# Contributions of KM3NeT to ICRC2023

**The KM3NeT Collaboration**

This document collects the contributions of the KM3NeT collaboration to the ICRC2023 conference, held from July 26 to August 3, 2023, in Nagoya, Japan. KM3NeT submitted 38 contributions to ICRC2023, on neutrino- and multimessenger astronomy, neutrino oscillation physics, cosmic ray physics, searches for dark matter and exotics, calibration, technical detector descriptions, and art. Proceedings are published in Proceedings of Science.

# Table of Contents

## Neutrino- and Multimessenger Astronomy

## Neutrino Oscillation Physics

## Cosmic Ray Physics

## Dark Matter and Exotics

## Calibration

## Other

## Full Authors List: The KM3NeT Collaboration

S. Aiello[a], A. Albert[b,bd], S. Alves Garre[c], Z. Aly[d], A. Ambrosone[f,e], F. Ameli[g], M. Andre[h], E. Androutsou[i], M. Anguita[j], L. Aphecetche[k], M. Ardid[l], S. Ardid[l], H. Atmani[m], J. Aublin[n], L. Bailly-Salins[o], Z. Bardačová[q,p], B. Baret[n], A. Bariego-Quintana[c], S. Basegmez du Pree[r], Y. Becherini[n], M. Bendahman[m,n], F. Benfenati[t,s], M. Benhassi[u,e], D.M. Benoit[v], E. Berbee[r], V. Bertin[d], S. Biagi[w], M. Boettcher[x], D. Bonanno[w], J. Boumaaza[m], M. Bouta[y], M. Bouwhuis[r], C. Bozza[z,e], R.M. Bozza[f,e], H.Brânzaş[aa], F. Bretaudeau[k], R. Bruijn[ab,r], J. Brunner[d], R. Bruno[a], E. Buis[ac,r], R. Buompane[u,e], J. Busto[d], B. Caiffi[ad], D. Calvo[c], S. Campion[g,ae], A. Capone[g,ae], F. Carenini[t,s], V. Carretero[c], T. Cartraud[n], P. Castaldi[af,s], V. Cecchini[c], S. Celli[g,ae], L. Cerisy[d], M. Chabab[ag], M. Chadolias[ah], A. Chen[ai], S. Cherubini[aj,w], T. Chiarusi[s], M. Circella[ak], R. Cocimano[w], J.A.B. Coelho[n], A. Coleiro[n], R. Coniglione[w] P. Coyle[d], A. Creusot[n], A. Cruz[al], G. Cuttone[w], R. Dallier[k], Y. Darras[ah], A. De Benedittis[e], B. De Martino[d], G. De Wasseige[bf], V. Decoene[k], R. Del Burgo[e], U.M. Di Cerbo[e], L.S. Di Mauro[w], I. Di Palma[g,ae], A.F. Díaz[j], C. Diaz[j], D. Diego-Tortosa[w], C. Distefano[w], A. Domi[ah], C. Donzaud[n], D. Dornic[d], M. Dörr[am], E. Drakopoulou[i], D. Drouhin[b,bd], R. Dvornický[q], T. Eberl[ah], E. Eckerová[q,p], A. Eddymaoui[m], T. van Eeden[r], M. Eff[n], D. van Eijk[r], I. El Bojaddaini[y], S. El Hedri[n], A. Enzenhöfer[d], J. de Favereau[bf], G. Ferrara[w], M. D. Filipović[an], F. Filippini[t,s], D. Franciotti[w], L.A. Fusco[z,e], J. Gabriel[ao], S. Gagliardini[g], T. Gal[ah], J. García Méndez[l], A. Garcia Soto[c], C. Gatius Oliver[r], N. Geißelbrecht[ah], H. Ghaddari[y], L. Gialanella[e,u], B.K. Gibson[v], E. Giorgio[w], I. Goos[n], D. Goupilliere[o], S.R. Gozzini[c], R. Gracia[ah], K. Graf[ah], C. Guidi[ap,ad], B. Guillon[o], M. Gutiérrez[aq], H. van Haren[ar], A. Heijboer[r], A. Hekalo[am], L. Hennig[ah], J.J. Hernández-Rey[c], F. Huang[d], W. Idrissi Ibnsalih[e], G. Illuminati[s], C.W. James[al], M. de Jong[as,r], P. de Jong[ab,r], B.J. Jung[r], P. Kalaczyński[at,be], O. Kalekin[ah], U.F. Katz[ah], N.R. Khan Chowdhury[c], A. Khatun[q], G. Kistauri[av,au], C. Kopper[ah], A. Kouchner[aw,n], V. Kulikovskiy[ad], R. Kvatadze[av], M. Labalme[o], R. Lahmann[ah], M. Lamoureux[bf], G. Larosa[w], C. Lastoria[d], A. Lazo[c], E. Le Guirriec[d], S. Le Stum[d], G. Lehaut[o], E. Leonora[a], N. Lessing[c], G. Levi[t,s], M. Lindsey Clark[n], F. Longhitano[a], J. Majumdar[r], L. Malerba[ad], F. Mamedov[p], J. Mańczak[c], A. Manfreda[e], M. Marconi[ap,ad], A. Margiotta[t,s], A. Marinelli[e,f], C. Markou[i], L. Martin[k], J.A. Martínez-Mora[l], F. Marzaioli[u,e], M. Mastrodicasa[ae,g], S. Mastroianni[e], J. Mauro[bf], S. Miccichè[w], G. Miele[f,e], P. Migliozzi[e], E. Migneco[w], S. Minutoli[ad], M.L. Mitsou[e], C.M. Mollo[e], L. Morales-Gallegos[u,e], C. Morley-Wong[al], A. Moussa[y], I. Mozun Mateo[ay,ax], J. M. Mulder[ab], R. Muller[r], M.R. Musone[e,u], M. Musumeci[w], L. Nauta[r], S. Navas[aq], A. Nayerhoda[ak], C.A. Nicolau[g], B. Nkosi[ai], B. Ó Fearraigh[ab,r], V. Oliviero[f,e], A. Orlando[w], E. Oukacha[n], D. Paesani[w], J. Palacios González[c], G. Papalashvili[au], V. Parisi[ap,ad], E.J. Pastor Gomez[c], A. M. Păun[aa], G.E. Păvălaş[aa], S. Peña Martínez[n], M. Perrin-Terrin[d], J. Perronnel[o], V. Pestel[ay], R. Pestes[n], P. Piattelli[w], C. Poirè[z,e], V. Popa[aa], T. Pradier[b], S. Pulvirenti[w], G. Quéméner[o], C. Quiroz[l], U. Rahaman[c], N. Randazzo[a], R. Randriatoamanana[k], S. Razzaque[az], I.C. Rea[e], D. Real[c], S. Reck[ah], G. Riccobene[w], J. Robinson[x], A. Romanov[ap,ad], A. Šaina[c], F. Salesa Greus[c], D.F.E. Samtleben[as,r], A. Sánchez Losa[c,ak], S. Sanfilippo[w], M. Sanguineti[ap,ad], C. Santonastaso[ba,e], D. Santonocito[w], P. Sapienza[w], J. Schnabel[ah], J. Schumann[ah], H. M. Schutte[x], J. Seneca[r], N. Sennan[y], B. Setter[ah], I. Sgura[ak], R. Shanidze[au], Y. Shitov[p], F. Šimkovic[q], A. Simonelli[e], A. Sinopoulou[a], M.V. Smirnov[ah], B. Spisso[e], M. Spurio[t,s], D. Stavropoulos[i], I. Štekl[p], M. Taiuti[ap,ad], Y. Tayalati[m], H. Tedjditi[ad], H. Thiersen[x], I. Tosta e Melo[a,aj], B. Trocmé[n], V. Tsourapis[i], E. Tzamariudaki[i], A. Vacheret[o], V. Valsecchi[w], V. Van Elewyck[aw,n], G. Vannoye[d], G. Vasileiadis[bb], F. Vazquez de Sola[r], C. Verilhac[n], A. Veutro[g,ae], S. Viola[w], D. Vivolo[u,e], J. Wilms[bc], E. de Wolf[ab,r], H. Yepes-Ramirez[l], G. Zarpapis[i], S. Zavatarelli[ad], A. Zegarelli[g,ae], D. Zito[w], J.D. Zornoza[c], J. Zúñiga[c], and N. Zywucka[x].

[a]INFN, Sezione di Catania, Via Santa Sofia 64, Catania, 95123 Italy
[b]Université de Strasbourg, CNRS, IPHC UMR 7178, F-67000 Strasbourg, France
[c]IFIC - Instituto de Física Corpuscular (CSIC - Universitat de València), c/Catedrático José Beltrán, 2, 46980 Paterna, Valencia, Spain
[d]Aix Marseille Univ, CNRS/IN2P3, CPPM, Marseille, France
[e]INFN, Sezione di Napoli, Complesso Universitario di Monte S. Angelo, Via Cintia ed. G, Napoli, 80126 Italy
[f]Università di Napoli "Federico II", Dip. Scienze Fisiche "E. Pancini", Complesso Universitario di Monte S. Angelo, Via Cintia ed. G, Napoli, 80126 Italy
[g]INFN, Sezione di Roma, Piazzale Aldo Moro 2, Roma, 00185 Italy
[h]Universitat Politècnica de Catalunya, Laboratori d'Aplicacions Bioacústiques, Centre Tecnològic de Vilanova i la Geltrú, Avda. Rambla Exposició, s/n, Vilanova i la Geltrú, 08800 Spain
[i]NCSR Demokritos, Institute of Nuclear and Particle Physics, Ag. Paraskevi Attikis, Athens, 15310 Greece
[j]University of Granada, Dept. of Computer Architecture and Technology/CITIC, 18071 Granada, Spain
[k]Subatech, IMT Atlantique, IN2P3-CNRS, Université de Nantes, 4 rue Alfred Kastler - La Chantrerie, Nantes, BP 20722 44307 France
[l]Universitat Politècnica de València, Instituto de Investigación para la Gestión Integrada de las Zonas Costeras, C/ Paranimf, 1, Gandia, 46730 Spain
[m]University Mohammed V in Rabat, Faculty of Sciences, 4 av. Ibn Battouta, B.P. 1014, R.P. 10000 Rabat, Morocco
[n]Université Paris Cité, CNRS, Astroparticule et Cosmologie, F-75013 Paris, France
[o]LPC CAEN, Normandie Univ, ENSICAEN, UNICAEN, CNRS/IN2P3, 6 boulevard Maréchal Juin, Caen, 14050 France
[p]Czech Technical University in Prague, Institute of Experimental and Applied Physics, Husova 240/5, Prague, 110 00 Czech Republic
[q]Comenius University in Bratislava, Department of Nuclear Physics and Biophysics, Mlynska dolina F1, Bratislava, 842 48 Slovak Republic
[r]Nikhef, National Institute for Subatomic Physics, PO Box 41882, Amsterdam, 1009 DB Netherlands
[s]INFN, Sezione di Bologna, v.le C. Berti-Pichat, 6/2, Bologna, 40127 Italy
[t]Università di Bologna, Dipartimento di Fisica e Astronomia, v.le C. Berti-Pichat, 6/2, Bologna, 40127 Italy
[u]Università degli Studi della Campania "Luigi Vanvitelli", Dipartimento di Matematica e Fisica, viale Lincoln 5, Caserta, 81100 Italy

[v]E. A. Milne Centre for Astrophysics, University of Hull, Hull, HU6 7RX, United Kingdom
[w]INFN, Laboratori Nazionali del Sud, Via S. Sofia 62, Catania, 95123 Italy
[x]North-West University, Centre for Space Research, Private Bag X6001, Potchefstroom, 2520 South Africa
[y]University Mohammed I, Faculty of Sciences, BV Mohammed VI, B.P. 717, R.P. 60000 Oujda, Morocco
[z]Università di Salerno e INFN Gruppo Collegato di Salerno, Dipartimento di Fisica, Via Giovanni Paolo II 132, Fisciano, 84084 Italy
[aa]ISS, Atomistilor 409, Măgurele, RO-077125 Romania
[ab]University of Amsterdam, Institute of Physics/IHEF, PO Box 94216, Amsterdam, 1090 GE Netherlands
[ac]TNO, Technical Sciences, PO Box 155, Delft, 2600 AD Netherlands
[ad]INFN, Sezione di Genova, Via Dodecaneso 33, Genova, 16146 Italy
[ae]Università La Sapienza, Dipartimento di Fisica, Piazzale Aldo Moro 2, Roma, 00185 Italy
[af]Università di Bologna, Dipartimento di Ingegneria dell'Energia Elettrica e dell'Informazione "Guglielmo Marconi", Via dell'Università 50, Cesena, 47521 Italia
[ag]Cadi Ayyad University, Physics Department, Faculty of Science Semlalia, Av. My Abdellah, P.O.B. 2390, Marrakech, 40000 Morocco
[ah]Friedrich-Alexander-Universität Erlangen-Nürnberg (FAU), Erlangen Centre for Astroparticle Physics, Nikolaus-Fiebiger-Straße 2, 91058 Erlangen, Germany
[ai]University of the Witwatersrand, School of Physics, Private Bag 3, Johannesburg, Wits 2050 South Africa
[aj]Università di Catania, Dipartimento di Fisica e Astronomia "Ettore Majorana", Via Santa Sofia 64, Catania, 95123 Italy
[ak]INFN, Sezione di Bari, via Orabona, 4, Bari, 70125 Italy
[al]International Centre for Radio Astronomy Research, Curtin University, Bentley, WA 6102, Australia
[am]University Würzburg, Emil-Fischer-Straße 31, Würzburg, 97074 Germany
[an]Western Sydney University, School of Computing, Engineering and Mathematics, Locked Bag 1797, Penrith, NSW 2751 Australia
[ao]IN2P3, LPC, Campus des Cézeaux 24, avenue des Landais BP 80026, Aubière Cedex, 63171 France
[ap]Università di Genova, Via Dodecaneso 33, Genova, 16146 Italy
[aq]University of Granada, Dpto. de Física Teórica y del Cosmos & C.A.F.P.E., 18071 Granada, Spain
[ar]NIOZ (Royal Netherlands Institute for Sea Research), PO Box 59, Den Burg, Texel, 1790 AB, the Netherlands
[as]Leiden University, Leiden Institute of Physics, PO Box 9504, Leiden, 2300 RA Netherlands
[at]National Centre for Nuclear Research, 02-093 Warsaw, Poland
[au]Tbilisi State University, Department of Physics, 3, Chavchavadze Ave., Tbilisi, 0179 Georgia
[av]The University of Georgia, Institute of Physics, Kostava str. 77, Tbilisi, 0171 Georgia
[aw]Institut Universitaire de France, 1 rue Descartes, Paris, 75005 France
[ax]IN2P3, 3, Rue Michel-Ange, Paris 16, 75794 France
[ay]LPC, Campus des Cézeaux 24, avenue des Landais BP 80026, Aubière Cedex, 63171 France
[az]University of Johannesburg, Department Physics, PO Box 524, Auckland Park, 2006 South Africa
[ba]Università degli Studi della Campania "Luigi Vanvitelli", CAPACITY, Laboratorio CIRCE - Dip. Di Matematica e Fisica - Viale Carlo III di Borbone 153, San Nicola La Strada, 81020 Italy
[bb]Laboratoire Univers et Particules de Montpellier, Place Eugène Bataillon - CC 72, Montpellier Cédex 05, 34095 France
[bc]Friedrich-Alexander-Universität Erlangen-Nürnberg (FAU), Remeis Sternwarte, Sternwartstraße 7, 96049 Bamberg, Germany
[bd]Université de Haute Alsace, rue des Frères Lumière, 68093 Mulhouse Cedex, France
[be]AstroCeNT, Nicolaus Copernicus Astronomical Center, Polish Academy of Sciences, Rektorska 4, Warsaw, 00-614 Poland
[bf]UCLouvain, Centre for Cosmology, Particle Physics and Phenomenology, Chemin du Cyclotron, 2, Louvain-la-Neuve, 1349 Belgium

## Acknowledgements

The authors acknowledge the financial support of the funding agencies: Agence Nationale de la Recherche (contract ANR-15-CE31-0020), Centre National de la Recherche Scientifique (CNRS), Commission Européenne (FEDER fund and Marie Curie Program), LabEx UnivEarthS (ANR-10-LABX-0023 and ANR-18-IDEX-0001), Paris Île-de-France Region, France; Shota Rustaveli National Science Foundation of Georgia (SRNSFG, FR-22-13708), Georgia; The General Secretariat of Research and Innovation (GSRI), Greece Istituto Nazionale di Fisica Nucleare (INFN), Ministero dell'Università e della Ricerca (MIUR), PRIN 2017 program (Grant NAT-NET 2017W4HA7S) Italy; Ministry of Higher Education, Scientific Research and Innovation, Morocco, and the Arab Fund for Economic and Social Development, Kuwait; Nederlandse organisatie voor Wetenschappelijk Onderzoek (NWO), the Netherlands; The National Science Centre, Poland (2021/41/N/ST2/01177); The grant "AstroCeNT: Particle Astrophysics Science and Technology Centre", carried out within the International Research Agendas programme of the Foundation for Polish Science financed by the European Union under the European Regional Development Fund; National Authority for Scientific Research (ANCS), Romania; Grants PID2021-124591NB-C41, -C42, -C43 funded by MCIN/AEI/ 10.13039/501100011033 and, as appropriate, by "ERDF A way of making Europe", by the "European Union" or by the "European Union NextGenerationEU/PRTR", Programa de Planes Complementarios I+D+I (refs. ASFAE/2022/023, ASFAE/2022/014), Programa Prometeo (PROMETEO/2020/019) and GenT (refs. CIDEGENT/2018/034, /2019/043, /2020/049. /2021/23) of the Generalitat Valenciana, Junta de Andalucía (ref. SOMM17/6104/UGR, P18-FR-5057), EU: MSC program (ref. 101025085), Programa María Zambrano (Spanish Ministry of Universities, funded by the European Union, NextGenerationEU), Spain; The European Union's Horizon 2020 Research and Innovation Programme (ChETEC-INFRA - Project no. 101008324); Fonds de la Recherche Scientifique - FNRS, Belgium.

# Search for cosmic neutrino point sources and extended sources with 6-21 lines of KM3NeT/ARCA

**Rasa Muller,**[a,*] **Thijs van Eeden**[a] **and Aart Heijboer**[a] **on behalf of the KM3NeT Collaboration**

[a] *Nikhef, Science Park 105, Amsterdam, Netherlands*

*E-mail:* rmuller@km3net.de, aart.heijboer@nikhef.nl, tjuanve@nikhef.nl

The identification of cosmic objects emitting high energy neutrinos provides new insights about the Universe and its active sources. The existence of cosmic neutrinos has been proven by the IceCube Neutrino Observatory, but the big question of which sources these neutrinos originate from remains largely unanswered. The KM3NeT detector for Astroparticle Research with Cosmics in the Abyss (ARCA), is currently being built in the Mediterranean Sea. It will have an instrumented volume of a cubic kilometre, and excel at identifying cosmic neutrino sources due to its unprecedented angular resolution (< 0.1 degree for muon neutrinos with E > 100 TeV). KM3NeT has a view of the sky complementary to IceCube, and is sensitive to neutrinos across a wide range of energies. Currently KM3NeT/ARCA is taking data with 21 detector lines. This contribution will present the results of point source and extended source searches with KM3NeT/ARCA with data from 2021 and 2022 taken with an evolving detector geometry.

*Speaker

## 1. Introduction

IceCube has identified TXS 0506+056 as a flaring neutrino source [1–3] and has recently identified NGC 1068 as a steady point source of high energy neutrinos ($4.2\sigma$) [4] (November 4th, 2022). In their respective energy ranges the sources NGC 1068 and TXS 0506+056 contribute no more than ~ 1% to the overall diffuse flux of astrophysical neutrinos. This indicates that there is still a lot of work to be done in identifying the sources that produce high energy neutrinos. This is where KM3NeT/ARCA will play an important role.

The KM3NeT/ARCA detector is a neutrino telescope currently under construction at the bottom of the Mediterranean Sea. It consists of a three-dimensional grid of optical modules that detect the Cherenkov light from neutrino interaction products. The full detector will have 230 detection lines of 18 optical modules and will instrument ~ 1 $\text{km}^3$ volume of sea water.

For 101 astrophysical objects it is tested whether they are high energy neutrino emitters. In order to identify a cosmic neutrino signal on top of the atmospheric background of muons and neutrinos, statistical methods are developed based on Monte Carlo pseudo experiments. With the KM3NeT/ARCA data taken with 6, 8, 19, and 21 detection units, the expected sensitivity of KM3NeT/ARCA to neutrino point and extended sources in our universe is calculated in a binned likelihood analyses. The methods and results are presented in this contribution.

## 2. KM3NeT/ARCA6-21 data and performance

### 2.1 Data sample

This analysis uses data from a period when KM3NeT/ARCA was running with 6 - 21 detector strings between May 2021, and December 2022. The effective data taking time of ~ 424 days contains: ~ 92 days with 6 lines (referred to as ARCA6), ~ 210 days with 7 - 8 lines (referred to as ARCA8), ~ 52 days with ARCA19 and ~ 69 days with ARCA21. The different periods have been studied individually, but unless stated differently, the plots shown in this contribution include the full period.

### 2.2 Background rejection, event selection

The event selection criteria are applied to reduce the atmospheric muon contamination and discard badly reconstructed events. Events are selected using cuts on the number of hits used in the reconstruction, the reconstructed direction, and the fit quality, which is based on the likelihood of the reconstruction. For the ARCA19-21 period an additional boosted decision tree model is trained to further discriminate signal from background.

The selection efficiencies for $\nu_\mu$ CC track events reconstructed within 1° from the true neutrino direction is 81% for the ARCA6-8 period and 95% for the ARCA19-21 period. A cosmic flux of $\phi^{\cos}_{\nu_i+\bar{\nu}_i} = 1.2 \cdot 10^{-8} \left(\frac{E_\nu}{\text{GeV}}\right)^{-2}$ $\text{GeV}^{-1}\ \text{cm}^{-2}\ \text{s}^{-1}\ \text{sr}^{-1}$ yields 16.5 events in the ARCA6-21 data sample.

In Figure 1, the reconstructed energy distribution of the event samples are shown for the ARCA6-8 and ARCA19-21 periods. There is a 13% overall underestimation of the data by the simulation, but this effect does not affect the analysis since the background estimation comes from scrambled data. As it can be seen in Figure 1 (left), the ARCA6-8 is still dominated by badly reconstructed atmospheric muons, while the ARCA19-21 sample has a higher neutrino purity (Figure 1 - right). It can be explained by the different size of the 2 detectors.

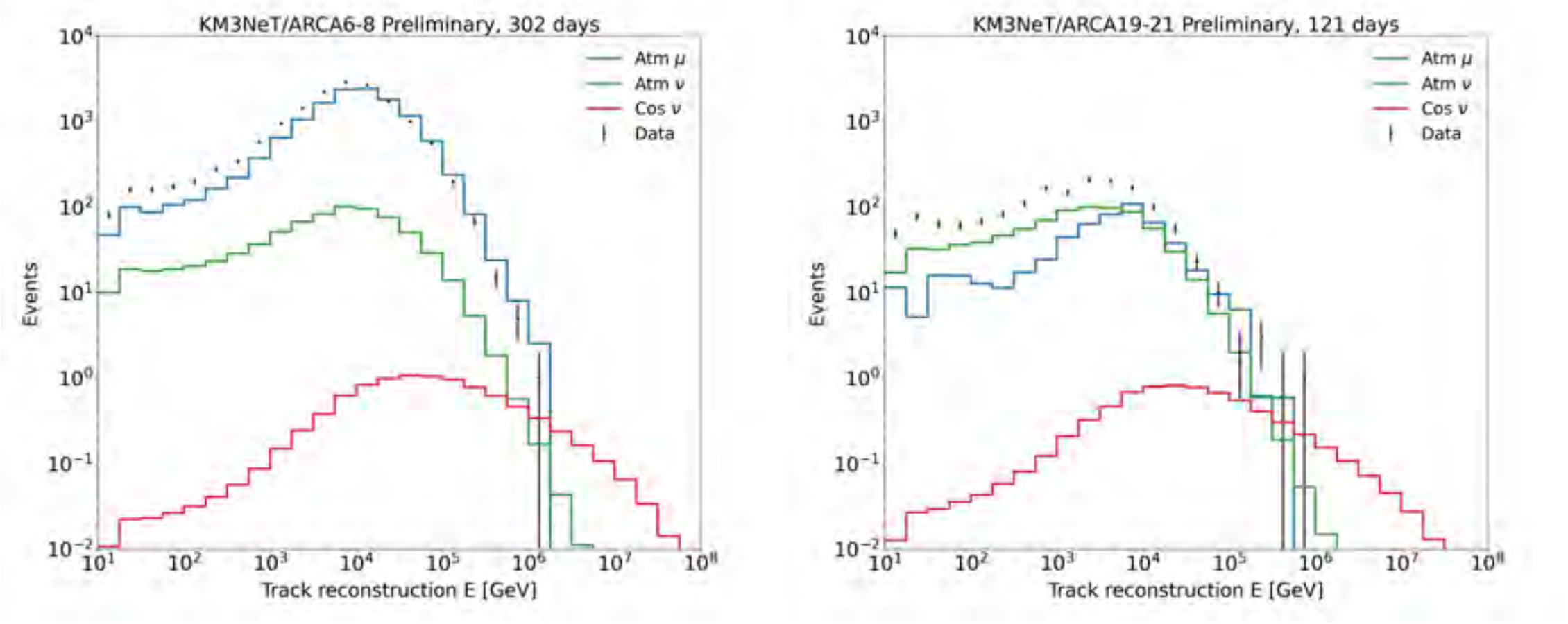

**Figure 1:** Reconstructed energy distributions after applying the event selections for the ARCA6-8 (left) and ARCA19-21 (right) period.

### 2.3 Background expectation from scrambled data

The background rate, as a function of reconstructed energy and declination ($\delta$), is data-driven. For the $\sin(\delta)$ a spline $F(\delta)$ is fitted, and the energy dependence $F(\log_{10}(E))$ is parameterised by a fit with two Gaussians. For each individual candidate source that is studied, the expected density of background events in $\mathrm{sr}^{-1}$ $\log_{10}(\mathrm{GeV})^{-1}$ is obtained by:

$$N_{\mathrm{bkg}} = n \times F(\delta) \times F(\log_{10}(E)) \tag{1}$$

where the normalisation $n$ is chosen such that the integral over the full solid angle of the sky and over $E$ gives exactly the total number of events in the data.

### 2.4 Detector response from Monte Carlo simulations

The response of the detector to a possible signal is modeled via detector response functions (acceptance, and resolutions), which are based on simulations [5–7]. Figure 2 (left) shows the effective area of ARCA6-21 compared with ANTARES for similar - yet not exactly the same - event selection, and the angular deviation (right) for selected track events from $\nu_\mu$ CC interactions for the ARCA6-8 and ARCA19-21 periods. The ARCA6-8 period reaches an angular deviation of $\sim 1°$ at 100 PeV while the ARCA19-21 period already reaches 0.2°. For the full detector this is expected to improve towards 0.06° [8].

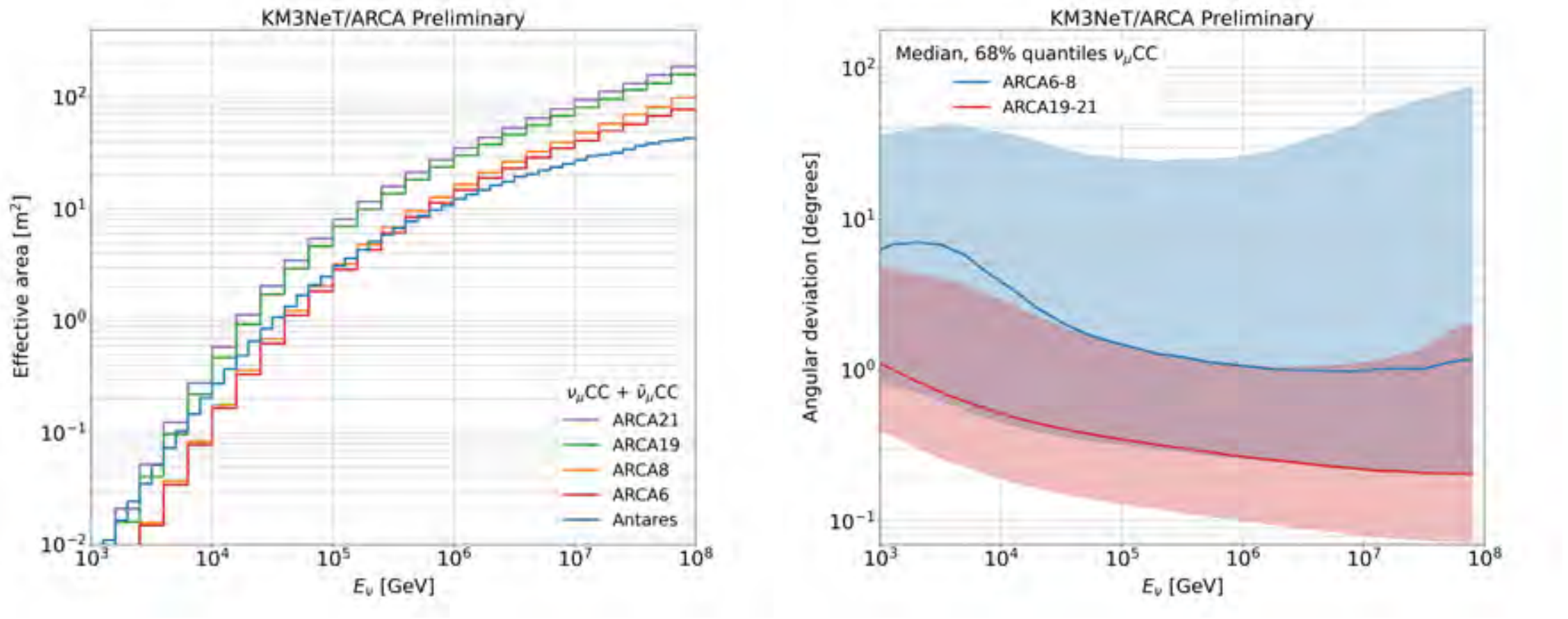

**Figure 2:** Effective area at selection level (left) for the different ARCA detectors for a flux of $\nu_\mu + \bar{\nu}_\mu$ that interact in the CC interaction. The effective areas are compared with the ANTARES effective area for upgoing events. The angular deviation (right) for the ARCA6-8 and ARCA19-21 periods with their corresponding 68% quantiles.

## 3. Method

### 3.1 Candidate sources

The 101 astrophysical objects[1], are selected based on GeV – PeV information from other neutrino experiments, cosmic ray experiments as well as electromagnetic measurements. Besides adding interesting sources from previous point source studies and real time alerts by IceCube and ANTARES, historically interesting sources were added as well as high-energy $\gamma$-ray source by the LHAASOO collaboration. Furthermore the $\gamma$-ray TeVCat is consulted to select interesting Galactic sources with a hint for a hadronic component, and active galactic nuclei were selected based on their maximal flux observed in radio. For the 10% of sources that are spatially extended in the sky, the detector point spread function is modified with a Gaussian with the spread ($\sigma$) equal to the corresponding extension ranging from 0.11 to 1 degrees.

### 3.2 Analysis method

A binned formalism is used where the compatibility of the data with a point source hypothesis is tested by means of 2D histograms of distance to the candidate source $\psi$ in the range $[0 - 5]$ in degrees, v.s. $\log_{10}(E_{\text{rec}})$ in the range $[1 - 8]$, in $\log_{10}(\text{GeV})$. For each bin $i$, there is an estimate of the number of signal events, $\mathcal{S}_i$, expected for a reference flux $\phi_{\text{ref}}$ and the number of background events, $\mathcal{B}_i$.

[1]The 101 analysed candidates are:
LMC N132D, HESS J1356-645 , SNR G318.2+00.1, IC-hotspot South hemisphere, HESS J1614-518, PKS 2005-489, HESS J1640-465, RX J0852.0-4622, HESS J1641-463, VelaX , PKS 0537-441, CentaurusA, PKS 1424-418, J0106-4034, RX J1713.7-3946, CTB 37A, PKS 1454-354, HESS J1741-302, J1924-2914, Galactic center, J2258-2758, J1625-2527, NGC 253, J0457-2324, J1833-210A, J0836-2016, J1911-2006, J0609-1542, SNR G015.4+00.1, J2158-1501, LHAASO J1825-1326 , QSO 1730-130, J1337-1257, J2246-1206, PKS 0727-11, TXS 1749-101, HESS J1828-099, J1512-0905, J0607-0834, QSO 2022-077, RS Ophiuchi, J0006-0623, 3C279, LHAASO J1839-0545 , J2225-0457, 4FGL J0307.8-0419, PKS 1741-038, LHAASO J1843-0338 , J0339-0146, J0423-0120, J0725-0054, LHAASO J1849-0003 , NGC 1068, J2136+0041, J1058+0133, J0108+0135, PKS 0215+015, J1229+0203, TXS 0310+022, 3C403, CGCG 420-01, J0433+0521, TXS 0506+056, HESS J0632+057, LHAASO J1908+0621 , PKS 2145+067, W 49B, OT 081, PKS 1502+106, J0242+1101, J2232+1143, J0121+1149, J1230+1223, J0750+1231, PKS 1413+135, J0530+1331, W 51, J2253+1608, PKS 0735+178, LHAASO J1929+1745 , J0854+2006, RGB J2243+203, LHAASO J0534+2202 , IC 443, PKS 1424+240, MG3 J225517+2409, 2HWC J1949+244, LHAASO J1956+2845 , J0237+2848, J1310+3220, J1613+3412, LHAASO J2018+3651 , J2015+3710, MGRO J2019+37, Mkn 421, J0927+3902, NGC 4151, Mkn 501, J1642+3948, J0555+3948, LHAASO J2032+41025.

The analysis is done for spectral index $\gamma = 2$ and 2.5. Furthermore a spectrum in line with the most recent NGC 1068 IceCube observation [4] ($\gamma = 3.2$) is tested for this particular source, and this particular source only.

#### 3.2.1 Likelihood formalism

For every (pseudo) dataset it is determined how compatible it is with the expected signal + background model (H1), and with the background-only model (H0). This is expressed in the log-likelihood ratio ($\lambda$) which serves as a test statistic:

$$\lambda = \log L(\mu = \hat{\mu}) - \log L(\mu = 0) \tag{2}$$

$$\log L = \sum_{i \in \text{bins}} N_i \log(\mathcal{B}_i + \mu \mathcal{S}_i) - \mathcal{B}_i - \mu \mathcal{S}_i. \tag{3}$$

where $\mu$ represents the signal strength and $\hat{\mu}$ the best fitted signal strength for a given (pseudo) dataset.

### 3.3 Systematic uncertainties

Systematic uncertainties are taken into account for two main parameters describing the detector performance: the angular resolution (0.5° Gaussian smearing) and the acceptance (30% Gaussian spread). The systematic uncertainty on the estimated angular deviation comes from the uncertainty on the absolute orientation of the detector around its z-axis and uncertainty on the tilts of the lines. The uncertainty on the acceptance takes into account the uncertainties on the water and PMT properties. They have been studied in detail in [9, 10].

### 3.4 Discovery potential and sensitivity

The sensitivity and discovery potential is obtained for a range of declinations with 50.000 pseudo-experiments for the H0 distribution, and 5000 pseudo-experiments for each H1 distribution. Table 1 summarises the results for the $E^{-2}$ and $E^{-2.5}$ spectra. For each of the candidate sources, the discovery potential and sensitivity are calculated with 30.000 pseudo-experiments for the H0 distribution, and 3000 pseudo-experiments for each H1 distribution. Depending on the declination, the sensitivity is at a flux level that would produce 2.2 – 3.2 (for an $E^{-2}$ spectrum), 2.4 – 3.6 (for an $E^{-2.5}$ spectrum) signal events per source in the full data taking period. A discovery would require 1.5 – 4.0 (for an $E^{-2}$ spectrum), 2.2 – 5.6 (for an $E^{-2.5}$ spectrum) signal events per source in the full data taking period.

## 4. Results

After unblinding the data, there were on average 31 events observed inside the 5 degree cone around each candidate source. For each dataset in the 5° cone, the p-value (significance) is computed to determine whether or not the candidate source is significantly detected.

All candidate sources are consistent with a background-only hypothesis, i.e. no candidate source is significantly observed. The most signal-like sources are:

$E^{-2.0}$

| $\delta$ (deg) | $N_{\text{sig}}$ ref.flux | $N_{\text{bg}}$ | med. $\lambda_{H0}$ | sens. ($N_{\text{sig}}$) | disc. ($N_{\text{sig}}$) |
|---|---|---|---|---|---|
| -90 | 1.25 | 51.05 | -0.91 | 3.13 | 3.96 |
| -70 | 1.30 | 51.12 | -0.95 | 3.05 | 3.82 |
| -50 | 1.43 | 51.65 | -1.07 | 2.90 | 3.30 |
| -30 | 0.88 | 29.85 | -0.65 | 2.68 | 3.17 |
| -10 | 0.78 | 24.44 | -0.58 | 2.70 | 2.88 |
| 10 | 0.72 | 28.00 | -0.54 | 2.60 | 2.62 |
| 30 | 0.66 | 28.35 | -0.51 | 2.47 | 2.26 |
| 50 | 0.51 | 21.98 | -0.40 | 2.32 | 1.83 |
| 55 | 0.33 | 13.75 | -0.26 | 2.24 | 1.59 |

$E^{-2.5}$

| $\delta$ (deg) | $N_{\text{sig}}$ ref.flux | $N_{\text{bg}}$ | med. $\lambda_{H0}$ | sens. ($N_{\text{sig}}$) | disc. ($N_{\text{sig}}$) |
|---|---|---|---|---|---|
| -90 | 13.18 | 51.05 | -8.89 | 3.58 | 5.38 |
| -70 | 13.17 | 51.12 | -8.93 | 3.59 | 5.53 |
| -50 | 12.91 | 51.65 | -8.90 | 3.57 | 5.16 |
| -30 | 9.02 | 29.85 | -6.13 | 3.46 | 4.82 |
| -10 | 7.65 | 24.44 | -5.23 | 3.19 | 4.38 |
| 10 | 6.17 | 28.00 | -4.28 | 3.21 | 3.87 |
| 30 | 4.65 | 28.35 | -3.37 | 2.78 | 3.35 |
| 50 | 2.74 | 21.98 | -2.05 | 2.62 | 2.66 |
| 55 | 1.74 | 13.75 | -1.31 | 2.47 | 2.23 |

**Table 1:** Summary of the discovery potential and sensitivity for the reference fluxes: $\phi^{\cos}_{\nu_i+\bar{\nu}_i} = 1.2 = 1 \cdot 10^{-4} \left(\frac{E}{\text{GeV}}\right)^{-2}$ (top), $1.8 \cdot 10^{-1} \cdot \left(\frac{E}{\text{GeV}}\right)^{-2.5}$ (bottom), in GeV$^{-1}$ cm$^{-2}$ s$^{-1}$. Listed are the total number of signal events for the concerned reference flux and the number of background events inside the 5 degree cone represented by the analysed 2D histograms, and the medians of the H0 test statistic ($\lambda$) distribution with the resulting number of events needed for a 90% confidence level exclusion or a $3\sigma$ discovery.

- For the $E^{-2.0}$ spectrum: The AGN J1512-0905 at right ascension 228.21°, declination −9.10° with a pre-trial p-value of 0.011. The post-trial p-value for the $E^{-2}$ analysis was 0.66.
- For the $E^{-2.5}$ spectrum: The BL Lac Mkn 421 at right ascension 166.11°, declination 38.21° with a pre-trial p-value of 0.020. The post-trial p-value for the $E^{-2.5}$ analysis was 0.56.
- For the $E^{-3.2}$ spectrum: NGC 1068 was the only source candidate studied. This active galactic nuclei is located at right ascension 40.7°, declination −0.01° ($\sin\delta = 0.00$) with a pre-trial p-value of 0.98. Since there was only one trial, the post-trial p-value for the $E^{-3.2}$ analysis is automatically also: 0.98.

Because no significant detection is made, upper limits are set on the flux. If $\lambda_{\text{obs}}$ is below the median expected $\lambda$ for the background-only test statistic distribution, $\lambda(\mu_{\text{true}} = 0)$ is used for the upper limit calculation. This means the limit will match with the sensitivity for such cases. The computed upper limits for each source are shown in Figure 3 together with the overall sensitivity.

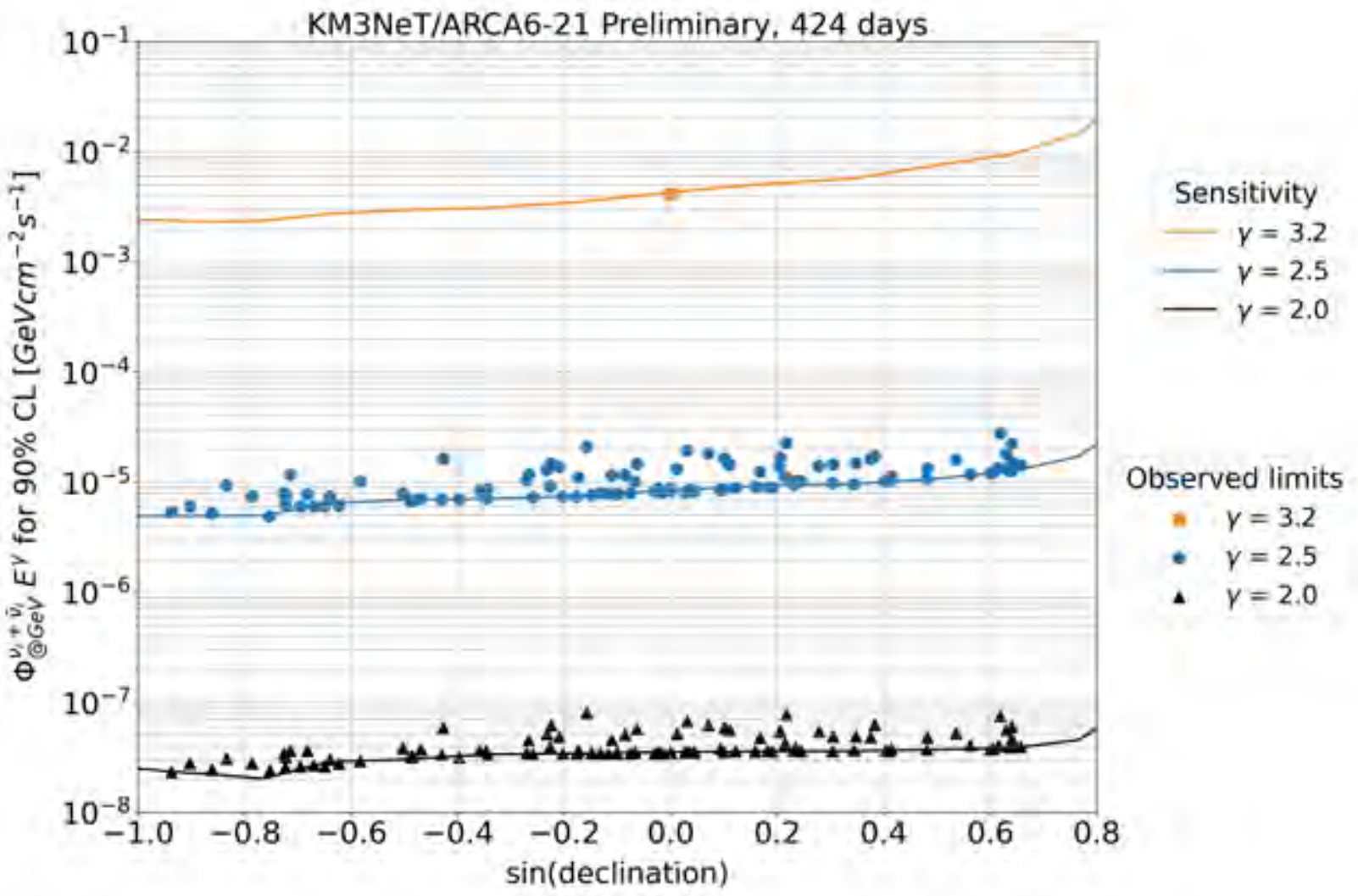

**Figure 3:** Sensitivity and observed limits for ARCA6-21 using 424 days of data. The $\gamma = 2.0$ and 2.5 analyses show the observed limits of 101 sources while the 3.2 analysis only looked at NGC 1068.

In Figure 4 the final ARCA6-21 $E^{-2}$ point source results are compared with the IceCube and ANTARES experiments [11–13], as well as the previous ARCA point source analyses, and with the expected sensitivity for the full ARCA detector comprising 2 building blocks [8]. The improvement in sensitivity is expected to accelerate due to the planned deployment of 9 detection lines before the end of 2023 and another 9 months of unprocessed ARCA21 data.

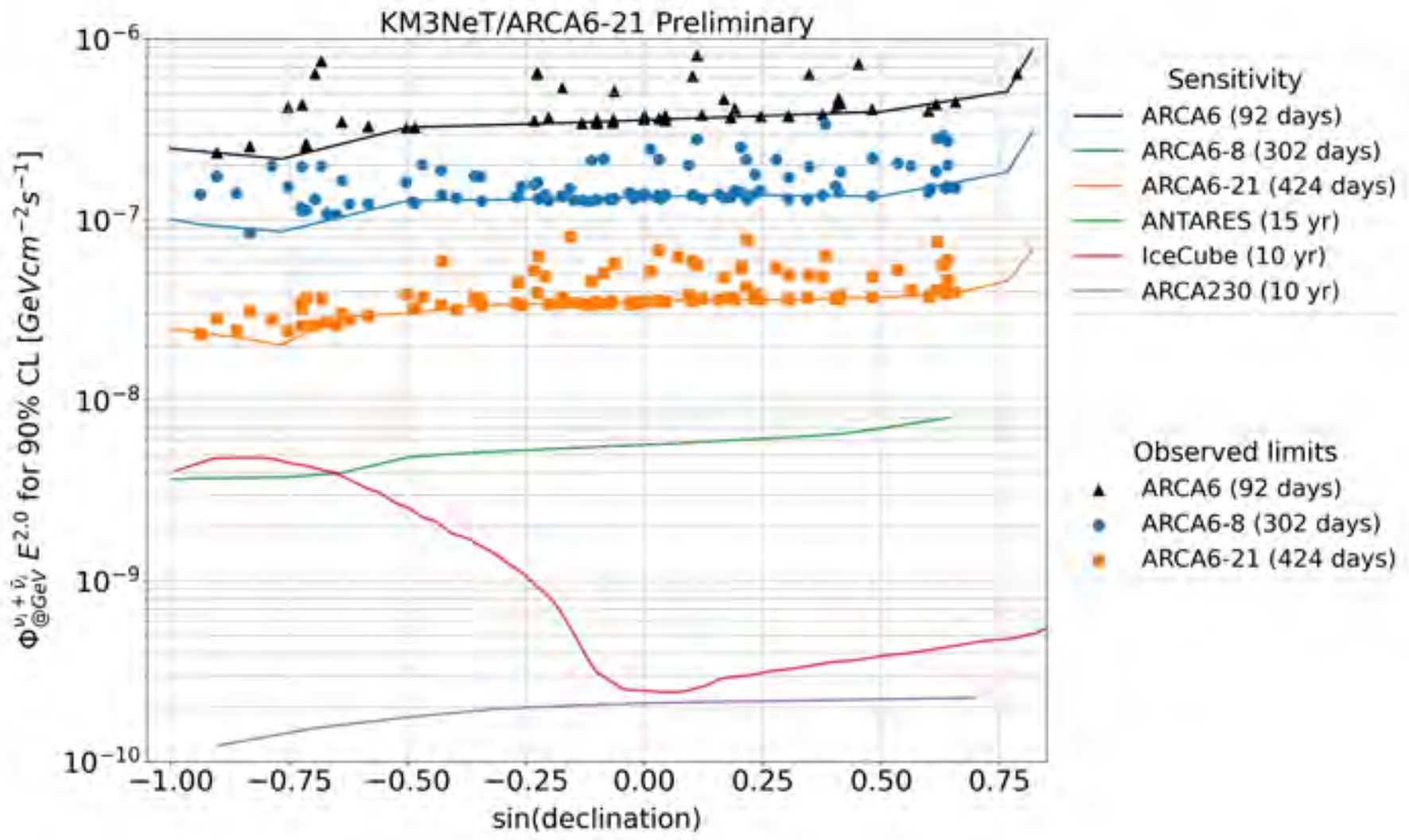

**Figure 4:** Comparison of the observed limits on the flux for the ARCA6-21 point source analysis assuming an $E^{-2}$ source spectrum as a function of $\sin(\delta)$, with earlier presented results of ANTARES 15 years [11] and IceCube [12, 13], as well as the ARCA230 10 years sensitivity [8].

# Event reconstruction and neutrino selection with KM3NeT/ARCA230

**Thijs Juan van Eeden**[a,*] **and Jordan Seneca**[a] **for the KM3NeT collaboration**

[a] *Nikhef,*
*Science Park 105, Amsterdam, Netherlands*

*E-mail:* tjuanve@nikhef.nl

The discovery of high-energy cosmic neutrinos has raised great interest in the field of neutrino astronomy. The KM3NeT/ARCA detector is currently under construction at the bottom of the Mediterranean Sea and is measuring high-energy neutrinos. The complete detector will instrument a cubic kilometre of seawater with photomultiplier tubes that detect the Cherenkov radiation from the products of neutrino interactions. The design is focused on discovering high-energy neutrino sources which requires an excellent angular resolution. This contribution covers the reconstruction algorithms deployed in the full KM3NeT/ARCA detector and their performances. The angular resolution reaches below 0.1° for tracks and below 1° for cascades and double cascades that originate from tau neutrinos. Additionally, the neutrino selections for the full KM3NeT/ARCA diffuse and point source studies are covered.

*Speaker

## 1. Introduction

The detection of high energy cosmic neutrinos by IceCube accelerated the field of neutrino astronomy [1]. IceCube detected a high-energy neutrino event coincident with a gamma-ray flare from TXS 0506+056 [2] and an excess of events from nearby active galaxy NGC 1068 [3] but there are many sources yet to be discovered.

The KM3NeT/ARCA detector is currently under construction offshore from Porto Palo di Capo Passero, Sicily, Italy [4]. Its primary purpose is to detect neutrinos with energies ranging from TeV and beyond. The field of the detector view covers the Southern sky in order to study potential Galactic sources. The detector consists of a three-dimensional grid of optical modules at the bottom of the Mediterrean sea [5]. Each optical module houses 31 photomultiplier tubes (PMTs) and data acquisition hardware. Each vertical detection string contains 18 modules that are mounted to the seabed. The final detector configuration will have 230 strings covering an instrumented volume of approximately 1 $km^3$.

The discovery and characterisation of new neutrino sources is enabled by reconstruction and selection of neutrino and background event. This contribution covers the full detector (KM3NeT/ARCA230) reconstruction algorithms and performance for track, shower and double shower topologies. This is followed by the track and shower event selections that are used for the current point source and diffuse flux analyses.

## 2. Simulation

Neutrino events are simulated using the gSeaGen simulation framework [6]. All neutrino flavours and interactions are simulated with energies from $10^2$ to $10^8$ GeV and weighted using different flux models. The atmospheric flux consists of a conventional and prompt component as described in [7].

The atmospheric muon events are generated using parametric formulas implemented in the MUPAGE software package [8]. Two samples of atmospheric muon are generated. One with a bundle threshold of 10 TeV and one of 50 TeV to increase statistics at high energies.

The generated events are processed by software packages for the light simulation and detector response. The track and shower reconstruction are applied to all events.

## 3. Event reconstruction

The event reconstruction algorithms covered in this contribution are based on a maximum likelihood search method.

### 3.1 Track reconstruction

The track reconstruction fits the energy and direction of high-energetic muons [9]. A coordinate transformation is applied to align a muon direction hypothesis with the z-axis. The muon trajectory can then be described by 5 parameters

- $\rho_i$: distance of closest approach muon track with PMT,

- $\theta_i, \phi_i$: PMT orientation angles,
- $\Delta t$: difference between and measured and expected hit time according to the Cherenkov hypothesis,
- $E$: energy of the muon.

The non-linearity of the problem requires a prefit where scattering and dispersion of light is neglected. This results in a linear fit that is performed over all directions with a 1° grid angle. The best 12 fits are stored and passed to the final fit. This fit maximises the likelihood

$$\text{Likelihood} = \prod_{\text{hit PMTs}} \frac{\partial P}{\partial t}(\rho_i, \theta_i, \phi_i, \Delta t, E) \tag{1}$$

where $\frac{\partial P}{\partial t}$ is a semi-analytical probability density function (PDF) which gives the expected scattered and unscattered photons from Chrerenkov radiation and energy losses of a muon. This PDF includes detector effects like the quantum efficiency and transit time spread of PMTs and background rates from K40 decays in seawater. Figure 1 shows the angular resolution of the track reconstruction for the track selection (described in section 4) $\nu_\mu$CC events using the KM3NeT/ARCA230 detector. The energy resolution is defined as half the difference between the 16th and 84th percentiles of the energy bias distributions (Energy bias $= 100\% \cdot \left(\frac{E_{rec}}{E_{visible}} - 1\right)$) and is shown in Figure 2(b).

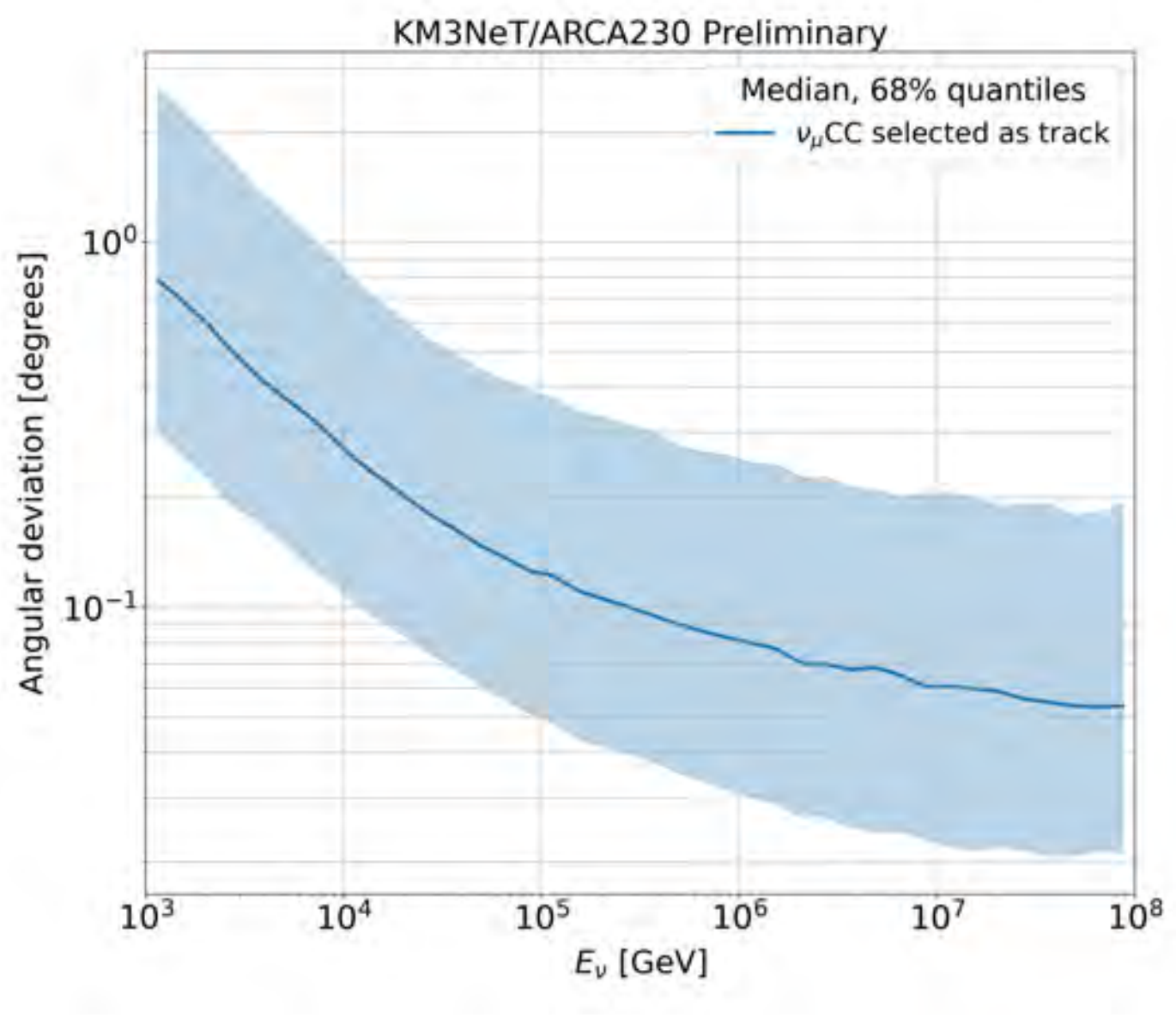

**Figure 1:** KM3NeT/ARCA230 angular resolution for $\nu_\mu$CC of the track selection described in section 4.

### 3.2 Shower reconstruction

The KM3NeT/ARCA230 shower reconstruction also follows a two-step procedure [9]. The first step selects coincident hits on the same optical module within 20 ns. Subsequently the vertex position and time is found by minimising

$$M = \sum_{i \in \text{hits}} \sqrt{1 + (t_i - \hat{t}_i)} \tag{2}$$

where $t_i$ is the measured hit time and $\hat{t}_i$ the expected hit time assuming a spherical light pattern emitted from the vertex. The final fit tests different direction hypotheses around the fitted vertex position. Twelve isotropic starting directions are chosen where the following likelihood is maximised

$$\log L = \sum_{i \in \text{hits}} \log(P_i^{\text{hit PMTs}}) + \sum_{i \in \text{no hit PMTs}} \log(P_i^{\text{no hit}}) \tag{3}$$

$$P_i^{\text{hit}} = 1 - P_i^{\text{no hit}} = 1 - \exp(-\mu_{\text{sig}}(r_i, z_i, a_i, E) - R_{\text{bg}} \cdot T) \tag{4}$$

where $\mu_{\text{sig}}$ is obtained from interpolating a PDF based on Monte Carlo simulations and $R_{bg}$ is expected background rate in time window $T$. The PDF depends on the distance from the vertex to the PMT ($r_i$), the angle between the neutrino direction and the vector between the vertex and the PMT ($z_i$), the angle between the normal vector of the PMT and the vector between the vertex and the PMT ($a_i$) and scales linearly with energy. The direction and energy resolution of $\nu_e$CC events selected as shower (described in section 4) are shown in figure 2.

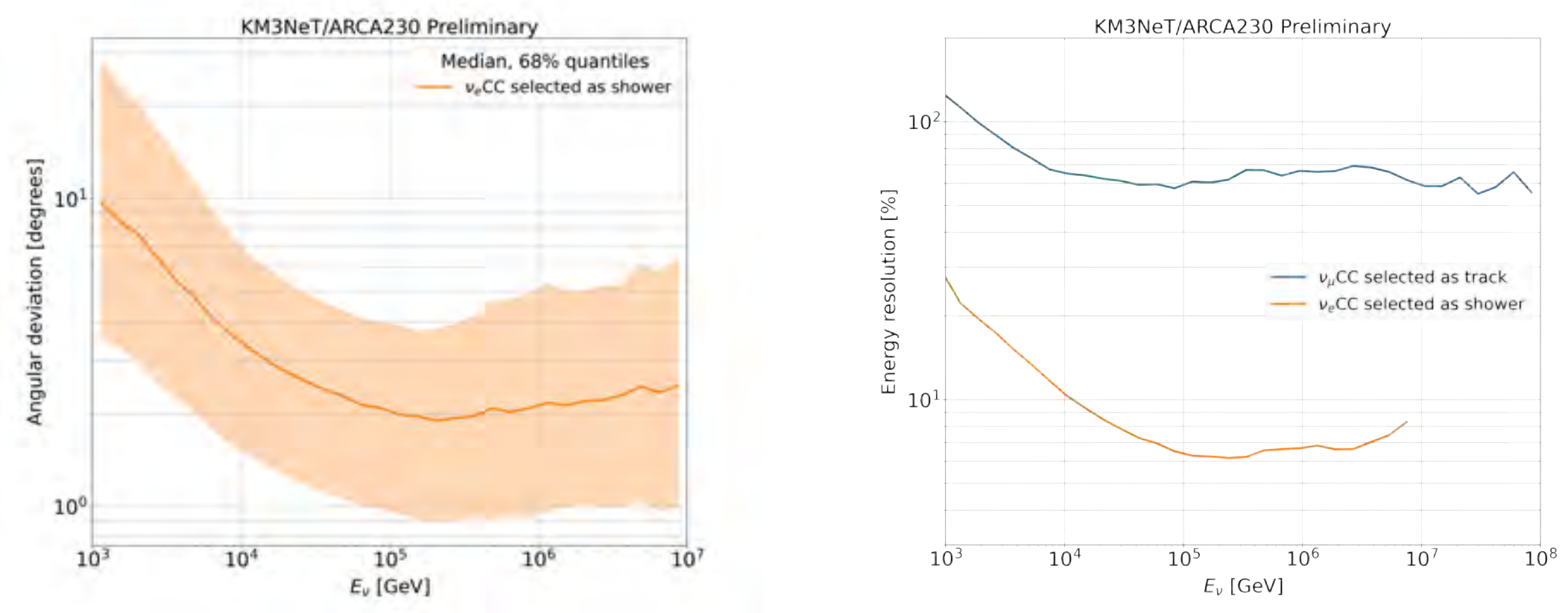

**(a)** **(b)**

**Figure 2:** KM3NeT/ARCA230 angular resolution (a) for $\nu_e$CC of the shower selection (described in section 4). The energy resolution (b) of $\nu_\mu$CC selected as track and $\nu_e$CC selected as shower.

### 3.3 Improvements

The shower reconstruction can be improved when including timing information in the fit procedure [10]. High-energy showers are elongated over several meters resulting in a small lever arm that helps the direction reconstruction. The algorithm is currently too computationally expensive to run over large datasets, but figure 3 (a) shows contained $\nu_e$CC events reconstructed with and without timing information. The resolutions drops from 2° to below 1°.

The inclusion of timing information in the reconstruction also opens up the door in reconstructing double shower events. These events are caused by $\nu_\tau$CC interactions where the $\tau$ decays into a electromagnetic or hadronic shower. The relativistic $\tau$ can travel approximately 5 m/100 TeV resulting in two separated showers. The double shower reconstruction fits a double shower model based on the hit time information [10]. The fit starting point is the traditional shower reconstruction result. This is followed by a prefit to get a first estimate of the first and second shower position. Finally a full fit is performed where the primary vertex, tau flight length, direction and two energies are fitted. Figure 3 (b) shows the angular resolution as a function of the tau length for contained events with $E_\nu$ > 100 TeV. Dedicated tau selections and analyses are in preparation.

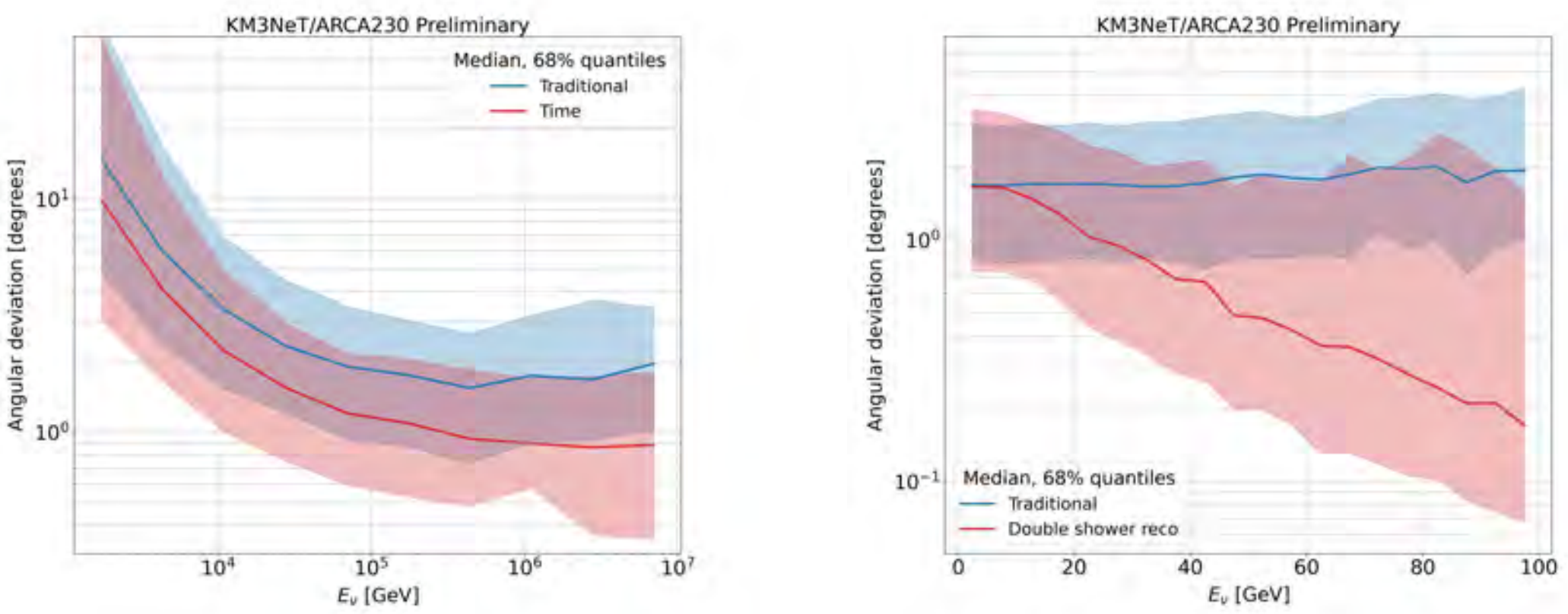

**(a)** **(b)**

**Figure 3:** Angular resolution for the shower reconstruction with timing information (a) and the double cascade reconstruction (b). Both results are compared with the traditional shower reconstruction. The time shower reconstruction shows contained $\nu_e$CC. The double cascade reconstruction contained $\nu_\tau$CC shower decay events with $E_\nu$ > 100 TeV.

## 4. Neutrino selections

The neutrino purity is increased by applying cuts to remove atmospheric muon events. This is done for two separate observation channels; tracks and showers. The variables used for rejecting background are output from the track and traditional shower reconstruction. Selection cuts are followed by the training of a boosted decision tree model to obtain the final selections. Selections based on the time shower fit and double cascade reconstruction are under preparation.

### 4.1 Track selection

The track selection is optimised to find upgoing track-like neutrino events that are reconstructed within 10° of the neutrino direction. Upgoing and horizontal events are selected for reconstructed zenith $\theta$ < 100°. Various variables of these events are used to train the boosted decision tree model including reconstructed energy, fit quality, direction error, track length and more. The final selection contains a more strict cut for horizontal (80° < $\theta$ < 100°) events due to higher mis-reconstructed muon contamination. Figure 4 shows the KM3NeT/ARCA230 event rate per year versus the reconstructed energy and zenith of the track selection for a cosmic neutrino flux of $\phi_{\nu_i+\bar{\nu}_i} = 1.2 \cdot 10^{-8} \left(\frac{E_\nu}{\text{GeV}}\right)^{-2}$ GeV$^{-1}$ cm$^{-2}$ s$^{-1}$ sr$^{-1}$.

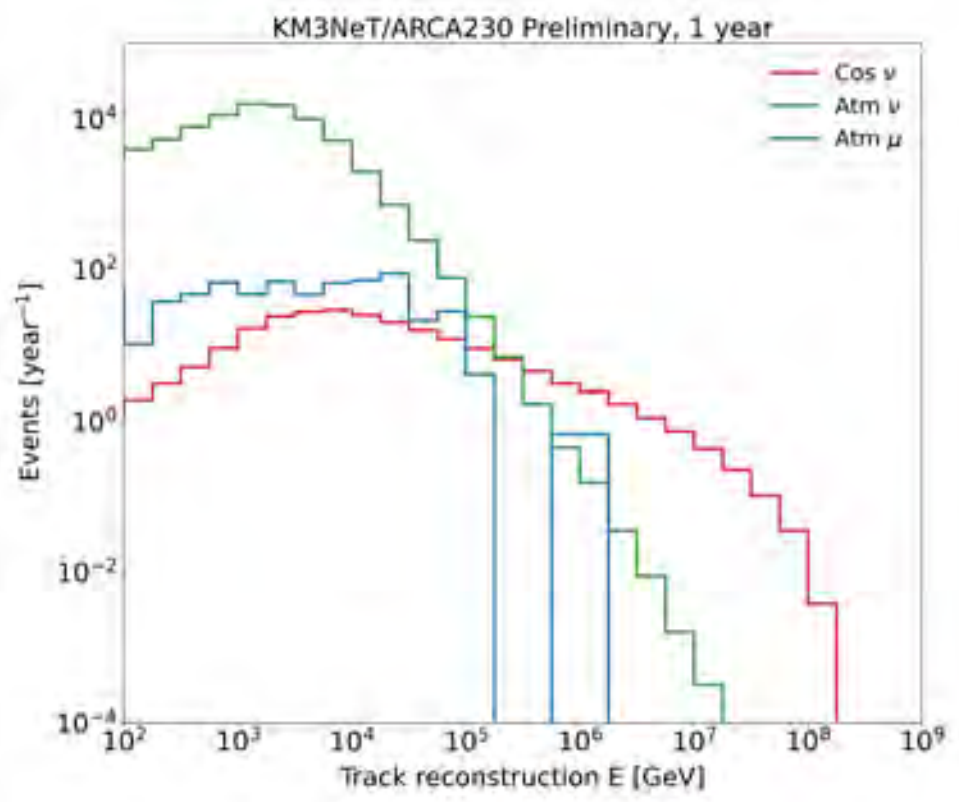

**(a)**

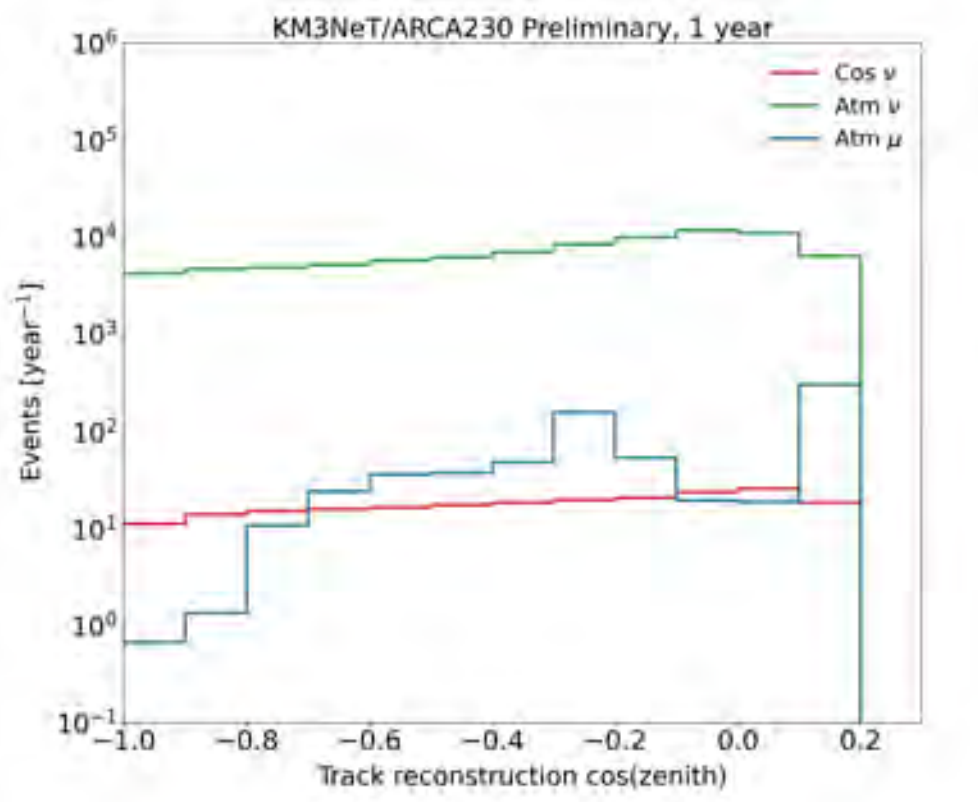

**(b)**

**Figure 4:** KM3NeT/ARCA230 event rate per year versus the reconstructed energy (a) and zenith (b) of the track selection for a cosmic neutrino flux of $\phi_{\nu_i+\bar{\nu}_i} = 1.2 \cdot 10^{-8} \left(\frac{E_\nu}{\text{GeV}}\right)^{-2}$ GeV$^{-1}$ cm$^{-2}$ s$^{-1}$ sr$^{-1}$.

### 4.2 Shower selection

The shower selection aims to improve the sensitivity to point sources and the diffuse neutrino flux by selecting as many neutrino events that not pass the track selection. Events are selected with a shower reconstruction vertex z position below the top layer of optical modules. This is followed by cuts on the number of hits that fulfill the Cherenkov hypothesis. Another boosted decision tree model is trained using variables like reconstructed position, direction, energy, fit quality and the inertia ratio of the hits. Figure 5 shows the KM3NeT/ARCA230 event rate per year versus the reconstructed energy and zenith of the shower selection.

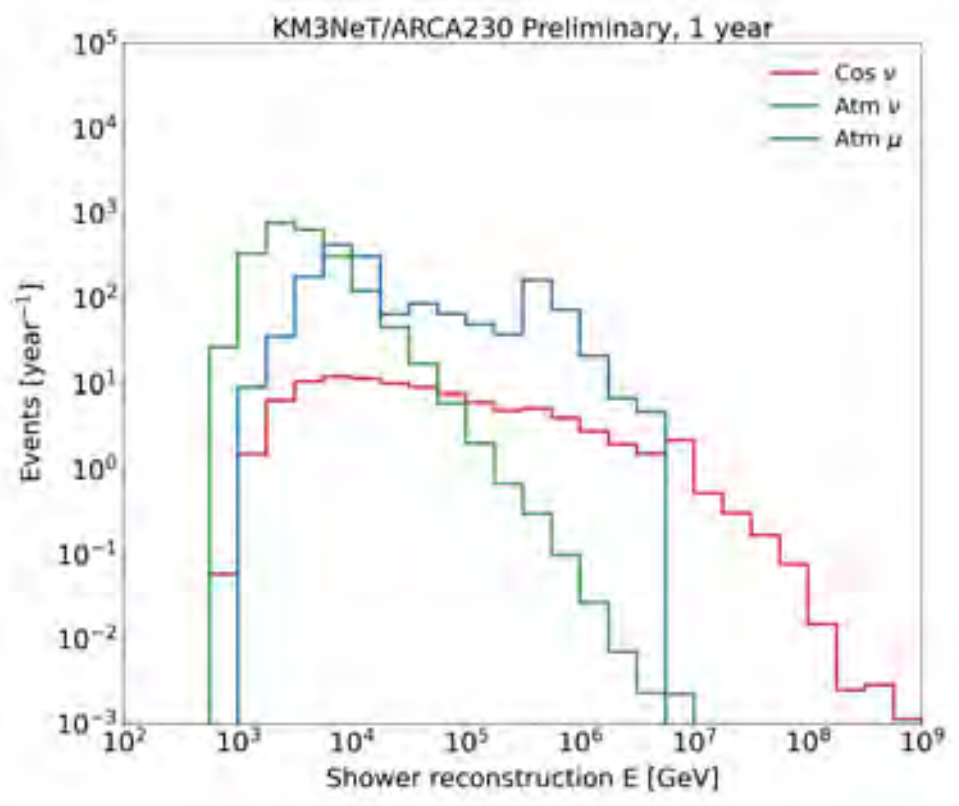

**(a)**

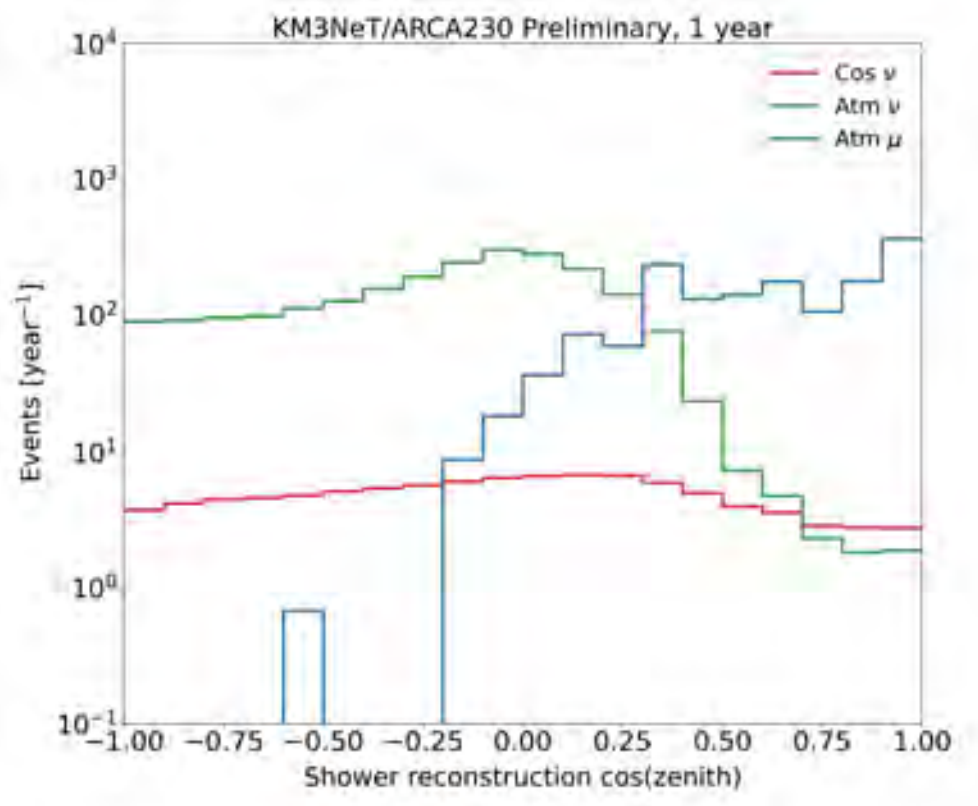

**(b)**

**Figure 5:** Event rates per year versus the reconstructed energy (a) and zenith (b) of the shower selection for a cosmic neutrino flux of $\phi_{\nu_i+\bar{\nu}_i} = 1.2 \cdot 10^{-8} \left(\frac{E_\nu}{\text{GeV}}\right)^{-2}$ GeV$^{-1}$ cm$^{-2}$ s$^{-1}$ sr$^{-1}$.

### 4.3 Performance

The effective area of the final selections can be found in figure 6.

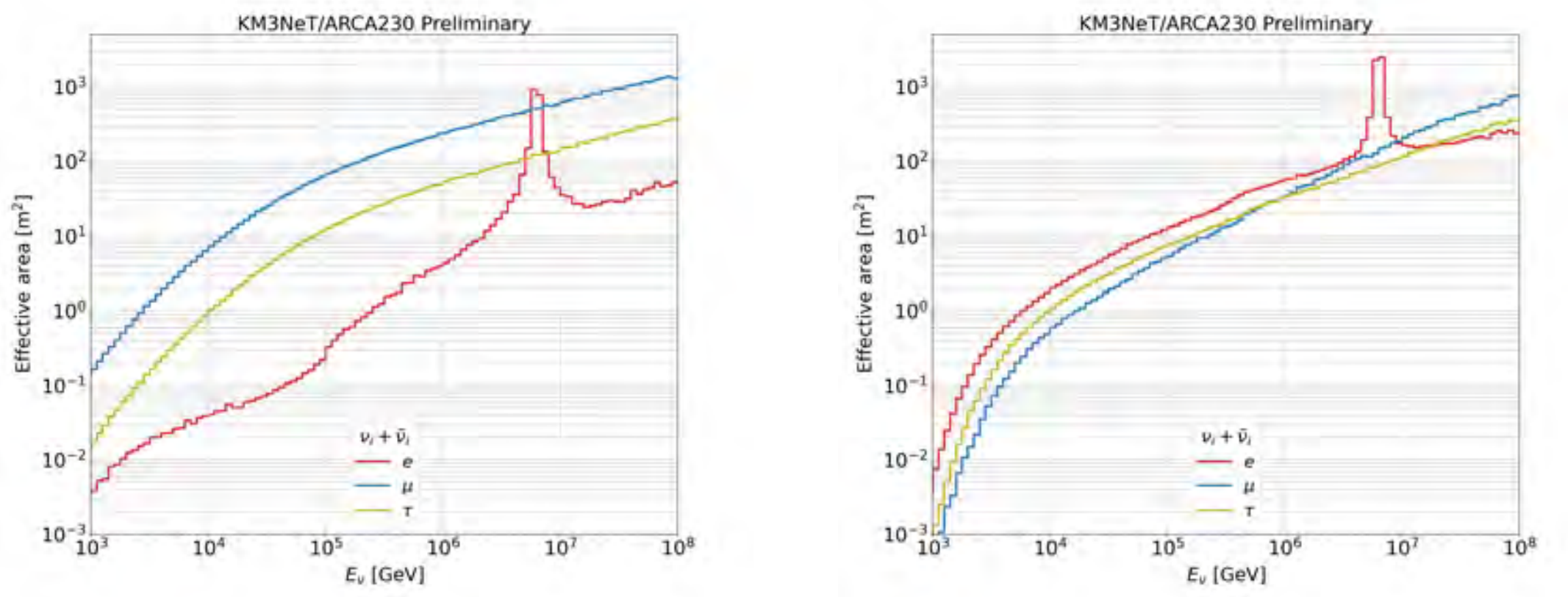

**(a)** **(b)**

**Figure 6:** Effective area for a flux of $\nu_i + \bar{\nu}_i$ for the track (a) and shower (b) selection for KM3NeT/ARCA230.

The final event rates per year for the full detector are shown in table 1.

| | Trigger | Track selection | Shower selection |
|---|---|---|---|
| Atmospheric $\mu$ | 80.6e6 | 713 | 1524 |
| Atmospheric $\nu$ | 19.1e4 | 85.2e3 | 2264 |
| Cosmic $\nu$ | 728 | 220 | 96 |
| Signal efficiency | | 94.5% | 69.7% |
| Neutrino purity | | 99% | 60.8% |

**Table 1:** Yearly event rate for KM3NeT/ARCA230 at trigger level and for the track and cascade selection for a cosmic neutrino flux of $\phi_{\nu_i+\bar{\nu}_i} = 1.2 \cdot 10^{-8} \left(\frac{E_\nu}{\text{GeV}}\right)^{-2}$ GeV$^{-1}$ cm$^{-2}$ s$^{-1}$ sr$^{-1}$.

The signal efficiency is calculated by dividing the signal events that pass the selections divided by upgoing signal events for tracks and all signal events for showers. The signal definition for tracks entails events with a muon and that have the direction reconstruction within 10°. Signal events for showers are events without a muon, that are contained within the detector and have the direction reconstruction within 10°.

## 5. Conclusions

This contribution summarises the KM3NeT/ARCA230 reconstruction software and neutrino selections that are used for the current point source and diffuse a nalyses. The track selection is a pure upgoing neutrino sample and the cascade selection includes as much extra signal as possible. Future improvements will include timing information in the shower reconstruction resulting in an angular resolution below 1° above several 100 TeV. A dedicated double shower algorithm and a dedicated neutrino selection is in preparation.

# Astronomy potential of KM3NeT/ARCA230

**Thijs Juan van Eeden**[a,*] **for the KM3NeT collaboration**

[a] *Nikhef,*
*Science Park 105, Amsterdam, Netherlands*

*E-mail:* tjuanve@nikhef.nl

The KM3NeT/ARCA neutrino detector is currently under construction at the bottom of the Mediterranean Sea. The main science objective is the detection of high-energy cosmic neutrinos and the discovery of their sources. This is achieved by instrumenting a cubic kilometre of seawater with photo-multiplier tubes that detect Cherenkov radiation from neutrino interaction products. In recent years, there have been advancements in the reconstruction algorithms for muon neutrinos, along with the development of techniques for identifying and accurately reconstructing electron and tau neutrinos. This contribution will present the updated prospects for cosmic sources using the track and cascade detection signatures for the full detector.

*Speaker

## 1. Introduction

The detection of high-energy neutrinos from TXS 0506+056 and NGC 1068 are the first discoveries of cosmic neutrino sources in our Universe [1, 2]. The KM3NeT/ARCA detector is currently under construction and is optimised for the discovery of cosmic neutrino sources. The detector consists of optical modules that are attached to vertical detection strings. Each string is mounted to the seabed of the Mediterranean sea and houses 18 optical modules. The final detector configuration will consist of two blocks with 115 strings each (KM3NeT/ARCA230).

Statistical methods have been developed to study the sensitivity and discovery potential of KM3NeT/ARCA to point sources and the diffuse cosmic neutrino flux. The KM3NeT collaboration is sharing first analyses with recent data in [citations], but this contribution covers updated sensitivity projections for the full detector using the combination of tracks and showers.

## 2. Simulation and selection

Neutrino events are generated over the full-sky with energies in the range $10^2 < E_\nu < 10^8$ GeV using the gSeaGen framework [3]. Atmospheric muons are simulated using the MUPAGE software [5]. All events are passed through the KM3NeT light simulation and detector response software packages. The track and shower reconstruction is applied as described in [6].

The track selection is optimised to select upgoing muons produced in neutrino interactions. The background from atmospheric muons and noise is rejected using a boosted decision tree model trained on reconstructed quantities like the reconstructed zenith angle and energy, the track length length and the fit quality. The shower selection covers the full-sky and also uses a boosted decision tree model to reject atmospheric muons. Events are selected based on the height of the reconstructed shower vertex using the top layer of optical modules as a veto. The effective area of both selections is given in Figure 1 (a). The angular resolution for $\nu_\mu$CC which are selected as track and for $\nu_e$CC selected as shower are shown in Figure 1 (b).

The energy resolution is defined as half the difference between the 16th and 84th percentiles of the energy bias distributions where

$$\text{Energy bias } = 100\% \cdot \left(\frac{E_{rec}}{E_{visible}} - 1\right). \quad (1)$$

The energy resolution is shown in Figure 2.

The detector simulations are stored in detector response functions containing the effective area, angular resolution and energy response for different flavours, interactions and observation channels. These functions are input to the analysis described in the next section. This contribution extends the track analysis [4] with the shower channel and uses updated Monte Carlo simulations.

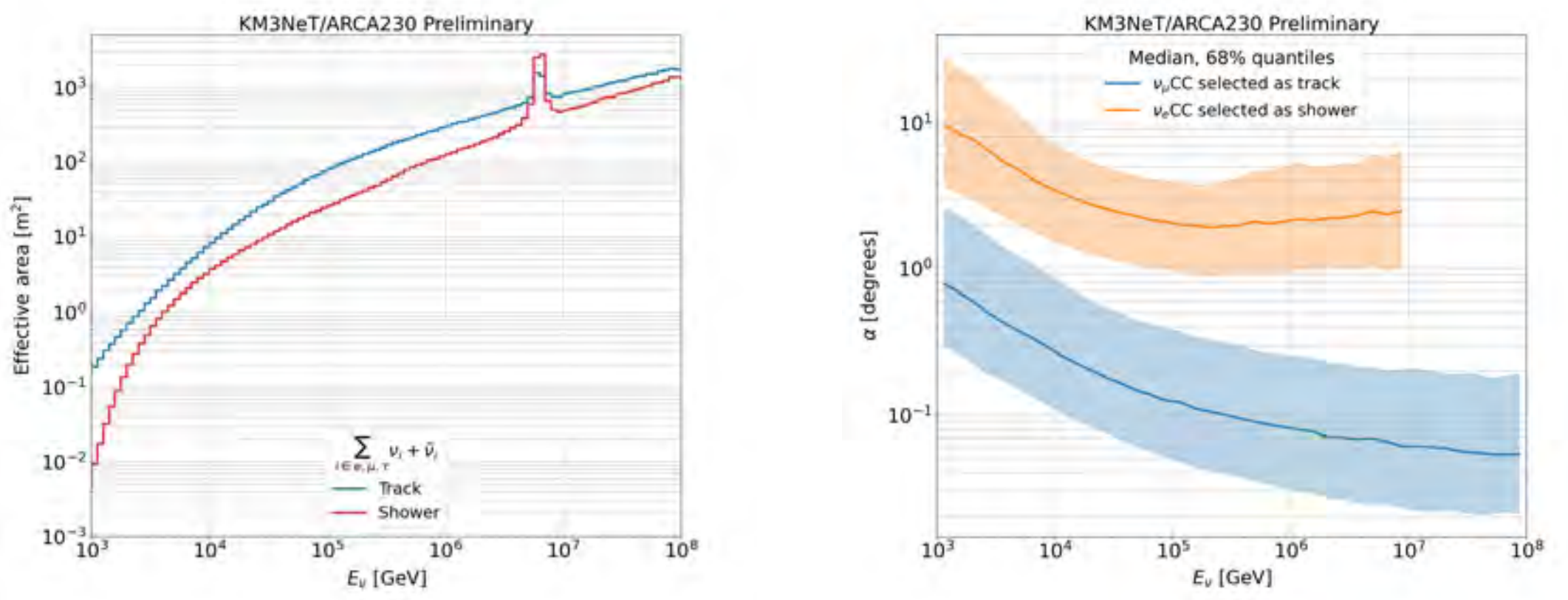

**Figure 1:** Effective area (a) for the track and shower channel. The effective area contains the sum of interactions from $\nu_e, \nu_\mu, \nu_\tau$ and averaged over $\nu, \bar{\nu}$. The angular resolution (b) for $\nu_\mu$CC selected as track and for $\nu_e$CC selected as shower.

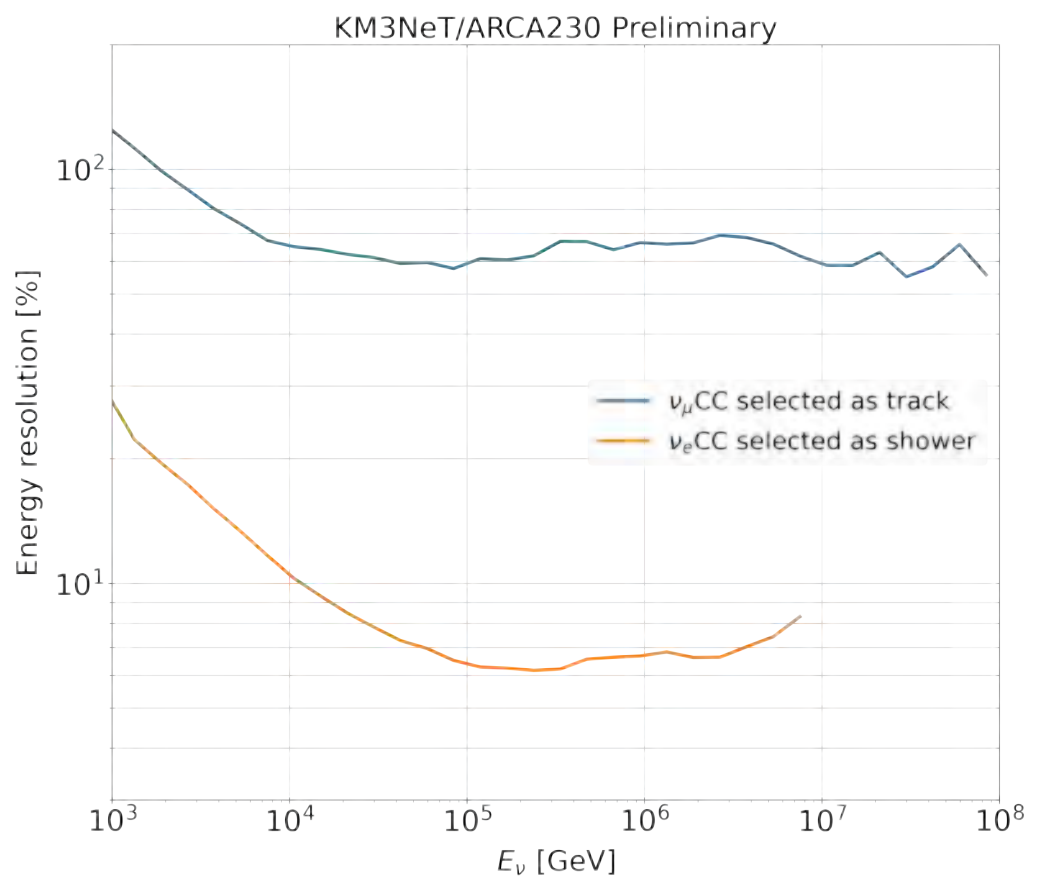

**Figure 2:** Energy resolution of $\nu_\mu$CC for the track selection (a) and $\nu_e$CC for the shower selection (b).

## 3. Method

The detector response functions are used to create expected distributions for signal and background for different flux models and observation times. The flux models are characterised by different spectral indices for point sources while the IceCube spectral index of [7] is adopted for the diffuse analysis. The flux models are convoluted with the effective area and detector response to obtain two-dimensional distributions for expected signal ($S_i$) and background ($B_i$). An arbitrary flux normalisation is scaled with a varying signal strength ($\mu$) to study the sensitivity and discovery potential of KM3NeT/ARCA230.

The expected event rate distributions for the point source analysis have bins in reconstructed energy and distance to source in degrees. An example distribution of signal and background events is shown in Figure 3. Both distributions are shown for events selected as tracks with a source at sin(declination) = 0.1, a spectral index $\gamma$ = 2 and 3 years of KM3NeT/ARCA230 operation.

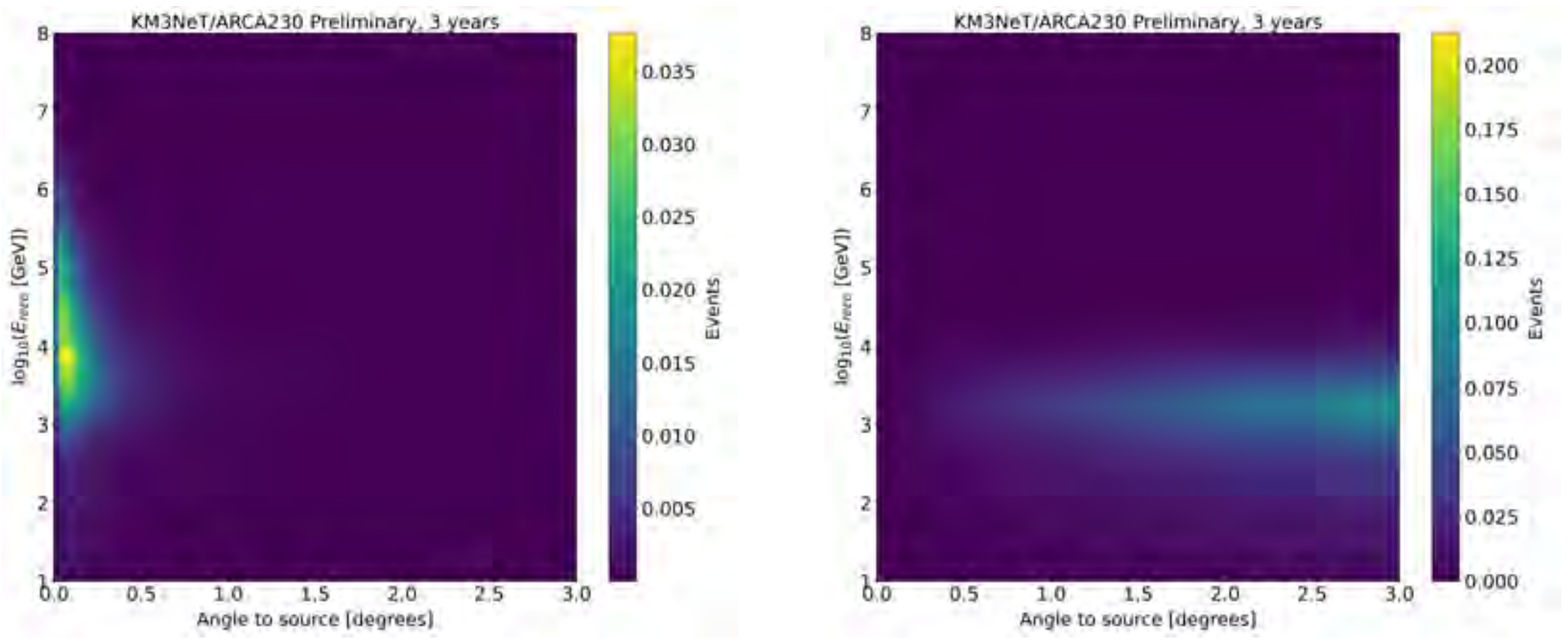

**(a)** **(b)**

**Figure 3:** Expected signal (a) and background (b) distributions for events selected as track for the point source analysis. The distributions were obtained for a source at sin(declination) = 0.1, spectral index $\gamma$ = 2 and 3 years of KM3NeT/ARCA230 operation.

The diffuse analysis uses two-dimensional distributions with bins in reconstructed energy and zenith. The corresponding signal and background distributions for events selected as showers are shown in Figure 4. Both distributions are obtained for 3 years of KM3NeT/ARCA230 operation.

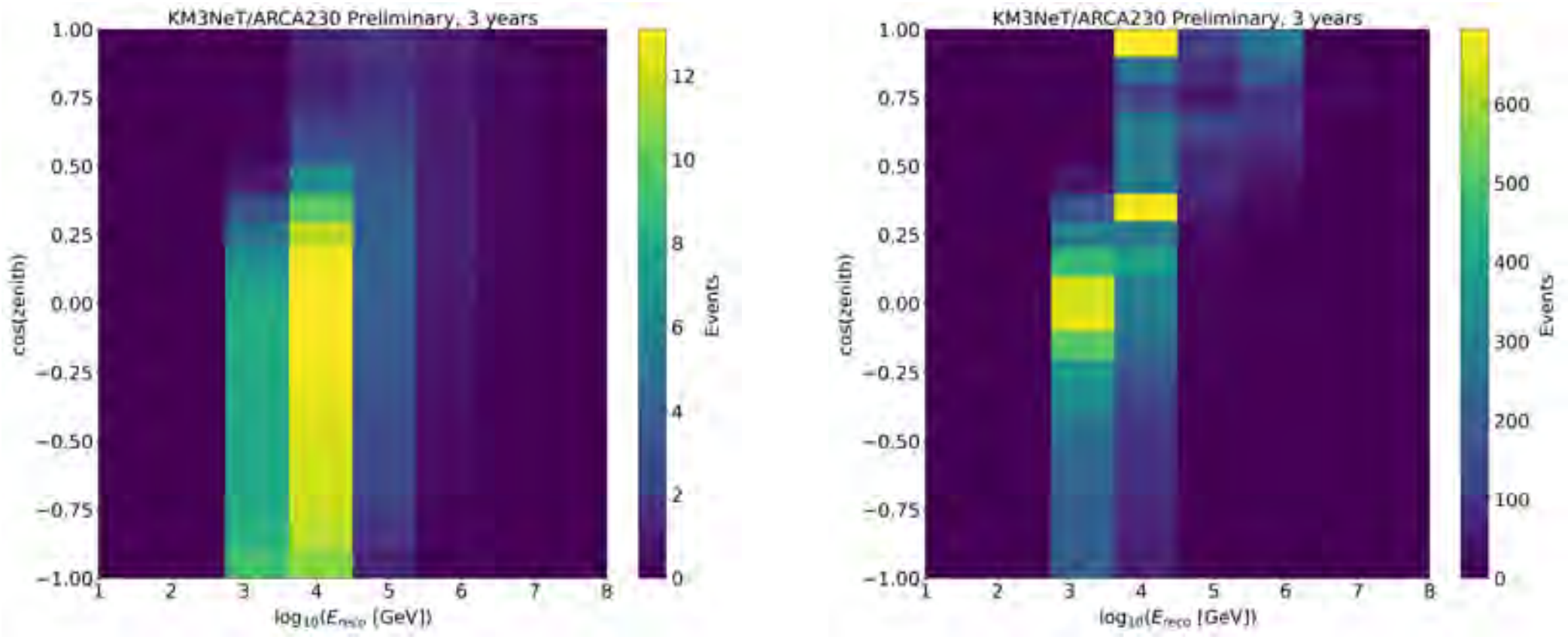

**(a)** **(b)**

**Figure 4:** Expected signal (a) and background (b) distributions for events selected as shower for the diffuse flux analysis. The distributions were obtained for 3 years of KM3NeT/ARCA230 operation.

Based on these expected signal and background distributions, pseudo experiments using Poissonian statistics are made where the expectation value is defined as

$$E(N_i) = B_i + \mu S_i. \tag{2}$$

The signal strength ($\mu$) is varied to obtain test statistic distributions for the statistical analysis. The definition of the likelihood for a pseudo dataset is

$$\log L = \sum_{i \in \text{bins}} N_i \log(B_i + \mu S_i) - B_i - \mu S_i \tag{3}$$

and the test statistic ($\lambda$) is a likelihood ratio defined as

$$\lambda = \log L(\mu = \hat{\mu}) - \log L(\mu = 0). \tag{4}$$

The test statistic distributions are used to calculate the confidence level and p-values. An example of the test statistic and confidence level for the point source analysis is shown in Figure 5.

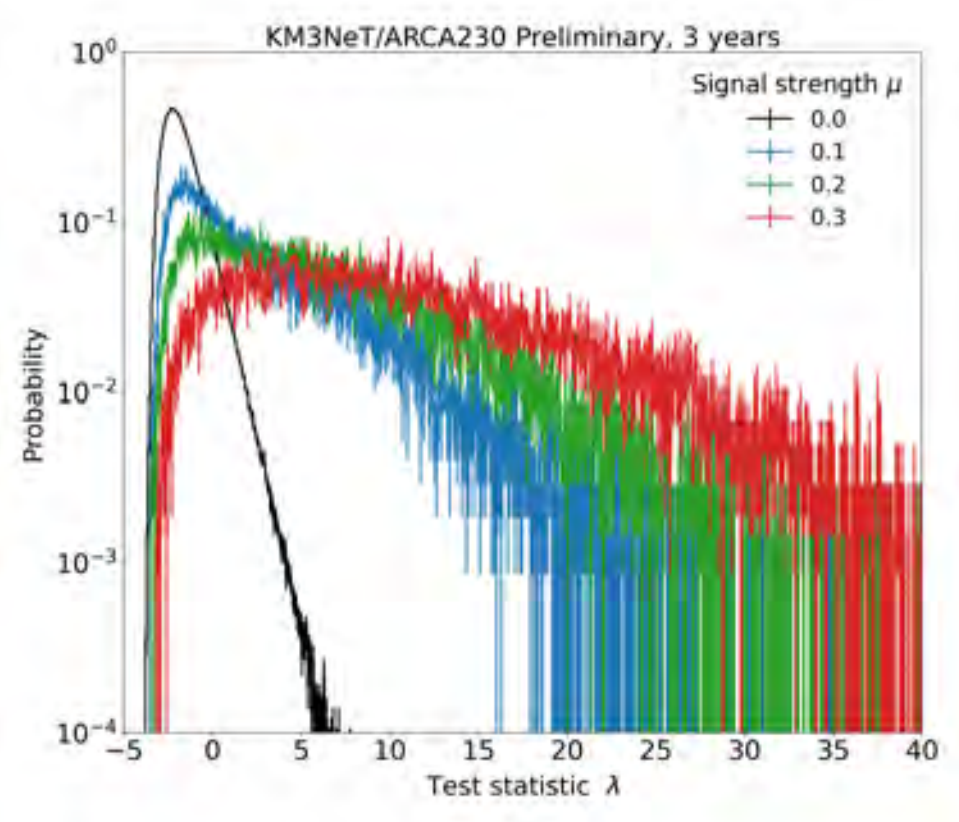

**(a)**

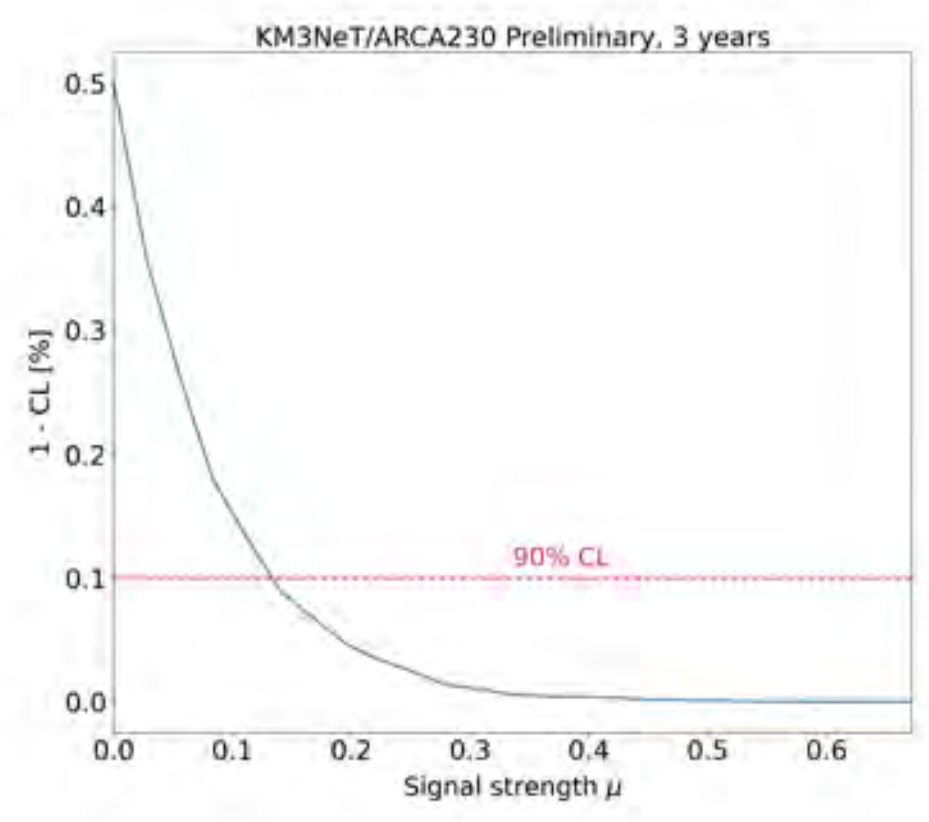

**(b)**

**Figure 5:** Test statistic distribution (a) of the point source analysis for various signal strengths for a source at sin(declination) = 0.1 and 3 years of KM3NeT/ARCA230 operation. The corresponding confidence levels and the 90% confidence level (b).

## 4. Results

### 4.1 Point sources

The point source sensitivity for different spectral indices is given in Figure 6 . For a spectral index of $\gamma = 2$ the KM3NeT/ARCA230 sensitivity is compared with the corresponding sensitivity of 15 years of Antares [8] and 7 years of IceCube [9].

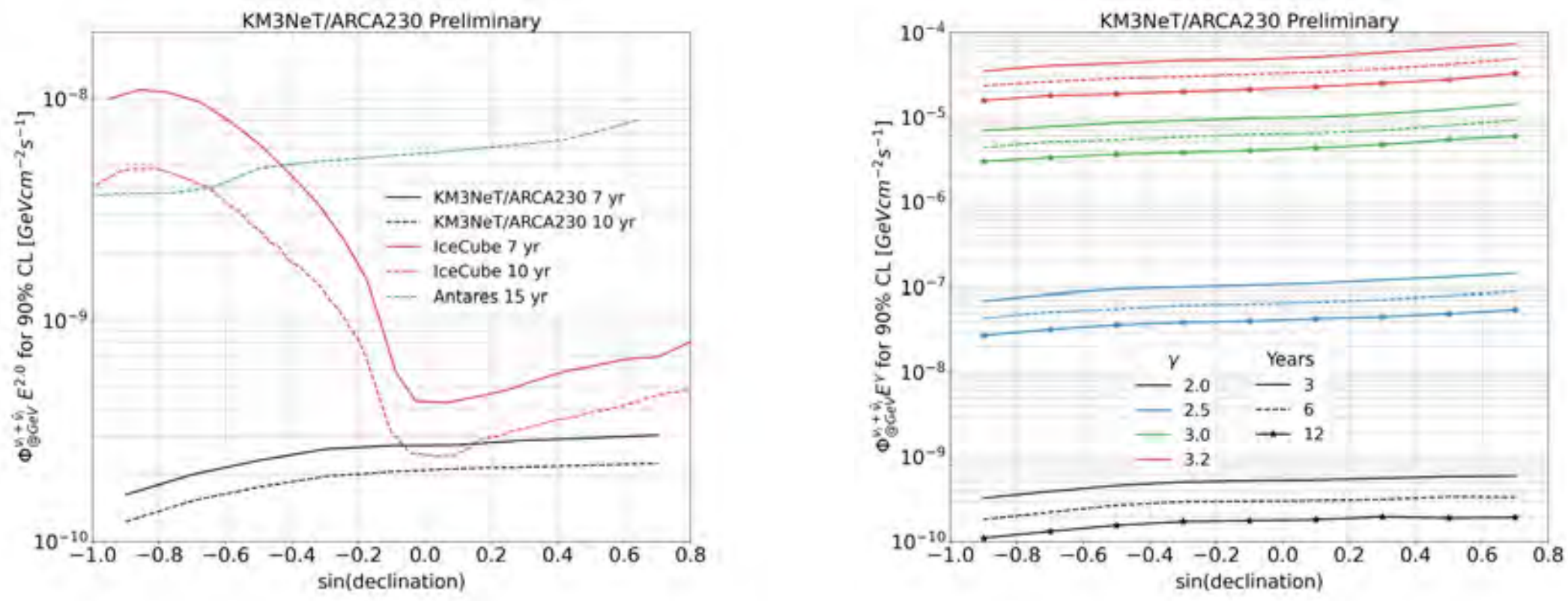

**(a)** **(b)**

**Figure 6:** Point source sensitivity for $\gamma = 2$ (a) and other spectral indices (b). The $\gamma = 2$ results are compared with 15 years of Antares and 7 years of IceCube [8, 9].

The discovery flux for point sources is given in Figure 7 with different spectral indices.

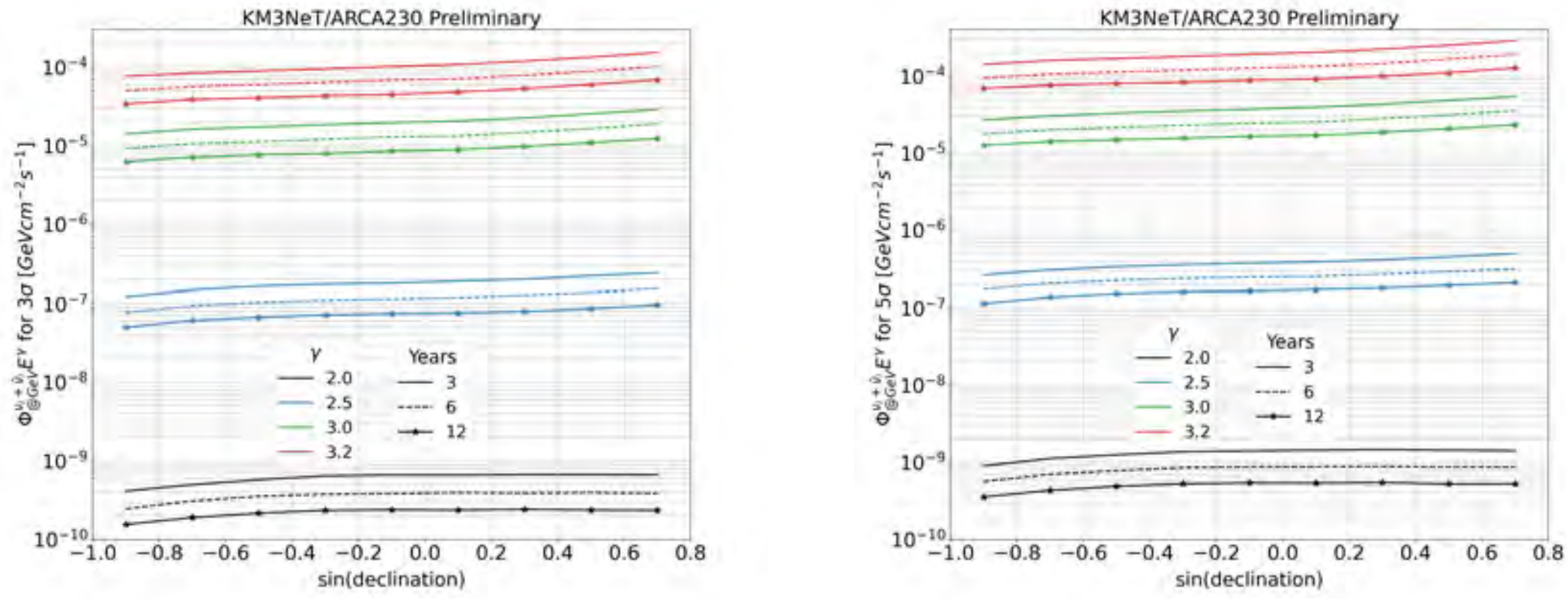

**(a)** **(b)**

**Figure 7:** Point source discovery flux for different spectral indices. The $3\sigma$ discovery flux is given in (a) and the $5\sigma$ is given in (b).

### 4.2 NGC 1068

The discovery potential for NGC 1068 was studied and is shown in Figure 8. A $5\sigma$ discovery can be claimed after 3 years of full KM3NeT/ARCA operation. The power law flux with $\gamma = 3.2$ was extended over the full energy range.

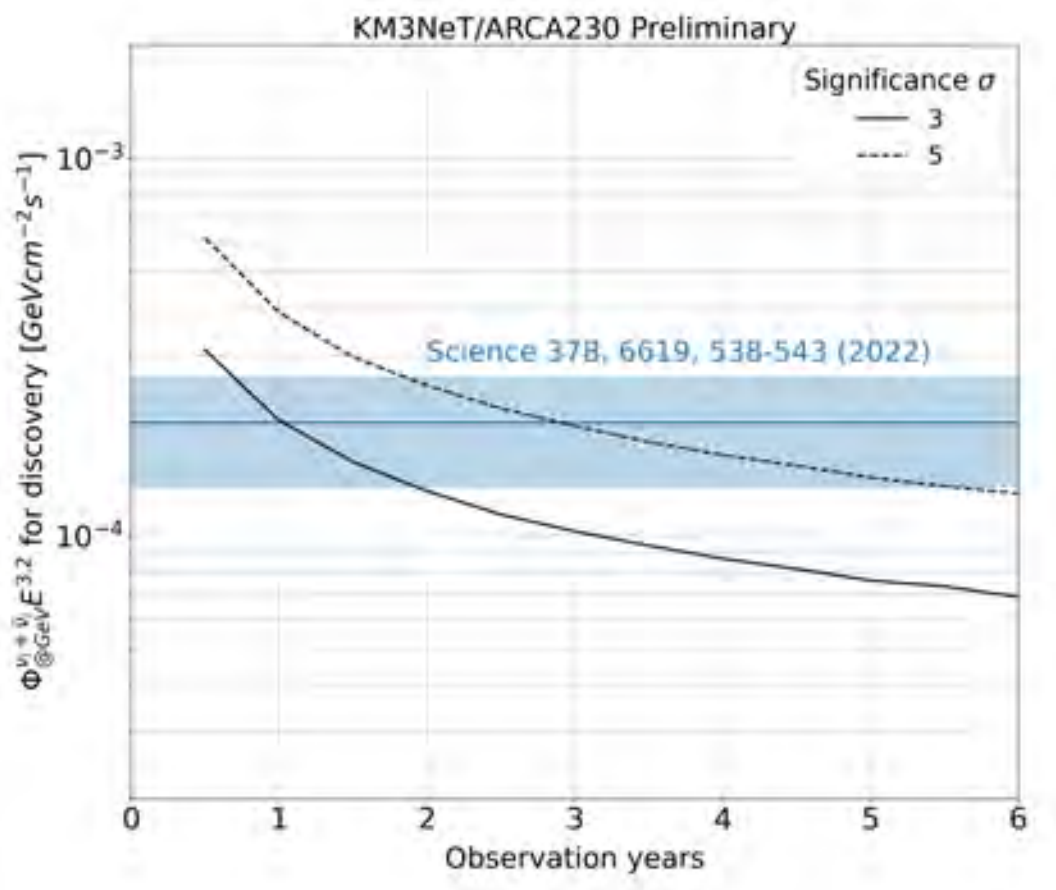

**Figure 8:** Discovery flux of KM3NeT/ARCA230 for NGC 1068 assuming a spectral index $\gamma$ = 3.2 as reported by IceCube [2]. The blue lines represent the fitted flux normalisation of IceCube including statistical and systematic uncertainties.

### 4.3 Diffuse flux

The 3 and 5 $\sigma$ discovery flux for KM3NeT/ARCA230 is shown in Figure 9 for a diffuse neutrino flux with $\gamma$ = 2.37 as reported by IceCube [7]. The fitted flux normalisation of IceCube can be discovered with 5 $\sigma$ within a half year of full KM3NeT/ARCA operation.

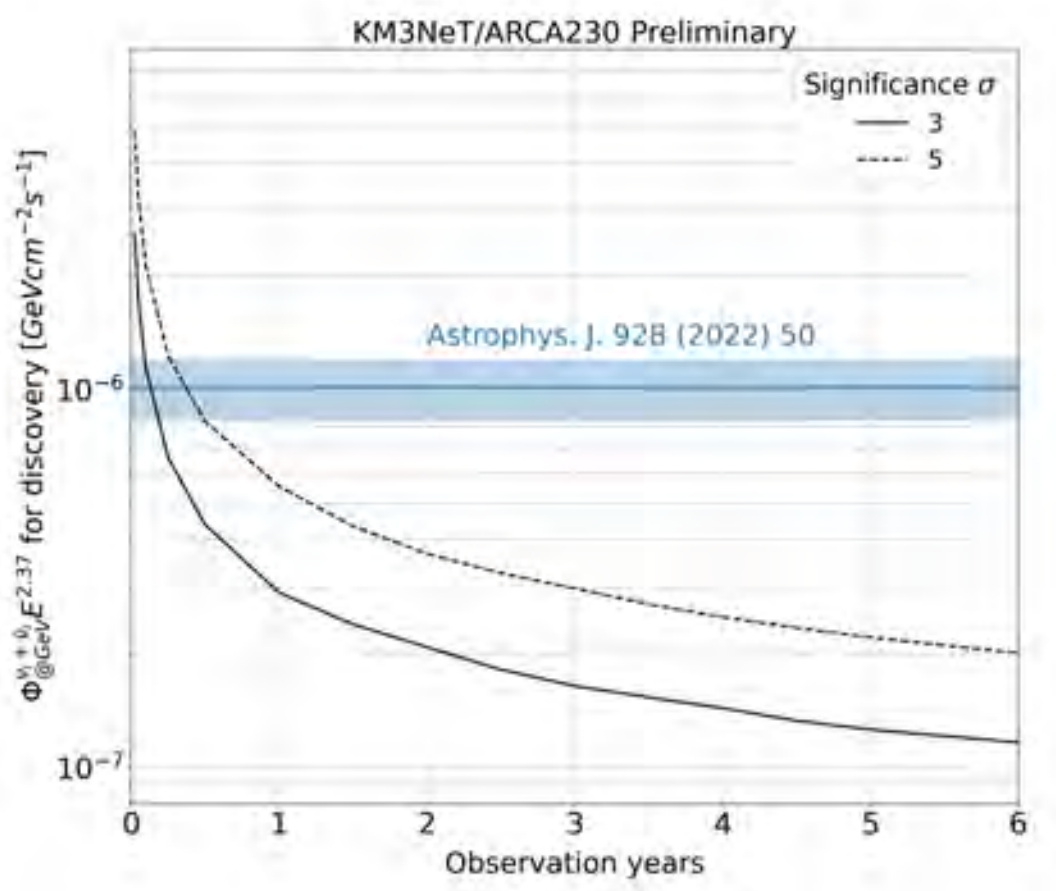

**Figure 9:** Discovery flux of KM3NeT/ARCA230 for the diffuse neutrino flux with spectral index $\gamma$ = 2.37 as reported by IceCube [7]. The blue lines represent the fitted flux normalisation of IceCube including statistical and systematic uncertainties.

## 5. Conclusions

The KM3NeT/ARCA detector is currently under construction and is already contributing to neutrino and multi messenger astronomy. The full detector will be competitive in discovering point sources and will be able to detect the diffuse and NGC 1068 neutrino fluxes within respectively 3 and 0.5 years. Improvements in the shower reconstruction described in [6] will bring up to 10% improvement in the discovery potential of point sources and for the diffuse neutrino flux. Systematic uncertainties in the absorption and scattering length of water (±10%) and the effective area of an optical module (±10%) influence the results with < 10%.

# The Real-Time Analysis Platform of KM3NeT and its first results

**S. Celli**[a,b,*] **and P. Demin, D. Dornic, F. Filippini, E. Giorgio, E. Le Guirriec, J. De Favereau de Jeneret, M. Lamoureux, M. Mastrodicasa, J. Palacios Gonzalez, S. Le Stum, J. Tanasijczuk, G. Vannoye, A. Veutro, A. Zegarelli for the KM3NeT Collaboration**

[a] *Sapienza Università di Roma,*
*Piazzale Aldo Moro 5, 00185, Rome, Italy*

[b] *Istituto Nazionale di Fisica Nucleare, Sezione di Roma,*
*Piazzale Aldo Moro 5, 00185, Rome, Italy*

*E-mail:* silvia.celli@roma1.infn.it

KM3NeT is a multi-site neutrino telescope under construction in the depth of the Mediterranean Sea, consisting of two Cherenkov telescopes, ARCA and ORCA, both of which are currently in data-taking. Among the primary scientific goals of KM3NeT are the observation of cosmic neutrinos and the investigation of their sources. ARCA and ORCA are optimized in complementary energy ranges, allowing for the exploration of neutrino astronomy from MeV to tens of PeV. The combination of an extended field of view and a high duty cycle of Cherenkov-based neutrino detectors is crucial for detecting and informing other telescopes about interesting neutrino candidates in a very short time. As emission from these sources can rapidly fade, the alerts need to be shared with low latency, in order to allow for a prompt follow-up in the multi-messenger and multi-wavelength domains, particularly for the detection of transient and variable sources. In the case of poorly localized triggers, such as gravitational waves, KM3NeT can provide refined pointing directions, representing a further advantage. This contribution reports on the status of the software architecture implemented in KM3NeT for a fast reconstruction and classification of events occurring in the detectors. Additionally, the results of the online processing of KM3NeT data in coincidence with GRB221009A will be presented.

*Speaker

## 1. Introduction

The goal of real-time alert systems is to enable multi-messenger (MM) observations allowing the localization and identification of sources, particularly in the presence of fading events. Thanks to its fast response, almost continuous duty cycle, and large field of view, KM3NeT is well suited to trigger observatories with limited solid angle visibility to quickly point their instruments in a well-defined direction of the sky. Other way around, KM3NeT can carry out follow-up observations of external triggers, thus enhancing the utility of joint observation campaigns. The rapid provision of alerts is crucial to detect transient and variable sources, such that fast algorithms coupled with automated notification systems have to be employed in the online processing of events.
The KM3NeT Collaboration has developed a Real-Time Analysis (RTA) platform for each of its two detectors: after validation in a six-month commissioning period, the system is currently implemented into the standard data flow. It is now regularly reconstructing and classifying events with a median latency below $\sim$ 10 s for both ARCA and ORCA data, as well as receiving and processing external alerts. The recent start of the fourth data taking run of the two LIGO Gravitational-Wave (GW) interferometers is hence accompanied by the KM3NeT monitoring of the neutrino sky, in analogy with the MM activities performed during the previous acquisition run of LIGO-Virgo [1]. This contribution describes the status of the RTA framework implemented in KM3NeT. In Sec. 2 the main features of the experimental apparatus are described, followed in Sec. 3 by details of the framework that has been developed for the real-time trigger and follow-up of MM alerts. This section also reports on the online analysis of GRB221009A, the most luminous Gamma-Ray Burst (GRB) ever detected, which occurred during the commissioning of the KM3NeT RTA system. Conclusions are drawn in Sec. 4.

## 2. KM3NeT

KM3NeT [2] is an European research infrastructure consisting of a network of deep-sea neutrino detectors located in the Mediterranean Sea: ARCA offshore Sicily (Italy) at 3.5 km depth, and ORCA offshore Toulon (France) at 2.5 km depth. These are three dimensional arrays of photosensors recording the time, position and charge deposit of Cherenkov hits induced by the passage in water of ultra-relativistic charged particles produced at neutrino interactions, thus allowing for the reconstruction of the incoming neutrino direction and energy. The two instruments profit of the same technology, based on the Digital Optical Modules (DOMs), glass spheres hosting 31 3$''$ Photo-Multiplier Tubes (PMTs) as well as the electronics for data acquisition and calibration [3]. 18 DOMs compose a Detection Unit (DU). At the time of writing, the two detectors are taking data, ARCA with 21 DUs and ORCA with 18 DUs, in the so-called ARCA21 and ORCA18 configurations, respectively. Besides the different location and depth, ARCA is the larger and more sparsely instrumented detector compared to ORCA. In ARCA, the DUs are horizontally separated from each other by an average distance of about 90 m and the vertical distance between DOMs belonging to the same DU is 36 m. In contrast, the DUs in ORCA are installed about 20 m apart and the vertical distance between DOMs is 9 m. The different geometries of ARCA and ORCA permit scientists to explore complementary energy ranges of the interacting neutrinos: ARCA has enhanced sensitivity in the multi-TeV domain, while ORCA is optimized for the detection of sub-TeV neutrinos. Moreover, searches for MeV neutrinos, e.g. those expected in Core Collapse SuperNova (CCSN) explosions, can be performed exploiting signals in individual DOMs, by looking for an increased rate in hit coincidences within its PMTs. The broad energy coverage of the two KM3NeT detectors hence enables analyses in the MeV to multi-PeV domain. In addition, the high duty cycle of Cherenkov detectors, coupled to the $4\pi$ field of view, guarantees continuous monitoring of the neutrino sky, which is of paramount importance in the context of transient and variable emissions from cosmic sources. To fully explore the time-domain neutrino astronomy, the KM3NeT Collaboration has implemented a real-time reconstruction and classification system of events, allowing for a fast selection of a high-purity neutrino sample as well as for follow up external triggers. The KM3NeT RTA system is described in the following, with a focus on the latency times for event processing and alert distribution to the external community.

## 3. The RTA framework

The KM3NeT telescopes follow the so-called *all data-to-shore* concept, i.e. all data are sent from the detectors to their respective control stations on the shores, where the Data Acquisition System (DAQ) applies an automated filtering in order to reduce the data output for storage. Before data writing, the collected hits and triggered events are propagated into the RTA platform, where they are sudden processed through different pipelines. Two RTA modules are running continuously in the detector shore stations, in order to identify interesting neutrino-induced events within KM3NeT data. One is the online processing pipeline for reconstruction and classification of events induced by the interaction of GeV-PeV neutrinos; the other is the MeV SN analysis, as detailed in Sec. 3.1. Data from each detector are then transferred to a common dispatcher, where analysis pipelines are activated: these include both auto-correlation searches as well as follow-up studies, starting automatically whenever an interesting external alert is received from the MM community. Currently, four follow-up analyses are in place, looking for space and time coincidences between KM3NeT events and GWs, GRBs, other transients and high-energy neutrinos from IceCube, as presented in Sec. 3.2. The described data flow is schematically illustrated in Fig. 1.

### 3.1 The online GeV-PeV event processing module

Neutrinos with energies in the GeV-PeV range can trigger multiple DOMs in the detector, leaving two distinct topological signatures according to their flavor and interaction channels, namely: 1) track-like events, mainly from the charged-current (CC) interactions of muon neutrinos $\nu_\mu$ and partially from the CC interactions of tau neutrinos $\nu_\tau$; 2) shower-like events, from the electron neutrino $\nu_e$ CC interactions, the neutral-current (NC) interactions of all neutrino flavors and the majority of tau neutrino CC interactions. Based on these two event signatures, KM3NeT reconstruction algorithms can be divided into track and shower reconstructions [4], the former providing the best angular resolution and the latter featuring a remarkable energy resolution, particularly for contained events. Both the online directional and energy reconstruction adopt the same algorithms as in offline, with the exception that some steps are avoided in the online chain in order to process data as fast as possible. A further difference among online and offline reconstructions is the fact that dynamical calibrations are not yet implemented in the real-time chain: while optimizations are in progress, the current track-like reconstruction of ARCA21 data can achieve sub-degree precision for neutrino energies above 10 TeV [5].
To handle the high trigger rate with low latency, parallel multi-core processing is adopted: Fig. 2 shows the rate of reconstructed events in a time window of few days with the detector configurations ARCA21 and ORCA18. For the same event topology, the rate of reconstructed events is different for the two detectors because of the different depth where they are located, the deeper location of ARCA implying a lower background from atmospheric muons.

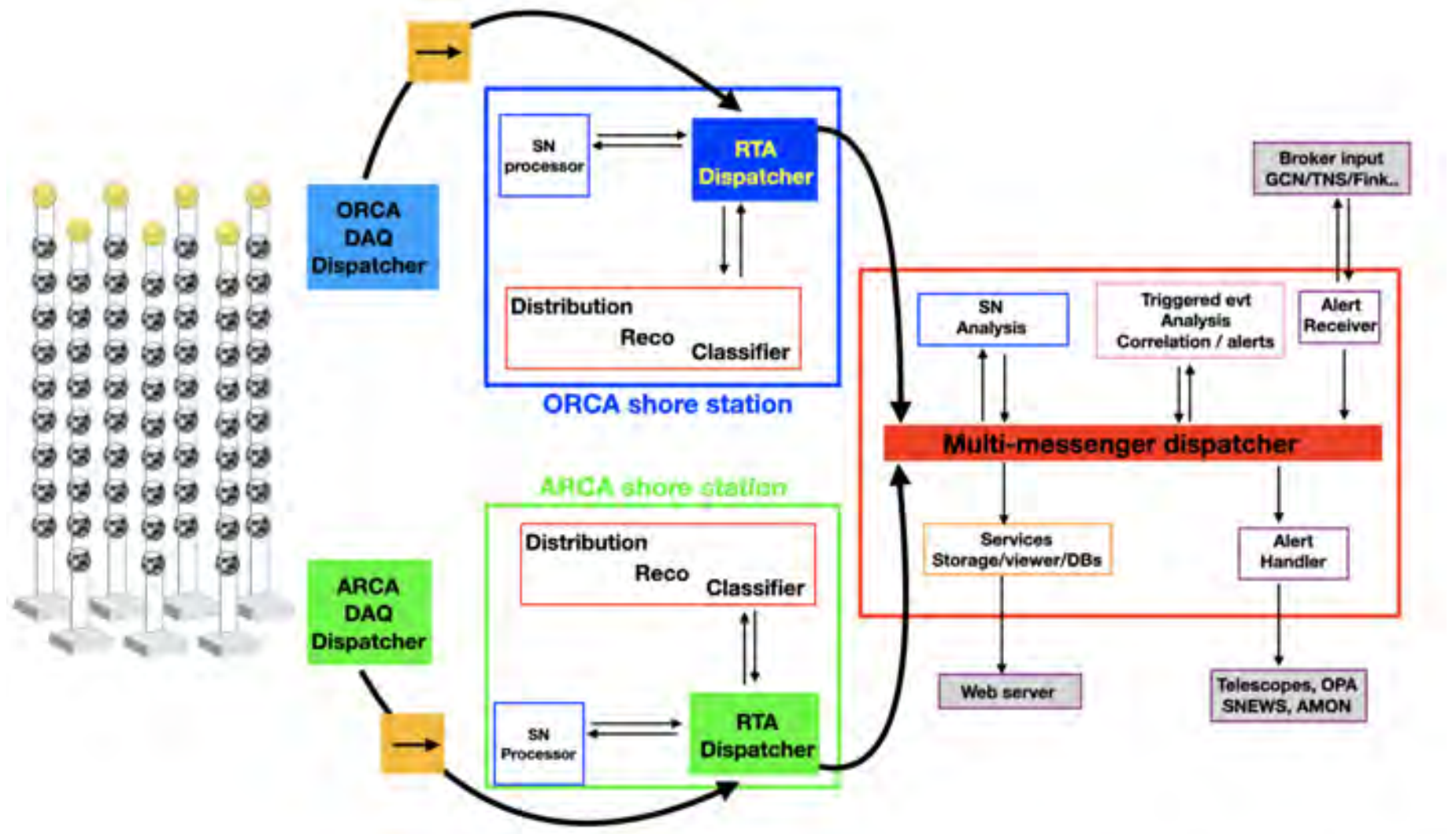

**Figure 1:** A schematic illustration of the real-time framework architecture of KM3NeT.

The event processing at the ARCA and ORCA shore stations is implemented through different architectures: for each ARCA event, topological reconstructions and classifications are run in parallel, while ORCA data are processed serially, namely a single client is in charge for both the track-like, the shower-like and the classification of an individual triggered event. Fig. 3 shows histograms of the resulting online processing times of individual events, in the two current detector configurations: in the case of ARCA21, a median delay of 3.6 s is achieved between the event triggering and its reconstruction by either the track-like or the shower-like algorithm (hence including the data filtering from background noise, its buffering, dispatching and finally reconstruction times), as visible in Figs. 3(a) and 3(b), respectively. An analogous timescale is obtained in the classification processing, as shown in Fig. 3(c), where the time shown includes again filtering, buffering, and dispatching of data before the actual classification. A median of 6.0 s is achieved for ORCA18 events, such a delay including filtering, buffering, dispatching, serial topological reconstruction and classification, as shown in Fig. 3(d).
With regards to the classification algorithms that are currently in place for both ARCA and ORCA, machine learning techniques are being adopted, both Boosted Decision Trees and Graph Neural Networks capable of fully exploiting the detector geometry. A first classifier aims at performing a fast neutrino classification to separate neutrinos from the large muon background: each event is evaluated with a classification score indicating the probability of it being a neutrino, its output being crucial in the process of neutrino selection in the following correlation search module. A second classifier, in turns, determines the most likely event topology, providing the likelihood for the event to belong to either the track or the shower sample. Currently, the processing time taken by each classifier amounts to $\sim 0.3$ s/event. After the event classification, ARCA and ORCA data streams are combined for common analysis and any subsequent alerts in the so-called MM dispatcher, shown in Fig. 1. At this level, a neutrino sample can be selected by choosing the appropriate cut on the classification scores depending on the analysis.

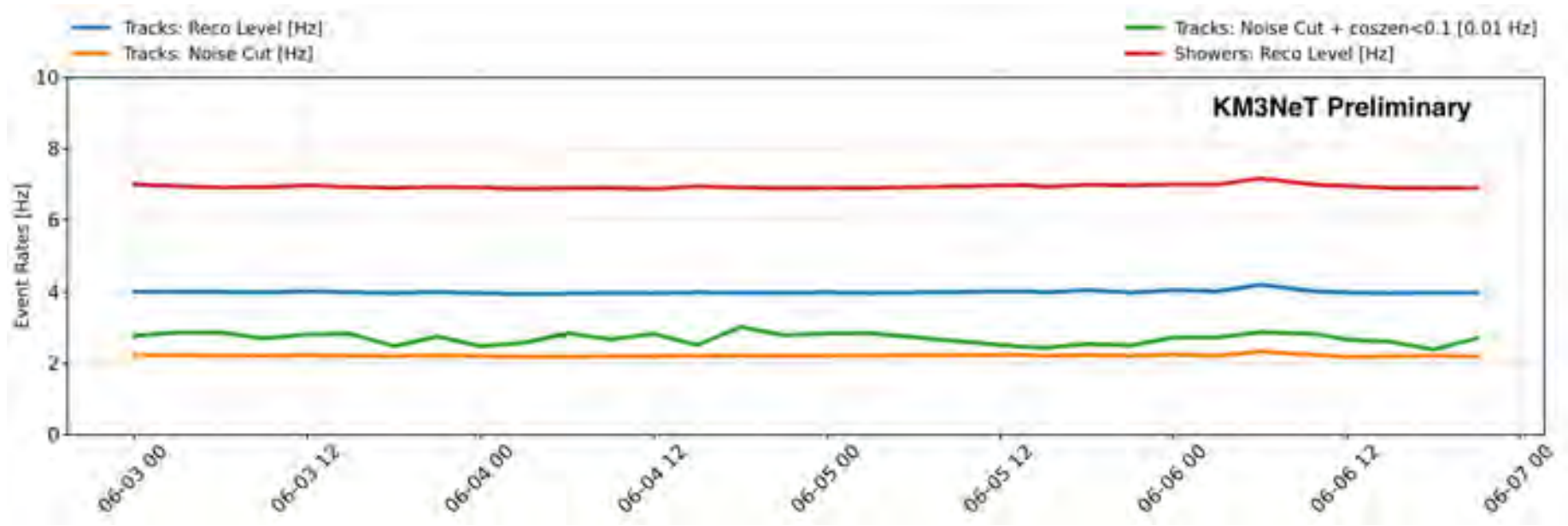

(a)

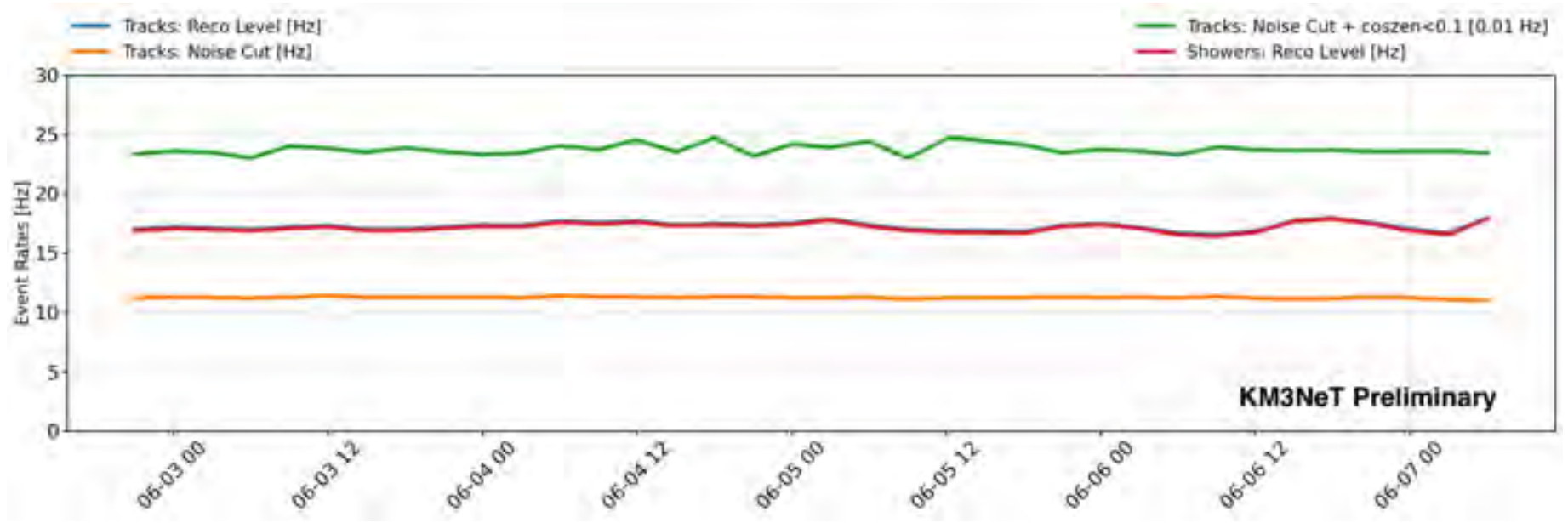

(b)

**Figure 2:** Event rate at (online) reconstruction level in ARCA21 (a) and ORCA18 (b) as a function of time, from June 03rd to June 07th 2023.

### 3.2 The MeV neutrino CCSN module

The main goal of the MeV CCSN pipeline is to provide early warning for optical telescopes for the observation of the next Galactic CCSN, as SN neutrinos arrive hours before the electromagnetic signals [6]. The KM3NeT real-time SN analysis takes as input the raw PMT data, such that each DOM acts as a standalone detector and coincidences among its PMT hits are investigated [7, 8]. The number of PMTs hit in a coincidence is defined as the *multiplicity*: the KM3NeT CCSN pipeline evaluates the multiplicity every 0.1 s in a 0.5 s sliding window, where the size of the window corresponds to the typical duration of the accretion phase of the $\bar{\nu}_e$ burst. Radioactive decays dominate at low multiplicities, while the contribution of atmospheric muons dominates above a multiplicity of 8. The latter background can be reduced by exploiting the fact that muon tracks typically produce correlated coincidences on multiple DOMs: as shown in Fig. 4(a), such a filtering permits ORCA to achieve lower background values than ARCA, thanks to the fact that the denser array can more efficiently identify low-energy atmospheric events. The same figure shows the expected event rates from simulated CCSN bursts, which increase with the mass of the progenitor's star. The simulations indicate that the KM3NeT sensitivity is maximal at high multiplicities and that more than 95% of the Galactic CCSNe are potentially observable, in the scenario with progenitor's mass of 11 $M_\odot$ and above. The KM3NeT MeV neutrino CCSN module is operational, and it is sending alerts with false alarm rate less than 1/week to SNEWS [9] with a latency shorter than 20 s.

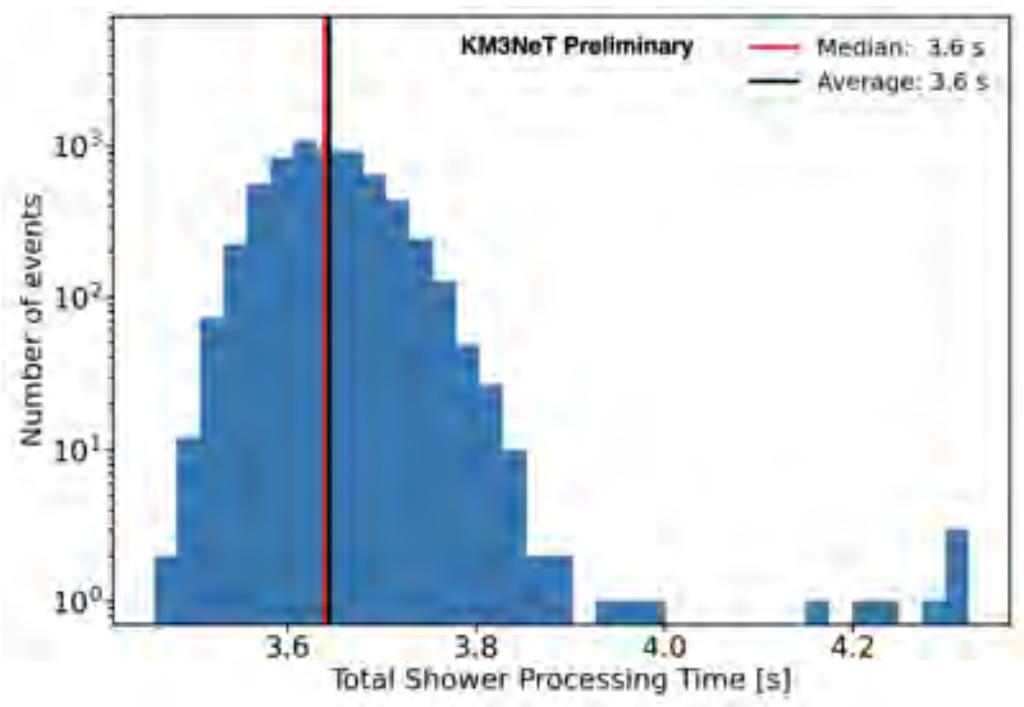

(a)

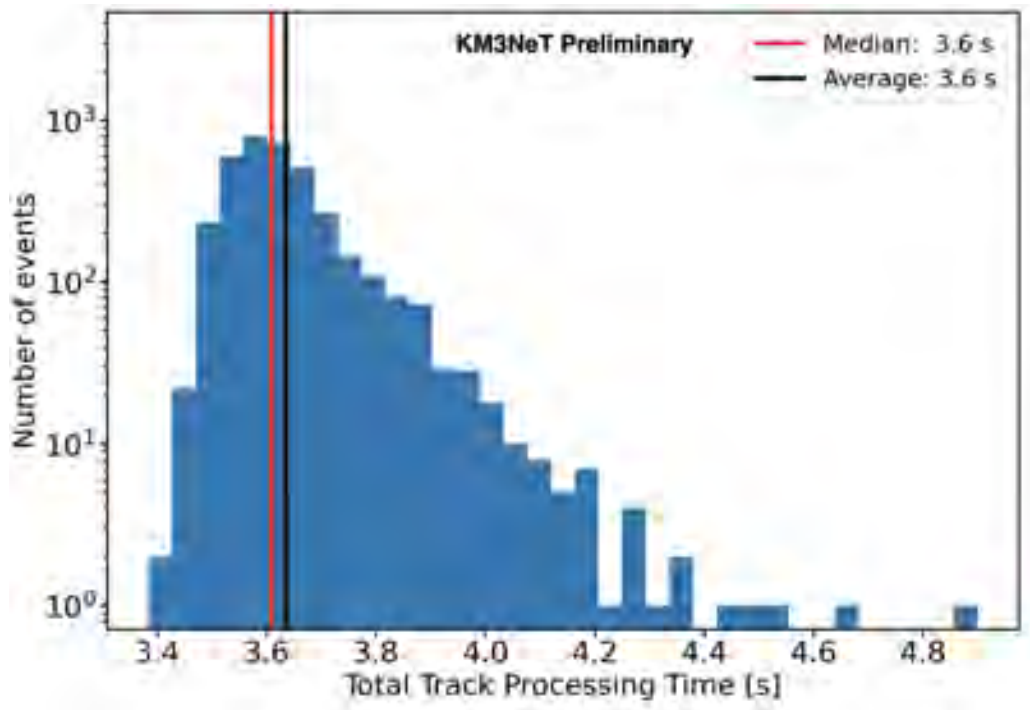

(b)

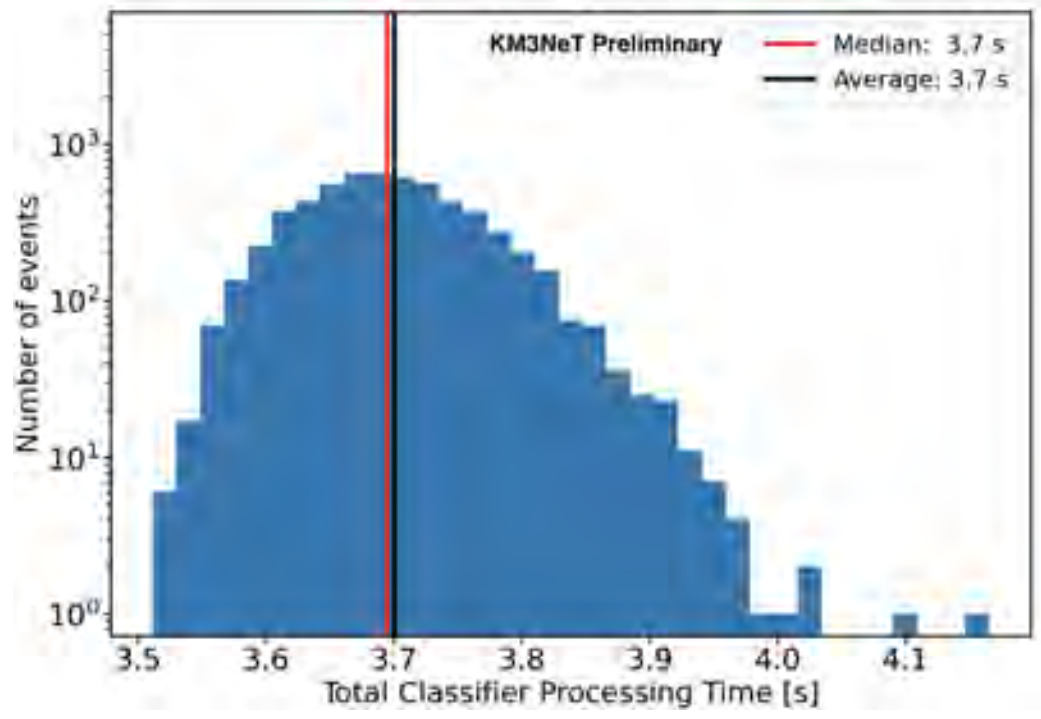

(c)

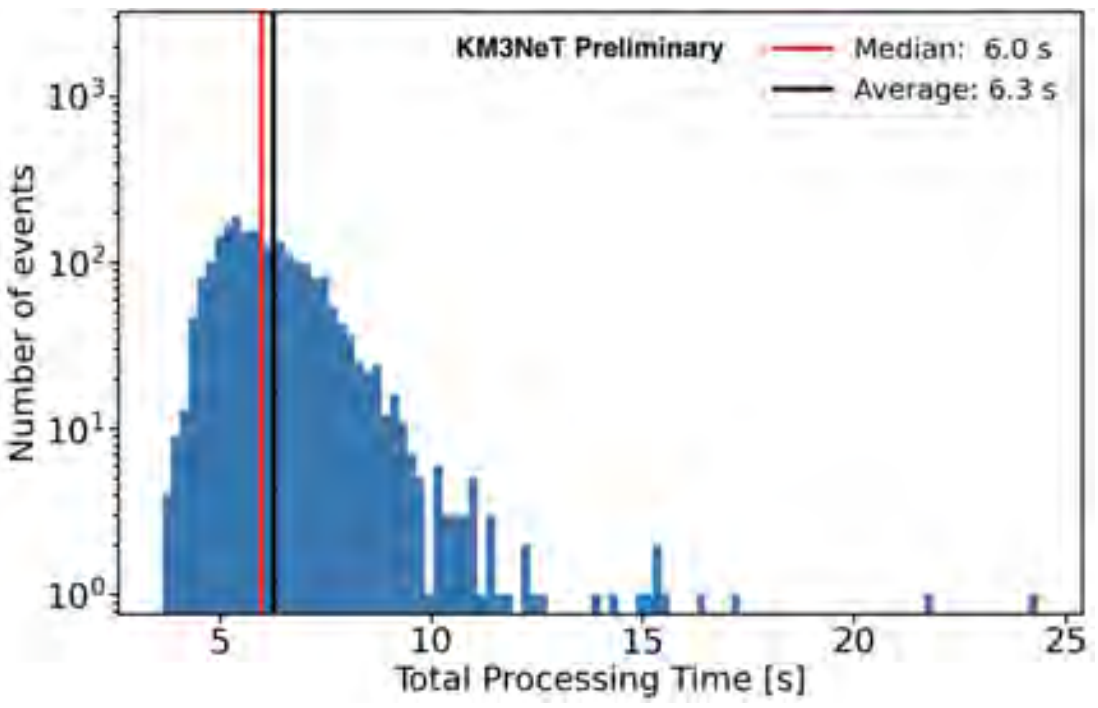

(d)

**Figure 3:** ARCA21 real-time processing times for event acquisition, distribution, reconstruction into track-like (a) and shower-like (b) topologies, and classification (c). (d) ORCA18 real-time processing times for event acquisition, distribution, reconstruction into both track-like and shower-like topologies, and classification.

### 3.3 Analysis pipelines for multi-messenger alert follow-up

At the reception of an external alert, analysis pipelines are triggered simultaneously both for ARCA and ORCA accordingly to the type of alert, namely a parser reads the incoming message and extracts relevant information to run the corresponding analysis pipeline [10]. A preliminary alert selection is performed at this stage, based on source visibility, quoted size of error box, time delay between event trigger and alert reception, and false alarm rate of the notice.
Currently, four kinds of real-time analyses are in place to look for temporal and spatial coincidences among the KM3NeT reconstructed events and either: i) GRBs, ii) GW extended region, iii) neutrinos identified by IceCube, and iv) transient events (e.g. flaring/variable objects). In addition to these, a pipeline for MeV neutrinos in coincidence with GWs is also in place, that similarly to the SN analysis counts the number of coincidences within single DOMs in a sliding window of 0.5 s.
Each analysis pipeline implements its own search strategy (upgoing/downgoing), gets the recon-structed events, applies the predefined event selection for track and shower events, performs the correlation search and produces the results, merging both event channels. Different iterations of the pipeline are performed with progressively extended time windows, possibly including the most refined coordinates of the alert. For the MeV neutrino analysis, the time window starts at the time of the alert $t_{\rm alert}$ with a width of 2 seconds afterwards. For the high-energy event search in coincidence with GWs, two iterations are run, the first from $t_{\rm alert} - 500$ s to $t_{\rm alert} + 500$ s, and the second up to 6 hours after the alert. Also for neutrinos two iterations are performed, but in different time windows, respectively $[t_{\rm alert} - 1\ {\rm h}; t_{\rm alert} + 1\ {\rm h}]$ and $[t_{\rm alert} - 24\ {\rm h}; t_{\rm alert} + 24\ {\rm h}]$. For GRBs and all other transients, in turn, four iterations are run, the first in the time window $[t_{\rm alert} - 24\ {\rm h}; t_{\rm alert}]$, and the following up to 3 hr, 6 hr and 24 hr. The expected background is estimated using a few days before the alert.

#### 3.3.1 The online follow-up of GRB221009A

On 2022 October 9th, the brightest GRB ever recorded triggered the gamma-ray instruments onboard of the Swift and Fermi satellites. At first, Swift-BAT reported a transient event at 14:10:17 UT, first cataloged as Swift J1913.1+1946 [11], at RA = 288.263°, DEC = +19.803°. Fermi-GBM triggered an event at 13:16:59 UT at a location consistent with Swift-BAT [12]. Later, LHAASO reported the observation of GRB221009A [13] with energy above 500 GeV by the LHAASO-WCDA (within 2000 seconds after T0) and with energy up to ∼ 10 TeV with LHAASO-KM2A, the highest energy radiation ever detected from GRBs. IceCube soon performed a search for track-like muon neutrino events, at first in a time window of -1 hour/+2 hours from the initial trigger reported by Fermi-GBM, and later in a time window of 2 days [14]. In both cases, data were consistent with background only expectations, which allowed for setting upper limits on the time-integrated muon-neutrino flux set respectively at $E^2 dN/dE = 3.9 \times 10^{-2}$ GeV cm$^{-2}$ for the first search and $4.1 \times 10^{-2}$ GeV cm$^{-2}$ at 90% CL (assuming an E$^{-2}$ power law for the neutrino spectrum).
A quick follow-up was also performed using KM3NeT online data: unfortunately, at the time of the alert, the source was above the KM3NeT horizon, as shown by the sky map in Fig. 4(b). The search for MeV neutrinos yielded a result compatible with background expectations. In addition, two high-energy analyses were run, based on an on/off technique: for ARCA, a region of interest radius of 2° was used; for ORCA, an angular search radius of 4° was adopted instead, the larger size being related to the higher kinematics angle between the parent neutrino and the emerging muon at low energies. No events were found in the on region of any of the online analyses, whose results have been published in [15]; a refined follow up analysis was also performed, confirming the lack of associated neutrinos in KM3NeT data, as reported in [16].

## 4. Conclusions

The construction and data taking of the KM3NeT infrastructure is ongoing at two designated sites, where the ARCA and ORCA telescopes are located. Each detector currently implements a real-time alert system, capable of fast reconstruction and classification of neutrino-induced events as well as of prompt correlation analyses in response to MM external alerts. A specific monitoring system ensures the stability and sanity of all processes involved in the RTA framework, supported by dedicated shift crews. At the time of writing, more than 300 alerts have been selected and processed, of which 191 with GRBs, 119 in coincidence with GW candidates, 50 with neutrinos, and 13 with other transients. A summary of these activities can be found at [10].

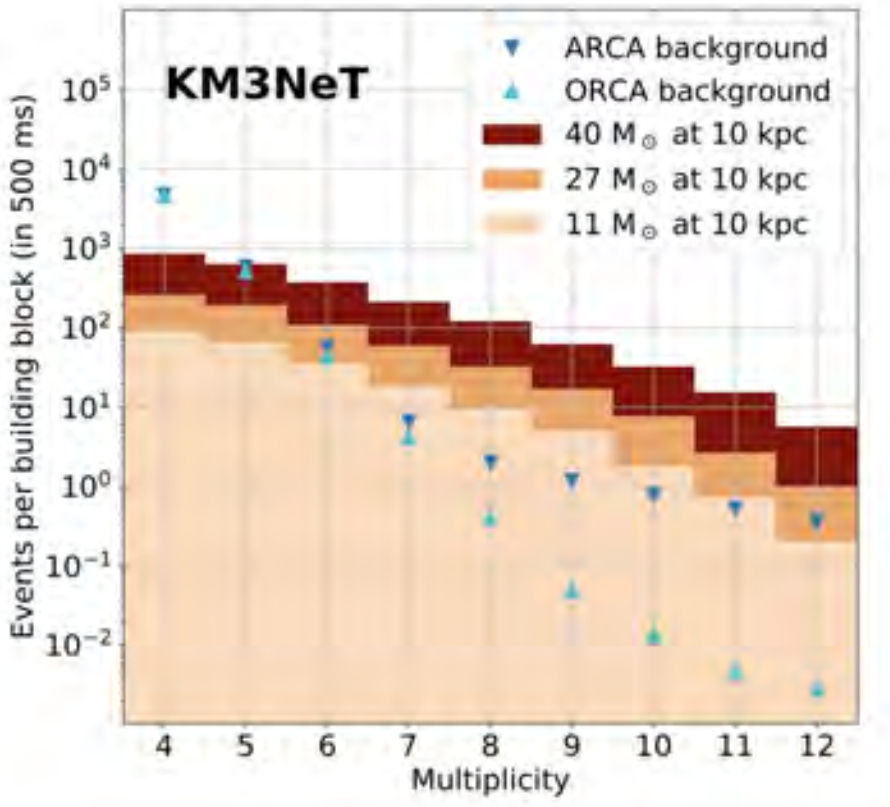

(a)

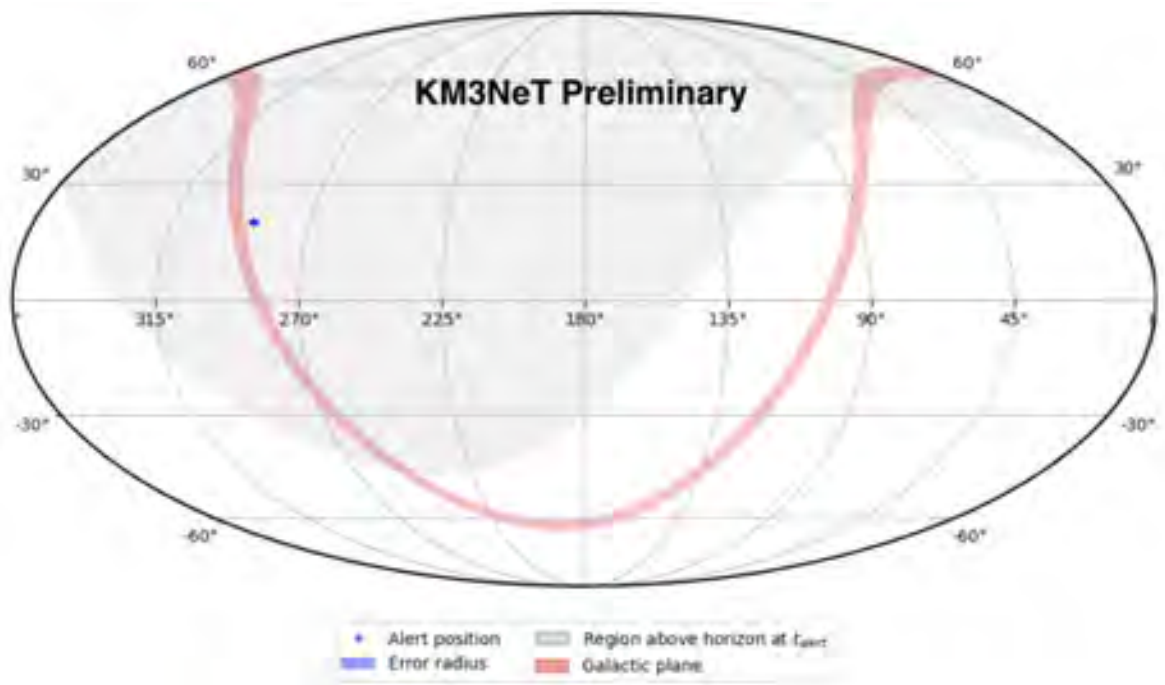

(b)

**Figure 4:** (a) Expected events in ARCA and ORCA (one BB) from a Galactic SN with different progenitor's mass as a function of multiplicity, compared with background rates. Figure from [6]. (b) Equatorial coordinate sky map showing the position of GRB221009A at its occurrence and the KM3NeT visibility.

# Combined KM3NeT-ARCA and ANTARES searches for point-like neutrino emission

**Barbara Caiffi,[a] Vladimir Kulikovskiy,[a] Vittorio Parisi,[b,a] Matteo Sanguineti,[b,a] Sandra Zavatarelli,[a] Sergio Alves Garre,[c] Damien Dornic,[d,*] Thijs van Eeden,[e] Aart Heijboer,[e] Giulia Illuminati[f] and Rasa Muller[e] on behalf of the KM3NeT and ANTARES collaborations**

[a] *INFN, Sezione di Genova, Via Dodecaneso 33, Genova, 16146 Italy*
[b] *Università di Genova, Via Dodecaneso 33, Genova, 16146 Italy*
[c] *IFIC - Instituto de Física Corpuscular (CSIC - Universitat de València), c/Catedrático José Beltrán, 2, 46980 Paterna, Valencia, Spain*
[d] *Aix Marseille Univ, CNRS/IN2P3, CPPM, 163, avenue de Luminy, Marseille, France*
[e] *Nikhef, National Institute for Subatomic Physics, PO Box 41882, Amsterdam, 1009 DB Netherlands*
[f] *INFN, Sezione di Bologna, v.le C. Berti-Pichat, 6/2, Bologna, 40127 Italy*

*E-mail:* km3net-astro-convenor@km3net.de

Neutrino telescopes are the leading-edge instrument for the detection of high energy cosmic neutrinos. The ANTARES detector operated offshore Toulon (France) for 16 years until 2022, while KM3NeT-ARCA is in construction in Southern Italy. The ANTARES telescope was composed of 12 strings of optical modules. Each optical module contained one 10" photomultiplier tube to detect the faint light produced by neutrinos interacting in the surrounding water. Similarly, the KM3NeT-ARCA detector will count 230 strings of 18 optical modules containing 31 3" photomultipliers. In recent years, there has been a growing interest in studying potential sources of neutrinos, as these sources can provide valuable information about the most extreme phenomena in the Universe. This contribution will showcase the analyses of the combined data sample of ANTARES and the first two years of KM3NeT-ARCA to detect high energy cosmic neutrinos from point-like sources.

*Speaker

## 1. Introduction

In the year 2013, a groundbreaking discovery was made by the IceCube Neutrino Observatory, discovering the presence of a flux of astrophysical neutrinos at exceptionally high energy levels [1], followed by the measurement of cosmic neutrino from blazar TXS 0506 +056 in 2018 [2] and from Messier 77 in 2022 [3]. Despite these achievements, the origins of the majority of these neutrinos remain unknown. In the following years a significant contribution in the investigation of these sources will be played by the KM3NeT/ARCA detector [4], which is currently under construction and the data taking is already in progress.

The detector is located in the Mediterranean Sea offshore Sicily at a depth of 3500 m . It will comprise two building blocks, each one equipped with an array of 115 strings and housing 18 digital optical modules per string with 31 3-inch photomultiplier tubes per module. This apparatus is mainly devoted to neutrino astrophysics thanks to its excellent sub-0.2° pointing resolution for muon neutrinos with energies above 10 TeV. The KM3NeT/ARCA detector has remarkable sensitivity across a vast energy spectrum, spanning from hundreds of GeV to PeV, and possesses a unique vantage point that complements the coverage of the IceCube detector, including a very good visibility of the galactic center.

The ANTARES detector [5], the predecessor of KM3NeT, operated from May 2008 until February 2022 in the depths of the Mediterranean Sea offshore Toulon ( France). The telescope consisted of 12 detection strings, each one with 25 storeys. Every storey counted a triad of 10-inch photomultiplier tubes housed within pressure-resistant glass spheres.

The neutrino detection mechanism of the above mentioned detectors is identical. The neutrino, interacting inside or in the surroundings of the detectors, produce charged particles. These particles, travelling at velocities surpassing that of light in the sea-water or ice medium, induce the emission of Cherenkov light. The detection of this radiation by the PMTs allows the reconstruction of the particle direction and energy using the temporal information and the spatial coordinates of these signals. In this way, this comprehensive reconstruction enables the determination of the region of the sky from where the neutrino has been produced.

In the pursuit of identifying cosmic neutrino signals among the background of atmospheric muons and neutrinos, efforts are being applied to develop statistical methodologies using Monte Carlo pseudo experiment approach. It is important to note that the methods presented in this contribution have been developed in the software framework of the KM3NeT collaboration that has been already presented in [6]. The main steps of the analysis frameworks are briefly summarized here below.

Once suitable criteria are established to enhance the signal-to-background ratio, the response functions of the detectors are derived. These distributions, including effective area, energy resolution, and angular resolution, are then transformed into probability density functions, serving as fundamental inputs for the binned likelihood analysis. Through this analytical approach, the sensitivity of the detectors to different predefined source locations can be evaluated. By checking the data with a background model one can estimate the significance of the observation and the upper flux limits in case significance is low. This is done for a set of locations in a source catalog comprising known gamma-ray and other multi-messenger emitters.

This contribution describes the effort to develop the analysis framework to combine the KM3NeT and ANTARES data sample.

## 2. Data sample

In order to harmoniously merge the diverse detector data in an already developed anlaysis framework the *data set* is introduced in a slightly different way. In particular, we consider that the data set refers to a particular detector period and a dedicated event selection. No overlaps between events are allowed.

The different data sets used in this analysis are

1. ANTARES tracks: This data set pertains to the ANTARES point source analysis, track events selection, incorporating data from the period spanning 2007 to 2020 [7];
2. ARCA6 tracks: KM3NeT/ARCA period with 6 working lines (around 100 days) [6];
3. ARCA8 tracks: KM3NeT/ARCA period with 8 working lines (around 200 days) [6].

Each data set is accompanied with its instrument response functions (IRF) needed for the signal simulation and the backgrounds simulated from the data itself as it will be described in the following.

Both ANTARES and KM3NeT/ARCA detectors provides also selection of the events reconstructed as cascades. The developed update for the KM3NeT framework can deal with these non-overlapping additional data sets, however, in this contribution we will not show the results on this yet.

## 3. Background and signal model

In this analysis the determination of the background rate, as a function of the reconstructed energy and declination, has been estimated with a data-driven approach; we hence do not rely on atmospheric neutrinos and muon simulations for this. In view of the limited statistical data available, the employed model applies a factorization technique to account for the dependence on both declination and energy. This factorization is expressed as follows:

$$N_{bg} = n \cdot F(\delta) \cdot F(E)\,\mathrm{sr}^{-1}, \tag{1}$$

Here, the normalization factor $n$ is chosen to ensure that the integration over the celestial sphere and energy range precisely yields the total number of events recorded in the data, for each respective data set. To justify the sampling of the background from the aforementioned function, previous ANTARES and ARCA6+8 analyses have confirmed the flatness of the data sets with respect to right ascension.

It is important to note that when constructing background histograms for parameters such as the distance from the source center and energy, solely the declination of the source center is considered. Although this approximation may introduce some degradation for bins situated farther away from the source center, it remains a practical approach for the purposes of analysis.

The model employed to represent the detector response for the signal includes several key components:

1. the effective area, dependent on true neutrino energy and zenith/declination. It serves as a measure of the detectors sensitivity to neutrinos at different energies and incoming angles;
2. the neutrino energy resolution estimated as the fraction of the events with a given reconstructed neutrino energy for a given bin of true neutrino energy and zenith/declination;
3. the point spread function estimated as the fraction of the events with a reconstructed angular distance from the true source center for a given bin of true neutrino energy.

The point spread function was summed over zenith angles in simulation in order to increase the statistics. The point spread function can be further convoluted with the source extension if it is provided in the source catalogue.

## 4. Likelihood formalism

The compatibility of the data with a background or signal hypothesis is quantified by filling the histograms of $\alpha$ (angular distance of the reconstructed event from the source center) and $\log_{10}(E_{\rm rec})$ (event energy estimation) with the events. In this framework each data set has its own range and number of bins. For the ANTARES track events 20 bins for $\alpha$ in the range $[0, 10]$ and 12 bins in $\log E[\rm GeV] \in [0, 12]$ are used, while for both ARCA data sets there are 50 bins in $\alpha \in [0, 5]$ and 14 bins for $\log E[\rm GeV] \in [0, 8]$.

For each data set the following histograms are created: the one with a number of observed/simulated events $N$, the one with an estimate of the number of signal events, $S$, expected for a reference flux $\Phi_0$ and the one with a number of background events, $B$. The log-likelihood is the Poisson probability of the bin-contents:

$$\log L = \sum_{\rm bins} \log\left[\exp(-B_i - \mu S_i)(B_i + \mu S_i)^{N_i}/N_i!\right], \tag{2}$$

where $\mu$ is the "signal strength", which is the scaling factor of the provided reference flux $\Phi_0$. Summing over bins of each data set and then summing $\log L$ is equivalent to summing over generic bin number $i$ through each data set histogram bins. Omitting all terms that do not depend on $\mu$:

$$\log L = \sum_{\rm bins} N_i \log(B_i + \mu S_i) - (B_i + \mu S_i). \tag{3}$$

Note, that for empty expectation, i.e. $B_i + \mu S_i = 0$, $B_i + \mu S_i = 1e - 8$ is explicitly assumed. For a given data in $N$ the best signal strength,$\hat{\mu}$, is determined by maximizing $\log L$.

The logarithm of the likelihood ratio is used as a test-statistic to quantify the compatibility of the data with the signal/background hypotheses and it is described as follows:

$$\lambda = \log L(\mu = \hat{\mu}) - \log L(\mu = 0) \tag{4}$$

For a true value of the signal strength $\mu_{\rm true}$, pseudo-experiments (PE) can be generated by randomly drawing each $N_i$ from a Poisson distribution with mean $B_i + \mu_{\rm true} S_i$. Each PE then undergoes the treatment described above (a maximum likelihood fit yielding $\hat{\mu}$ which is then used to compute the test statistic $\lambda$ for that PE. In this way, distributions of $\lambda(\mu_{\text{true}})$ are obtained. In the current update of the KM3NeT framework the $\lambda$ distributions for pseudo-experiments are stored as vectors of $\lambda$ values instead of histograms which improves precision on the median values calculation.

The $\lambda(\mu_{\text{true}})$ distributions are used to extract Neyman upper limits. For a given data, $N$, one has $\lambda_{obs}$ that is applied as threshold to calculate the confidence level, which is a fraction of pseudo-experiments with a given $\mu_{\text{true}}$ and $\lambda(\mu_{\text{true}}) > \lambda_{\text{obs}}$. The maximum value of $\mu_{\text{true}}$ reaching requested confidence level is the upper limit on $\mu$. The sensitivity is defined as the upper limit on $\mu$ at 90% confidence level, $\mu_{90}$, for the median of the pure background pseudo-experiments, $\lambda_{\text{obs}} = \tilde{\lambda}(\mu_{\text{true}} = 0)$, i.e. median upper limit. This limit is converted to the flux as follows:

$$\Phi_{90} = \mu_{90}\Phi_0. \tag{5}$$

Of course, one can use real data to build the histograms of $N_i$ and use them to fit $\mu$ ($\mu_{\text{obs}}$). The value of $\lambda(\mu_{\text{obs}})$ can be used to calculate the observed upper limit at 90% confidence level. Additionally, $p$-value can be calculated as $P(\lambda \geq \lambda(\mu_{\text{obs}}))$ for distribution of pure background pseudo-experiments. The $p$-value further can be converted to significance in number of $\sigma$ following 1-sided convention.

## 5. Results

The sensitivities for ARCA6 and ARCA8 track data set, the ANTARES 2007–2020 track data set and for their joint analysis are estimated within this framework. The results for the sensitivity on a grid of declinations is reported in Table 1 and plotted in Figure 1.

From the number of signal and background events of both detectors that are also shown in Figure 2 one can see that despite different detector geometry, KM3NeT/ARCA is already running in a similar signal to noise ratio to the ANTARES detector. The event reconstruction quality is expected to improve with bigger detector configuration and with higher statistics more stringent event selection can be applied. This will further improve signal to noise ratio for the KM3NeT/ARCA detector.

As one can see the sensitivity of the joint data analysis is dominated by the ANTARES 2007-2020 data set. The ARCA6 and ARCA8 contribution improves the sensitivity by 0–10%; the fluctuations are mostly dominated by the statistical fluctuations and limited number of pseudo-experiments. With the rapidly growing ARCA detector configurations and more statistics, the joint analysis will exploit data from ARCA and ANTARES detectors at the best during the upcoming years. The KM3NeT collaboration runs ARCA with 19–21 strings since summer 2022. For this contribution no data was unnecessary spoiled and both collaborations will present the standalone analysis updates [8, 9]. In order to perform the real data analysis this analysis will be completed with the cascade events, the full ANTARES data set (2007–2022) and bigger ARCA data sets in the near future.

**Table 1:** Summary of the sensitivity studies with ARCA6, ARCA8 and ANTARES 2007–2020 track data sets. The total number of signal events for the reference one flavour flux $d\Phi_{\nu+\bar{\nu}}/dE = 10^{-4}\,(E[\text{GeV}])^{-2}$ $\text{m}^{-2}\text{s}^{-1}\text{GeV}^{-1}$ in the 10° search cone, the number of background events in the search cone and the median signal strength of the reference flux, $\tilde{\mu}_{90}$.

| decl.[deg] | signal ARCA | signal ANTARES | background ARCA | background ANTARES | sensitivity ARCA | sensitivity ANTARES | sensitivity joint |
|---|---|---|---|---|---|---|---|
| -90 | 0.54 | 4.14 | 45.95 | 165.79 | 7.87 | 0.53 | 0.54 |
| -80 | 0.54 | 4.14 | 45.96 | 165.73 | 7.96 | 0.56 | 0.54 |
| -70 | 0.56 | 4.06 | 46.03 | 165.30 | 7.26 | 0.56 | 0.53 |
| -60 | 0.58 | 4.02 | 46.32 | 163.48 | 7.06 | 0.55 | 0.53 |
| -50 | 0.61 | 4.01 | 46.67 | 158.41 | 6.18 | 0.55 | 0.50 |
| -40 | 0.43 | 3.56 | 39.88 | 141.31 | 9.32 | 0.57 | 0.55 |
| -30 | 0.37 | 2.95 | 25.91 | 113.25 | 9.63 | 0.69 | 0.66 |
| -20 | 0.35 | 2.70 | 22.98 | 100.44 | 9.91 | 0.71 | 0.67 |
| -10 | 0.33 | 2.56 | 21.58 | 89.57 | 9.86 | 0.72 | 0.66 |
| 0 | 0.32 | 2.33 | 23.48 | 76.51 | 9.89 | 0.75 | 0.73 |
| 10 | 0.31 | 2.16 | 25.35 | 62.59 | 10.55 | 0.78 | 0.77 |
| 20 | 0.30 | 1.97 | 24.39 | 50.63 | 10.65 | 0.89 | 0.81 |
| 30 | 0.29 | 1.64 | 26.36 | 41.71 | 10.39 | 1.02 | 0.97 |
| 40 | 0.27 | 1.42 | 28.75 | 28.63 | 11.30 | 1.12 | 1.05 |

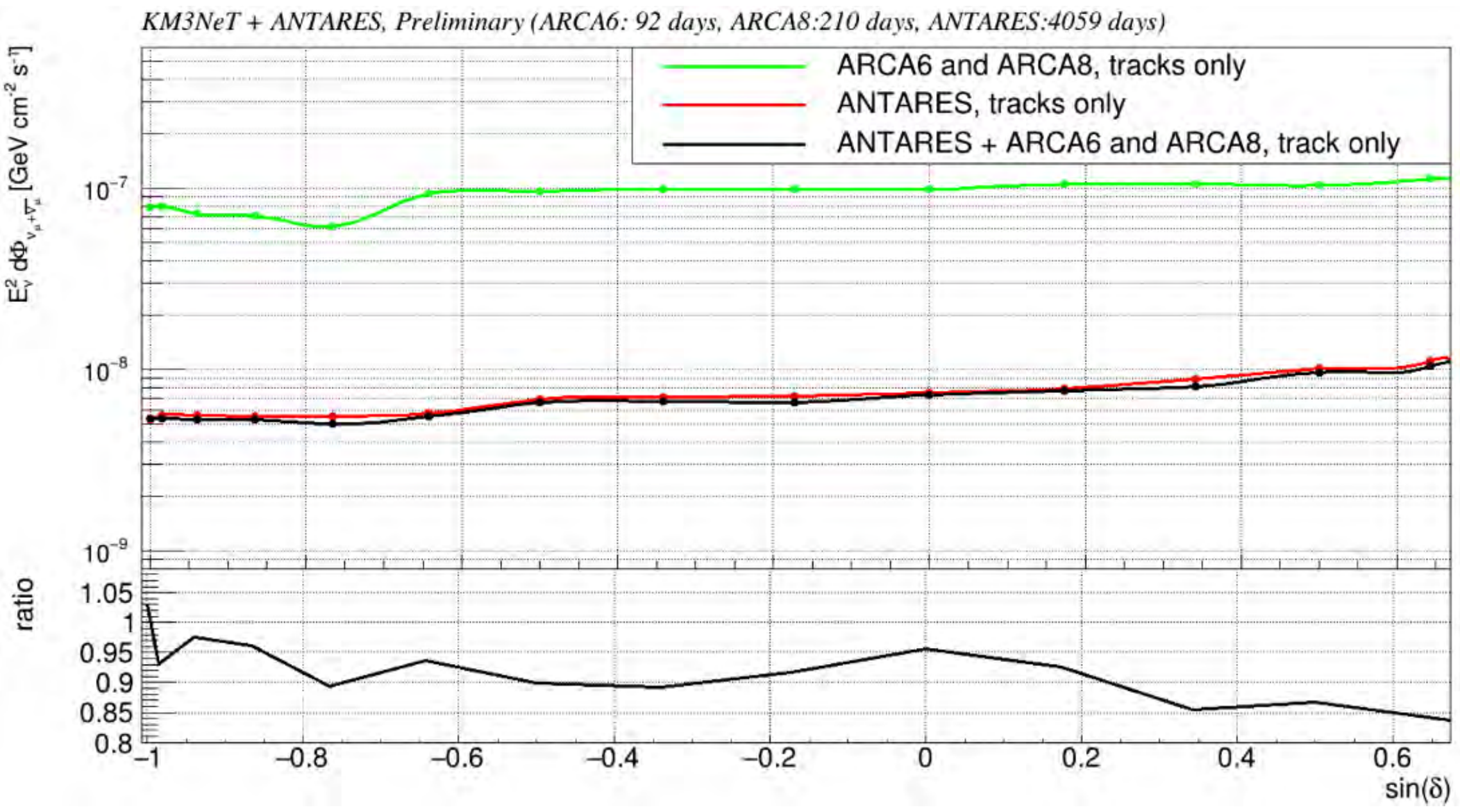

**Figure 1:** Sensitivity as a function of source declination for ARCA6 and ARCA8 track data sets with about 300 days (green), ANTARES track data set for 2007–2022 period (blue) and ARCA6/ARCA8/ANTARES combined analysis (red). The ratio of the latter two sensitivities is shown in the bottom plot.

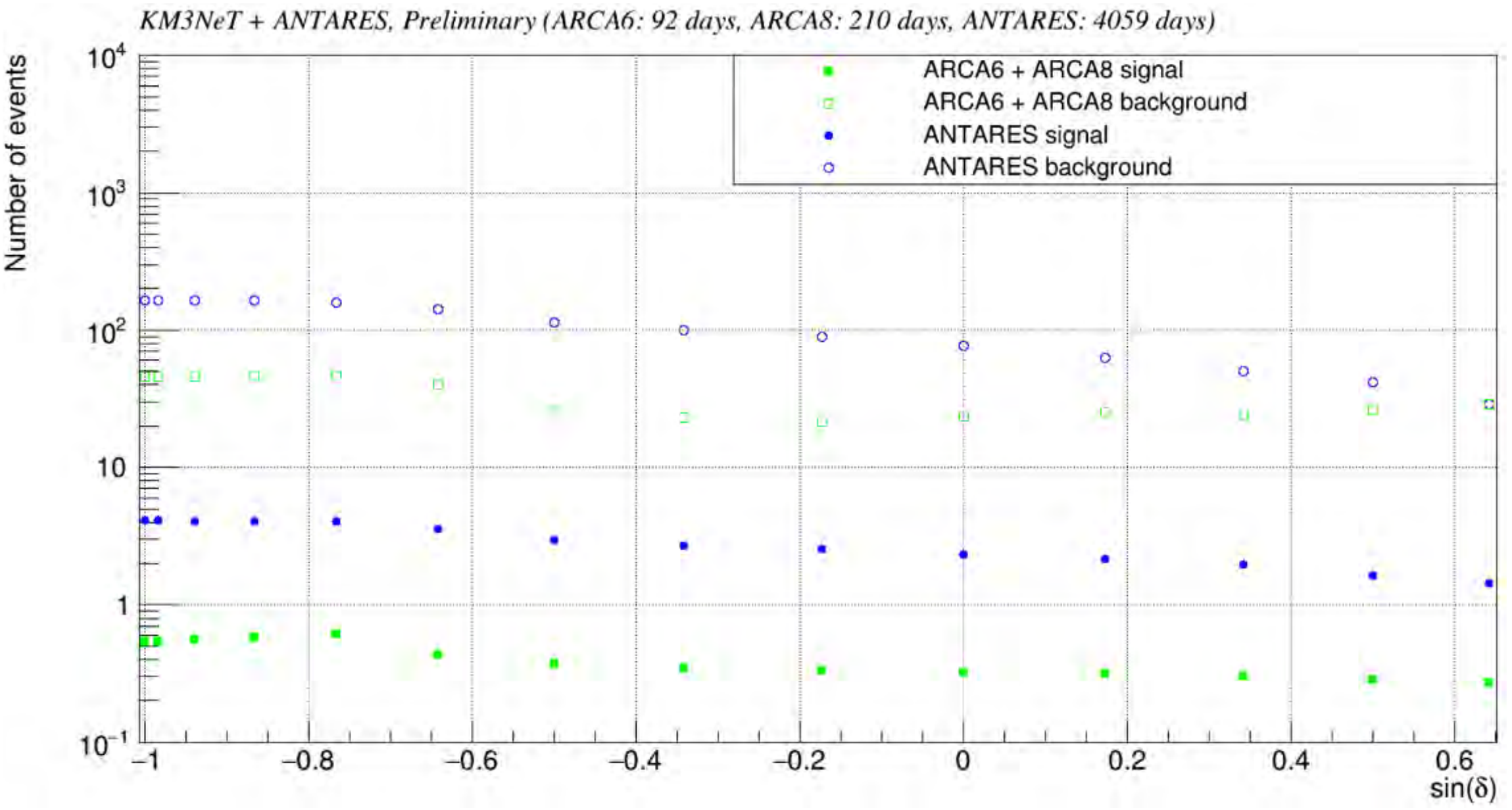

**Figure 2:** Mean number of signal and background events for ARCA6/ARCA8 and ANTARES track data sets.

# Energy-Dependent Expectations for the Full KM3NeT/ARCA Detector: Application to Starburst Galaxies

**W. I. Ibnsalih,**[b,*] **A. Ambrosone,**[a,b] **A. Marinelli,**[a,b,e] **G. Miele,**[a,b,d] **P. Migliozzi**[b] **and M. R. Musone**[b,c] **for the KM3NeT collaboration**

[a] *Dipartimento di Fisica "Ettore Pancini",*
*Università degli studi di Napoli "Federico II", Complesso Univ. Monte S. Angelo, I-80126 Napoli, Italy*
[b] *INFN - Sezione di Napoli,*
*Complesso Univ. Monte S. Angelo, I-80126 Napoli, Italy*
[c] *Università degli studi della Campania Luigi Vanvitelli*
[d] *Scuola Superiore Meridionale,*
*Università degli studi di Napoli "Federico II", Largo San Marcellino 10, 80138 Napoli, Italy*
[e] *INAF-Osservatorio Astronomico di Capodimonte,*
*Salita Moiariello 16, I-80131 Naples, Italy*

*E-mail:* walid.idrissiibnsalih@unicampania.it, aambrosone@km3net.de

The energy-dependent expectations for the full KM3NeT/ARCA detector, both for diffuse and point-like signals are presented. For the diffuse analysis, the 90% C.L. quasi-differential sensitivity for 10 years of the full detector operations is computed, and compared with the IceCube measured diffuse spectrum. The whole energy range observable by ARCA $\sim$ (1 TeV – 100 PeV) is taken into account, selecting upgoing track-like as well as all-sky cascade-like events. For the point-like analysis, the differential sensitivity for several declinations is computed, selecting upgoing tracks $\nu_\mu$ and $\bar{\nu}_\mu$ charge-current interactions. Some particular starburst galaxies (SBGs) are considered such as NGC 1068, the Small Magellanic Cloud (SMC) and Circinus Galaxy. The resulting sensitvities demonstrate the importance of KM3NeT/ARCA to link star-forming processes with high-energy neutrino production within a few years of operation.

*Speaker

## 1. Introduction

The KM3NeT/ARCA is a water Cherenkov neutrino telescope under construction in the Mediterranian sea [1]. It will consist of two building blocks with 115 detector units each, with a total effective volume of $\sim$ 1 Km$^3$. Given its median-latitude position, the detector will provide a complementary view of the sky with respect to IceCube. KM3NeT is expected to have full visibility for the southern hemispher and therefore for the galactic center [2]. In this proceeding, the energy-dependent expected sensitivities of the full KM3NeT/ARCA are reported, both for diffuse and point-like spectra after 10 years of operation. For the diffuse spectrum analysis, a binned likelihood ratio technique is employed in order to compute the 90% C.L. quasi-differential sensitivity, considering both upgoing tracks and all-sky cascades. A dedicated boosted decision trees (BDTs) is used in order to reject the background from atmospheric muons. All the neutrino flavours are taken into account for charged and neutral current interactions (see [2, 3] for details). For the point-like analysis, a cut & count technique is used to calculate the 90% C.L. quasi-differential sensitivity for sources on several declinations, showing the potential of KM3NeT/ARCA to constrain neutrino emission from point-like and extended SBG emission. In order to reach such a purpose, only upgoing $\nu_\mu$ and $\bar{\nu}_\mu$ events are considered. Finally, some particular sources such as Starburst Galaxies are considered. Indeed, the expectations are presented for NGC 1068, whose neutrino emission evidence has recently increased up to 4.2$\sigma$ over the null hypothesis [4]. Also, the Small Magellanic Cloud (SMC) and the Circinus Galaxy are considered, since they are situated in positions of the sky where KM3NeT/ARCA has full visibility and expected neutrino fluxes constrainable by KM3NeT/ARCA [5]. The resulting sensitivity demonstrate that KM3NeT/ARCA, in few years of operation, will be able to constrain the considered scenarios.

## 2. Diffuse Analysis: Simulations and Event-Selection

The analysis is based on the latest simulations of the Full KM3NeT/ARCA detector Monte Carlo files. In particular, neutrino simulations are performed through the gSeaGen code [6] and muons through the MuPage code [7]. The generated events are subject to the simulation of the light generation and the response of the detector and are reconstructed with the same algorithm used for actual data (see also [8] for further details). For the track-like and cascade events, the same selection explained in Ref. [3] are employed. In particular, for the tracks only those with reconstructed zenith angle $\theta \leq 100°$ are selected, in order to use the Earth as a shield to reduce the contamination of downgoing atmospheric muons. Furthermore, a dedicated BDT is employed in order to reduce the contamination of mis-reconstructed muons (muons which are downgoing but which are mis-reconstructed as upgoing). This allows for a selection with a reliable track reconstruction (see [8] for details about the performance of the event selection). For the the full-sky cascade selection, events satisfying the track selection requirement are removed and, in addition, only events contained withn the instrumented volume of the detector are considered(containment). A BDT has been employed also for the final step of the cascade selection (see [3] for details) in order to obtain a sample of events which are reliably reconstructed.

### 2.1 Likelihood Framework and Results

In order to evaluate the sensitivity, a binned maximum likelihood ratio framework is used. The likelihood is defined following Ref. [9] as:

$$\mathcal{L}(\lambda) = \prod_i P\left(n_i; \lambda\mu_i^{\text{Sig}} + \mu_i^{\text{Back}}\right) \tag{1}$$

where $i$ runs over the 140 reconstructed energy bins. $P(n, \mu)$ is the Poisson PDF to observe n events with an expected value $\mu$ [9]. $\lambda$ is a free-parameter which represents the signal strength (or normalization), $\mu_i^{\text{Sig}}$ is the expected number of signal events for a given $i^{th}$ bin (in the case $\lambda = 1$) and finally $\mu_i^{\text{Back}}$ represents the expected number of background events for each bin. $10^4$ pseudo experiments (PEs) are generated for different signal strengths and the test-statistics (TS) is defined as:

$$\text{TS} = \log \frac{\mathcal{L}(\tilde{\lambda})}{\mathcal{L}(\lambda = 0)} \tag{2}$$

where $\tilde{\lambda}$ is the signal strength which maximizes the likelihood for each pseudo experiment. We determine the PDF distribution of the TS for each signal strength. The Model rejection factor (MRF) is defined as the value of the signal strength $\lambda_{90}$ for each:

$$\int_{TS_m}^{+\infty} d(\text{TS}|\lambda_{90})\text{dTS} = 90\% \tag{3}$$

where $TS_m$ is the median distribution in the null hypothesis (only background) and $d(\text{TS}|\lambda_{90})$ is the PDF distribution for the TS for a given $\lambda_{90}$ value. In other words, the sensitivity is defined as the signal strength for which 90% of the signal is above the median of the background-only distribution [9]. The TS is corrected (allowing for negative values of $\tilde{\lambda}$) in order to get always TS $> 0$. The final sensitivity is given by:

$$\phi_{90}(E) = \lambda_{90} \cdot \phi^{\text{Sig}}(E) \tag{4}$$

where $\phi^{\text{Sig}}(E)$ is the signal flux corresponding to $\lambda = 1$ (the final sensitivity does not depend on the normalization of the input signal spectrum, but only on its shape). For this contribution, the signal is considered as a $E^{-2}$ spectrum binned over half-decade true energy bins. Fig. 1 shows the differential sensitivity after 10 years of operation for 2BB both for upgoing tracks and all-sky cascades. The sensitivities refer to one flavour of neutrinos + antineutrinos. They are compared with the corresponding IceCube measured spectra. In particular, the track sensitivity is compared with the 9.5 yr through-going muon flux [10] and the cascade sensitivity is compared with the 6 yr cascade flux [11]. Interestingly, the track sensitivity peaks around 1 PeV where there is a significant contribution from horizontal events. On the other hand, the cascade sensitivity peaks at 100 TeV, since the containment requirement harshly reduces the amount of signal events above this energy. For this reason, at high energies a better sensitivity is obtained for tracks compared to cascades, while at low energies the sensitivity is better for cascades due to the reduced background rate. The Result demonstrates the potential of KM3NeT/ARCA to constrain a diffuse spectrum outside the energy ranges where IceCube has measured it. Indeed, the spectrum is not well known below $\sim 10 - 15$ TeV and above $\sim 1$ PeV. Therefore, it is crucial to understand if it can be extrapolated also for lower energies as well as if it changes its shape. Indeed, as highlighted in Ref. [12], some tension is reported between the gamma-ray and neutrino data have tension due to the electromagnetic cascade contribution.

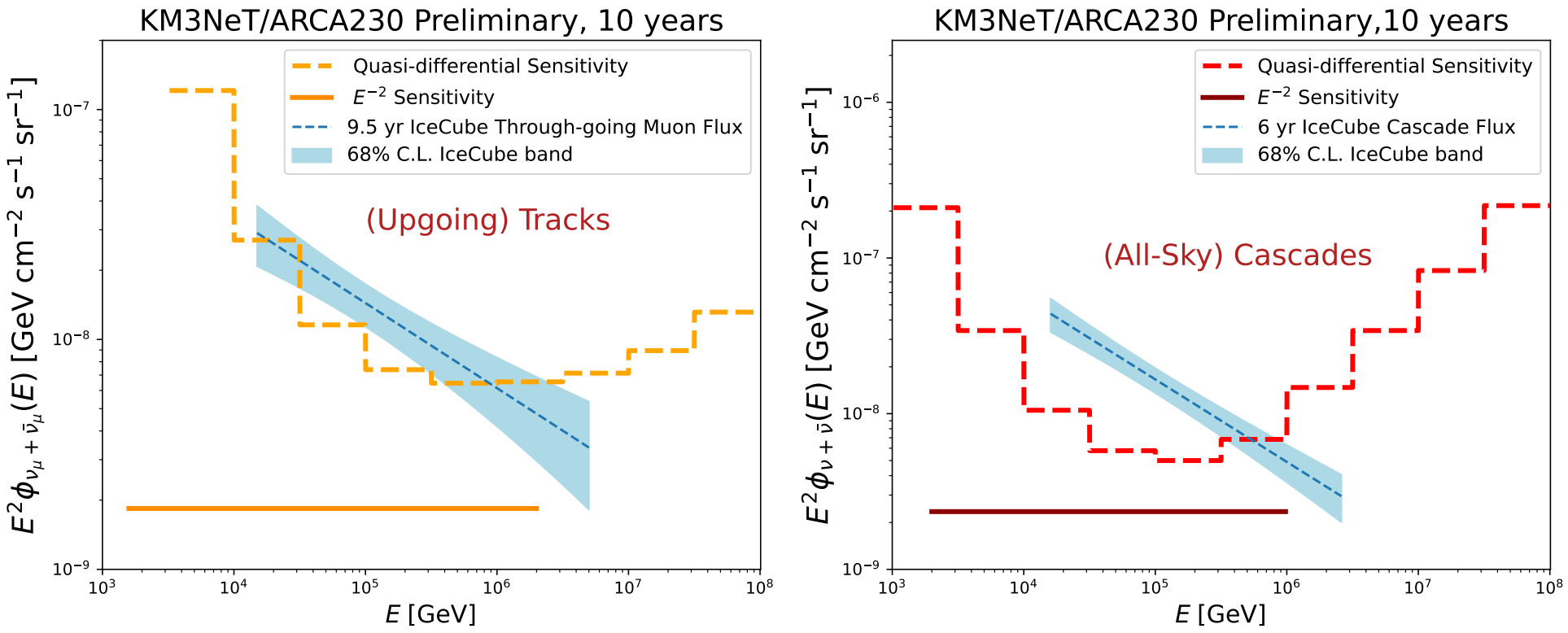

**Figure 1: Left**: 90% C.L. quasi-differential sensitivity (per neutrino flavour) for KM3NeT/ARCA after 10 years of operation and 2BB for a diffuse spectrum for upgoing tracks (dashed orange line). The continuous orange line shows the integrated sensitivity for a $E^{-2}$ diffuse spectrum. The sensitivities are compared with the corresponding 9.5 yr diffuse spectrum measurement by the IceCube Collaboration in the through-going muon channel [10]. **right**: 90% C.L. quasi-differential sensitivity (per neutrino flavour) for KM3NeT/ARCA after 10 years of operation and 2BB for a diffuse spectrum for the all-sky contained cascades (dashed red line). The integrated sensitivity for a $E^{-2}$ diffuse spectrum is shown with the continuous dark red line. The sensitivities are compared with the corresponding 6 yr cascade flux measured by the IceCube collaboration [11].

## 3. Point-Like Analysis

Regarding the Point-Like Analysis, particular SBGs are considered in order to to evaluate the KM3NeT/ARCA expectations: NGC 1068, SMC and Circinus Galaxy. For such a purpose, a simple cut & count technique is used in order to optimize the event selection. The same event selection explained in Ref. [3] is employed. In particular, the sensitivity is defined as the average upper limit evaluated through the 90% C.L. Felmand and Cousins upper limits [13]. Given an astrophysical flux $\phi^{\mathrm{Sig}}$, the sensitivity is

$$\phi_{90} = \frac{\mu_{90}}{n_s} \cdot \phi^{\mathrm{Sig}} \tag{5}$$

where $n_s$ is the expected number of signal events induced by the astrophysical flux and $\mu_{90}$ is the average upper limit. $\mu_{90}/n_s$ is defined as the Model Rejection Factor (MRF). The minimization of the MRF is performed over the cone angle ($\alpha$) and track reconstruction related variables such as the likelikood ($\Lambda$), the angular error ($\beta$) and the track length (Len). A final optimization over a reconstructed energy range $[E_{\mathrm{min}}, E_{\mathrm{max}}]$ is used to further optimize the selection of signal events.

### NGC 1068

NGC 1068 is a nearby SBG located 10 − 14 Mpc [15] away from the Earth and it also shows Seyfert Galaxy activity [16]. The IceCube collaboration has recently found compelling evidence for neutrino emission along the direction of this source measuring 79 astrophysical neutrino events with a significance at $4.2\sigma$ [4]. The sensitivity and the discovery potential for this source is shown in Fig. 2.

Firstly, on the left, the 90% C.L. quasi-differential sensitivity (orange line) is compared with the IceCube measured band [4]. Given the importance of this source, the model discovery potential (MDP) is reported, defined as the minimum flux needed for a $5\sigma$ discovery in the 50% of cases (see [2] for more details). In order to compute it, the event selection is optimized in order to produce the minimum MDP for each bin. The integrated MDP for a $E^{-3.2}$ spectrum (dashed dotted dark red line) is reported. For this particular quantity, it has been calculated both with Cut & Count and binned likelihood ratio (see [14] for details). The integrated MRF and MDP are shown as functions of the KM3NeT/ARCA operation time. The result demonstrates that, after few years, KM3NeT/ARCA is expected to discover the IceCube flux. Indeed, the binned likelihood ratio method provides a strong improvement with respect to the cut & count method. So, further improvements are also expected for the energy-dependent sensitivity. Finally, the result shows that the energy-dependent sensitivity peaks at $\sim$ 100 TeV for this particular position in the sky.

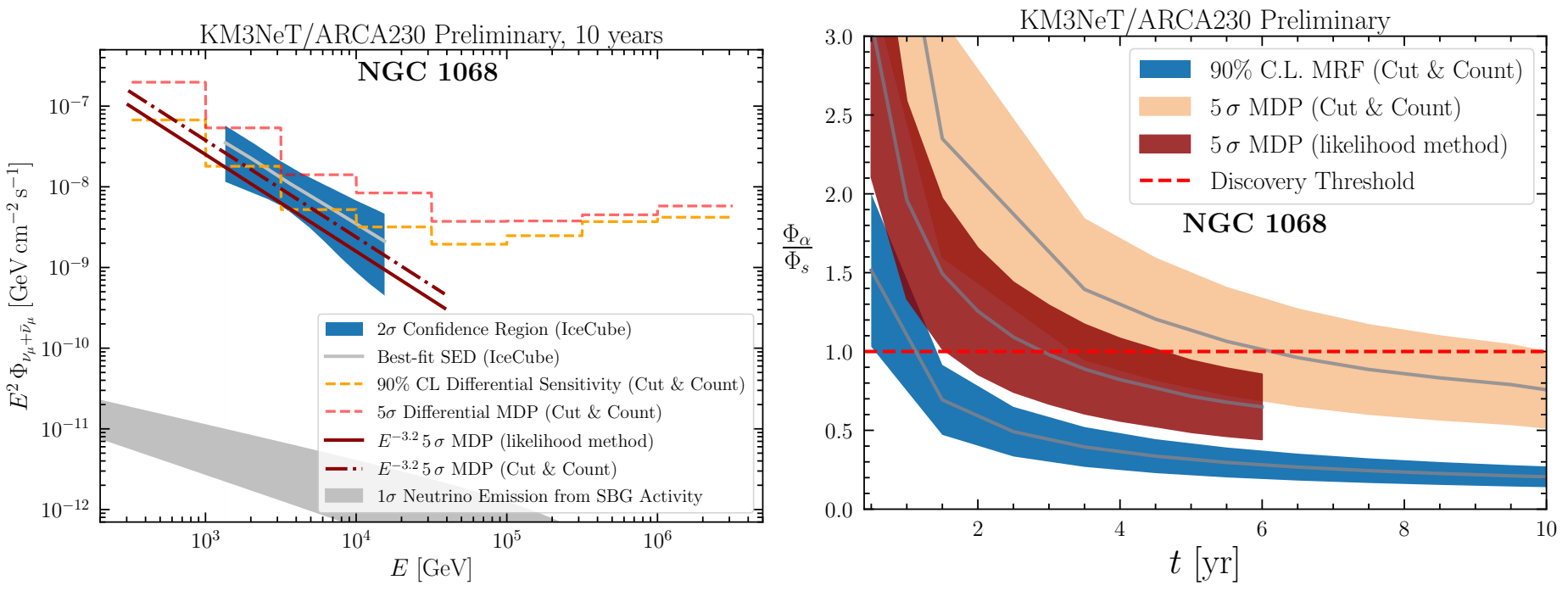

**Figure 2: Left**: 90% C.L. quasi-differential sensitivity (dashed orange line), the 5 $\sigma$ differential MDP (dashed red line) and finally the integrated 5 $\sigma$ MDP for a $E^{-3.2}$ spectrum, obtained both with the Cut & count approach (dashed dotted dark red line) and with the Binned Likelihood Ratio (continous dark red line). It is also shown the the IceCube $2\sigma$ region [4]. Furthermore, the neutrino expectations from SBG activity are shown according to the analysis of Ref. [5]. **Right**: The integrated MRF (blue band) and the MDP (orange band for cut & count and dark red band for the Binned likelihood ratio) are shown as functions of the KM3NeT/ARCA operation time. The bands correspond to the $1\sigma$ uncertainty provided by the normalization uncertainty provided by the IceCube fit ([4]).

## The Small Magellanic Cloud (SMC)

SMC is a very nearby star-forming galaxy which exhibits a hard gamma-ray spectrum up to 1 TeV, which could be attributed to its star forming activity [16]. The importance for studying this source with KM3NeT/ARCA and the potential to significantly constrain the theoretical models has been pointed out in Ref. [5]. In this contribution, in order to evaluate the sensitivity, the source is simulated as a disk of radius $r = 0.5°$ (therefore a total extension of $\sim$ 1°) which is consistent also with the Fermi-LAT observations [16]. The expected sensitivity is shown in Fig. 3 (for 2 BB after 20 years of operation) compared with theoretical neutrino expectations according to [5]. The integrated MRF is shown as a function of time. In this case, it will take a couple of decades in order to produce an upper limit. It is worth mentioning that if the neutrino emission is concentrated in a smaller region of the galaxy effectively, the expectations will improve. Even in case no neutrino excess will be found, it will be important to constrain the hadronic budget of this source at 1−10 TeV.

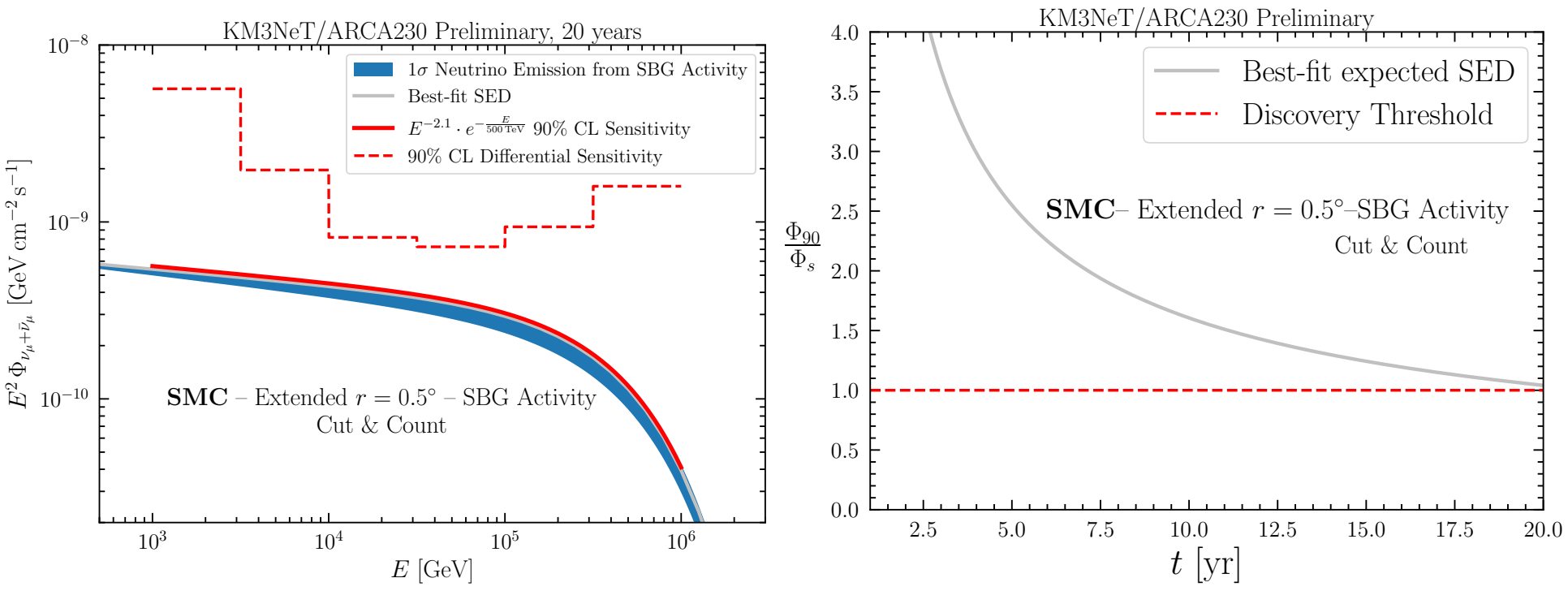

**Figure 3: Left**: Theoretical neutrino expectations according to Ref. [5] with the corresponding integrated and differential sensitivity after 20 years of operation with 2BB. **Right**: The integrated MRF is shown as a function of the KM3NeT/ARCA operation time (for 2BB).

Furthemore, further analysis using a likelihood framework is also expected to significantly improve the expectations.

## Circinus Galaxy

Circinus Galaxy is a nearby SBG situated 4 Mpc away from the Earth [15] in a region of the sky where KM3NeT/ARCA is expected to have full visibility [3]. It has also AGN actvity just like NGC 1068 with Seyfert Galaxy activity. Therefore, it might be the perfect candidate not only to study SBG emissions, but also to understand if other seyfert galaxies have similar emission as NGC 1068. It is also noteworthy to remark that this particular source is nearer than NGC 1068, thereby providing better prospects for future discovery for this source. The resulting sensitivity is shown in Fig. 4. Analogously to SMC, the neutrino expectations from SBG activity ([5]) are compared with the corresponding integrated and differential sensitivities. The expectations are less promising since this source has a lower flux compared to SMC. However, a potential AGN component might significantly increase the normalization of the flux, leading to much better expectations.

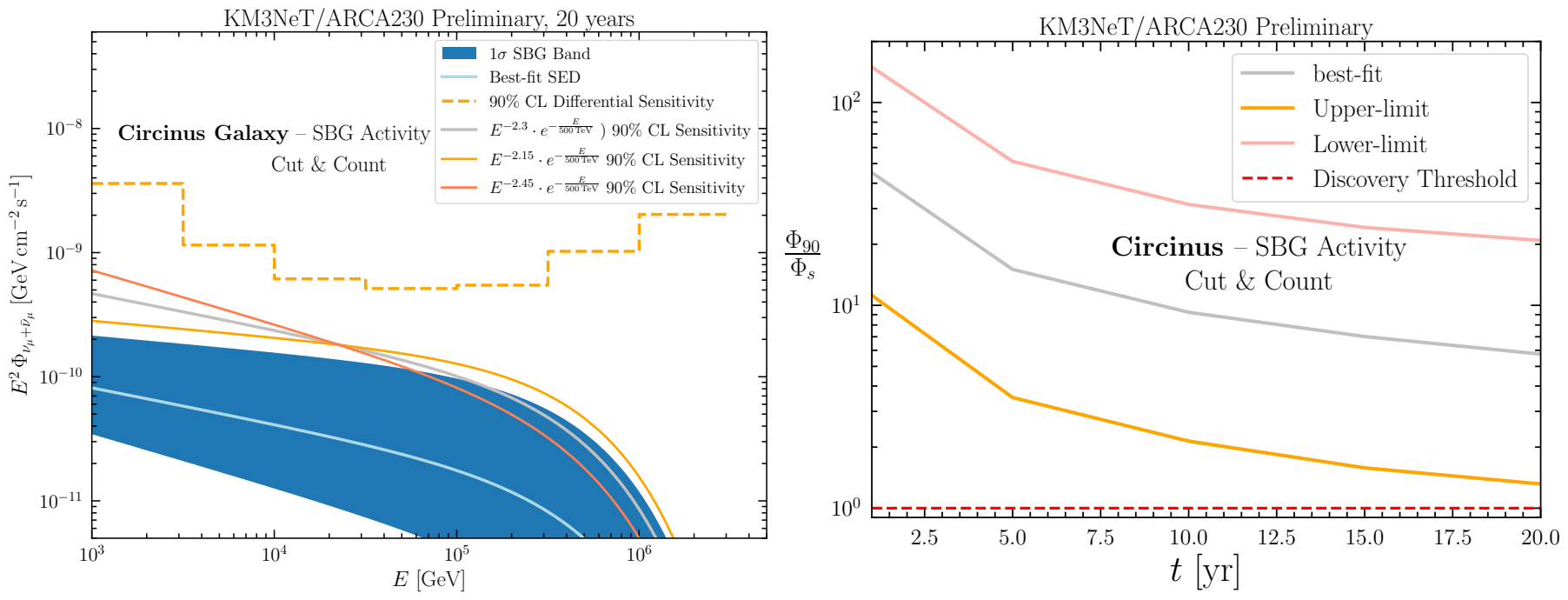

**Figure 4: Left**: Theoretical neutrino expectations from the SBG activity according to Ref. [5] compared with the corresponding integrated and differential sensitivity after 20 years and 2BB. **Right**: The integrated MRF is shown as a function of the KM3NeT/ARCA operation time (for 2BB).

### 3.1 Sensitivities for Different Positions in the Sky

The differential sensitivity after 10 years and 2BB for different declinations is shown in Fig. 5. The result highlights how the KM3NeT/ARCA expectations vary along different position in the sky.

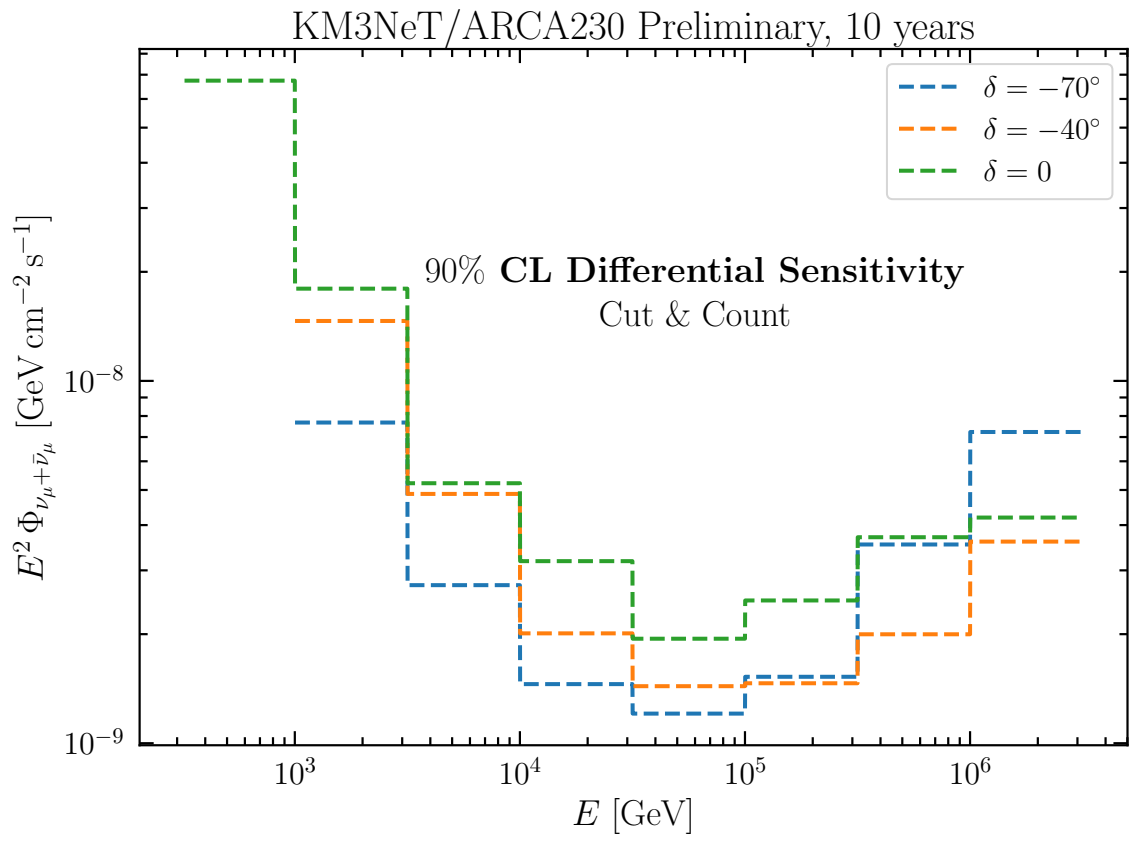

**Figure 5:** 90% C.L. Quasi-differential sensitivity for point-like emission for several positions in the sky. The cut & count method is used and the sensitivity refers to one flavour of neutrinos and antineutrinos.

## 4. Conclusions

The potential of the full KM3NeT/ARCA to discover diffuse and point-like spectra has been presented in this contribution. The result highlights that the upcoming detector will be able to strengthen IceCube's observations, confirming measured spectral features in few years of operation. KM3NeT/ARCA will also be able to detect neutrinos from nearby starburst galaxies. It is worth emphasizing that the expectations are going to improve with future event selections able to better reject the background and with more sophisticated statistical treatment of the data. For instance, a maximum likelihood ratio framework might improve the point-like sensitivity, thereby reducing the time needed for discovery for NGC 1068 and for constraints on the other sources. Nonetheless, already with cut & count method, it is clear that KM3NeT/ARCA will pave the way for a new era of the multi-messenger astronomy.

# Searching for Core-Collapse Supernova neutrinos at KM3NeT

**Isabel Goos**[a,*] **, Sonia El Hedri, Corinne Donzaud, Meriem Bendahman, Vladimir Kulikovskiy, Godefroy Vannoye and Damien Dornic on behalf of the KM3NeT Collaboration**

[a] *Particles Group, Astroparticle and Cosmology Laboratory (APC)*
*10 Rue Alice Domon et Léonie Duquet, 75013 Paris, France*

*E-mail:* goos@apc.in2p3.fr, elhedri@apc.in2p3.fr

The discovery of 25 neutrinos coming from the SN1987A core-collapse supernova (CCSN) by the Super-Kamiokande, IMB and Baksan experiments marked the beginning of neutrino astronomy. A new observation of supernova neutrinos with current or upcoming experiments could provide key insight into the underlying mechanism of CCSNe, which is currently poorly understood. Due to the low interaction rate of neutrinos, experiments are however only sensitive to close-by supernovae. Since these events are quite rare, it is crucial to optimize the detection channels of all available experiments. In this contribution, a study of the backgrounds for CCSN searches at the KM3NeT neutrino telescope, currently under construction and taking data in the Mediterranean Sea, is presented. Using new dedicated observables, signatures from radioactive decays and atmospheric muons are modelled, and these results are compared to data to assess the quality of the modelling of the detector response to supernova neutrinos. Finally, based on these results, an improvement of 23% in KM3NeT's distance horizon to a CCSN is presented.

*Speaker

## 1. Introduction

Core-Collapse supernovae (CCSNe) are the end of life of heavy stars ($8M_{\odot}$ and above), whose core collapses in a fraction of a second, often leading to an extremely powerful explosion whose mechanism is not yet completely understood. The detection of 25 neutrinos from supernova 1987A has demonstrated that CCSNe are associated with an extremely powerful neutrino emission, which could crucially affect the dynamics of this explosion. If another CCSN occurs in or near our Galaxy, the detection of the resulting $O(10)$ MeV neutrino burst would provide invaluable information on the CCSN mechanism and neutrino properties. Moreover, this burst could be intense enough to be visible at experiments targeting higher-energy neutrinos, such as KM3NeT.

KM3NeT [1] is a cubic-kilometer neutrino observatory currently under construction in the Mediterranean Sea. It is based on the detection of Cherenkov light induced by product particles of neutrino interactions in seawater. This light is detected by Digital Optical Modules (DOMs) arranged in groups of 18 along vertical *detection lines*. KM3NeT is composed of two detectors: ORCA, a densely instrumented 115-line array aimed at the characterization of $O(10)$ GeV neutrino oscillations, and ARCA, a kilometer-cube detector with two 115-line arrays aimed at the detection of TeV to PeV astrophysical neutrinos. To date, 21 lines have been installed at ARCA and 18 at ORCA. By the end of the year, ARCA will have 29 lines and ORCA 24 lines. In this study, these near-future ARCA29 and ORCA24 configurations are therefore considered.

KM3NeT's current CCSN search strategy [2] makes the detector's final configuration sensitive to 96% of Galactic CCSNe by leveraging its unique DOM structure. The present study shows how to further exploit this structure to improve KM3NeT's low-energy detection potential.

## 2. Low-energy neutrino signatures at KM3NeT

Interactions of $O(10)$ MeV neutrinos, such as CCSN neutrinos, will generally activate at most one DOM and hence are below KM3NeT's reconstruction threshold. If their single-DOM signatures were not resolved, low-energy neutrinos would be indistinguishable from KM3NeT's main backgrounds: bioluminescence, ambient radioactivity, and atmospheric muons as well as muons from neutrino interactions. However, each KM3NeT DOM is composed of 31 small photomultipliers (PMTs), grouped into a sphere of 21.6 cm radius, as shown in Figure 1, left [3]. This small but dense PMT array can be used to characterize low-energy neutrino signatures.

In particular, KM3NeT's current CCSN analysis uses the *multiplicity*, defined as the number of PMT hits in a DOM within a 10 ns window, to distinguish CCSN neutrinos from ambient backgrounds [2, 4]. An example of a multiplicity 4 signature is shown in Figure 1, left. The multiplicity distribution for a 6 hour period of ORCA6 is shown in Figure 1, right. Simulations show that the low multiplicity region up to a value of 7 is dominated by radioactive decays in seawater, while muons dominate the background at values above 7. This contribution presents a proposal for improving KM3NeT's sensitivity to low-energy neutrinos by considering not only the multiplicity of a single-DOM signal but also other single-DOM observables, such as the signal's position on the DOM, and the time and space correlations between activated PMTs.

After considering a wide array of possible single-DOM observables, four weakly-correlated quantities are selected. $\cos\theta$ is the zenith angle of the average direction of the activated PMTs and thus indicates the position of the signal on the DOM. The spatial concentration of activated PMTs is captured by the $|R|$ observable, which is the magnitude of the average direction of activated PMTs. The last two observables considered are the total time over threshold (ToT), which reflects the intensity of the signal, and $\Delta t$, which is the mean time difference between the first hit of a coincidence and the rest of the hits and indicates thus the temporal spread of the signal. The distributions of these four observables for simulated events at the ORCA detector with a 6-line configuration are shown in Figure 2, selecting events with multiplicity 8 that are not associated with KM3NeT high-energy triggers. GEANT4-based simulations of the two dominant KM3NeT backgrounds for this multiplicity are also shown: radioactivity, notably from $^{40}$K in the seawater, and muons. The quality of these simulations is validated by comparing the shapes of the resulting distributions to data. These shapes are found to be similar between data and simulations. The remaining differences will be accounted for in subsequent analyses as systematic uncertainties.

The shapes of the $|R|$, $\cos\theta$, and $\Delta t$ distributions differ significantly between radioactivity and muon backgrounds, showing the excellent discriminating potential of these observables. In particular, these new quantities capture essential differences between radioactivity and muons, which do not necessarily translate into multiplicity differences, as shown in the $\cos\theta$ and $|R|$ distributions in Figure 2. Long downward-going muon tracks leave spread-out signatures (low |R|), mostly on the upper half of the DOMs (large $\cos\theta$). Conversely, low-energy radioactivity signals occur more often near the bottom of the DOM (low $\cos\theta$), which is more instrumented, and are emitted close to the DOM, thus activating small groups of close-by PMTs (large $|R|$). In addition to these observables, for higher multiplicities where muon backgrounds dominate, the total ToT also becomes an essential background characterization tool.

In the following section, the single-DOM observables presented above are incorporated into a multivariate analysis, in order to improve KM3NeT's sensitivity to CCSN neutrinos.

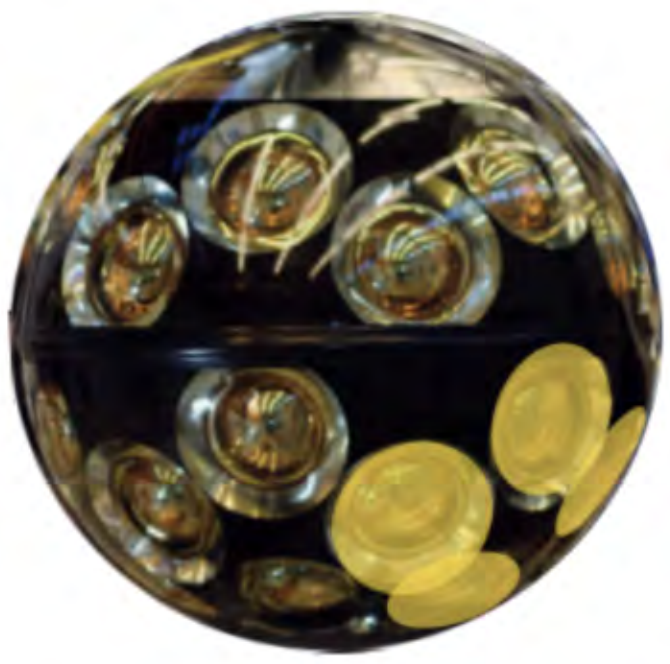

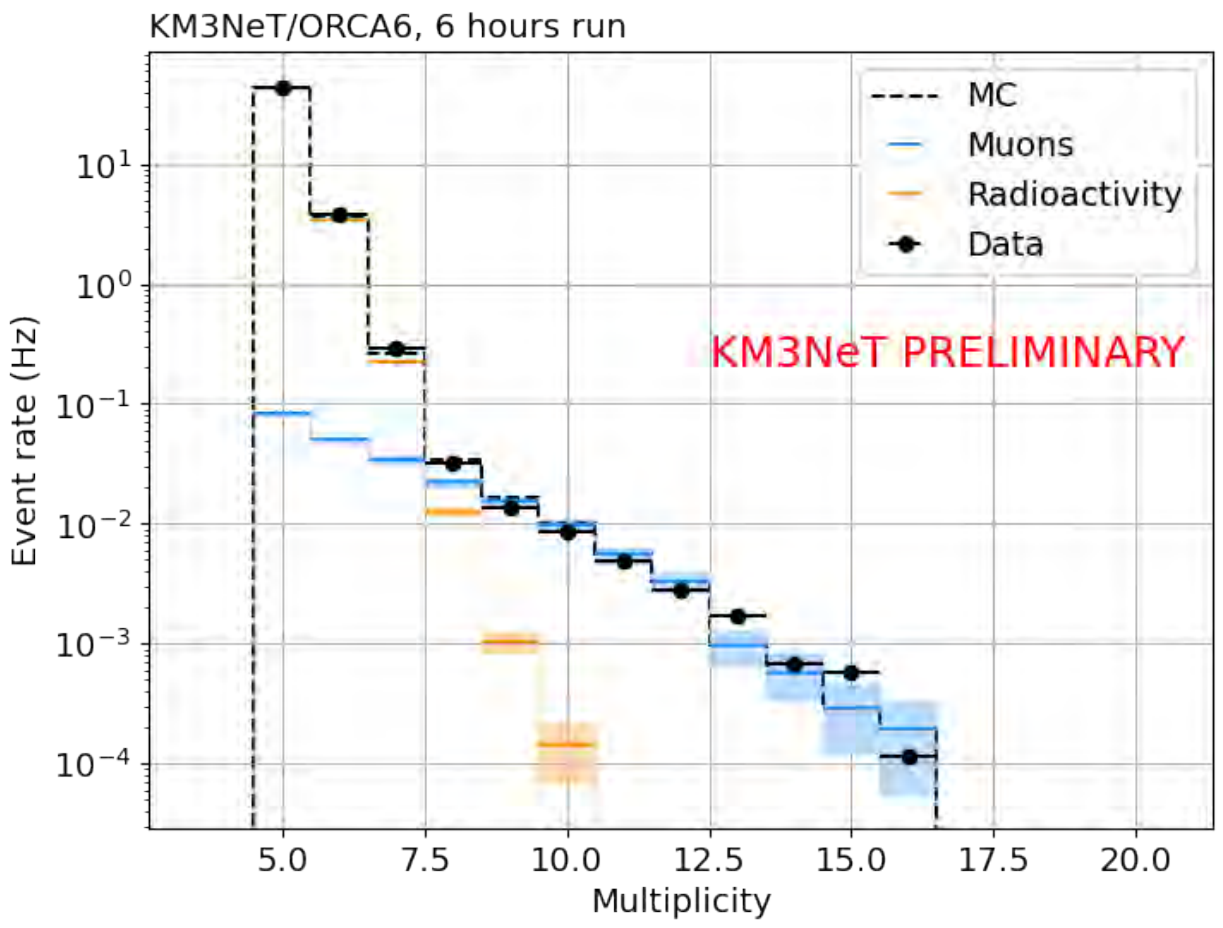

**Figure 1:** Left: image of a DOM with 4 out of the 31 PMTs highlighted to show an example of a multiplicity 4 low-energy neutrino signature. Right: Multiplicity distribution for a 6 hour period of ORCA6 (full black) compared to simulations. Simulated muons are shown in blue, simulated radioactive decays in orange and the total simulated background in dashed black.

## 3. Searching for CCSN neutrinos

Neutrinos emitted by a Galactic CCSN could leave signatures in individual KM3NeT DOMs, as discussed in the previous section. If the neutrino burst is intense enough, these signatures could register as a rise of the number of recorded single-DOM events at ORCA and ARCA, lasting for about 0.5 s. To identify this rise, an online analysis system [4] processes ORCA and ARCA data in realtime, selects suitable CCSN candidates among single-DOM events based on their multiplicity, and evaluates the expected background level. This background is dependent of the number of active PMTs, and its characterization is described in PoS(2023)1223. The number of candidates in a sliding 0.5 s window is then computed and, if the false alarm rate is less than 0.125 per day, a CCSN trigger is issued and an alert is sent to the Supernova Early Warning System (SNEWS, [5]). This contribution focuses on improving the CCSN candidate selection to increase the sensitivity of this CCSN trigger.

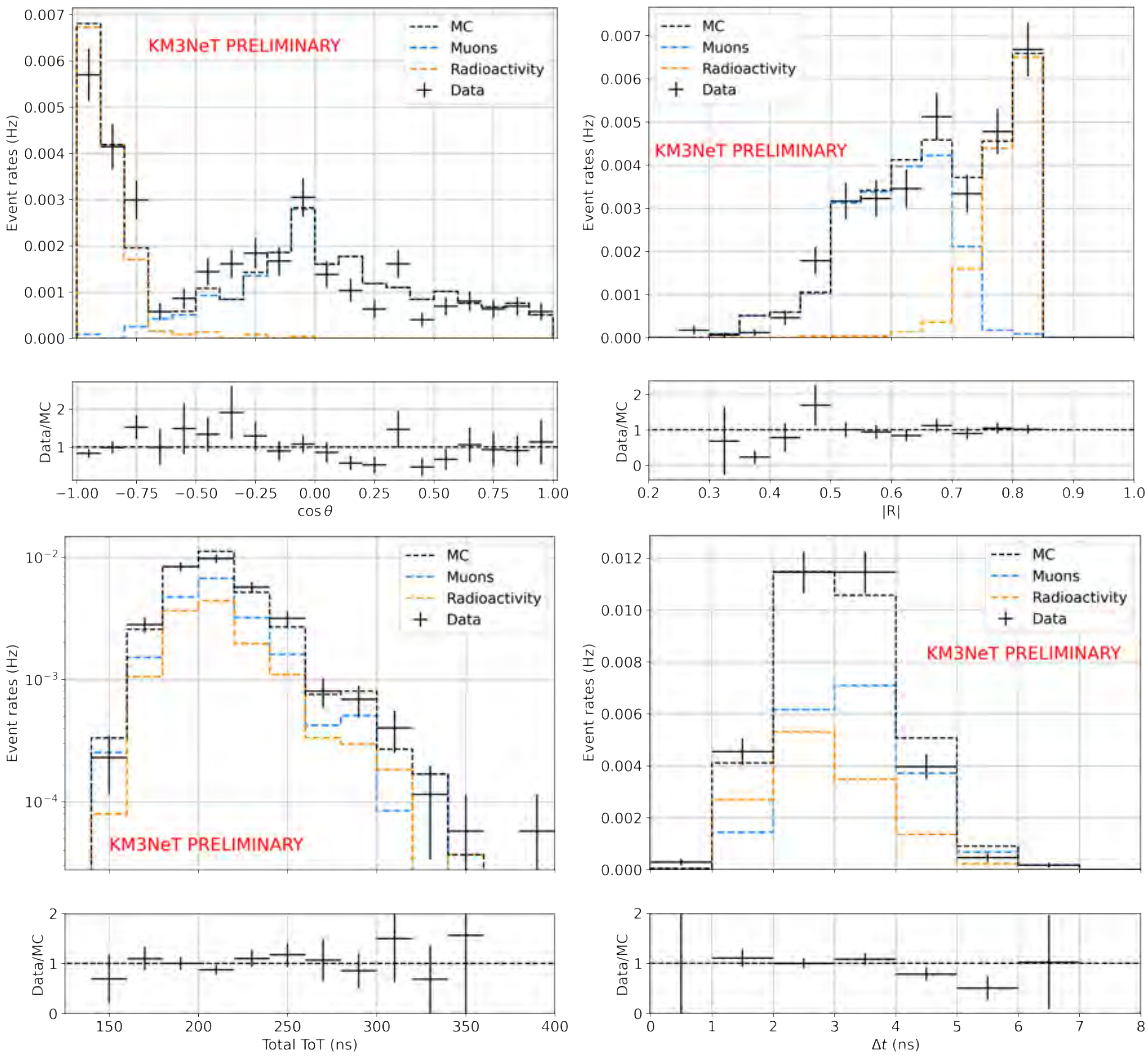

**Figure 2:** Distributions of the most relevant new single-DOM observables for ORCA6 data compared to simulations: $\cos\theta$ (top left), $|R|$ (top right), total ToT (bottom left) and $\Delta t$ (bottom right). ORCA6 data is shown with full black lines, while MC is shown with dashed black lines (muons in blue and radioactivity in orange).

The present analysis relies on two event selection steps. First, a muon veto based on current high-energy KM3NeT triggers is applied to reduce the muon background. The expected number of events in ORCA24 and ARCA29 after applying the muon veto and as a function of the multiplicity is shown in Figure 3, left. The muon veto is more efficient for ORCA than ARCA since the ORCA array is denser. The simulated signal from a CCSN expected at ORCA24 & ARCA29 for a distance of 8 kpc is also shown in this figure for different progenitor masses. These simulations are based on the models from [6–8] for the progenitors of $11\text{M}_\odot$, $27\text{M}_\odot$ and $40\text{M}_\odot$, respectively. The total CCSN rate expected at a water Cherenkov detector is then computed using the SNEWPY software [9] and the result is rescaled to the size of ORCA24 & ARCA29. As can be seen in Figure 3, the signal can be clearly identified above the background contribution at intermediate multiplicity values.

Second, a method to select signal-like single-DOM events is developed, making use of the observables described in Section 2. For this purpose, Boosted Decision Trees (BDTs) are used. BDTs are a machine learning technique which is used in this analysis to assess how signal-like an event is, based on the values of the single-DOM observables. Since the background and signal distributions are different for different multiplicities, one BDT is trained for each multiplicity between 5 and 9 (below 5 the background is too high and above 9 the statistics of our data and simulation samples becomes too low). In addition, since the muon veto affects the ORCA and ARCA backgrounds differently, BDTs for ORCA and ARCA are trained using ORCA6 data (livetime of 463 h) and ARCA8 data (livetime of 293 h) as backgrounds, respectively. During the training process, the $11M_\odot$ model is used for the signal, since it is the most probable supernova scenario considered here.

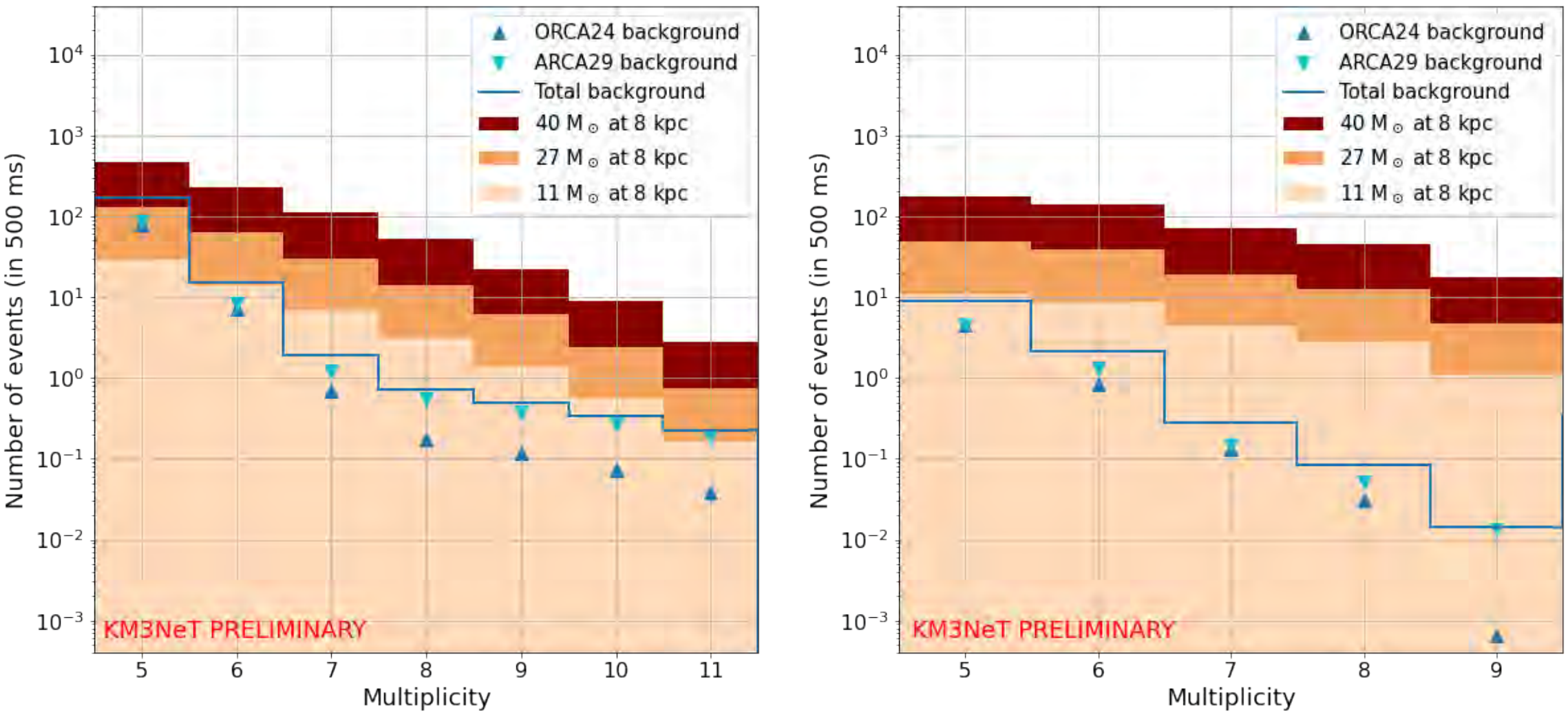

**Figure 3:** Expected number of events in ORCA24 and ARCA29 as a function of the multiplicity before (left) and after (right) applying the BDT cuts (where only the multiplicities for which a BDT cut is applied are displayed). In both cases, the muon veto is applied. The background is shown separately for ORCA24 (light blue markers) and ARCA29 (dark blue markers); the sum of the two is shown with a blue line. The signal of a CCSN at a distance of 8 kpc is represented with coloured bars in orange shades for different m odels: light for $11\text{M}_\odot$, intermediate for $27\text{M}_\odot$, and dark for $40\text{M}_\odot$.

BDTs are trained using different subsets of the single-DOM observables described in Section 2 as input features. The distribution of the discriminant for a BDT trained on ORCA6 events and restricted to multiplicity 8 is shown in Figure 4. The jagged shape of the curves is due to the discrete segmentation of the DOM. For each multiplicity, the BDT selection is then optimized by minimizing the number of signal events needed for a $5\sigma$ discovery, using the Rolke method [10]. This is a frequentist method that takes uncertainties in the background and event selection efficiency into account. In this analysis, background rates are estimated using large data samples collected with the most recent detector configurations, as well as GEANT4-based simulations for the CCSN signal. Systematic uncertainties associated with the signal simulation are modelled using the relative discrepancies between the data and background simulations introduced in Section 2. Uncertainties on the DOMs' efficiencies and the number of activated PMTs are taken to be of 11%, following the estimates from [2].

Applying the optimal BDT cuts in addition to the muon veto to the background and simulated signal distributions leads to the distributions shown in Figure 3, right. With this procedure, a considerable reduction of the background with respect to the signal is achieved. In particular, after the cut on the BDT score, the signal at multiplicity 6 is now above the background even for the lightest progenitor.

## 4. Distance horizons

After applying the muon veto and the optimal BDT cuts to background and signal events, the optimal multiplicity range is computed for the combined detector ORCA24 & ARCA29, scanning multiplicities from 5 to 11 and again minimizing the number of signal events corresponding to a $5\sigma$ discovery. The $5\sigma$ distance horizon for the most conservative case of an $11M_\odot$ progenitor is shown in Figure 5, for all considered multiplicity ranges, before (left) and after (right) applying the BDT cuts. The best horizon without applying BDT cuts is at 8.1 kpc, selecting events in the multiplicity range 6-11. The use of BDT cuts leads to a 23% increase of this horizon, up to 10.0 kpc, using the multiplicity range 6-9. Hence, for the low-emission CCSN model considered, the use of single-DOM observables makes it possible for current detectors to probe a significant fraction of the Galactic bulge. In the cases of the $27M_\odot$ and $40M_\odot$ models, the $5\sigma$ distance horizons making use of the BDT cuts are at 21.1 kpc and 40.4 kpc, respectively.

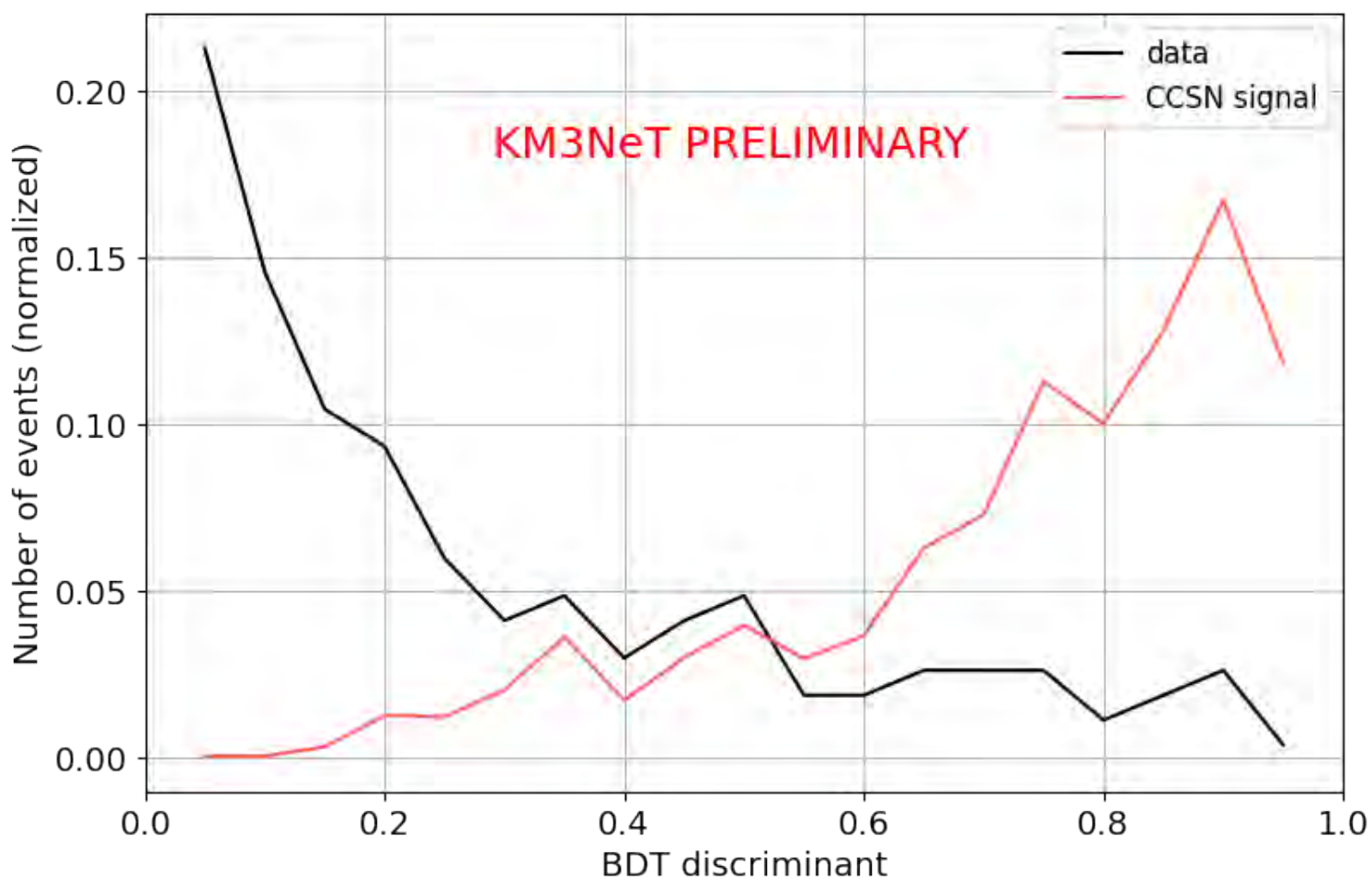

**Figure 4:** BDT discriminant distribution for a BDT trained on ORCA6 data, using $|R|$, $\cos\theta$ and the total ToT as input features, for events with multiplicity 8. The jagged shape of the curves is due to the discrete segmentation of the DOM.

It is also instructive to compute the sensitivity (in units of sigma) to a CCSN signal as a function of the distance from it. The result is shown in Figure 6 for the three progenitors considered and using the combined ORCA24 & ARCA29 detector and the optimal multiplicity range. The distance to Betelgeuse, which has a mass of close to $11M_\odot$, is also displayed as a reference, since it is a good candidate for the next Galactic supernova [11]. For this scenario, the detection of associated neutrinos would be certain. With the upcoming detector configuration considered, the analysis proposed here can reach the Galactic Center and, for a sufficiently heavy progenitor, a CCSN close to the most distant Milky Way edge could be detected.

## 5. Conclusions

The design of the DOMs at KM3NeT makes it possible to detect a CCSN signal despite the fact that CCSN neutrinos have energies below the detector's energy threshold. Here, a thorough study of single-DOM observables, which capture different characteristics of a MeV-scale signal, is performed. BDTs are used to discriminate signal from background events. With this method, KM3NeT's distance horizon is increased by 23%. With the upcoming ORCA24 and ARCA29 detector configurations it will thus be possible to probe a significant fraction of the Galactic bulge, even for the case of a light progenitor.

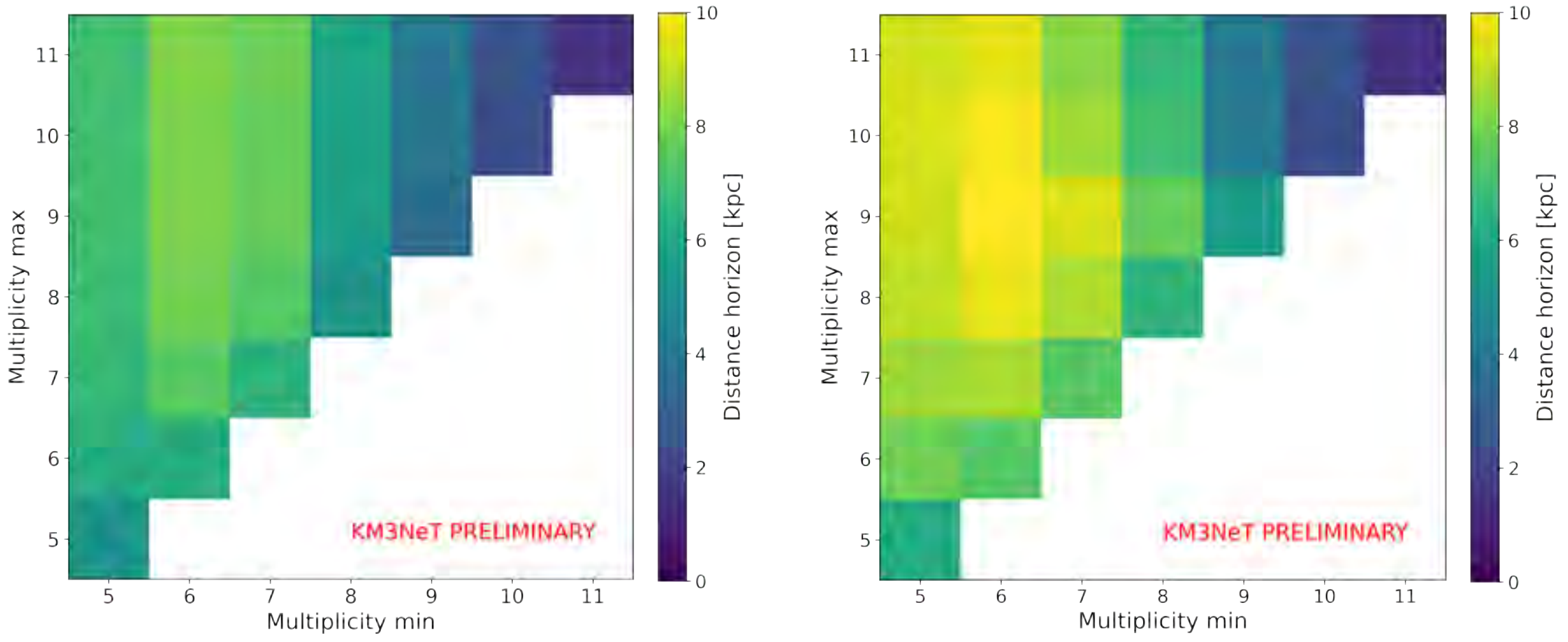

**Figure 5:** Distance horizons at $5\sigma$ obtained by selecting different multiplicity ranges, without BDT cuts (left) and with the optimal BDT cuts (right). The horizons for these two scenarios are of 8.1 kpc and 10.0 kpc, respectively.

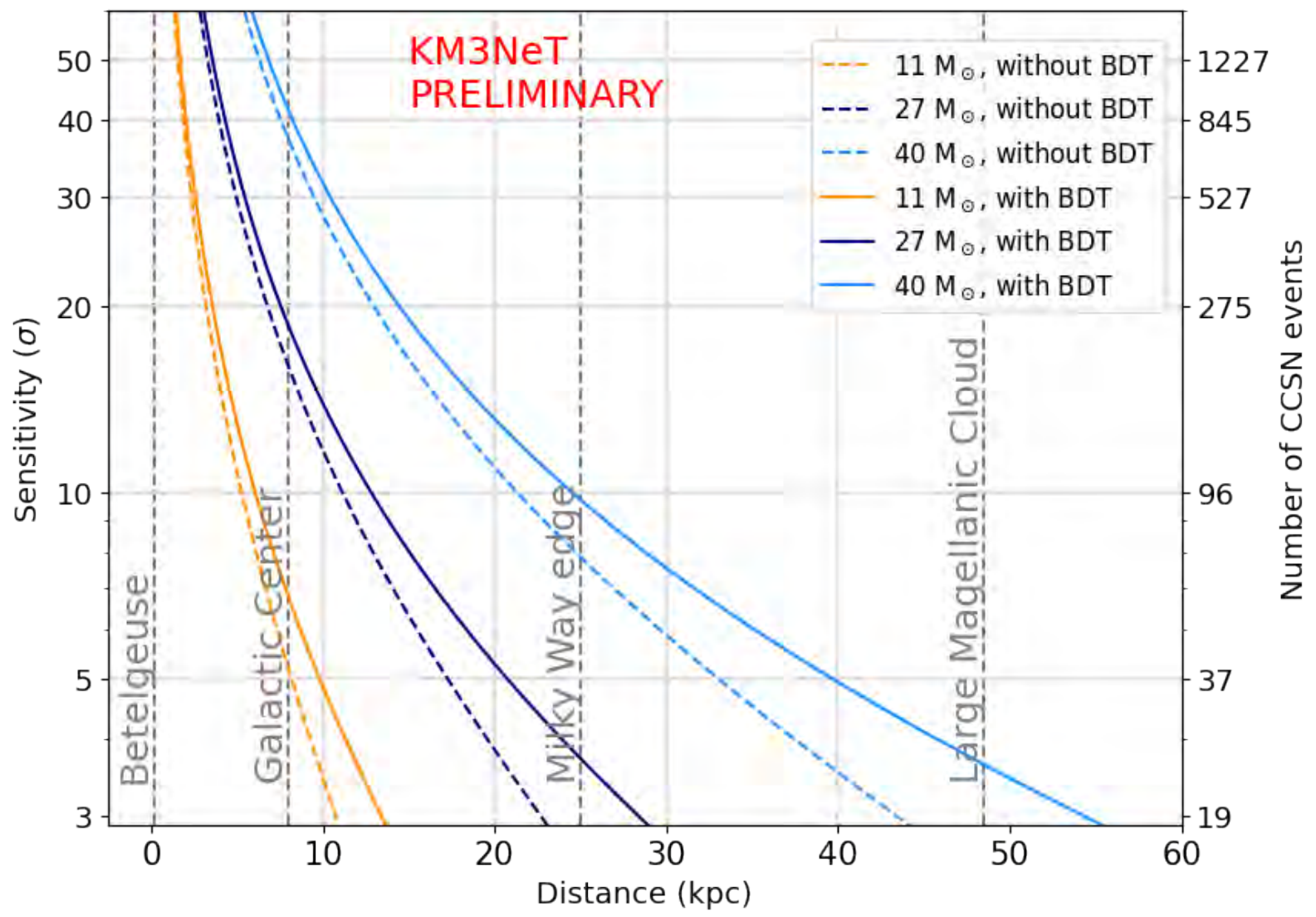

**Figure 6:** KM3NeT detection sensitivity as a function of the distance from the CCSN for the three progenitors considered: $11M_\odot$ (orange), $27M_\odot$ (dark blue) and $40M_\odot$ (light blue).

## Acknowledgments

This work is supported by LabEx UnivEarthS (ANR-10-LABX-0023 and ANR-18-IDEX-0001) and Paris Region (DIM ORIGINES).

# Search for a diffuse astrophysical neutrino flux from the Galactic Ridge using KM3NeT/ARCA data

**Francesco Filippini**[a,b,*] **on behalf of the KM3NeT collaboration**

[a] *Università di Bologna, Dipartimento di Fisica e Astronomia, v.le C. Berti-Pichat, 6/2, Bologna, 40127, Italy*

[b] *INFN, Sezione di Bologna, v.le C. Berti-Pichat, 6/2, Bologna, 40127, Italy*

*E-mail:* francesco.filippini9@unibo.it

Several theoretical models predict and describe the properties of part of the diffuse neutrino flux, originating from the interaction of Galactic cosmic rays with the interstellar medium matter located in the centre of our Galaxy. This neutrino flux is expected to be of the same order of magnitude as the diffuse $\gamma$-ray flux measured by Fermi-LAT, close to the Galactic plane. Recently hints by the ANTARES Collaboration, at a significance level over 2 $\sigma$, and the observation by the IceCube Collaboration, at a significance level of 4.5 $\sigma$, of a high-energy neutrino emission from the Galactic plane have been reported. For these reasons, and considering the privileged position of the KM3NeT/ARCA telescope, being located in the Northern hemisphere, data have been analysed searching for a possible excess of events coming from an extended region, with Galactic coordinates $|l| < 30°$ and $|b| < 2°$, namely the Galactic Ridge. In this contribution, the result of the analysis exploiting data gathered with KM3NeT/ARCA in the period in which comprises 6, 8, 19 or 21 active detection units is presented, showing the capabilities and performance of KM3NeT.

*Speaker

# 1. Introduction

## 1.1 Diffuse neutrino emission from the Galactic Ridge

The Galactic plane is the most evident source in the sky in all the electromagnetic wavelengths. These photons can be produced either in specific point sources, positioned on the Galactic plane, or by the interaction of cosmic rays (CRs) with the interstellar medium matter (ISM), located in the center of our Galaxy. A significant fraction of energy released during these interaction processes is transformed into short living particles, that can decay into $\gamma$-rays and neutrinos. While photons can be produced by energy-losses of electrons, interacting with magnetic fields or ISM, neutrinos can be produced only in the decays of particles produced in hadronic interactions, representing therefore a unique probe of acceleration and interaction sites of CRs. Several theoretical models have been developed in the past years, to try to constrain the flux of neutrinos originated inside the Galactic plane [1–3]. The strict relation that link $\gamma$-rays and neutrinos, produced via hadronic interactions, allows to constrain the expected neutrino fluxes thanks to measured CRs local properties and to $\gamma$-rays measurements themselves, performed both by space-based telescopes and by detectors located on Earth. In the specific, the predicted neutrino flux is expected to be of the same order of the $\gamma$-ray one, and in the innermost part of the Galactic plane, defined here $|l| < 30°$ and $|b| < 2°$, namely the Galactic Ridge, the CRs spectrum should be described by a harder spectral index with respect to the one locally measured at Earth. The ANTARES Collaboration reported an excess of events coming from the Galactic Ridge incompatible with the background expectation at $\sim 96\%$ confidence level [4], and at the end of June 2023, IceCube Collaboration has reported the first observation of high-energy neutrinos from the Galactic plane, with a statistical significance of 4.5 $\sigma$ [5].

## 1.2 KM3NeT neutrino telescope

KM3NeT[1] is the next-generation neutrino telescope project that will instrument, in its final configuration, an overall volume of several cubic kilometres of sea water [6]. KM3NeT comprise two different instrumented regions, placed in separate locations: KM3NeT/ARCA[2], off-shore the Sicilian coast at a depth of about 3500 m, optimised for searching of neutrinos from astrophysical sources, and KM3NeT/ORCA[3], off-shore the French south coast, that will study neutrino properties exploiting atmospheric neutrinos. The two detectors share the same technology and neutrino detection principle: namely a 3D array of photosensors, called digital optical modules (DOM) [7], capable to detect the Cherenkov light produced by relativistic particles emerging from neutrino interactions. The main difference between the two detectors consists in the density of photosensors, which is optimised for the study of neutrinos in the few-GeV region for ORCA and the TeV-PeV energy range for ARCA. The detectors are built with detection units (DUs), which stand on the sea bottom and comprise 18 DOMs each. In its final configuration the KM3NeT/ARCA telescope will consist of two building blocks of 115 vertical detection units each. Both sites however are in the Northern hemisphere at a latitude between 36° and 43° North, allowing to observe upgoing events coming from most of the sky. In fact, looking at the KM3NeT sky coverage, reported in Figure 1, we can see that most of the galactic plane, including the Galactic Ridge, is fully visible through upgoing events.

[1] KM3NeT is an acronym for 'Cubic Kilometre Neutrino Telescope'
[2] ARCA stands for Astroparticle Research with Cosmics in the Abyss
[3] ORCA stands for Oscillation Research with Cosmics in the Abyss

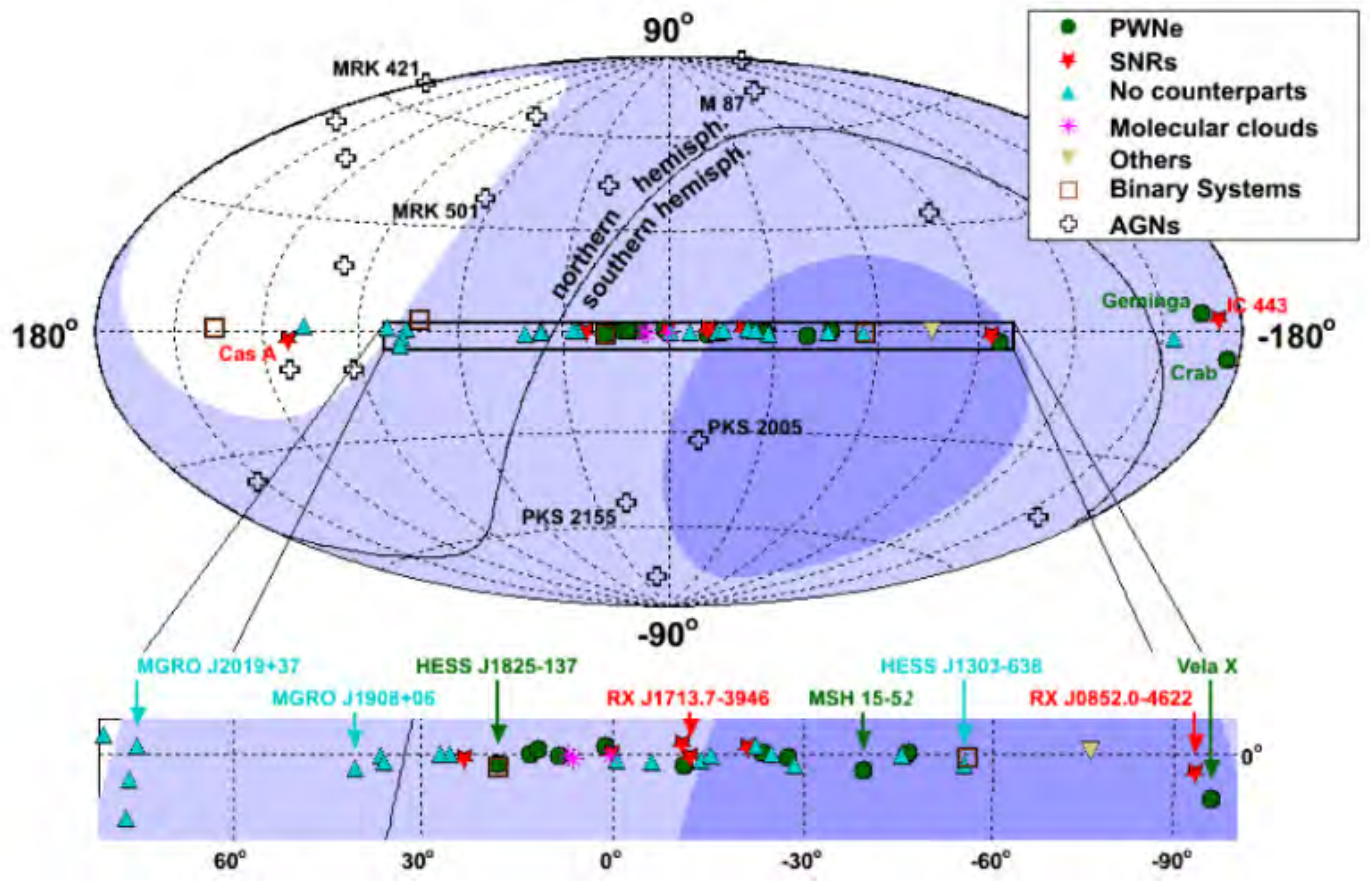

**Figure 1:** Sky coverage of a neutrino telescope located in the Mediterranean sea, in galactic coordinates. Dark (light) blue shaded areas represent visibility over 75% (25%). Some known astrophysical objects are marked. Figure taken from [6].

## 2. Analysis

In the following sections, a data analysis to search for a diffuse neutrino flux coming from the Galactic Ridge is reported. In the specific, data gathered with 6, 8, 19 and 21 active DUs of the KM3NeT/ARCA detector have been exploited, for a total lifetime of 432 days.

### 2.1 Event selection

The events considered in the analysis are track candidate events, generated by $\nu_\mu$ charged-current interactions, that have passed quality selection criteria. To further reduce the surviving background, represented by atmospheric muons, Boost Decision Trees (BDTs) have been trained on the different detector configurations, in order to classify and reject these events (developed in synergy with [8]). In Figure 2 the BDT score data-Monte Carlo distribution is shown, evaluated in this specific case on KM3NeT/ARCA8 data. An optimal event selection is then derived, through the minimization of the Model Rejection Factor (MRF) [9], specifically for a signal flux with spectral index $\Gamma_\nu = 2.4$. The optimal region, from which to require that reconstructed directions of the track events should come from, has been found to be $|l| < 31°$ and $|b| < 5°$ for KM3NeT/ARCA6-8 and $|l| < 31°$ and $|b| < 4°$ for KM3NeT/ARCA19-21, due to the angular resolution of the detector in the considered configurations. The whole event selection has been carried out following the blinding policies of the KM3NeT Collaboration.

### 2.2 Method

The methodology adopted for the analysis is an on-off technique, similarly to what has been done in [4] and [10]. The on-region is defined for each detector configuration from the optimization procedure described in section 2.1. The background expectation instead is directly estimated from off-regions in the data, which have been selected so as to have the same sky coverage of the on-region and are shifted in right ascension, while avoiding also the region of the Fermi Bubbles. The expected neutrino signal has been simulated following the standard Monte Carlo chain developed within the KM3NeT collaboration [11], assuming the signal to be originated inside the Galactic Ridge. Each simulated event has been then weighted assuming a single power-law spectrum, of the form $\Phi_\nu(E) = \frac{dN_\nu}{dE_\nu} = \Phi_0 \times \left(\frac{E_\nu}{E_0}\right)^{-\Gamma_\nu}$ with a normalization energy $E_0$ set at 40 TeV, for convenience. The selected events are binned in function of the reconstructed energy, and the statistical analysis, based on the same procedure developed in [4], adopts the following binned likelihood approach:

$$\mathcal{L}(N_i; S_i^{\Gamma_\nu}, B_i, \Phi_0) = \prod_{i=1}^{12} Poisson(N_i, B_i + S_i^{\Gamma_\nu}) \quad (1)$$

where $N_i$ is the number of events observed in the on-region in each energy bin, $B_i$ the background derived from the off-regions and $S_i^{\Gamma_\nu}$ is the signal expectation, derived from Monte Carlo simulations for a given spectral index $\Gamma_\nu$ and normalization flux $\Phi_0$. A posterior probability distribution is derived, following a Bayesian approach, in order to include statistical and systematic uncertainties, and assuming a flat prior for the two parameter of i nterest: the neutrino spectral index $\Gamma_\nu$ and normalization flux $\Phi_0$. The combination of different detector configurations is performed multiplying the respective posterior probabilities. From the profiled posterior probability then upper limits and sensitivities are derived.

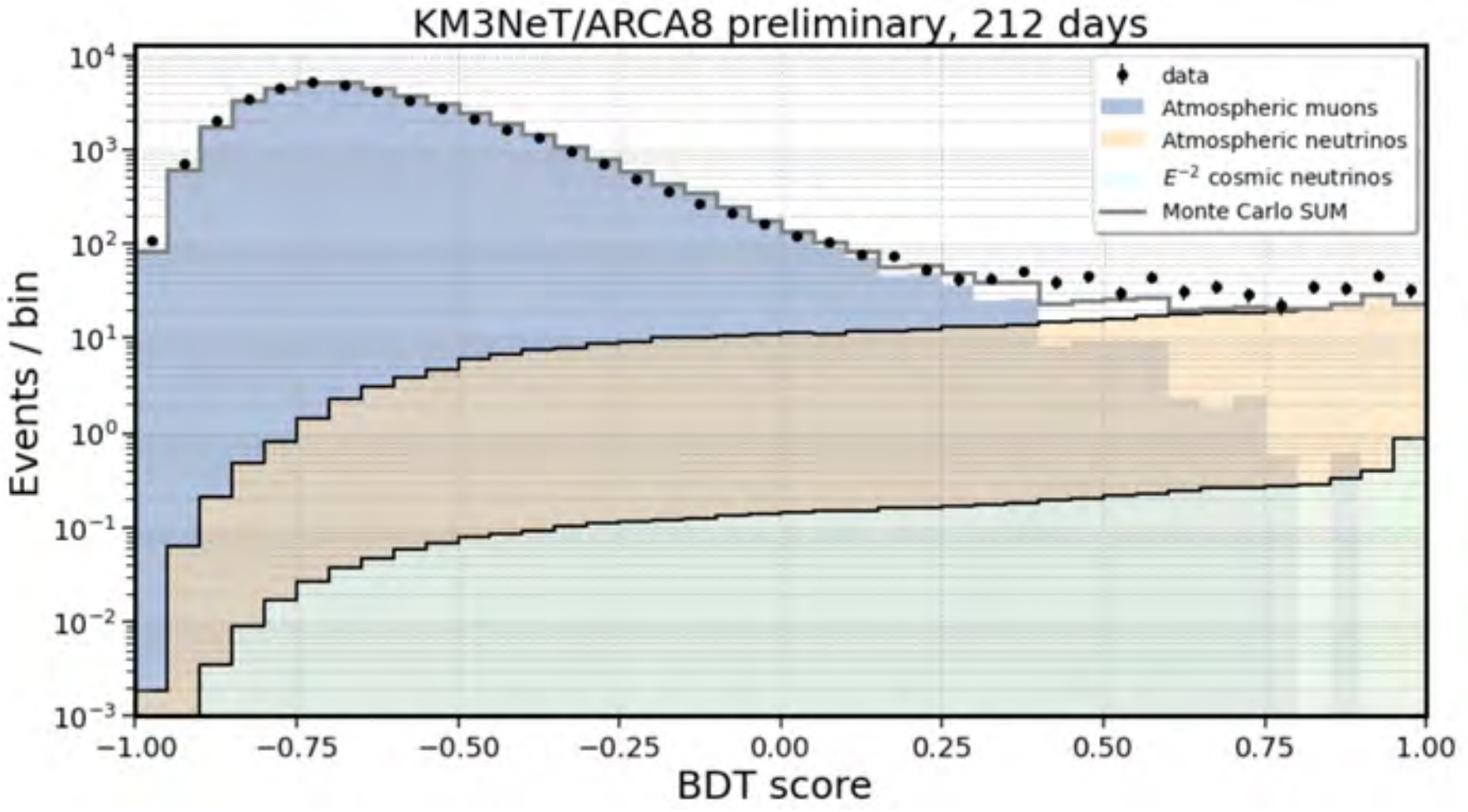

**Figure 2:** Data-Monte Carlo comparison for Boost Decision Tree output score, evaluated on test sample. Classification among atmospheric muons (blue shaded distribution) and neutrinos (orange and green shaded area).

### 2.3 Results

In Figure 3 the unblinded energy distributions are shown for respectively KM3NeT/ARCA6, KM3NeT/ARCA8, KM3NeT/ARCA19 and KM3NeT/ARCA21.

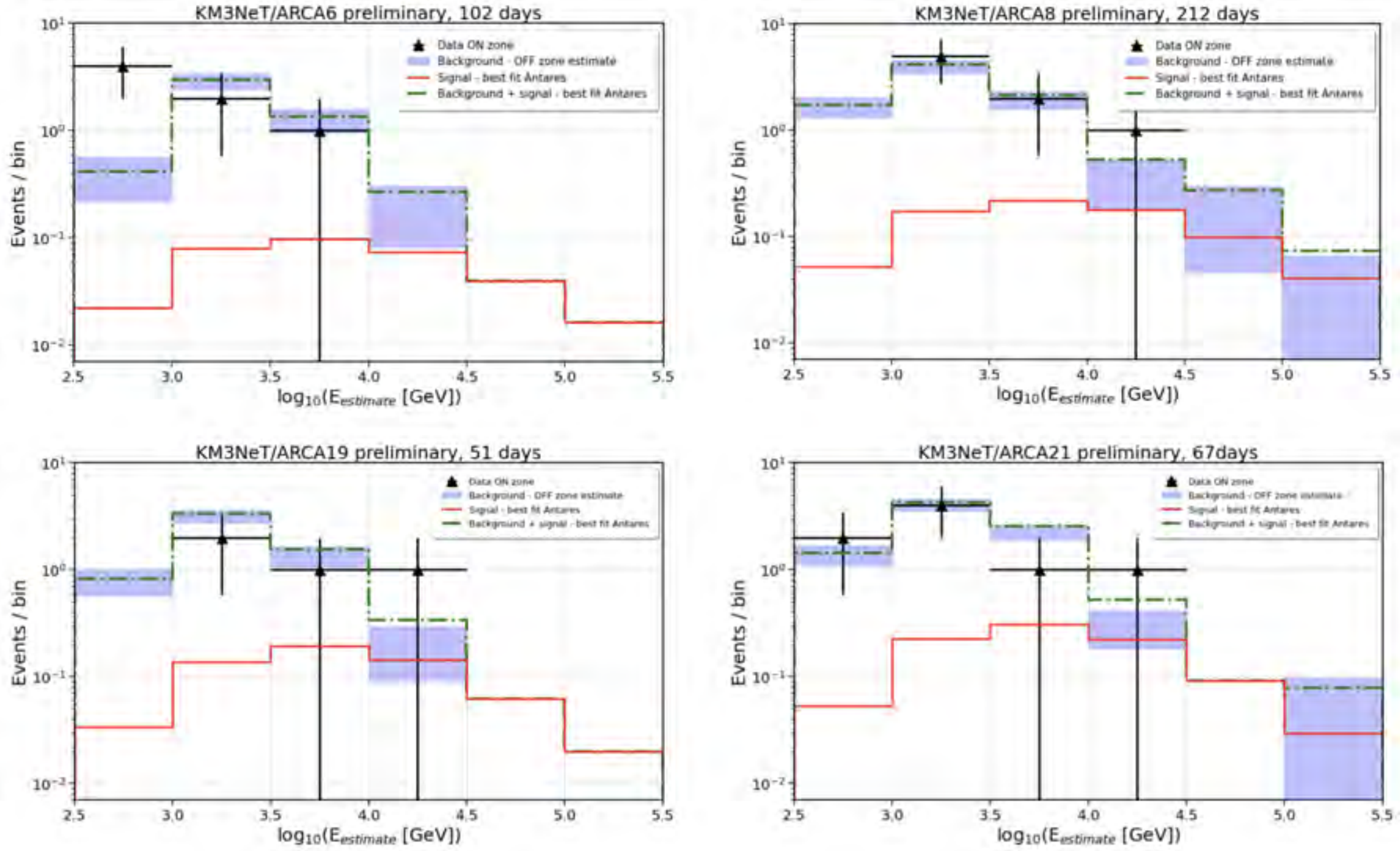

**Figure 3:** Energy distribution of selected events for KM3NeT/ARCA with 6 (**top left**), 8 (**top right**), 19 (**bottom left**) and 21 (**bottom right**) active detection units. The blue bands represent the background estimation derived from the off-regions, with the associated statistical error. The black points represent the data found in the on-region while the red solid lines represent the numbers of events expected in each data sets, assuming the ANTARES best-fit flux found in [4]. The green dashed lines show the sum of the expected signal and background.

No excess of events has been found with respect to the background expectation. Therefore the 90% C.L. upper limits have been calculated and reported in Table 1. Limits for each single detector configuration and for the combination KM3NeT/ARCA6+8 and KM3NeT/ARCA6+8+19+21 are reported, in order to highlight the impact of including data sets gathered with 19 and 21 DUs, despite the limited lifetime.

| **90% C.L. upper limits** | | | | | | |
|---|---|---|---|---|---|---|
| $\Gamma_\nu$ | ARCA6 | ARCA8 | ARCA6+8 | ARCA19 | ARCA21 | ARCA6+8+19+21 |
| 2.2 | $8.6 \times 10^{-5}$ | $4.5 \times 10^{-5}$ | $3.4 \times 10^{-5}$ | $4.9 \times 10^{-5}$ | $3.4 \times 10^{-5}$ | $1.9 \times 10^{-5}$ |
| 2.3 | $2.7 \times 10^{-4}$ | $1.3 \times 10^{-4}$ | $1.1 \times 10^{-4}$ | $1.5 \times 10^{-5}$ | $1.0 \times 10^{-4}$ | $5.8 \times 10^{-5}$ |
| 2.4 | $8.2 \times 10^{-4}$ | $3.9 \times 10^{-4}$ | $3.0 \times 10^{-4}$ | $4.1 \times 10^{-4}$ | $2.8 \times 10^{-4}$ | $1.7 \times 10^{-4}$ |
| 2.5 | $2.3 \times 10^{-3}$ | $1.1 \times 10^{-3}$ | $9.0 \times 10^{-4}$ | $1.1 \times 10^{-3}$ | $7.8 \times 10^{-4}$ | $4.8 \times 10^{-4}$ |
| 2.6 | $6.5 \times 10^{-3}$ | $2.9 \times 10^{-3}$ | $2.5 \times 10^{-3}$ | $2.8 \times 10^{-3}$ | $2.1 \times 10^{-3}$ | $1.3 \times 10^{-3}$ |
| 2.7 | $1.7 \times 10^{-2}$ | $7.4 \times 10^{-3}$ | $6.8 \times 10^{-3}$ | $7.1 \times 10^{-3}$ | $5.5 \times 10^{-3}$ | $3.5 \times 10^{-3}$ |

**Table 1:** 90% C.L. upper limits under a single power-law assumption, referred to in the text as $\Phi_\nu$, for a reference energy $E_0 = 1$ GeV and with spectral indices $\Gamma_\nu$ ranging from 2.2 to 2.7 for KM3NeT/ARCA6, KM3NeT/ARCA8, KM3NeT/ARCA19, KM3NeT/ARCA21 and the combined data sets. All the results are expressed in units of $\mathrm{GeV}^{-1}\,\mathrm{cm}^{-2}\,\mathrm{s}^{-1}\,\mathrm{sr}^{-1}$.

In Figure 4 the best-fitting flux obtained from the IceCube Collaboration [5] for two templates models is reported, together with the ANTARES results [4] and KM3NeT upper limits. Since the analysis methodology adopted by IceCube is based on a full-sky template search, the ANTARES and KM3NeT limits has been integrated over the solid angle extension of the Galactic Ridge.

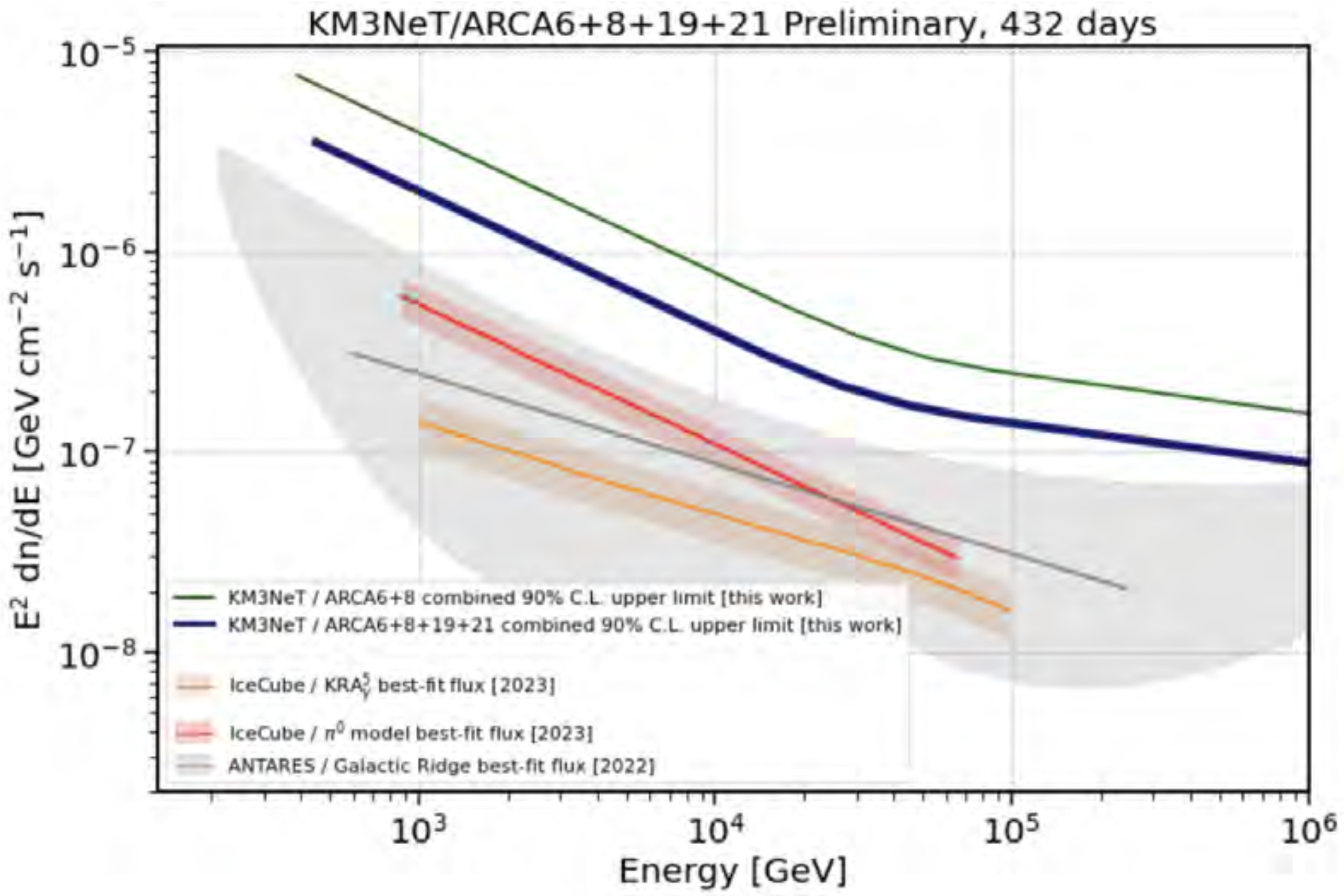

**Figure 4:** KM3NeT/ARCA6+8 combined (green solid line) and KM3NeT/ARCA6+8+19+21 combined (blue solid line) 90% C.L. upper limits to a diffuse neutrino emission from the Galactic Ridge, for a range of spectral indices $\Gamma_\nu \in [2.2, 2.7]$. The ANTARES limits and best-fitting flux are also reported (grey shaded area and grey solid line) for the same type of search, derived from [4]. The KM3NeT and ANTARES limits have been integrated over the solid angle, spanned by the Galactic Ridge, to be compared to the best fitting fluxes (red and orange lines) reported by IceCube analysis, which are based on full-sky template method [5].

### 2.4 Conclusions and outlook

The discovery of a neutrino emission, recently reported by IceCube collaboration, following a hint reported by the ANTARES Collaboration, has opened a new perspective on the possibility to study the properties of our Galaxy through neutrinos [12]. The analysis illustrated in this contribution, searching for a diffuse neutrino flux originated from the Galactic Ridge region, has been performed exploiting data collected by KM3NeT/ARCA with 6, 8, 19 and 21 active detection units, for a total lifetime of 432 days. No excess of events has been found with respect to the background estimation. Currently the KM3NeT/ARCA detector comprises 21 active detection units, for an effective area which is three times higher than the one of ARCA6/8. The first period of KM3NeT/ARCA21 has been included in this analysis, but further 6 months of data gathered with this configuration geometry are currently under analysis. A further expansion of the detector with ~ 10 more detection units is planned for the coming autumn. The limits shown in this work for this type of search are not yet competitive with the results reported by ANTARES and IceCube, but the fast growth planned for the KM3NeT detectors in the near future will allow soon to complement those observations and to further constrain the neutrino emission from the center of our Galaxy. Moreover multi-messenger observations, combining information from gamma-rays observatories like CTA, LHAASO, SWGO and from second generation neutrino telescopes (KM3NeT, GVD, IceCube Gen-2, P-One, TRIDENT) will be fundamental to identify individual sources of CRs and to deeply study the CR properties inside our galaxy.

## A. Evaluation of systematic uncertainties

In order to evaluate the systematic uncertainties on the signal acceptance, dedicated Monte Carlo simulations have been performed, allowing to identify a limited number of parameters as the main contributors to the systematic uncertainties. In order to disentangle the different contributions to the overall uncertainty, a dedicated simulation has been produced, in which only one parameter was varied at a time. In Figure 5 the relative difference of the modified Monte Carlo production with respect to the standard one has been calculated in bins of reconstructed neutrino energy. This percentage variation is assumed to be the systematic uncertainty associated with the specific parameter variation. The PMT quantum efficiency has been modified by a 5% (*red dotted line*), while the uncertainty on the water properties are taken into account by modifying the light absorption length by a factor 10%, coherently to what was previously done by ANTARES Collaboration [13] (*green line*). These two different contributions are then summed in quadrature (*blue solid line*) to derive an overall uncertainty, which has been added to the statistical analysis described in section 2.2.

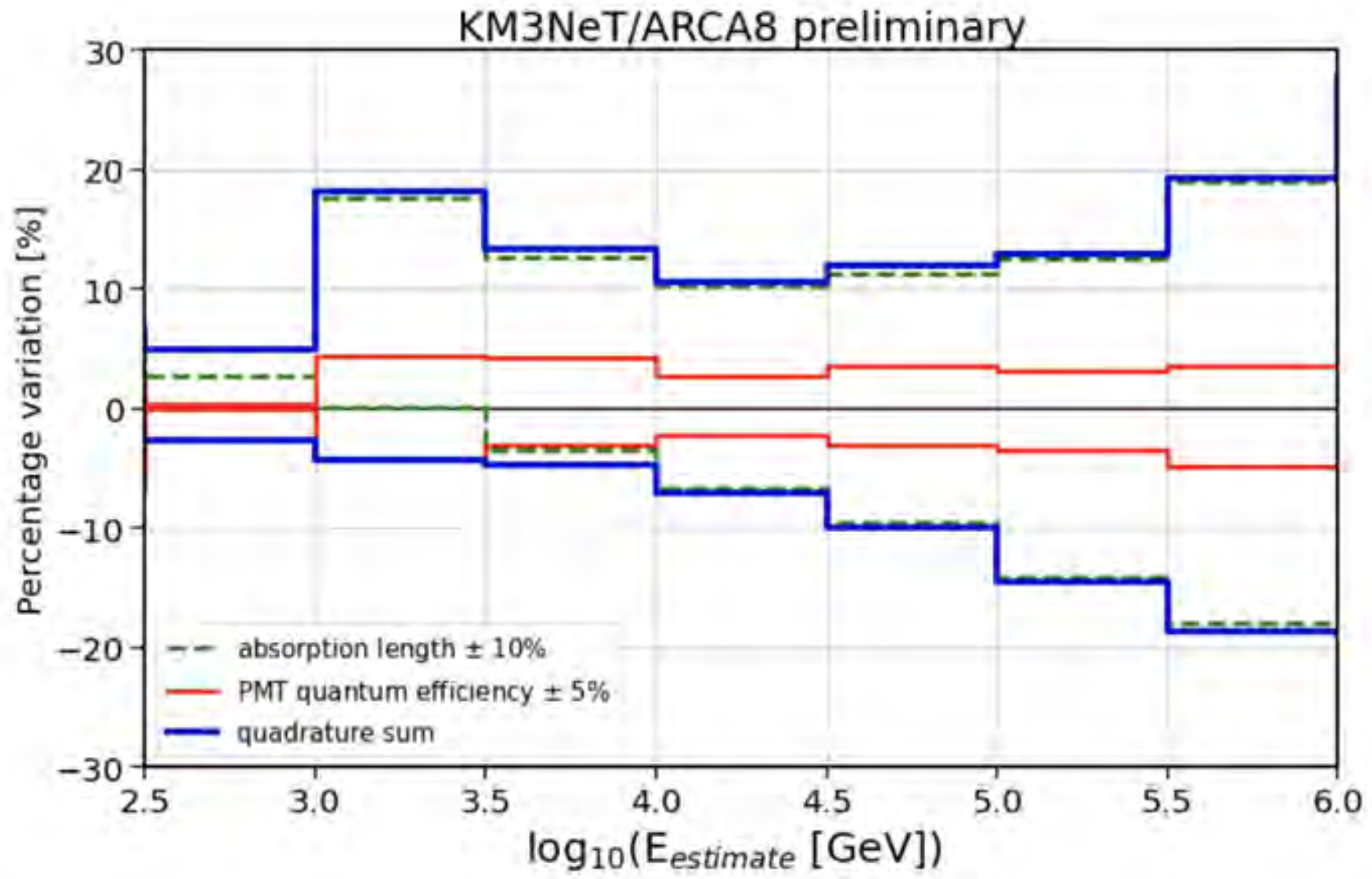

**Figure 5:** Percentage variation of the modified Monte Carlo simulations, for each parameter modification (specified in the legend) with respect to standard Monte Carlo simulation, as a function of reconstructed neutrino energy.

# Data reconstruction and classification with graph neural networks in KM3NeT/ARCA6-8

**Francesco Filippini,**[a,b,*] **Eleni Androutsou,**[c] **Alba Domi,**[d] **Bernardino Spisso,**[e,f,g] **and Evangelia Drakopoulou**[c] **on behalf of the KM3NeT collaboration**

[a] *Università di Bologna, Dipartimento di Fisica e Astronomia, v.le C. Berti-Pichat, 6/2, Bologna, 40127, Italy*

[b] *INFN, Sezione di Bologna, v.le C. Berti-Pichat, 6/2, Bologna, 40127, Italy*

[c] *NCSR Demokritos, Institute of Nuclear and Particle Physics, Ag. Paraskevi Attikis, Athens, 15310, Greece*

[d] *Erlangen Centre for Astroparticle Physics of Friedrich-Alexander-Universität, Nikolaus-Fiebiger-Straße 2, 91058 Erlangen*

[e] *INFN, Sezione di Napoli, Complesso Universitario di Monte S. Angelo, Via Cintia ed. G, Napoli, 80126, Italy*

[f] *Gruppo Collegato di Salerno, Dipartimento di Fisica, Via Giovanni Paolo II 132, Fisciano, 84084, Italy*

*E-mail:* francesco.filippini9@unibo.it, androutsou@inp.demokritos.gr, Alba.Domi@ge.infn.it, Spisso@na.infn.it, drakopoulou@inp.demokritos.gr

KM3NeT is a research infrastructure hosting two large-volume Cherenkov neutrino detectors which are currently under construction in the Mediterranean Sea. The KM3NeT/ARCA detector is optimised for the detection of high-energy neutrinos from astrophysical sources in the TeV-PeV energy range. Once completed, the detector will consist of 230 detection units. Here, we present a Deep Learning method using graph neural networks that is trained and applied to events gathered with 6 and 8 active detection units of KM3NeT/ARCA. Graph neural networks have been trained for classification and regression tasks, showing very promising performances in a range of different tasks like neutrino-background identification, neutrino event topology classification, energy and direction reconstruction, and also in the study of properties of muon bundles.

*Speaker

## 1. Introduction

Charged particles produced in neutrino interactions with water induce Cherenkov light, which is detected by 3" photomultiplier tubes (PMTs) hosted in pressure resistant glass spheres. 31 PMTs are stored in a single digital optical module (DOM), and 18 DOMs are fixed to a long vertical detection unit (DU) that is anchored to the sea floor. Each of the building blocks of KM3NeT will consist of 115 DUs [1]. The KM3NeT detector can therefore be modelled as a 3D array of photosensors capable of registering the arrival time and time over threshold of photons impinging on a photomultiplier tube (hit) contained inside a DOM. The detected light on each PMT can be used to reconstruct the particle properties, such as their energy or direction. Each hit has information about the time and the xyz-position of the PMT that recorded the light, as well as its pointing direction. All the hit information of an event serves as input to reconstruction and classification algorithms. Since data recorded by KM3NeT closely resembles point clouds, graph neural networks (GNNs) are a natural choice as architecture for deep neural networks employed for the data analysis.

## 2. Graph neural networks

In the past, Deep Learning methods have acquired huge popularity on image recognition tasks, exploiting convolution and pooling operations on images encoded as a fix grid of pixels. These techniques have also been developed in the context of the KM3NeT experiment [2]. However, the fixed pixel structure has shown limitations in its capability to represent data collected by the telescope. The high dimensionality and sparse signal registered in the detector can be much better encoded in graphs. Other advantages of graphs with respect to image based methods are linked to the limited resolution in position and time that can be achieved through images/fixed grid pixels. At the same time, DOMs in the KM3NeT detector move under the effect of the sea current: this information is completely lost within the position bin size. The most natural way to encode information of events into a graph is to represent every photon hit as a node. Therefore, each node has a 7-dimensional feature space represented by: 3 spatial coordinates, 3 PMT directions, and time. To create the final graph structures then, nodes are connected to each other, based on Euclidean metric. For memory usage optimization and for keeping the number of connections under control, each node is connected to its *k-nearest neighbours*. For the model architecture adopted in the next sections, the ParticleNet architecture has been exploited [3]. This architecture was originally designed for point cloud applications, and used for jet tagging at LHC, showing outstanding performances with respect to image convolutional techniques.

## 3. Event reconstruction and classification

### 3.1 Neutrino energy regression

The architecture explained in Section 2 has been exploited using a last fully-connected layer with a linear activation function in order to produce an estimation of neutrino energy. The training was performed with about 4 million simulated events of ARCA, with 6 active DUs, equally divided among track-like ($\nu_\mu$ charged current interaction) and shower-like events ($\nu_\mu$ neutral current, $\nu_e$ charged current, $\nu_e$ neutral current interaction), including the corresponding antiparticles. The validation data set was composed of 200k events. The Monte Carlo (MC) simulated energy is used as truth reference value for this learning task. A $r^2$ score (coefficient of determination[1] ) is computed on a test data set, reaching a value of 0.835. For comparison, the $r^2$ score achieved by the standard reconstruction algorithm, based on maximum likelihood, for the same set is 0.353.

Track-like and shower-like event topologies are characterized by a different spatial evolution inside the detector, hence in Figure 1 the performances are reported separately.

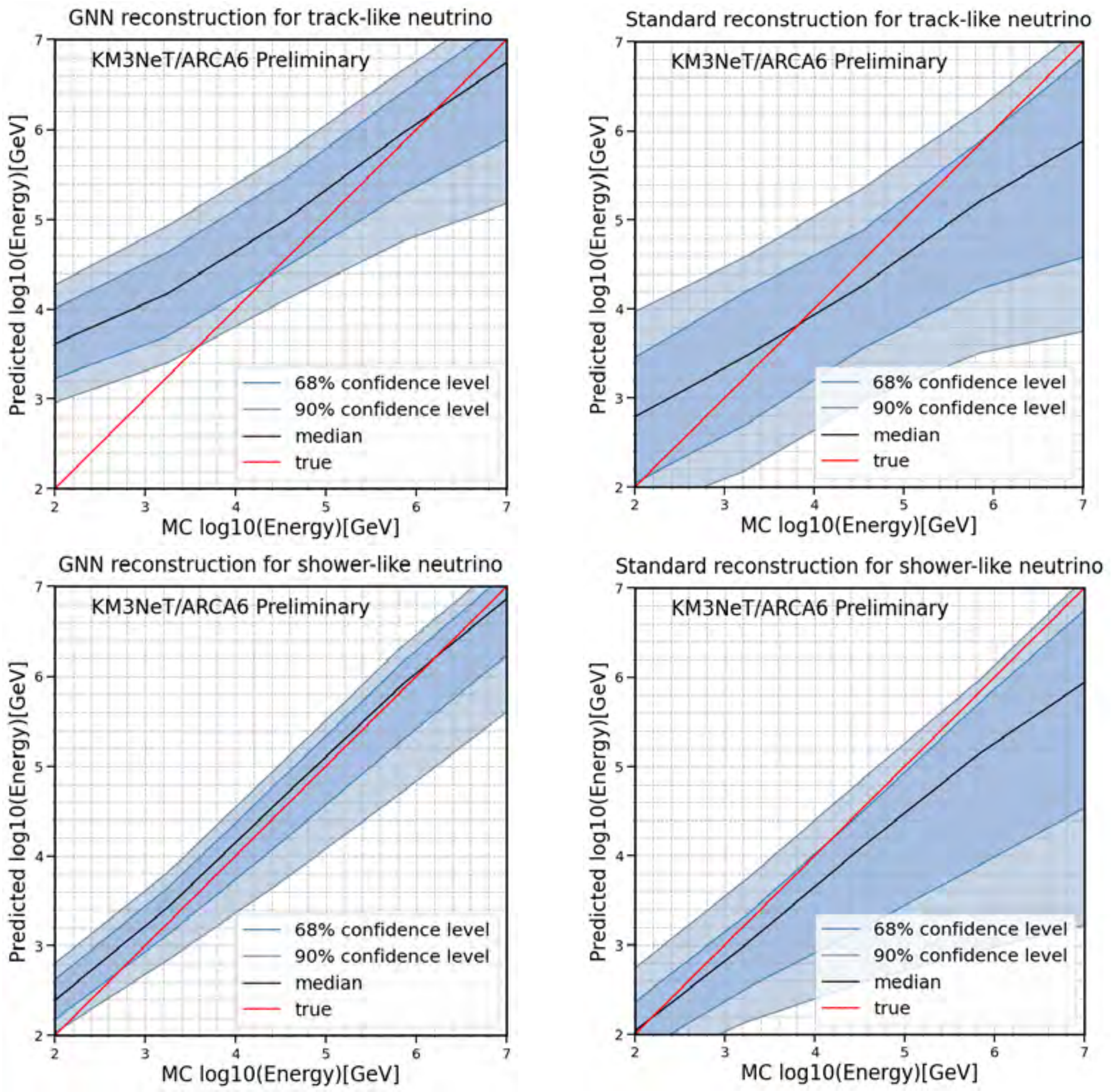

**Figure 1: Top:** predicted energies versus true MC energies for track-like events in the GNN (*top left*) and standard reconstruction case (*top right*). **Bottom:** predicted energies versus true MC energies for shower-like events in the GNN (*top left*) and standard reconstruction case (*top right*).

It is worth to notice the better performances of the shower-like topologies: in fact the GNN $r^2$ for shower-like and track-like events is respectively 0.895 and 0.628. This behaviour is probably due to the better event containment for showers.

[1] $r^2$ is calculated by using the following formula: $r^2 = 1 - \frac{SS_{res}}{SS_{tot}}$ where $SS_{res}$ is the residual sum of squares $\sum_{i=1}^{n}(y_i^{true} - y_i^{pred})^2$ and $SS_{tot}$ is the total sum of squares $\sum_{i=1}^{n}(y_i^{true} - \bar{y}^{\,true})^2$.

### 3.2 Neutrino direction regression

The reconstruction of the neutrino direction is performed by a similar GNN as adopted for the energy estimation but followed in this case by three parallel layers, one for each component of the neutrino direction $CosX$, $CosY$, $CosZ$, with custom cosine activation function. The training, validation and test events are the same as in the energy regression case.

In Figure 2 the performances for the reconstruction of the neutrino zenith with the GNN and the standard reconstruction algorithms are reported. The GNN reconstructions show a better behavior compared to the standard method for both the shower-like component ($r^2_{GNN}$ = 0.74 vs. $r^2_{STD}$ = 0.58) and the track-like component ($r^2_{GNN}$ = 0.96 vs. $r^2_{STD}$ = 0.78). The track-like component plot for the GNN case exhibits a narrow distribution around the trues values respect the standard reconstruction. However, the median of the distribution produced by the standard reconstruction shows a better agreement with the true values. The other directions show similar behavior, so they will be omitted here.

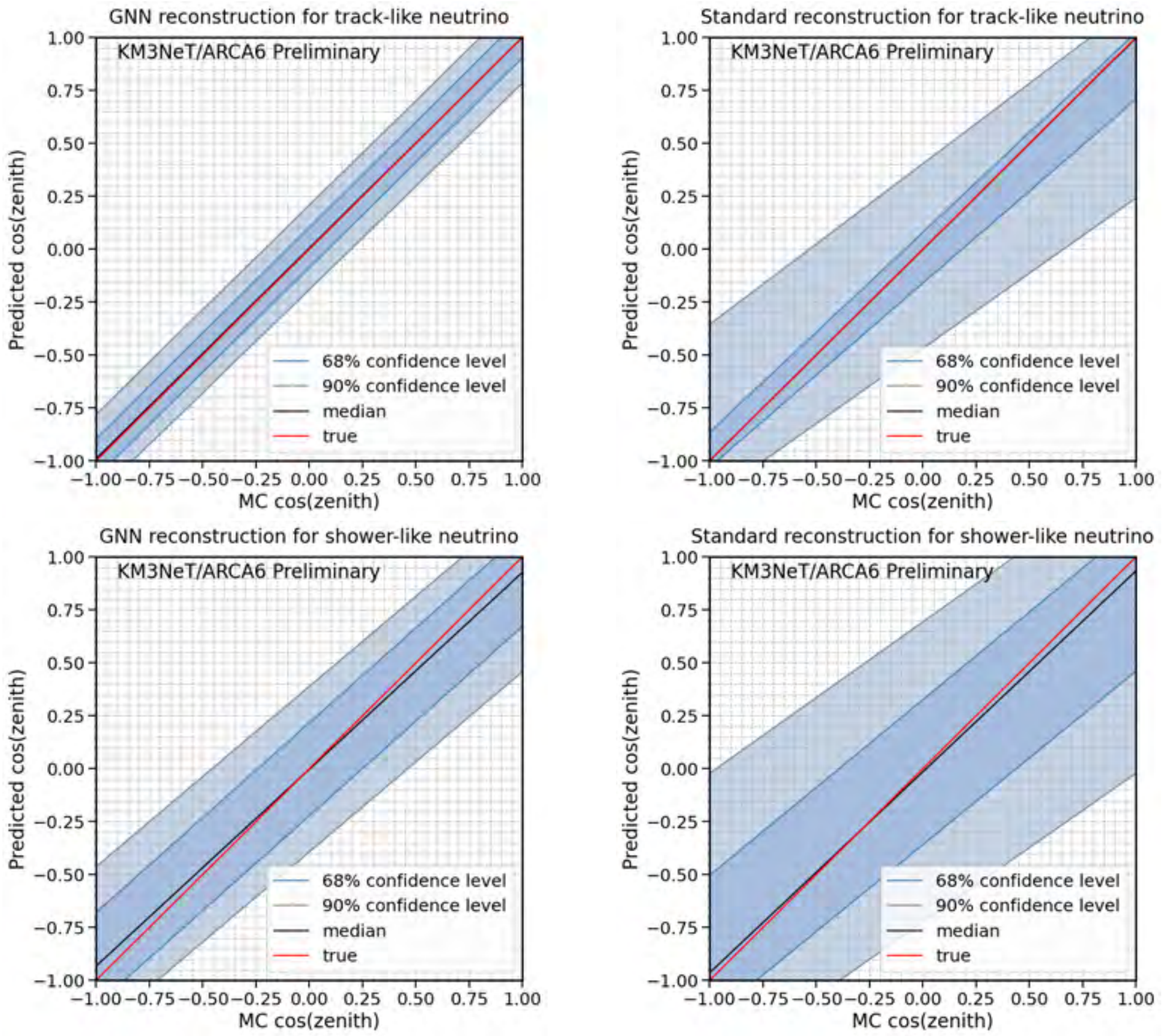

**Figure 2: Top:** predicted cosine of the zenith versus true MC value for the GNN (*top left*) and the standard reconstruction case (*top right*) for the track-like component. **Bottom:** predicted cosine of the zenith versus true MC for the GNN (*bottom left*) and standard reconstruction case (*bottom right*) for the shower component.

The trained models have also been used for inference on real data: Figure 3 shows the data to Monte Carlo comparison for energy and zenith regression.

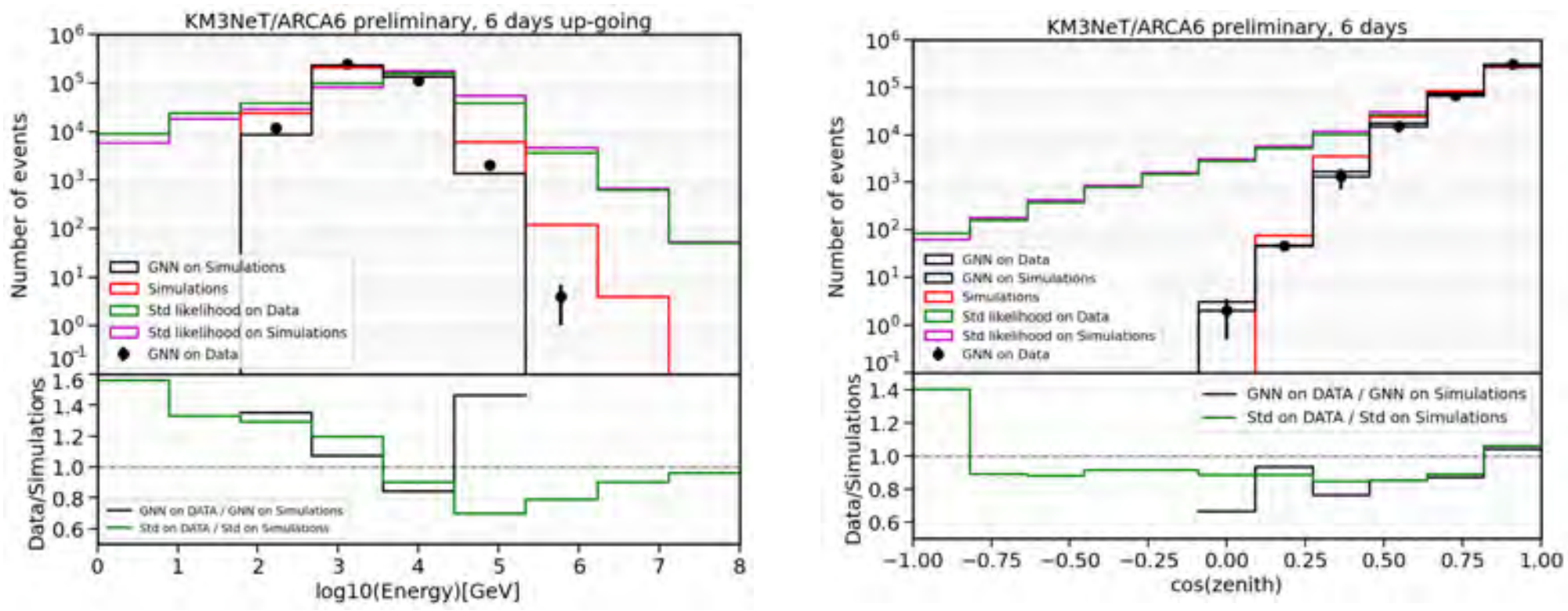

**Figure 3:** Comparisons of the inferred energies (left) and cos(zenith) (right), varying the inference test data set and reconstruction methods.

The previous plots have been obtained applying a cut on the number of triggered hits for each event in order to suppress events generated by environmental optical background, mainly due to $^{40}K$ decay. The good agreement between the GNN inference on real data (back dots), GNN inference on Monte Carlo (black lines), and the Monte Carlo simulated value (red lines) for both energy and zenith is clearly noticeable.

## 4. Signal/background classification

A classification model has been trained to distinguish between atmospheric muons and neutrinos. The classifier produces a score for each event, ranging from 0 to 1, that represents the probability of the event to be of a certain class. During the training phase the GNN takes as input graphs created from MC simulations of the recorded hits for each event: specifically, each hit represents a node of the graph and causality relations between hits represent the edges of the graph. The training process exploits approximately 90% of the data set for training and 10% for validation. Three training sessions were conducted, one for each different detector configuration geometry: one for ARCA with 6 active DUs (ARCA6), one using ARCA7 and one employing ARCA8. The training sets consist of 800k events for ARCA6, 500k for ARCA7, and 1 million events for ARCA8, all with a fraction of 50% atmospheric muons and 50% neutrinos. The networks are trained with k-nearest neighbors equal to $k = 16$, ReLU activation function and Adam optimizer ($\beta_1 = 0.9$, $\beta_2 = 0.999$ and $\varepsilon = 0.1$).
The inference of the network trained on ARCA6 has been performed on a total lifetime of 45 days, for the ARCA7 trained GNN, a period of 25.5 days has been used, while for the network trained on ARCA8, a period of 22.2 days was examined. In total 93 days have been analysed. The analysis results are depicted in Figure 4, where the probability for each event to be classified as neutrino is reported. In order to exclude events mainly due to $^{40}K$ decay, the same event selection described in Section 3.2 has been applied. It can be noted a peak of events with very high neutrino score in the data, compatible with an excess of atmospheric neutrinos in that region of the neutrino score. The data-Monte Carlo comparison is compatible with values obtained in other KM3NeT analyses, exploiting other selection methodologies [4].

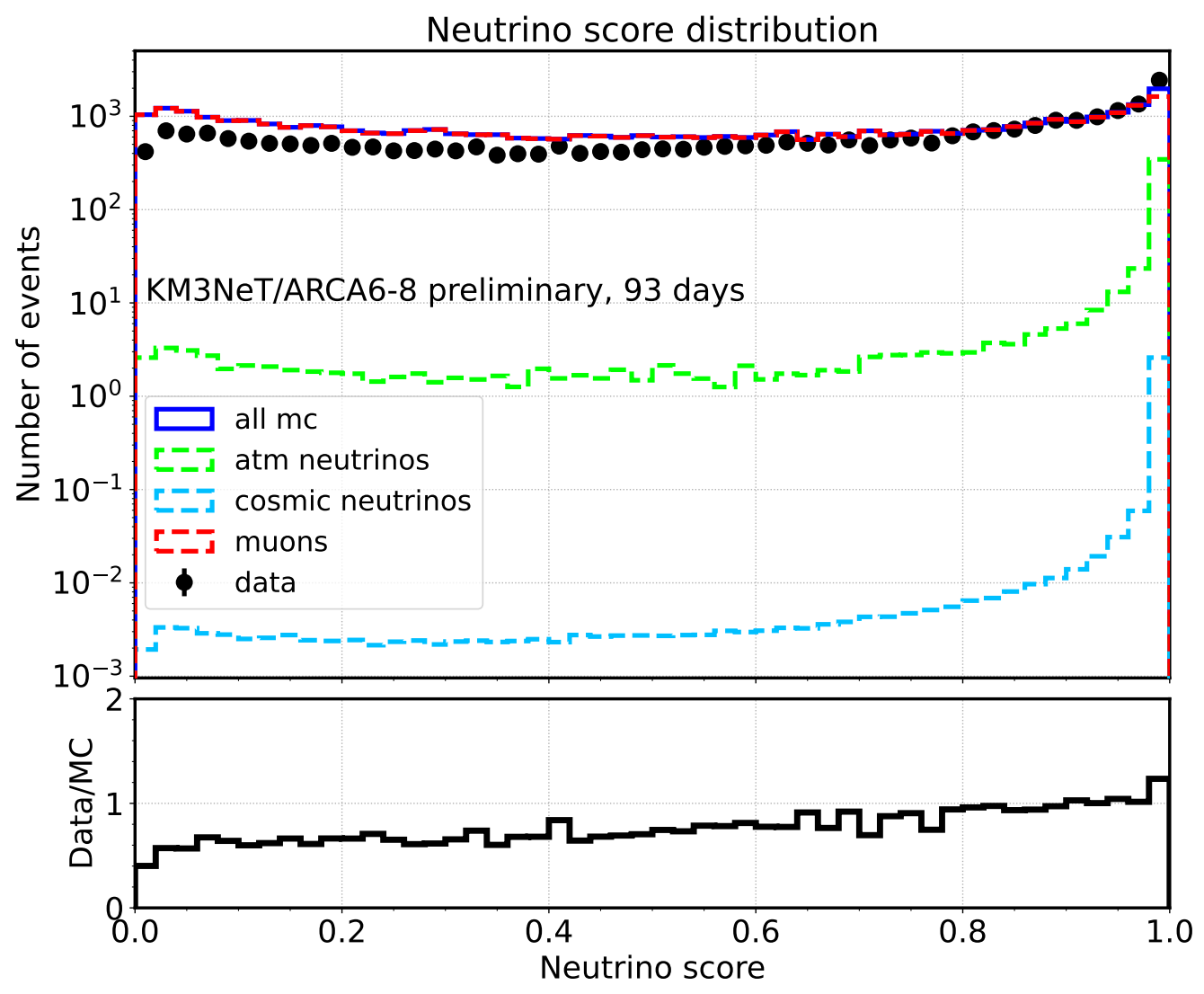

**Figure 4:** Probability of the events to be classified as neutrino for ARCA6-8.

## 5. Muon bundles reconstruction

The bulk of events measured in a neutrino telescope corresponds to atmospheric muons, induced by the interaction of cosmic rays (CRs) with the nuclei in the atmosphere. A common technique of background rejection is to use the Earth as a shield, selecting therefore only upgoing particles. Atmospheric muons can be exploited for a variety of physics cases and detector checks, like primary CR composition studies and validation of detector performances and calibrations. A great challenge is the reconstruction of the properties of atmospheric muons, especially if multiple muons traverse the detector simultaneously (muon bundles) [5].

### 5.1 Direction reconstruction

A regression model has been trained on Monte Carlo events, in order to infer the direction (zenith and azimuth) of the incoming bundle. Each muon inside the bundle, even if having slightly different zenith angles, is simulated with a parallel direction, therefore the two targets for the regression are uniquely defined. The architecture adopted for this task follows the ParticleNet structure described in Section 2. The difference is in the output activation function, for which a normal distribution is assumed in order to measure also an uncertainty associated to the reconstruction. In Figure 5, the 2D distribution of predicted cosine of zenith as a function of the true simulated value is reported, after applying a cut on the uncertainty value itself.

The training sample was composed of ~2 million events, and the validation set was chosen to be 10% of the training one.

An important aspect of Deep Learning algorithms is their robustness, when applied to data. The performance of the GNN algorithm is compared to the standard reconstruction which is based on the maximization of a likelihood function. The data-Monte Carlo comparison for both the algorithms is reported in Figure 6.

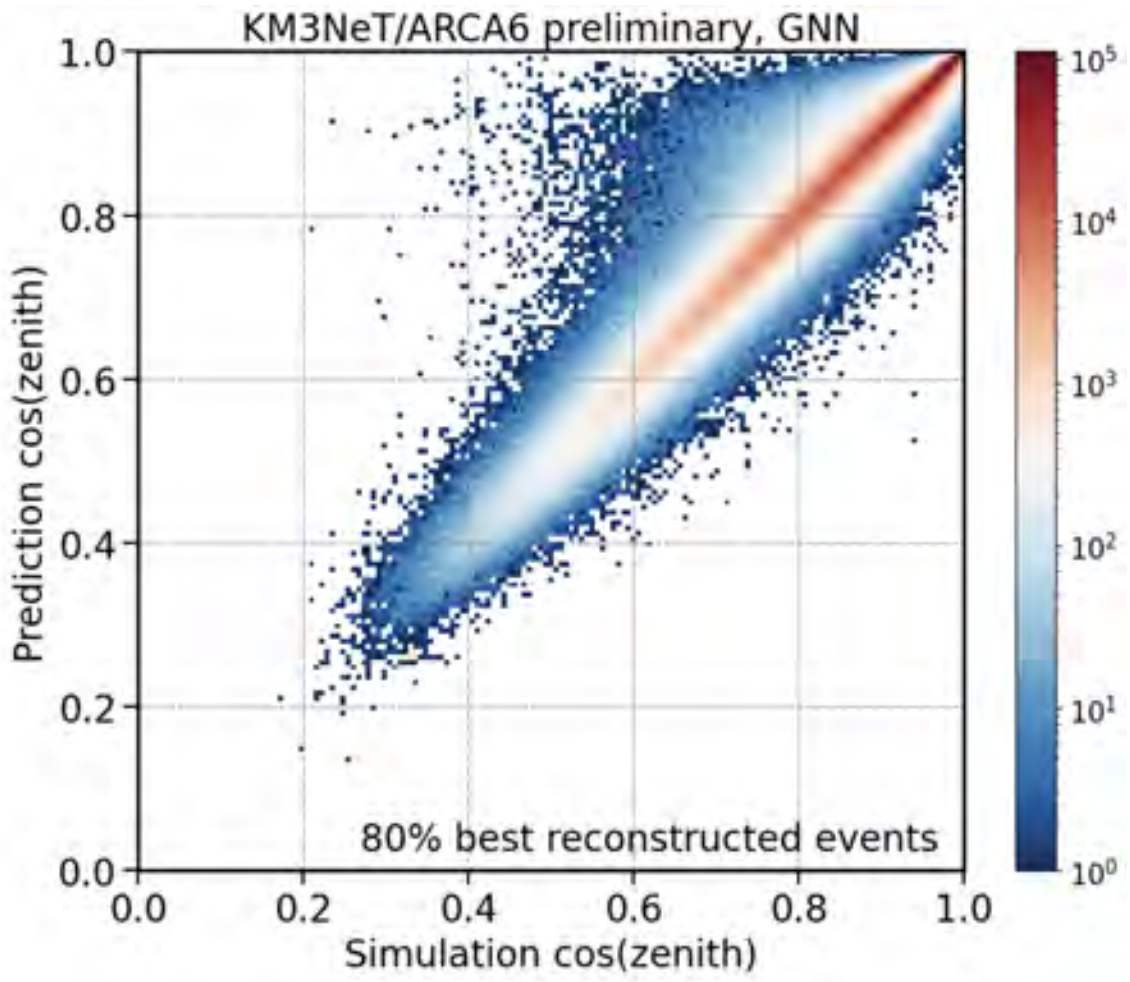

**Figure 5:** 2D distribution of the predicted cosine of zenith as a function of the true MC one, selecting the best 80% reconstructed events, cutting on the uncertainty estimation.

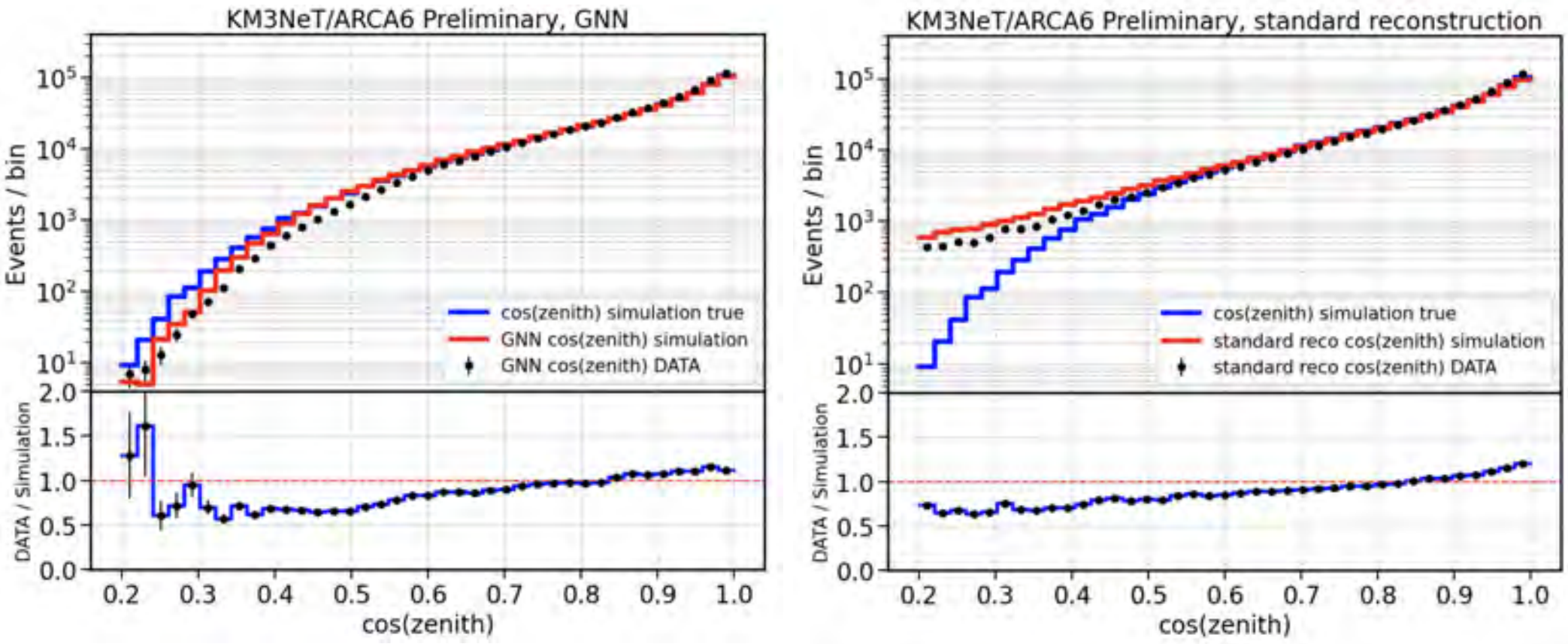

**Figure 6:** Data-Monte Carlo comparison of the reconstructed cosine of the zenith angle for the GNN (**left**) and for the standard algorithm (**right**). In both plots the Monte Carlo true cosine of the zenith is also reported.

### 5.2 Muon bundle multiplicity reconstruction

A graph neural network was also trained in order to infer the number of muons in a bundle traversing simultaneously the detector, called *muon multiplicity*. The muon multiplicity is a very important parameter since it allows various studies of CRs and their interactions, and in particular to estimate the mass of the CR primary particle. At the same time, the light yield produced by muon bundles, when traversing the detector, is collected and reconstructed by classical algorithms as a single muon, with energy equal to the sum of the energies of the muons in the bundle.

The choice on the architecture was done similarly to the direction regression task, even if the muon multiplicity is a discrete quantity, and bound from below. An example of the network capabilities is shown in Figure 7: the 2D histogram represents the predicted multiplicity in function of the true one. Also in this case the network is capable to provide not only the inferred multiplicity on a given event but also the error on the estimate itself. In Figure 7 also the data-Monte Carlo comparison for the predicted muon multiplicity is reported showing a general agreement, especially at mid to high muon multiplicities.

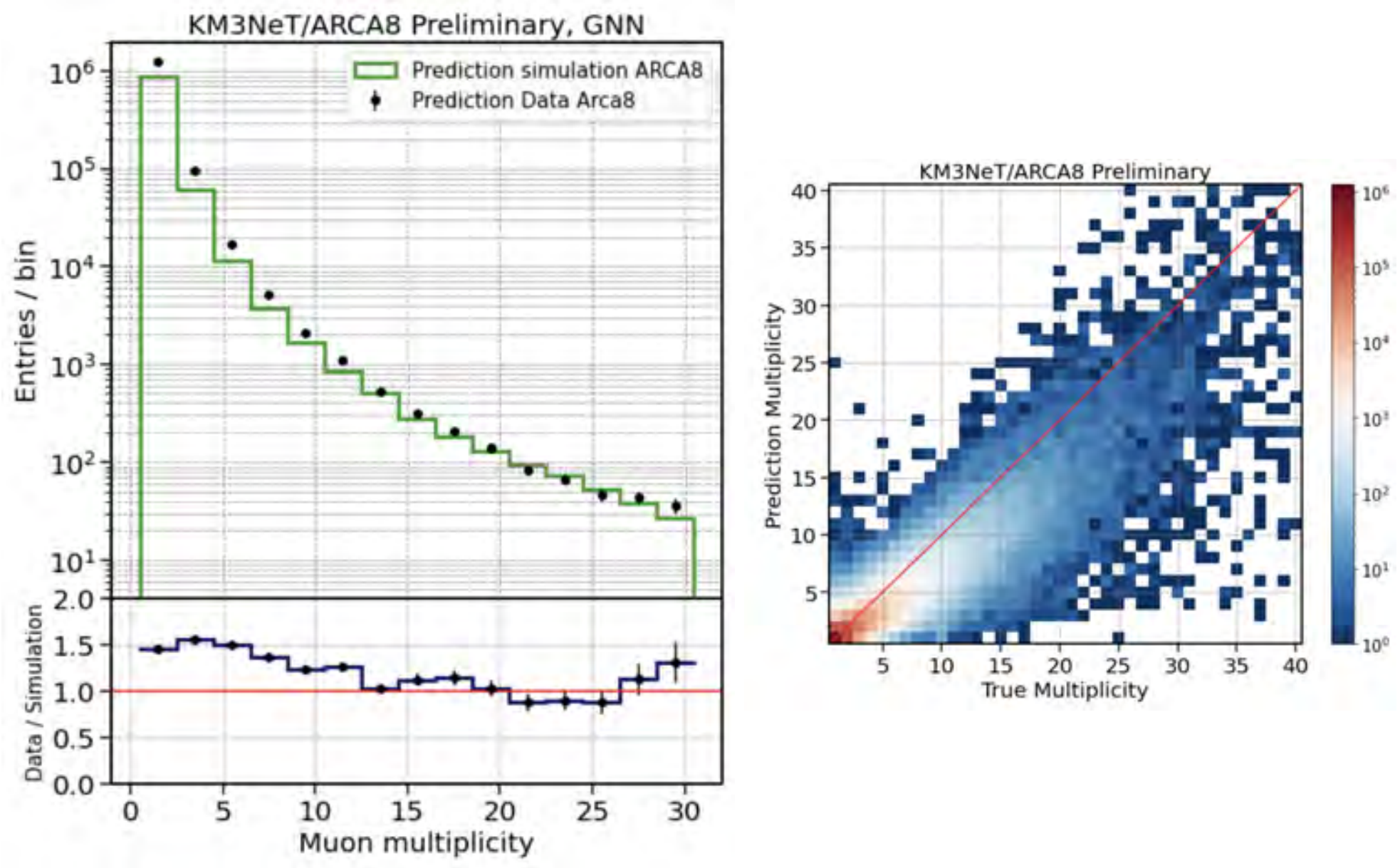

**Figure 7: Left:** Data Monte Carlo comparison on ARCA8 data set for GNN predicted muon multiplicity. **Right:** 2D histogram showing the predicted muon multiplicity in function of the true one for ARCA8 data set.

## Acknowledgments

A. Domi acknowledges the support by the HORIZON-MSCA-2021-PF-01 project QGRANT No. 101068013.

# Search for a diffuse astrophysical neutrino flux with KM3NeT/ARCA using data of 2021-2022

**Vasileios Tsourapis,**[a,b,*] **Evangelia Drakopoulou,**[a] **Christos Markou,**[a] **Anna Sinopoulou**[c] **and Ekaterini Tzamariudaki**[a] **on behalf of the KM3NeT collaboration**

[a] *NCSR "Demokritos", Institute of Nuclear and Particle Physics, 15310 Ag. Paraskevi Attikis, Athens, Greece*

[b] *National Technical University of Athens, School of Applied Mathematical and Physical Sciences, Zografou Campus, 9, Iroon Polytechniou str, 15780 Zografou, Athens, Greece*

[c] *INFN Sezione di Catania, Via Santa Sofia,64 - 95123 Catania, Italy*

*E-mail:* tsourapis@inp.demokritos.gr

KM3NeT is a research infrastructure hosting two neutrino detectors which are currently under construction in the Mediterranean Sea. The KM3NeT/ARCA detector focuses on the detection of high energy neutrinos (>TeV) from astrophysical sources. In this contribution, an analysis of the data obtained during 2021-2022, using machine learning techniques, is presented. The potential of KM3NeT/ARCA to detect a diffuse flux of astrophysical neutrinos at the early stages of the detector construction is reported. The results of this analysis are presented.

*Speaker

## 1. Introduction

One of the main goals of modern astronomy is to resolve the particle acceleration mechanisms at place at distant astrophysical sources. The measurement of a diffuse flux of cosmic neutrinos can provide an insight to this endeavor. Responsible for the diffuse astrophysical neutrino flux observed at Earth are: a) the high energy cosmic rays which produce neutrinos through $pp$ and $p\gamma$ interactions while traveling at cosmic distances and b) the collection of emitters that can not be detected individually.

At source, the ratio of the neutrino flavours is typically assumed to be $\nu_e : \nu_\mu : \nu_\tau = 1 : 2 : 0$. The probability of a certain flavour $\alpha$ oscillating to a different flavour $\beta$ depends on the distance traveled (L) and the neutrino energy (E): $P_{\alpha\to\beta} \sim L/E$. Neutrinos traveling over cosmic distances are expected to reach the Earth with a ratio $\nu_e : \nu_\mu : \nu_\tau = 1 : 1 : 1$. The energy spectrum of these neutrinos is typically modeled as a power-law, $\Phi = \Phi_0 E^{-\gamma}$. In this work, the following parametrization is assumed:

$$\Phi_{\nu+\bar{\nu}}(E) = \frac{dN}{dE} = \phi_0 \cdot 10^{-18} \cdot \left(\frac{E}{100\ \text{TeV}}\right)^{-\gamma} \left[\text{GeV}^{-1}\text{cm}^{-2}\text{s}^{-1}\text{sr}^{-1}\right] \tag{1}$$

## 2. Detection method

### 2.1 KM3NeT detector

The KM3NeT [1] infrastructure currently hosts two Cherenkov neutrino detectors under construction in the Mediterranean Sea: KM3NeT/ARCA, offshore Capo Passero (Italy) and KM3NeT/ORCA, offshore Toulon (France). KM3NeT/ORCA is designed for studying atmospheric neutrino oscillations, while KM3NeT/ARCA is optimized for detecting neutrinos of astrophysical origin.

ARCA is located at a depth of 3500 m and, when complete, will consist of 2 Building Blocks of 115 Detection Units (DUs) each. Every DU is a vertical, string-like, structure anchored to the seabed and carrying 18 Digital Optical Modules (DOMs). A DOM [2] is the main detector element, housing 31 3"-PMTs (photo-multiplier tubes). DOMs are spaced by 36 m along a DU, with an horizontal spacing between DUs of 90m. Both ORCA and ARCA share the same design, only differing in the spacing between DUs (20 m for ORCA) and DOMs along a DU (9 m for ORCA), since the two detectors are optimised for the detection of neutrinos of different energy ranges.

Currently, the ARCA detector consists of 21 DUs and the ORCA detector of 18 DUs.

### 2.2 Data taking periods

Data taking periods refer to the number of deployed DUs at running time. A stable data taking period starts after the new DUs are commissioned and calibrated. In this study, the full data sample of the ARCA6/ARCA8 running periods is analyzed. ARCA6 ran from May 2021 through September 2021 for a total of 102 analyzed days and ARCA8 from September 2021 through June 2022 for a total of 212 analyzed days. The analysis has been then extended for the recent detector configurations, ARCA19 (July 2022 - September 2022) [51 analyzed days] and ARCA21 (September 2022 – December 2022) [1] [67 analyzed days] and the final results are presented. The event selection for those two periods is described in Section 5.

## 3. Event selection

A collection of photons that impinge on PMTs (hits) that occur within a certain distance and a specific time window, constitute an event. The underwater environment produces noise hits to the PMTs, mainly through $^{40}K$ decays and bioluminescence. These contributions can easily be eliminated by applying appropriate constraints on the reconstruction variables. Signal events come from neutrinos and anti-neutrinos of astrophysical origin of all 3 flavours and both interaction channels (CC and NC) reconstructed as tracks. Atmospheric (anti)neutrinos and muons represent the physics background.

### 3.1 Preselection

Extensive data-MC comparisons have been performed to evaluate the description of the detector performance. Several reconstruction variables are used during the preselection in order to remove the contribution of noise, as well as of events with a poor reconstruction performance, based on the previous analysis on the diffuse astrophysical neutrino flux [3]. Constraints are applied on the basis of the goodness of fit of the event reconstruction, the length of the track in the detector volume, the angular error estimate from the event reconstruction, the number of triggered DOMs, and the value of the energy estimate of the event.

**Upgoing** events, with zenith > $90^o$, are selected, in order to suppress the background from atmospheric muons since only (anti)neutrinos can survive passing through the Earth.

### 3.2 Usage of a Boosted Decision Tree

A Boosted Decision Tree (BDT) classifier, using ROOT TMVA [4], was developed, based on the previous diffuse analysis [3], in order to better reject the atmospheric muon background. Since only upgoing events are selected, all remaining atmospheric muons are wrongly-reconstructed events.

The BDT is trained using as *class 0* well reconstructed (WR) $\overset{(-)}{\nu}_\mu$ CC interactions (the dominant track-like channel) and as *class 1* badly reconstructed (BR) muons; this definition relies on the amplitude of the angular difference, $\Delta\Omega$, between the neutrino direction from Monte Carlo (MC) simulations and the reconstructed track.

The model was trained using a combination of a special MC production dedicated for machine learning and a 10% of the standard MC production of the ARCA8 dataset. The rest of the MC was used for the evaluation. As input to the BDT, 15 variables extracted from the output of the event reconstruction were used, as the ones with the best separation power and smallest correlation with each other. Using the Grid Search technique, the best combination of hyper-parameters for the BDT was selected. After thorough tests, the classifier was deemed suitable to be applied also to the ARCA6 dataset.

[1]ARCA21 detector configuration is expected to complete its uptime on September 2023, when new detection units will be added. The dates here refer to the time period analyzed in this contribution.

A **blind policy** has been followed, in order to avoid any biases, according to which all cuts are optimized only on the MC and 10% of the data.

The performance of the classifier can be seen in Figure 1. For atmospheric (anti)neutrinos the Honda [5] and Enberg [6] models were used, once a correction for the presence of the cosmic ray knee was applied. For all the optimizations, the cosmic neutrino flux used, following the normalization of eq. 1, is the one presented by IceCube [7] with $\phi_0 = 1.44$ and spectral index $\gamma = 2.37$.

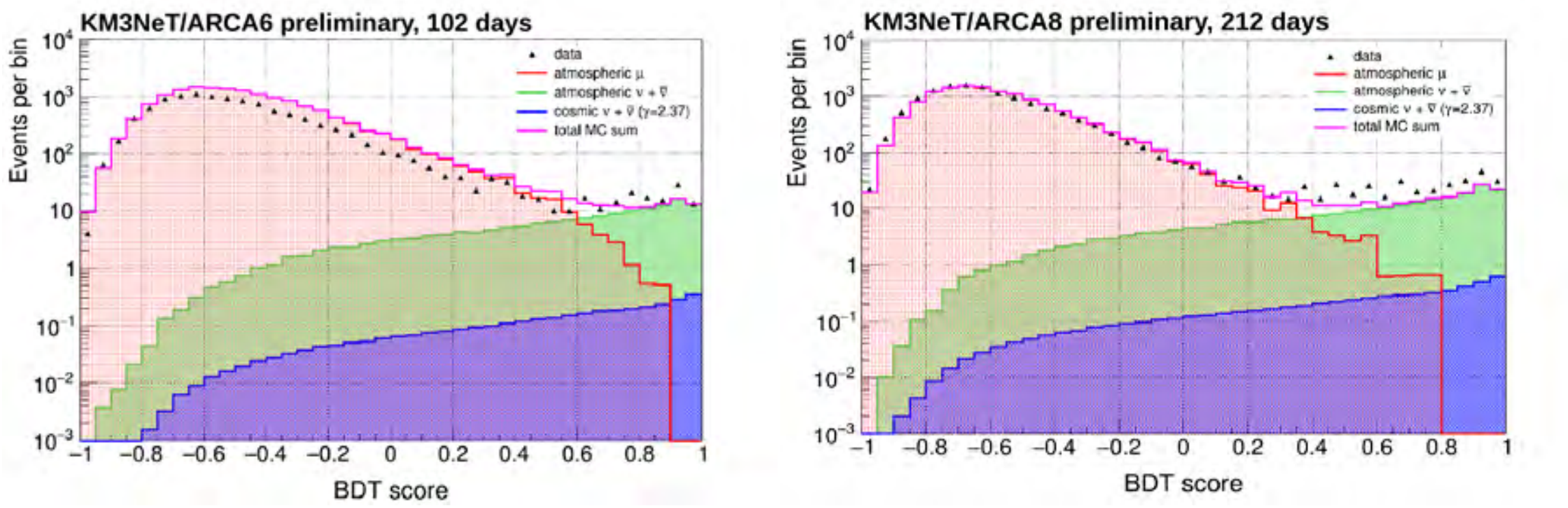

**Figure 1:** The BDT score distribution for all events for the ARCA6 and ARCA8 periods. 100% of the data is shown in this Figure.

The choice of the optimal BDT score cut was made after *evaluating* on a 3-flavour signal, based on the requirement that an eciency more than 70% is reached, and that the Model Rejection Factor (MRF) is optimised [8]. The MRF procedure, based on the Feldman-Cousins [9] upper limit estimations, is also used to select the optimal energy cut value that selects the cosmic signal.

In order to account for the low statistics of the atmospheric muon MC simulation for high BDT scores and high muon energies, the atmospheric muon contribution is extracted by fitting the estimated energy spectrum *after* the BDT score selection. Also, *after* the BDT selection, the atmospheric muon contributions were scaled up by 40% to compensate for the data-MC differences. Consequently, a systematic error of 40% was assigned as referenced in Section 4.2.

The energy resolution for events surviving the BDT score cut is shown in Figure 2.

### 3.3 Final sample

The BDT score for ARCA6 dataset was required to be more than 0.35 and for ARCA8 more than 0.27. The optimal threshold on the logarithm of the energy estimate is 4.20 and 4.04 for ARCA6 and ARCA8, respectively.

The number of the events surviving the final selection for each category, after *unblinding*, is shown in Table 1 and the corresponding energy distributions in Figure 3.

## 4. Analysis method

### 4.1 Binned-Likelihood method

Data are analyzed using a binned-likelihood method, described in [10]. Based on the Bayesian interpretation of probability, the possibility of a certain hypothesis to be true, given some experi-

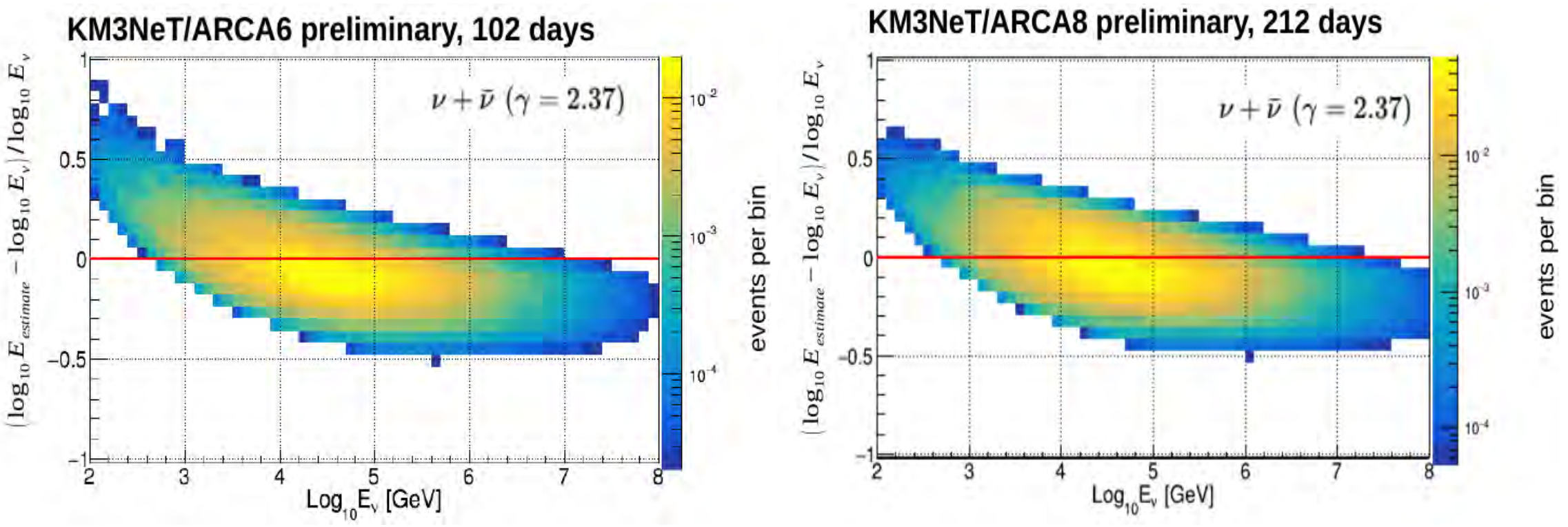

**Figure 2:** Energy resolution for cosmic neutrinos, after the BDT score cut is applied for ARCA6 (score > 0.35) and ARCA8 (score > 0.27), versus the true (MC) neutrino energy.

| | ARCA6 | | ARCA8 | |
|---|---|---|---|---|
| | BDT score > 0.35 | $\log E_{estimate} > 4.20$ GeV | BDT score > 0.27 | $\log E_{estimate} > 4.04$ GeV |
| atm. $\nu + \bar{\nu}$ | 117.68 | 16.08 | 185.18 | 33.04 |
| atm. $\mu$ | 150.91 | 39.07 | 49.12 | 25.39 |
| all atmospheric | 268.59 | 55.15 | 234.30 | 58.43 |
| cosmic $\nu + \bar{\nu}$ | 2.54 | 1.40 | 4.76 | 3.06 |
| data | 223 | 26 | 365 | 61 |

**Table 1:** Events at final selection level. Cosmic $\overset{(-)}{\nu}$ with $\phi_0 = 1.44$ and $\gamma = 2.37$

mental data, is given by:

$$p(\vec{\theta}; x) = \frac{L(x; \vec{\theta})\pi(\vec{\theta})}{\int L(x; \vec{\theta})\pi(\vec{\theta})d\vec{\theta}} \tag{2}$$

with $x$ being the experimental data, $\vec{\theta}$ a set of parameters describing the hypothesis under test, $\pi(\vec{\theta})$ the *prior* distribution and $L(x; \vec{\theta})$ the *likelihood* function, which is the joint probability of the experimental data corresponding to the the hypothesis. The maximum of the likelihood will give the best estimation of the parameters for the hypothesis.

In our case:

$$L(N; S(\gamma), B, \phi_0) = \prod_i Poisson(N_i, B_i + \phi_0 S_i(\gamma)) \tag{3}$$

where, $N_i$ is the number of data events in bin i, $S_i$ is the number of expected signal events for a given spectral index $\gamma$, with $\phi_0 = 1$ and $B_i$ is the number of expected background events. The product runs over all bins of the energy estimate.

The marginalization of the posterior probability, that is to integrate out the *nuisance* parameters, gives:

$$p(\phi_0, \gamma; N) = \int L(N; S(\gamma), B, \phi_0) \cdot \pi(B_i) \cdot \pi(S_i(\gamma)) \cdot \pi(\phi_0, \gamma) \cdot \prod_i dB_i dS_i(\gamma) \tag{4}$$

The prior distributions incorporate the uncertainties of the MC estimates. Here: $\pi(B_i) = Gaussian(\mu = B_i, \sigma = \sigma_{B_i})$, $\pi(S_i(\gamma)) = Gaussian(\mu = 1, \sigma = \sigma_{S_i})$ are used, and $\pi(\phi_0, \gamma) = 1$, since no previous knowledge of $\phi_0$ and $\gamma$ is assumed.

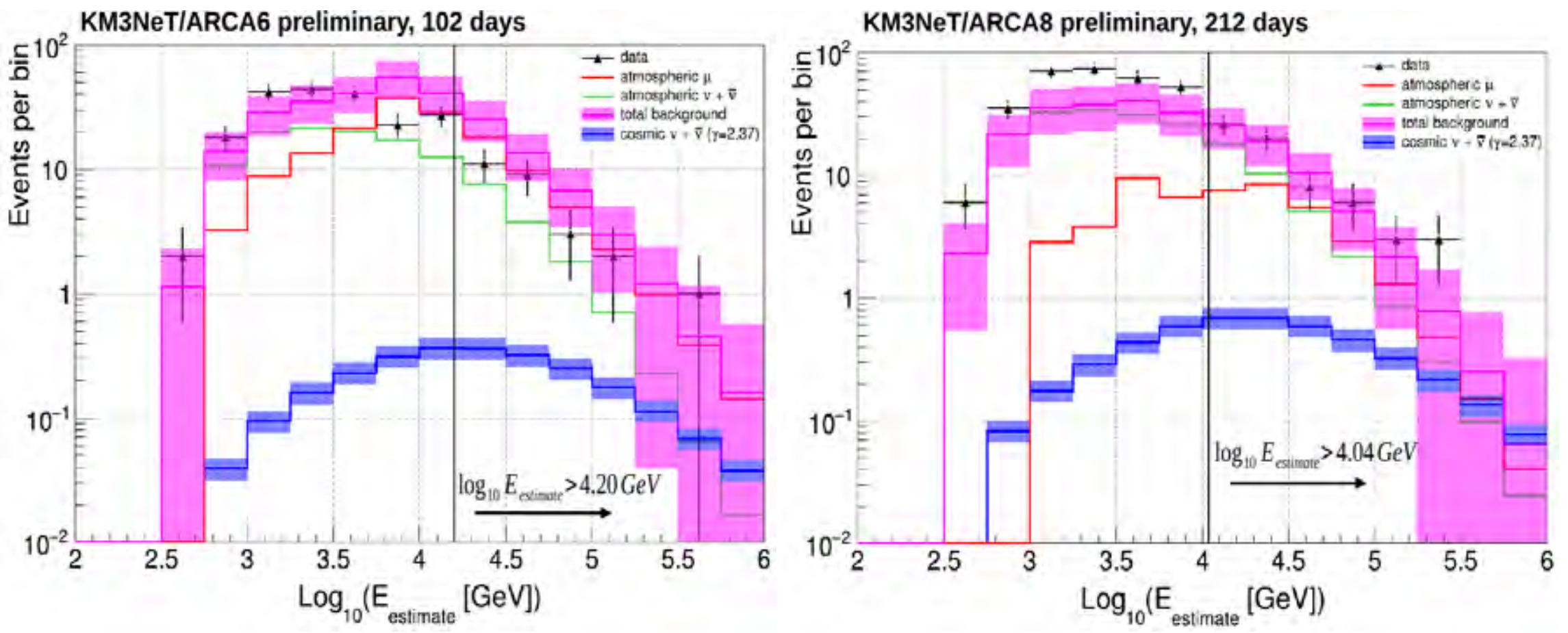

**Figure 3:** Distributions of the energy estimate for ARCA6 and ARCA8 after applying the BDT score cut. The total background of atmospheric neutrinos and muons (purple) and signal (blue) are shown with their total uncertainties. The vertical line in each plot shows the optimal threshold of the energy estimate.

### 4.2 Uncertainties

Statistical and systematic uncertainties are taken into account through the gaussian priors of eq. 4. For both the ARCA6 and ARCA8 running periods, the same uncertainties were considered. For the *signal*, an overall uncertainty of 20% for every energy bin is assumed. For the *background* the total uncertainty, for every energy bin, is given as:

$$\sigma_{B_i} = \sqrt{\sigma^2_{M_i} + \sigma^2_{A_i}} = \sqrt{\sigma^2_{stat,(M_i)} + \sigma^2_{data/mc,(M_i)} + \sigma^2_{stat,(A_i)} + \sigma^2_{simulation,(A_i)}} \tag{5}$$

since the background (B) consists of of atmospheric $\nu + \bar{\nu}$ (A) and atmospheric $\mu$ (M).

Here, $\sigma_{stat,(M_i)}$ and $\sigma_{stat,(A_i)}$ is the statistical error for atmospheric muons and atmospheric (anti)neutrinos respectively. $\sigma_{data/mc,(M_i)}$ is the systematic uncertainty for the atmospheric muon background estimation equal to 40% due to the difference between data and the expected events from MC simulation. $\sigma_{simulation,(A_i)}$ is the collective uncertainty on the atmospheric (anti)neutrino simulation, equal also to 40%. Uncertainties on the (anti)neutrino simulation come from the effect of uncertainties on the water properties, the uncertainty on the PMTs efficiency and the atmospheric (anti)neutrino flux normalization [11].

## 5. ARCA19/ARCA21 samples

A similar procedure, as the one described in previous sections, was followed for the ARCA19 and ARCA21 data samples. After some tests, the same *Preselection* as for ARCA6/8 periods was applied to the ARCA19/21. A special MC production dedicated to be used in machine learning (ML) techniques was made for both the ARCA19 and the ARCA21 periods. A BDT was trained on the combination of the two ML samples with the same *class 0* and *class 1* as for ARCA6/8. The model used the same hyper-parameters and similar input variables. An additional constraint based on a different way of calculating the length of the track was applied for ARCA19/21. The optimization of the BDT and the MRF procedure returned a BDT score > 0.45 and $\log E_{estimate}$ > 4.12 GeV, and BDT score > 0.40 and log$E_{estimate}$ > 4.36 GeV for ARCA19 and ARCA21 respectively. The same statistical analysis strategy was followed for those two periods. After applying the BDT score cut, a fit to the atmospheric muon energy distribution was performed for both periods. *No* other modifications to the MC were made (in contrast to ACA6/8). Following eq. 5, a *conservative* systematic uncertainty of 40% was applied to the atmospheric muon contribution, and also a 40% systematic uncertainty on the atmospheric (anti)neutrino simulation. A 20% total uncertainty was applied to the signal. The same binned-likelihood method was exploited for the ARCA19/21 periods and their combination with ARCA6/8.

## 6. Results

For the computation of the Likelihood, 8 bins for the log$E_{estimate}$ ∈ [4,6] were used for each period. The spectral index was scanned in the range [1,4], and the flux normalization, $\phi_0$, in [0.001,10].

By profiling the posterior (eq. 4) to specific spectral indices, the upper limits (U.L.s) to the diffuse astrophysical neutrino flux are found. In Table 2, the 90% confidence level (C.L.) U.L.s are given for our baseline $\gamma = 2.37$ and also for $\gamma = 2.0$ and 2.5 in order to compare with the values reported by ANTARES' all-flavour search for a diffuse flux with 9 years of data [12].

| | ARCA6+8 | ARCA19+21 | ARCA6+8+19+21 | ANTARES | 5% quantile | 95% quantile |
|---|---|---|---|---|---|---|
| $\gamma = 2.0$ | 5.11 | 3.13 | 2.09 | 4.0 | 15.07 TeV | 11.71 PeV |
| $\gamma = 2.37$ | 6.92 | 4.68 | 3.06 | | 5.88 TeV | 1.73 PeV |
| $\gamma = 2.5$ | 6.76 | 4.94 | 3.12 | 6.8 | 4.43 TeV | 1.03 PeV |

**Table 2:** 90% C.L. upper limits for ARCA6+ARCA8 and ARCA19+ARCA21 periods and their combination compared with 9 years of ANTARES, computed for the central 90% energy range of the signal events. Numbers reported are $\times 10^{-18}[\text{GeV}^{-1}\text{cm}^{-2}\text{s}^{-1}\text{sr}^{-1}]$ in accordance to eq. 1. Energy quantiles refer to the combined ARCA6+8+19+21 period.

The convolution of sensitivities and U.L.s for selected spectral indices in the range [2.1,2.5] in terms of energy-flux $E^2\Phi(E)$ versus the (anti)neutrino energy, calculated for the central 90% energy range of the signal events, can be seen in Figure 4.

## 7. Conclusions and outlook

In this work, an analysis of the full dataset collected from the first KM3NeT/ARCA configuration of appreciable volume, namely ARCA6 & ARCA8, and also of the recent configuration ARCA19 & ARCA21 was presented, focusing on the search for a 3f diffuse astrophysical neutrino flux. Calculations were performed using as a baseline the one-flavour IceCube neutrino flux with $\Phi_0 = 1.44 \cdot 10^{-18}$ and $\gamma = 2.37$ at 100 TeV normalization. An event selection was applied for all periods with a similar preselection and by considering only *upgoing* events. A BDT was used to further discriminate signal from background. The MRF procedure was applied for selecting the BDT score and the cut on the optimal energy estimate for each period. A binned likelihood method was used for the statistical analysis of the signal properties. Statistical and systematic uncertainties were taken into account. The 90% upper limits set on the cosmic neutrino flux for various spectral indices are reported, and compared with the ones obtained with 9 years of ANTARES data. The results are within expectations and demonstrate the potential of KM3NeT/ARCA for measuring a diffuse astrophysical neutrino flux.

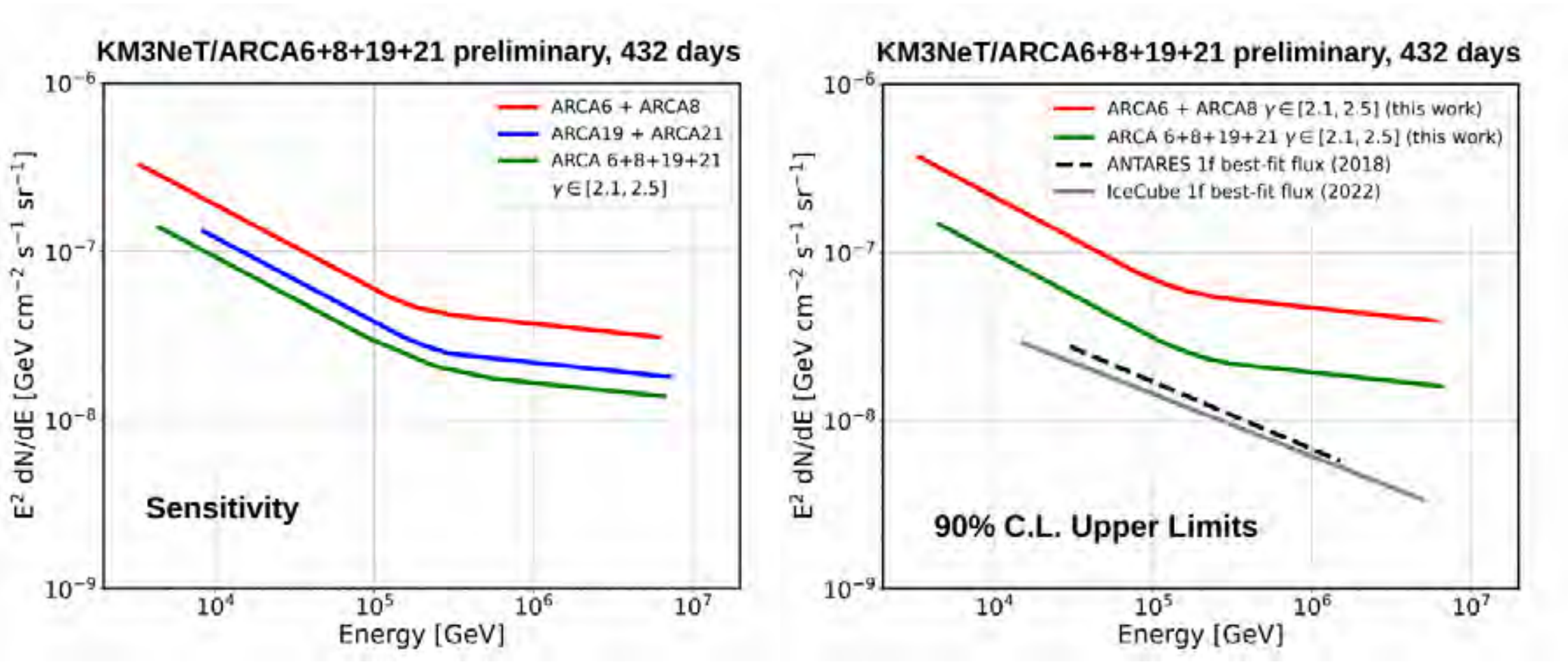

**Figure 4: (Left)** Convolution of sensitivities at 90% C.L. for selected $\gamma$ in range [2.1,2.5] for ARCA6+ARCA8, ARCA19+ARCA21 and their combination, computed for the central 90% energy range of the signal events. **(Right)** Convolution of U.L.s at 90% C.L. for the same spectral indicies and energy ranges. As a comparison, IceCube's and ANTARES' 1-flavour best fit flux is drawn.

## Acknowledgements

Vasileios Tsourapis acknowledges the support of the Hellenic Foundation for Research and Innovation (HFRI) under the 3rd Call for HFRI PhD Fellowships (Fellowship Number: 5403).

H.F.R.I. Hellenic Foundation for Research & Innovation

# Implementation of the KM3NeT Online Core-Collapse Supernova neutrino search

**C. Donzaud,[b] D. Dornic,[a,*] S. El Hedri,[b] I. Goos,[b] V. Kulikovskiy[c] and G. Vannoye[a,*] for the KM3NeT collaboration**

[a] *Aix Marseille Univ, CNRS/IN2P3, CPPM, Marseille, France*
[b] *Université de Paris, CNRS, Astroparticule et Cosmologie, F-75013 Paris, France*
[c] *INFN, Sezione di Genova, Via Dodecaneso 33, Genova, 16146 Italy*

*E-mail:* dornic@cppm.in2p3.fr, vannoye@cppm.in2p3.fr

The detection of a neutrino burst from the next Galactic Core-Collapse Supernova (CCSN) will provide us invaluable information on this extreme phenomenon. Furthermore, the detection of its gravitational waves and electromagnetic signals would give us a complete picture of all emitted messengers. KM3NeT is a neutrino telescope consisting of two detectors, ORCA and ARCA, currently under deployment in the Mediterranean Sea. By looking for an excess of coincidence events above the optical background, it will be able to detect low-energy neutrinos from CCSN. A sensitivity to Galactic and near-Galactic events is expected when data from the two infrastructures is combined. With its integration in the SNEWS global alert network and the ongoing work to compute and combine the neutrino light-curves of different detectors, KM3NeT will play a key part in notifying other telescopes before the arrival of the other messengers. In this contribution, we present the real-time detection capabilities of KM3NeT, the additional information that can be brought by light-curve computations and the follow-up of external alerts.

*Speaker

## 1. Detection of CCSN neutrinos with KM3NeT

KM3NeT [1] is a distributed water Cherenkov neutrino telescope currently in construction in the Mediterranean sea. It is made of two detectors, ORCA offshore Toulon (France) at 2500 km depth and ARCA offshore Sicily (Italy) at 3500 km d epth. Its detection principle is the reconstruction of neutrinos from the light induced from secondary interaction products using three dimensional arrays of photo-multipliers. The base component of these arrays is the Digital Optical Module (DOM) [12]: it is a pressure-resistant sphere containing 31 photo-multiplier tubes (PMTs) and embedded electronics to digitize the signal of the PMTs, as well as orientation sensors used to locate the DOM in real time. The DOMs are installed in groups of 18 in a vertical string called Detection Unit (DU) anchored to the seabed, with different spacing according to the d etector. ORCA is more compact, optimized for the detection of tens of GeV neutrinos (mainly atmospheric neutrinos) and ARCA is a more voluminous, less dense detector, whose main goal is to discover astrophysical neutrino sources in the TeV to PeV energy range. KM3NeT is currently taking data with 16 DUs for ORCA and 21 DUs for ARCA and new DUs are continuously added to reach the objective of respectively 115 and 230 DUs by 2028.

Core-collapse supernovae are explosive phenomena ending the life of massive stars (above 10 solar masses) releasing a massive amount of energy (about $3 \times 10^{53}$ erg) and causing a luminous stellar explosion. During the collapse, 99 % of the energy is released as low energy neutrinos ($\sim$ 10 MeV), emitted during a time window of a few hundreds of ms to one second. As the medium surrounding a proto-neutron star stays opaque to electromagnetic radiation for a few hours after the first emission of neutrinos, it is possible to detect those neutrinos on Earth before the CCSN becomes visible to light telescopes. This phenomenon allows for the possibility of alert systems: neutrino detectors could detect the neutrinos emitted during the core-collapse, eventually locating the source of those neutrinos, so that telescopes could observe all electromagnetic radiation emitted by a CCSN.

The main detection channel of KM3NeT for CCSN neutrinos is through inverse beta decay (IBD) of electron anti-neutrinos on free protons in the water ($\bar{\nu}_e + p \rightarrow e^+ + n$). As KM3NeT is optimized for the detection of neutrinos above the GeV scale, the individual positron trajectories cannot be reconstructed. Instead, the strategy adopted is to search for an excess of coincidences between PMTs in single DOMs above the expected background of the detector [2]. Apart from the PMT dark rate (about 1 kHz to 2 kHz) there are three main sources of optical background with KM3NeT: radioactive decay in the sea water (mostly $^{40}$K), bioluminescence and atmospheric muons [10]. Radioactive decay causes a constant background in every PMT with a hit (PMT voltage rising above 0.3 photo-electrons) rate of about 7 kHz. Bioluminescence leads to a localized increase of the hit rates up to the MHz range, causing the need to veto rates above 20 kHz with the embedded electronics of the DOMs [11]. Finally, atmospheric muons generate long, downgoing tracks, inducing Cherenkov radiation detected by multiple DOMs, allowing to identify them.

In order to reject sources of optical background, a coincidence is defined as at least four hits within one DOM and with PMTs within 90 degrees of each other, with all the hits in a time window of 10 ns. The multiplicity of a given coincidence is the number of unique PMTs involved. Some of those coincidences are caused by atmospheric muons, but can be discarded using the muon reconstruction capabilities of the detector. The multiplicity distribution of the number of coincidences due to optical background in 500 ms can be seen in Figure 1 left: in light blue markers for ORCA and dark blue markers for ARCA, shown for a fully built detector. On this plot, the number of coincidences that would be measured in the event of CCSN with progenitors of different masses at 10 kpc can be seen.

The optimal sensitivity is obtained for a multiplicity range between 6 and 10 and the coincidence level is defined as the number of coincidences in 500 ms within this multiplicity range. The right plot of Figure 1 shows the sensitivity of KM3NeT as a function of the distance of the CCSN. In the most conservative scenario (progenitor of mass 11 $M_\odot$), KM3NeT is able to observe 95 % of the Milky Way, and can go up to the Large Magellanic Cloud.

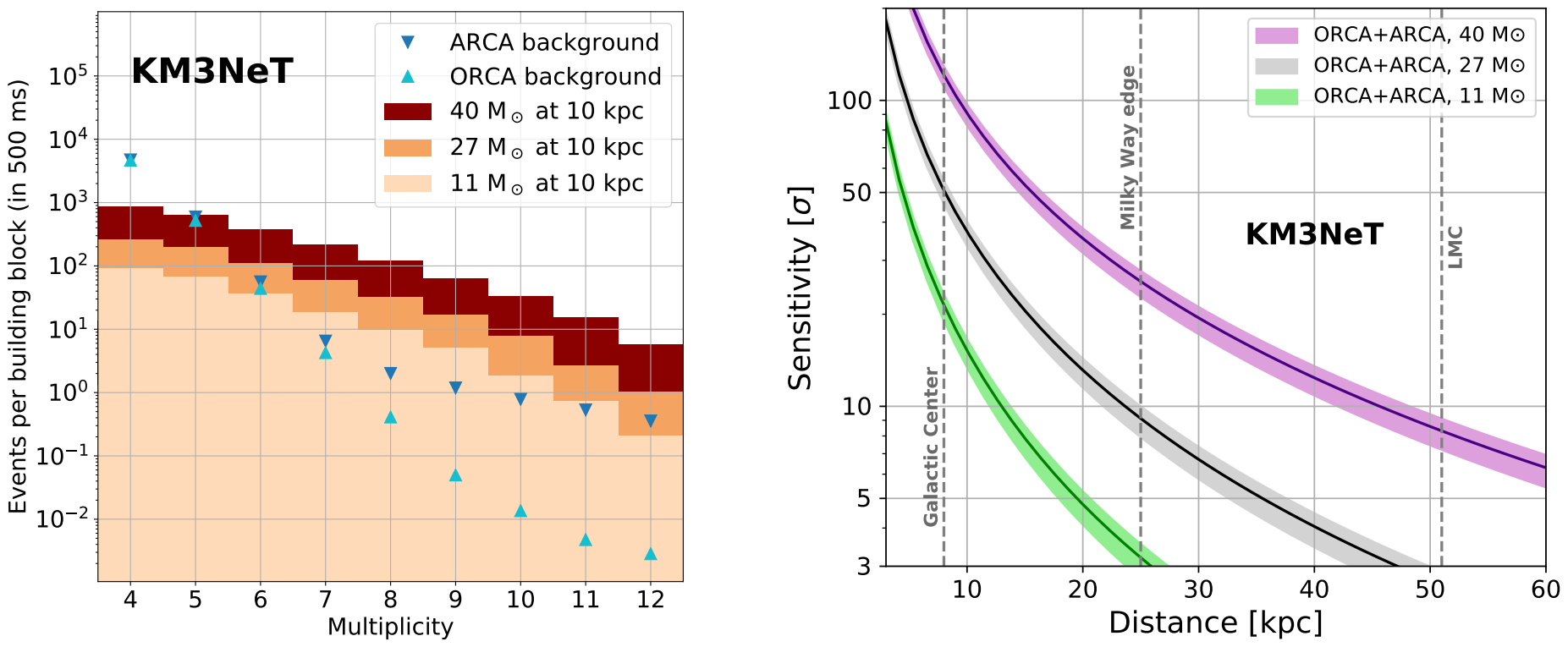

**Figure 1:** Left: expected number of events in a KM3NeT building block as a function of the multiplicity. The background is shown with markers in light blue for ORCA and dark blue for ARCA. The signal is represented with colored bars in orange shades for the different models: light for 11 $M_\odot$, intermediate for 27 $M_\odot$, dark for 40 $M_\odot$. Right: KM3NeT detection sensitivity as a function of the distance to the CCSN for the three progenitors considered: 11 $M_\odot$ (green), 27 $M_\odot$ (gray) and 40 $M_\odot$ (purple). The error bars include the systematic uncertainties. Figures from [2].

Those results describe the expected sensitivity with the analysis currently implemented within the current KM3NeT online system. Another method, which uses new observables to characterize the optical background, is also presented during this conference [3]. This new selection improves the overall sensitivity to a CCSN by ~ 23 % and will be implemented in the online system before the end of 2023.

As the expected Galactic CCSN rate is 1.5 per century and the typical emission duration is a few hundred ms, a fast online system is needed to be ready for the next CCSN.

## 2. Description of the CCSN online pipeline

The CCSN online pipeline is part of the real-time multi-messenger analysis platform of KM3NeT [7], whose main goals are CCSN monitoring, which is presented in this contribution, searching for neutrinos correlated with external electromagnetic, gravitational waves (GW) or neutrinos alerts, which is presented in [8] and sending all-sky all-flavor neutrino alerts to external observatories for follow-up.

An overview of the CCSN online pipeline is given in Figure 2. It consists of multiple modules (small independent programs running on a common server to both detectors), which are organized in three sections: the real-time CCSN search pipeline, in blue, analyzing the data collected by both KM3NeT detectors in order to evaluate a time-dependent significance that can be exploited to identify a CCSN signal and generate alerts; the quasi-online pipeline, in red, which is activated on special occasions to obtain and analyze data containing all detected coincidences, without selection on multiplicity or angle between PMTs; the triggered analysis pipeline, in yellow, made to follow-up external multi-messenger alerts in order to provide a fast response using a semi-automated analysis [6]. Those modules communicate through a common dispatcher, which is also running on the common server. The different networks connected to the pipeline to send or receive alerts and data are also represented in green.

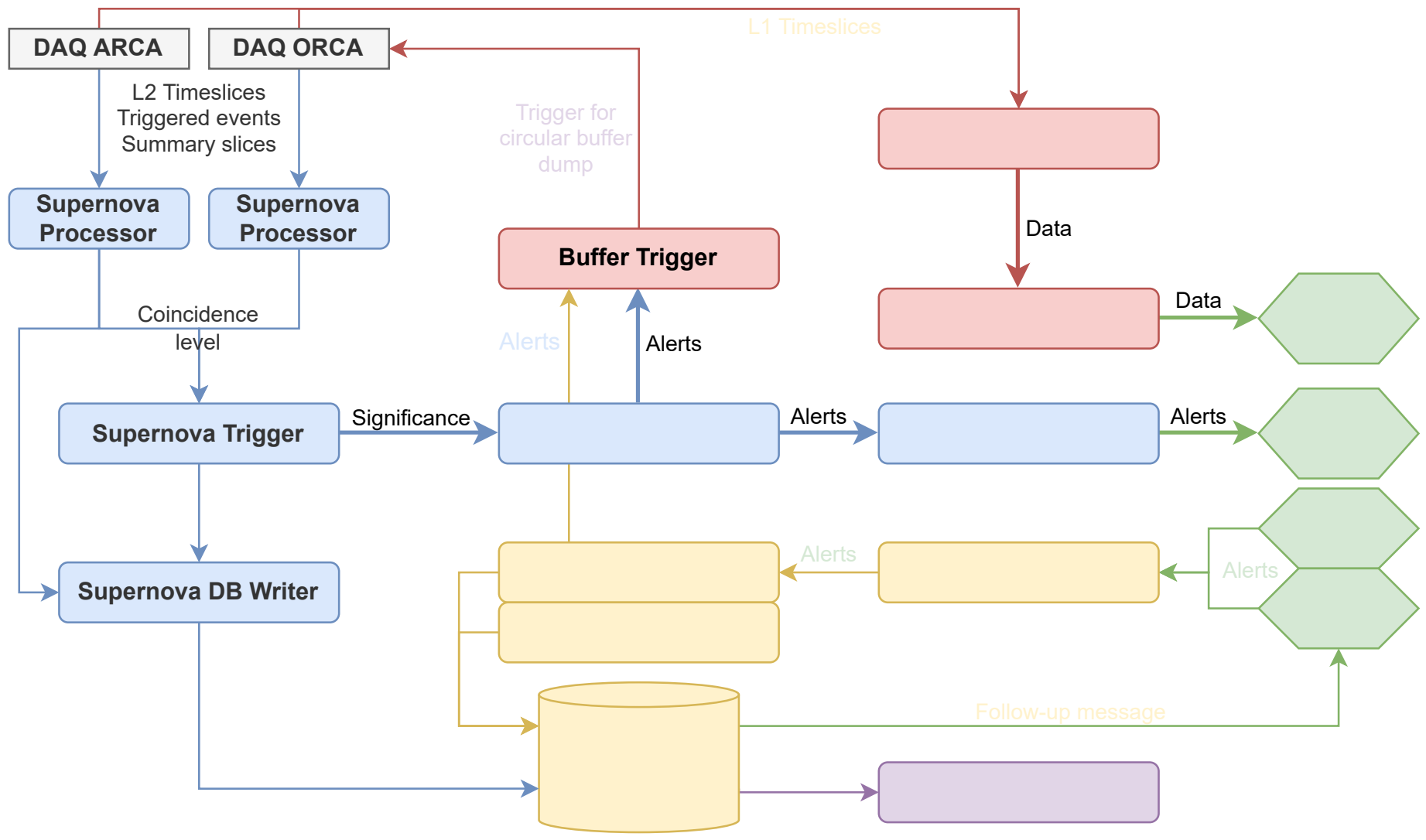

**Figure 2:** Scheme describing the dataflow of the CCSN online pipeline. The colors describe the three sections of the pipeline: in blue, the real-time CCSN search pipeline, to identify a CCSN signal and generate alerts; in red, the quasi-online pipeline, to extract parameters from all detected coincidences during an alert; in yellow, the triggered analysis pipeline, to follow-up external multi-messenger alerts.

The goal of the real-time CCSN search pipeline is to evaluate a time-dependent significance used to identify a CCSN and generate alerts that can be shared with networks such as SNEWS [4].

The first module of the chain is called Supernova Processor and computes every 100 ms the coincidence level. One Supernova Processor is running on each shore station, receiving low-level data in the form of timeslices (pack of segments of 100 ms of data) from the Data Acquisition (DAQ) system. As a reminder, the coincidence level is the number of coincidences between multiple PMTs with a maximal opening angle between them in single DOMs in a given time window length on which only coincidences with an optimal multiplicity are selected, before applying a veto to remove muon induced coincidences and finally summing the number of coincidences over 500 ms.

Coincidence levels from both detectors are sent to a common server, where the Supernova Trigger computes the combined Gaussian significance and the False Alarm Rate (FAR). The combined significance is obtained using a weighted linear combination of the ORCA and ARCA significance, where the weights are chosen as the detection sensitivity at a reference distance of 10 kpc for a benchmark flux model. In order to insure that the significance is computed correctly also in the case of an event during the downtime of one detector and to accommodate for the variations of delay between the two detectors, the incoming data are queued for an configurable maximum amount of time, at the end of which the module computes the significance for only one detector. If data from the second detector arrive after this delay, they are stored for a potential triggered follow-up analysis, but are not considered in the real-time computation of the significance. A description of the method to compute the significance can be found in section 3.

The Supernova Alert module selects messages whose significance is above a threshold before triggering the last module, the SNEWS handler, which transmits the alert to the SNEWS network. Four thresholds are currently defined: `test`, alerts with an hourly FAR, used for internal monitoring purposes; `snews-low`, alerts with a FAR below one per day, sent as sub-threshold alerts; `snews`, alerts with a FAR below one every eight days, sent to SNEWS for alerting purposes; `high-Z`, alerts with a FAR below one per year, aimed at public advertising via the Gamma-ray Coordinates Network (GCN).

The quasi-online pipeline aims to retrieve all hits belonging to coincidences without selection in order to be able to perform further analyses based on the neutrino light-curve characterization, such as fit of the precise arrival time or detection of the standing accretion shock instability [2]. As this data cannot be stored continuously because of storage limitations, a circular buffer is implemented in the DAQ system to continuously cache 10 minutes of data. On request, a file containing the circular buffer is written on each shore station, and recovered to the common server. Currently, the events considered significant enough to trigger this writing are the following: `snews` alert sent by KM3NeT; reception of an external SNEWS alert; external gravitational wave observation of a compact binary merger involving at least one neutron star; GW detection of a significant burst event.

Once the data containing all coincidences have been written and retrieved using the Buffer Trigger module, the analysis is performed by the Buffer Analyzer module in order to build the neutrino light-curve and to fit it with given models. From this fit, the signal arrival time called $T_0$ is precisely determined, and transmitted along with the light-curve to the SNEWS2.0 network [5].

At the reception of an external alert, either through the GCN for the follow-up of GW or through the SNEWS system, the follow-up pipeline is automatically triggered. The principle of the search is to look for the maximum significance during a given time-window, which differs according to the type of alert: $[0, 2\,\mathrm{s}]$ for GW detection and $[-5\,\mathrm{s}, 5\,\mathrm{s}]$ for external CCSN alerts.

## 3. Adaptive background expectation

As written previously, bioluminescence in the sea may lead to a localized increase of the hit rates up to the MHz range per PMT, causing the need to veto rates above kHz within the embedded electronics of the DOMs. This leads to a non-constant number of active PMTs over the whole detector, which also causes variation in the expected background. This veto occurs more often with the ORCA detector due to the common bioluminescence in its location, thus the following text focuses on ORCA data.

If a constant expected background $b$ and a coincidence level $n$ for a given period of 500 ms is considered, computing the one sided p-value $p$, associated Gaussian significance $\mathrm{zscore}$ and daily FAR is straightforward using the following definition: $p = P_{\mathrm{Poisson}}(X \geq n; b)$, $p = \int_{\mathrm{zscore}}^{+\inf} P_{\mathrm{Gaussian}}(x)\,\mathrm{d}x$ and FAR [per day] $= 10 \times 86400 \times p$. The FAR is calculated as the product of the p-value and the update frequency of the search window, corresponding to $100\ \mathrm{ms}^{-1}$.

For a non-constant expected background, the relationship between $b$ and the number of active PMTs needs to be properly established. Two previous determinations of the expected background as a function of the number of active PMTs have been done in the past, respectively for ORCA4 and ORCA6 [6, 9]. The first one, shown in Figure 3 left, considers a variable called "instrumentation efficiency", defined as the ratio between the expected background measured with a given fraction of active PMTs and the expected background measured with a fully active detector. The second one, which can be seen in Figure 3 right, considers directly the expected background (mean rate on the plot) as a function of active PMTs. For both plots, a binning is applied to the fraction of active PMTs and a linear relationship is found between expected background and number of active PMTs.

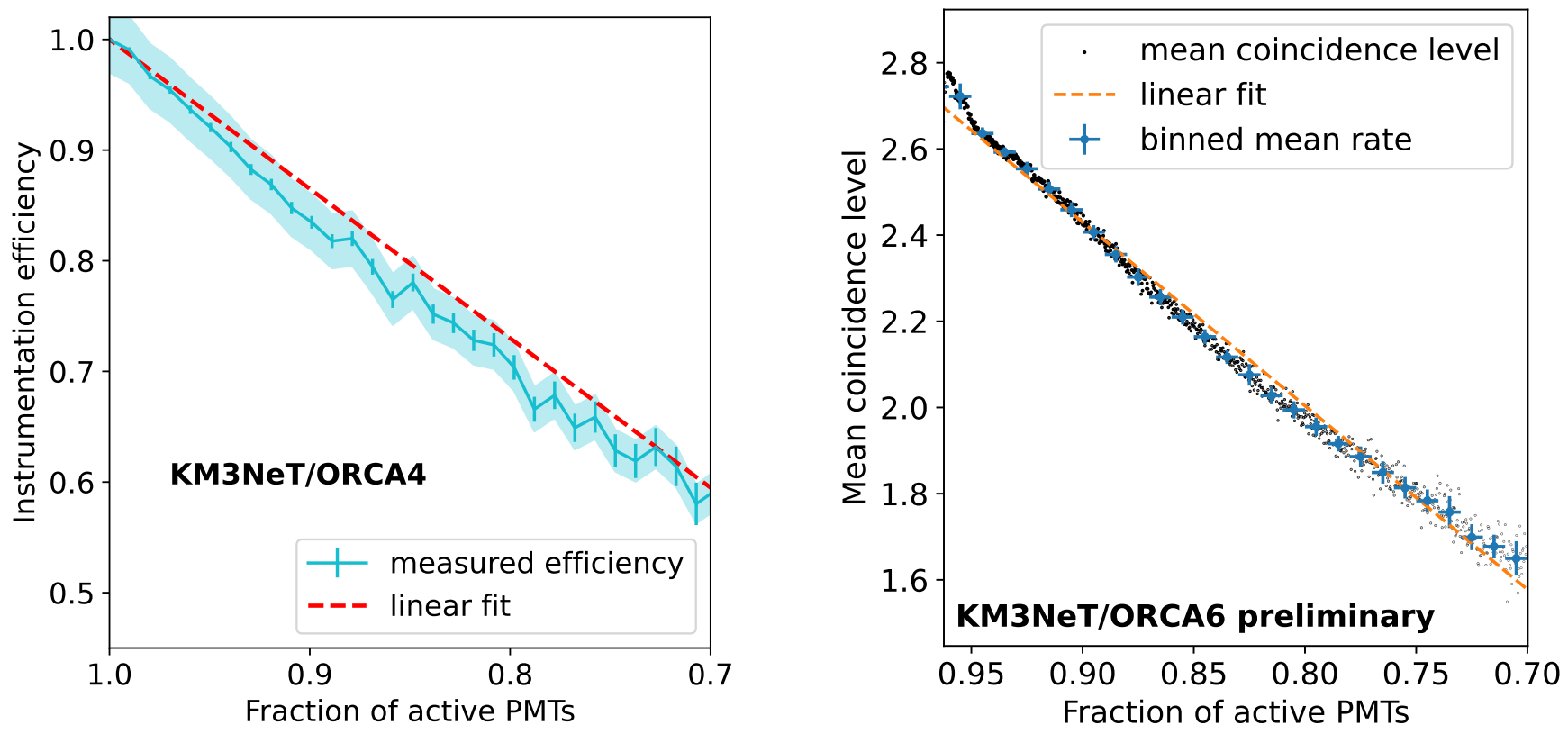

**Figure 3:** Left: Measured instrumentation efficiency as a function of the fraction of active PMTs in ORCA4 detector. The blue line shows the averaged measured efficiency with bins of width 0.01, monthly variability is covered by the shaded area and the dashed red line shows the linear fit applied to the d ate. Figure from [6]. Right: Mean rate of the coincidence level as a function of the fraction of active PMTs for ORCA6. Every black dot represents the computed mean rate averaged for the whole detector, for a given fraction of active PMTs. The blue crosses indicate averaged values in bins of width 0.05 and the orange dashed line is a linear fit of the binned d ata. Figure made to characterize the background for the MeV analysis of [9].

The issue with those two previous plots arises when drawing a scatter plot of values averaged over groups of ten minutes, as shown in the left of Figure 4, instead of applying a binning to the number of active PMTs. On this scatter plot, the mean number of active PMTs as a function of the mean expected background (*ie* coincidence level) is displayed, with the color representing the mean number of active DOMs during those 10 minutes. The relationship is no longer linear but seems to be quadratic, with multiple behaviors depending on the number of active DOMs.

The behavior seen in the left plot of Figure 4 can be easily explained by the fact that the detector with all lines and half of the PMTs active does not behave the same way as the detector with half of the lines and all PMTs active. The solution is to normalize both the coincidence level and the number of active PMTs by the number of active DOMs, as done in the right plot of Figure 4. On this figure, it can be seen that the detector follows a single quadratic function. A non-negligible variability is however still visible and needs to be taken into account. The goal of the following analysis is to show that a new quadratic fit needs to be performed frequently to factor in this variability and to insure a good estimation of the significance.

Data from the beginning of ORCA10 (December 2021) to ORCA18 (middle of June 2023) were processed using two different methods. In the first method, a single fit is performed with the first 30 days of data. For every timeslice, the expected background is computed from the fit, which is used to compute a zscore and an expected zscore distribution. In the second method, multiple fits are performed on groups of 30 days of data, and the expected background for a given timeslice is computed from the fit of the previous group. The choice of the 30-day period is a good compromise between the need to have enough data to be able to have a variation in the number of active PMTs to perform the fit and the need for the most recent behavior of the detector to be taken into account. The two distributions are computed in the same way as in the first method.

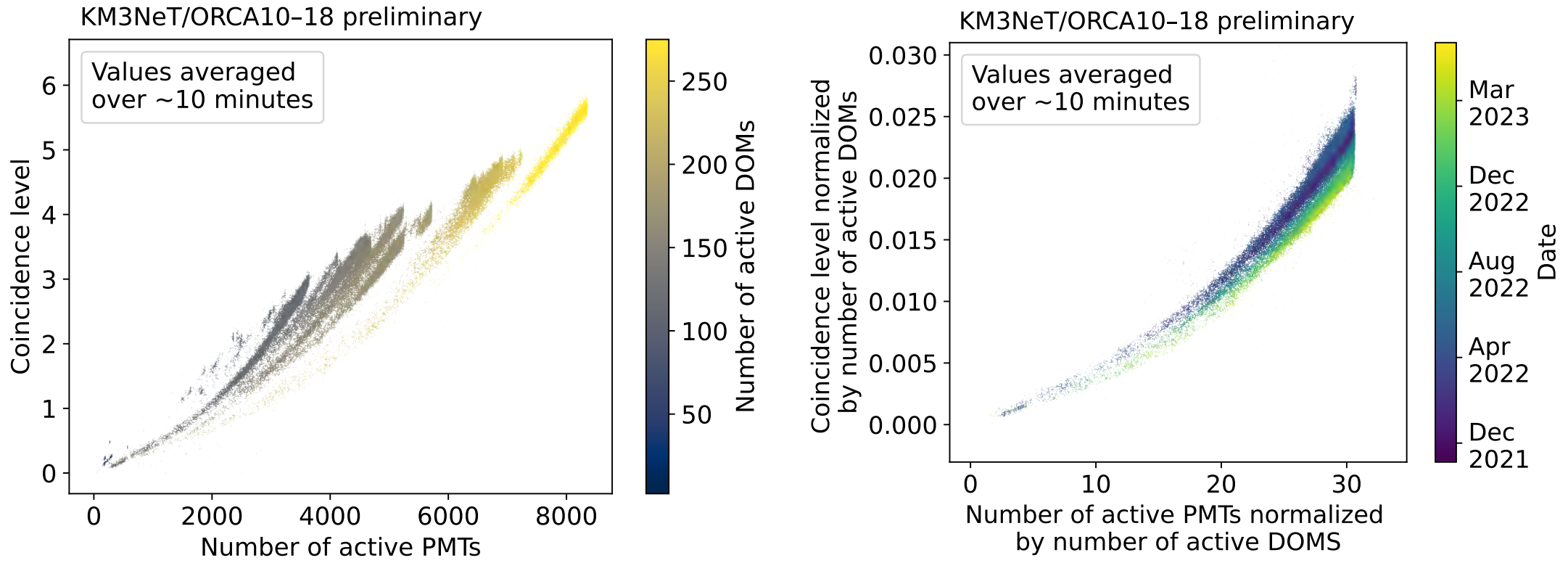

**Figure 4:** Left: Scatter plot of average over groups of 10 minutes of coincidence level vs number of active PMTs. The colors represent the average number of active DOMs during the 10 minutes period. Right: Scatter plot of average over groups of 10 minutes of coincidence level vs number of active PMTs both normalized by number of active DOMs. Time of the 10 minutes group is shown as a color. Both plots are made with data from ORCA10 to ORCA18.

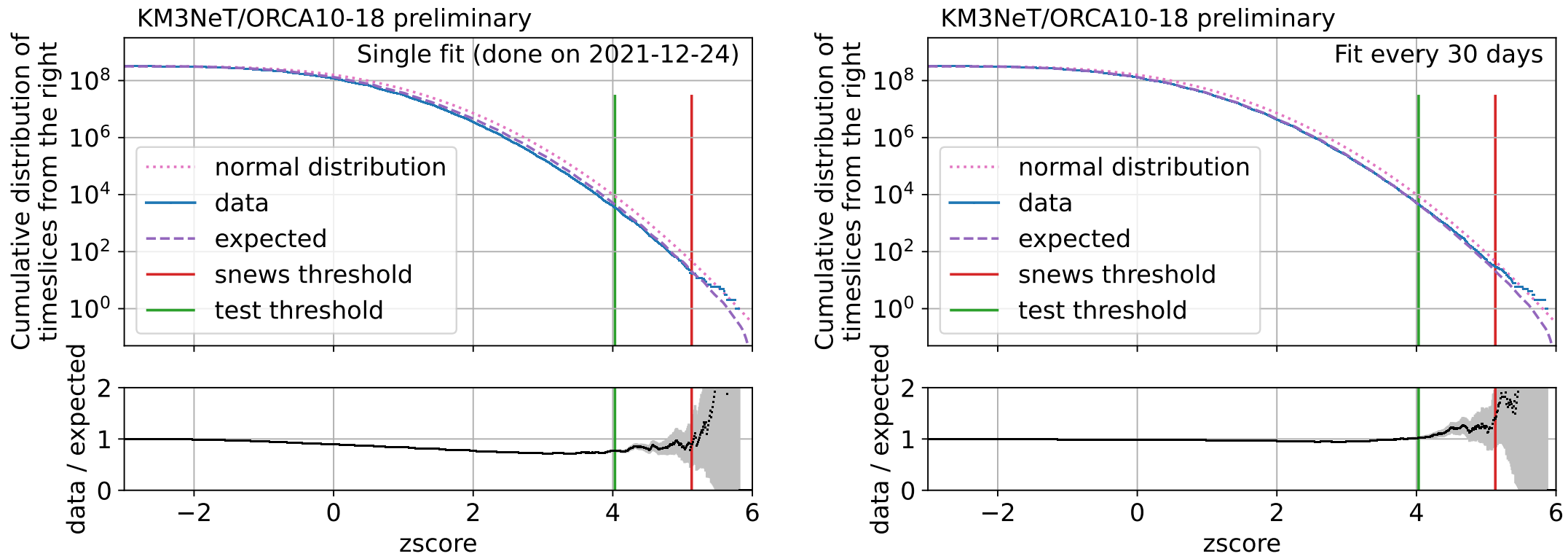

**Figure 5:** Cumulative distribution of timeslices from the right as a function of zscore computed for data from ORCA10 to ORCA18. The distribution of a normal variable is drawn in a dotted pink line; the expected number of timeslices in a purple dashed line; the number of timeslices obtained from data in a full blue line. Two threshold levels, `snews` (1 per 8 days) in red and `test` (1 per hour) in green are displayed for reference. A comparison between data and expected number of timeslices is also shown at the bottom. Left: distribution for a single fit (done from the first 30 days of da ta). Ri ght: distribution for fits performed every 30 days.

The cumulative distributions from high zscore to low zscore for both methods are displayed on Figure 5 (left for single fit, right for multiple fits). This distribution gives the number of alerts that would be sent for a given zscore threshold. A good agreement between the zscore distribution obtained from data (full blue line) and the expected zscore distribution (dashed purple line) would show that the expected background is properly estimated. The bottom plots of Figure 5 showing the relative comparison between them clearly indicate that for zscore below 4, the agreement between data and expected is much better using the second method. For zscore above 4, the statistical uncertainty starts to be predominant.

## 4. Conclusions

The CCSN online pipeline of KM3NeT is currently running with data from ORCA18 and ARCA21. The real-time CCSN search pipeline, looking for an excess of coincidences above the optical background, will be able to detect the next Galactic CCSN with a $5\sigma$ discovery potential. With the quasi-online pipeline, it is possible to perform further studies from the neutrino light-curve, such as the fit of the time of arrival of the neutrino burst. The triggered follow-up analysis can provide fast response to external alerts, whether from GCN or the SNEWS network. Finally, an improved estimation method for the expected background was also presented.

# Improving the sensitivity of KM3NeT to MeV-GeV neutrinos from solar flares

**Jonathan Mauro[a,*] and Gwenhaël de Wasseige[a] on behalf of the KM3NeT Collaboration**

[a] *Centre for Cosmology, Particle Physics and Phenomenology - CP3, Universite Catholique de Louvain, B-1348 Louvain-la-Neuve, Belgium*

*E-mail:* jonathan.mauro@uclouvain.be, gwenhael.dewasseige.@uclouvain.be

The detection of MeV-GeV neutrinos from astronomical sources is a long-lasting challenge for neutrino experiments. The low flux predicted for transient sources, such as solar flares, and their low-energy signature, requires a detector with both a large instrumented volume as well as a high density of photomultiplier tubes (PMTs). We discuss how KM3NeT can play a key role in the search for these neutrinos. KM3NeT is a Cherenkov neutrino telescope currently under deployment, located at the bottom of the Mediterranean Sea. It consists of two arrays of Digital Optical Modules (DOMs): KM3NeT/ORCA and KM3NeT/ARCA, which are optimised for the detection of GeV neutrinos for oscillation studies, and higher-energy astronomical neutrinos respectively. We exploit the multi-PMT configuration of KM3NeT's DOMs to develop the techniques that allow the disentangling of the MeV-GeV neutrino signature from the atmospheric and environmental background. Comparing data with neutrino simulations we identify the variables with discriminating power, and by applying hard cuts we are able to reject a large fraction of background. We present a graph neural network approach to classify signal from background. To further improve the sensitivities compared to previous studies, we will make use of the Hierarchical Graph Pooling with Structure Learning algorithm and will use graph-structured data to reproduce the hit geometry on the DOM. This will allow for stronger constraints on the hits and reduce the fraction of background that survives the selection.

*Speaker

## 1. Introduction

KM3NeT is a neutrino telescope currently operating at the bottom of the Mediterranean Sea. It consits of an array of Detection Units (DUs), to each of which are attached 18 Digital Optical Modules (DOMs), which contain 31 photomultipliers tubes (PMTs) each. These are used to detect the Cherenkov light emitted from charged particles originating from interactions of neutrinos with the sea water [1]. KM3NeT consists of two separate blocks: KM3NeT/ORCA, which is located in the South of France near Toulon, and is optimised to detect GeV neutrinos, and KM3NeT/ARCA, which is close to Capo Passero in Sicily, and more sensitive to neutrinos with energies between several tens of GeV to PeV. The detector is currently under deployment, and it is going to result in an instrumented volume of about 1 cubic kilometre for KM3NeT/ARCA and $3.6 \times 10^6 \mathrm{m}^3$ for KM3NeT/ORCA.

Although the main objective of KM3NeT is to investigate neutrinos in the multiple-GeV to TeV energy range, its large volume and relative high-density of instrumentation present a unique opportunity to study transient neutrino sources, such as solar flares, at sub-GeV energies. The current sensitivity of KM3NeT to astrophysical sources in the few-GeV energy range comes from the modified Neutrino-Mass-Ordering (NMO) selection in KM3NeT/ORCA, which is described in [2]. However, this selection still relies on multi-DOM coincidences, thus it's not optimal to investigate sub-GeV neutrinos as they are not expected to trigger multiple DOMs.

For neutrinos with energies of a few MeV, an event selection that uses coincidences at the single-DOM level has been implemented in KM3NeT [3]. This is mainly used to investigate Core Collapse Supernovae (CCSN), and it demonstrates the potential of KM3NeT to resolve low-energy signatures thanks to its unique DOM design. The CCSN neutrino flux is however expected to be high at energies well below the GeV, and the dedicated selection is then not well suited to probe lower fluxes at higher energies.

In its rawest form, KM3NeT data contains groups of two or more PMT-hits recorded in coincidence of 10 ns on the same DOM. Each hit has an associated time and time-over-treshold (ToT), where ToT is the duration of the pulse above approximately 0.3 photo-electrons, and can loosely be interpreted as a charge measurement. This type of data is referred to as L1, it is dominated by coincidences caused by decay of $^{40}$K, naturally present in sea water, while other major contributions come from bioluminescent species and atmospheric muons [1].

Solar flares are extremely energetic events that occur in the solar atmosphere, they are observed as peaks in the gamma-ray flux and/or in the X-ray flux, and they are often associated with other interesting phenomena, such as Coronal Mass Ejections (CMEs) and production of Solar Energetic Particles (SEPs), i.e., high-energy charged particles. These observations describe solar flares as sites of particle acceleration, and therefore, as perfect candidates to be neutrino sources. Moreover, there is most likely a link between gamma rays and neutrinos, as discussed in [4]: gamma-ray observations and spectral analysis prove pion production to happen in the most energetic solar flares, implying that flares with a strong gamma-ray flux should have an associated neutrino flux in the MeV-GeV range, as a result of pion decay.

Estimates of the solar-flare neutrino flux can be found in the literature, we refer the interested reader to some notable examples [4–6]. Solar-flare neutrino flux predictions are inferred by estimating the accelerated proton's energy fraction that goes into pion production. As such predictions are highly dependent on the assumptions made on the accelerated proton's beam, and on the sun's density profile, there are significant differences between the proposed models. We highlight that the optimistic prediction of Fargion's model [6] has the expected neutrino flux almost within the sensitivities of current neutrino telescopes, and strict constraints have already been set by IceCube in the high energy range [7], and by Super-Kamiokande at lower energies [8]. Being able to cover a broad energy range, KM3NeT will be the ideal instrument to further constrain neutrino production in solar flares.

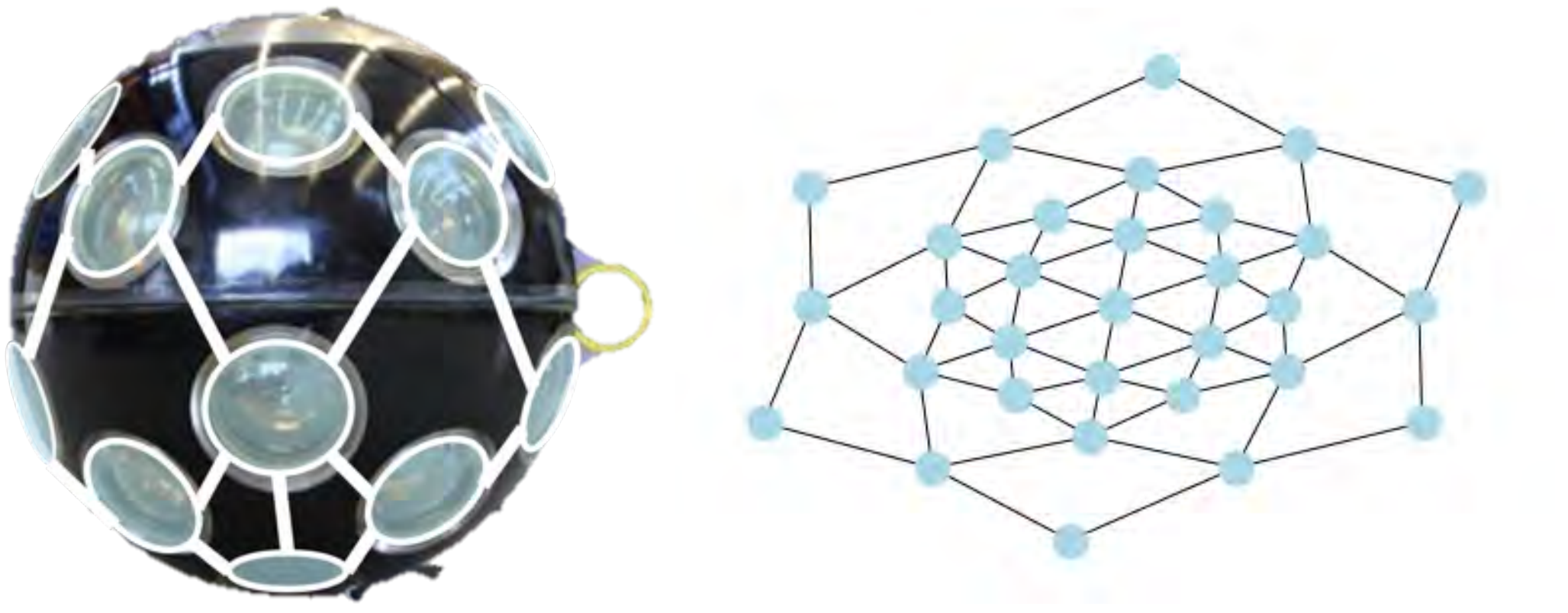

(a) (b)

**Figure 1:** (a) DOM-graph superimposed on a picture of a KM3NeT DOM. (b) DOM-graph as represented by NetworkX's graphic tools.

While machine learning has been largely implemented in all applications of particle physics, recently graph neural networks (GNNs) have caught the attention of the experimental neutrino physics community as well. The reason for this is that the architecture of the detectors, and thus the structure of the data, can be naturally encoded into graphs by exploiting the spatial and temporal relations between the hits recorded by multiple PMTs. In this context GNNs can be used for regression and classification tasks, examples of which are found in [9, 10].

In this analysis we will use a simple GNN to perform classification on single-DOM events. Since the predicted energies of solar-flare neutrinos are too low to apply classification methods that use the large-scale structure of the detector, our aim is to disentangle the neutrino signature from environmental and atmospheric background by looking at coincident hits on a single DOM. This will allow to further expand the energy range probed by KM3NeT to sub-GeV level.

## 2. Method

The core of this analysis is to perform classification on a small-graph dataset, where our graphs correspond to single-DOM events. These are fixed shape graphs where each node represent one of the 31 PMTs on a DOM, while the edges connect PMTs that are physically close to each other. Most traditional GNN classifiers architecture can be described as a combination of a graph-convolutional model, a readout function, and a multilayer perceptron (MLP). The model that we use for this analysis has been adapted from [11]. This model introduce the hierarchical pooling with structure learning (HGP-SL) operator, which is used to obtain a lower dimensional representation of a graph preserving the original graph-structure. This algorithm has shown outstanding performances for graph classification on a variety of small-graph benchmark datasets, which have similar features to our MeV-GeV-neutrino dataset.

Our dataset only comprises two classes: the MeV-GeV-neutrino signal, and background. Our signal is made of events built from 0.1-3 GeV muon-neutrino and electron-neutrino simulations, which reproduce the detector response. These simulations were produced using gSeaGen [12] for event generation, and KM3Sim [13] to simulate light propagation in water. The background sample events are instead built using the actual data recorded from the detector in absence of astrophysical transient events. Data was randomly obtained from the KM3NeT/ORCA detector in the period starting on 18 May 2020 at 12:00 and finishing at 18:00 of the same day.

## 2.1 Single-DOM Events in KM3NeT

### 2.1.1 Precuts

Precuts have to be applied to L1 data in order to reject the largest portion of background, which is mostly generated by K40 decay coincidences. These cuts also make sure that the classifier avoids training on signal events that are indistinguishable from background. The precuts are performed during the building process of the single-DOM events, i.e., while grouping the coincident hits. For the precuts, the following variables are considered: ToT of individual hits, time offset between hits, and the number of coincident hits in each event. Starting from L1 data and neutrino simulation, we first reject all of the hits with ToT below a threshold. We then look at the hits recorded on the same DOM, and group them in the same event until the time offset between consecutive hits is larger than a maximum time offset. Finally, we discard the events with less than a given number of hits.

In order to optimise these three cuts we study the increments in the fraction of background and signal that is rejected. As there is a clear hierarchy in the impact that these cuts have on the rejected fractions, we investigate them independently. The major rejection is done by constraining the number of hits in the events. The optimal values are found to be:

- Minimum ToT per hit: 6 ns.
- Maximum time offset between hits: 30 ns.
- Minimum number of hits per events: 3.

Roughly 46% of the total (summed) ToT in our simulations survives these cuts, and we have an average event rate of around 56 Hz per DOM in KM3NeT/ORCA at the time period considered for this analysis.

### 2.1.2 DOM graphs

The graph structured data are processed to naturally reproduce the geometrical configuration of the DOMs. The graphs are thus made of 31 nodes corresponding to the 31 PMTs of a DOM. The edges are drawn between PMTs whose relative distance is smaller than a given threshold. This threshold is chosen to be the smallest value that produces a connected graph for each DOM (with a small margin to account for the inexact cylindrical symmetry of the DOMs). The resulting graphs are shown as processed with NetworkX [14] graphical tools in Fig. 1, together with a representation of the same graph drawn on a DOM. The graphs have a higher density of edges around the nodes that correspond to the lower hemisphere of the DOM, as in this region PMTs are closer together, e.g., in Fig. 1b the node in the centre of the graph represents the lowest PMT on the DOM.

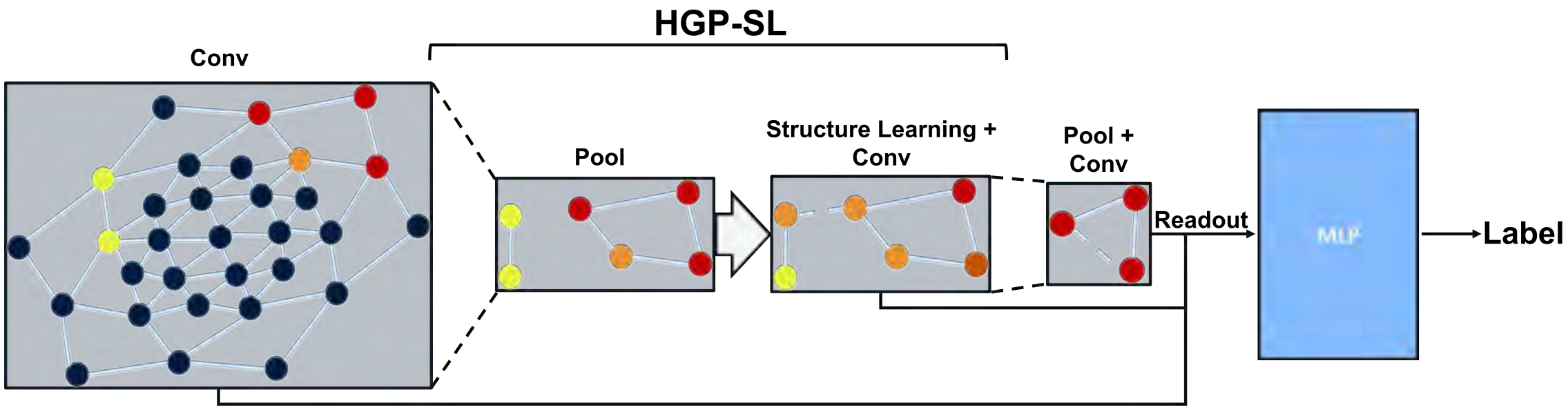

**Figure 2:** Diagram of the model architecture. Figure inspired by [11].

Using full-DOM graphs as opposed to only-hit graphs is fundamental to make our data representative of the spatial distribution of the hits in the events. Each graph is completed by assigning to each node a label corresponding to hit/no-hit, and the measured ToT. When multiple hits occur on the same PMT, in the same event, their ToTs are summed over. Moreover, we compute the standard deviation of the timing of the hits weighted by the corresponding ToT and use it as a graph attribute, this variable describes the spread over time of the deposited charge.

### 2.2 Model Architecture

For graph classification, we use the model proposed in [11], in the configuration used for classification on the PROTEIN dataset. The original code is provided by the authors and it is publicly available[1]. The model is built on PyTorch [15] and PyTorch Geometric (PyG) [16]. It consists of a series of three Graph Convolutional (GCN) layers, alternated with the hierarchical pooling operator. Starting from the second iteration of convolution and pooling the adjacency matrix of the graph is substituted by the output of the structure learning layer. A readout function is used to combine the output of the three convolutional layers into a vector that is fed into a 3-Layers dense MLP for label-prediction. The full model architecture is shown in Fig. 2. The GCN layers use the GCNConv operator for message-passing, the specifics of which can be found in [17]. This is a common choice for graph convolution, and this layer is readily available in PyG.

The pooling operation is based on the concept of 'information score'. The information score can be interpreted as the distance between the true node representation and the prediction obtained from its neighbours, where a node will receive a lower score if it can be well reconstructed from its neighbours. The idea is that a node with a low score can be removed from the graph without losing much of the information encoded in the original graph. This allows to select the subset of most significative nodes, i.e. with the highest score, for the pooled graph. The number of nodes in the pooled graph is specified by the 'pooling ratio' hyper-parameter. The pooling ratio is defined as the fraction of the number of nodes in the pooled graph and the number of nodes in the graph before pooling. It is set to 0.5 such that graphs halve in size each time pooling is performed.

[1]Code is available at `https://github.com/cszhangzhen/HGP-SL`

Structure learning is the other main component of the HGP-SL operator. It is used to deal with the problem of the highly-disconnected graphs that can arise from the pooling operation. As the subset of nodes selected during pooling is likely composed of non-neighbouring nodes, this can hinder the effectiveness of the following convolution. This is an artifact of the way that the pooling is performed, but its effect can be mitigated by virtually drawing additional edges in the pooled graph. The main goal of structure learning is to reproduce the structure of the original graph in the pooled graph to allow for better message passing. This is done by creating a substitute to the adjacency matrix, which is obtained by learning an optimal similarity between the nodes of the original graph. The training process is biased to give high similarity to directly connected nodes.

The readout function implemented in this model uses a concatenation of mean pooling and max pooling to obtain an equal-size graph-representation from the outputs of the convolutional layers. The final representation is obtained simply by summing over the individual readouts of the three convolutional layers.

Ultimately, the representation obtained with the readout function is fed into a softmax MLP classifier. This is a linear model made of three dense layers with a pyramidal structure. The number of neurons in the first layer is equal to the readout representation size and double the size of the nodes' hidden representations, which is set to 128. The hidden layer is half the size of the input one, while the output layer is made of two neurons. To train our model we minimize for a cross-entropy loss function over both classes.

### 2.3 Dataset and training

As mentioned in section 2.1 our dataset is composed of fixed-size graphs with 31 nodes. For this analysis we use a balanced dataset of 1750 events, meaning that half of our graphs have been built from neutrino simulations and labeled as signal, while the other half is built from calibrated

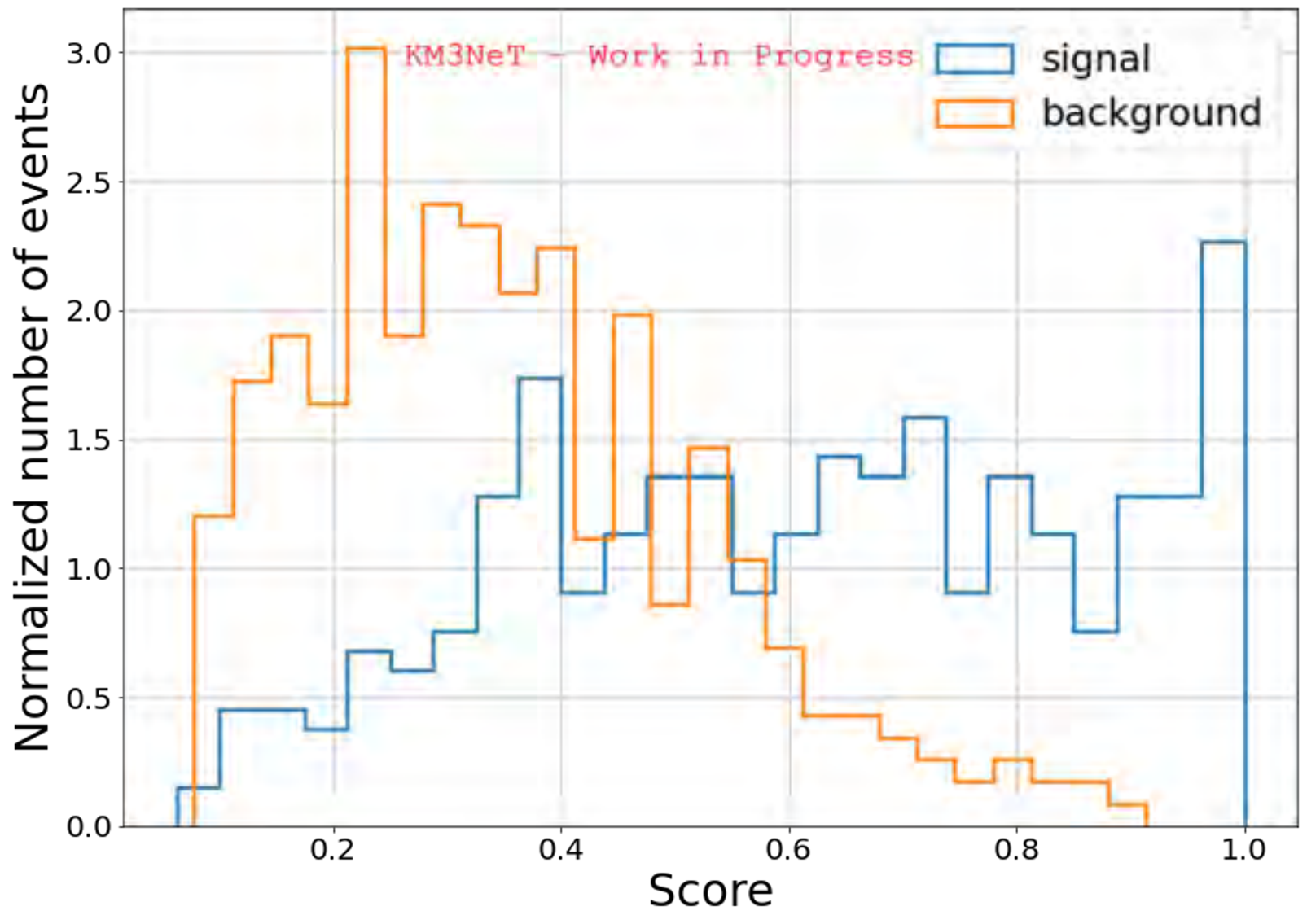

**Figure 3:** Score distribution of the test sample for the best model on validation. Here signal corresponds to MeV-GeV neutrinos, while background is dominated by environmental noise

L1 data and labeled as background. The reduced size of the dataset is due to the number of neutrino simulations available at the time.

This dataset has been split into three subsets used for training, validation and testing. These subsets correspond respectively to 50%, 10% and 40% of the full sample. this subdivision has been chosen to allow for higher statistics in the testing sample in order to reduce statistical fluctuations in the score distribution shown in Fig. 3.

As per common practice, the model is trained by minimizing a cross-entropy loss function on the training dataset, where at the end of each epoch the model is tested on the validation sample, and training is interrupted when the loss on validation stops decreasing. The best model is chosen to be the one with the lower loss on validation, which implies that 15 more epochs have been tested afterwards without producing better results.

## 3. Conclusions and prospects

We presented here a method to use KM3NeT for solar-flare neutrino searches in the MeV-GeV energy range. We propose to use single-DOM events to identify the MeV-GeV-neutrino signature enhancing the sensitivity of the detector to low-energies transient sources. By investigating low-level data and neutrino simulations, and implementing hard cuts on the ToT, time coincidence, and number of hits we are able to partially reject atmospheric and environmental background. Starting from the hits that survive these cuts, we build a graph dataset that encodes the information on timing, deposited charge, and geometrical distribution of coincident hits on the same DOM. We conclude our event selection by training a GNN to perform classification on this dataset. Despite the limiting low statistics, preliminary investigations show promising performances.

The single-DOM nature of this selection allows it to be used on both KM3NeT/ORCA, and KM3NeT/ARCA data, making use of the full detector volume. Further investigations will focus on including a high-energy veto using multiple DOMs coincidences, and deriving the sensitivity of KM3NeT to solar-flare neutrinos and other transient events.

# Refined neutrino follow-up analysis of GRB 221009A with KM3NeT ARCA and ORCA detectors

**J. Palacios González,[a,*] S. Le Stum,[b] D. Dornic,[b] F. Filippini,[c] G. Illuminati,[c] F. Salesa Greus,[a] A. Sánchez Losa,[a] G. Vannoye[b] and A. Zegarelli[d] for the KM3NeT collaboration**

[a] *Instituto de Física Corpuscular (CSIC-UV). Valencia, Spain.*
[b] *Aix Marseille University, CNRS/IN2P3, CPPM, Marseille, France.*
[c] *Dipartimento di Fisica e Astronomia dell' Università, Bologna, Italy, and INFN, Sezione di Bologna, Bologna, Italy.*
[d] *Dipartamento di Fisica, Universita' Sapienza. Roma, Italy, and INFN, Sezione di Roma, Roma, Italy.*

*E-mail:* Juan.Palacios@ific.uv.es, lestum@cppm.in2p3.fr

On October 9th 2022, the Swift-BAT telescope detected a spectacular transient event, soon classified as a Gamma-Ray Burst (GRB), based on the Fermi-GBM observation performed one hour earlier. Photons up to TeV energies were observed from such GRB by LHAASO, corresponding to the highest energy ever detected from a GRB. Just after this detection, a large number of observatories detected and characterized the multi-wavelength and multi-messenger emissions of this GRB, in one of the largest worldwide follow-up campaigns ever.

The KM3NeT neutrino telescope was one of the experiments that participated in the follow-up effort. KM3NeT is currently being built in the Mediterranean Sea and is composed of two detectors: ORCA, optimized for the detection of signals induced by neutrinos in the GeV-TeV range, and ARCA, mainly focused in neutrinos at the TeV-PeV range. MeV neutrinos can also be detected by looking for rate coincidences of Photo-Multipliers Tubes signals in both detectors. A first fast analysis was performed using data from the online reconstruction chain. In this contribution, we present a refined follow-up analysis, where new offline features are added together with improved calibration and optimized event selection.

*Speaker

## 1. GRB 221009A

On 2022 October 9, at 13:16:59.0 UT ($T_0$ from now on) the GBM instrument onboard the Fermi satellite triggered an extraordinarily bright transient phenomenon [1]. Shortly later, at 14:10:17 UT, the Swift-BAT telescope also detected a transient event consistent with the location of Fermi-GBM (but with better accuracy) and with candidate counterparts by Swift-XRT and Swift-UVOT [2]. The event was located at RA=288.263° and DEC= +19.803° (J2000) with an uncertainty of 3 arc mins. The transient was quickly identified as a long Gamma-Ray Burst (GRB), probably as a consequence of the collapse of a supermassive star.

GRB 221009A is one of the brightest gamma-ray bursts ever detected. Notably, the LHAASO gamma-ray observatory reported the observation of this event at unprecedented TeV energies [3]. Fermi-LAT also detected photons with energies up to $\sim$ 99 GeV, being the highest energies ever detected by such instrument [4].

This GRB is a relatively nearby event, with a redshift of $z = 0.151$ and an isotropic energy of at least $2 \cdot 10^{54}$ erg based on the gamma-ray fluence measured by Fermi-GBM [5]. The proximity, together with the fact that the jet emission is thought to be very collimated [6], could explain the extraordinary brightness of this event.

GRB 221009A has been proposed to be a collapsar, i.e. a very energetic supernova that results from an extreme core-collapse scenario [7]. However, there has not been observed clear evidence for a supernova signal [8]. Results from this long GRB have been reported by various multi-wavelength facilities such as MAXI/GSC [9], INTEGRAL SPI/ACS [10] or HAWC [11], among others. Close to fifty entries were published in the Gamma-ray Coordinate Network (GCN) [12] during the three days after the event, which proves the relevance of the event for the astrophysical community and motivates multi-messenger campaigns.

The peculiar characteristics of this GRB made it very interesting also for neutrino astronomy studies. Indeed, there are several models that predict the emission of TeV-PeV neutrinos from GRBs as a result of hadronic interactions of protons with high-density matter or radiation field photons [13]. The IceCube Neutrino Observatory reported the results of a fast follow-up one day after the event [14]. The analysis was based on two track-like muon neutrino searches:

- Using a time window $[T_0 - 1\text{ h}, T_0 + 2\text{ h}]$: no track-like events were found in coincidence. The upper limit of the time-integrated muon-neutrino flux was set at $E^2 dN/dE = 3.9 \cdot 10^{-2}$ GeV cm$^{-2}$ at 90% CL (assuming an E$^{-2}$ power law).

- During a time window of $T_0 \pm 1$ day: a p-value of 1.0 was reported, consistent with background expectations. In this case, the time-integrated muon-neutrino flux upper limit was set at $E^2 dN/dE = 4.1 \cdot 10^{-2}$ GeV cm$^{-2}$ at 90% CL (assuming an E$^{-2}$ power law).

A refined search was published months later by the IceCube Collaboration, including restrictive upper limits in the neutrino emission from GRB 22100A in a broad energy range [15].

The KM3NeT Collaboration also reported results for a quick follow-up search three days after the event [16]. Indeed, the Online KM3NeT framework [17, 18] for multi-messenger studies was in the commissioning period at that time. Three different real-time analyses were performed:

- A low energy analysis in the MeV range, based on the search for the maximum number of 10 ns coincidences between Photo-Multipliers Tubes (PMTs) in single modules during 500 ms, computed every 100 ms. A post-trial p-value of 0.9 was reported, compatible with background expectations.
- Two high-energy searches (one for KM3NeT/ORCA and one for KM3NeT/ARCA) based on binned techniques [18]. In the case of ARCA (ORCA) a search cone with a radius of 4° (2°) centered around the GRB position and the time window [T0−50 s, T0+5000 s] were used, with zero events observed and ∼ 0.1 events expected from the atmospheric background.

The present work includes a refined search for neutrino emission from GRB 221009A using KM3NeT data, including new features such as improved calibrations and dedicated Monte Carlo (MC) simulations. Section 2 includes a detailed description of the KM3NeT detectors. Section 3 describes the method used to perform the follow-up analysis. The results and the conclusions are provided in Sections 4 and 5, respectively.

## 2. The KM3NeT detectors

KM3NeT [19] (Cubic Kilometre Neutrino Telescope) is an international collaborative project currently deploying two deep-sea detectors in the Mediterranean Sea. These detectors consist of three-dimensional arrays of PMTs, able to detect the Cherenkov light emitted by particles resulting from neutrino interactions in seawater. The construction comprises two separate arrays: ORCA (Oscillation Research with Cosmics in the Abyss) and ARCA (Astroparticle Research with Cosmic in the Abyss).

On one hand, ORCA, located 40 km from Toulon at a depth of 2.5 km, is designed to study atmospheric neutrino oscillations and the neutrino mass hierarchy. On the other hand, ARCA, located 100 km from Portopalo di Capo Passero, Sicily, at a depth of 3.5 km, is optimized for studying high-energy neutrinos originating from astrophysical sources. The complementarity of the two detectors allows to study neutrinos from the MeV range up to PeV energies.

The main components of the KM3NeT detectors are the Digital Optical Modules (DOMs) [20], pressure-resistant glass spheres, each housing 31 PMTs. This multi-PMT approach, as opposed to using a single large PMT in each module, offers advantages such as the ability to identify physical signals through coincident hits on the same DOM. Each vertical string of 18 DOMs is called a Detection Unit (DU), and a grouping of 115 DUs forms a building block.

ORCA has a higher DOM density that enables it to study neutrinos in the GeV energy range. In contrast, ARCA has a lower density of DOMs, allowing it to cover a broader energy range, spanning from the sub-TeV range up to a few PeVs. This complementary between the two detectors motivates studies across a wide energy spectrum. Additionally, the high-duty cycle (> 95%) and the good angular resolution of the detectors (below one degree for $E > 10$ TeV) make them excellently suited instruments to perform multi-messenger studies such as the present work.

During October 2022, when GRB 221009A took place, ARCA had 21 DUs in operation, while ORCA counted 10 DUs. These partial configurations were taking good-quality data, monitoring the sky searching for neutrino signals.

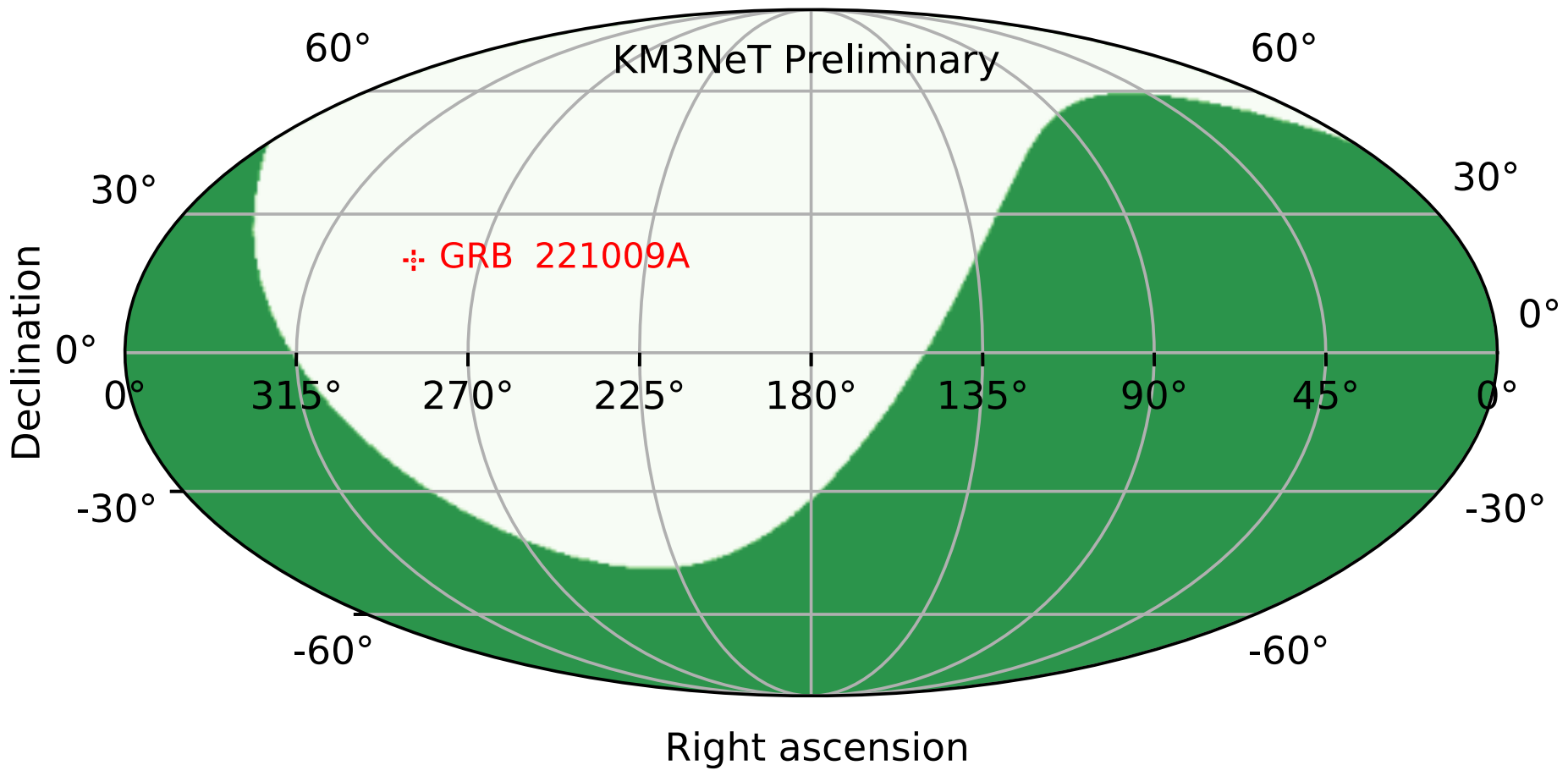

**Figure 1:** Skymap with the position of GRB 221009A in equatorial coordinates. The green shadowed region indicates the instantaneous visibility for upgoing events of KM3NeT/ARCA at the moment of the event (similar to the one of KM3NeT/ORCA).

## 3. Search method

As mentioned in Section 1, the data used in this analysis are at a refined calibration level with respect to the first, fast analyses performed earlier. An acoustic dynamical correction has been implemented to estimate the positions of the DOMs after the lines displacement due to currents in the seabed [21]. Additionally, a dedicated MC has been produced to compute the expected signal and set upper limits in the fluence emission.

As GRB position, the one given by Swift-BAT has been considered [2] (due to its better angular resolution) using as $T_0$ the trigger time by Fermi-GBM [1]. Three different time windows have been investigated using both ARCA and ORCA independently:

- One short-duration search in the range $[T_0 - 50\text{ s}, T_0 + 5000\text{ s}]$, as the online follow-up, with a downgoing track event selection (i.e. events reconstructed as not crossing the Earth).
- Two long-duration searches in the range $T_0 \pm 1$ day: one using an upgoing track event selection (i.e. events reconstructed as crossing the Earth) and another with a downgoing one.

The format of these searches is motivated by the fact that GRB 221009A was in the downgoing sky of the KM3NeT detectors at the time of the event, as shown in Figure 1. The analyses have been conducted by selecting only the corresponding upgoing or downgoing events during the period of the time windows considered. Indeed, for 45.2% of the time during one day, the location in the sky of GRB 221009A was found to be in the upgoing sky of the KM3NeT detectors.

The search method is based on a binned ON/OFF technique [22]. The ON region is defined as the area of the sky where the signal is expected to dominate over the atmospheric background, while the OFF region is defined as a region comparable to the ON region where only background is expected. We use as ON region a circular cone centered in the GRB position, which represents the Region of Interest (RoI) of the analysis. The OFF region is a declination/elevation band (for time windows above/below the day range) that is re-scaled in solid angle and time to the ON region size.

The event selection is based on the reconstruction variables of track-like events, i.e. events with a straight-line pattern that are originated in charged-current interactions from muon-flavour neutrinos and some tau-flavour. These events represent the ones with the best angular resolution. Cascade-like events, i.e. spherical-pattern events that emerge from electron-neutrino and some tau-flavour charged-current interactions, and neutral-current neutrino interactions, are not considered in this analysis.

The selection has been optimized in order to achieve a background level where one single event in the ON region provides a $3\sigma$ excess with respect to the expected background. Using a 2-sided convention, this corresponds to an expected background $\leq 2.7 \cdot 10^{-3}$ events. From all the event selections that fulfill this condition, we select the one that provides the largest expected signal, which has been computed from MC simulations assuming a neutrino flux $\phi \propto E^{-2}$. Systematic effects have been taken into account in these expectations computed from simulations, such as the uncertainty in the reconstruction direction of the data events.

For KM3NeT/ORCA, a data livetime of ~41 days has been used to estimate the expected background in the OFF region, re-scaled in time to the ON region time window. Consideration has been given to ensure stable data-taking conditions during the period studied. The event selection optimization is done on a machine learning classification score [23] that aims to reduce the large background of atmospheric muon events. After some minimal quality cuts, the $3\sigma$-1event optimization has been performed to determine the optimum values for the classification score and the RoI radius.

For KM3NeT/ARCA, a similar stability study has been conducted using a sample with ~70 days of livetime. This analysis applies straightforward selections on the reconstruction variables of the track-like events. In concrete, the optimization focuses on the quality of the event reconstruction, the estimated angular uncertainty of the event, the number of hits used in the reconstruction, and the estimated length of the track event in meters. The latter variable, which is useful to reject atmospheric muons misreconstructed as upgoing, is substituted in the downgoing analyses by a cut in the energy estimator in order to address the high atmospheric muon contamination in this part of the sky.

## 4. Results

After the optimization procedure described in Section3, the KM3NeT data was unblinded. No candidate neutrino event has been found in any of the searches performed. The expected signal events (from MC simulations) together with the expected background events (from data in the OFF region) are provided in Table 1.

Given a null result in the correlation analyses, upper limits (UL) in the neutrino emission from GRB 221009A were determined. The 90% CL UL in the flux normalization factor is defined as

$$\Phi_0^{\mathrm{UL}}\ (90\%\ \mathrm{CL}) \equiv \frac{\mu_{90}^{FC}(n_b)}{Acc}, \tag{1}$$

where $\mu_{90}^{FC}(n_b)$ denotes the 90% CL UL in the number of events by Feldman-Cousins [24]. $Acc$ in eq. 1 stands for the detector acceptance, a quantity defined as the proportionality constant that relates the number of expected signal events with a given flux normalization.

| **ANALYSIS** | RoI radius | Expected signal events | Expected background events | Events in ON region |
|---|---|---|---|---|
| ARCA upgoing $T_0 \pm 1$ day | 1.7° | $4.7 \cdot 10^{-3}$ | $(2.7 \pm 0.2) \cdot 10^{-3}$ | 0 |
| ARCA downgoing $T_0 \pm 1$ day | 1.0° | $1.2 \cdot 10^{-3}$ | $(2.6 \pm 0.1) \cdot 10^{-3}$ | 0 |
| ARCA downgoing $T_0[-50\text{s}, +5000\text{s}]$ | 1.2° | $4.4 \cdot 10^{-5}$ | $(2.66 \pm 0.03) \cdot 10^{-3}$ | 0 |
| ORCA upgoing $T_0 \pm 1$ day | 1.2° | $3.5 \cdot 10^{-4}$ | $(2.7 \pm 0.3) \cdot 10^{-3}$ | 0 |
| ORCA downgoing $T_0 \pm 1$ day | 1.0° | $1.7 \cdot 10^{-5}$ | $(2.7 \pm 0.3) \cdot 10^{-3}$ | 0 |
| ORCA downgoing $T_0[-50\text{s}, +5000\text{s}]$ | 5.4° | $6.9 \cdot 10^{-8}$ | $(2.7 \pm 0.3) \cdot 10^{-3}$ | 0 |

**Table 1:** Results of the analyses performed, both for KM3NeT ARCA and ORCA searches. The RoI used in each search is shown, together with the expected number of background and signal events. No candidate event has been found inside the ON region for any of the searches.

Since we are dealing with a transient event, it is also interesting to set an UL on the radiant fluence which is defined as the energy flux (per flavour) integrated over a certain emission period of interest. It can be computed as

$$\mathcal{F}^{\text{UL}} = \Delta T \int_{E_{min}}^{E_{max}} E\Phi_0^{\text{UL}} \left(\frac{E}{E_0}\right)^{-\gamma} dE, \tag{2}$$

where $\Delta T$ is the time window covered, $E_0$ is a reference energy level, in our case 1 GeV, and $E_{min}$ and $E_{max}$ correspond respectively to the 5% and 95% energy quantiles in the energy range of the detectable neutrino flux. The UL results for the six searches performed are provided in Table 2.

These results can be compared with the ones obtained by the IceCube Neutrino Observatory in similar searches. Figure 2 shows an adaptation of Figure 1 in [15], where the results from the present contribution have been added alongside the ones of IceCube. In concrete, ULs on $E^2F(E) = \Delta T \times \Phi_0^{UL}$, the energy-scaled per-flavour neutrino flux, integrated in time, have been computed to include the results in the mentioned figure. Only the ULs derived for a $\gamma = 2$ spectral index are shown. Note that a direct comparison is not straightforward, as each UL is computed according to different time window assumptions. The results derived in these searches are coherent with the ones obtained by IceCube, considering that we have worked with partial configurations of both KM3NeT detectors.

## 5. Conclusions

In this contribution, we have summarised the results of the offline search for neutrino emission from GRB 221009A performed using KM3NeT data. This analysis, together with the ones presented in [18], represents the first multi-messenger study performed by the KM3NeT Collaboration. Moreover, this is the first analysis where data from ARCA in the 21-line configuration, and ORCA in the 10-line one, have been analyzed.

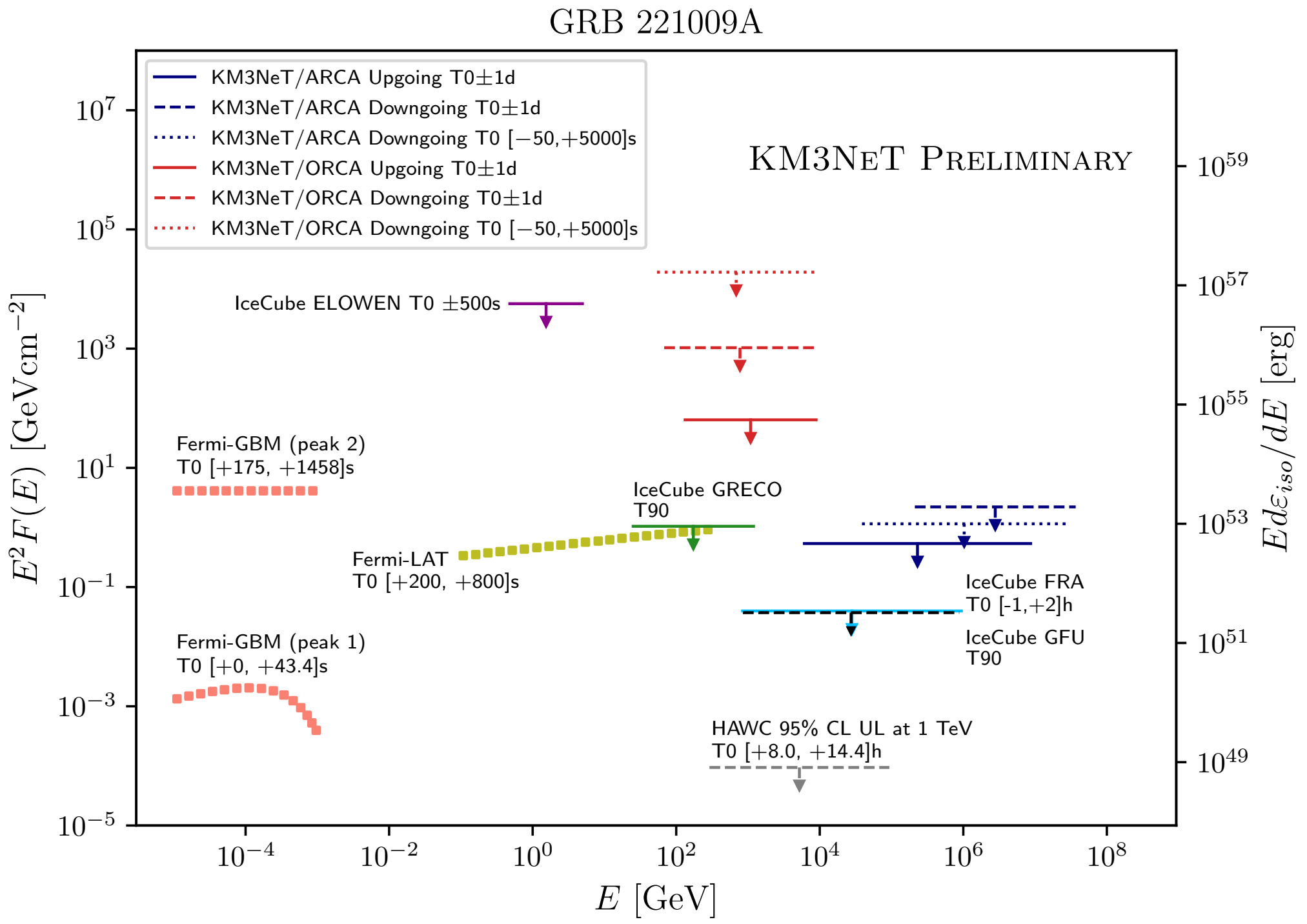

**Figure 2:** Comparison of the upper limits derived for IceCube and KM3NeT on the energy-scaled time-integrated neutrino emission from GRB 221009A (left y-axis), together with several gamma-ray observations. The right y-axis shows the differential isotropic equivalent energy. All the neutrino ULs shown are derived for a $\gamma = 2$ spectral index. This figure has been adapted from Figure 1 in [15], incorporating the results obtained in this contribution. More information about the IceCube results and the electromagnetic results can be consulted in [15].

| **ANALYSIS** | $\Phi_0$ UL [GeV$^{-1}$cm$^{-2}$s$^{-1}$] | 5% energy quantile | 95% energy quantile | Radiant fluence $\mathcal{F}^{UL}$ [GeVcm$^{-2}$] | $E^2F(E)$ UL [GeVcm$^{-2}$] |
|---|---|---|---|---|---|
| ARCA upgoing $T_0 \pm 1$ day | $6.2 \cdot 10^{-6}$ | 6.1 TeV | 8.7 PeV | 7.8 | $5.4 \cdot 10^{-1}$ |
| ARCA downgoing $T_0 \pm 1$ day | $2.6 \cdot 10^{-5}$ | 211.1 TeV | 36.8 PeV | 22.6 | 2.2 |
| ARCA downgoing $T_0[-50\text{s}, +5000\text{s}]$ | $2.3 \cdot 10^{-4}$ | 38.8 TeV | 27.3 PeV | 7.6 | 1.2 |
| ORCA upgoing $T_0 \pm 1$ day | $7.4 \cdot 10^{-4}$ | 133 GeV | 9.8 TeV | $5.4 \cdot 10^{2}$ | $6.4 \cdot 10^{1}$ |
| ORCA downgoing $T_0 \pm 1$ day | $1.6 \cdot 10^{-2}$ | 68 GeV | 8.8 TeV | $1.3 \cdot 10^{4}$ | $1.0 \cdot 10^{3}$ |
| ORCA downgoing $T_0[-50\text{s}, +5000\text{s}]$ | 3.8 | 54 GeV | 8.7 TeV | $9.8 \cdot 10^{4}$ | $1.9 \cdot 10^{4}$ |

**Table 2:** UL results in the neutrino emission from GRB 221009A for the analyses performed, both for KM3NeT ARCA and ORCA. $\mathcal{F}^{UL}$ stands for the energy-integrated per-flavour neutrino flux integrated over the emission period, while $E^2F(E)$ represents the energy-scaled per-flavour neutrino flux, also integrated in time. The 90% sensitivity energy range is also provided. Note that in the case of KM3NeT/ARCA, the energy range goes up to the few PeV range.

Despite no neutrino candidate has been found, upper limits have been set in the flux normalization factor and in the fluence neutrino emission of GRB 221009A. The next step is to reduce the latency of this kind of search and be able to introduce typically offline features (dynamic calibrations, UL computations, etc.) in the online follow-ups. For that purpose, the KM3NeT Online Framework [17] will include analysis methods derived from this contribution.

# Searching for neutrinos from microquasar flares with ANTARES and KM3NeT

**Sébastien Le Stum,**[a,*] **Damien Dornic**[a] **and Sergio Alves Garre**[b] **for the ANTARES and KM3NeT collaborations**

[a] *Aix-Marseille Univ., CNRS/IN2P3, CPPM, Marseille, France*

[b] *IFIC, Instituto de Fisica Corpuscular (CSIC - Universitat de Valencia), Valencia, Spain*

*E-mail:* lestum@cppm.in2p3.fr

Very high energy emissions are believed to happen through particle acceleration in the jet originating from microquasars. This hypothesis was corroborated by the recent detections from the HAWC collaboration of very high energy gamma-ray events in the vicinity of SS433 and V4641Sgr microquasars. Through hadronic processes, neutrinos are produced together with these gamma-rays. We present results from a search of a neutrino signal from 13 microquasars which, according to multiwavelength observations, exhibited events of fast bulk ejecta in the context of outbursts characterized by strong X-Ray flares and spectral state transitions. This search is focused on X-Ray flaring and transitioning periods determined from RXTE/ASM, MAXI/GSC and Swift/BAT public data monitoring. The analysis is performed using data from the ANTARES neutrino telescope, which was dismantled during 2022 after 16 years of steady data acquisition, as well as with its successor KM3NeT, in construction in the Mediterranean Sea, with data from the KM3NeT/ORCA detector in its 6 lines configuration. No significant excess was found and upper limits on the neutrino fluxes are computed in the total observed flaring times and in specific spectral states.

*Speaker

## 1. Introduction

This study aims to search for a neutrino signal correlated with X-Ray flares from microquasars in the ANTARES and KM3NeT neutrino detectors. Microquasars are binary objects - with a compact object and a star - that exhibit phenomena of accretion/ejection in the form of an accretion disk/jet system. The matter is ejected with relativistic bulk speed, which could lead to the acceleration of particles to very high energy via shock formation. In the case of a significant hadronic contribution in the jet, a detectable neutrino flux could be emitted. The multiwavelength behaviour of these sources shows high levels of variability, with periods of intense flaring mostly detected in radio and X-Rays. This study focuses on the flaring periods, as these exhibit higher ejection rates, which lead to preferential conditions for neutrino production.

Sources are selected for this study that present the following behaviour: periods of clear X-Ray flaring with X-Ray spectra following the characteristic hysteresis cycle as shown in [1]. During this cycle a source typically exhibits an increase in luminosity followed by a transition from a hard X-Ray spectrum to a soft, thermal, spectrum. This cycle is the focus of the search due to its correlation with radio bright ejecta with relativistic bulk speeds, as seen in [2] for MAXI J1820+070.

## 2. Detectors

ANTARES and KM3NeT are undersea neutrino telescopes built with photo-multiplier tubes (PMT) aiming to detect Cherenkov light induced by secondary particles created in the interaction of neutrinos with sea water. These interactions generate in the detector two main event topologies: track-like from muons produced in the charged-current interaction of muon and anti-muon neutrinos and cascade-like from other kinds or interactions. The now-dismantled ANTARES telescope [3] was located 2500 meters deep near Toulon (France) and ran for 16 years before stopping data-taking in February 2022. It was composed of 12 vertical lines containing 885 PMTs and was optimised for the detection of high-energy neutrinos with energy ranging from a few hundreds GeV to a few PeV. KM3NeT [4] is its successor and is currently being built on two sites: ORCA, located near the ANTARES site and optimised for the detection of lower-energy neutrinos (GeV to TeV) and ARCA, located offshore of Sicily (Italy) and optimised for TeV to PeV neutrinos. The planned KM3NeT detector will be composed of 115 lines on the ORCA site and 230 lines on the ARCA site.

This analysis uses data from the ANTARES detector in most of its data taking duration, from January 2007 to February 2022 and data from the KM3NeT/ORCA detector in its 6-lines configuration (ORCA6), from January 2020 to November 2021. In both detectors, quality controls for the data taking condition are performed, resulting in a total live time of 4541 days for ANTARES and 555 days for KM3NeT/ORCA6.

## 3. Search periods determination

The flaring periods of the candidate sources are determined using publicly available X-Ray lightcurve data from the following telescopes: RXTE/ASM (2-10 keV), MAXI/GSC (2-20 keV) and Swift/BAT (15-150 keV).

An estimation of the baseline rates and standard deviations for each source and telescope is performed with a Gaussian fit of the daily averaged rates of the lightcurves. Data points are retained that verify $F - \Delta F > \mu_{\mathrm{BL}} + 8\,\sigma_{\mathrm{BL}}$, with $F$ the flux and $\Delta F$ its corresponding error, $\mu_{\mathrm{BL}}$ and $\sigma_{\mathrm{BL}}$ the mean and standard deviation of the Gaussian fitted baseline. To account for possible instrumental effects, data points verifying this criterion are removed if they are not accompanied by another selected point in a ± 5-days time window. The resulting high-significance data points are used as seeds to search for flaring time windows: starting from each point, the flaring period is defined by the time window in which the flux verifies $F - \Delta F > \mu_{\mathrm{BL}} + 2\,\sigma_{\mathrm{BL}}$ in at least 1 point in a 5-day sliding time window starting from each high-significance data point. A neutrino search is performed if a flaring period is found in any telescope. The lightcurves for GX339-4 with the corresponding flaring periods are shown in Figure 1. From the initial source list, 3 microquasars exhibit flaring in X-Ray during the KM3NeT/ORCA6 period: GX339-4, AqlX-1 and 4U1630-472, with observation livetimes of respectively 250, 143 and 83 days.

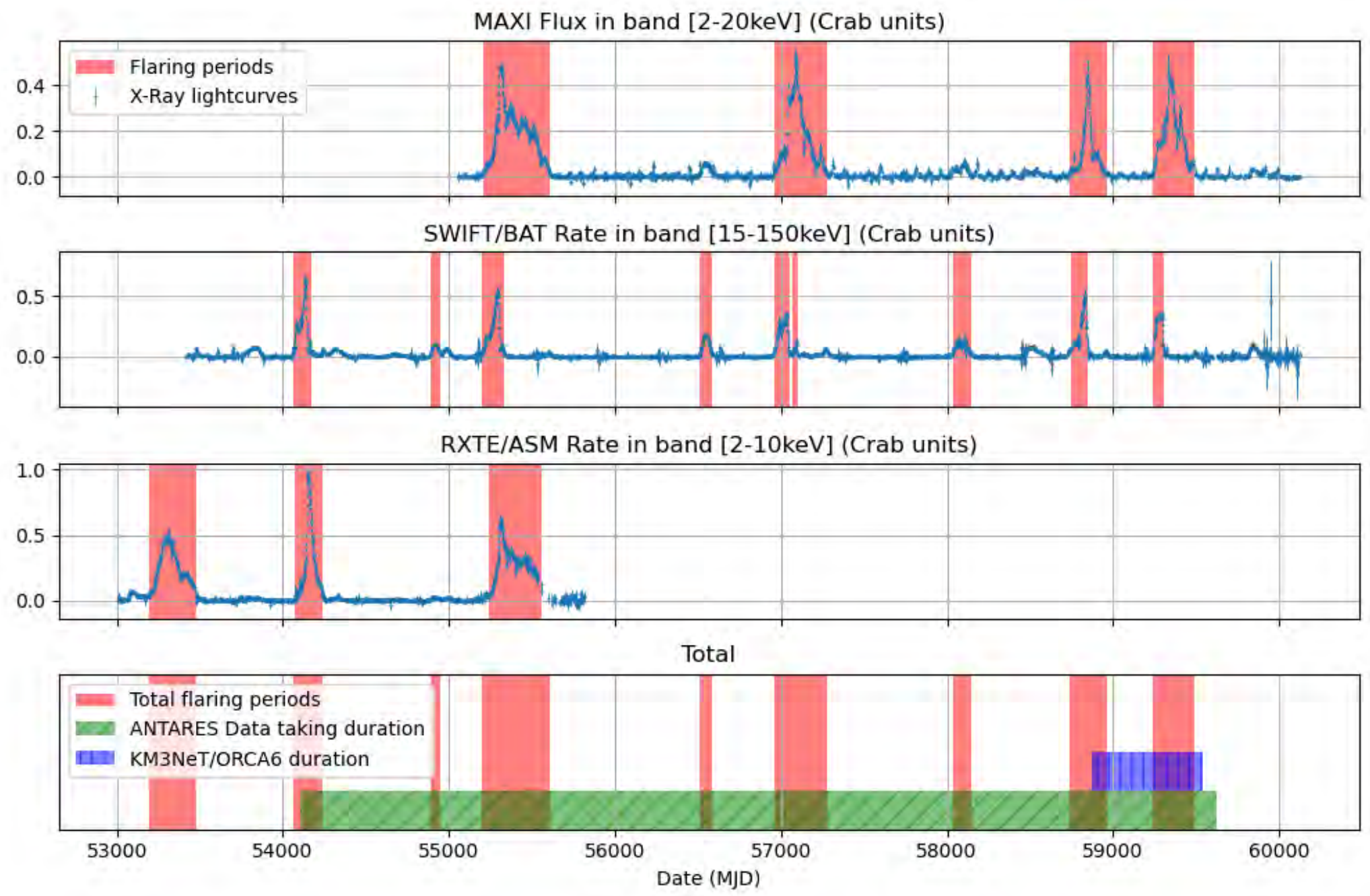

**Figure 1:** X-Ray lightcurves from GX339-4, recorded by MAXI/GSC, Swift/BAT and RXTE/ASM. Flaring periods are highlighted in red. The ANTARES data-taking duration is represented as a green line and the KM3NeT/ORCA6 data-taking duration is represented as a blue line. The total flaring periods are a stacking of the flaring periods in each telescope and define the neutrino signal search window.

Additionally, for sources MAXIJ1820+070 and GX339-4 the search is also performed in 3 subdivisions of the flaring periods, each corresponding to a given X-Ray spectral s tate: the Hard State, Hard to Soft state transitions, and Soft State. This is motivated by the fact that these sources exhibit different multi-wavelength properties in each state, and neutrino fluxed are expected to also differ.

For MAXIJ1820+070, periods are taken from [5], rounded to the floor/ceiling Modified Julian Day (MJD) for the beginning/end of the period to account to ensure a flat Right Ascension background distribution. They are as follow, given in MJD: Hard State from 58189 to 58304, State Transition from 58304 to 58311 and Soft State from 58311 to 58380.

GX339-4 exhibits regular flares with multiple state transitions. The spectral states where divided using X-Ray hardness ratios from RXTE/ASM and MAXI/GSC, with total livetimes of 248, 30 and 634 days for the Hard States, State transitions, and Soft States respectively.

## 4. Analysis methods

### 4.1 ANTARES

The signal search in ANTARES data is a standard unbinned likelihood analysis. The likelihood PDF is built using spatial, time and energy distributions obtained from simulations, taking into account event rate fluctuations over the detector life time [6]. Pseudo-experiments are performed to obtain the distribution of test statistic (TS) used to compute flux sensitivities for each source. The TS is defined as the ratio of the signal plus background likelihood, for which the number of signal events is fitted, over the background-only likelihood.

The search is performed in the time windows obtained in the previous section. They are defined as a box-shaped PDF term in the likelihood. In the case of discontinuous flaring periods the analysis is performed only once, effectively stacking the multiple flares on a single search. The analysis takes into account track-like and cascade-like event topologies. The events are selected on the quality of their reconstruction to reduce the background coming from atmospheric muons and to improve the angular resolution. Cuts on these quality parameters are then chosen by maximizing the probability of a $5\sigma$ discovery assuming an incoming neutrino spectrum of $\phi(\mathrm{E}) = 10^{-7}\ \mathrm{E}^{-2}\ \mathrm{GeV}^{-1}\ \mathrm{cm}^{-2}\ \mathrm{s}^{-1}$

### 4.2 KM3NeT

The analysis is a binned ON/OFF region search, similar to the one used in [7] and [8]. It is a counting experiment in which the rate of events in an OFF (control) region is used to estimate the number of background events in an ON (signal) region, which is the search region. These ON/OFF regions are defined as follows: the ON region is a circle with a radius to be optimised (Region of Interest, RoI), centered to the coordinates of the source, and the OFF region is a band in equatorial coordinates, +/- 10° around the source, minus the RoI defining the ON region. Here the OFF events are recorded at the same time as the ON events, to obtain a reasonable estimate of background event rate in the search region. The expected $N_{\mathrm{ON}}$ coming from background is then: Exp.Bkg = $\mathrm{N_{OFF}} \times \Omega_{\mathrm{ON}}/\Omega_{\mathrm{OFF}}$ with $\Omega$ the solid angles of the corresponding regions. In this search only events reconstructed as upgoing or horizontal are taken into account, with events selected with a reconstructed zenith angle $\theta$ such as $\cos(\theta) > -0.1$. This allows to significantly reduce the number of background events coming from atmospheric muons, with the remaining background being misreconstructed muons and atmospheric neutrinos. Moreover, the analysis only uses events reconstructed as a muon track-like, typically associated with muons. Tracks originating from neutrinos are separated from tracks originating from misreconstructed muons using a boosted decision tree (BDT) classifier, based on gradient boosting [9]. The training is done using simulated events, with all-flavor $\nu$ events as signal and atmospheric muons as background. The final selection criteria are determined by minimizing the upper limit expected from the background and a neutrino signal modelled by $\phi(\mathrm{E}) \propto \mathrm{E}^{-2}$ [10]. The two free parameters for this minimization are the classification score and the radius of the ON region (RoI). The optimization is performed using data from the OFF region defined above to estimate the background level.

## 5. Results

In the ANTARES analysis, the likelihood fit yielded no signal event for each source and time period. Upper limits are given as the 90% confidence level sensitivities derived from pseudo-experiments. The upper limits on the fluence for the flaring periods are shown as a function of the source galactic longitude Figure 2 and as a function of the search livetimes Figure 3. For the different spectral states of MAXIJ1820+070 and GX339-4, the upper limits on the fluence and flux normalisation are given in Figure 4. Due to the smaller observation time windows, the flux is less constrained during the state transition periods, but it should be highlighted that these periods are the most favourable for neutrino emission in the case of luminous ejecta with high bulk Lorentz factor [11].

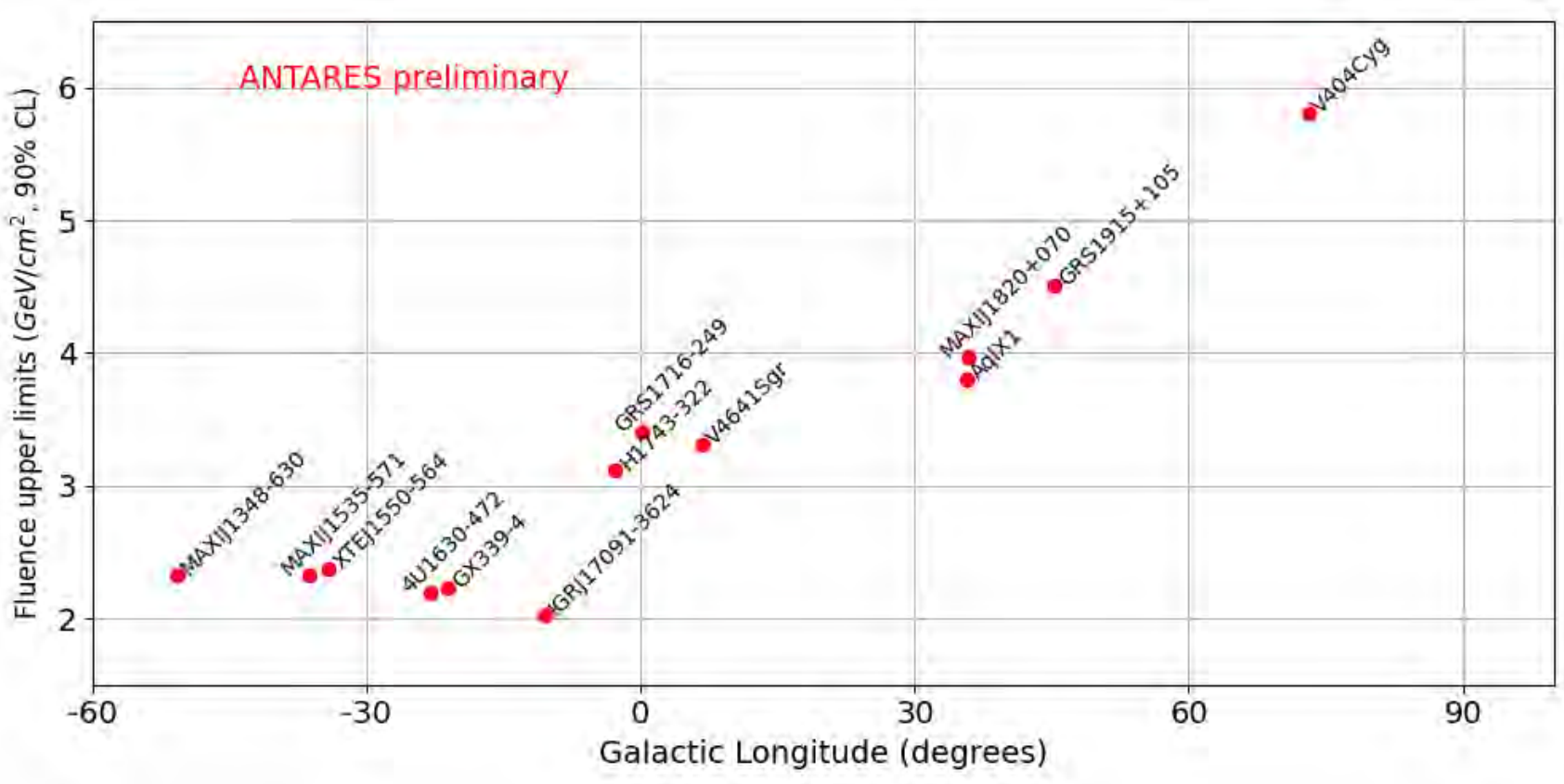

**Figure 2:** Neutrino fluence upper limits with ANTARES against the galactic longitudes of the studied sources.

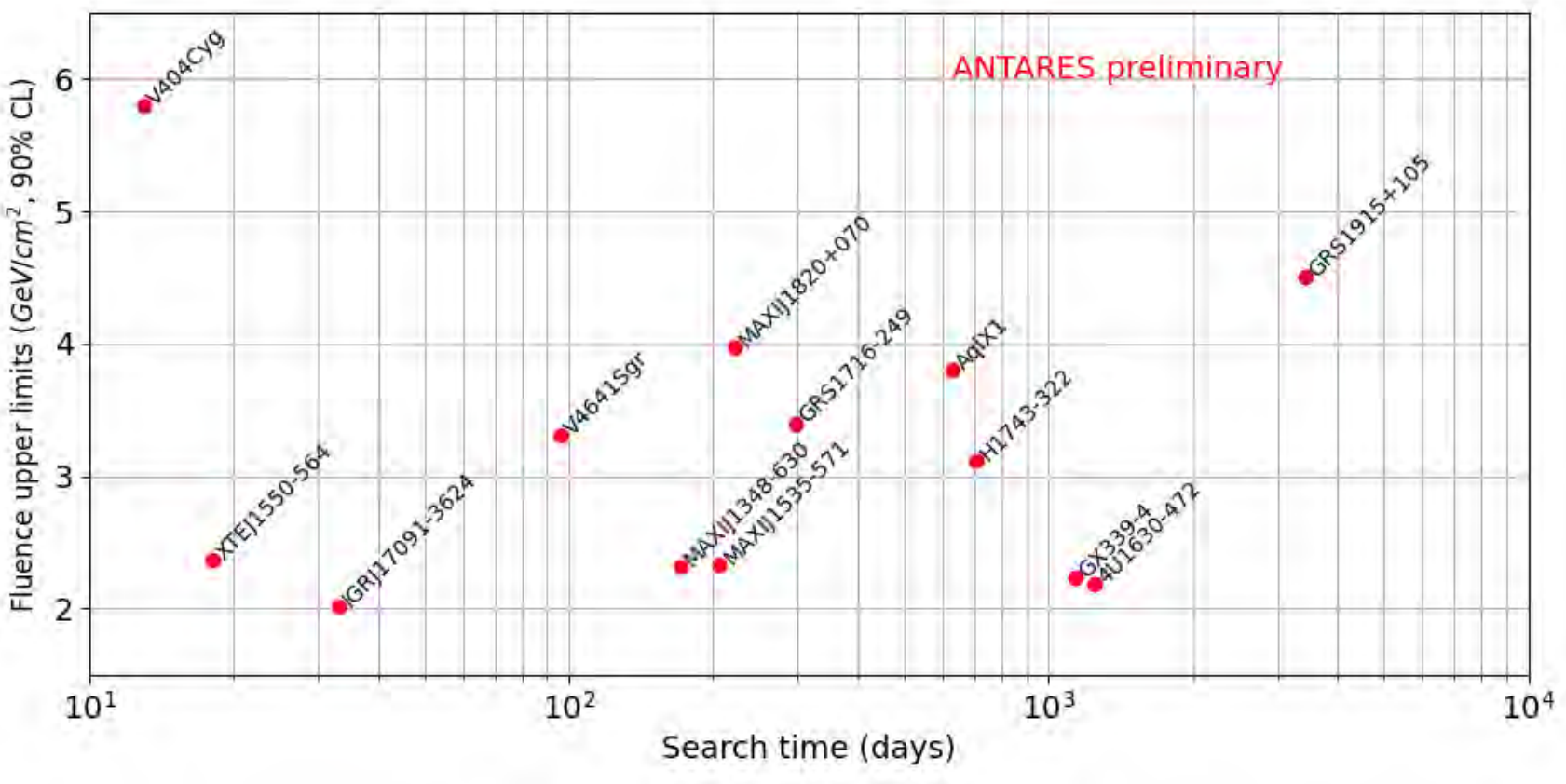

**Figure 3:** Neutrino fluence upper limits with ANTARES against the search time. This duration corresponds to the flaring time of each source under good data-taking conditions.

The fluence is defined as :

$$\mathcal{F} = \Delta T \int_{E_{5\%}}^{E_{95\%}} E \frac{dN}{dE} dE \tag{1}$$

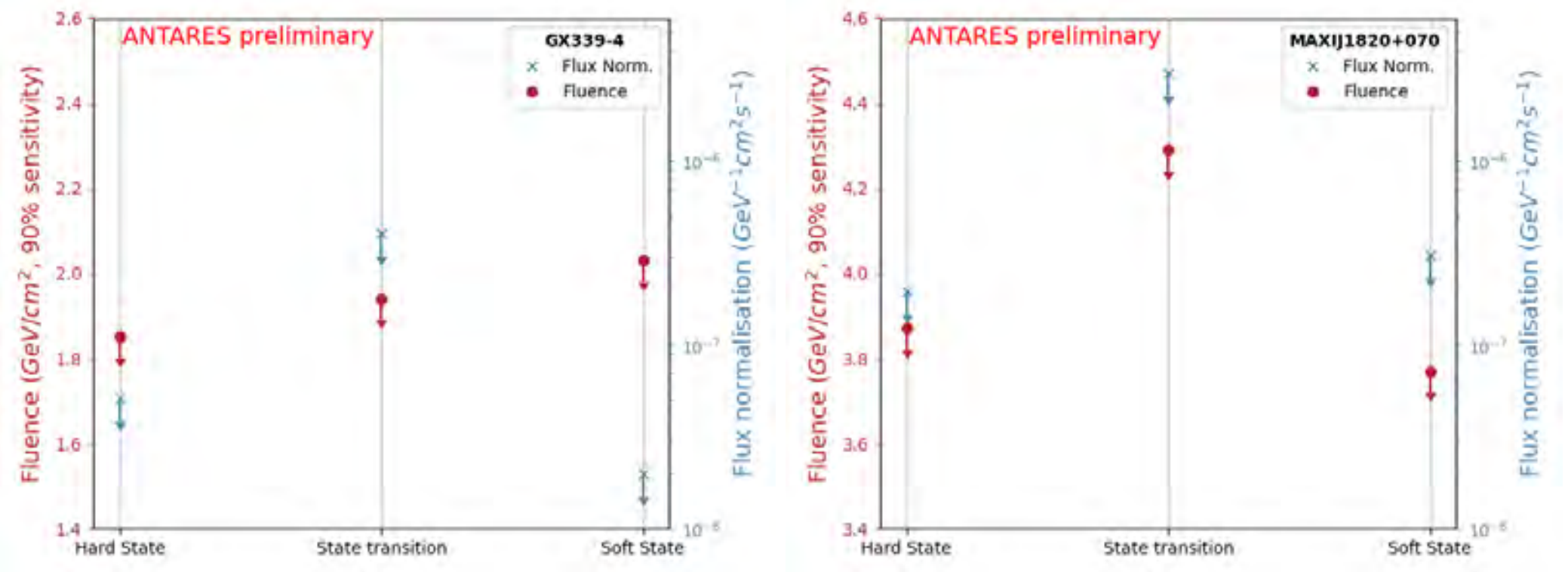

**Figure 4:** Upper limits with ANTARES on neutrino fluence (in GeVcm$^{-2}$) and flux normalisations (in GeV$^{-1}$cm$^{-2}$s$^{-1}$), assuming an incoming neutrino flux with an energy spectrum $\phi$(E) $\propto$ E$^{-2}$. Limits are given in 3 X-Ray spectral states: Hard State, State Transition, and Soft State for GX339-4 (left) and MAXIJ1820+070 (right).

with $\Delta T$ the observation livetime and with $E_{5\%}$ and $E_{95\%}$ the 5th and 95th energy percentiles of the signal events energy distribution, assuming a power law spectral distribution: $\frac{\mathrm{d}N}{\mathrm{d}E} \propto E^{-2}$. The energy ranges are as follows: for the KM3NeT/ORCA6 detector, $E_{5\%}$ = 100 GeV and $E_{95\%}$ = 9 TeV while for the ANTARES detector, $E_{5\%}$ = 3 TeV and $E_{95\%}$ = 3 PeV.

For the KM3NeT/ORCA6 analysis, the results can be found in Table 1. No significant excess was found, with the lowest p-value being 32% for 4U1630-472 assuming the background follows Poisson statistics. Upper limits are computed using Rolke confidence intervals [12], with a 90% confidence level. In this computation the background is assumed to follow Poisson statistics and the detection efficiency is modeled as a Gaussian with a relative standard deviation of 30% to conservatively take into account systematic effects. The flux and fluence upper limits are computed assuming the incoming neutrino flux has an energy spectrum $\phi$(E) $\propto$ E$^{-2}$ and are evaluated using simulations. For the 3 sources that could be observed in a flaring state with KM3NeT/ORCA6 and ANTARES, upper limits on the flux as a function of energy can be found in Figure 5. The KM3NeT/ORCA6 upper limit is less constraining due to the smaller detector and the shorter observation times, but provides complementary coverage in a lower energy range with respect to ANTARES.

## 6. Conclusion

This contribution presents a search for a neutrino signal correlated with X-Ray flares from 13 microquasars with ANTARES with its full life time, and with KM3NeT/ORCA in its 6 lines configuration. No significant signal was observed in either case and upper limits on the neutrino fluxes were derived.

In a recent analysis from the Icecube neutrino detector [13] that included a similar search that the one presented here, the microquasar V404 Cyg yielded a signal with a pre-trial p-value of 1.4% , while taking into account trial correction led to results fully compatible with background expectations. The complete KM3NeT detector will be able to further study galactic objects as potential neutrino sources thanks to its location in the northern hemisphere allowing a good view of the galactic center.

| Source | | 4U1630-472 | | GX339-4 | | Aql X-1 | |
|---|---|---|---|---|---|---|---|
| $N_{\mathrm{ON}}$ | Time (MJD)<br>Dist. from source | **2** | 59495.887, 59526.473<br>1.77°, 1.80° | **1** | 58956.383<br>1.35° | **1** | 59128.48<br>1.52° |
| $N_{\mathrm{OFF}}$ | | 474 | | 648 | | 291 | |
| Exp.Bkg | | 1.15 | | 1.35 | | 0.60 | |
| p-value | | 0.32 | | 0.74 | | 0.45 | |
| $\Phi_0^{\mathrm{UL}}$ ($\mathrm{GeV^{-1}cm^{-2}s^{-1}}$) | | $6.7 \times 10^{-6}$ | | $2.2 \times 10^{-6}$ | | $8.8 \times 10^{-6}$ | |
| $\mathcal{F}^{\mathrm{UL}}$ ($\mathrm{GeVcm^{-2}}$) | | $3.3 \times 10^{2}$ | | $1.9 \times 10^{2}$ | | $2.5 \times 10^{2}$ | |

**Table 1:** KM3NeT/ORCA6 search results. $N_{\mathrm{ON}}$ and $N_{\mathrm{OFF}}$ are respectively the number of events in the ON and OFF regions after selection. The times and distances from the source are given for events in the ON region. Exp.Bkg. is the expected number of background events in the search region. The p-value is computed assuming a Poisson-distributed background. $\Phi_0^{\mathrm{UL}}$ and $\mathcal{F}^{\mathrm{UL}}$ are respectively the upper limits of the flux normalisation and the fluence, assuming an incoming neutrino flux with an energy spectrum $\phi(\mathrm{E}) \propto \Phi_0 \mathrm{E}^{-2}$. Upper limits are given with a 90% confidence level.

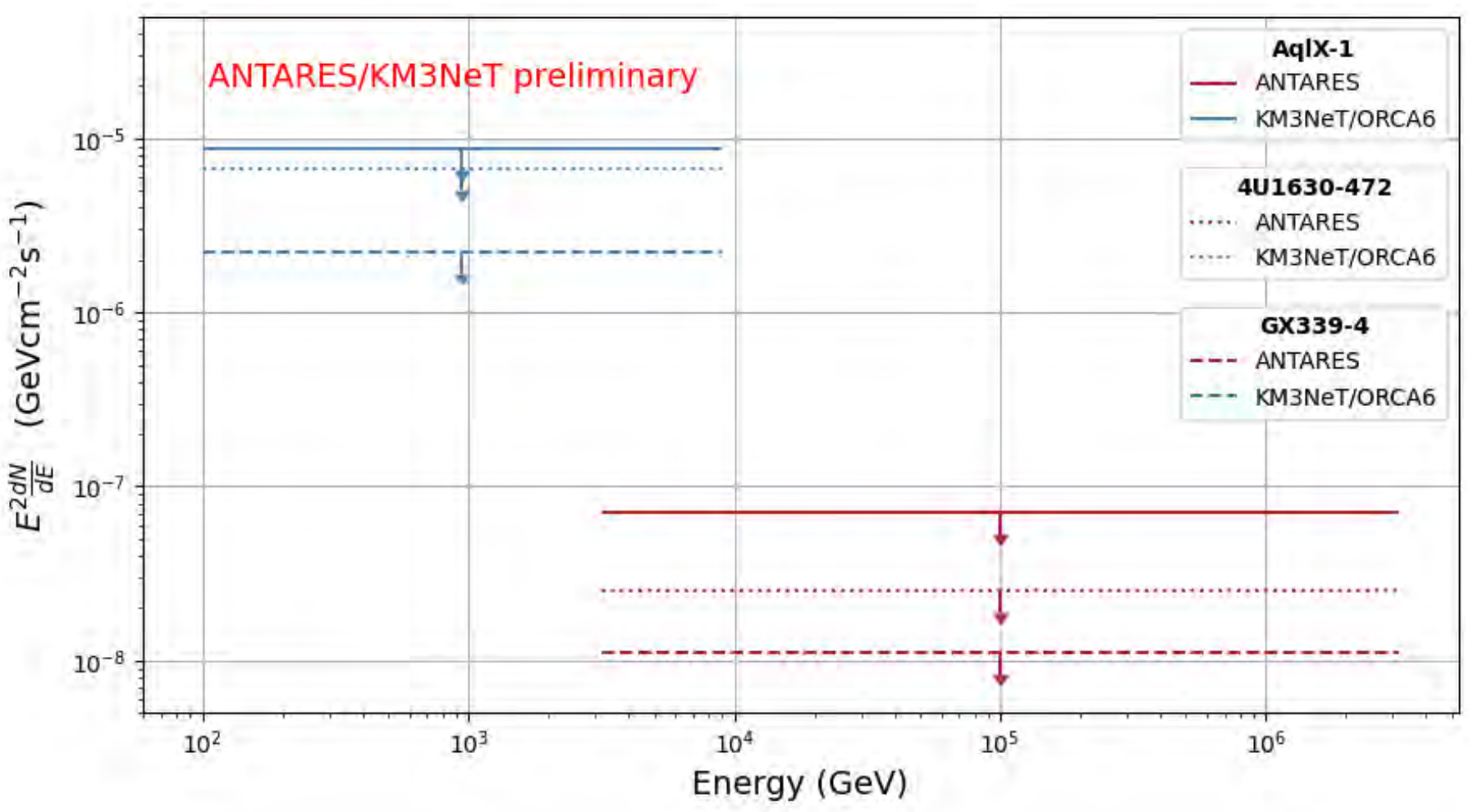

**Figure 5:** Neutrino flux upper limits for AqlX-1 (solid), 4U1630-472 (dotted) and GX339-4 (dashed) in KM3NeT/ORCA (blue) and ANTARES (red) energy ranges. Limits are given as $E^2\frac{dN}{dE}$, (in $\mathrm{GeVcm^{-2}s^{-1}}$), assuming an incoming neutrino flux with an energy spectrum $\phi(\mathrm{E}) \propto \mathrm{E}^{-2}$.

The KM3NeT detector, which is under construction, is already showing performances competing with ANTARES with 18 lines in ORCA and 21 in ARCA as of summer 2023. Plus, the KM3NeT/ORCA detector is demonstrating capabilities for neutrino astronomy with its lower energy range, with already 6 out of the 115 planned lines.

# Follow-up of O3 gravitational wave events with neutrinos in ANTARES and KM3NeT telescopes

**Mathieu Lamoureux,**[a,b,*] **Damien Dornic,**[c] **Sébastien Le Stum,**[c] **Godefroy Vannoye**[c] **and Gwenhaël de Wasseige**[a] **for the ANTARES and KM3NeT collaborations**

[a] *Centre for Cosmology, Particle Physics and Phenomenology - CP3, Université Catholique de Louvain, B-1348 Louvain-la-Neuve, Belgium*
[b] *Université Paris Cité, CNRS, Astroparticule et Cosmologie, F-75013 Paris, France*
[c] *Aix Marseille Univ, CNRS/IN2P3, CPPM, Marseille, France*

*E-mail:* mathieu.lamoureux@uclouvain.be

Astrophysical neutrinos may be produced during the coalescence of compact objects, in particular those involving neutron stars. Such mergers have been identified through gravitational wave detections by the LIGO and Virgo collaborations and reported in published catalogs. The ANTARES and KM3NeT deep-sea neutrino telescopes are sensitive to neutrino interactions in a wide range of energies, from MeV to PeV. The under-construction KM3NeT telescope covers this energy range with two detectors: ORCA for neutrinos below the TeV and ARCA for TeV–PeV, extending the capabilities of the now-decommissioned ANTARES telescope. This contribution presents the search for neutrinos in time and space correlation with the gravitational wave signals reported during the Third Observing Run of LIGO/Virgo. The ANTARES analysis uses track-like and shower-like events originating from high-energy neutrino interactions. It focuses on a ±500-second time window centered on the time of the merger given by the gravitational wave signal. Two KM3NeT studies are carried out using the data from the partial KM3NeT/ORCA detector: a search for upgoing tracks induced by GeV-TeV neutrinos in the same window as above; and a search for a MeV neutrino signal in a shorter 2-second time window. The results are provided in terms of upper limits on the incoming neutrino flux in the various energy ranges and the total isotropic energies emitted in neutrinos. High-energy observations are also stacked to probe the typical neutrino emission from different populations of mergers. The complementarity of ANTARES and KM3NeT results is also explored.

*Speaker

## 1. Introduction

Since 2015, compact binary mergers have been identified by the detection of gravitational waves with the LIGO/Virgo interferometers [1]. These astrophysical sources, involving stellar black holes and/or neutrons stars may also emit neutrinos. High-energy (GeV–PeV) neutrinos may be produced in hadronic processes occurring in the dense environment surrounding the source [2, 3]; MeV neutrinos may also be produced, especially in the case of binary neutron star mergers [4].

In 2019–2020, the network formed by LIGO and Virgo interferometers identified 83 significant GW sources, reported in three catalogs: GWTC-2, GWTC-2.1, and GWTC-3 [5]. This contribution reports the search for neutrino counterparts from these objects with the ANTARES and KM3NeT deep-sea neutrino telescopes.

The ANTARES detector was located at 2500 m depth, offshore from Toulon (France), and was composed of 12 vertical strings, for a total of 885 photomultiplier tubes (PMTs) and an instrumented volume of about 0.01 $km^3$ [6]. The spacings between the strings and PMTs were optimized for the detection of high-energy neutrinos with 100 GeV–100 PeV. The data taking lasted for 15 years before the experiment was decommissioned in February 2022.

The KM3NeT telescope is the successor of ANTARES and is currently under construction [7]. It is split into two sites: ORCA is located close to the site of ANTARES and its final configuration will have 115 lines covering a volume of 0.007 $km^3$; ARCA, located near Sicily (Italy), will consist of 230 lines instrumenting 1 $km^3$. The geometry of the ORCA (ARCA) detector is optimized for the detection of GeV–TeV (TeV–PeV) neutrinos. During O3, only ORCA was taking exploitable data, specifically with four lines from July 2019 to January 2020 (ORCA4) and six lines from January to March 2020 (ORCA6).

This contribution summarises the O3 follow-up results presented separately from the ANTARES and KM3NeT collaborations [8, 9], as well as combined limits and prospects.

## 2. Method

### 2.1 Event selections

The analyses performed by ANTARES and KM3NeT covered neutrino energies ranging from MeV to PeV. For energies above few GeV (referred to as "high-energy" in the following), a neutrino interaction induces Cherenkov light in the water that may then be reconstructed as a track-like event (mostly induced by muon neutrinos) or a shower-like event (produced by electron neutrinos and neutral current interactions). The neutrino direction can be reconstructed with dedicated algorithms and it is notably possible to separate upgoing from downgoing events, the latter being more likely to be contaminated by atmospheric muon backgrounds. Low-energy (MeV) neutrino interactions produce secondary particles that emit only a few Cherenkov photons, corresponding to a faint signal indistinguishable from optical noises due to bioluminescence and $^{40}K$ decays. Therefore, such neutrinos can only be detected through a global increase of the rate across the whole detector.

In the case of high-energy follow-ups, the search is performed in a time window of ±500 s centered on the GW time $t_{GW}$ and exploiting the spatial correlation between the GW localization and neutrino directions. More precisely, only events with a reconstructed direction within the region $\mathcal{R}_{90}^{+\alpha}$ containing 90% of the GW probability (as it can be uniquely defined from the provided skymap), extended by an angle $\alpha$ to cover for detector angular resolution, are considered. The KM3NeT low-energy search concentrates on a shorter time window $[t_{GW}, t_{GW} + 2\,\text{s}]$ and does not apply any spatial constraints as the original neutrino directions are not available.

The ANTARES search focuses on four event categories: upgoing tracks, downgoing tracks, upgoing showers, and downgoing showers. These only cover energies above 100 GeV and no MeV selection is performed. The expected background is estimated independently for each category and follow-up as it may vary strongly from one GW to another due to the variability of bioluminescence in the sea. The cuts on the track/shower parameters and the value of the extension angle $\alpha$ are optimized to ensure a background level of $2.7 \times 10^{-3}$ events, such that any non-null observation would be associated with a $\geq 3\sigma$ excess while maximizing the expected number of signal events for an $E^{-2}$ spectrum. More details on the selection procedure are presented in [8].

The KM3NeT high-energy search only considers upgoing tracks, while other topologies are not considered. The event selection mainly relies on a Boosted Decision Tree (BDT) classifier to separate mis-reconstructed atmospheric muon background events from upgoing neutrino signals. The ON region consists of upgoing track events in the ±500 s time window and within the region $\mathcal{R}_{90}^{+\alpha}$ and the OFF region is defined by selecting events reconstructed within the same region in local coordinates but outside the time window. For building the latter, data over several months are considered but only during periods with similar data-taking conditions as at the time of the GW (same number of lines and similar event rates). For each GW, the cut on the BDT score is then optimized by minimizing the model rejection factor (MRF) for an $E^{-2}$ neutrino spectrum. The extension $\alpha$ is fixed to 30° as the small size of ORCA4 and ORCA6 configurations does not allow precise direction reconstruction.

The KM3NeT MeV analysis follows the methods originally developed for core-collapse supernova searches and described in [10]. A coincidence is defined by the detection of four close hits within a single DOM in a time window of 15 ns and the coincidence level $C$ is the total number of such occurrences in 500 ms (computed every 100 ms). In the presence of optical noise, the latter is expected to follow a Poisson distribution with a rate $\bar{b}_C$, while a signal from several neutrino interactions across the detector would be associated with an over-fluctuation. The maximum coincidence level in the 2 s time window $C_{\text{max}}$ is thus compared with the expectation, taking into account the number of trials due to the multiple 500 ms slices in this window.

### 2.2 Statistical analysis

The observations are converted into an upper limit on the incoming neutrino time-integrated flux at Earth and on the total energy emitted in neutrinos by the source (correcting for the distance). The standard scenario for the reported limits in high-energy analyses is using the following assumptions:

- The source is assumed to be located within the region containing 90% of the GW probability.
- The neutrino spectrum is assumed to follow a single power law with a spectral index of 2. The flux can be written as $\frac{\mathrm{d}n}{\mathrm{d}E} = \phi \left(\frac{E}{\mathrm{GeV}}\right)^{-2}$, where $\phi$ is the flux normalisation (in GeV$^{-1}$ cm$^{-1}$).
- The neutrinos are equipartitioned between the different flavors so that $\left.\frac{\mathrm{d}n}{\mathrm{d}E}\right|_x = \frac{1}{6} \times \left.\frac{\mathrm{d}n}{\mathrm{d}E}\right|_{\text{tot}}$ for $x = \nu_e, \nu_\mu, \nu_\tau, \bar{\nu}_e, \bar{\nu}_\mu, \bar{\nu}_\tau$. The reported limits are then on the all-flavor emission.

- The neutrino emission is assumed to be isotropic around the GW source. The total energy emitted in neutrinos is then simply related to the flux at Earth: $E^{\text{iso}}_{\text{tot},\nu} = 4\pi D_L^2 \int_{E_{\min}}^{E_{\max}} E \times \frac{\mathrm{d}n}{\mathrm{d}E}\, \mathrm{d}E$, where $D_L$ is the source luminosity distance and $E_{\min}, E_{\max}$ are the bounds of the spectrum.

**High-energy searches** For each event category $c$ (four for ANTARES, one for KM3NeT), the detector acceptance $a^{(c)}$ is estimated as a function of the direction $\Omega$. A Poisson likelihood is then defined to convert the observed and expected number of events $N^{(c)}$ and $B^{(c)}$ into a limit on $\phi$:

$$\mathcal{L}\left(\{N^{(c)}\}; \{B^{(c)}\}, \{a^{(c)}(\Omega)\}, \phi, \right) = \prod_c \text{Poisson}\left(N^{(c)}; B^{(c)} + \phi \cdot a^{(c)}(\Omega)\right), \tag{1}$$

where the product runs over available event categories. A Bayesian analysis is then applied to account for the priors on $\Omega$ (as extracted from the GW skymap), on the background (covering for statistical and systematic uncertainties), on the normalization of the acceptance (systematic uncertainty), and on the parameter of interest $\phi$ (a flat prior is a ssumed). The upper limit is simply obtained by marginalizing all the nuisance parameters and finding the range containing 90% of the marginalized posterior probability. The same strategy is employed for obtaining a limit on $E^{\text{iso}}_{\text{tot},\nu}$ or on the ratio $f^{\text{iso}}_{\nu} = E^{\text{iso}}_{\text{tot},\nu}/E_{\text{tot,GW}}$, where $E_{\text{tot,GW}}$ is the energy radiated in GWs. The method is described at length in [8]. For KM3NeT, as only upgoing tracks are considered in the selection, the obtained constraints are given in the additional assumption that the source is visible in the upgoing sky of the detector at the time of the alert.

**KM3NeT MeV search** For KM3NeT MeV results, pseudo-experiments are generated using the expected mean rate, to estimate how often one would expect to achieve the measured $C_{\max}$. This is then converted to an upper limit on the number of signal events contributing to the observation. Assuming a quasi-thermal neutrino spectrum with an average energy of 13 MeV and the method described in [10], one can then obtain upper limits on the total neutrino flux and on the total energy released in such MeV neutrinos.

## 3. Results

### 3.1 Results from individual searches

Out of the 83 significant GW sources reported by LIGO and Virgo during the O3 run, the ANTARES detector has performed a follow-up for 80 of them. Concerning KM3NeT searches, only GWs after July 1, 2019 are considered, and strict cuts are applied to ensure analysis quality. Therefore, only 50 (55) follow-ups have been done for the KM3NeT high-energy (MeV) analysis.

No significant excess has been detected in any of these searches, such that the main results are upper limits on the neutrino emission. Figure 1 summarises the results in terms of the total energy emitted in neutrinos assuming isotropic emission. The limits range from $10^{54}$ to $10^{59}$ erg for high energies while the MeV analysis yields constraints of the order of $10^{60}$–$10^{63}$ erg.

### 3.2 Stacking analyses

One may also consider the stacking of individual follow-ups into a global constraint on the typical emission from subpopulations of sources, assuming that they all behave the same. This has

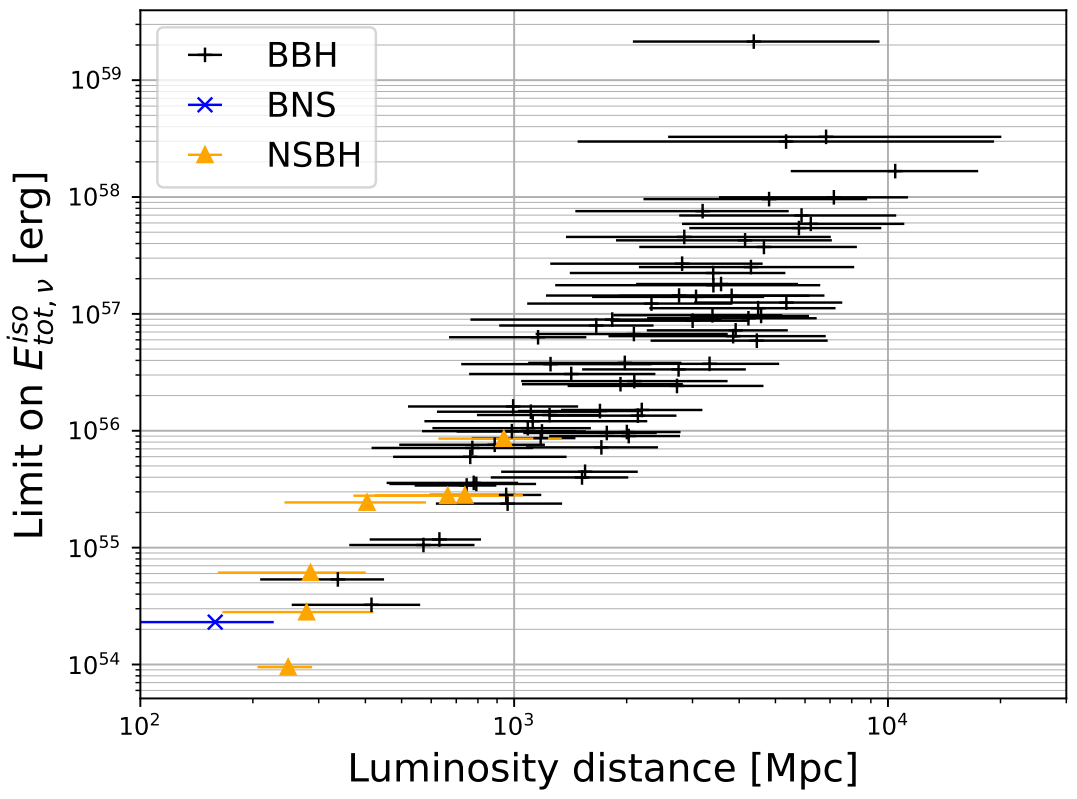

**(a)** ANTARES analysis [8]

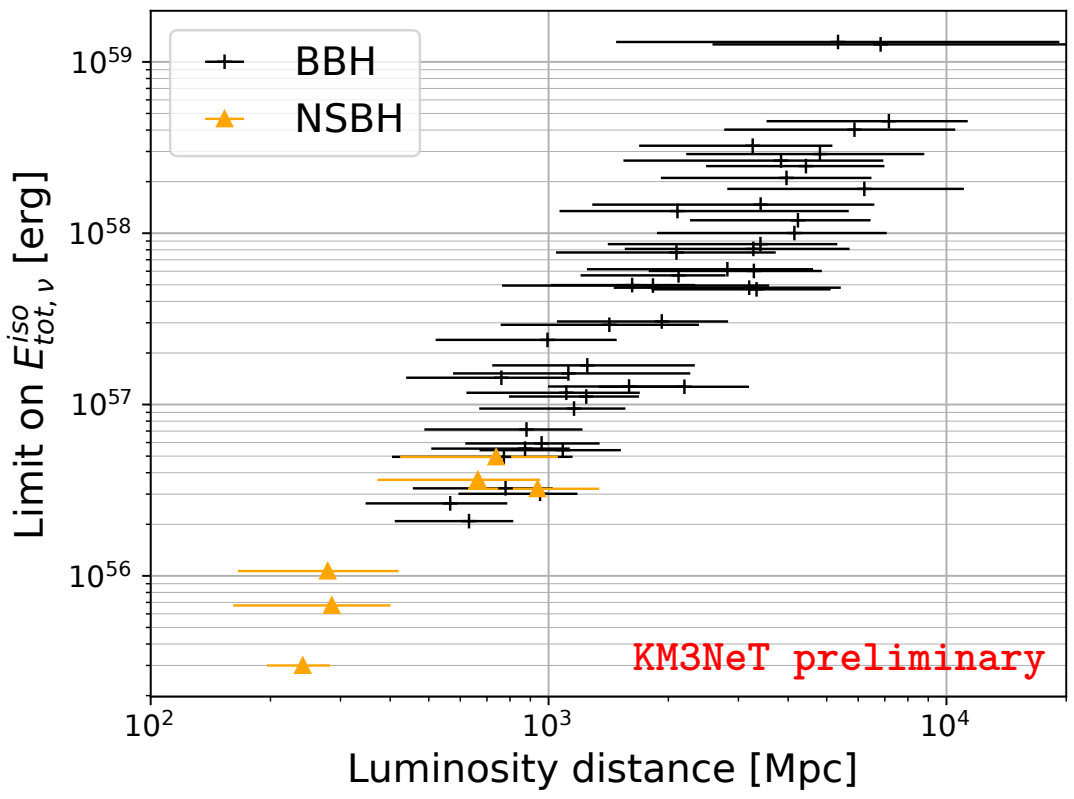

**(b)** KM3NeT high-energy analysis [9]

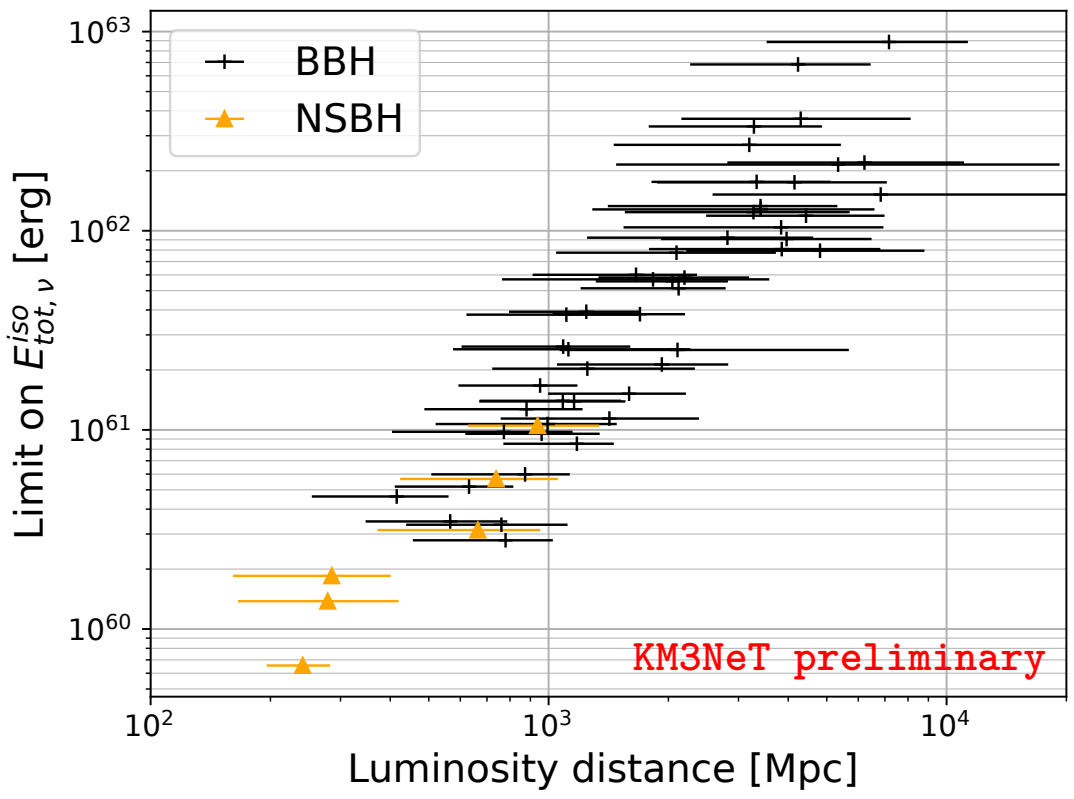

**(c)** KM3NeT MeV analysis [9]

**Figure 1:** 90% upper limits on the total energy $E^{\mathrm{iso}}_{\mathrm{tot},\nu}$ emitted in neutrinos of all flavor as a function of the source luminosity distance, assuming isotropic emission and an $E^{-2}$ (quasi-thermal) spectrum for high-energies (MeV) samples, for the different analyses. The horizontal bars indicate the 5 – 95% range of the luminosity distance estimate, and the markers/colors correspond to the different source categories. The integration range is fixed to $E_{\mathrm{min}} - E_{\mathrm{max}} = 5$ GeV – 100 PeV (1 GeV – 100 PeV) for ANTARES (ORCA).

only been performed for high-energy searches. The GWs have been separated into two categories based on the reported source masses: Binary Black Hole mergers (BBH) if both masses are above $3\,\mathrm{M}_\odot$; Neutron Star Black Hole mergers (NSBH) if one of the masses is below this threshold. In the case of ANTARES, it is straightforward to compute stacked limits by simply multiplying the individual likelihoods. For KM3NeT, the visibility is defined as the probability of the source being in the upgoing sky at the time of the alert, and stacking pseudo-experiments are performed where each GW is considered with a probability equal to its visibility. The final reported stacking limit is the median upper limit from these pseudo-experiments, as shown in Table 1.

For BBH mergers, the combination can improve the constraints by a factor of 5 – 10 with respect to event-by-event follow-ups, thanks to the large number of such sources in the catalogs. Given the low number of NSBH candidate sources, the corresponding stacking does not yet bring significant improvements with respect to individual results. The ANTARES publication [8] also presents stacking limits for other emission scenarios, including softer spectra and jetted emission.

| Category | Experiment | Number of sources | Best individual limits | | Stacking limits | |
|---|---|---|---|---|---|---|
| | | | $E^{\rm iso}_{{\rm tot},\nu}$ [erg] | $f^{\rm iso}_{\nu}$ | $E^{\rm iso}_{{\rm tot},\nu}$ [erg] | $f^{\rm iso}_{\nu}$ |
| BBH | ANTARES | 72 | $3.2\times10^{54}$ | $2.4\times10^{0}$ | $3.8\times10^{53}$ | $1.4\times10^{-1}$ |
| | KM3NeT | 44 | $2.1\times10^{56}$ | $5.7\times10^{1}$ | $3.0\times10^{55}$ | $1.2\times10^{1}$ |
| NSBH | ANTARES | 7 | $9.5\times10^{53}$ | $2.2\times10^{0}$ | $3.2\times10^{53}$ | $8.8\times10^{-1}$ |
| | KM3NeT | 6 | $3.0\times10^{55}$ | $6.8\times10^{1}$ | $1.9\times10^{55}$ | $4.6\times10^{1}$ |

**Table 1:** Comparison of stacking upper limits for ANTARES and KM3NeT high-energy analyses in the BBH and NSBH populations with the best individual upper limits within the same category.

### 3.3 Combination of ANTARES and KM3NeT observations

As the ANTARES and KM3NeT high-energy searches employ very similar selection techniques and statistical approaches, it is natural to compare the obtained results and consider combining them into a single limit. The Figure 2 shows the comparison between the ORCA effective areas (ORCA4 and ORCA6) and the ANTARES effective area considering only the upgoing track selections (ANTARES shower sample is not considered to allow direct comparison). These translate to the differential sensitivities reported on the same figure, where it is clear that, despite being only 3/5% of the total number of lines to be deployed, the ORCA4/6 configurations already provide better constraints for energies below 100 GeV.

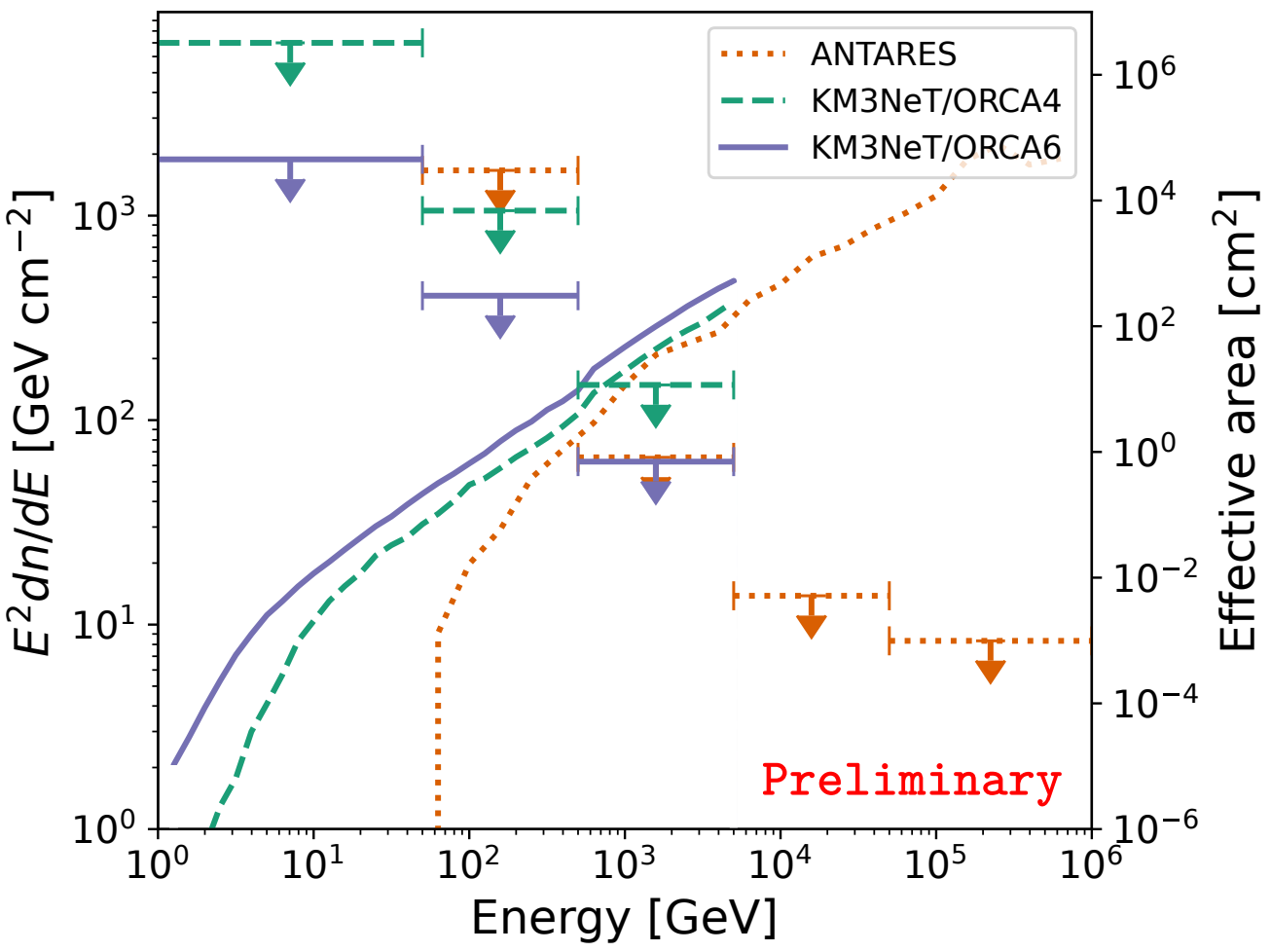

**Figure 2:** Comparison of effective areas of upgoing track selections for ANTARES, KM3NeT/ORCA4, and KM3NeT/ORCA6, computed for GWs with similar sky coverage. The horizontal lines show the corresponding differential sensitivities for various energy ranges.

As the presented KM3NeT selection solely focuses on upgoing tracks, such a combination has only been tested for GW whose localization is fully above the detector horizons at the time of the alert. Only three GWs analyzed both by ANTARES and KM3NeT pass this condition: GW190814, GW190925_232845, GW200208_130117. The first two GWs are during the ORCA4 period while the last one is during ORCA6. The combination is performed by defining the likelihood in Equation 1 with the product running on both ANTARES and KM3NeT event categories and following the same procedure as in the ANTARES-only and KM3NeT-only analyses. Different spectral indices $\gamma$ ranging from 2 (value used above) to 3 are considered.

The results are shown in Figure 3. Even though the integrated ORCA limits are worse than the ANTARES ones for an $E^{-2}$ spectrum, they become competitive for softer spectral indices (towards $\gamma = 3$). For $\gamma > 2.5$, the combination of ANTARES and KM3NeT results improves the constraints by a factor 1.5 – 2 with respect to single-detector limits.

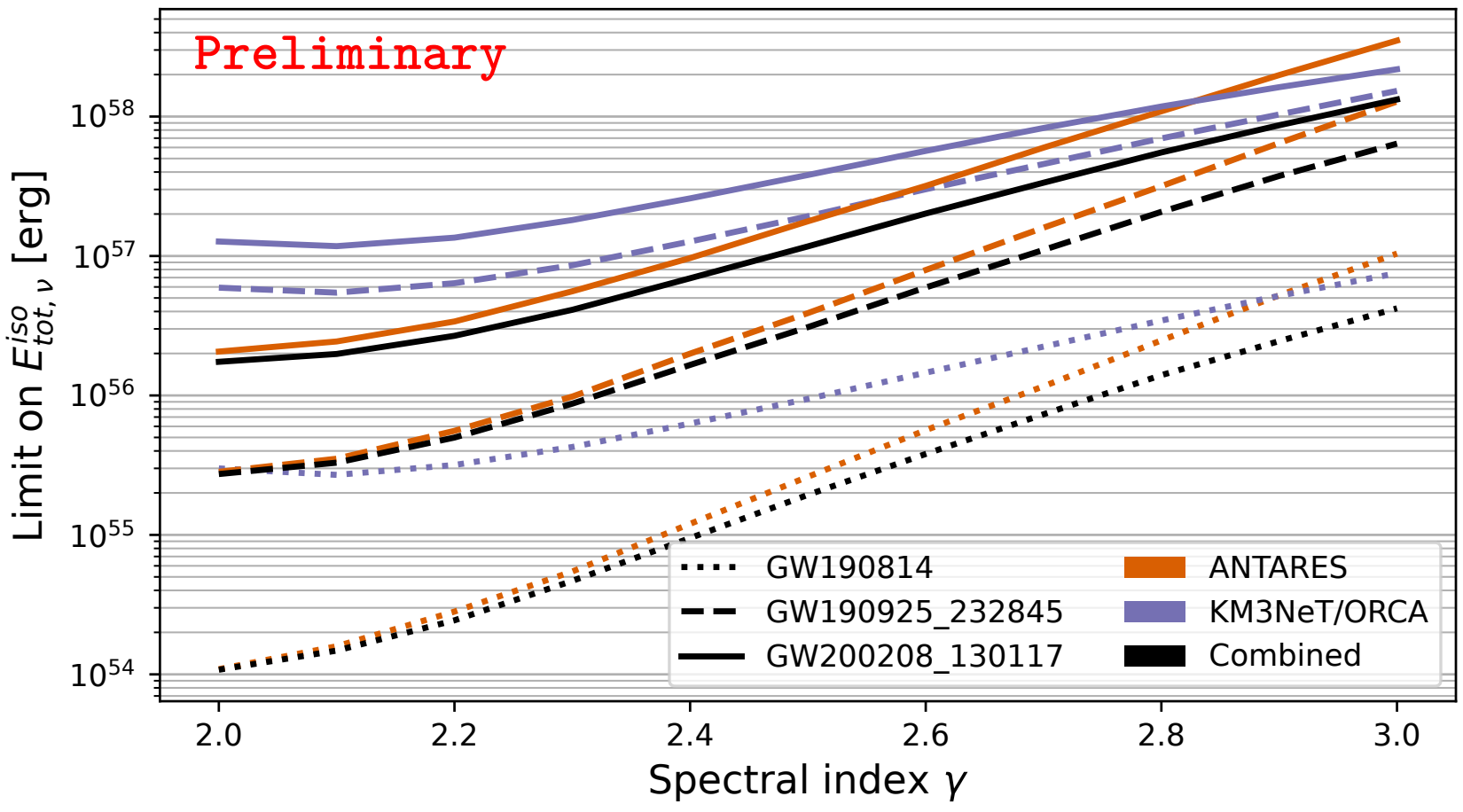

**Figure 3:** Comparison of the upper limits on $E^{\rm iso}_{\rm tot,\nu}$ as a function of the assumed spectral index, for ANTARES, KM3NeT/ORCA, and the combination. The different line styles correspond to the three considered GW sources. The integration range is fixed to $E_{\rm min} - E_{\rm max} = 1\,{\rm GeV} - 100\,{\rm PeV}$ for both analyses.

## 4. Conclusion and outlooks

This contribution summarises the methods and results presented by the ANTARES and KM3NeT collaborations in [8] and [9], respectively. Overall, the searches for neutrinos from GW-emitting sources have yielded no significant excess, such that upper limits on the incoming flux and on the total energy emitted in neutrinos have been computed. The large number of objects allows population studies to constrain the typical emission.

The new observation period from the LIGO, Virgo, and KAGRA collaborations started in May 2023, with enhanced sensitivities. The ANTARES experiment having been decommissioned, it is now up to KM3NeT to take the lead in the search for neutrino counterparts in the depths of the Mediterranean Sea. Since the end of O3, the detector has considerably increased in size, with 18 lines on the ORCA site and 21 lines for ARCA. The latter will participate for the first time in the follow-ups. As it has a complementary energy coverage from TeV to PeV, the joint analyses will greatly enhance KM3NeT sensitivity thanks to the important lever arm between ORCA and ARCA. Such a joint study has been illustrated here with the example of ANTARES+KM3NeT during O3, meaning that the tools are already in place to perform it. Concerning MeV neutrino searches, the gain in sensitivity is mostly proportional to the total number of active lines, meaning that an improvement by a factor > 6 – 10 is expected with respect to ORCA4/6 analyses.

In the near future, using the online framework presented in [11], follow-up results with KM3NeT will be made available a few minutes after the GW public alert, so that an eventual neutrino detection may help in better localizing a potential joint source and guide electromagnetic observations. The first results obtained within this system are available in [12]. In the following years, when new GW source catalogs will be made available, refined searches will be performed. These include notably population studies taking benefit of the expected large number of reported astrophysical objects in these catalogs.

**Acknowledgments**

M.L. is a Postdoctoral Researcher of the Fonds de la Recherche Scientifique - FNRS.

# Follow-up of multi-messenger alerts with the KM3NeT ARCA and ORCA detectors

**J. Palacios González,[a,*] S. Celli, D. Dornic, F. Filippini, E. Le Guirriec, J. De Favereau, G. Illuminati, M. Lamoureux, M. Mastrodicasa, R. Muller, F. Salesa Greus, A. Sánchez Losa, S. Le Stum, G. Vannoye, A. Veutro and A. Zegarelli for the KM3NeT collaboration**

[a] *Instituto de Física Corpuscular (CSIC-UV), Parque Científico, C/ Catedrático José Beltrán, 2. Valencia, Spain.*

*E-mail:* Juan.Palacios@ific.uv.es

The strength of multi-messenger astronomy comes from its capability to increase the significance of a detection through the combined observation of events coincident in space and time. This is particularly valuable for transient events, since the use of a narrow time window can allow a reduction of background of the search.

In KM3NeT, we are actively monitoring and analysing a variety of external triggers in real-time, including alerts like IceCube neutrinos, HAWC gamma-ray transients, LIGO-Virgo-KAGRA gravitational waves, SNEWS neutrino alerts, and others.

In this contribution, we present the follow-up of various external alerts using the complementary capabilities of the two KM3NeT detectors, ORCA (covering the few GeV to few TeV energy range) and ARCA (ranging from sub-TeV energies up to tens of PeV). Both detectors were collecting high-quality data with partial configurations during the period of the studied alerts, which goes from December 2021 until June 2023.

*Speaker

## 1. Multi-messenger astronomy

Multi-messenger (MM) astronomy is based on the principle that detecting various cosmic messengers in spatial and temporal coincidence, originating from the same source, enhances the potential for discovery compared to single-messenger observations. This approach has become critical in the quest to identify the sources of the most energetic cosmic rays.

Different messengers can provide diverse information about the astrophysical cosmic-ray accelerators, each with its own advantages and disadvantages. For example, photons are very abundant particles and easy to detect, but at high energies they are attenuated by the extra-galactic background light, limiting observations to nearby sources. In addition, charged particles like cosmic rays lose directional information once they are deflected by with magnetic fields.

Gravitational waves, discovered in 2015 [1], opened a new window to the Universe. The case of GW170817, with the electromagnetic counterpart quickly identified by Fermi GBM, and followed-up by observations in all wavelengths of light and in all cosmic messenger, presents an excellent example of the potential behind MM astronomy [2]. The recent start of O4 run of gravitational wave observatories motivates follow-up campaigns with instruments all around the world.

Cosmic neutrinos, first hinted at IceCube in 2013 [3], play an important role in MM astronomy. Being electrically neutral particles, neutrinos can provide unique insights into the properties of sources that are not accessible through other cosmic messengers. Traveling undeviated through very large distances, they offer a unique opportunity to investigate the most energetic non-thermal phenomena in the Universe. Given their small interaction cross section with matter, they can also travel through the Universe without any significant absorption.

However, their weakly-interacting nature – together with the large atmospheric backgrounds in cosmic neutrino searches – hinders the possibility of obtaining large statistical samples of cosmic events in neutrino telescopes. This limitation can be addressed with a MM approach, by conducting real-time searches for neutrino signals triggered by transient events detected in other messengers.

In this context, the KM3NeT neutrino observatory is currently setting up an Online Framework system [4] that will take care of performing real-time follow-up of external triggers, reporting relevant results, as well as sending to external observatories alerts of neutrino events having high probability of an astrophysical origin. The aim is to build a system that is as automatic as possible and that can react quickly as required in the MM approach. The structure of the system stems from the online activities performed in the context of the predecessor of KM3NeT, the ANTARES neutrino telescope [5].

This contribution summarizes the results of various MM analyses conducted by the KM3NeT Collaboration. Section 2 describes the performances of the KM3NeT detectors. Section 3 presents the follow-up of IceCube alerts potentially correlated with blazars, one of the first MM analyses performed by the KM3NeT Collaboration. In Section 4, a review of the recent online searches is summarized. Finally, in Section 5, future perspectives are presented.

## 2. The KM3NeT detectors

KM3NeT [6] is a research infrastructure currently deploying two deep-sea neutrino telescopes in the Mediterranean Sea. Two separate apparatuses are currently being deployed: ORCA (Oscilla-tion Research with Cosmics in the Abyss) for the study of atmospheric neutrino oscillations and the neutrino mass hierarchy, and ARCA (Astroparticle Research with Cosmics in the Abyss) for identifying high-energy neutrinos from astrophysical sources. The detectors consist of three-dimensional arrays of photo-multiplier tubes (PMTs) that are able to collect the Cherenkov light emitted as a result of neutrino interactions in seawater.

The main difference between the two detectors is the density of the Digital Optical Modules (DOMs) [7], where the PMTs are hosted. The DOMs are embedded in vertical lines, called Detection Units (DU). The higher DOM density of ORCA is optimised to study the GeV range, while the lower DOM density of ARCA allows it to cover the energy range from sub-TeV to PeV. This complementarity between the two detectors motivates astrophysical studies using data from both over a broader energy range.

The detectors are currently taking data with partial configurations. The high-duty cycle (>95%) together with the full-sky coverage (with better sensitivity to sources in the Galactic plane) makes ARCA and ORCA well-suited detectors to perform MM studies. In addition, the good angular resolution (better than 1° for $E > 10$ TeV for events selected in the analyses) is a crucial feature to perform correlation analyses.

## 3. Follow-up of IceCube alerts

One of the first MM analyses performed by the KM3NeT Collaboration was the follow-up of selected alerts sent by the IceCube Neutrino Observatory, potentially correlated to blazars, between December 2021 and May 2022. Results from these analyses were presented in previous conferences (see e.g. [8]). The selection of potentially-correlated blazars is performed on the basis of multiple criteria, such as the distance to the IceCube best-fit coordinates or the blazar's flaring state.

The search technique is based on a binned ON/OFF method [9]. The ON region, where the signal is expected to dominate over the background, is defined as the region where the angular distance to the source position is lower than the radius of an optimised Region of Interest (RoI). The OFF region is defined as an area of the sky for which the detector has a detection efficiency that is comparable to that of the ON region, while not including it, and in which only background is expected. A declination band centered on the blazar position has been used in the later case. Figure 1 illustrates the definition of these two regions for the case of the blazar candidate PKS 0215+015.

In order to increase the statistics, the OFF region can be extended both in declination size and in time, provided that later it is re-scaled to the size and time span of the ON region. These factors can introduce systematic effects in the analysis that must be taken into account. A declination band width of 30°, centered at the position of the blazar, has been chosen after evaluating its limited effects on the analysis results. Similarly, the extension of the OFF time window covers only periods of times when the detector was in similar data-taking conditions as the time of the alert under study.

At the time of the alerts, KM3NeT/ARCA was taking data with eight active detection lines, while KM3NeT/ORCA had ten. The analyses have been performed using only track-like events, i.e. events with a hit pattern in the detector compatible with a straight line, mainly induced by muon neutrino charged-current interactions. Additionally, only upgoing events have been considered, namely events that have been reconstructed as crossing the Earth, to reduce in our analyses the huge amount of background events due to atmospheric muons.

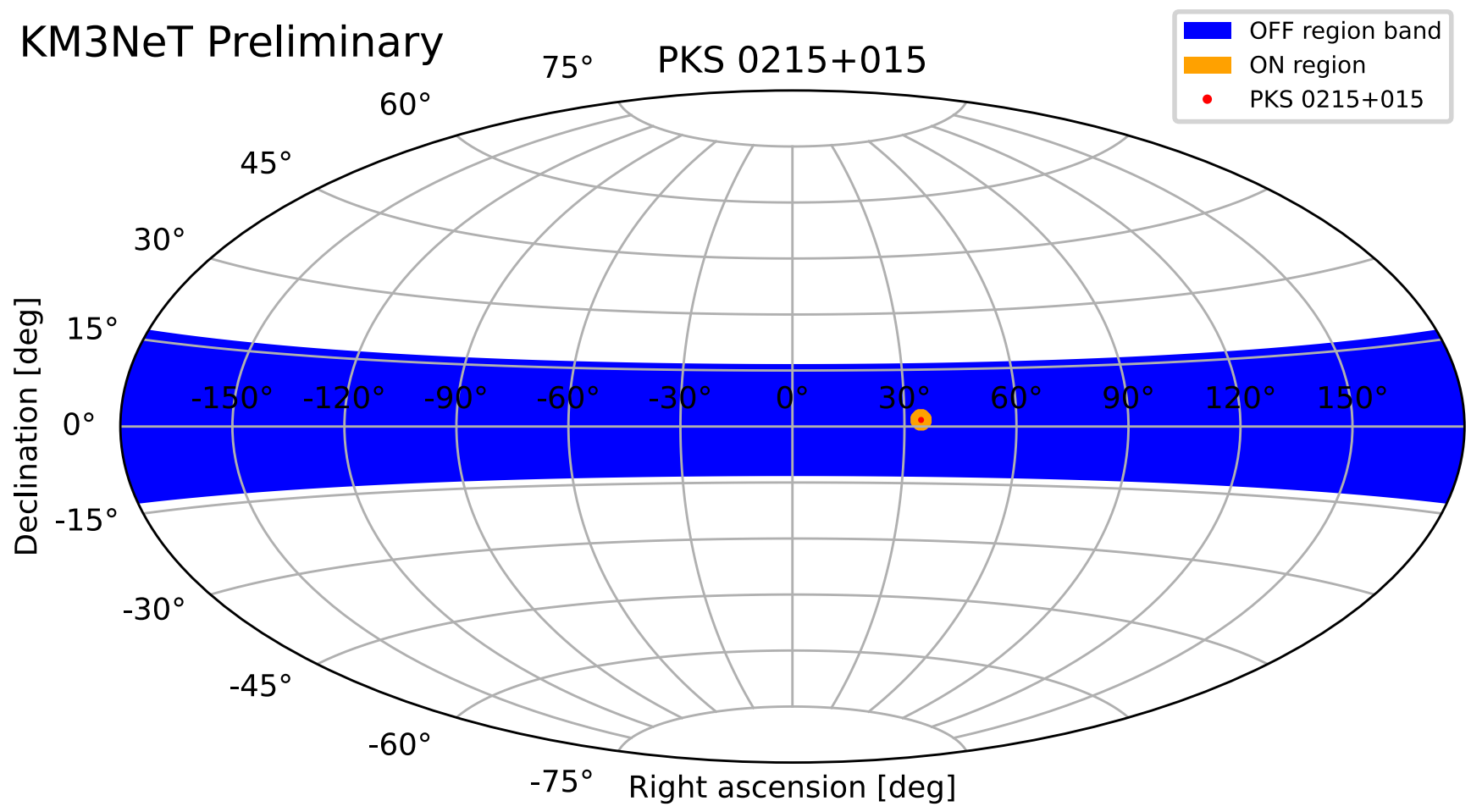

**Figure 1:** Skymap in equatorial coordinates showing the ON and OFF regions for PKS 0215+015 (potentially correlated with IC-220225A).

The event selection has been optimised by means of Monte Carlo simulations. First, a preliminary set of cuts in the reconstruction variables is derived in order to reduce the atmospheric contamination for each alert. Then, an optimum RoI radius is determined both for ARCA and ORCA according to the Model Discovery Potential (MDP) method [10]. The MDP is defined as MDP $= n_\alpha\left(n_{bckg}\right)/n_{sig}$, where the function $n_\alpha$ is defined in our analysis as the signal strength that leads to a $p$-value smaller than $\alpha = 2.7 \cdot 10^{-3}$ (2-sided convention) with a 50% statistical power. The expected signal is computed from MC simulations, using a neutrino flux $\phi \propto E^{-2}$. The optimum value of the RoI radius is determined by the minimum of the MDP.

Finally, concerning the time windows, we have performed the follow-ups in a ± 1 day range for all the alerts, centered around the trigger time provided by IceCube. The results are reported in Table 1 for ARCA and Table 2 for ORCA. In the particular case of IC211208A, we performed also a one-month search. This follow-up was motivated by the fact that the corresponding blazar PKS 0735+17 was found to be in a gamma-ray enhanced emission period during the month of December 2021 [11]. The results for this extended search are reported in Table 3.

No significant deviation from the expected background has been found for any of the analysed alerts. In the case of IC211208A, the 1-month follow-up by KM3NeT/ARCA resulted in one event in the ON region with an estimated energy of ~ 18 TeV. However, this event is compatible with the background expectations, with an associated $p$-value of 0.14 (pre-trial). The results for the follow-up of this alert were published in a dedicated ATel entry [12].

## 4. Online searches

As mentioned in Section 1, the natural step is to advance from an offline search, such as the one described in the previous section, to an online follow-up in real-time. For that purpose, the KM3NeT Collaboration is currently establishing an Online Framework system dedicated to perform multi-messenger related tasks, with a team of dedicated shifters overseeing the correct behaviour of the different analysis steps and f unctionalities. More information about the architecture of the system can be found in a dedicated contribution [4].

| IceCube alert | Potential blazar | Sky location (RA, DEC) | Optimum RoI | Expected background | Expected signal | Events in ON region |
|---|---|---|---|---|---|---|
| IC211208A | PKS 0735+17 | (114.5°, +17.7°) | 1.4° | $(4.7 \pm 0.7) \cdot 10^{-2}$ | $8.9 \cdot 10^{-3}$ | 0 |
| IC220205B | PKS 1741-03 | (266.1°, −3.9°) | 1.9° | $(4.9 \pm 0.9) \cdot 10^{-2}$ | $9.7 \cdot 10^{-3}$ | 0 |
| IC220225A | PKS 0215+15 | (34.5°, +1.7°) | 3.0° | $(2.9 \pm 0.4) \cdot 10^{-3}$ | $1.4 \cdot 10^{-2}$ | 0 |
| IC220304A | TXS 0310+022 | (48.3°, +2.5°) | 2.9° | $(2.6 \pm 0.4) \cdot 10^{-3}$ | $1.4 \cdot 10^{-2}$ | 0 |

**Table 1:** KM3NeT/ARCA results for the follow-up of IceCube alerts potentially correlated with blazars, selected from the period between December 2021 and May 2022, when ARCA was taking data with eight lines. A time window of ± 1 days is applied. The location of the blazars has been taken from the 4FGL-DR3 Fermi-LAT catalogue [13]. No event has been found inside the ON region for any of the searches.

| IceCube alert | Potential blazar | Sky location (RA, DEC) | Optimal RoI | Expected background | Expected signal | Events in ON region |
|---|---|---|---|---|---|---|
| IC211208A | PKS 0735+17 | (114.5°, +17.7°) | 4.2° | $(9 \pm 2) \cdot 10^{-2}$ | $8.6 \cdot 10^{-4}$ | 0 |
| IC220205B | PKS 1741-03 | (266.1°, −3.9°) | 3.6° | $(9 \pm 1) \cdot 10^{-2}$ | $6.7 \cdot 10^{-4}$ | 0 |
| IC220225A | PKS 0215+15 | (34.5°, +1.7°) | 4.0° | $(8 \pm 1) \cdot 10^{-2}$ | $6.5 \cdot 10^{-4}$ | 0 |
| IC220304A | TXS 0310+022 | (48.3°, +2.5°) | 4.0° | $(9 \pm 1) \cdot 10^{-2}$ | $6.3 \cdot 10^{-4}$ | 0 |

**Table 2:** KM3NeT/ORCA results for the follow-up of IceCube alerts potentially correlated with blazars, selected from the period between December 2021 and May 2022, when ORCA was taking data with ten lines. A time window of ± 1 days is applied. The location of the blazars has been taken from the 4FGL-DR3 Fermi-LAT catalogue [13]. No event has been found inside the ON region for any of the searches.

| Detector | Optimal RoI | Expected background | Expected signal | Events in ON region | $p$-value |
|---|---|---|---|---|---|
| KM3NeT/ORCA | 2.3° | $(2.3 \pm 0.2) \cdot 10^{-1}$ | $1.0 \cdot 10^{-2}$ | 0 | 1.0 |
| KM3NeT/ARCA | 1.4° | $(6.6 \pm 0.3) \cdot 10^{-1}$ | $1.2 \cdot 10^{-1}$ | 1 | 0.14 |

**Table 3:** Results for the dedicated follow-up of IC2111208A (potentially correlated to PKS 0735+17) using a time window of 1 month. One event in the ON region has been found in the case of ARCA. This event is not statistically significant, being compatible with the expected background (the $p$-value is 0.14).

From the incoming flux of external alerts received by the system, only the ones that satisfy some minimum conditions are analyzed. These criteria are based on multiple verifications, such as the nature of the event, the visibility conditions, or the reported false alarm rate. Each selected alert is then tagged according to its type: **GRB** (for Gamma-Ray Burst triggers such as the ones provided by Fermi-GBM), **TRANSIENT** (for general transitory phenomena), **GW** (for gravitational wave candidates provided by the LIGO-Virgo-KAGRA Collaborations), and **NEUTRINO** (for high energy events reported by the IceCube Collaboration). Figure 2 shows the rate of incoming alerts from January 2023 to June 2023. Note that, on average, one or two alerts are received every day, which justifies the need for a platform as automatized as possible.

For each type of alert, a different analysis pipeline is used to perform the correlation analysis. Each pipeline incorporates a specific neutrino event selection, optimised according to the nature of the alert to follow up. In addition to it, two main inputs are considered: the time window to be used during the correlation analysis, and the RoI to be studied.

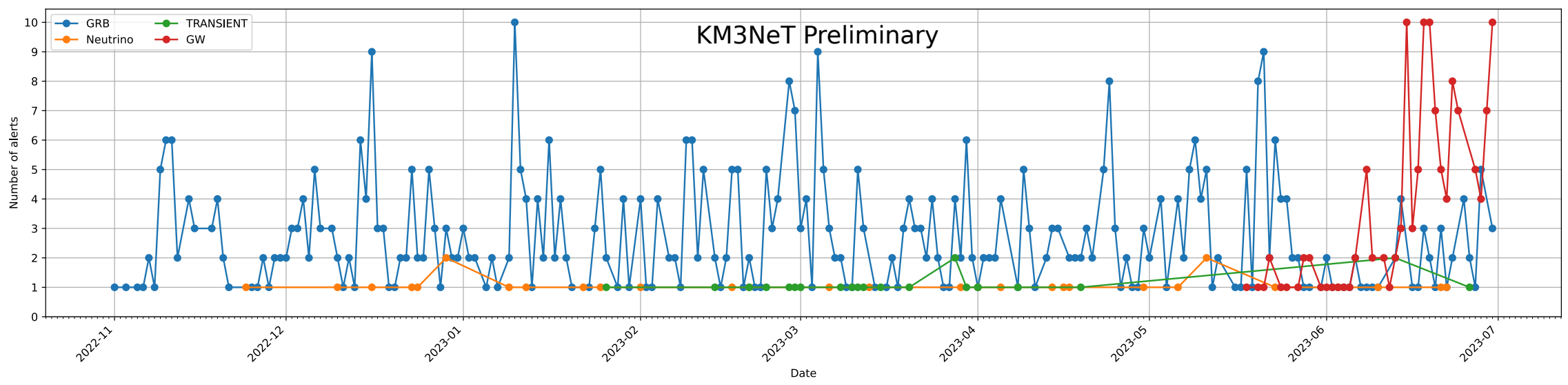

**Figure 2:** Rate of incoming alerts per day during the period November 2022 to June 2023. The most common type of alert is GRB. The start of run O4 for the gravitational waves at the end of May can be clearly seen.

Each pipeline can use multiple time windows that are defined with respect to the trigger time of the alert ($T_0$). The search is performed iteratively, extending the time window to include more data and updating the alert information in order to account for possible refinements from the external observatory. The time windows covered in each case are:

- For **GRB**, four time windows are considered: from $T_0 - 1$ day up to $T_0$, $T_0 + 3$ hours, $T_0 + 12$ hours and $T_0 + 1$ day.
- For **TRANSIENT** and **NEUTRINO**: $[T_0 - 1$ h, $T_0 + 1$ h$]$ and $[T_0 - 1$ day, $T_0 + 1$ day$]$.
- For **GW**: $[T_0 - 500$ s, $T_0 + 500$ s$]$ and $[T_0 - 500$ s, $T_0 + 6$ h$]$. Additionally, a dedicated analysis pipeline searches for MeV neutrinos detected in a short time window of $[T_0, T_0 + 2$ s$]$.

Concerning the RoI, the search method is based on binned techniques, as the analyses described in Section 3. The ON region is determined on the basis of the error box region in the sky associated with the alert being studied, including considerations for the angular uncertainty of the detector. It is, in general, a circular region; in the case of GWs, the contour probability of the gravitational wave event is used, expanded by a few degrees to take into account the angular uncertainty of the detector. These ON regions are defined in equatorial coordinates. In the case of analyses with a time window shorter than one day, the movement of the region in local coordinates is considered.

The event selection is optimised for each alert reducing the expected background to the minimum possible taking into account the shape of the RoI. This expected background is computed from an OFF region defined for each alert, using various days before the alert trigger time in a region with similar coverage in local coordinates as the ON region. Checks on the stability conditions during the ON and the OFF period ensure a stable data-taking flow.

Current analyses consider only track-like events, which are the ones with the best angular resolution. Work is ongoing to include shower-like events (coming from electron neutrino charge-current interactions and all-flavor neutral-current interactions), that have a better energy resolution as they are completely contained in the detector. In addition, currently, the event selection is mainly focus on upgoing events. Progress is being made in the inclusion of downgoing events, to ensure a full sky coverage of the system, with the first tests already performed for GW alerts.

Finally, it is worth mentioning that a dedicated module is being developed to monitor the detection of MeV neutrinos emitted during a Core-Collapse Supernova event (CCSN). This pipeline focuses on the search for an increased rate of coincident hits in the PMTs of the KM3NeT DOMs [14].

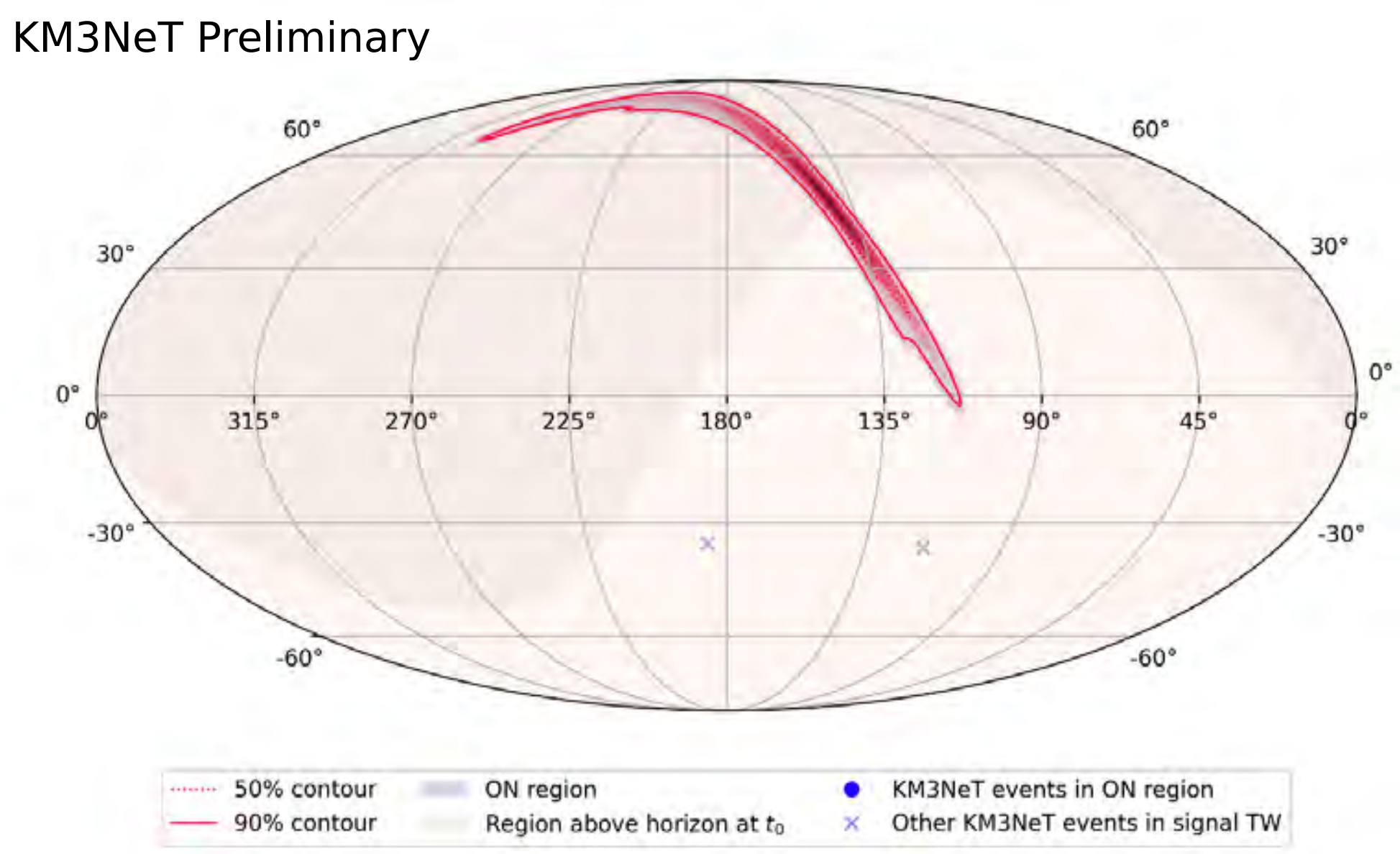

**Figure 3:** Skymap in equatorial coordinates for the GW S230628ax. The light region denotes the upgoing visibility of KM3NeT at the trigger time of the alert.

In the period from October 2022 to June 2023, around 300 alerts have been analysed. About 170 of these are GRB alerts, where high-significance GRB triggers have been considered. In the case of NEUTRINO alerts, around 50 alerts sent by the IceCube Collaboration have been studied, while TRANSIENT alerts accounted 13 times. Concerning GW events, up to mid-June, all the alerts sent by the LIGO Collaboration have been analysed, including the ones from the engineering run ER15. Starting from mid-June, only high-significance GW alerts are examined. In total, more than 100 GW alerts have been followed up, 15 of which corresponding to significant GW events

No significant excess has been found in any of the analyses performed, since the number of track-like candidates that have been found inside the ON region has always been compatible with the atmospheric background expectations. An illustrative example of analysis can be the case of the GW alert S230628ax, a likely BBH merger event detected by LIGO-Livingston and LIGO-Hanford on June 28. Once the event selection has been optimised, in the case of ARCA, an atmospheric background of $2.6 \cdot 10^{-3}$ events is expected for the short time window ($2.3 \cdot 10^{-3}$ for ORCA), being $1.8 \cdot 10^{-2}$ in the case of the longer one ($4.8 \cdot 10^{-2}$ for ORCA). No significant deviation from this background has been observed, as can be seen in Figure 3 for ARCA. In the case of the MeV analysis, the results are also compatible with the expected background. These online analyses will be complemented by offline searches using refined calibrations and reconstructions.

## 5. Conclusions and perspectives

The main results of the follow-up analyses performed with the KM3NeT detectors, based on binned sky searches, have been reviewed in this contribution. For IceCube neutrino alerts, the evolution from offline searches (Section 3) to real-time correlation studies (Section 4) has been outlined. For GRB alerts, offline follow-ups such as the one performed for GRB 221009A [4, 15] are now conducted automatically by the Online Framework as described. Moreover, the real-time follow-up of GWs alerts using KM3NeT data has been detailed.

Although no significant neutrino excess has been found in any of the searches, the increasing interest in multi-messenger campaigns motivates to continue monitoring the sky using the Online Framework. Indeed, the complementary hemispheres covered by KM3NeT and IceCube allows a global coverage of the sky. Work is ongoing to automatise the dissemination of analysis results in the case of interesting alerts, also providing upper limits on the neutrino emission −foreseen by Fall 2023−. Furthermore, the implementation of a module for sending KM3NeT neutrino alerts of likely astrophysical origin is currently in progress and is expected to be completed by the end of the year.

# Measuring atmospheric neutrino oscillations with KM3NeT/ORCA6

**V. Carretero**[a,*] **on behalf of the KM3NeT Collaboration**

[a] *IFIC (UV-CSIC),*
*Carrer del Catedràtic José Beltrán Martinez, 2, 46980, Valencia, Spain*

*E-mail:* vcarretero@km3net.de

The KM3NeT Collaboration is constructing two water-Cherenkov neutrino detectors at the bottom of the Mediterranean Sea: ARCA, which is designed for neutrino astronomy in the TeV to PeV range, and ORCA, optimised for GeV neutrino detection. The ORCA detector will comprise 115 string-like vertical Detection Units arranged in a cylindrical array. Its main objectives are to determine the neutrino mass ordering and measuring atmospheric neutrino oscillations. During 2020 and 2021, an early configuration of the detector with six lines was in operation. A high-purity neutrino sample covering 433 kton-years of exposure was extracted using optimised reconstruction algorithms and machine learning classifiers. In this contribution, the measurement of the neutrino oscillation parameters $\sin^2\theta_{23} = 0.51^{+0.06}_{-0.07}$ and $\Delta m^2_{31} = 2.14^{+0.36}_{-0.25} \cdot 10^{-3}\mathrm{eV}^2$, as well as the sensitivity to determine the neutrino mass ordering based on this data sample, will be presented.

*Speaker

## 1. Introduction

KM3NeT is a research infrastructure for neutrino experiments located on the seabed of the Mediterranean Sea and currently undergoing construction [1]. The infrastructure comprises two detectors that use the same technology but are designed to achieve distinct physics objectives based on their respective spatial configurations. ARCA (Astroparticle Research with Cosmics in the Abyss) is being installed at the KM3NeT-It site, situated 100 km off the coast of Sicily near Capo Passero, Italy, at a depth of 3500 m. ARCA is dedicated to the search for high-energy neutrinos originating from astrophysical sources. ORCA (Oscillation Research with Cosmics in the Abyss) is being built near the coast of Toulon, France, positioned 40 km offshore at a depth of 2500 m. The primary objective of ORCA is to investigate the Neutrino Mass Ordering (NMO) by detecting the neutrino flux generated in the Earth's atmosphere [2]. With a planned instrumented volume of about one cubic kilometer of seawater, ARCA will encompass a mass of one gigaton, while ORCA will cover approximately 7 megatons.

The detection system of ORCA consists of Digital Optical Modules (DOMs) which are pressure-resistant glass spheres housing 31 photomultiplier tubes (PMTs) and corresponding readout electronics [3]. These DOMs are arranged along vertical flexible strings called Detection Units (DUs), anchored to the sea floor and maintained in a vertical position through the buoyancy of the DOMs and a submerged buoy at the top. ORCA will comprise 115 DUs, each equipped with 18 DOMs, with a vertical spacing of 9 m and a horizontal separation of about 20 m. Currently, 18 DUs have been deployed, while the presented results correspond to the initial data acquired using the configuration with 6 DUs referred to as ORCA6.

The ORCA detector employs the Cherenkov effect as its detection principle. Charged particles exceeding the local speed of light induce Cherenkov radiation, which is then registered by the DOMs. This mechanism allows for the reconstruction of interaction parameters, including the interaction vertex, and event topology. By harnessing these capabilities, ORCA enables the exploration of the neutrino properties. Notably, the resulting event topology exhibits distinct light emission characteristics appearing as track-like patterns for GeV muons produced in $\nu_\mu$-CC interactions and shower-like patterns for other neutrino channels.

## 2. Data taking and selection

The ORCA detector has been continuously measuring in the deep sea since mid-2019, while it was extended. From January of 2020 to November 2021, the detector was taking data with a 6-DU configuration. A run selection has been applied to ensure strict data quality criteria resulting in a total of 510 days of data. The time distribution of the exposure is shown in figure 1 in terms of kton-years, taking into account the instrumented volume of seawater, which is corrected by removing non-working PMTs and PMTs with very high rates induced by enviromental optical background. The current dataset corresponds to 433 kton-years. A previous preliminary analysis was carried out with a part of this dataset, in particular using a sample equivalent to 296 kton-years [4]. Several improvements have been made in the selection and analysis with respect to the previous analysis: shower reconstruction was added, selection is now based in machine learning algorithms and energy reconstruction was improved.

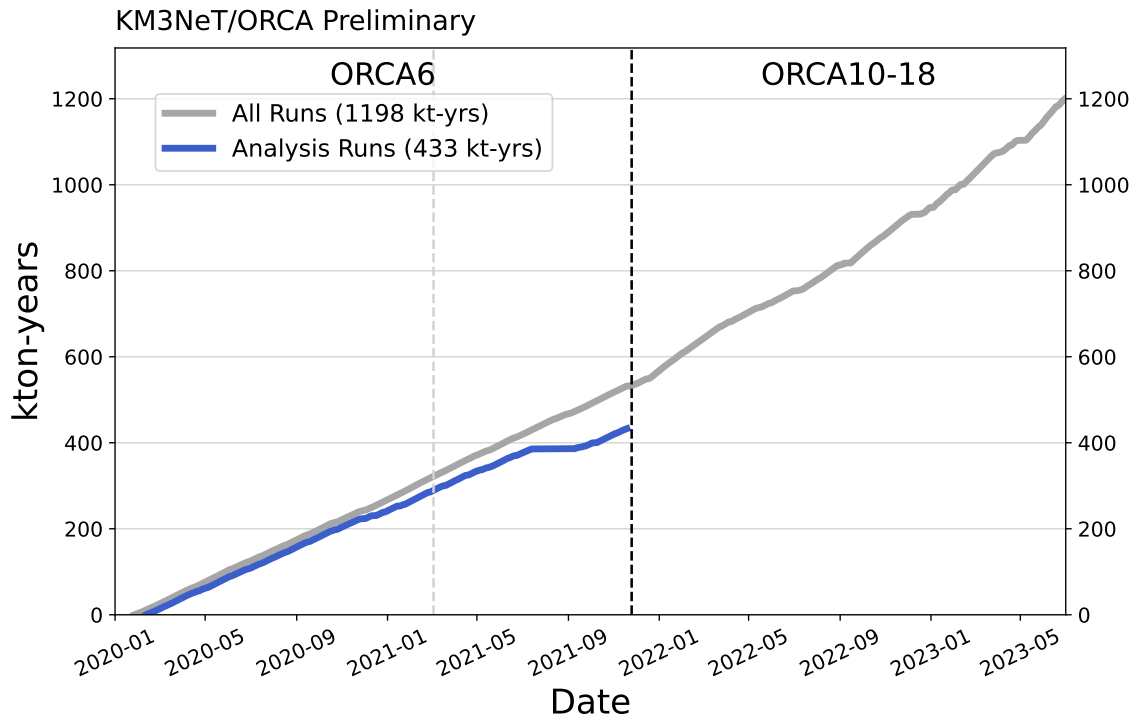

**Figure 1:** Detector exposure shown as a cumulative livetime as a function of time and configuration in units of kton-years.

The overwhelming majority of recorded events in ORCA are atmospheric muons and pure noise, which have to be rejected in the selection process. To remove these events, first selection is applied by requesting a direction reconstruction with a high quality score, a number of triggered hits above a certain threshold and the reconstructed direction to be up-going, since no atmospheric muons are expected in the up-going direction. A Boosted Decision Tree (BDT) machine learning algorithm is then used to assign particle identification scores to the events that allow to remove remaining atmospheric muons and to discriminate between the two possible event topologies: track-like events and shower-like events. To this end, two scores are calculated which will be called atmospheric muon score and track score, respectively.

After implementing that pre-selection and excluding events with high muon scores, the remaining events are initially categorised into two classes based on the track score: tracks and showers. Subsequently, the track class is further divided into two subclasses using the muon score, which serves as a quality parameter to distinguish well-reconstructed tracks from poorly reconstructed tracks that may be misidentified as atmospheric muons. As a result, three distinct classes of events are formed: High Purity Tracks, Low Purity Tracks, and Showers.

To refine the event selection, an additional selection is applied to the reconstructed energy, but with different thresholds for the shower and track classes. Events with a reconstructed energy exceeding 1 TeV for showers and above 100 GeV for tracks are removed. This specific threshold is employed to mitigate the impact of the migration of high-energy events (above 10 TeV) that are not simulated. Based on this selection process, a total of 5830 events are expected from Monte-Carlo (MC) simulations to meet the criteria while a total of 5828 events were actually observed.

Figure 2 displays the compatibility of the data and MC simulations for the described selection process for both scores. Pre-selection is applied to both data and MC distributions. The additional cut on the muon score to reject the atmospheric muon background is already applied in the track score distribution.

The MC distributions are constructed using events generated with gSeaGen [5] and MUPAGE [6], followed by propagation, triggering, and reconstruction using custom KM3NeT software. MC distributions are simulated with oscillation parameters based on NuFit 5.0 [7] (with Super-Kamiokande data) and normal ordering (NO). The measured data and the modelled MC exhibit

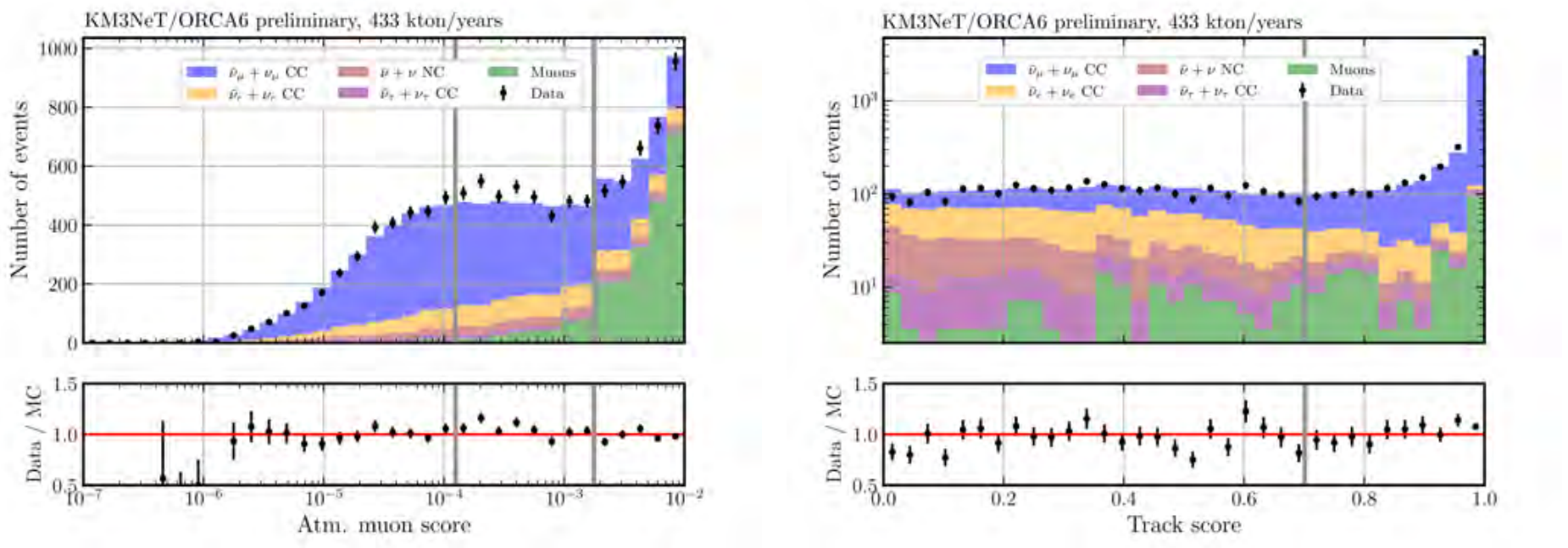

**Figure 2:** Data/MC distributions as a function of the BDT score, atmospheric muon score (left) and track score (right). Pre-selection is applied to both distributions, an additional cut to remove events with atmospheric muon score higher than $1.8 \cdot 10^{-3}$ is applied. Both show an excellent agreement of data and MC.

good agreement for all the distributions.

## 3. Analysis

The analysis is based on 2-dimensional distributions of the reconstructed energy and reconstructed cosine of the zenith angle for each of the three event classes. These distributions are obtained through the implementation of Swim [8], an analysis framework developed for KM3NeT which employs MC simulations to model the detector response. The true energy and zenith angle distributions are calculated for each (anti)neutrino interaction type by taking into account cross sections, neutrino fluxes, the interaction volume, and the oscillation probabilities.

Neutrinos are simulated across a wide true energy range, from 1 GeV to 10 TeV, and from all directions. However, only reconstructed up-going directions are used for the analysis, with 10 bins in the reconstructed cosine of the zenith angle. The range of reconstructed energies spans from 2 to 1000 GeV, employing 15 non-equally spaced bins chosen in a way to ensure sufficient statistics per bin. Note that the 15th bin, ranging from 100 to 1000 GeV, exclusively contains shower events, as tracks with reconstructed energies above 100 GeV are rejected.

To account for the detector resolution, a response matrix which is evaluated by reconstructing MC events is employed. This matrix establishes the relationship between the true and reconstructed variables utilised in the analysis. For each interaction channel $\nu_x$ and classification $i$, a 4-dimensional response matrix $R^{[\nu_x \to i]}(E_{\text{true}}, \theta_{\text{true}}, E_{\text{reco}}, \theta_{\text{reco}})$ is defined. Each entry within the matrix represents the detection efficiency, classification, and reconstruction probability for a given bin of true energy and zenith angle $(E_{\text{true}}, \theta_{\text{true}})$. The number of reconstructed events for a given class $i$ is determined by multiplying the expected number of interacting events with these corresponding efficiencies.

$$n^{i}_{\text{reco}}(E_{\text{reco}}, \theta_{\text{reco}}) = \sum_{x} n^{x}_{\text{int}}(E_{\text{true}}, \theta_{\text{true}}) \times R^{[\nu_x \to i]}(E_{\text{true}}, \theta_{\text{true}}, E_{\text{reco}}, \theta_{\text{reco}}), \quad (1)$$

The resulting effective mass for different neutrino flavours and interaction types is shown in figure 3 as a function of neutrino energy. Selection cuts are applied. The effective masses continue growing as energy grows due to the contribution of events starting outside of the instrumented volume.

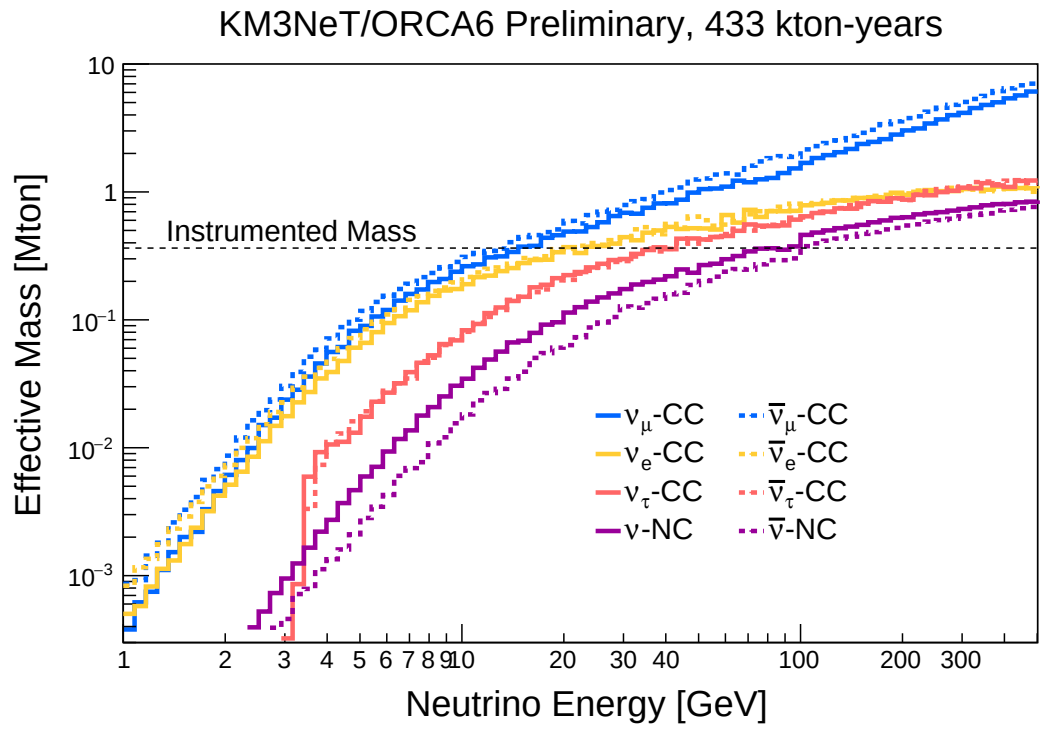

**Figure 3:** Effective mass for different neutrino flavours and interaction types as a function of the true neutrino energy. Selection cuts are applied. The ORCA6 instrumented mass is shown with a vertical line.

The analysis procedure to constrain the oscillation parameters is based on the maximisation of a binned likelihood for the 2-dimensional distribution of events in $\log_{10}(E_{\rm reco}/{\rm GeV})$ and $\cos\theta_{\rm reco}$, comparing the observed data to a model prediction. The sensitivities are computed using the Asimov approach, in which the observed data is replaced by a representative dataset using the expected values of the null hypothesis for each bin [9]. This analysis is not sensitive to $\theta_{13}$, $\theta_{12}$, $\Delta m^2_{21}$ and $\delta_{CP}$, so they are fixed to the NuFit 5.0 [7] values. The log-likelihood is modelled as a combination of Poisson distributions for the expected number of events in each bin and Gaussian distributions associated with the nuisance parameters:

$$-2\log L(\vec{\theta}) = \min_{\vec{\epsilon}} \left\{ 2\sum_{i,j}\left[ (N^{\rm mod}_{ij}(\vec{\theta};\vec{\epsilon}) - N^{\rm dat}_{ij}) + N^{\rm dat}_{ij}\log\left(\frac{N^{\rm dat}_{ij}}{N^{\rm mod}_{ij}(\vec{\theta};\vec{\epsilon})}\right)\right] + \sum_k \left(\frac{\epsilon_k - \langle\epsilon_k\rangle}{\sigma_k}\right)^2 \right\}. \quad (2)$$

$N^{\rm mod}_{ij}$ and $N^{\rm dat}_{ij}$ represent the number of reconstructed events in bin $(i,j)$ expected by the model and the number of observed events, respectively. The parameters of interest, $\vec{\theta}$, are $\theta_{23}$ and $\Delta m^2_{31}$ in the standard oscillation analysis. The rest of the oscillation parameters is kept fixed. The parameters of the model that characterise the distributions ($\vec{\epsilon}$) are composed by nuisance parameters which are related to systematic uncertainties. Some of these parameters are constrained with priors representing constraints from other experiments. Specifically:

1. Normalisations: the overall normalisation as well as the relative normalisations of the High Purity Track and Shower classes are allowed to vary with no constraints. A 20% prior uncertainty is applied to the normalisation of neutral current (NC) and $\tau$-CC events. At high energies, further approximations are included in the light propagation simulation in KM3NeT/ORCA. A 50% relative normalisation uncertainty is applied to events simulated in this regime.

2. Flux: the spectral index of the neutrino flux energy distribution, as $\phi \times E^s$, is allowed to vary from $s = 0$ with a standard deviation of 0.3. The ratio of electron neutrinos to electron antineutrinos is allowed to vary with a 7% prior uncertainty. The ratio of muon neutrinos to muon antineutrinos is allowed to vary with a 5% prior uncertainty. The ratio of muon neutrinos to electron neutrinos is allowed to vary with a 2% prior uncertainty. The ratio of vertical to horizontal neutrinos, introduced as $1 + r_{\mathrm{h/v}} \cos\theta$, is allowed to vary from $r_{\mathrm{h/v}} = 0$ with a standard deviation of 0.02.

3. The absolute energy scale of the detector is allowed to vary with a 9% prior uncertainty. The energy scale is related to the uncertainty on water optical properties and on the knowledge of the PMT efficiencies.

## 4. Results

The ORCA6 dataset has been studied to determine the neutrino oscillation parameters. Specifically, the focus in this section is on constraining the oscillation parameters $\Delta m^2_{31}$ and $\theta_{23}$.

The model is fitted to the dataset using 2-dimensional histograms on the reconstructed energy and direction, as illustrated in figure 4. 5 shows the results of the fit transformed to the L/E (path length over neutrino energy) ratio and normalised with respect to the "non-oscillations" hypothesis for illustration purposes.

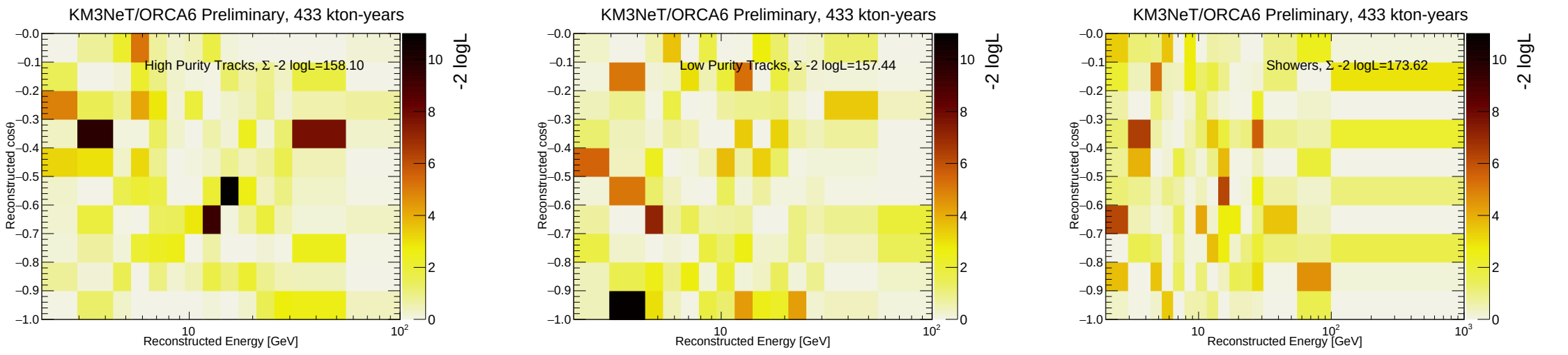

**Figure 4:** Negative log-likelihood landscape as a function of the reconstructed cosine of the zenith angle and energy for the three classes, High Purity Tracks (left), Low Purity Tracks (middle) and Showers (right). Total negative log-likelihood is reported per class.

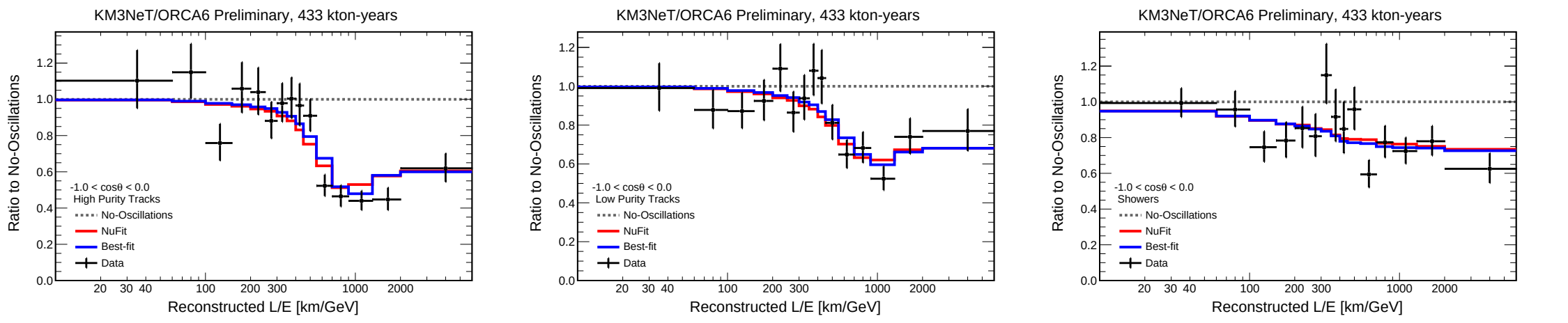

**Figure 5:** Ratio to non-oscillations as a function of the reconstructed path length over reconstructed neutrino energy, $L/E$, for data (black), the best-fit (blue), and NuFit (red) for the three c lasses: High Purity Tracks (left), Low Purity Tracks (middle), and Showers (right). Non-oscillations and NuFit hypotheses are computed taking the best-fit and fixing the oscillation parameters to the corresponding hypothesis.

The fit deviates from the expectations derived from the NuFit model by less than $1\sigma$. There is a preference for NO over IO: $-2\log\left(\frac{L_{NO}}{L_{IO}}\right) = 0.9$. The best-fit values for the parameters are $\sin^2\theta_{23} = 0.51^{+0.06}_{-0.07}$ and $\Delta m^2_{31} = 2.14^{+0.25}_{-0.36}\cdot 10^{-3}\text{eV}^2$.

The space of oscillation parameters has been scanned and profiled in terms of the negative log-likelihood, in order to provide 1-dimensional scans and a 2-dimensional contour of the sensitivity of the ORCA detector for neutrino oscillations. Figure 6 shows the 1-D scans of the profiled likelihood with 68% and 90% CL bands, computed by generating pseudo-experiments with the best-fit values of oscillation and nuisance parameters. The 90% CL contour for both parameters constrained simultaneously is shown in figure 7 in comparison to other experiments.

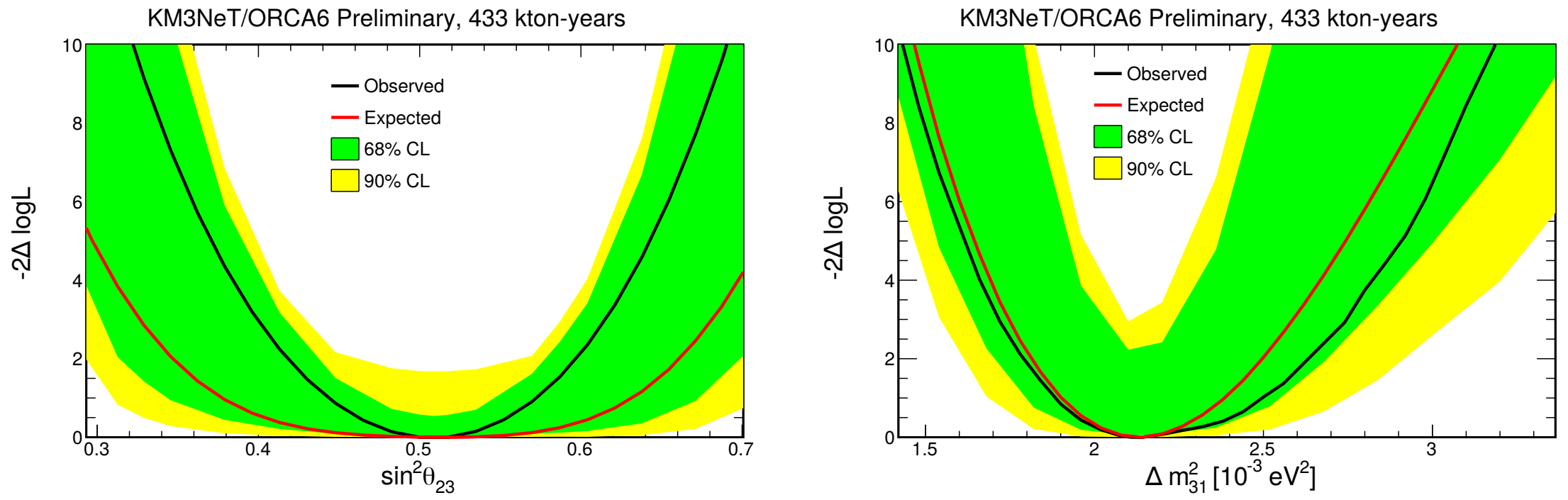

**Figure 6:** Profiled negative log-likelihood scan of the oscillation parameters, $\sin^2\theta_{23}$ (left) and $\Delta m^2_{31}$ (right). Observed limits are compatible within the 68% CL bands.

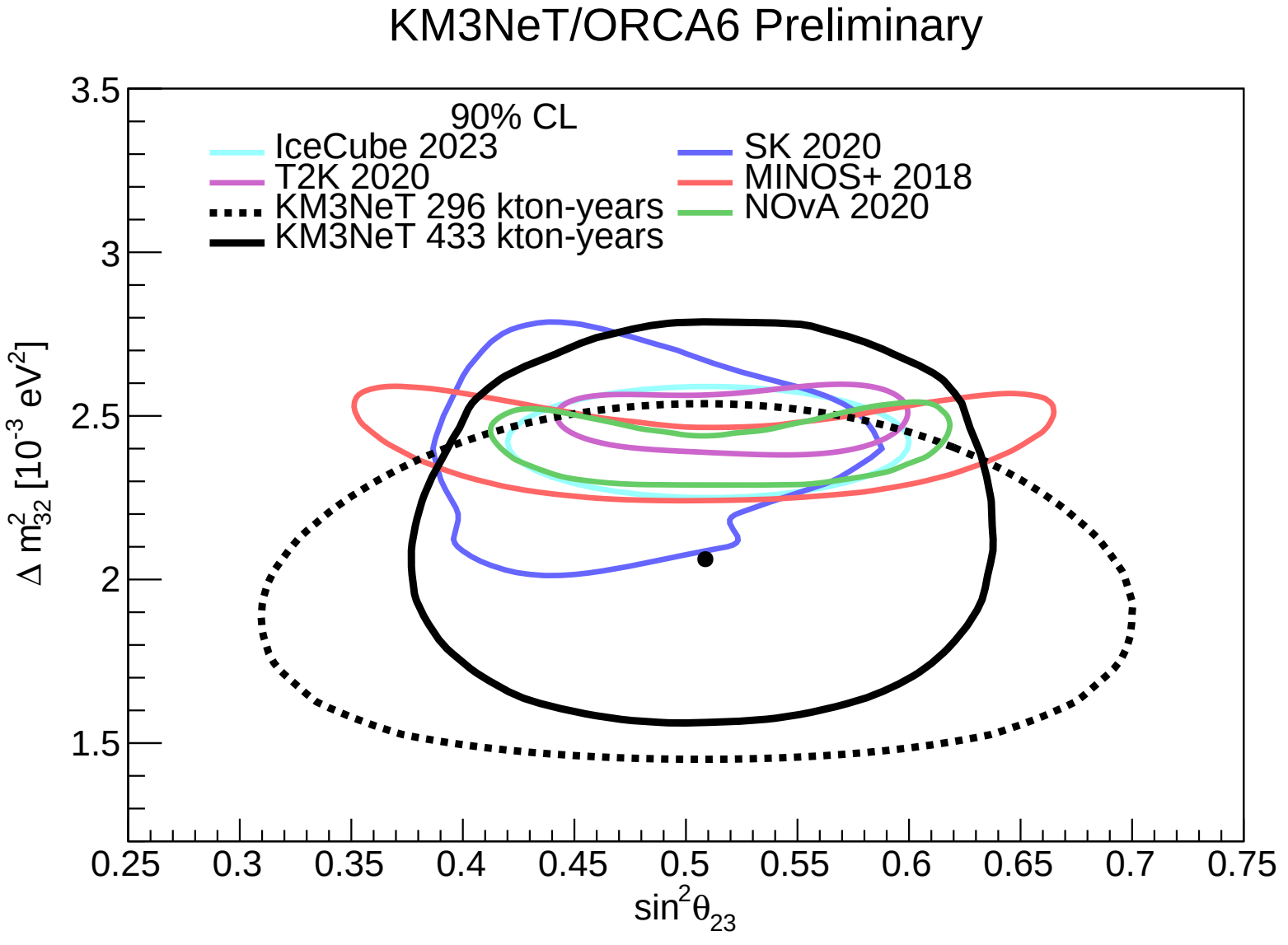

**Figure 7:** Contour at 90% CL of ORCA6 for the oscillation parameters $\sin^2\theta_{23}$ and $\Delta m^2_{31}$ compared with other experiments. A previous result obtained with a first ORCA6 dataset is included [4].

Figure 8 shows the effect of the different systematic uncertainties. The impact is computed comparing the nominal best-fit value of $\theta_{23}$ and $\Delta m^2_{31}$ with the result of the fit when fixing the considered nuisance parameter shifted by ± its post-fit uncertainty. Black dots with error bars represent the pulls of the parameter's best-fit. The error bars are computed as the post-fit uncertainty divided by the pre-fit uncertainty (priors). If the parameter was unconstrained, the pull is computed based on the post-fit uncertainty and the error bar is 1.

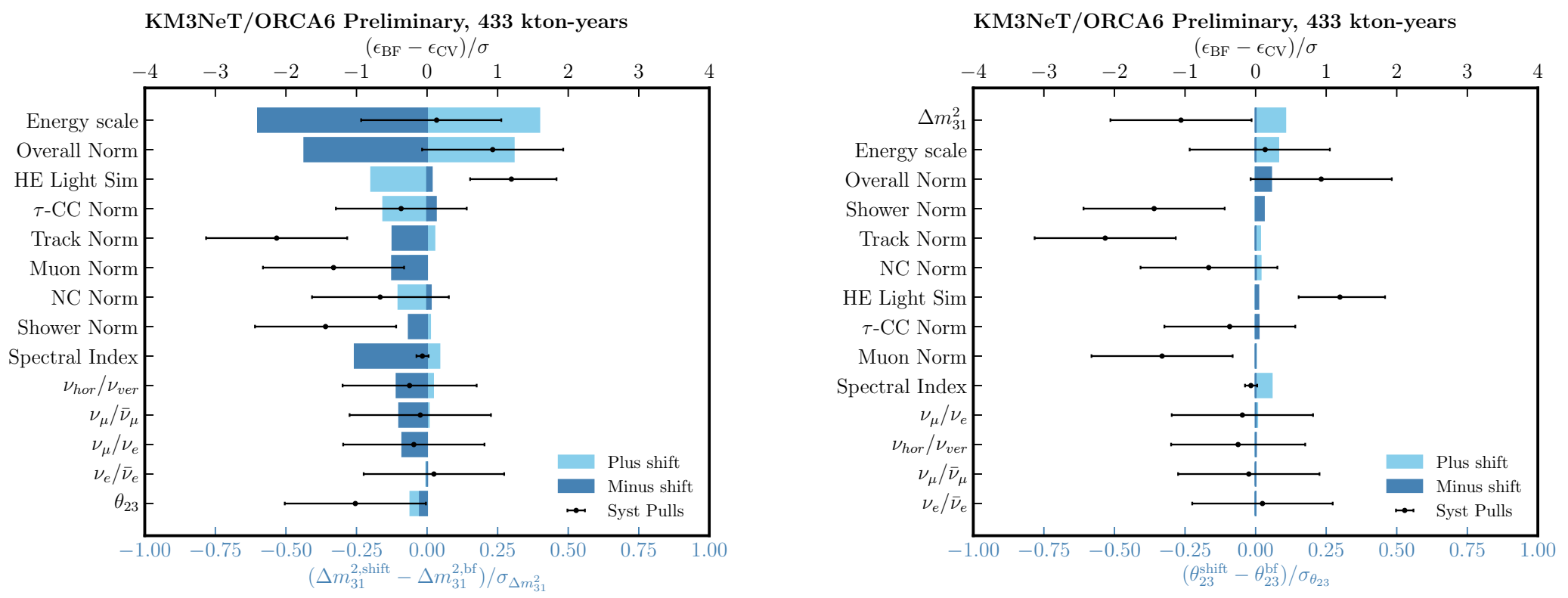

**Figure 8:** Impact of the different systematic parameters computed as a shift in the parameter of interest, $\Delta m^2_{31}$ (left) and $\theta_{23}$ (right) when the nuisance parameter is shifted and fixed. Systematic pulls are reported as black dots.

This test provides insights into the parameter correlations. Specifically, $\Delta m^2_{31}$ exhibits strong correlations with the systematic due to energy scale, energy spectral index, and the overall normalisation. On the other hand, correlations involving $\theta_{23}$ are generally small, although non-negligible contributions are observed from the energy scale, $\Delta m^2_{31}$, and the overall normalisation nuisance parameters.

## 5. Conclusions

With only 5% of its final configuration, the ORCA detector starts to contribute to the measurement of atmospheric neutrino oscillations. The best fit values for the parameters are $\sin^2\theta_{23} = 0.51^{+0.06}_{-0.07}$ and $\Delta m^2_{31} = 2.14^{+0.36}_{-0.25}\cdot 10^{-3}\text{eV}^2$ with a preference for NO: $-2\log\left(\frac{L_{NO}}{L_{IO}}\right) = 0.9$. The detector deployment is progressing continuously and these measurements will gain in precision as the detector volume increases and the reconstruction and selection efficiencies are improved.

# Searches for invisible neutrino decay with KM3NeT/ORCA6

**V. Carretero**[a,*] **on behalf of the KM3NeT Collaboration**

[a] *IFIC (UV-CSIC),*
*Carrer del Catedràtic José Beltrán Martinez, 2, 46980, Valencia, Spain*

*E-mail:* vcarretero@km3net.de

In the era of precision measurements of the neutrino oscillation parameters, upcoming neutrino experiments will also be sensitive to physics beyond the Standard Model. ORCA is an atmospheric neutrino detector currently being built at the bottom of the Mediterranean Sea, that will measure atmospheric neutrino oscillation parameters with high precision and probe new physics at GeV energies. The final ORCA configuration of 115 string-like vertical detection units will be able to probe several theories beyond the Standard Model in neutrino physics. In this work, a three-flavour neutrino oscillation scenario in which the third neutrino mass state, $\nu_3$, decays into an undetectable state, e.g., a sterile neutrino, is investigated with the first configuration of ORCA with six detection lines, ORCA6. A refined high-purity neutrino sample corresponding to 433 kton-years of data taking has been analysed and optimised for the search of this phenomenon. This contribution presents the bounds obtained in the decay parameter, $\alpha_3 = m_3/\tau_3$, and future sensitivity perspectives with ten years of data taking with the future ORCA configuration of 115 detection units.

*Speaker

## 1. Introduction

KM3NeT is an ongoing research infrastructure situated on two sites of the Mediterranean Sea's seabed [1]. ARCA (Astroparticle Research with Cosmic in the Abyss) is positioned 100 km off the Sicilian coast near Capo Passero, Italy, at a depth of 3500 m. With a planned instrumented mass of 1 Gton of seawater, it is dedicated to the search for high-energy neutrinos from astrophysical sources. ORCA (Oscillation Research with Cosmic in the Abyss) is situated near Toulon, France, 40 km offshore at a depth of 2500 m and will cover approximately 7 Mton of seawater. ORCA aims to investigate neutrino oscillations and determine the neutrino mass ordering (NMO) by detecting neutrinos produced in the Earth's atmosphere [2].

The detection system consists of an array of pressure-resistant glass spheres denominated Digital Optical Modules (DOMs) housing 31 photomultiplier tubes (PMTs) and the corresponding readout boards [3]. The DOMs are supported by verticle flexible strings known as Detection Units (DUs), which are firmly anchored to the seabed. ORCA will have a total of 115 DUs, each containing 18 DOMs, vertically spaced by 9 m, while the horizontal separation between DUs is 20 m. Presently, 18 DUs have been deployed. The results presented herein correspond to the initial data collection using the 6-DU configuration referred to as ORCA6.

The detection principle employed by the ORCA detector relies on the Cherenkov effect, which occurs when charged particles surpass the speed of light in the medium and emit Cherenkov radiation. This detection mechanism allows for the accurate reconstruction of essential interaction parameters such as the interaction vertex, energy, direction, and event topology of the interacting neutrino. The resulting event topology exhibits specific characteristics, with track-like patterns for GeV muons produced in $\nu_\mu$-CC interactions and shower-like patterns for other neutrino channels.

The ORCA detector, currently under construction, has been continuously collecting data in the deep sea since mid-2019. From January 2020 to November 2021, data acquisition was carried out using the 6-DU configuration. A careful run selection process was implemented to ensure stringent data quality standards. Following this selection, a total of 510 days of high-quality data were included in the analysis, which, taking into account the current instrumented volume corrected by the working PMTs, corresponds to an exposure of 433 kton-years. Notably, a preliminary analysis searching for invisible neutrino decay was previously conducted using a subset of this dataset, comprising a total of 296 kton-years [4]. Several enhancements have been implemented in both the selection process and the subsequent analysis. Anti-noise cuts and a Boosted Decision Tree (BDT) machine learning algorithm are used to remove the atmospheric muon background (based on an atmospheric (atm.) muon score) and to discriminate between the two possible event topologies: track-like events and shower-like events, based on a track score.

After applying the pre-selection and removing events with high atm. muon scores, the remaining events are divided into two classes: tracks and showers. The track class is then subdivided into two subclasses based on the muon score, distinguishing well-reconstructed tracks from potentially misidentified atmospheric muons. This results in three distinct classes: High Purity Tracks, Low Purity Tracks, and Showers.

To further enhance the event selection, an additional criterion is implemented by imposing distinct thresholds for the reconstructed energy in the shower and track classes. Showers with a reconstructed energy exceeding 1 TeV and tracks with reconstructed energy above 100 GeV are excluded. This specific threshold is chosen to mitigate the influence of unaccounted migrations from higher energies (above 10 TeV) in the simulation. Following this selection, it is expected that 5830 events will satisfy the criteria, whereas a total of 5828 events were observed in the dataset.

## 2. Invisible neutrino decay

Neutrino decay can be mathematically represented by a depletion factor, given by $D = e^{-\frac{t}{\tau_i}}$, where $\tau_i$ corresponds to the rest-frame lifetime of the mass state $m_i$, and $t$ represents the proper time [5]. For relativistic neutrinos in the laboratory frame, the depletion factor can be expressed as $D = e^{-\frac{m_i L}{\tau_i E}}$, where $L$ denotes the distance travelled by the neutrino and $E$ represents its energy. This equation quantifies the fraction of neutrinos with a specific energy that remains intact after traversing a given distance. All three neutrino mass states can decay in principle, but in this analysis we focus on the third neutrino mass state, of which the invisible decay is not constrained from solar and supernova data, as it happens for $\nu_1$ and $\nu_2$. In order to allow for the invisible neutrino decay, a new term must be included in the Hamiltonian:

$$H_{\text{Total}} = \frac{1}{2E}\left[U\begin{pmatrix} 0 & 0 & 0 \\ 0 & \Delta m^2_{21} & 0 \\ 0 & 0 & \Delta m^2_{31}\end{pmatrix}U^\dagger + U\begin{pmatrix} 0 & 0 & 0 \\ 0 & 0 & 0 \\ 0 & 0 & -i\alpha_3\end{pmatrix}U^\dagger\right] + \begin{pmatrix} V & 0 & 0 \\ 0 & 0 & 0 \\ 0 & 0 & 0\end{pmatrix}, \tag{1}$$

where $E$ is the neutrino energy, $U$ is the Pontecorvo-Maki-Nakagawa-Sakata (PMNS) neutrino mixing matrix, $V = \pm\sqrt{2}N_e G_F$ being the matter potential, $N_e$, the electron density in matter and $G_F$, the Fermi constant. Essentially, the only change in the Hamiltonian is a shift in the mass basis term, from $\Delta m^2_{31}$ to $\Delta m^2_{31} - i\alpha_3$.

As a consequence of neutrino decay, the mixing matrix becomes non-hermitian. Consequently, the total sum of neutrino oscillation probabilities deviates from unity,

$$P_{\beta e} + P_{\beta\mu} + P_{\beta\tau} = 1 - P_D(\beta) \qquad \beta = e, \mu, \tau, \tag{2}$$

where $P_D(\beta)$ is the decay probability for flavour $\beta$.

The impact of neutrino decay on the oscillation pattern can be observed in figure 1, depicting the decay effects on the survival ( resp. transition) probability of muon ( resp. electron) neutrinos. The oscillation parameter values used are obtained from NuFit 5.0 [8]. Notably, the decay effects induced by $\alpha_3$ has a more pronounced impact on channels associated with the muon flavor, owing to the relatively higher contribution of $\nu_3$ in the $\nu_\mu$ component. Regardless of the mass ordering, the channel that experiences the most substantial effects of neutrino decay is $P_{\mu\mu}$.

The correlation between $\alpha_3$ and $\theta_{23}$ exhibits a subtle yet distinct behavior in the oscillation and survival channels. Extensive investigations on the interplay between $\theta_{23}$ and $\alpha_3$ have been conducted for specific baselines in references [6, 7]. However, in atmospheric neutrino experiments encompassing a wider range of baselines, this correlation becomes more complex. Figure 2 displays the muon neutrino survival probability (electron neutrino oscillation to muon) for four scenarios characterised by different $\theta_{23} - \alpha_3$ values. The decrease in the survival probability $P_{\mu\mu}$ at the oscillation maxima, induced by neutrino decay effects, can be partially compensated by reducing the value of $\theta_{23}$ to the lower octant. However, such compensation results in an increased probability in the energy range where matter effects play a significant ro le. On the other hand, for the transition probability $P_{e\mu}$, a higher value of $\theta_{23}$ serves as a counterbalance to the decay-induced decrease. This will partially affect to the possibility to constrain both parameters at the same time.

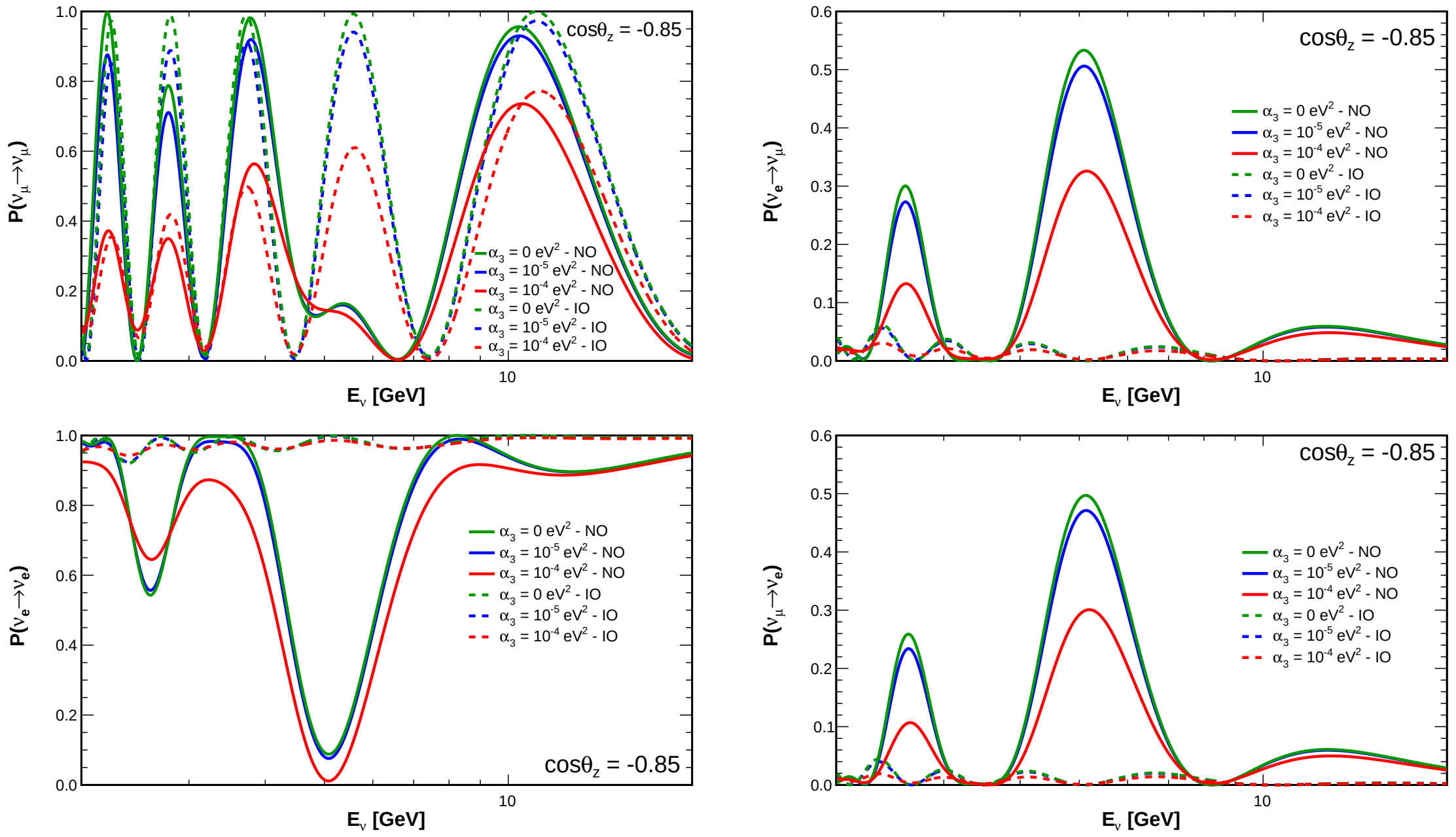

**Figure 1:** Probability for muon neutrino survival (top left), electron-to-muon transition (top right), electron neutrino survival (bottom left) and muon-to-electron transition (bottom right) as a function of energy at a cosine of the zenith angle $\cos\theta_z = -0.85$. Three values of the decay constant are considered: $\alpha_3 = 0$ (green), $\alpha_3 = 10^{-5}$ eV$^2$ (blue) and $\alpha_3 = 10^{-4}$ eV$^2$ (red). The solid (dashed) curves are for normal ordering, NO (inverted ordering, IO). Antineutrino probabilities can be described by the same curves but swapping the orderings.

## 3. Analysis

The analysis relies on 2-dimensional distributions of the reconstructed energy and reconstructed cosine of the zenith angle for each of the three event classes. These distributions are obtained through Swim [9], an analysis framework developed for KM3NeT which incorporates Monte Carlo (MC) simulations to model the detector response. Cross sections, neutrino fluxes, the interaction volume, and oscillation probabilities are taken into account when calculating the true energy and zenith angle distributions for each type of (anti)neutrino interaction.

Neutrinos are simulated from all angles and over a broad true energy range, from 1 GeV to 10 TeV. For the analysis, 10 bins in the reconstructed cosine of the zenith angle are used, and only reconstructed upward directions are taken into account. The reconstructed energies range from 2 to 1000 GeV, employing 15 unevenly spaced bins to ensure that each bin had enough statistics. Note that tracks with reconstructed energies above 100 GeV are rejected, so the 15th bin, which ranges from 100 to 1000 GeV, only contains shower events.

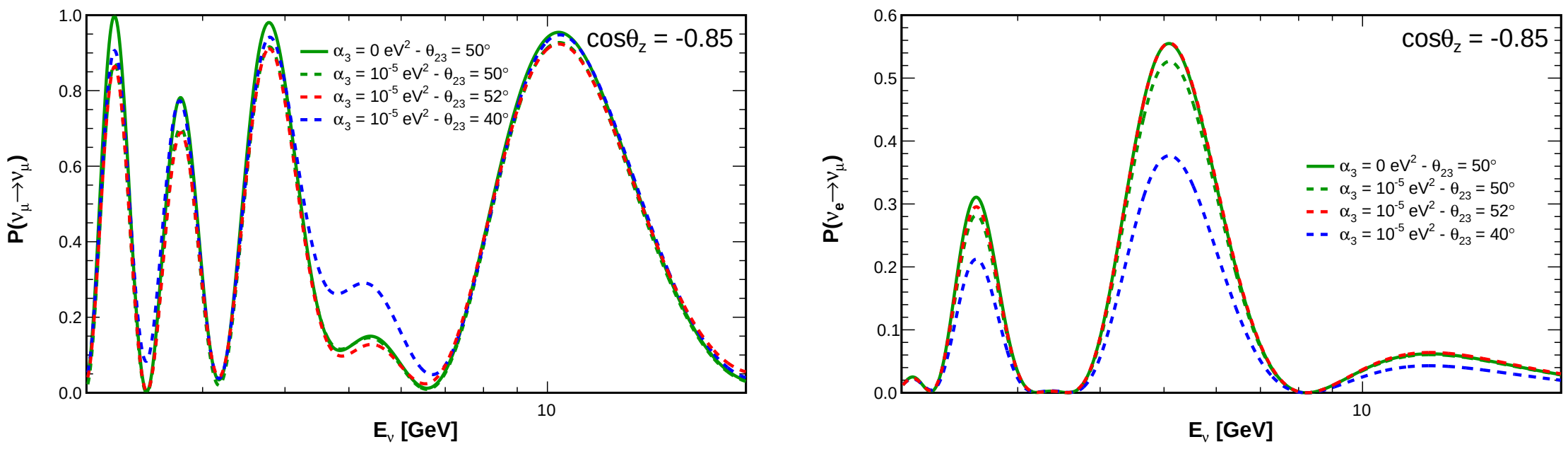

**Figure 2:** Probability for muon neutrino survival (left) and electron-to-muon transition (right) as a function of energy at a cosine of the zenith angle $\cos\theta_z = -0.85$ assuming NO. Four cases are shown: $\alpha_3 = 0$ with $\theta_{23} = 50°$ (solid green), $\alpha_3 = 10^{-5}$ eV$^2$ with the same value of $\theta_{23}$ (dashed green) and with two different values: $\theta_{23} = 52°$ (dashed red) and $\theta_{23} = 40°$ (dashed blue).

A response matrix is used to account for the detector resolution and it is evaluated by reconstructing MC events that relates true and reconstructed variables. For each interaction channel $\nu_x$ and class $i$, a 4-dimensional response matrix $R^{[\nu_x \to i]}(E_{\text{true}}, \theta_{\text{true}}, E_{\text{reco}}, \theta_{\text{reco}})$ is defined. Each entry within the matrix represents the efficiency of detection, classification, and reconstruction probability for a specific true bin $(E_{\text{true}}, \theta_{\text{true}})$. The expected interacting events are multiplied by these corresponding efficiencies to yield the reconstructed events of a given class $i$.

$$n^i_{\text{reco}}(E_{\text{reco}}, \theta_{\text{reco}}) = \sum_x n^x_{\text{int}}(E_{\text{true}}, \theta_{\text{true}}) \times R^{[\nu_x \to i]}(E_{\text{true}}, \theta_{\text{true}}, E_{\text{reco}}, \theta_{\text{reco}}), \tag{3}$$

The analysis method used to constrain the invisible decay parameter, $\alpha_3$, is based on the maximization of a binned log-likelihood of the 2-dimensional distribution of events in $\log_{10}(E_{\text{reco}}/\text{GeV})$ and $\cos\theta_{\text{reco}}$, which compares the observed data to a model prediction. The sensitivities are computed using the Asimov approach, where the observed data is replaced by a representative dataset defined as the one which provides the expected values of the null hypothesis in each bin [10]. This analysis is not sensitive to $\theta_{13}$, $\theta_{12}$, $\Delta m^2_{21}$ and $\delta_{CP}$, so they are fixed to the NuFit 5.0 [8] values. Poisson distributions for the expected number of events in each bin and Gaussian distributions related to the nuisance parameters are used to model the log-likelihood:

$$-2\log L(\vec{\theta}) = \min_{\vec{\epsilon}} \left\{ 2 \sum_{i,j} \left[ (N^{\text{mod}}_{ij}(\vec{\theta}; \vec{\epsilon}) - N^{\text{dat}}_{ij}) + N^{\text{dat}}_{ij} \log\left( \frac{N^{\text{dat}}_{ij}}{N^{\text{mod}}_{ij}(\vec{\theta}; \vec{\epsilon})} \right) \right] + \right. \tag{4}$$

$$\left. + \sum_k \left( \frac{\epsilon_k - \langle \epsilon_k \rangle}{\sigma_k} \right)^2 \right\}.$$

$N^{\text{mod}}_{ij}$ and $N^{\text{dat}}_{ij}$ represent the number of reconstructed events expected by the model and the number of events observed, respectively, in the bin $(i,j)$. The parameter of interest, $\vec{\theta}$, is $\alpha_3$ in this analysis. The parameters of the model that characterise the distributions ($\vec{\epsilon}$) are composed by nuisance parameters which are related to systematic uncertainties. Some of these parameters are constrained with priors representing constraints from other experiments. Specifically:

1. Normalisations: the overall normalisation as well as the relative normalisations of the High Purity Track and Shower classes are allowed to vary with no constraints. A 20% prior uncertainty is applied to the normalisation of neutral current (NC) and $\tau$-CC events. At high energies (above 500 GeV), as a result of further approximations in the light propagation simulation in KM3NeT/ORCA, a 50% relative normalisation uncertainty is applied to events simulated in this regime.
2. Flux: the spectral index of the neutrino flux energy distribution, as $\phi \times E^s$, is allowed to vary from $s = 0$ with a standard deviation of 0.3. The ratio of electron neutrinos to electron antineutrinos is allowed to vary with a 7% prior uncertainty. The ratio of muon neutrinos to muon antineutrinos is allowed to vary with a 5% prior uncertainty. The ratio of muon neutrinos to electron neutrinos is allowed to vary with a 2% prior uncertainty. The ratio of vertical to horizontal neutrinos, introduced as $1 + r_{\text{h/v}} \cos\theta$, is allowed to vary from $r_{\text{h/v}} = 0$ with a standard deviation of 0.02.
3. The absolute energy scale of the detector is allowed to vary with a 9% prior uncertainty. The energy scale is related to the uncertainty on water optical properties and on the knowledge of the PMT efficiencies.
4. Oscillation parameters: $\Delta m^2_{31}$ and $\theta_{23}$ are allowed to vary without constrains.

## 4. Results

The ORCA6 dataset has been studied to constrain the invisible decay parameter, $\alpha_3$. The model is fitted to the dataset using 2-dimensional histograms on the reconstructed energy and direction. The results of the fit are subsequently transformed to the L/E (path length over neutrino energy) variable for the purpose of visualising the outcomes in figure 3. This is done for the standard case (stable scenario), decay ($\alpha_3$ is freely fitted) and a high decay scenario to illustrate its effects ($\alpha_3 = 1.1 \times 10^{-3}\text{eV}^2$, nuisance parameters are fixed to the decay best-fit).

The best-fit values for the parameters are $\sin^2\theta_{23} = 0.51^{+0.06}_{-0.07}$ and $\Delta m^2_{31} = 2.14^{+0.36}_{-0.25} \cdot 10^{-3}\text{eV}^2$ for the standard case and $\sin^2\theta_{23} = 0.52^{+0.07}_{-0.07}$, $\Delta m^2_{31} = 2.21^{+0.33}_{-0.24} \cdot 10^{-3}\text{eV}^2$ and $\alpha_3 = 1.08^{+1.4}_{-0.7} \cdot 10^{-4}\text{eV}^2$ for the decay best-fit. The significance of the data preference to invisible neutrino decay compared to the stable scenario is estimated at $1.8\sigma$.

The invisible neutrino decay parameter has been scanned and profiled in terms of the negative log-likelihood to provide a 1-dimensional scan in figure 4 (left) in comparison with the results of other experiments (T2K+NO$\nu$A combination [12], T2K+MINOS combination [13] and SK+K2K+MINOS combination [14]). The 2-dimensional scan to constrain $\theta_{23}$ and $\alpha_3$ at the same time is shown as a 90% CL contour in figure 4 (right) compared to the previous ORCA6 result [4]. At every point in the contour, the log-likelihood is minimised relative to all nuisance parameters and the 90% CL contour is drawn through the recovered likelihood landscape.

Figure 5 shows the impact of the different systematic uncertainties is computed comparing the nominal best-fit value of $\alpha_3$ with the result of the fit when fixing the considered nuisance parameter shifted $\pm$ its post-fit uncertainty. Black dots with error bars represent the pulls of the best-fit parameters. The error bars are computed as the post-fit uncertainty divided by the pre-fit uncertainty (priors). If the parameter was unconstrained, the pull is computed based on the post-fit uncertainty and the error bar is 1. This test provides insights into the parameter correlations. Specifically, $\alpha_3$ exhibits strong correlations with the normalisations, the spectral index, the horizontal to vertical flux ratio and $\theta_{23}$.

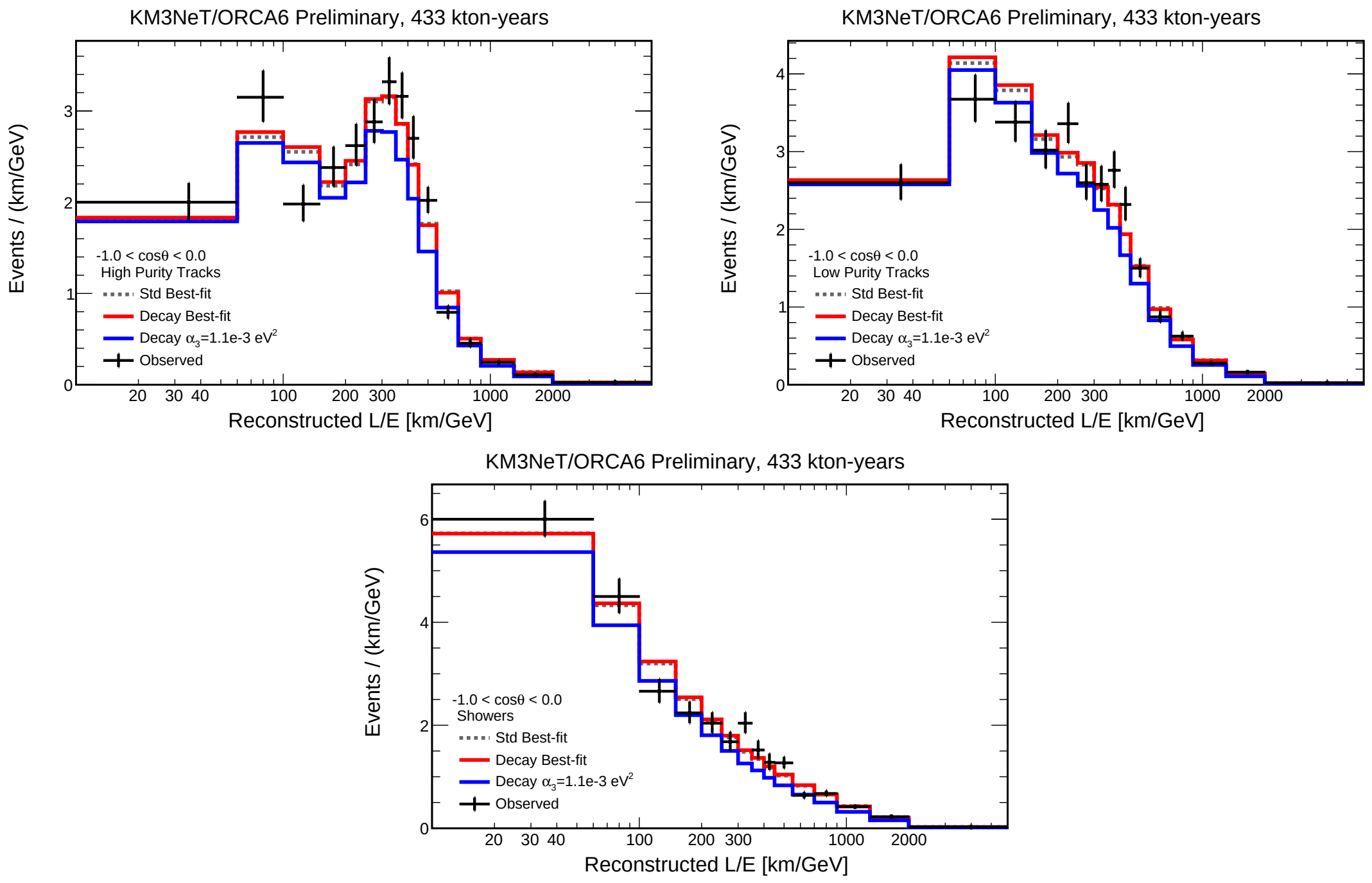

**Figure 3:** Distributions of the events divided by the bin width as a function of the reconstructed path length over neutrino energy, $L/E$, for the Standard bestfit (gray), decay bestfit (red), high decay scenario (blue) and data (black) for the three classes, High Purity Tracks (top left), Low Purity Tracks (top right) and Showers (bottom).

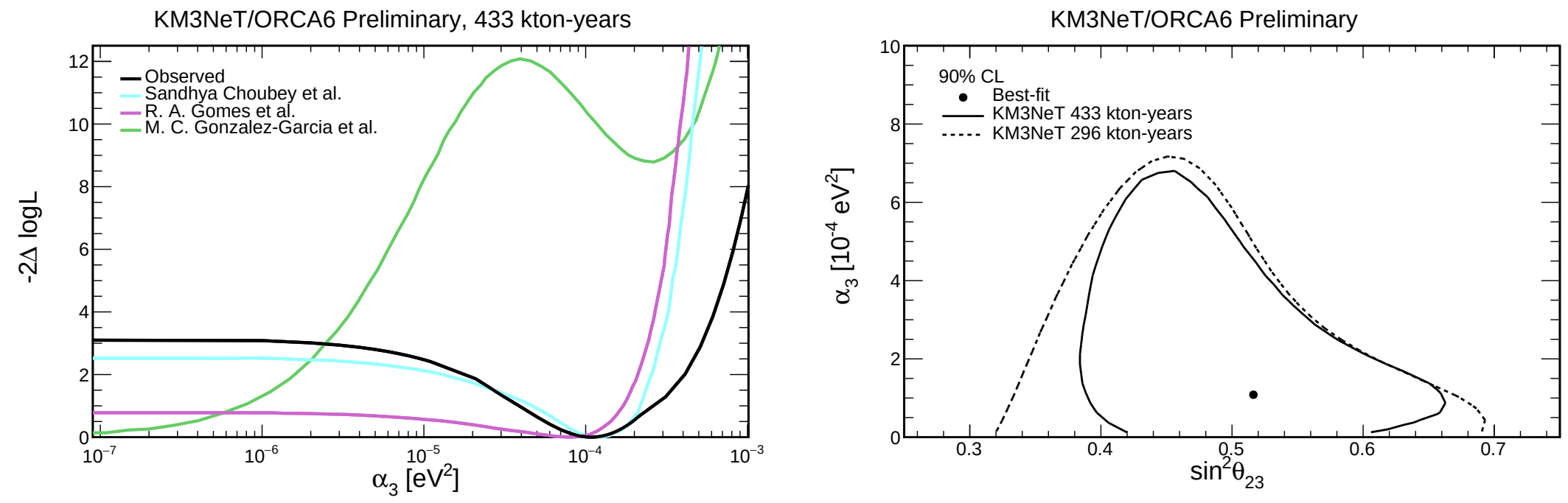

**Figure 4:** Profiled negative log-likelihood scan of the invisible neutrino decay parameter, $\alpha_3$ (left) and 90% CL $\theta_{23} - \alpha_3$ contour (right).

## 5. Conclusions

With only 5% of its final configuration, the ORCA detector is starting to provide competitive results probing physics beyond the Standard Model such as the invisible neutrino decay. The best fit value is $\alpha_3 = 1.08^{+1.4}_{-0.7} \cdot 10^{-4}\mathrm{eV}^2$ with a preference for invisible neutrino decay of $1.8\sigma$ with respect to the stable scenario. The detector deployment is progressing steadily and these measurements will improve as the detector volume gets larger and reconstruction and selection efficiencies are enhanced. Specifically, after 10 years of the complete ORCA115 detector, the 90% CL sensitivity is expected to improve by two orders of magnitude [15].

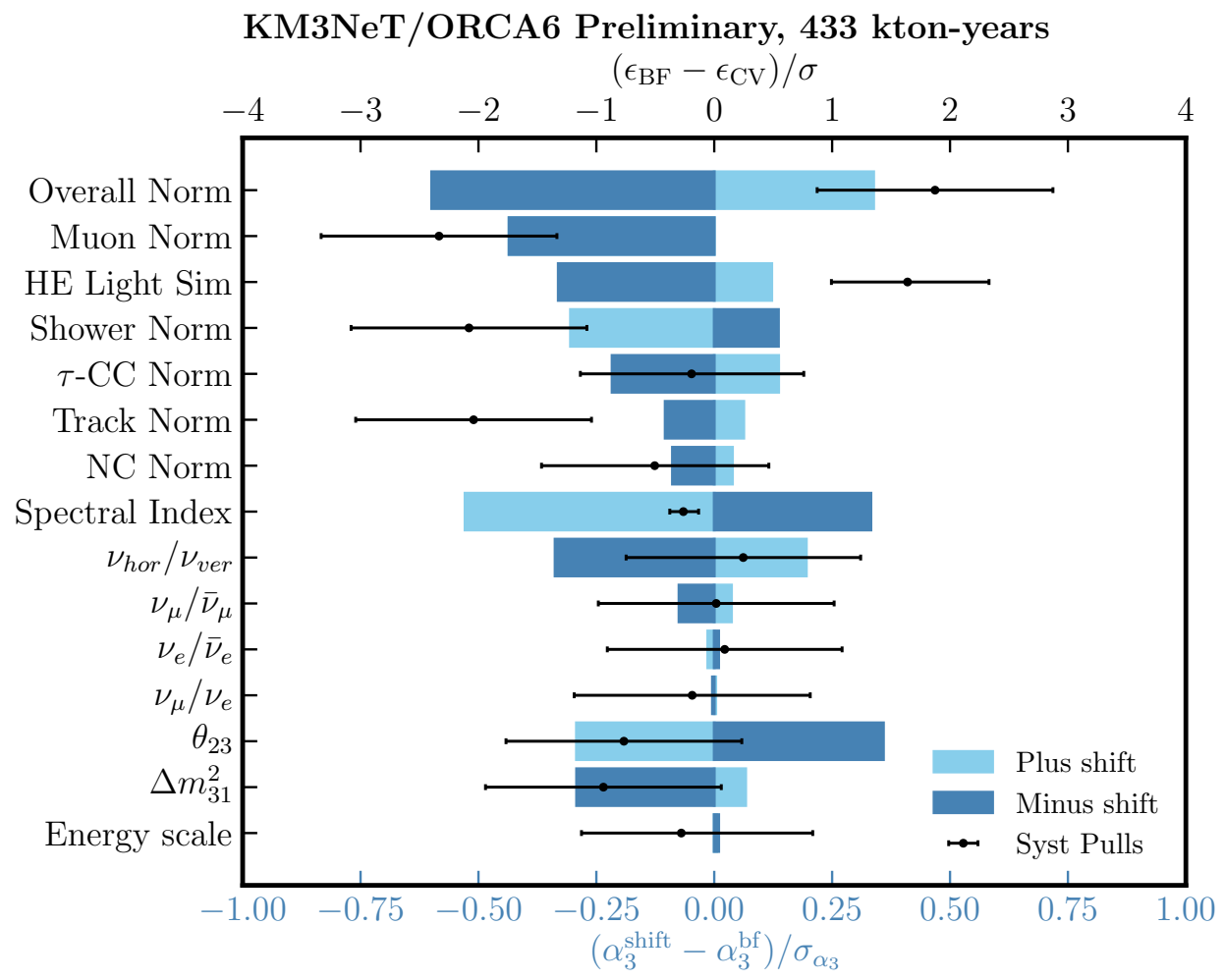

**Figure 5:** Impact of the different systematic parameters computed as a shift in the parameter of interest, $\alpha_3$ when the nuisance parameter is shifted and fixed. Systematics Pulls are reported as black dots.

# Updated results on neutrino Non-Standard Interactions with KM3NeT/ORCA6

**Alfonso Lazo Pedrajas on behalf of the KM3NeT Collaboration**[a,*]

[a] *IFIC - Instituto de Física Corpuscular (CSIC - Universitat de València), c/Catedrático José Beltrán, 2, 46980 Paterna, Valencia, Spain*

*E-mail:* Alfonso.Lazo@ific.uv.es

KM3NeT/ORCA is an underwater neutrino telescope currently under construction in the Mediterranean Sea, with the goal of measuring atmospheric neutrino oscillation parameters and determining the neutrino mass ordering. KM3NeT/ORCA can additionally provide constraints on physics beyond the Standard Model which could appear through strong matter effects, such as the neutrino Non-Standard Interactions (NSIs). This work reports on the results of the NSIs search with the final dataset of ORCA6, the first configuration of ORCA with six detection units, which uses 433 kton-years of exposure and improved calibration, reconstruction and selection methods compared to previous works. The obtained bounds at 90% CL, $|\varepsilon_{\mu\tau}| \leq 5.5 \cdot 10^{-3}$, $|\varepsilon_{e\tau}| \leq 7.8 \cdot 10^{-2}$, $|\varepsilon_{e\mu}| \leq 5.8 \cdot 10^{-2}$ and $-0.015 \leq \varepsilon_{\tau\tau} - \varepsilon_{\mu\mu} \leq 0.016$, are comparable to the current most stringent limits on any NSIs parameter.

*Speaker

# 1. Introduction

## 1.1 Neutrino Non-Standard Interactions

The Neutrino Non-Standard Interactions (NSIs) appear naturally in several extensions of the Standard Model (SM) proposing mechanisms for the origin of neutrino masses. NSIs are incorporated through effective four-fermion interactions which lead to both charged-current (CC) and neutral-current (NC) interactions. NC NSIs affect the coherent forward scattering of neutrinos on fermions in matter, ultimately leading to modifications of neutrino oscillation probabilities in matter potentials. In the ultrarelativistic limit, neutrino propagation in matter is governed by the effective Hamiltonian $\mathcal{H}_{\text{eff}} = \mathcal{H}_{\text{vac}} + \mathcal{H}_{\text{SI}} + \mathcal{H}_{\text{NSIs}}$ given by [2]

$$\mathcal{H}_{\text{eff}} = \frac{1}{2E}\mathcal{U}\begin{bmatrix} 0 & 0 & 0 \\ 0 & \Delta m^2_{21} & 0 \\ 0 & 0 & \Delta m^2_{31} \end{bmatrix}\mathcal{U}^{+} + A(x)\begin{bmatrix} 1+\varepsilon_{ee} & \varepsilon_{e\mu} & \varepsilon_{e\tau} \\ \varepsilon^*_{e\mu} & \varepsilon_{\mu\mu} & \varepsilon_{\mu\tau} \\ \varepsilon^*_{e\tau} & \varepsilon^*_{\mu\tau} & \varepsilon_{\tau\tau} \end{bmatrix}, \qquad (1)$$

where $A(x) = \sqrt{2}G_F n_e(x)$ is the standard matter potential for the electron number density at a given point, and $\mathcal{U}$ is the three-flavour PMNS matrix. The NSIs strength is therefore measured relative to the standard electroweak interaction, and is parameterised by six independent parameters: the complex off-diagonal terms induce flavour-changing (FC) neutral current interactions ($\nu_\alpha + f \to \nu_\beta + f$), whereas the real diagonal parameters cause non-universal couplings of the different neutrino flavours to fermions.

Neutrino propagation in matter is sensitive to the vector part of the NC-NSIs lagrangian, and to the incoherent sum of the scattering amplitudes of neutrinos on the three types of fermions found in matter [2]. For this reason, NSIs are customarily parameterised as

$$\varepsilon_{\alpha\beta} = \varepsilon^{eV}_{\alpha\beta} + \frac{N_u}{N_e}\varepsilon^{uV}_{\alpha\beta} + \frac{N_d}{N_e}\varepsilon^{dV}_{\alpha\beta}, \qquad (2)$$

where the fractions denote the relative abundance of electrons, $u$- and $d$-quarks inside neutral Earth matter. In the following, only the NSIs coupling to the $d$ quark is considered, since other choices can be derived by a simple rescaling given by the relative abundance of the other fermions.

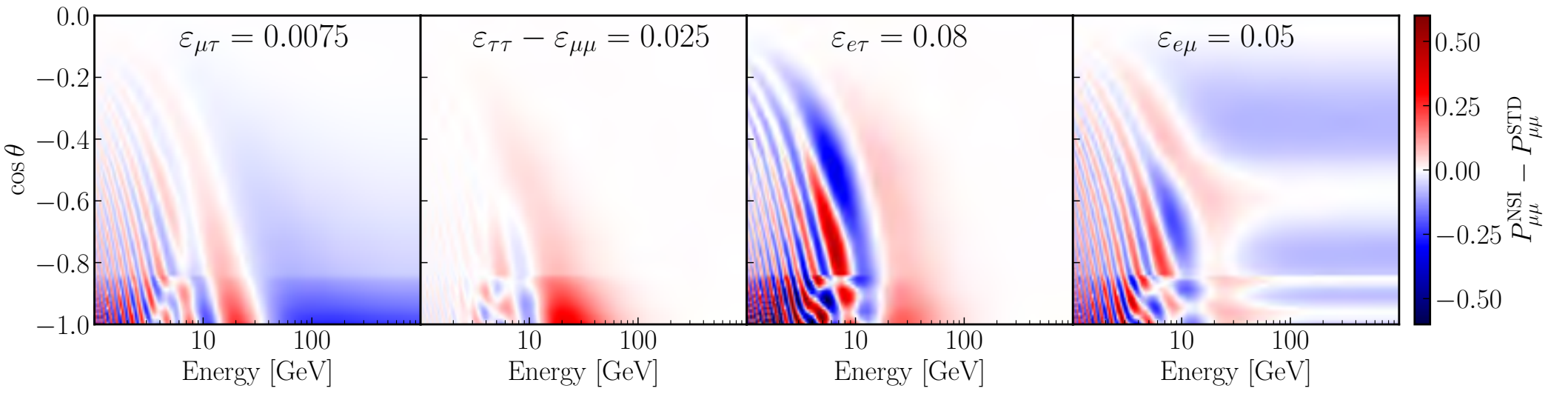

**Figure 1:** Difference in muon neutrino survival probability between the standard interaction case (STD) and four different non-zero NSIs hypothesis. Oscillation probabilities are computed numerically with the OscProb package [3].

Figure 1 presents the change induced by non-zero NSIs parameters (one at a time) to the muon neutrino survival probabilities. These are weighted according to the atmospheric flux composition and cross section ratio of $\nu_\mu/\bar{\nu}_\mu \sim 2$, with oscillation parameters fixed at the NuFit 5.0 values [4]. Given ORCA's current instrumented volume, NSIs effects are mostly accessible through modifications in the position and amplitude of the oscillation valley between 10 and 40 GeV. For $\varepsilon_{\mu\tau}$ and $\varepsilon_{e\mu}$, there are additional signatures at high energies (above 50 GeV) which are also observable with the current detector configuration.

### 1.2 The KM3NeT/ORCA detector

KM3NeT/ORCA is a water Cherenkov detector currently under construction in the Mediterranean Sea, 40 km offshore Toulon at a depth of around 2450 m [1]. Its main goal is the determination of the neutrino mass ordering (NMO) by studying the oscillations of few-GeV atmospheric neutrinos produced by cosmic ray interactions in the atmosphere. Neutrino interactions near or inside the instrumented volume produce secondary charged particles, whose Cherenkov light yield is detected by a three-dimensional grid of Digital Optical Modules (DOMs), housing 31 photomultiplier tubes (PMTs) each, in order to reconstruct the parent neutrino energy and direction based on the deposited light. By inspecting from horizontal to vertical neutrino arrival directions, KM3NeT/ORCA can probe baselines ranging from $\sim 10^2$ km to $O(10^4$ km) traversing different Earth layers, while being sensitive to energies between 1-100 GeV.

The detector is undergoing a modular construction with a final foreseen configuration of 115 vertical Detection Units (DUs), each housing 18 DOMs uniformly spaced every 9 m. DUs are deployed at the seabed following a layout with 20 m of inter-DU spacing and around 200 m of vertical length. Benefiting from this dense instrumentation, ORCA will be able to probe the low-energy range of the atmospheric neutrino spectrum (1-10 GeV) with unprecedented statistics. Since January 2020, the first configuration with six DUs (ORCA6) uninterruptedly took data until November 2021, when it was expanded with more detection units.

## 2. Event sample and selection

The dataset used in this work covers 433 kton-years of exposure of ORCA6, which were selected according to strict quality criteria on the environmental conditions and stability of the data-taking. The event filtering starts with a directional cut to keep up-going reconstructed events, followed by trigger and reconstruction quality requirements, efficiently rejecting most of the pure-noise background. The next level of event filtering uses the output of a Boosted Decision Tree (BDT), which was trained on reconstruction algorithm features to discriminate between atmospheric muons and neutrino-induced signal. The BDT cut allows to keep the atmospheric muon contamination in the sample below 5%, while retaining 60% of the neutrino signal from the previous selection level. Finally, a second BDT output discriminates between track-like topologies ($\nu_\mu$ CC and $\nu_\tau$ with subsequent $\tau$ muonic decays) and shower-like ones ($\nu_e$ CC, $\nu$ NC and remaining $\tau$ decays). The final sample comprises 5828 observed events split into three classes: high purity tracks with negligible atmospheric muon contamination and estimated 95% $\nu_\mu$-CC purity, low purity tracks with 4% muon contamination and 90% $\nu_\mu$-CC purity, and a shower class with 46% of muon neutrinos.

While the three are approximately equally populated, the high purity track class enhances the $\nu_\mu$-disappearance signal and renders the dataset the most sensitive to $\varepsilon_{\mu\tau}$ and $\varepsilon_{\tau\tau} - \varepsilon_{\mu\mu}$, whereas the shower class opens the possibility to constrain the electron NSIs sector ($\varepsilon_{e\tau}$ and $\varepsilon_{e\mu}$).

## 3. Analysis method

The analyses in ORCA proceed by comparing our observation with Monte Carlo (MC) templates weighted according to the hypothesis being tested. We use a Maximum Likelihood Estimator (MLE) of the NSIs and nuisance parameters, which uses the Poisson likelihood ($\mathcal{L}$) of observing the data given our MC expectation, binned in reconstructed zenith angle (baseline) versus reconstructed energy. Confidence intervals of the parameters are constructed by scanning the negative Log-Likelihood Ratio ($-2\log(\mathcal{L}_{NSIs}/\mathcal{L}_{bf}) = -2\Delta\log\mathcal{L}$) of each point in the NSIs space, computed using $\mathcal{L}$ at fixed NSIs over the likelihood at the global best fit. Fifteen nuisance parameters are profiled over in the MLE of the parameters. They model a variety of systematic uncertainties of the atmospheric neutrino flux (composition, energy and directional dependence), detector-related uncertainties (water properties, light propagation), interaction cross-section uncertainties (NC, $\nu_\tau$ and individual class normalisations) and background modelling. Table 1 presents a summary of the nuisance and oscillation parameters with their assumed prior uncertainties, when applicable. The nominal oscillation parameter values are taken from NuFit 5.0 [4].

| | **Nominal value** | **Syst. unc.** |
|---|---|---|
| $\Delta m^2_{31} \cdot 10^{-3}$ [eV$^2$] | 2.517 (NO) /−2.424 (IO) | free |
| $\Delta m^2_{21} \cdot 10^{-5}$ [eV$^2$] | 7.42 | fixed |
| $\theta_{23}$ [°] | 49.2 (NO) / 49.3 (IO) | free |
| $\theta_{21}$ [°] | 33.44 | fixed |
| $\theta_{31}$ [°] | 8.57 (NO) / 8.60 (IO) | fixed |
| High purity Normalisation | 1.0 | free |
| Overall Normalisation | 1.0 | free |
| Shower Normalisation | 1.0 | free |
| Atm. Muon Normalisation | 1.0 | free |
| HE Light Sim | 1.0 | ±50% |
| Energy Scale | 1.0 | ±9% |
| Flux energy slope | 0.0 | ±0.3 |
| Flux zenith slope | 0.0 | ±2% |
| $\nu_\tau$ Norm | 1.0 | ±20% |
| $\nu$ NC normalisation | 1.0 | ±20% |
| $\nu_\mu/\bar{\nu}_\mu$ | 0.0 | ±5% |
| $\nu_e/\bar{\nu}_e$ | 0.0 | ±7% |
| $\nu_\mu/\nu_e$ | 0.0 | ±2% |

**Table 1:** Summary of the systematic uncertainties considered in this study.

## 4. Results

The one-by-one best fit NSIs parameters and 90% CL allowed regions, extracted from the $-2\Delta\log\mathcal{L}$ scans to the data assuming Wilks' theorem [5], are presented in table 2. No significant deviation from the standard interaction hypothesis was found for any NSIs parameter, with p-values ranging from 0.23 to 0.90.

| **Hypothesis** | **Best fit** $\lvert\varepsilon_{ij}\rvert$, $\delta_{ij}$ | **p-value** | **Real-valued 90% CL limit** | **Complex 90% CL limit** |
|---|---|---|---|---|
| $\varepsilon_{\mu\tau}$ | $0.001^{+0.003}_{-0.001}$, $0^{+360°}_{-0}$ | 0.66 | $[-0.0047, 0.0052]$ | $\leq 0.0055$, $\delta_{\mu\tau} \in [0°, 360°]$ |
| $\varepsilon_{\tau\tau} - \varepsilon_{\mu\mu}$ | $0.00 \pm 0.01$ | 0.90 | $[-0.015, 0.016]$ | — |
| $\varepsilon_{e\tau}$ | $0.04 \pm 0.03$, $190 \pm 70°$ | 0.23 | $[-0.077, 0.028]$ | $\leq 0.078$, $\delta_{e\tau} \in [0°, 360°]$ |
| $\varepsilon_{e\mu}$ | $0.03 \pm 0.02$, $140 \pm 70°$ | 0.25 | $[-0.056, 0.043]$ | $\leq 0.058$, $\delta_{e\mu} \in [0°, 360°]$ |

**Table 2:** Best fit results and 90% CL limits for the NSIs parameters from the fits to the data.

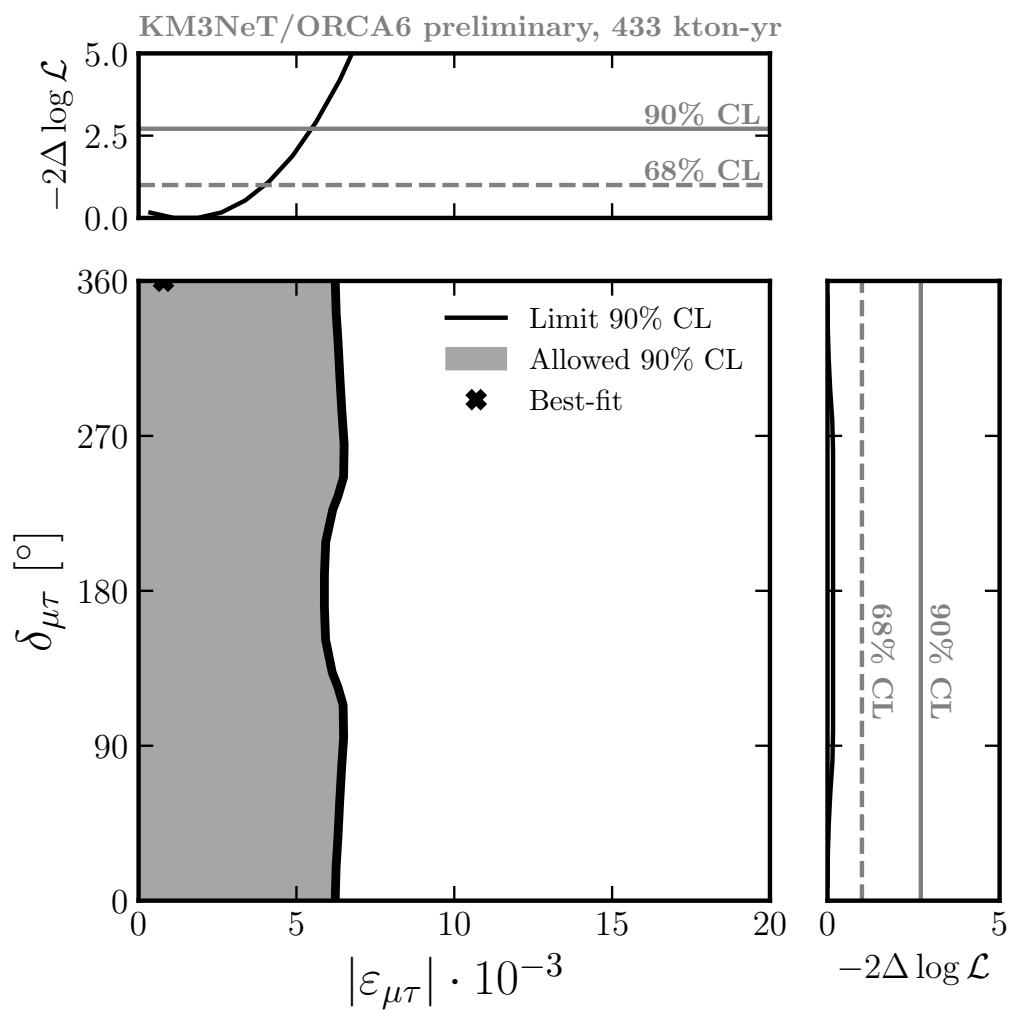

**(a)** $-2\Delta \log \mathcal{L}$ contour of $|\varepsilon_{\mu\tau}|$ vs $\delta_{\mu\tau}$ .

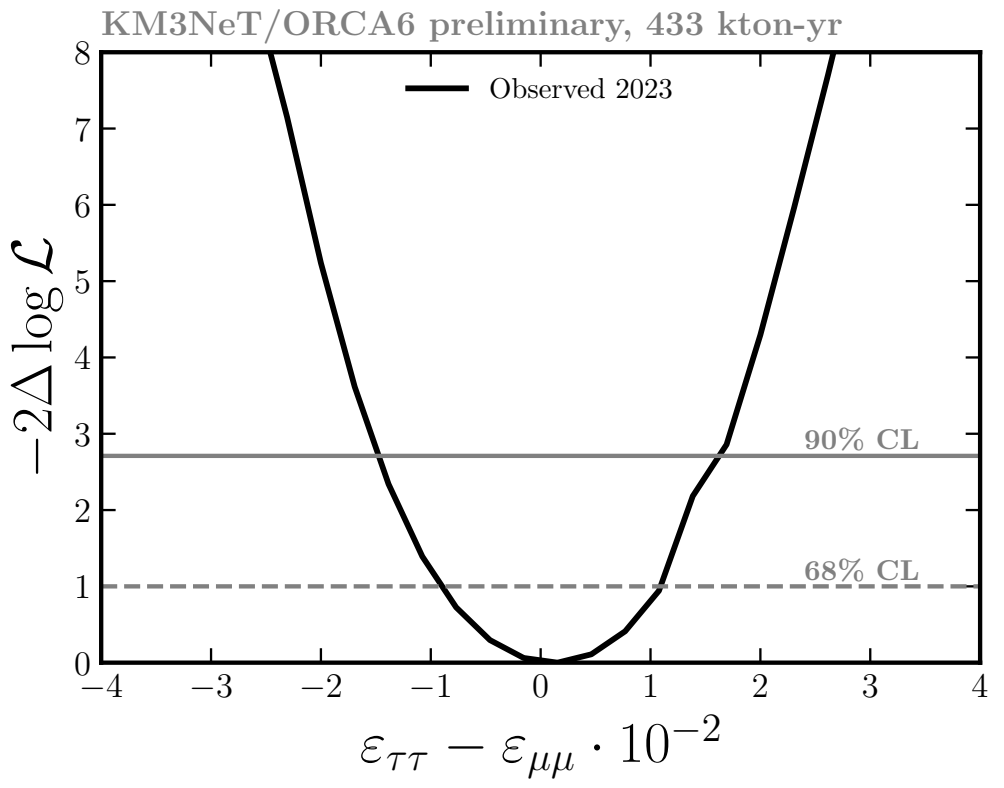

**(b)** $-2\Delta \log \mathcal{L}$ scan of $\varepsilon_{\tau\tau} - \varepsilon_{\mu\mu}$.

**Figure 2:** Likelihood ratio contour at 90% CL of the complex parameter $\varepsilon_{\mu\tau}$ (2a) and profile of the real parameter $\varepsilon_{\tau\tau} - \varepsilon_{\mu\mu}$. In 2a, the top and side plots show the projections of $-2\Delta \log \mathcal{L}$ when the other variable is profiled over.

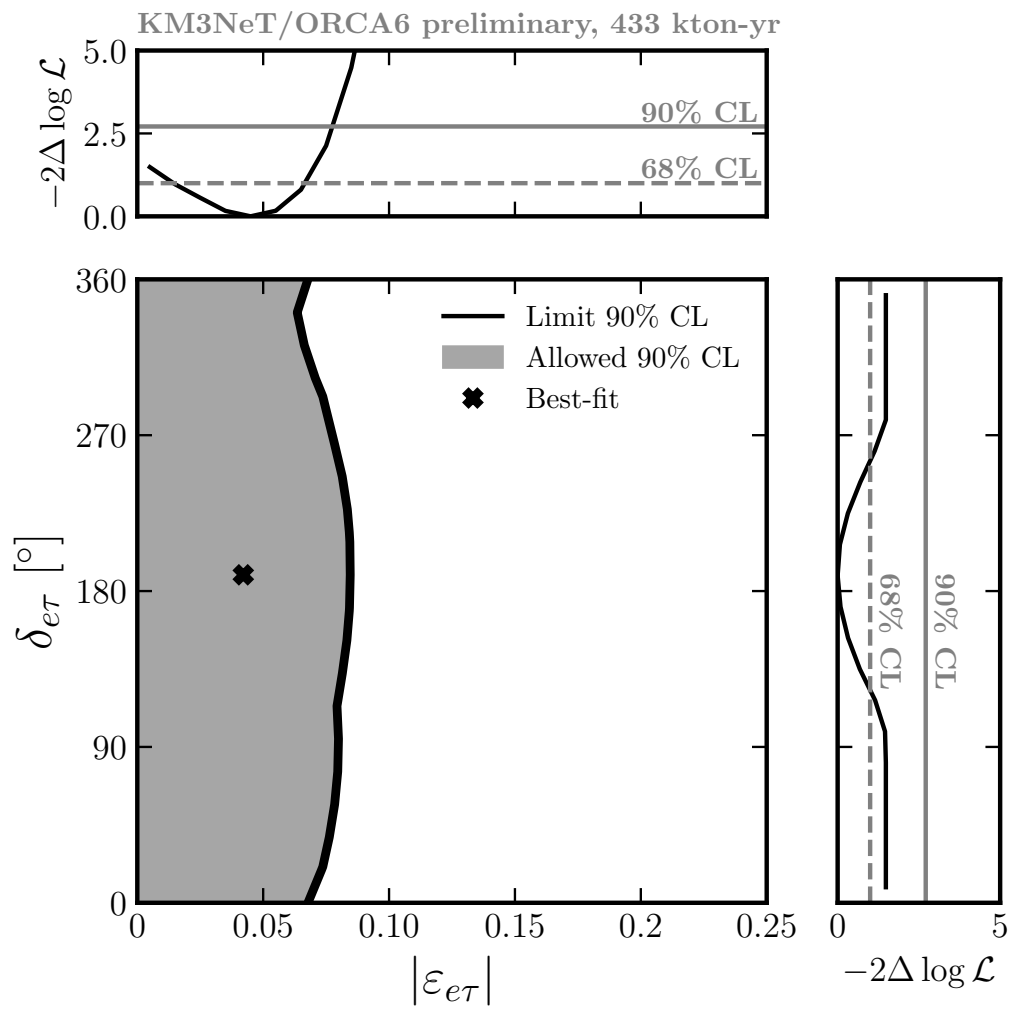

**(a)** $-2\Delta \log \mathcal{L}$ contour of $|\varepsilon_{e\tau}|$ vs $\delta_{e\tau}$ at 90% CL.

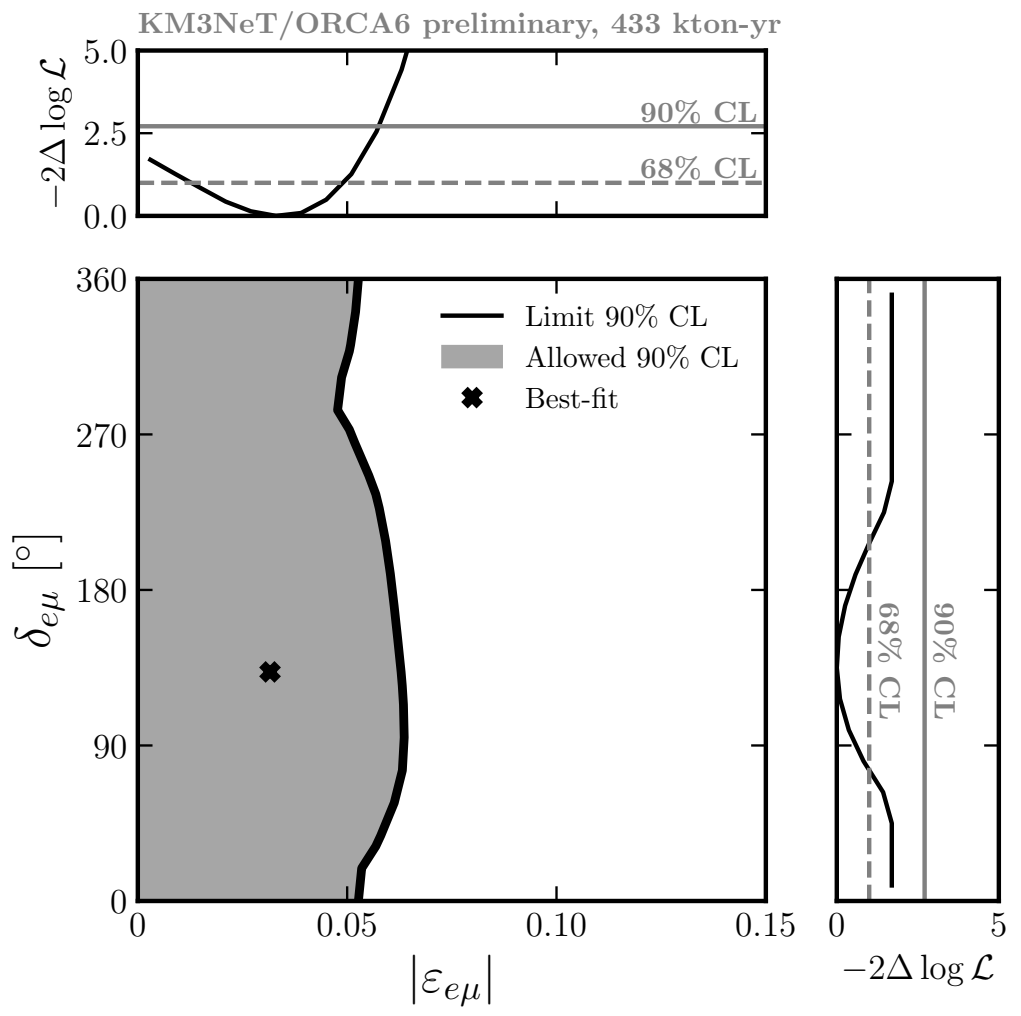

**(b)** $-2\Delta \log \mathcal{L}$ contour of $|\varepsilon_{e\mu}|$ vs $\delta_{e\mu}$ at 90% CL.

**Figure 3:** Likelihood ratio contour at 90% CL of the complex parameters $\varepsilon_{e\tau}$ (3a) and $\varepsilon_{e\mu}$ (3b). The top and side plots show the projections of $-2\Delta \log \mathcal{L}$ when the other variable is profiled over.

Figures 2a, 2b, 3a and 3b show the allowed regions obtained from this study. No sensitivity at 90% CL was observed for the complex phases of the flavour-violating NSIs coupling strengths, although somewhat stronger bounds are placed on $|\varepsilon_{e\tau}|$ and $|\varepsilon_{e\mu}|$ when their corresponding complex phases are restricted outside the range $[90°, 270°]$ and $[45°, 240°]$, respectively. Given the reduced sensitivity of ORCA6 to the NMO and $\theta_{23}$-octant, $-2\Delta \log \mathcal{L}$ is profiled over the two assumed neutrino mass orderings and the two octants as starting points for the fits.

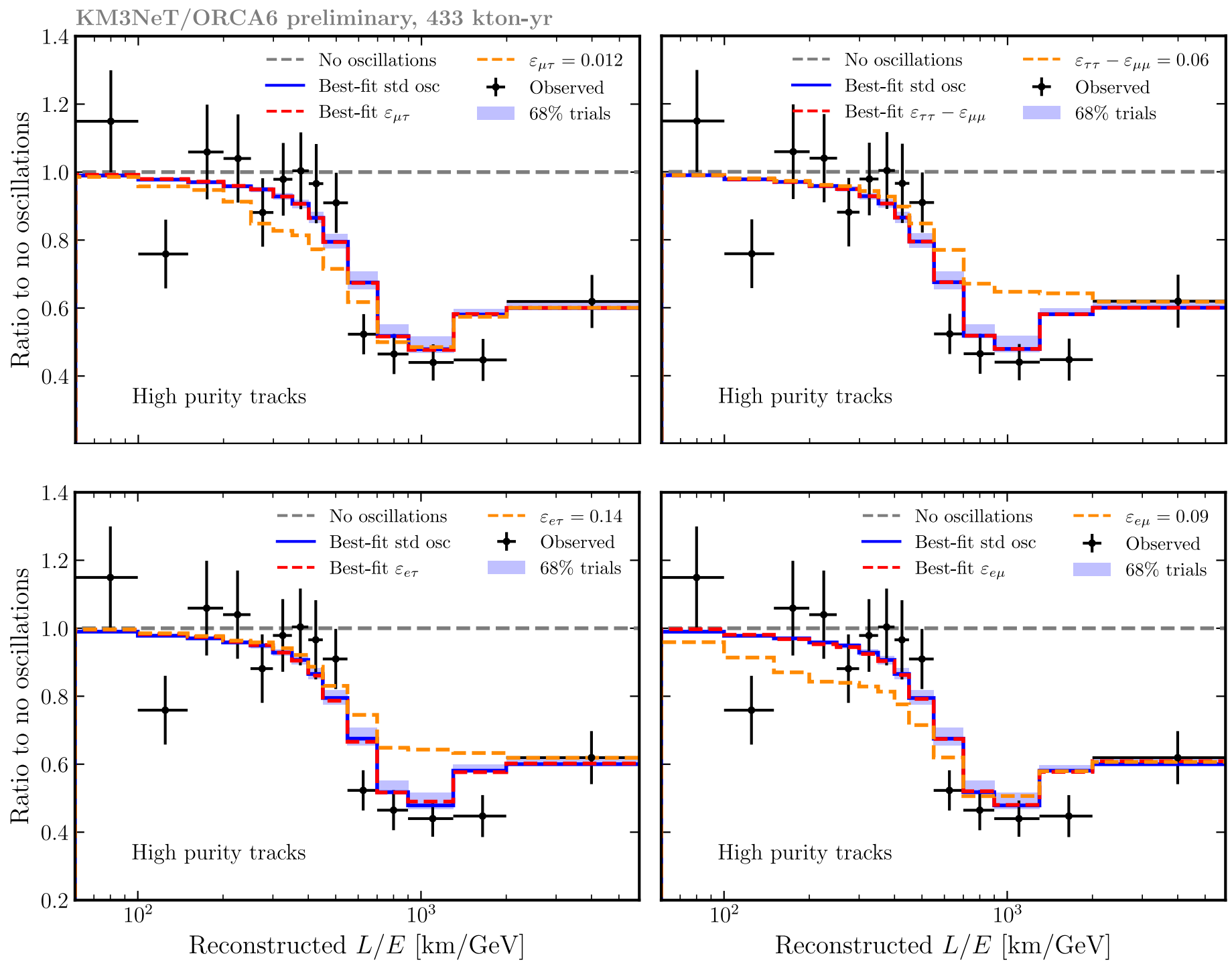

**Figure 4:** Ratio of events respect to the no oscillation hypothesis as a function of the baseline over reconstructed energy bin ($L/E$). The observed ratios are shown together with the prediction from the fits to five different hypothesis: the standard oscillation best fit (all), $\varepsilon_{\mu\tau}$ best fit, $\varepsilon_{\tau\tau} - \varepsilon_{\mu\mu}$ best fit, $\varepsilon_{e\tau}$ best fit and $\varepsilon_{e\mu}$ best fit. The 68 % trials band is drawn by fitting the standard oscillation hypothesis to 1000 pseudo-experiments, generated from Poisson fluctuations of the data best-fit MC template under standard oscillations. Additionally, the yellow lines overlaid on every plot show the expectation from shifting each NSIs parameter to its $5\sigma$ limit, while keeping all systematic values at their best fit.

Figure 4 presents the number of observed and expected events as a function of L/E (baseline over reconstructed energy) for different oscillation best-fit scenarios, normalised to the expectation for the no-oscillation case. All distributions are shown only for the high purity track class, which is the most sensitive set to neutrino oscillations out of the three classes fit in the analysis. The oscillation valley between 300 and 2000 km/GeV is clearly visible in data and all expectations. For the four NSIs best fits, the resulting distributions follow closely the standard oscillation case, since no significant pull was observed on any coupling strength. For comparison, yellow dashed curves show the expectation from shifting each NSIs parameter alone at its $5\sigma$ threshold, while keeping all systematic values at their corresponding best fit. The effects observed in these shifts are fully compatible with the oscillation probabilities computed numerically in figure 1.

In the following, the impact of each systematic on the individual NSIs coupling strengths is studied, in correlation to the rest of the systematics set. The approach followed consists in shifting the value of one systematic at a time $\pm 1\sigma$ away from its best fit value for the given NSIs parameter, fix it and fit the remaining systematics together with the parameter of interest (PoI, the NSIs parameters in our case). The impact of the systematic shift is reflected as a deviation in the PoI from its best fit to the data, divided by the $1\sigma$ uncertainty obtained in the real-valued profile of the PoI. Only deviations in the absolute value of the PoI are considered, since the NMO profiling during the fits can lead to sign flips in the PoI. This procedure yields the bar plots shown in figures 5a, 5b, 5c and 5d. On top of the bars, the black dots show the pulls exhibited by the systematics in the NSIs fits, where all nuisance parameters were free to vary. For this, the central values of the oscillation parameters were those of table 1 for NO , since $\Delta m^2_{31} > 0$ was preferred for the four NSIs fits. As can be seen, all constrained nuisance parameters were found well within their expectation.

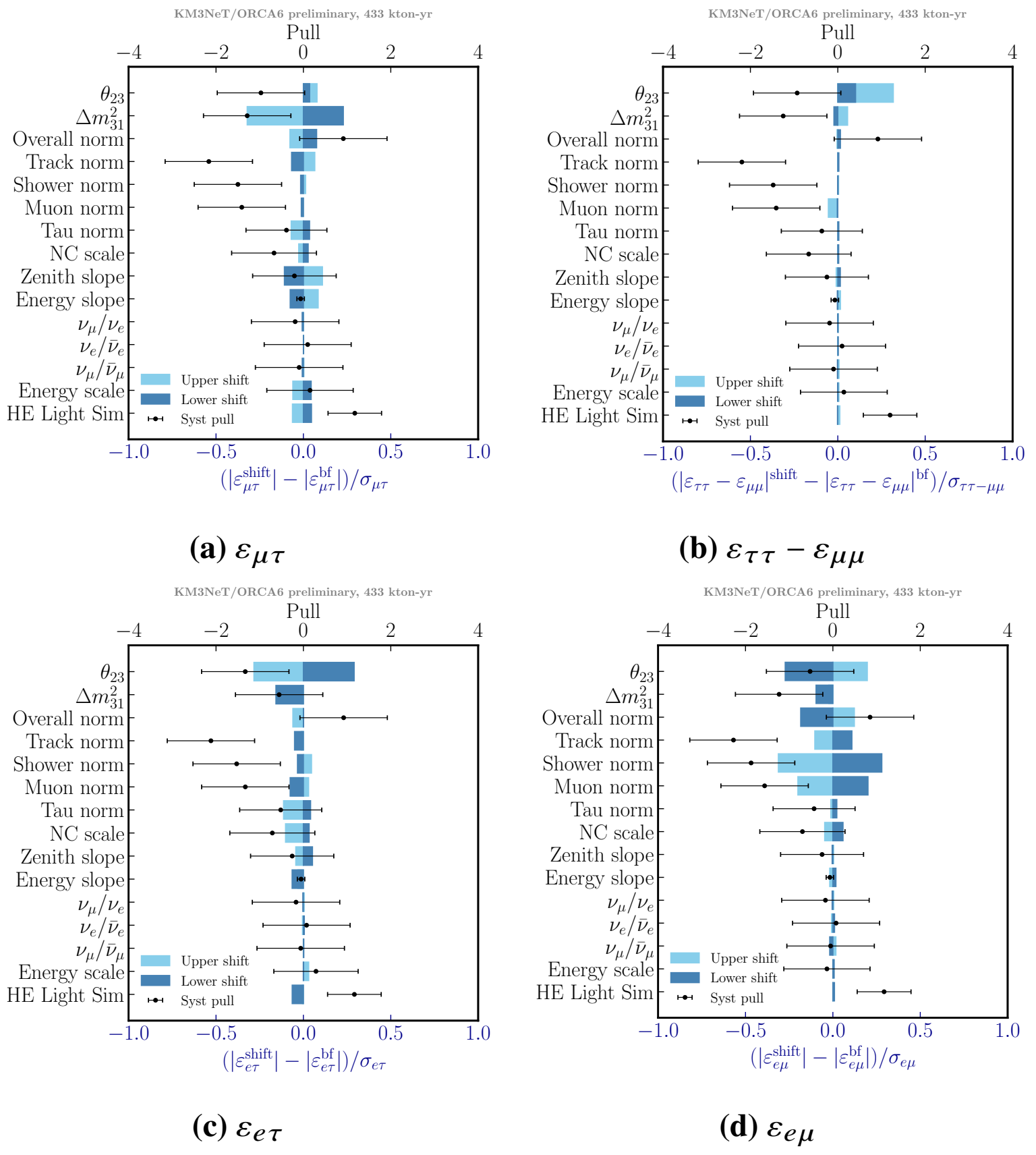

**(a)** $\varepsilon_{\mu\tau}$ **(b)** $\varepsilon_{\tau\tau} - \varepsilon_{\mu\mu}$

**(c)** $\varepsilon_{e\tau}$ **(d)** $\varepsilon_{e\mu}$

**Figure 5:** Systematics shift plot for the one-by-one NSIs coupling strengths. The light and dark blue bars reflect the shift exerted on the PoI by the upper and lower shift of the systematic, respectively. The bars are to be read with the lower blue axis. The overlaid black dots are the pulls experienced by the systematics in the NSIs fit, $(\text{syst}^{\rm bf} - \text{syst}^{\rm nom})/\sigma^{\rm syst}$, being $\text{syst}^{\rm nom}$ the nominal values from table 1, and $\sigma^{\rm syst}$ is the prior width for constrained systematics, or the post fit $1\sigma$ uncertainty for unconstrained ones. The pulls are to be read with the upper black axis. The horizontal error bars on the pulls are the ratio of the post fit uncertainty to the prior width. For unconstrained parameters which do not have priors, the error bars are set to one unit.

Figure 5 then shows that the oscillation parameters $\Delta m^2_{31}$ and $\theta_{23}$ have the biggest impact among all nuisance parameters, followed by the class normalisations, based on their shifts induced on the PoI. In particular, $\Delta m^2_{31}$ appears as the most important systematic for $\varepsilon_{\mu\tau}$, due to both parameters being able to shift the position of the oscillation valley, whereas $\theta_{23}$ is the one for the remaining NSIs parameters, driven by the similar effects produced on the oscillation valley amplitude by varying the value of $\theta_{23}$ and $\varepsilon_{\tau\tau} - \varepsilon_{\mu\mu}$ or $\varepsilon_{e\tau}$ . Finally, $\varepsilon_{e\mu}$ is most impacted by the shower class normalisation.

## 5. Summary and conclusions

This work reports on the results from the NSIs search with 433 kton-year of ORCA6. No significant deviation from standard interactions was found by measuring atmospheric neutrino oscillations with a sample of 5828 events, split into high purity tracks, low purity tracks and showers. The analysis has constrained NSIs coupling strengths at the 90% CL, assuming couplings to down quarks, to be $|\varepsilon_{\mu\tau}| \leq 5.5 \cdot 10^{-3}$, $|\varepsilon_{e\tau}| \leq 7.8 \cdot 10^{-2}$, $|\varepsilon_{e\mu}| \leq 5.8 \cdot 10^{-2}$ and $-0.015 \leq \varepsilon_{\tau\tau} - \varepsilon_{\mu\mu} \leq 0.016$. The resulting bounds are similar to the most stringent ones reported up to date, which are shown for real-valued NSIs parameters in figure 6, and still offer room for improvement in the near future, coming mainly from the extensions of ORCA's instrumented volume with more detection units.

Finally, the authors want to acknowledge the financial support from Generalitat Valenciana and the European Social Fund through the ACIF/2021/233 grant, and the support from Ministerio de Ciencia e Innovación (MCINN) through the programme PID2021-124591NB-C41.

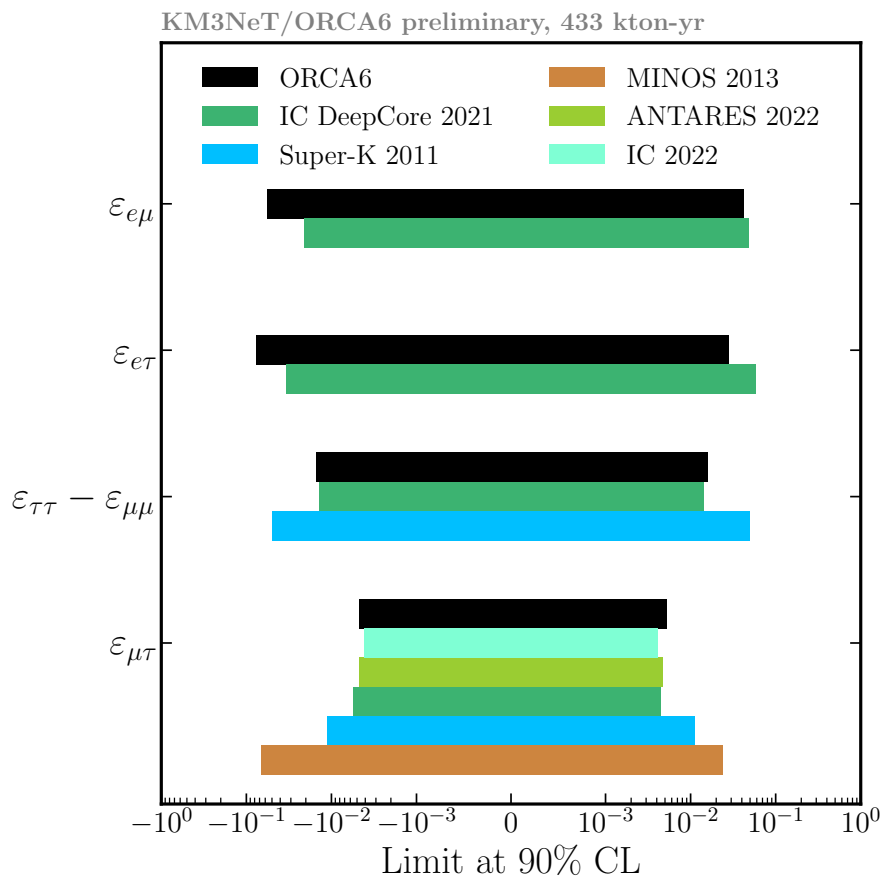

**Figure 6:** Comparison of the 90% CL limits reported in this work and DeepCore [6], IceCube [7], Super-K [8], ANTARES [9] and MINOS [10]. DeepCore and MINOS results were rescaled to match the down-quark coupling used by the other experiments, using the factor deduced from [6].

# Search for Quantum Decoherence in Neutrino Oscillations with KM3NeT ORCA6

**Nadja Lessing**[a,*] **on behalf of the KM3NeT Collaboration**

[a] *Instituto de Física Corpuscular IFIC (CSIC - Universidad de Valencia), Parque Científico Catedrático José Beltrán 2, Paterna, Spain*

*E-mail:* Nadja.Lessing@ific.uv.es

Quantum decoherence of neutrino states is an effect that is proposed in various theories of quantum gravity. It is envisaged to emerge from interactions of the neutrino as a quantum system with the environment and may destroy the superposition of neutrino mass states, which leads to a modification of neutrino oscillation probabilities.

KM3NeT/ORCA, a neutrino telescope designed to detect various neutrino flavours across a broad spectrum of energies, can probe this phenomenon. ORCA is a water Cherenkov detector that is currently under construction in the Mediterranean Sea. It was designed to detect atmospheric neutrinos in the GeV energy range. The main objective of ORCA is the precise measurement of neutrino oscillations. Therefore, the detector provides the possibility to investigate various beyond Standard Model scenarios that may alter the oscillation pattern.

This contribution reports on first constraints on the strength of decoherence effects with KM3NeT. We provide upper limits on the decoherence parameters $\gamma_{21}$ and $\gamma_{31}$. The high-purity neutrino sample used for this analysis was collected with a six Detection Units configuration of ORCA with an exposure of 433 kton-years.

*Speaker

## 1. The KM3NeT detectors

The KM3NeT research infrastructure comprises two water Cherenkov detectors which are currently under construction in the Mediterranean sea. These detectors, known as KM3NeT/ARCA and KM3NeT/ORCA are composed of 3D arrays of Digital Optical Modules arranged in vertical Detection Units. Each of the modules hosts 31 Photomultiplier tubes which collect Cherenkov light emitted by the charged particles produced in neutrino interactions. ORCA is optimized to detect atmospheric neutrinos in the GeV energy range and to precisely measure the neutrino oscillation parameters. In the final configuration the ORCA detector will encompass 115 Detection Units yielding a total instrumented volume of 7 Mton of sea water. The data analyzed for this proceeding was recorded between February 2020 and November 2021 with a configuration of 6 Detection Units, which we refer to as ORCA6 in the following. The total lifetime is 510 days which results in an exposure of 433 kton-years.

## 2. Theory of decoherence in neutrino oscillations

Neutrino oscillations are considered coherent because the different mass eigenstates maintain their relative phase relationships as they propagate. However, neutrino eigenstates may loose their quantum superposition in a phenomenon known as neutrino decoherence. In this analysis, we consider *non-standard* decoherence, induced by the coupling of the neutrino as a quantum system with the environment. The time evolution of the neutrino system including decoherence effects is described by adding a non-unitary term to the Liouville–von Neumann equation [1]

$$\frac{d}{dt}\rho = -i[H,\rho] - \mathcal{D}[\rho]\,, \tag{1}$$

where $\rho$ is the density matrix of the neutrino and $H$ is the full Hamiltonian in matter. With the condition of complete positivity of the time evolution and trace conservation, the decoherence term can be written in the so-called *Lindblad* form [2]

$$\mathcal{D}[\rho] = \sum_m (\{\rho, D_m D_m^\dagger\} - 2D_m \rho D_m^\dagger)\,, \tag{2}$$

where the $D_m$ are complex matrices, with the index $m \in [1,\, N^2 - 1]$ where $N$ is the dimension of the $SU(N)$ Hilbert space of the neutrino system. The $D_m$ are required to be Hermitian which implies increasing entropy. The condition of complete positivity ensures that the eigenvalues of $\rho$ remain positive which is crucial for their interpretation as probabilities. Assuming energy conservation in the effective mass basis, the $D_m$ and $H$ can be diagonalized simultaneously with $\tilde{H} = \mathrm{diag}\,(\tilde{E}_1, \tilde{E}_2, \tilde{E}_3)$ describing the energy in the effective mass basis. The solution of Equation 1 for three neutrino families in uniform matter is given by

$$\tilde{\rho}_{ij} = \tilde{\rho}_{ij}(0)\, e^{-i\Delta\tilde{E}_{ij}t - \gamma_{ij}t}\,,\ \ i,j = 1,2,3 \tag{3}$$

where $\tilde{\rho}$ is the density matrix in the effective mass basis, $\tilde{\rho}_{ij}(0)$ is determined by the initial conditions and $\Delta\tilde{E}_{ij} = \tilde{E}_i - \tilde{E}_j$ describes the difference in the total neutrino energies. Additionally, we introduced the decoherence parameter defined by

$$\gamma_{ij} = \sum_m (d_m^i - d_m^j)^2\,, \tag{4}$$

where the $d_m^i$ are the entries of the diagonalized matrices $D_m$. We assume that the $d_m^i$ are constant and independent of matter effects, following the approach in [2]. It is important to note that this does not imply that decoherence effects themselves are independent of matter effects. For this analysis we use atmospheric neutrinos that traverse the Earth before being detected. In order to apply Equation 3 the Earth is modeled by 15 layers of constant density. In presence of decoherence effects the oscillation probabilities are calculated via

$$P(\nu_\alpha \to \nu_\beta) = \sum_{i,j} \tilde{U}_{\alpha i} \tilde{U}^*_{\beta i} \tilde{U}^*_{\alpha j} \tilde{U}_{\beta j} e^{-i\Delta \tilde{E}_{ij} t - \gamma_{ij} t} , \tag{5}$$

where $\tilde{U}$ is the diagonalized Pontecorvo-Maki-Nakagawa-Sakata matrix. The only difference to standard neutrino oscillations is the presence of a damping term $e^{-\gamma_{ij} t}$. The common approach in decoherence studies is to assume that the decoherence parameter may depend on the neutrino energy as

$$\gamma_{ij} = \gamma_{ij}^0 \left( \frac{E}{\text{GeV}} \right)^n . \tag{6}$$

In this work we consider $n = -2, -1$ since these cases affect lower energies that ORCA is most sensitive to. As can be seen from Equation 4 the decoherence parameters are not independent of each other. Consequently, we provide upper limits on $\gamma_{21}$ and $\gamma_{31}$ as, under the most conservative assumptions for our experiment, the third parameters is fixed by [3]

$$\gamma_{32} = \gamma_{31} + \gamma_{21} - 2\sqrt{\gamma_{21}\gamma_{31}} . \tag{7}$$

## 3. Analysis methods

This analysis employs a binned log-likelihood minimization technique to compare the data to the Monte Carlo expectation. The nominal values of the oscillation parameters as given in Table 1 are provided by Nu-Fit 5.0 [5]. The data is reconstructed in 2D event histograms of the energy and zenith angle with a binning scheme that ensures an expectation of at least two events per bin. The negative likelihood function is minimized for $\theta_{23}$, $\Delta m^2_{31}$, and a set of nuisance parameters described in detail in [6]. The nuisance parameters along with their corresponding prior uncertainties are summarized in Table 2. The prior uncertainties are assumed to be Gaussian distributed.

| **Parameter** | **Nominal value NO** | **Nominal value IO** | **Treatment** |
|---|---|---|---|
| $\Delta m^2_{31}$ [eV$^2$] | $2.517 \cdot 10^{-3}$ | $-2.424 \cdot 10^{-3}$ | free |
| $\Delta m^2_{21}$ [eV$^2$] | $7.42 \cdot 10^{-5}$ | $7.42 \cdot 10^{-5}$ | fixed |
| $\theta_{12}$ [°] | 33.44 | 33.45 | fixed |
| $\theta_{13}$ [°] | 8.57 | 8.60 | fixed |
| $\theta_{23}$ [°] | 49.2 | 49.3 | free |
| $\delta_{CP}$ [°] | 197 | 282 | fixed |

**Table 1:** Oscillation parameters for normal ordering (NO) and inverted ordering (IO) with their treatment in the minimization.

Only data from periods characterized by high stability in environmental conditions as well as data acquisition were used in this analysis, which results in an exposure of 433 kton-years.

Cuts based on the trigger rate and the reconstruction quality efficiently reject background events from pure noise. The analysis is restricted to up-going events which significantly reduces the atmospheric muon background. Two Boosted Decision Trees (BDT) are employed to effectively discriminate against atmospheric muons and to differentiate between track-like and shower-like events. Additional cuts on the atmospheric muon BDT score further reduce the atmospheric muon contamination to below 5 %. This result in a total of 5828 observed events. These events are divided into three classes in order to optimize the sensitivity: a high purity tracks class with negligible muon contamination, a low purity tracks class and a showers class. The tracks classes contain events with reconstructed energies between 2 GeV and 100 GeV, whereas the showers class ranges from 2 GeV to 1 TeV.

More information on the selected data set can be found in the proceedings for measuring atmospheric neutrino oscillation with KM3NeT/ORCA6 [7].

| Parameter | Prior |
|---|---|
| Spectral index | ±0.3 |
| Energy scale | ±9 % |
| $\nu_{\text{hor}}/\nu_{\text{ver}}$ ratio | ±2 % |
| $\nu_e/\bar{\nu}_e$ ratio | ±7 % |
| $\nu_\mu/\bar{\nu}_\mu$ ratio | ±5 % |
| $(\nu_\mu+\bar{\nu}_\mu)/(\nu_e+\bar{\nu}_e)$ ratio | ±2 % |
| High-energy light simulation | ±50 % |
| NC normalization | ±20 % |
| $\tau$-CC normalization | ±20 % |
| Muon normalization | free |
| Track normalization | free |
| Shower normalization | free |
| Overall normalization | free |

**Table 2:** Nuisance parameters along with their prior uncertainties.

## 4. Results

Fitting the data, no significant deviation was found with respect to the standard oscillation analysis. The fitted values of $\gamma_{21}$ and $\gamma_{31}$ are consistent with zero. Figure 1 shows the difference in the log-likelihood ratios of decoherence and standard oscillations in dependence of the reconstructed energy and zenith angle at the best fit values $\gamma_{ij,\text{BF}}$ for the high purity tracks class. For $\gamma \propto E^{-2}$ (left) the red bins give a better fit for decoherence while the blue bins are in better agreement with standard oscillations. For $\gamma \propto E^{-1}$ (right) there is no difference with respect to the standard oscillations best fit.

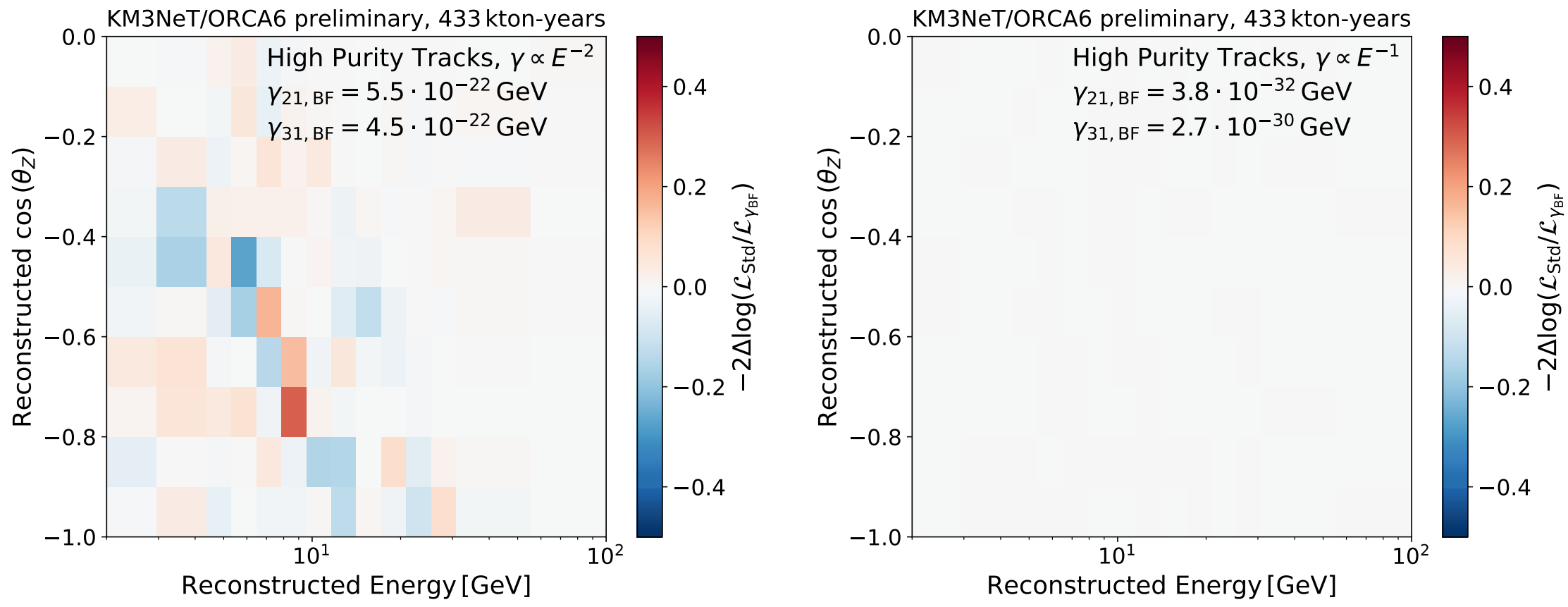

**Figure 1:** Difference in the log-likelihood ratios of decoherence and standard oscillations at the best fit for the high purity tracks class for $\gamma \propto E^{-2}$ (left) and $\gamma \propto E^{-1}$ (right).

The upper limits on $\gamma$ are determined by computing the log-likelihood ratio with respect to the global best fit as a function of one decoherence parameter while leaving the other parameter free in the minimization. By exploring the full parameter space and allowing for non-zero values of all three decoherence parameters, the most conservative limits are obtained. Furthermore, the minimization is performed for both octants of $\theta_{23}$ and for both neutrino mass orderings.

The solid curves in Figure 2 show the log-likelihood ratio as a function of $\gamma_{21}$ and $\gamma_{31}$ for each of the energy dependencies $\gamma \propto E^{-2}$ (left) and $\gamma \propto E^{-1}$ (right). Additionally, the blue dashed line shows the log-likelihood ratio for $\gamma_{21}$ assuming normal ordering (NO). For small values of $\gamma_{21}$ normal ordering is preferred whereas for large values of $\gamma_{21}$ inverted ordering (IO) gives the minimal log-likelihood ratio. The upper limits at the 95% CL are presented in Figure 3 for both, NO and IO. This figure illustrates how the limits strongly depend on the considered mass ordering which emphasizes the importance of performing the minimization for both cases.

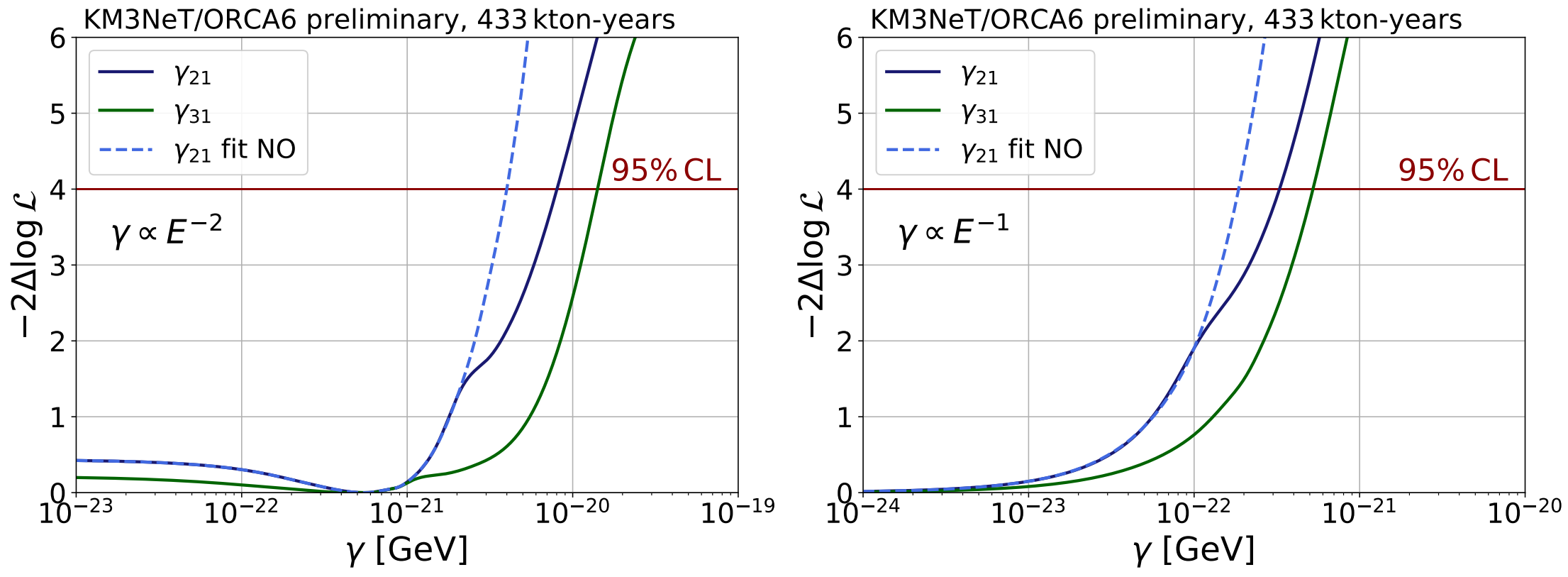

**Figure 2:** Log-likelihood ratio with respect to the global best fit for $\gamma \propto E^{-2}$ (left) and $\gamma \propto E^{-1}$ (right) as a function of the decoherence parameters $\gamma_{21}$ and $\gamma_{31}$. The solid lines were obtained fitting both, normal and inverted ordering, as well as both octants of $\theta_{23}$ and keeping the overall best fit. The blue dashed line was obtained assuming normal ordering.

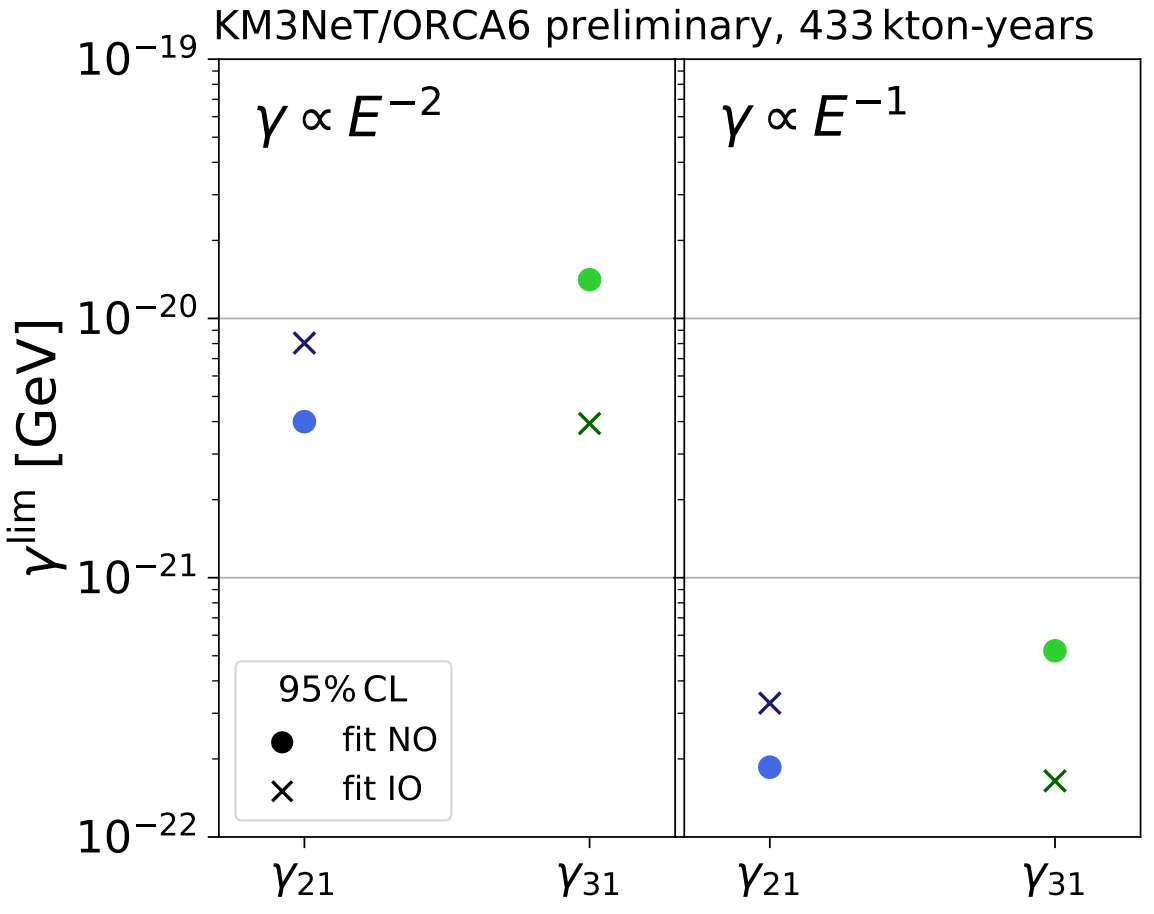

**Figure 3:** Upper limits on $\gamma_{21}$ and $\gamma_{31}$ at the 95% CL assuming NO (dots) and IO (crosses).

Figure 4 shows CL contours which serve to constrain both decoherence parameters $\gamma_{21}$ and $\gamma_{31}$ at the same time. The log-likelihood ratio is small along the diagonal since for $\gamma_{21} = \gamma_{31} \Rightarrow \gamma_{32} = 0$. The irregular shape of the curves is caused by a flip in the best fit mass ordering. This is evident from Figure 5 which depicts the 95 % confidence level contour assuming either NO or IO as well as considering both mass orderings. As already seen in the one-dimensional profiles, the upper limits differ greatly depending on the mass ordering.

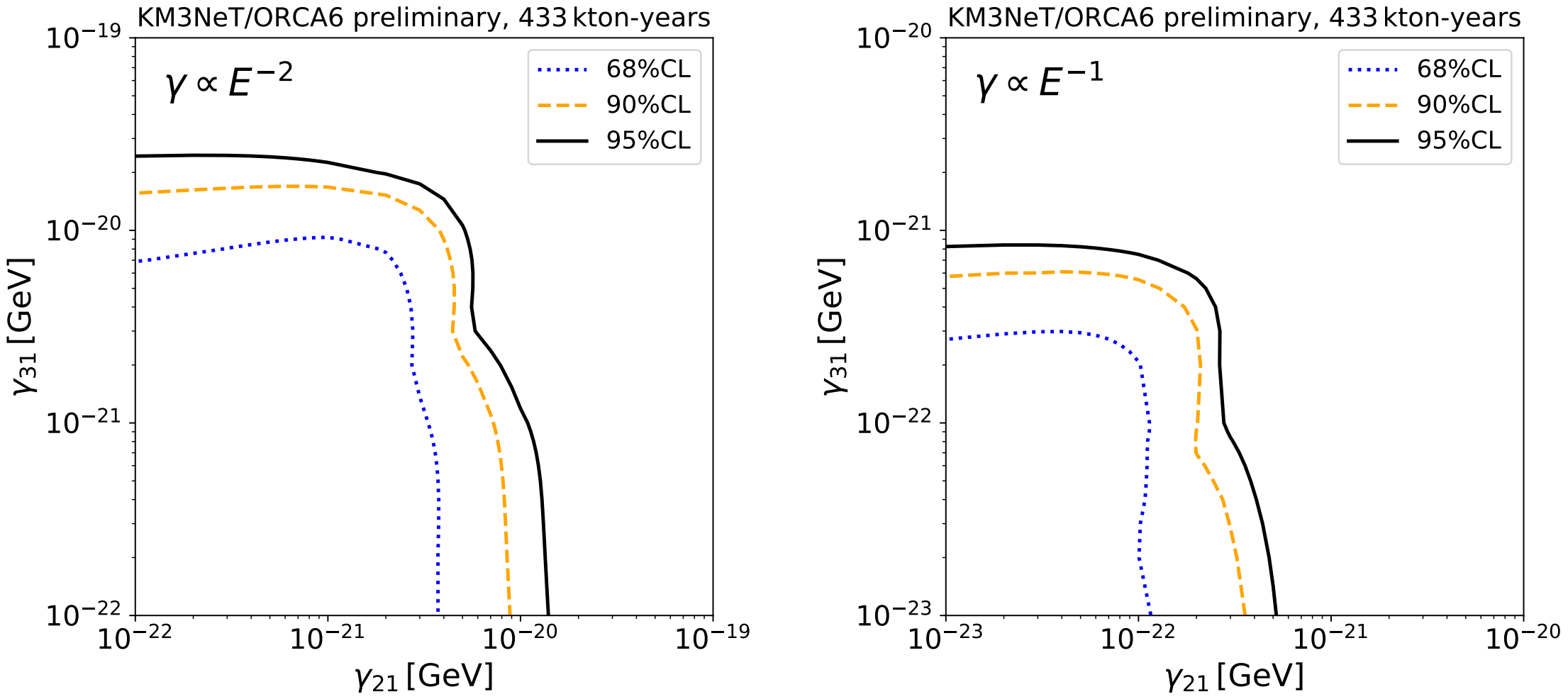

**Figure 4:** Confidence level contours of $\gamma_{31}$ and $\gamma_{21}$ for $\gamma \propto E^{-2}$ (left) and $\gamma \propto E^{-1}$ (right). The upper right part of the plane is excluded for each model at the corresponding CL.

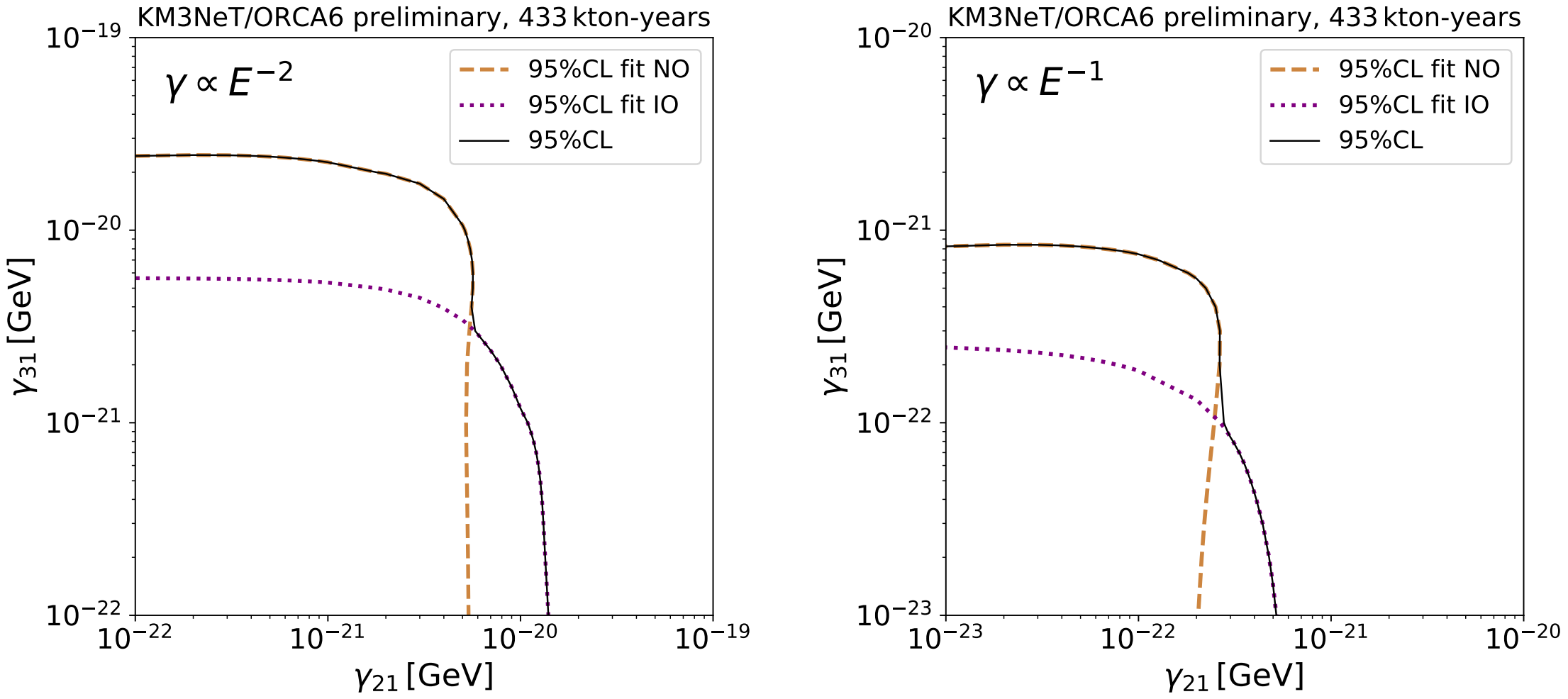

**Figure 5:** 95% CL contours of $\gamma_{31}$ and $\gamma_{21}$ assuming NO (brown dashed), assuming IO (purple dotted) and fitting both mass orderings (black solid).

Figure 6 shows the difference in the log-likelihood ratio of decoherence and standard oscillations for $\gamma \propto E^{-2}$ (left) and $\gamma \propto E^{-1}$ (right) with the decoherence parameters at the respective 95% CL upper limit for each model, and with $\gamma_{21} = \gamma_{31}$. The blue bins correspond to a better fit of standard oscillations than decoherence at the given values of $\gamma$ and therefore serve to exclude decoherence effects. It can be seen that the most relevant bins are centered around 10 GeV for both models but extend towards slightly higher energies for the $\gamma \propto E^{-1}$ model.

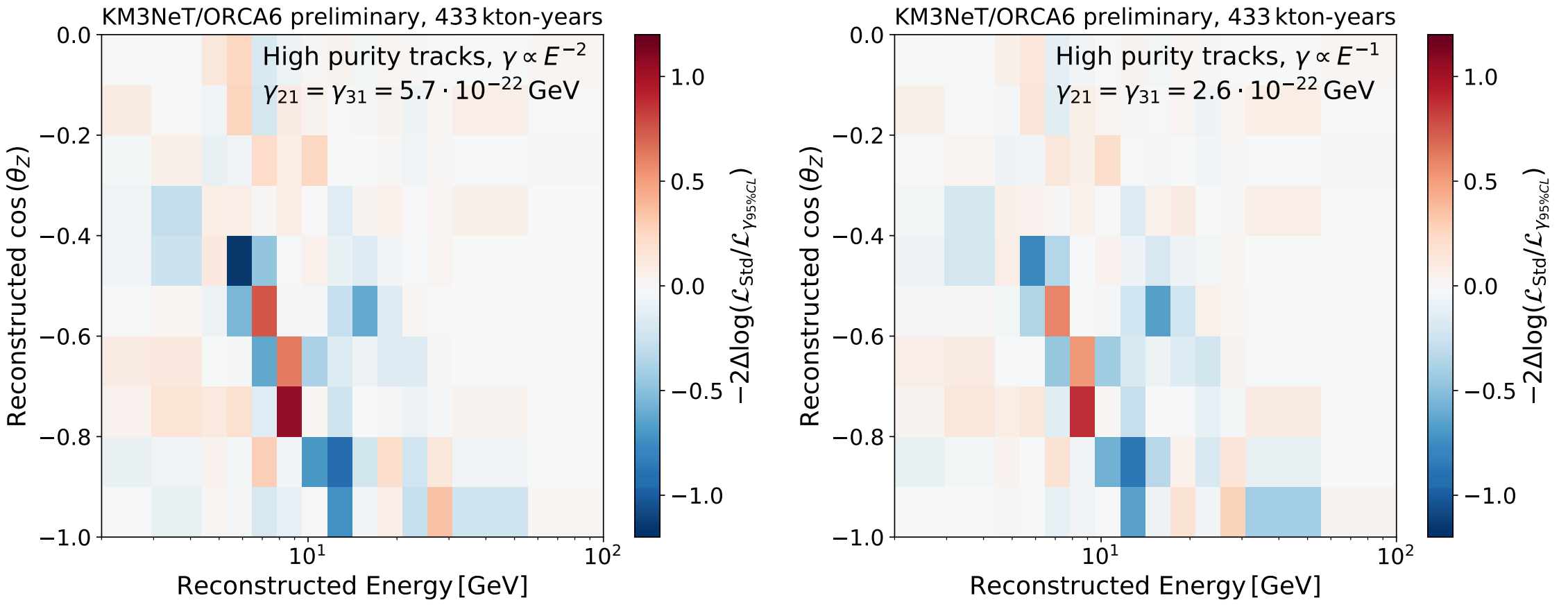

**Figure 6:** Difference in the log-likelihood ratios of decoherence and standard oscillations for $\gamma \propto E^{-2}$ (left) and $\gamma \propto E^{-1}$ (right) with the decoherence parameters fixed at the respective 95% CL upper limit with $\gamma_{21} = \gamma_{31}$. The blue bins serve to exclude decoherence since they correspond to a better fit of standard oscillations than decoherence.

| Upper limits in GeV | $\gamma \propto E^{-2}$ | | $\gamma \propto E^{-1}$ | |
|---|---|---|---|---|
| ORCA6 (this work) | NO | IO | NO | IO |
| $\gamma_{21}$ | $3.0 \cdot 10^{-21}$ | $5.2 \cdot 10^{-21}$ | $1.3 \cdot 10^{-22}$ | $1.8 \cdot 10^{-22}$ |
| $\gamma_{31}$ | $10.3 \cdot 10^{-21}$ | $2.9 \cdot 10^{-21}$ | $3.5 \cdot 10^{-22}$ | $1.0 \cdot 10^{-22}$ |
| $\gamma_{21} = \gamma_{31}$ | $4.5 \cdot 10^{-21}$ | $3.2 \cdot 10^{-21}$ | $2.1 \cdot 10^{-22}$ | $1.2 \cdot 10^{-22}$ |
| Reported in [4] | NO | IO | NO | IO |
| $\gamma_{21} = \gamma_{32}$ | $7.9 \cdot 10^{-27}$ (KL) | - | $1.8 \cdot 10^{-24}$ (KL) | - |
| $\gamma_{31} = \gamma_{32}$ | $6.9 \cdot 10^{-25}$ (R) | - | $2.1 \cdot 10^{-23}$ (T2K) | - |
| $\gamma_{21} = \gamma_{31}$ | $7.9 \cdot 10^{-27}$ (KL) | - | $1.8 \cdot 10^{-24}$ (KL) | - |

**Table 3:** Upper limits at the 90 % CL for ORCA6 in comparison with the most constraining upper limits reported in [4] using data from KamLAND (KL), RENO (R) and T2K. Note that in [4] one of the decoherence parameters is set to zero whereas in this work for the limit on $\gamma_{21}$ ($\gamma_{31}$) the other parameter $\gamma_{31}$ ($\gamma_{21}$) is left free in the fit in order to obtain the most conservative limits.

In general, comparing to results from other analyses is not straightforward since the assumptions on the model and the procedure used to obtain upper limits differ. Nevertheless, Table 3 summarizes recent upper limits reported in [4] using data from KamLAND, RENO and T2K along with the values obtained in this analysis for ORCA6. It is important to note that in our analysis, when setting a limit on $\gamma_{21}$ ($\gamma_{31}$) the other parameter $\gamma_{31}$ ($\gamma_{21}$) is left free in the fit whereas [4] considers limiting cases of decoherence where one of the parameters is set to zero, so the other two are equal. The same applies for the upper limits reported in [2] using three years of DeepCore data. Still, our limit on $\gamma_{21}$ ($\gamma_{31}$) is approximately comparable to the case $\gamma_{21} = \gamma_{32}$ ($\gamma_{31} = \gamma_{32}$) since the decoherence parameters that is left free in the fit tends to be at least one order of magnitude smaller.

| Upper limits in GeV | $\gamma \propto E^{-2}$ | | $\gamma \propto E^{-1}$ | |
|---|---|---|---|---|
| ORCA6 (this work) | NO | IO | NO | IO |
| $\gamma_{21}$ | $4.0 \cdot 10^{-21}$ | $8.0 \cdot 10^{-21}$ | $1.9 \cdot 10^{-22}$ | $3.3 \cdot 10^{-22}$ |
| $\gamma_{31}$ | $14.6 \cdot 10^{-21}$ | $3.9 \cdot 10^{-21}$ | $5.2 \cdot 10^{-22}$ | $1.6 \cdot 10^{-22}$ |
| $\gamma_{21} = \gamma_{31}$ | $5.7 \cdot 10^{-21}$ | $3.9 \cdot 10^{-21}$ | $2.6 \cdot 10^{-22}$ | $1.6 \cdot 10^{-22}$ |
| DeepCore | NO | IO | NO | IO |
| $\gamma_{21} = \gamma_{32}$ | $7.5 \cdot 10^{-21}$ | $5.0 \cdot 10^{-20}$ | $3.5 \cdot 10^{-22}$ | $2.3 \cdot 10^{-21}$ |
| $\gamma_{31} = \gamma_{32}$ | $4.3 \cdot 10^{-20}$ | $1.4 \cdot 10^{-20}$ | $2.0 \cdot 10^{-21}$ | $5.8 \cdot 10^{-22}$ |
| $\gamma_{21} = \gamma_{31}$ | $1.2 \cdot 10^{-20}$ | $8.3 \cdot 10^{-21}$ | $5.4 \cdot 10^{-22}$ | $3.6 \cdot 10^{-22}$ |

**Table 4:** Upper limits on the decoherence parameters at the 95 % CL for ORCA6 in comparison with results reported in [2] using DeepCore data.

## 5. Conclusions

We provided first constraints on neutrino decoherence effects using data from the KM3NeT/ORCA detector. Both octants of $\theta_{23}$ and both neutrino mass orderings were taken into account to derive the most conservative upper limits on the decoherence parameters. Also, we allowed all three decoherence parameters to take non-zero values and explored the full parameter space by computing confidence level contours of $\gamma_{21}$ and $\gamma_{31}$. Our results have demonstrated that even with only six Detection Units, the ORCA detector is capable of providing limits on decoherence effects that are comparable to those of other analyses. This allows to anticipate the potential of the full ORCA detector which will consist of 115 Detection Units and will be able to give more stringent limits.

# Deep Neural Networks for combined neutrino energy estimate with KM3NeT/ORCA6

**Santiago Peña Martínez**[a,*] **on behalf of the KM3NeT Collaboration**

[a] *Université Paris Cité, CNRS, Astroparticule et Cosmologie, F-75013 Paris, France*

*E-mail:* spenam@apc.in2p3.fr

KM3NeT/ORCA is a large-volume water-Cherenkov neutrino detector, currently under construction at the bottom of the Mediterranean Sea at a depth of 2450 meters. The main research goal of ORCA is the measurement of the neutrino mass ordering and the atmospheric neutrino oscillation parameters. Additionally, the detector is also sensitive to a wide variety of phenomena including non-standard neutrino interactions, sterile neutrinos, and neutrino decay.

This contribution describes the use of a machine learning framework for building Deep Neural Networks (DNN) which combine multiple energy estimates to generate a more precise reconstructed neutrino energy. The model is optimized to improve the oscillation analysis based on a data sample of 433 kton-years of KM3NeT/ORCA with 6 detection units. The performance of the model is evaluated by determining the sensitivity to oscillation parameters in comparison with the standard energy reconstruction method of maximizing a likelihood function. The results show that the DNN is able to provide a better energy estimate with lower bias in the context of oscillation analyses.

*Speaker

## 1. Introduction

KM3NeT is a neutrino telescope under construction in the Mediterranean Sea [1]. It consists of two water-Cherenkov detectors: ARCA and ORCA. The principle of detection uses the Cherenkov light induced by charged particles produced in neutrino interactions in the water. The emitted light is collected by an array of photomultiplier tubes (PMTs) housed in glass spheres to form digital optical modules (DOMs). Each DOM holds 31 PMTs arranged to provide directional information of the incoming photons. The ORCA detector will consist of 115 detection units (DUs) instrumented with 2070 DOMs, with a total volume of about 7 Mton. The main goal of ORCA is to determine the neutrino mass ordering and the atmospheric neutrino oscillation parameters. For this analysis, Monte Carlo (MC) simulations from an early detector configuration with 6 DUs are used.

The reconstruction of the neutrino energy is a key aspect for the oscillation analysis of KM3NeT/ORCA. Current methods for the energy reconstruction are based on the maximization of a likelihood function, which relies on a hypothesis for the distribution of light from a neutrino interaction. This distribution is either track-like when it is induced by muons or shower-like when it is generated by an electromagnetic or hadronic cascade of charged secondary particles. However, this approach has some limitations. For the case of the track-like topology, it is assumed for the reconstruction that the muon propagates in a straight line without scattering. For the shower-like topology, the produced lepton is e.g. an electron which will induce an electromagnetic cascade that is assumed to induce light emission in a sperically symmetric way. Both of the hypotheses do not cover the full picture of the interaction. When a neutrino interacts with a nucleus it produces not only an outgoing lepton but also a shower of hadrons that will induce light.

The present study aims to make use of the current energy estimates which rely on the track and shower hypotheses, and combines them with extra auxiliary information, such as the number of triggered hits, to consider the full information of the interaction topology. By using this ancillary information and the energy estimates from each hypothesis (track-like, shower-like, and the track length as an energy proxy) as an input, a Deep Neural Network (DNN), which outputs an improved energy estimate, is built. The training of the DNN is done using MC simulations of atmospheric neutrinos. In order to create the network, a set of hyperparameters has to be selected. The hyperparameters consist of configuration settings that the network cannot learn by itself, such as the number of layers, the number of neurons per layer, the activation function, and the loss function. The values of all of these hyperparameters have to be optimized in order to obtain the best possible network for the given problem.

The best set of hyperparameters will be determined by the performance of the network on validation data with respect to a chosen metric. For standard regression problems in machine learning, the metric used is the mean squared error (MSE). In general, the network should be trained for the lowest possible MSE. However, this might not be the best approach for studying neutrino oscillations. Ideally, the evaluation metric and the network loss function should be the sensitivity of the analysis to the oscillation parameters. However, it is not a simple problem to define a loss function that incorporates this into an event-by-event basis. Therefore, the sensitivity given by the $\Delta\chi^2$ between two oscillation hypotheses has to be computed when a training is done. The result is a $\Delta\chi^2$ value for each set of hyperparameters, and the set of hyperparameters that gives the lowest $\Delta\chi^2$ is the one that is selected.

Once the selection of the network hyperparameters is done, a full fit of the systematics for the oscillation analysis has to be done. This will be the final step to validate the energy estimate given by the network. This is done by applying the training model on the MC event sample containing neutrinos and atmospheric muons.

## 2. Methodology

### 2.1 Building the network

To perform the regression problem for energy estimation, a neural network using Keras with a TensorFlow backend is built [2]. The network consists of an input layer, followed by several dense hidden layers with an activation function, and a single output layer.

| **Track-like** | **Shower-like** | **Trigger information** | **PID features** |
|---|---|---|---|
| Energy | Energy | Triggered hits | Track score |
| Fit likelihood | Fit likelihood | Triggered DOMs | Background score |
| 3d event direction | 3d event direction | Triggered DUs | |
| 3d event position | 3d event position | | |
| Track length | | | |

**Table 1:** Training features from the topology reconstructions, information on number of triggered hits and PID classification scores.

The inputs of the network are stated in Table 1, these include the standard energy estimates for the track hypothesis, the shower hypothesis, and the track length as energy proxy, information about the number of triggered hits, DOMs, and DUs, and scores from the particle identification (PID) tasks. The values of the energy estimates have a range from 1 GeV to 1 TeV, which is a very wide range of values for the network to learn from. This may cause difficulties to the network when reconstructing true energies around the energy range of interest for oscillations (5 - 40 GeV). The reason for this is that the network will focus more on the higher energies, as the loss function will be dominated by the higher energies. To avoid this, the energy estimates were input in logarithmic scale and the maximum energy for the training sample is set to be 100 GeV, this allows the network to learn from a wider range of values and to have a better performance for the energies in the region of interest. We simulate Charged Current (CC) and Neutral Current (NC) interacting events. The event types used for training are $\nu_\mu$ CC, $\bar{\nu}_\mu$ CC, $\nu_e$ CC, $\bar{\nu}_e$ CC, $\nu_\tau$ CC, $\bar{\nu}_\tau$ CC, $\nu$ NC, and $\bar{\nu}$ NC events. Additionally, at each layer, a batch normalization layer is applied to the input data. This is to normalize the input data to have a mean of zero and a standard deviation of one.

### 2.2 Network architecture

As a starting point, the number of nodes per layer was set to 32 and the number of hidden layers to 12. Without much tuning, the network was able to learn the energy distribution of the events. However, the network did not show any improvement over the standard methods when computing the sensitivity to the oscillation parameters. This was expected since the network was not trained to learn the oscillation effects, but rather to learn the energy distribution of the events. Therefore, the events with the highest sensitivity to oscillations were not given more importance than the others and could eventually be regressed poorly by the network.

A skip connection architecture was implemented. This refers to an architecture breaking the sequential models by allowing layer outputs to skip connections and serve as input at a different layer on a later stage. The motivation behind this is to allow the network to learn the residual between the input and the output of the network, instead of learning the whole function. In our case, the input and output of the network are both energy estimates, and the residual between them is the energy resolution. Thus, we want the network to learn the energy resolution instead of the energy itself. This is done by concatenating the input energy estimates to the output of the network, and then train the network to learn the residual between them. With this method, the information about the energy estimates will not be washed out by mixing it with the other information in the network, and the network will be able to learn the energy resolution. As the energy estimate tends to perform differently depending on the type of event evaluated, it is a natural choice to include in the residuals the features related to the class selection given by the PID.

### 2.3 Weighting the events

In order to introduce some oscillation awareness during training, a novel method to set up the weights for training events was implemented. The method consists in giving more importance to events sensitive to oscillation effects. The usual weight of an event is given by $w = w(\Delta m^2_{31}, \theta_{23})$, where $\Delta m^2_{31}$ and $\theta_{23}$ are the oscillation parameters we want to be sensitive to. One could compute the weights for events having a different set of oscillation parameters $w(\Delta m^2_{31}{}', \theta'_{23})$. The difference $\Delta w = |w(\Delta m^2_{31}, \theta_{23}) - w(\Delta m^2_{31}{}', \theta'_{23})|$ corresponds to the difference in oscillation effects for each event. For neutral current events the value of $\Delta w$ will be zero, since they are unaffected by oscillations. For charged current events, $\Delta w$ will be non-zero and will have a strong dependence on the direction and energy of the event.

During training, each event is given the following weight:

$$w_{osc} = K\Delta w + w. \tag{1}$$

Where $K$ is a hyperparameter which sets the importance to the oscillation weights during the training. If $K$ is large, the network will give more importance to events with oscillation sensitivity. If $K$ is small, the most common events will be prioritized independent of their sensitivity oscillations.

### 2.4 Hyperparameter optimization

To further improve the performance of our neural network, the procedure of hyperparameter optimization to search for the optimal values of the hyperparameters was implemented. The choice for this purpose was the Optuna package in Python [3], which implements a Bayesian optimization algorithm (TPE) to efficiently search the hyperparameter space.

The selection of the hyperparameters to optimize are the following:

- Number of hidden layers: In the range of 8 to 64 with steps of 4.
- Number of neurons per hidden layer: In the range of 16 to 128, with steps of 16.
- Batch size: In the range of 32 to 254 with steps of 32.

- Learning rate: In the range of $10^{-7}$ to $10^{-5}$, using a log-uniform distribution.
- Activation function for each hidden layer: The options to select the optimal activation function were PReLU, ReLU, LeakyReLU and ELU.
- The options for selecting the optimal loss function were mean squared error, mean absolute error and log cosh.
- Training features were divided into groups and the one performing the better was selected, groups of features are shown in Table 2.

| | **G1** | **G2** | **G3** | **G4** | **G5** | **G6** | **G7** |
|---|---|---|---|---|---|---|---|
| **Track-like energy** | ✓ | ✓ | ✓ | ✓ | ✓ | ✓ | ✓ |
| **Track-like likelihood** | ✓ | ✓ | ✓ | ✓ | ✓ | ✓ | ✓ |
| **Track-like track length** | ✓ | ✓ | ✓ | ✓ | ✓ | ✓ | ✓ |
| **Track-like x,y,z positions** | | ✓ | | ✓ | | | ✓ |
| **Track-like x,y,z directions** | | | ✓ | | ✓ | ✓ | ✓ |
| **Shower-like energy** | ✓ | ✓ | ✓ | ✓ | ✓ | ✓ | ✓ |
| **Shower-like likelihood** | ✓ | ✓ | ✓ | ✓ | ✓ | ✓ | ✓ |
| **Shower-like x,y,z positions** | | ✓ | | ✓ | | | ✓ |
| **Shower-like x,y,z directions** | | | ✓ | | ✓ | ✓ | ✓ |
| **Triggered hits** | ✓ | ✓ | ✓ | ✓ | ✓ | ✓ | ✓ |
| **Triggered DOMs** | | | | ✓ | ✓ | ✓ | ✓ |
| **Triggered DUs** | | | | ✓ | ✓ | ✓ | ✓ |
| **Track score** | | | | | | ✓ | ✓ |
| **Background score** | | | | | | ✓ | ✓ |

**Table 2:** Group of features tested for the hyperparameter optimization procedure.

The hyperparameters were optimized using the MSE as a metric. The optimization was done using the training sample containing 1.48M neutrino CC and NC events, and the performance of the network was evaluated using the validation sample of 396k neutrino CC and NC events.

| **Hyperparameter** | **Value** |
|---|---|
| **Activation function** | ELU |
| **Osci. *w* normalization K** | 103188.39 |
| **Learning rate** | 8.009e-06 |
| **Batch size** | 64 |
| **Number of nodes per layer** | 128 |
| **Number of hidden layers** | 20 |
| **Loss function** | Mean Squared Error |
| **Feature group** | G6 |

**Table 3:** Table of Hyperparameters of the models trained and selected with the highest $\Delta\chi^2$ value.

It is difficult to assess if a reconstruction is good for computing sensitivities to oscillation parameters. The standard way to compare energy reconstructions is to look at the distribution of reconstructed energy versus true energy i.e. the energy response function. However, this is not enough to tell whether the reconstruction is good or not, since the distribution could be centered around the diagonal but with a large spread. As the response function is not a single number, it cannot be used to define a metric for our purposes. In particular, we care about having a correct energy reconstruction for oscillating events, since they are the ones used to compute the sensitivity to oscillation parameters. The best energy estimate may depend on the specific task one might want to accomplish. Therefore, we choose the sensitivity to oscillation as our ultimate metric of performance for this work. This is done computing the sensitivity to oscillation parameters using the energy estimate given by the DNN, and comparing it with the sensitivity using the standard energy estimates.

The standard way to do this is by computing the $\Delta\chi^2$ between two sets of oscillation parameters using our internal oscillation analysis framework [4]. The direct use of this framework during the training as part of a loss function is not possible. Also loading the data into the framework after each training epoch is not feasible, since it would be too slow. For this reason, we use a custom approximate implementation of the sensitivity calculation which can compute $\Delta\chi^2$ values directly from the obtained reconstructed energies at the end of each training. This allows to compare different models, and select the one with the best performance. The $\Delta\chi^2$ sensitivity is computed for different topology classes of events (High Purity Tracks, Low Purity Tracks and Showers) [5]. The model selected is the one with the best performance for the sum of the $\Delta\chi^2$ of the three classes.

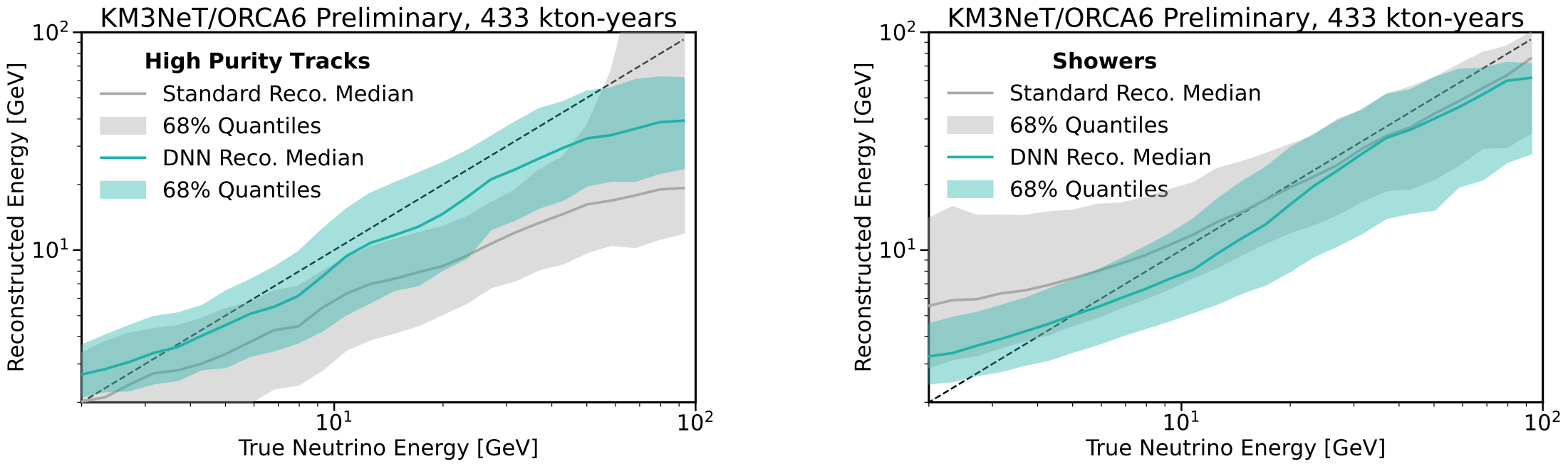

**(a)** High Purity Tracks class **(b)** Showers class

**Figure 1:** Energy response functions for different classes of neutrino events. Events are reconstructed using the standard method and the DNN to compare.

## 3. Results and discussion

After training the neural network and selecting the optimal hyperparameters, the performance of the network is evaluated.

### 3.1 Energy response function

The energy response function of the neural network is compared with the standard reconstruction method in Figure 1, where we show the energy resolution for the different c lasses of events.

The energy estimate given by the DNN shows a distribution with less bias than the standard reconstruction method. Additionally, the energy for the track events saturates eventually at a given energy for both energy estimates. This is a consequence of the finite size of the detector which may not contain the full energy deposition of the events. The muon produced in the interaction will leave the detector without fully depositing all its energy. For the case of the DNN, we see this saturation to occur at a higher reconstructed energy, possibly meaning the DNN is able to get information from the hadronic shower given by the triggered hits to correctly reconstruct the energy of the event. This effect is particularly useful to prevent the presence of events with high true energy from polluting the sample of reconstructed energies around the oscillation regime.

### 3.2 Sensitivity to oscillation parameters

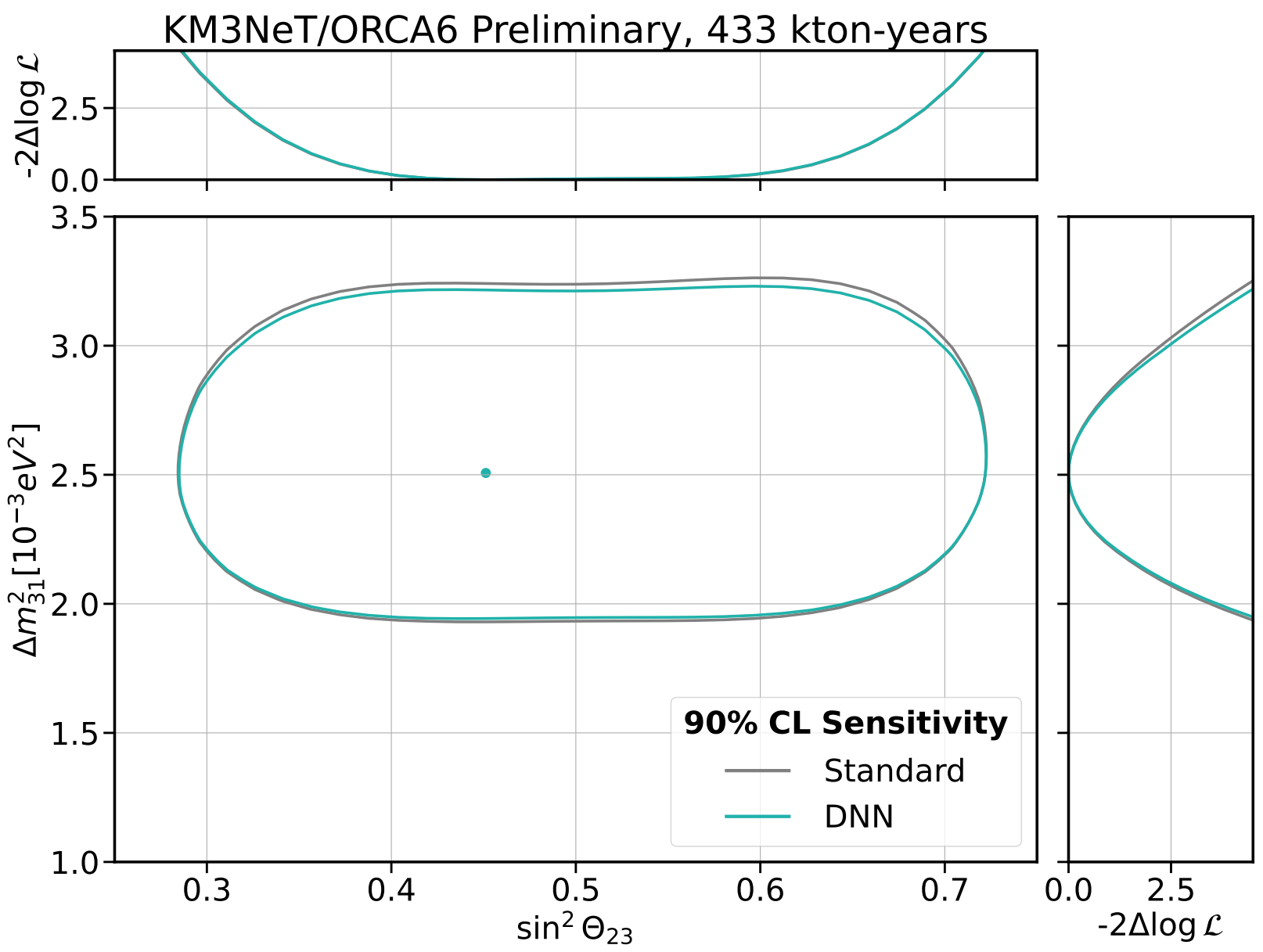

**Figure 2:** Contour at 90% CL of ORCA6 sensitivity to oscillation parameters $\theta_{23}$ and $\Delta m^2_{32}$ using the DNN reconstructed energy compared with the standard energy estimates.

The performance of the network is evaluated by computing the sensitivity to oscillation parameters as a $\Delta\chi^2$ using the reconstructed energy and by comparing with the standard energy estimates. Results are shown in Figure 2. At every point in the contour the log-likelihood is minimized relative to all nuisance parameters defined for the main oscillation analysis of the experiment [5]. The figure shows that the network leads to a better performance in constraining the value of $\Delta m^2_{32}$. For the case of $\theta_{23}$, the gain is negligible.

In order to assess the gain of using the DNN for the analyses, we compute the exposure needed to attain the same values of the errors on the oscillation parameters. Results are shown in Figure 3. The figure shows that the better precision of the DNN for the $\Delta m^2_{32}$ parameter is equivalent to having 12% more exposure, while for the $\sin^2(\theta_{23})$ parameter it is equivalent to having 2% more exposure.

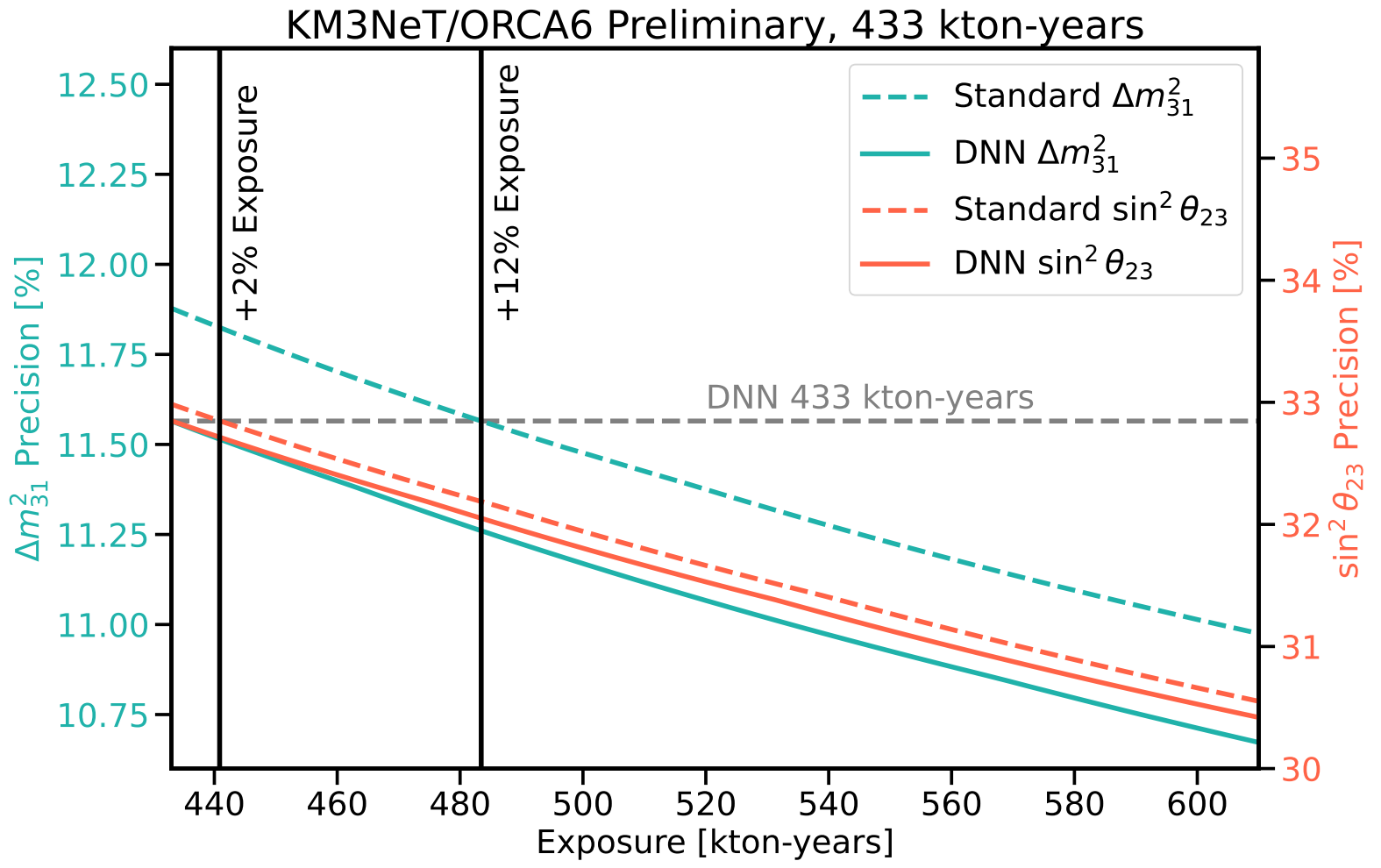

**Figure 3:** Precision to oscillation parameters $\sin^2\theta_{23}$ and $\Delta m^2_{32}$ as a function of exposure in kton-years for the DNN and the standard energy estimates.

### 3.3 Summary

Using a Deep Neural Network combining the energy estimates from three standard reconstruction methods plus additional topological information of the events allows to improve the performance of the energy reconstruction of an event. This improvement is translated into a better sensitivity to oscillation parameters, and a gain in precision for the same exposure. Additionally, the reconstructed energy has less bias than the standard reconstruction methods. This method highlights the importance of having a reconstruction that considers the different topologies of a neutrino interaction. This tool will continue to be improved and tested on the different physics analyses, including BSM scenarios within the collaboration. Future work will test different sample selections and a wider scan of training features to look for improvements on the sensitivity to oscillations.

# Lorentz Invariance Violation with KM3NeT/ORCA115

**Alba Domi,**[a,*] **on behalf of the KM3NeT Collaboration**

[a] *Erlangen Centre for Astroparticle Physics of Friedrich-Alexander-Universität, Nikolaus-Fiebiger-Straße 2, 91058 Erlangen*

*E-mail:* alba.domi@fau.de

Lorentz invariance (LI) underlies both the Standard Model of particle physics and General Relativity, and it represents our understanding of the nature of spacetime. It is therefore of fundamental interest to test its validity in every accessible regime as this would allow us to probe the microscopic structure of space-time and to constrain quantum gravity models. Lorentz Invariance Violation (LIV) would modify the observed energy and zenith angle distributions of atmospheric neutrinos that can be detected by neutrino telescopes such as KM3NeT. KM3NeT/ORCA115 is a next-generation neutrino telescope under construction in the Mediterranean sea, and is optimised for atmospheric neutrino oscillations studies. In this contribution, the sensitivity of ORCA115 to the presence of LIV is presented.

*Speaker

## 1. Introduction

Lorentz invariance underlies both the Standard Model (SM) of particle physics and General Relativity (GR), and it represents our understanding of the nature of spacetime. This symmetry guarantees that physical phenomena are observed to be the same by all inertial observers. Violations of this symmetry at or below the Planck scale, $m_P \sim 10^{19}$ GeV, have been predicted in a variety of quantum gravity (QG) models attempting to unify quantum field theory (QFT) and GR [1]. Indeed, many QG models involve some form of discretisation of spacetime, which is hard to reconcile with e.g. Lorentz boost invariance. It is actually possible to define a QG model which has LI sub-Planckian structure, however, this does not guarantee the preservation of LI at intermediate scales.

It is therefore of fundamental interest to test the validity of LI in every accessible regime as this would allow us to probe the microscopic structure of space-time and to constrain QG models.

The comprehensive effective field theory incorporating the SM and GR and characterizing general Lorentz and CPT violation is the Standard Model Extension (SME) [1]. In the SME, each Lorentz-violating term is formed by contracting a Lorentz-violating operator of a given mass dimension $d$ with a controlling coefficient that can be experimentally constrained.

Specifically, in the neutrino sector, deviations from standard neutrino oscillations can be expected in case of LIV. Such contributions would modify the observed energy and zenith angle distributions of atmospheric neutrinos that can be detected by neutrino telescopes.

## 2. Isotropic LIV with neutrinos

As discussed in Ref. [1], the general effective Hamiltonian describing neutrino propagation and mixing in the presence of LIV operators of renormalizable dimension contains four types of coefficients, leading to many novel effects that can be revealed in suitable experiments. The phenomenological approach to LIV assumes two separate cases: in one case, the rotational symmetry is preserved, and this is referred to as *isotropic* LIV. Other cases assume a breaking of the rotational symmetry, referred as to *sidereal* LIV. The analysis here presented focuses on isotropic LIV: the Lorentz symmetry is broken in the time coordinate.

To calculate the effect of isotropic LIV in the evolution of a neutrino system, we start from an effective Hamiltonian derived from the SME, which can be written as [1]:

$$H = H_0 + H_I + H_{LIV} \tag{1}$$

with

$$H_0 = \frac{1}{2E}\begin{pmatrix} 0 & 0 & 0 \\ 0 & \Delta m^2_{21} & 0 \\ 0 & 0 & \Delta m^2_{31} \end{pmatrix} \tag{2}$$

being the standard oscillation Hamiltonian that applies to the neutrino mass states,

$$H_I = \pm\sqrt{2}G_F \begin{pmatrix} N_e & 0 & 0 \\ 0 & 0 & 0 \\ 0 & 0 & 0 \end{pmatrix} \tag{3}$$

being the Hamiltonian that applies to the neutrino flavour states which accounts for matter effects in the regime of coherent scattering, and

$$H_{LIV} = \begin{pmatrix} \mathring{a}^{(3)}_{ee} & \mathring{a}^{(3)}_{e\mu} & \mathring{a}^{(3)}_{e\tau} \\ \mathring{a}^{(3)*}_{e\mu} & \mathring{a}^{(3)}_{\mu\mu} & \mathring{a}^{(3)}_{\mu\tau} \\ \mathring{a}^{(3)*}_{e\tau} & \mathring{a}^{(3)*}_{\mu\tau} & \mathring{a}^{(3)}_{\tau\tau} \end{pmatrix} - E \begin{pmatrix} \mathring{c}^{(4)}_{ee} & \mathring{c}^{(4)}_{e\mu} & \mathring{c}^{(4)}_{e\tau} \\ \mathring{c}^{(4)*}_{e\mu} & \mathring{c}^{(4)}_{\mu\mu} & \mathring{c}^{(4)}_{\mu\tau} \\ \mathring{c}^{(4)*}_{e\tau} & \mathring{c}^{(4)*}_{\mu\tau} & \mathring{c}^{(4)}_{\tau\tau} \end{pmatrix} + E^2\mathring{a}^{(5)} - E^3\mathring{c}^{(6)} + E^4\mathring{a}^{(7)} - E^5\mathring{c}^{(8)} + \ldots \tag{4}$$

being the isotropic Lorentz-violating Hamiltonian: the $\mathring{a}$ coefficients are CPT-odd, whereas the $\mathring{c}$ coefficients are CPT-even. From the Hamiltonian formulation, every dimension coefficient has a different impact in neutrino oscillations, which is summarised in Tab. 1. Specifically, the oscillation effect of $H_0$ is $\propto L/E$, which means that, fixing longer baselines $L$ and higher neutrino energies $E$ allow to probe higher dimension coefficients.

**Table 1:** LIV coefficients: for a comparison, the oscillation effect of $H_0$ is $L/E$.

| Coefficient | Unit | CPT | Oscillation effect |
|---|---|---|---|
| $\mathring{a}^{(3)}$ | GeV | odd | $\propto L$ |
| $\mathring{c}^{(4)}$ | - | even | $\propto LE$ |
| $\mathring{a}^{(5)}$ | GeV$^{-1}$ | odd | $\propto LE^2$ |
| $\mathring{c}^{(6)}$ | GeV$^{-2}$ | even | $\propto LE^3$ |
| $\mathring{a}^{(7)}$ | GeV$^{-3}$ | odd | $\propto LE^4$ |
| $\mathring{c}^{(8)}$ | GeV$^{-4}$ | even | $\propto LE^5$ |

## 3. LIV analysis with ORCA115

The analysis presented here follows the same procedure of Ref. [2]. Specifically, the analysis is based on detailed Monte Carlo (MC) simulations, accounting for neutrino interactions, secondary particles production and Cherenkov light emission and propagation. The atmospheric neutrino flux is computed from the Honda model [3] for the Gran Sasso site without mountain over the detector, assuming minimum solar activity. Atmospheric muons are generated with MUPAGE [2].
Event reconstruction is performed via a maximum likelihood fit to shower and track hypotheses. Background events arising from noise and atmospheric muons are rejected with two independent Random Decision Forests (RDF) trained on MC simulations. A third RDF was used to separate neutrino candidates into three topology classes defined by the output score of the RDF, trained to identify track-like events. Events with a track score larger than 0.7 are labelled as track-like, track scores less than 0.3 are labelled as shower-like, and other values are labelled as an intermediate topology. Moreover, as in Ref. [4], only upgoing events are considered in order to get rid of the atmospheric muon contamination.
Instead of using parametrised response functions as in Ref. [4], the analysis reported here is based on the aforementioned MC simulations to directly model the detector response. The two approaches have been compared and found consistent.
The MC-based modelling of the detector response is implemented in the KM3NeT framework

Swim [5]. The detector response is represented by a 4-dimensional matrix, as a function of true and reconstructed neutrino energy $E$, $E'$, and zenith angle $\theta$, $\theta'$, for each interaction channel $\nu_x$, $R^{[\nu_x \to i]}(E, \theta, E', \theta')$. Each entry of this matrix summarises in a single dimensionless coefficient the efficiency of detection, classification and probability of reconstruction for a given true bin $(E, \theta)$. Therefore, $R$ incorporates all the effects related both to the detector and to the event selection, which, in this analysis, uses atmospheric neutrino events with reconstructed energy up to 20 GeV.

The values of the standard neutrino parameters used in this analysis is taken from the NuFit v5.2 global fit result with Super-Kamiokande (SK) data [6] and are summarised in Tab. 2 for normal ordering (NO). Oscillation probabilities are evaluated with the software package OscProb [7], and to account for Earth's matter effects the PREM model [8] with 44 layers is used.

| | $\sin^2\theta_{12}$ | $\sin^2\theta_{23}$ | $\sin^2\theta_{13}$ | $\delta_{CP}$ | $\Delta m^2_{21}(\mathrm{eV}^2)$ | $\Delta m^2_{31}(\mathrm{eV}^2)$ |
|---|---|---|---|---|---|---|
| NO | 0.303 | 0.451 | 0.02225 | 232° | $7.41\times10^{-5}$ | $2.507\times10^{-3}$ |

**Table 2:** Benchmark oscillation parameters for NO, taken from the NuFit v5.2 result [6].

The above information can be used to define the distinguishability $\Delta\chi^2$, as a quick estimator of sensitivity of measurements, with the goal of illustrating the impact of LIV in the event distributions, as

$$\Delta\chi^2 = \frac{(N_{\mathrm{LIV}} - N_{\mathrm{Std}})|N_{\mathrm{LIV}} - N_{\mathrm{Std}}|}{N_{\mathrm{LIV}}}, \tag{5}$$

where $N_{\mathrm{LIV}}$ and $N_{\mathrm{Std}}$ are the number of events, as a function of reconstructed energy and zenith angle, in the LIV and standard hypothesis respectively. Fig. 1 shows the distinguishability distribution for LIV assuming dimension 3 coefficient.

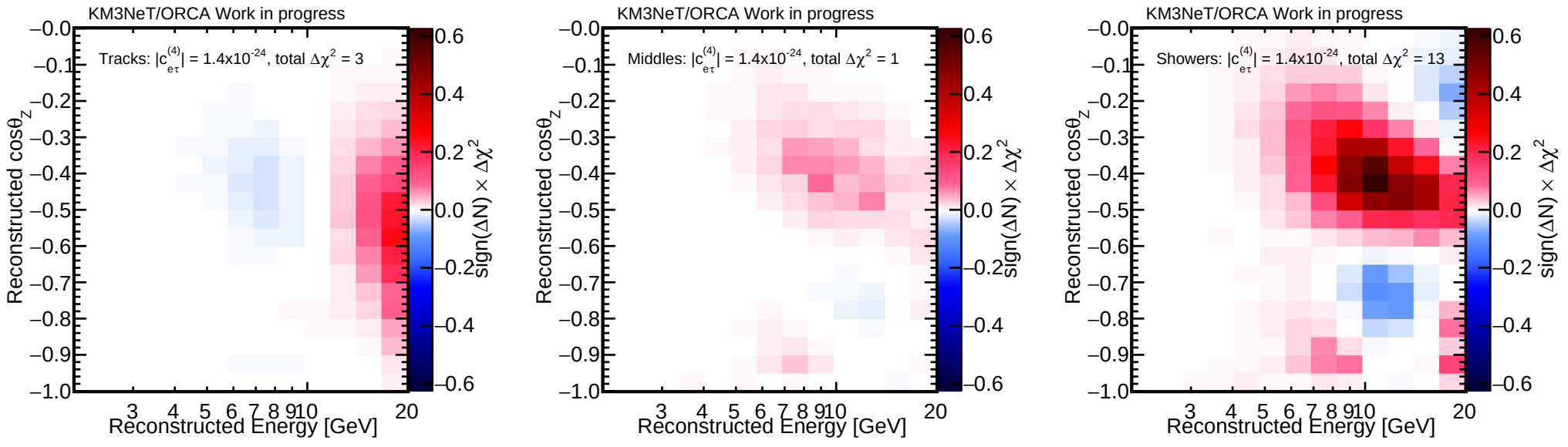

**Figure 1:** $\Delta\chi^2$ distribution of the three topologies considered in the analysis (tracks, intermediates and showers) assuming three years of data taking. The colour scale denotes the $S_\sigma$ value for each bin, whereas the total $S_\sigma$ is reported on top of the plots. The LIV parameters are $\left|\mathring{c}^{(4)}_{e\tau}\right| = 1.4\times10^{-24}$.

The same figure can be produced by fitting all the parameters of the analysis in order to see the impact of systematics. This is shown in Fig. 2.

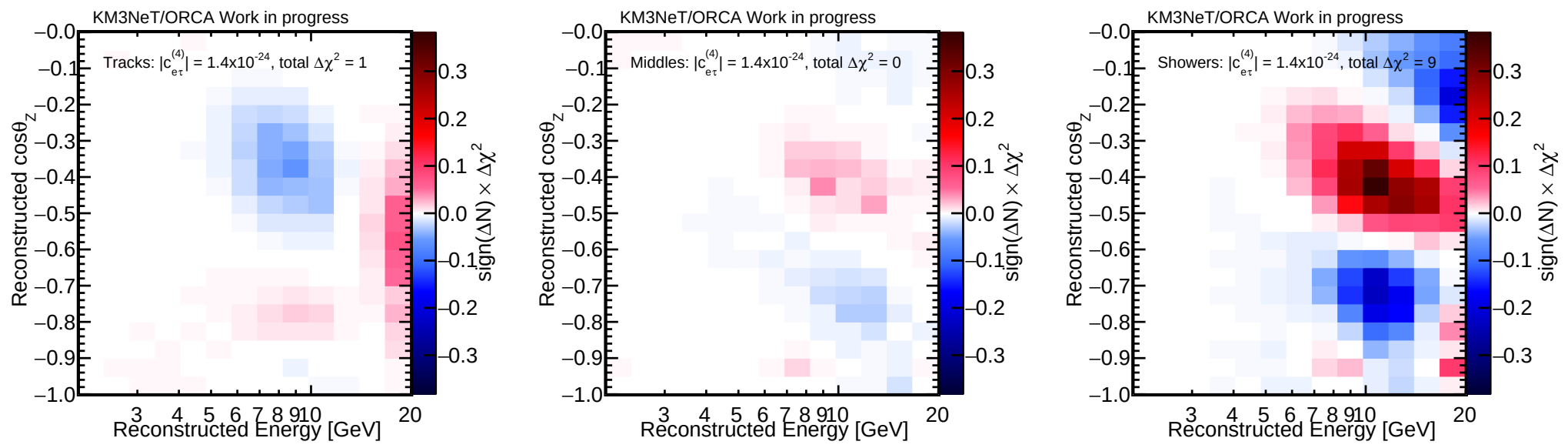

**Figure 2:** Same as Fig. 1 but by fitting all the analysis parameters.

The sensitivity evaluation is based on the minimisation of a negative log-likelihood function describing the agreement between a model prediction and observed data. This is done with the Asimov approach [2] assuming the negative log-likelihood follows a chi-squared distribution. Specifically, the negative log-likelihood function is defined as:

$$\chi^2 = -2\log L = \chi^2_{\text{stat}} + \chi^2_{\text{syst}} = 2\sum_{i=1}^{N_{E'}}\sum_{j=1}^{N_{\cos\theta'}}\sum_{t=1}^{3}\left[N^{\text{model}}_{ijt}(\eta) - N^{\text{data}}_{ijt} + N^{\text{data}}_{ijt}\log\left(\frac{N^{\text{data}}_{\text{ijt}}}{N^{\text{model}}_{\text{ijt}}(\eta)}\right)\right] + \sum_{k=1}^{N_{\text{Syst}}}\left(\frac{\eta'_k - \langle\eta'_k|\eta'_k\rangle}{\sigma_{\eta'_k}}\right)^2, \tag{6}$$

where $N^{\text{model}}_{ijt}$ and $N^{\text{data}}_{ijt}$ represent the number of expected and measured events in bin $(i, j)$ respectively and the sum over $t$ runs over the three event topologies: tracks, intermediates and showers. $\eta$ represents the model parameters, which comprise both the oscillation parameters listed in Tab. 2, and nuisance parameters $\eta'$, which are related to systematic uncertainties. The second sum runs over the nuisance parameters and $\langle\eta'_k|\eta'_k\rangle$ is the assumed prior of the parameter $k$ and $\sigma_{\eta'_k}$ its uncertainty. The set of free parameters considered in this analysis, together with the assumed gaussian priors with mean $\mu$ and standard deviation $\sigma$, is summarised in Tab. 3. More details can be found in Ref. [2].

## 4. Results

Fig. 3 shows the sensitivity of KM3NeT/ORCA115 to isotropic LIV coefficients up to dimension 4 which is represented by the area of excluded region of the parameters space. ORCA115 sensitivity is compared with current upper limits from 12 years of SK atmospheric neutrino data [9], two years of IceCube atmospheric neutrino data [10] and DUNE sensitivity assuming 7 years of data taking [11].

Since this analysis is limited to events up to 20 GeV, and as discussed in Sec. 2, the best sensitivity to higher dimension coefficients is reached with high energy neutrinos, currently the ORCA115 results do not extend to dimensions $> 4$. An update of this work, with events > 20 GeV is foreseen, which will include also higher dimension coefficients.

Current results show that with three years of data taking ORCA115 will allow to probe regions of the parameter space not yet probed by current analyses.

| **Parameter** | **Gaussian Prior** ($\mu \pm \sigma$) |
|---|---|
| $\nu_e/\bar{\nu}_e$ | $0 \pm 0.07$ |
| $\nu_\mu/\bar{\nu}_\mu$ | $0 \pm 0.05$ |
| $\nu_e/\nu_\mu$ | $0 \pm 0.02$ |
| NC Scale | No prior |
| Energy Scale | $1 \pm 0.05$ |
| Energy Slope | No prior |
| Zenith Angle Slope | $0 \pm 0.02$ |
| Track Normalisation | No Prior |
| Intermediate Normalisation | No Prior |
| Shower Normalisation | No Prior |
| $\Delta m^2_{31}$ | No prior |
| $\theta_{13}$ | $\theta_{13} \pm 0.13°$ |
| $\theta_{23}$ | No prior |

**Table 3:** List of fitted values and relative gaussian priors considered in this analysis. $\theta_{13}$ refers to the values listed in Tab. 2

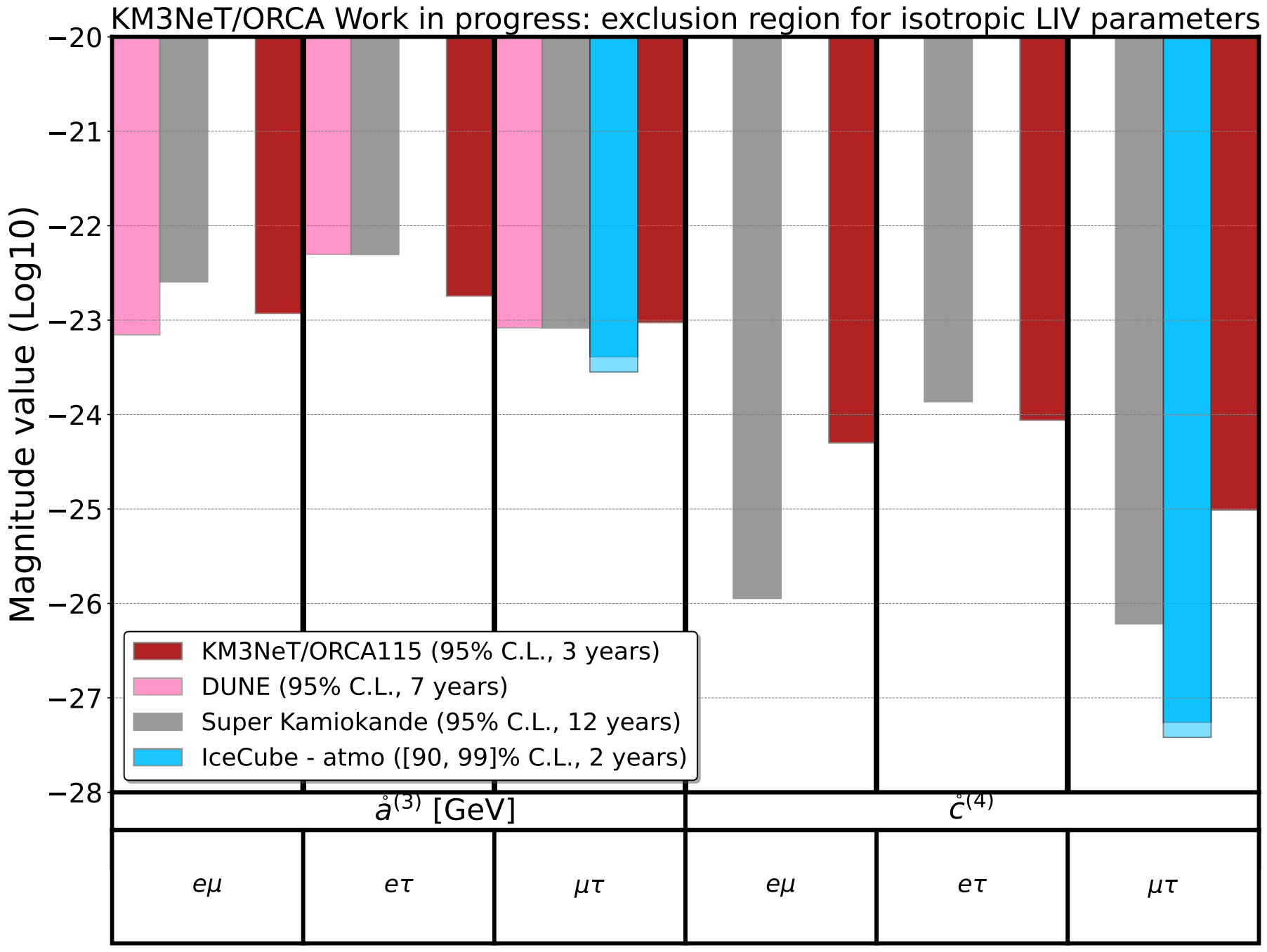

**Figure 3:** Two-dimensional sensitivity of KM3NeT/ORCA115 at the 95% C.L. for the real and imaginary parts of the isotropic coefficients of dimension 3 $a_{e\mu}$, $a_{\mu\tau}$, $a_{e\tau}$, for three years of data taking. Sensitivity results are compared with DUNE sensitivity [11] and current upper limits from 12 years of SK [9], two years of IC-atmospheric neutrino analysis [10].

**Acknowledgments**

This project has received funding from the European Union's HORIZON-MSCA-2021-PF-01 programme under the Marie Skłodowska-Curie grant agreement QGRANT No. 101068013.

# First measurement of tau appearance with KM3NeT/ORCA6

**Nicole Geißelbrecht**[a,*] **for the KM3NeT collaboration**

[a] *Friedrich-Alexander-Universität Erlangen-Nürnberg (FAU), Erlangen Centre for Astroparticle Physics, Nikolaus-Fiebiger-Straße 2, 91058 Erlangen, Germany*

*E-mail:* nicole.geisselbrecht@fau.de

KM3NeT/ORCA is an underwater Cherenkov neutrino detector currently being built in the Mediterranean Sea. The detector is optimised for the detection of atmospheric neutrinos in the energy range from a few GeV to 100 GeV in order to study neutrino oscillations and to determine the neutrino mass ordering.
The observation of oscillations of atmospheric electron and muon neutrinos into tau neutrinos is a primary physics goal during the ongoing detector construction phase, with a partially instrumented volume. The tau neutrino flux at the detector can be determined in a first step by identifying a statistical excess in the shower-like event topology compared to the expectation without oscillations. This measurement will allow to probe the standard three-flavour neutrino oscillation model.
This contribution will present the sensitivity of KM3NeT/ORCA to tau neutrino appearance and report a first measurement with ORCA6, an early 5% detector sub-array for an exposure of 433 kton-years.

*Speaker

## 1. Introduction - Tau appearance with KM3NeT/ORCA

The KM3NeT collaboration is currently building two neutrino telescopes in the Mediterranean Sea [1]. KM3NeT/ARCA is located near Sicily and is designed for neutrino astronomy. KM3NeT/ORCA is instead located offshore Toulon, France, and is optimised for the detection of atmospheric neutrinos in the energy range from a few to 100 GeV. The main goal is the study of neutrino oscillations and finally the determination of the neutrino mass ordering. Both detectors are three-dimensional arrays of photomultiplier tubes (PMTs) which detect the Cherenkov light that is emitted by the charged particles produced in neutrino interactions inside or close to the instrumented detector volume. KM3NeT/ORCA will consist of 115 vertical strings or detection units (DUs) anchored to the sea floor, and with an average distance of 20 metres. A DU comprises 18 so-called digital optical modules (DOMs) spaced 9 metres, and each DOM houses 31 3-inch PMTs in a 17-inch glass sphere. Data taking is already ongoing with preliminary sub-arrays of both detectors.

Sensitivity studies have shown that KM3NeT/ORCA has large potential for the observation of tau neutrinos [2]. The full detector will detect more than 3000 charged current tau neutrino ($\nu_\tau$ CC) events per year. These tau neutrinos must be a product of neutrino oscillations (tau appearance) since the atmospheric neutrino flux below 100 GeV is initially almost entirely composed by electron and muon neutrinos from the decays of charged pions and kaons. The dominant production channel for tau neutrinos in KM3NeT/ORCA is the transition $\nu_\mu \rightarrow \nu_\tau$. Figure 1 shows the oscillation probability at the detector level, i.e. the bottom of the Sea, dependant on the neutrino energy and direction, where the most prominent oscillation maximum corresponds to vertically up-going neutrino events with energies between 20 and 30 GeV. Matter effects, leading to the discontinuity in the probability pattern, affect the transition probabilities mostly at lower energies and are not expected to have a sizeable impact on this analysis.

However, it is not possible to detect tau neutrinos in KM3NeT/ORCA on an event-by-event basis but only statistically since the event topology is too similar to other neutrino interactions. Charged current muon neutrino ($\nu_\mu$ CC) interactions lead to a so-called track-like event signature in the detector. On the other hand, $\nu_\tau$ CC interactions are mostly shower-like. As a consequence, tau appearance can be observed as a statistical excess of shower-like events.

Tau appearance is quantified through the so-called tau normalisation parameter, $n_\tau$, which is defined as the ratio of the measured tau neutrino flux to the tau neutrino flux which is expected in the standard three-flavour neutrino oscillation scenario using the most updated measurements of the parameters that describe neutrino physics. There are non-negligible uncertainties on the charged current tau neutrino cross section [3] and on the unitarity of the PMNS matrix, especially in the tau sector [4], which could cause a deviation from the expected value. These two scenarios are respectively addressed by considering a scaling only on the CC component (CC-only), or by additionally scaling the expected fraction of neutral current (NC) events produced by tau neutrinos (CC+NC).

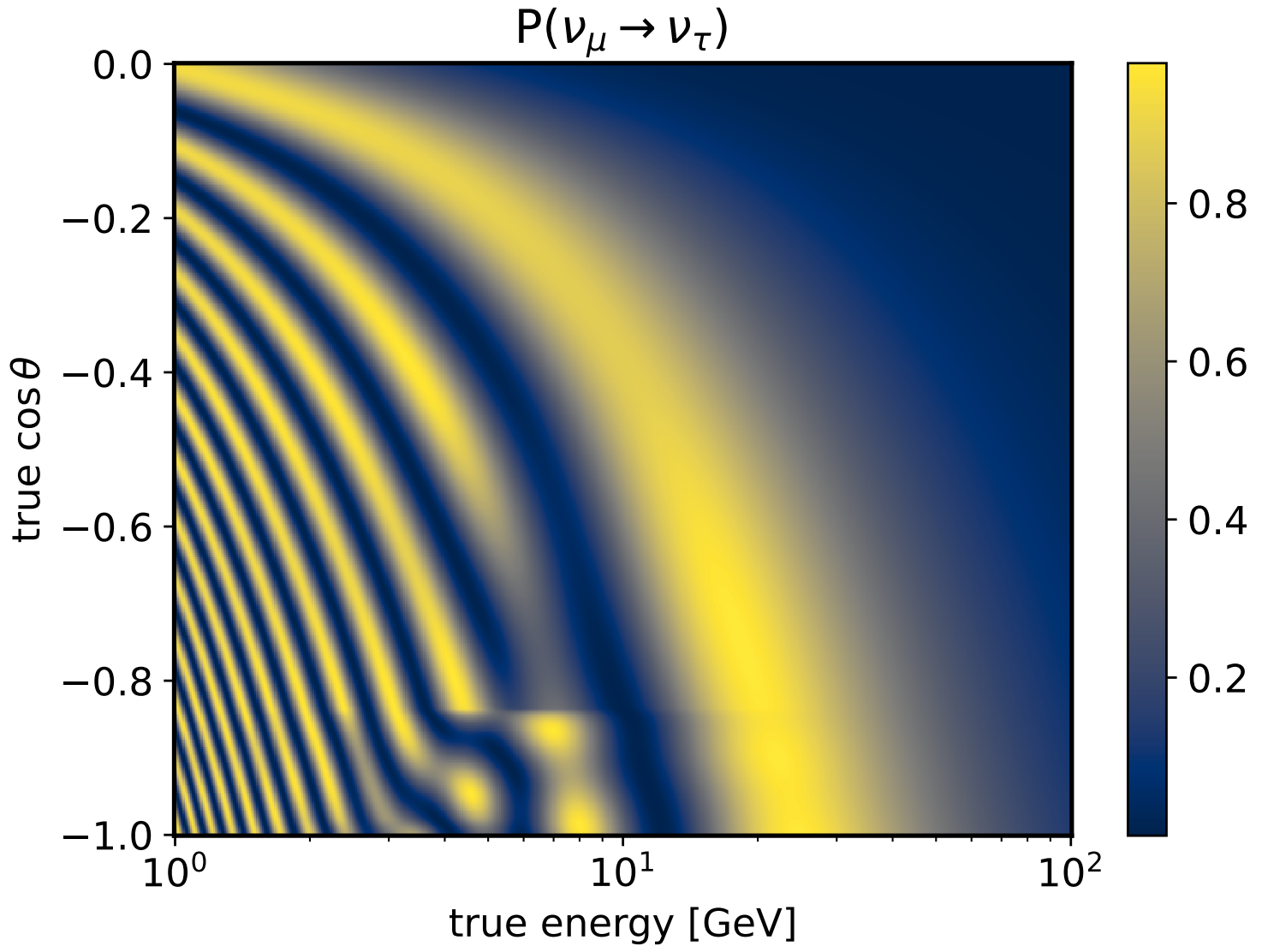

**Figure 1:** $\nu_\mu \rightarrow \nu_\tau$ oscillation probability at the detector level assuming normal ordering for true neutrino energy and direction relevant for KM3NeT/ORCA. Note that $\cos\theta$ equal to 0 (-1) refers to events that enter the detector horizontally (from below).

## 2. Data sample

This analysis is done with an early 5% sub-array of the full KM3NeT/ORCA detector composed of six detection units (KM3NeT/ORCA6). The total livetime is 510 days which corresponds to an exposure of 433 kton-years.

In order to obtain a clean neutrino sample, first, a pre-selection based on simple cuts which aims to discard pure noise events, is done. In a second step, atmospheric muon events are rejected by means of an event classifier based on boosted decision trees. Finally, the remaining events are further classified by a second set of boosted decision trees into track- and shower-like events. In order to minimise the impact of misclassified atmospheric muons and enhance the sensitivity, the track-like class is subdivided into a lower purity class (Low Purity Tracks) which contains the major amount of misclassified muons, and a higher purity class (High Purity Tracks) which contains only a very small fraction of muons.

The measurement of the tau normalisation parameter is carried out by fitting a model which includes the tau normalisation as a free parameter, to the observed event distribution in the two-dimensional space defined by the reconstructed energy and direction. To do this, histograms are created for each event class where the reconstructed direction is divided into 10 equally spaced bins in the up-going range of $\cos\theta$, where $\theta$ is the reconstructed zenith angle of an event. This is supposed to further reject the main background in KM3NeT/ORCA which is given by atmospheric muons that only enter the detector from above. Since tau neutrinos are only expected to come from below, this cut is not expected to have a negative impact on this analysis. Furthermore, 14 reconstructed energy bins are used between 2 and 100 GeV. In case of the shower class, one additional bin for events with reconstructed energies between 100 GeV and 1 TeV is introduced. After optimising the event selection and classification as described above, about 200 $\nu_\tau$ CC events are expected to be measured in 433 kton-years.

| Oscillation parameter | Prior |
|---|---|
| $\theta_{12}$ | fixed |
| $\theta_{13}$ | fixed |
| $\theta_{23}$ | free |
| $\Delta m^2_{31}$ | free |
| $\Delta m^2_{21}$ | fixed |
| $\delta_{CP}$ | fixed |

| Nuisance parameter | Prior |
|---|---|
| Spectral Index | ± 0.3 |
| $\nu_{\mathrm{hor}}/\nu_{\mathrm{ver}}$ | ± 2% |
| $\nu_\mu/\bar{\nu}_\mu$ | ± 5% |
| $\nu_e/\bar{\nu}_e$ | ± 7% |
| $\nu_\mu/\nu_e$ | ± 2% |
| NC Normalisation | ± 20% |
| Energy scale | ± 9% |
| High-energy Light Simulation | ± 50% |
| Overall Normalisation | free |
| Track Normalisation | free |
| Shower Normalisation | free |
| Muon Normalisation | free |

**Table 1:** All systematic uncertainties and their treatment in the fit.

## 3. Measurement

### 3.1 Fit and systematic uncertainties

The tau normalisation is fitted by the minimisation of a negative log-likelihood function $\mathcal{L}$. The fit is performed in the two-dimensional plane of reconstructed energy and direction for all event type classes.

$$-2\log\mathcal{L} = 2\sum_{ij}\left(n_{ij}^{\mathrm{model}} - n_{ij}^{\mathrm{data}} + n_{ij}^{\mathrm{model}}\log\left(\frac{n_{ij}^{\mathrm{data}}}{n_{ij}^{\mathrm{model}}}\right)\right) + \sum_{\epsilon}\left(\frac{\epsilon_{\mathrm{exp}} - \epsilon_{\mathrm{obs}}}{\sigma_\epsilon}\right)^2 \tag{1}$$

However, the fit does not only take $n_\tau$ into account but also the neutrino oscillation parameters, as well as other nuisance parameters as systematic uncertainties. The different parameters of the model as well as their treatment in the fit, i.e. if they are fitted with or without constraints, can be found in table 1. For each parameter $\epsilon$ with a prior $\sigma_\epsilon$, a term is added to equation 1 which penalises fitted values outside the prior range. The fixed oscillation parameters are taken from NuFIT v5.0 (with SK atmospheric data) [5] with both mass orderings tested. The remaining nuisance parameters account for uncertainties in the atmospheric neutrino flux and cross section. The priors are taken from [6]. The energy scale is a single parameter that combines different uncertainties of the detector. The high-energy light simulation parameter addresses the different light generators that are used in the simulations for low- and high-energy neutrinos. Finally, the global normalisation (Overall Normalisation) and the normalisations concerning the Showers (Shower Normalisation), the High Purity Tracks (Track Normalisation) and the atmospheric muons (Muon Normalisation) are fitted without constraints.

### 3.2 Results

The tau appearance analysis has been performed for CC-only and CC+NC scaling. The tau normalisation and the corresponding $1\sigma$ uncertainty were found to be $0.50^{+0.46}_{-0.42}$ ($0.67^{+0.37}_{-0.33}$) for CC-only (CC+NC). The $\Delta \log \mathcal{L}$ profile is shown in figure 2. It presents the difference between the log-likelihoods of the best fit and models with a tau normalisation fixed to values between 0 and 2. Even though the measurement deviates from the expected value of 1, it is still consistent with $n_\tau = 1$ on a $1.1\sigma$ ($0.9\sigma$) level in case of CC-only (CC+NC). For CC+NC, no tau appearance, i.e., $n_\tau = 0$, is disfavoured with $2.2\sigma$, whereas in the CC-only case, the measured value of $n_\tau$ agrees with no tau appearance within a $1.2\sigma$ level.

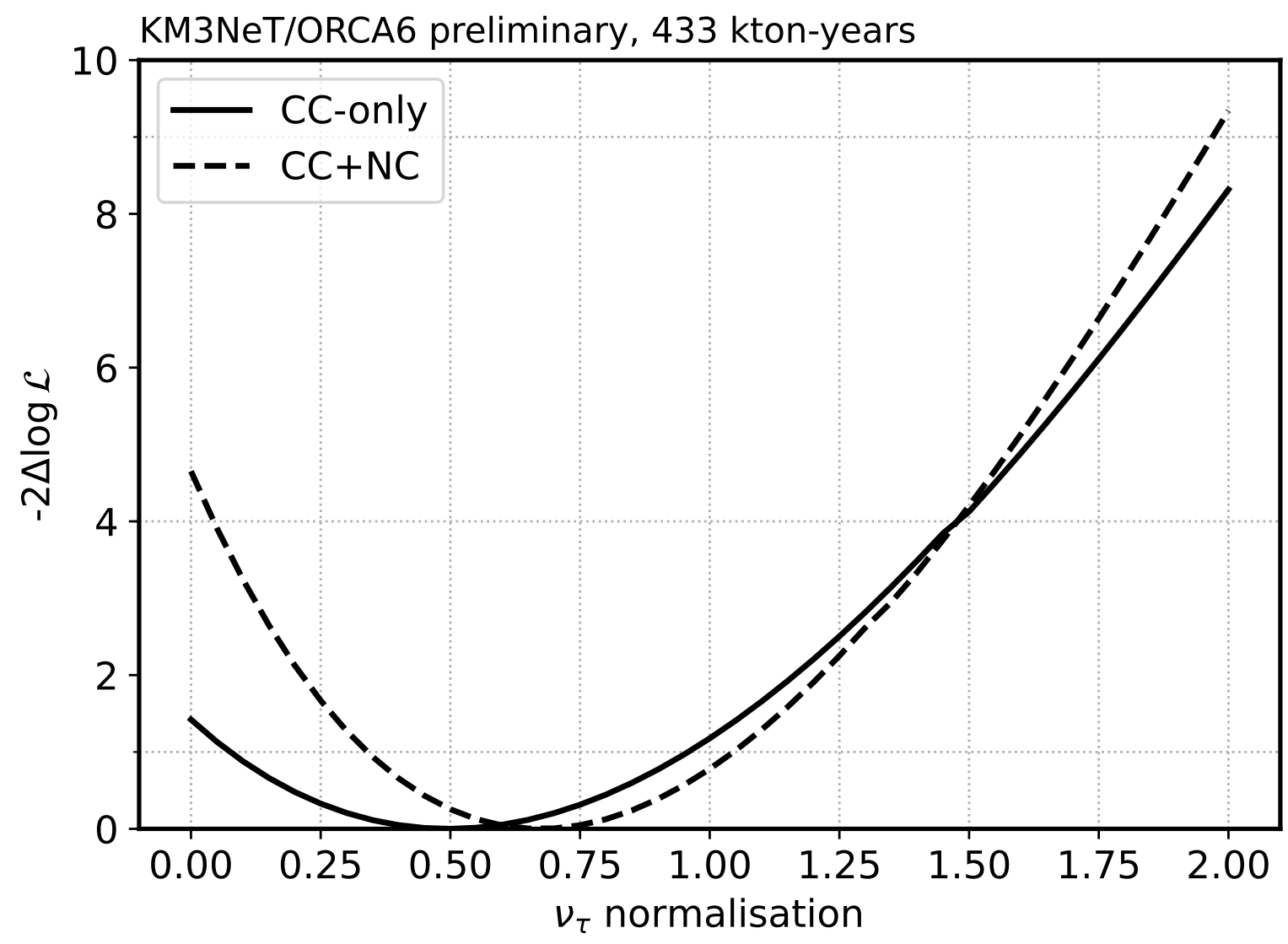

**Figure 2:** Measured $\Delta$ log-likelihood profile for CC-only (solid) and CC+NC (dashed).

As presented in figure 1, the oscillation probability and hence tau appearance depends on $L/E$ where $L$ is the neutrino baseline which depends on $\cos\theta$ and $E$ is the energy. Figure 3 shows the observed $L/E$ distribution of shower-like events, compared to the best fit with respect to no tau appearance for CC-only and CC+NC, respectively. Additionally, the distribution for a model with $n_\tau = 1$ is shown.

Figure 4 shows the impacts and pulls of all fitted oscillation and nuisance parameters. In order to study the impact that a variation on a certain model parameter has on the fitted tau normalisation, two additional fits are performed where the parameter is fixed at its best fit ± the $1\sigma$ MINOS errors calculated by the MINUIT package [7], while the rest of the parameters are left free. The impact on the tau normalisation is then calculated as $\left(n_\tau^{\text{shift}} - n_\tau^{\text{bf}}\right)/\sigma_{n_\tau}$. The coloured bars on figure 4 summarise the impact of the different parameters on the tau normalisation. For both scalings, the normalisation factors for the different light generators and the shower class as well as the zenith slope have large impacts. Slight differences between CC-only and CC+NC are a result of different correlations between the respective tau normalisation and the systematic uncertainties.

The black markers and their associated error bars on figure 4 represent the so-called pulls, which are defined as the difference of the best fit value and the expected value with respect to its uncertainty $(\epsilon_{\text{BF}} - \epsilon_{\text{CV}})/\sigma$. Here, $\sigma$ represents the pre-fit uncertainty in the parameter whenever available. For those parameters without pre-fit uncertainty, $\sigma$ represents the post-fit uncertainty. The error bars are calculated as the ratio $\sigma_\epsilon^{\text{post-fit}}/\sigma_\epsilon^{\text{pre-fit}}$ or set to 1 for those parameters without pre-fit uncertainties. As can be seen, the systematics with the largest pulls are the normalisations and the oscillation parameters. This can be expected since these are the parameters that are fitted without prior.

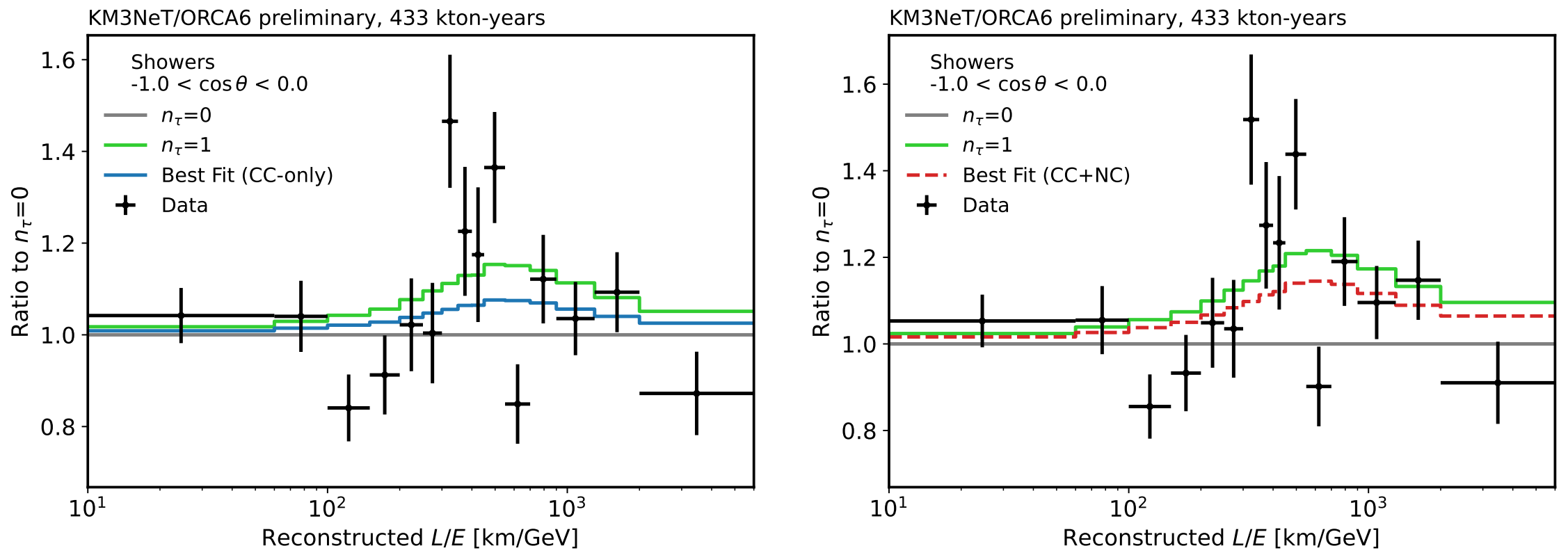

**Figure 3:** $L/E$ distributions for data, the best fit and $n_\tau = 1$ of the shower class with respect to no tau appearance for CC-only (left) and CC+NC (right).

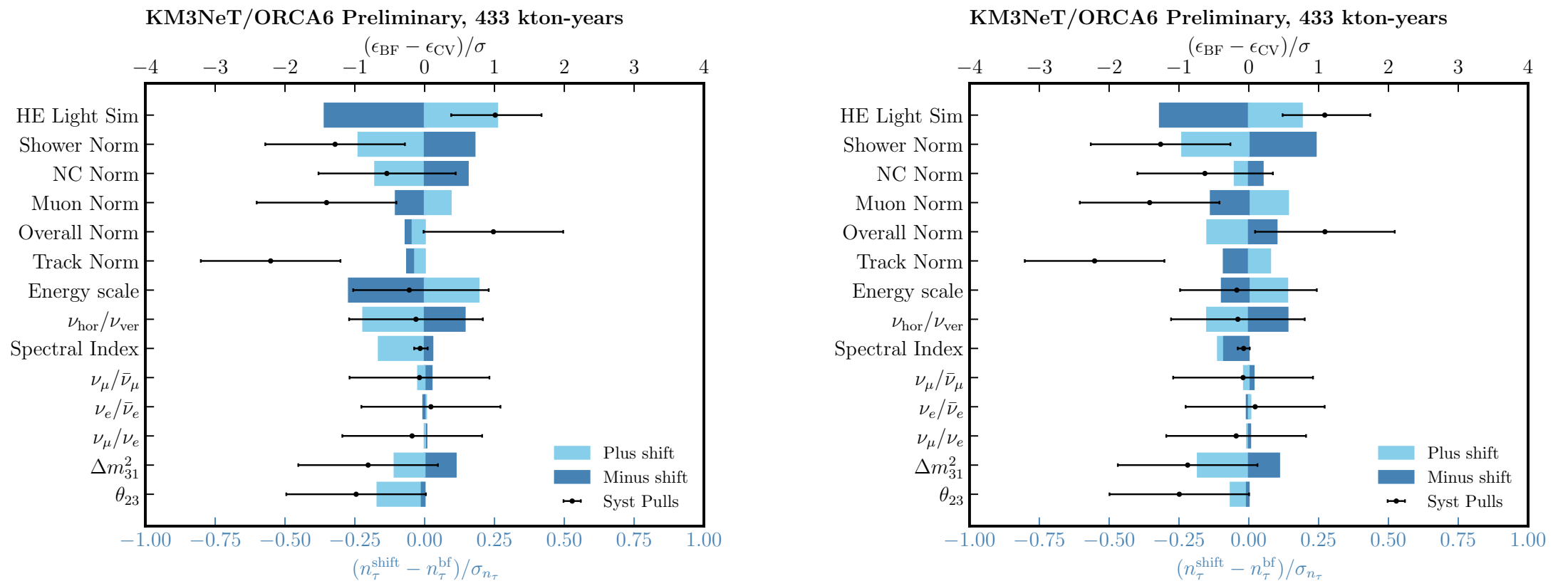

**Figure 4:** Impact on the tau normalisation of fixing oscillation and nuisance parameters at their best fit value $\pm 1\sigma$ (post-fit) uncertainty with respect to its uncertainty (coloured bars, lower axis). Additionally, the pulls and constraints (black markers with error bars, upper axis) are shown. Left: CC-only. Right: CC+NC.

Figure 5 presents the comparison of the results of KM3NeT/ORCA6 with the measurements of the tau normalisation that have been performed so far by OPERA [9], Super-Kamiokande [10] and IceCube/DeepCore [8]. All measurements agree with the Standard Model within their reported uncertainties. The first two experiments observed a tau normalisation larger than expected. However, a direct comparison of the results is non-trivial: on one hand, the differences with respect to OPERA could be related to the different neutrino sources used by both experiments. On the other hand, Super-Kamiokande's measurement was performed in an energy range where deep inelastic scattering contributes a 41% to the neutrino cross section, while being the main interaction channel in the KM3NeT/ORCA measurement. A direct comparison to the results of IceCube/DeepCore would be more straightforward because both detectors use the same neutrino source and both measurements are conducted in the same energy range. The measured tau normalisations from KM3NeT/ORCA6 and IceCube/DeepCore agree within a $1\sigma$ level. As can be seen in figure 5, none of the so far performed measurements can rule out a tau normalisation equal to 1.

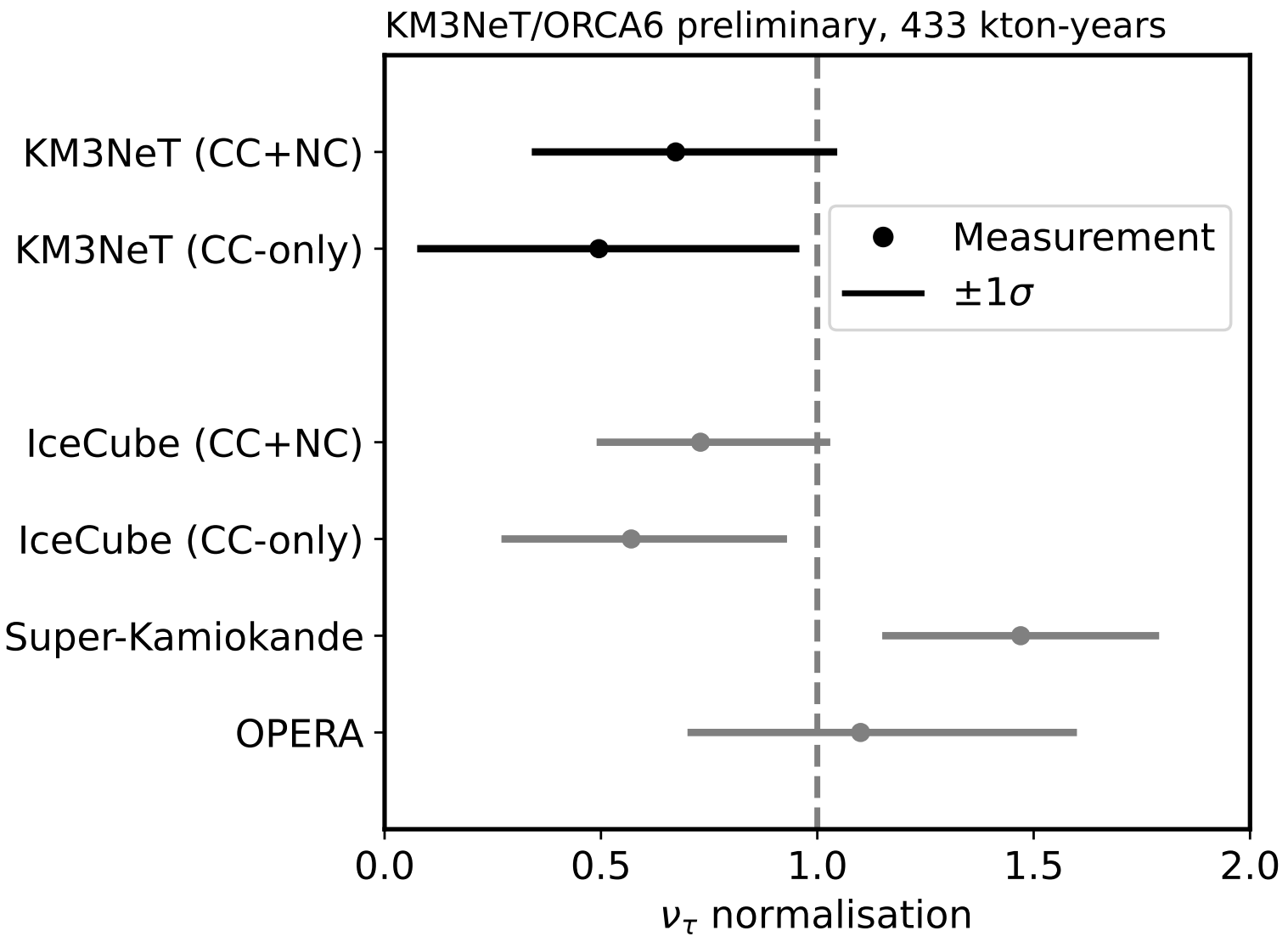

**Figure 5:** Comparison of the results for the tau normalisation from the different experiments. The errorbars show the reported $1\sigma$ uncertainties. Results are taken from [9] (OPERA), [10] (Super-Kamiokande) and [8] (IceCube/DeepCore).

## 4. Conclusion and Outlook

The first measurement of the tau normalisation for KM3NeT/ORCA6 has been performed for a data set with an exposure of 433 kton-years. The results are consistent with the expectations of a tau normalisation equal to 1, even though they indicate a slightly lower value than expected, namely $n_\tau = 0.50^{+0.46}_{-0.42}$ ($0.67^{+0.37}_{-0.33}$) for CC-only (CC+NC). The uncertainties of this measurement are on the same order of magnitude as the previous measurements from other experiments. However, the uncertainty is expected to be reduced in the future with extended data sets and larger detector configurations. This will decrease the impact of the statistical as well as of the systematic uncertainties. For an exposure of 21 Mton-years, which corresponds to 3 years of data taking with the full KM3NeT/ORCA detector, the tau normalisation will be constrained to ± 7% on a $1\sigma$ level (CC-only) [2]. On the long term, the measurement of $n_\tau$ can help to constrain elements of the PMNS matrix, as well as to constrain the $\nu_\tau$ CC deep inelastic scattering cross section.

# Particle identification in KM3NeT/ORCA

**L. Cerisy,[a,*] A. Lazo,[b] C. Lastoria,[a] M. Perrin-Terrin,[a] J. Brunner[a] and V. Dabhi[a,1] on behalf of the KM3NeT collaboration**

[a] *Aix Marseille Univ, CNRS/IN2P3, CPPM, Marseille, France*

[b] *IFIC - Instituto de Física Corpuscular (CSIC - Universitat de València), c/Catedrático José Beltrán, 2, 46980 Paterna, Valencia, Spain*

[1] *Not a KM3NeT member, contributed during his master thesis*

*E-mail:* cerisy@cppm.in2p3.fr

One of the main goals of KM3NeT/ORCA is to measure atmospheric neutrino oscillation parameters with competitive precision. To achieve this goal, good discrimination between track-like and shower-like events is necessary, with particular focus on the measurement of the tau neutrino normalisation. The track-like signal is mainly carried by muon neutrinos from charged current interactions, while the shower-like signal comes from charged current interactions of electron and tau neutrinos, and neutral current interactions of all flavours. A Random Grid Search algorithm is optimised to separate these channels and its performance is compared with machine learning methods using boosted decision trees. This contribution will report on the technical aspects of the algorithm and the performance of the particle identification with data recorded in 2020 and 2021 using an early six-lines configuration of the ORCA detector (ORCA6).

*Speaker

## 1. Introduction

Particle identification plays a crucial role in most neutrino studies aimed at measuring flavour oscillations with high precision. This contribution reports on the track-shower separation to identify neutrino flavours using the KM3NeT/ORCA6 data, with a particular focus on the optimisation of the $\nu_\tau$ appearance signal. The track-like signal refers to the muon from the charged current muon neutrino interaction. The shower-like signal refers to the electro-magnetic(EM) shower from the charged current electron neutrino interaction, the EM/hadronic shower from the charged current tau neutrino interaction, and the hadronic shower from the neutral current channel for all-flavors. With a still size-limited detector, no significant separability between electro-magnetic and hadronic shower is expected.

Sophisticated methods are being explored to perform particle identification such as Boosted Decision Trees (BDTs), used in the official KM3NeT/ORCA results, and most recently Deep Neural Networks (DNN). While these approaches show a strong potential to find the purest classification, new ideas to improve the robustness against mismodelling effects from the simulations and the understanding of the parameters used for particle identification, are described in this contribution.

The Random Grid Search (RGS) algorithm consist in a transparent and robust approach that relies on a combination of cuts in one or two dimension to separate two (or more) populations of events. A strong advantage of this method is the ability to look for the reasons why a particular data sample was chosen. By using a combination of cuts that involve a few features (typically 4 or 5) one can investigate the data/MC agreement and interpret the physical meaning of the features.

## 2. Detector and data sample

KM3NeT is an undersea Cherenkov neutrino telescope currently under construction at the bottom of the Mediterranean Sea off-shore the Italian Sicily coast (KM3NeT/ARCA) and 40 km off-shore Toulon, France (KM3NeT/ORCA). The two detectors are optimised for different neutrino energy ranges. They are composed of vertical Detection Units (DUs), each consisting of 18 Digital Optical Modules (DOMs). A DOM is housing 31 photomultiplier tubes (PMTs) and the corresponding readout electronics. Six DUs were operational in KM3NeT/ORCA when the data used in this analysis were acquired. A hit consists of a time stamp and a time over threshold. An event is created when the trigger algorithm identifies a series of causally-connected hits. The vertex position, the time and the direction of the event is determined by using a maximum-likelihood method based on a set of causally-connected hit times and positions. For the track reconstruction the hits are fitted under the assumption of a Cherenkov-light-emitting muon. The muon is assumed to follow a long, straight trajectory and to propagate practically at the speed of light in vacuum through water. For the shower reconstruction the signal is searched in all direction within 80m from the shower vertex pre-fit position, the hits are fitted to find the brightest point of the shower, expected a few meters from the neutrino vertex position depending on energy.

The data used in this analysis were collected between mid-February 2020, and mid-November 2021 for a total of 510 days or 433 kton-years. Quality cuts on the number of used hits >= 15, the likelihood >= 40 and the direction (up-going) of the track reconstruction were applied to remove poorly reconstructed events, noise events from K40 decay, and most of the atmospheric muon background. A cut keeping events below 100GeV in the energy from the track reconstruction is applied to the Tracks class (defined in Section 3) and 1TeV from the shower reconstruction energy to the Showers class to remove high energy events migration.

## 3. BDT Performance

A BDT based algorithm is used in the official oscillation analysis and relies on the training of the classifier on 45 different features using unweighted MC events. Those features are related to the energy, likelihood of the track/shower reconstruction, direction, hits of each reconstructed event. The trained classifier applied to a sample predicts a Track score and an atmopsheric muon (Atm. muon) score that determines the likeliness of an event to be associated to a **muon neutrino** or an **atmospheric muon**, respectively. In Figure 1 the weighted distribution of the scores and the data/MC ratio illustrates the good agreement between data and expectation for those variables.

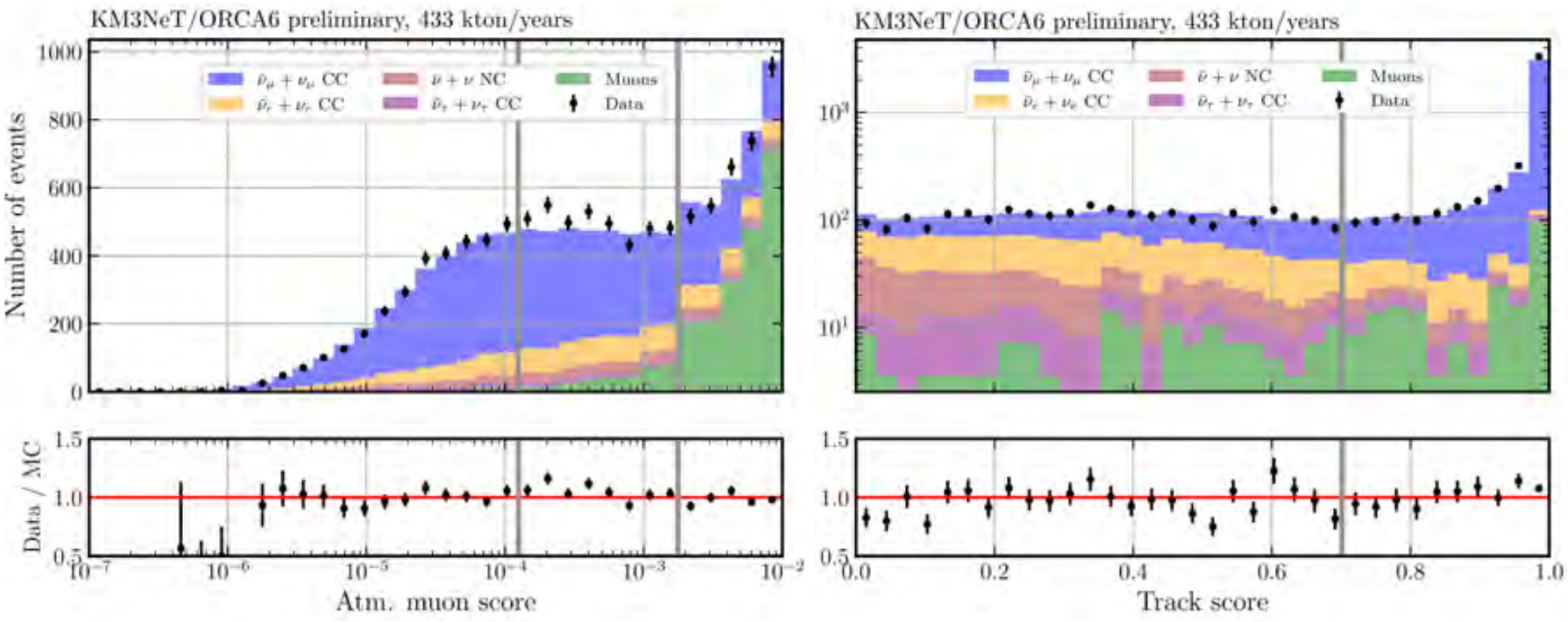

**Figure 1:** The Atmospheric muon score and Track score stacked distribution for all neutrino flavors is drawn in colors with atmospheric muons in green, $\nu_\tau + \bar{\nu}_\tau$ CC in purple, $\nu + \bar{\nu}$ NC in brown, $\nu_e + \bar{\nu}_e$ CC in orange and $\nu_\mu + \bar{\nu}_\mu$ CC in blue. The Data is shown with black dots and error bars. On the bottom part the data/MC ratio shows the agreement between the expected number of events and the collected data. The grey bands represents the cut values used to defined the classes.

To measure neutrino oscillations, a clean neutrino sample is produced by further removing the atmospheric muon background that passed the upgoing direction selection with a cut on the Atm. muon score at $1.8 * 10^{-3}$. In this approach the BDT outputs are also used to separate the data into three classes. The events that have a Track score below 0.7 are classified as Showers, while the events that fall above 0.7 are further divided into 2 classes the High and Low Purity Tracks by cutting on the Atm. muon score at $1.1 * 10^{-4}$. The cuts in the BDT scores have been optimised for performance to measure neutrino oscillation parameters [1]. Each class is defined to ensure enough statistics per class and per bin. The Table 1 shows the statistics contained in each class, the % of muons, $\nu_\mu/\bar{\nu}_\mu$ CC and $\nu_\tau/\bar{\nu}_\tau$ CC, for 433 kton-years and 296 kton-years samples.

| Selection | All events | Atm. muons | $\nu_\mu/\bar{\nu}_\mu$ CC | $\nu_\tau/\bar{\nu}_\tau$ CC |
|---|---|---|---|---|
| High Purity Tracks | 1870 | 7 | 1779 | 20 |
| Low Purity Tracks | 2001 | 83 | 1792 | 18 |
| Showers | 1959 | 21 | 908 | 130 |
| 433 kton-years | 5830 | 111 | 4480 | 169 |
| 296 kton-years | 1250 | 38 | 900 | 65 |

**Table 1:** Summary of the number of MC events for the different classes defined by the BDT separation. The exposure for ICRC 2023 is 433 kton-years, while the exposure for ICRC21 is 296 kton-years. The muon contamination is given as the number of atmospheric muons divided by the number of total events. The % of $\nu_\mu/\bar{\nu}_\mu$ CC events and number of $\nu_\tau/\bar{\nu}_\tau$ is also written to illustrate the purity of the classes.

## 4. Random Grid Search

The Random Grid Search algorithm was introduced in 1995 during the search of the top quark at FermiLab [2], and is still used in multiple experiments like in the search for supersymmetry at LHC [3].

The RGS procedure starts with the ranking of the features using weighted events from the track and shower reconstructions based on their 2D separability that measures the overlap between the track ( $\nu_\mu + \bar{\nu}_\mu$ CC) and the shower ($\nu_e + \bar{\nu}_e$ CC) distribution. As a second step the 2D asymmetry and the data/MC agreement for the best ranked features are investigated to verify the understanding of the features. Then the RGS algorithm is applied, with the idea to search for cuts where they are likely to be useful, i.e. in the expected signal region. If events $E_0 ... E_n$ represent neutrino interactions, with for instance a direction and and position which are called features $X$ and $Y$ then $(E_0(X_0), E_0(Y_0))...(E_n(X_n), E_n(Y_n))$ can be used to cut the sample in two. A combination of consecutive cuts for a given event $E_i$ with $0 \leq i \leq n$ in 1D like $((E_i(X_i), >), (E_i(Y_i), <))$ is called a set of cuts. By keeping track of the signal events (true positive rate) and background events (false positive rate) that passed the set of cuts, the performance graph shown in Figure 2 is produced. The RGS tests all the sets of cuts in 2D based on a fixed set of f eatures. After comparing the best sets of cuts for many different sets of features, the best set of features is fixed.

Once the best set of features is fixed, the optimisation stage starts in order to find the best cut values which would give the highest sensitivity for a given study. In this work the purpose was to increase the sensitivity to $\nu_\tau$ appearance [4]. It was found that the best classes division for $\nu_\tau$ appearance observation was into 3 classes: track, mixed and showers without separating in this approach the track class into 2 as this has no effect on the $\nu_\tau$ appearance s ensitivity. In this approach the events that have an Atm. muon score below $3 * 10^{-3}$ are kept. In Figure 3 the red lines represents the optimised 2D combination A & B that define the RGS track c lass. The mixed+showers region is further divided in two classes thanks to the intersection of the 2D combination C & D & F written in Table 2.

A Cherenkov hit is defined as a hit whose closest distance from the track is below 100 meters and whose time is within ±15 ns range from the expected time of the hit following the Cherenkov hypothesis. The definition of the features shown in Figure 4 are reported here from left to right and top to bottom. 1 - number of reconstructed tracks within 1° from the best track. 2 - the furthest Cherenkov hit distance to the start of the track in meters, and zero for the events that do not have any Cherenkov hits. 3 - the mean of the absolute value of the time residuals of the hits within 10° around the Cherenkov angle assuming the best track direction from the shower reconstruction. 4 - the log of the distance between shower reco vertex and the start of the track propagated in the track direction from a distance corresponding to the time difference of the two reconstructions at the speed of light in vacuum. 5 - the distance between the shower reco pre-reconstructed vertex and the position of the brightest point of the shower after the position fit.

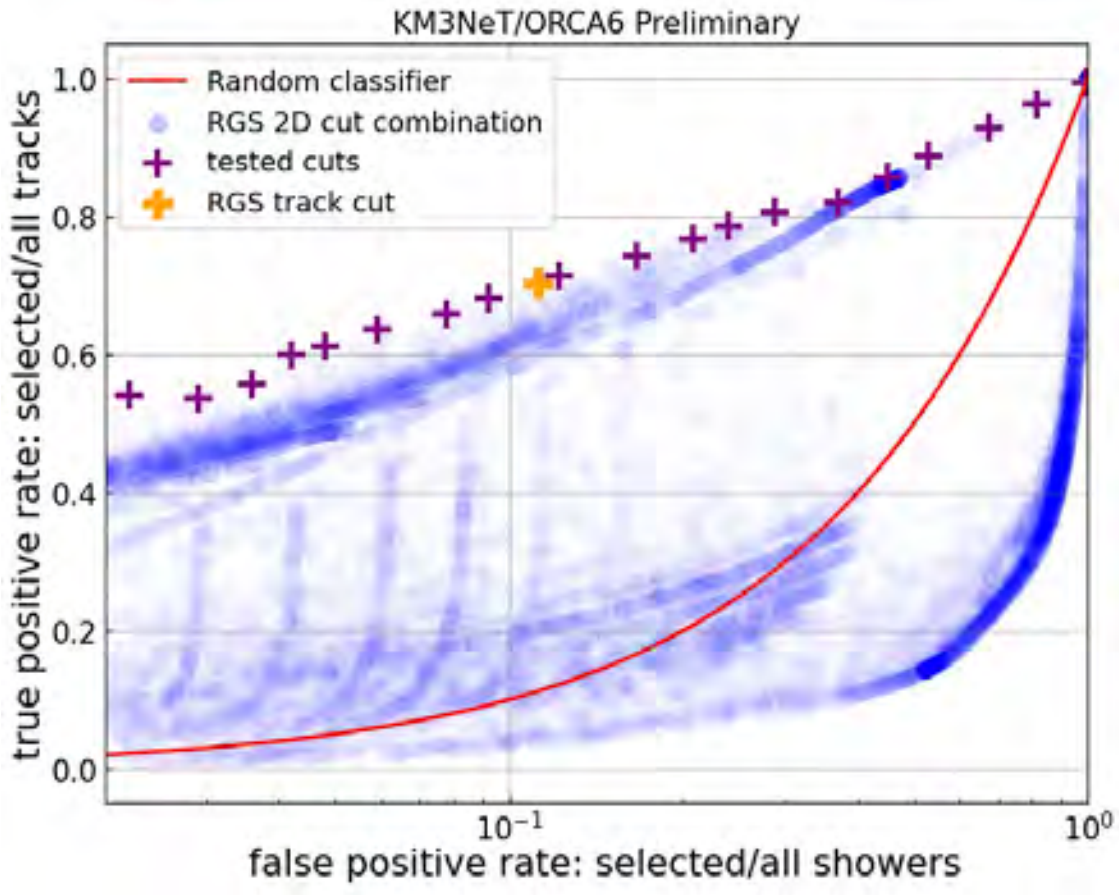

**Figure 2:** Efficiency and purity of each of the RGS sets of cuts. Those are shown with blue dots, while the tested cut values for optimisation are shown with purple crosses. The orange cross represents the chosen value for the RGS cut for track isolation after optimisation of the sensitivity to $\nu_\tau$ appearance. A random classifier is shown with a red line for comparison.

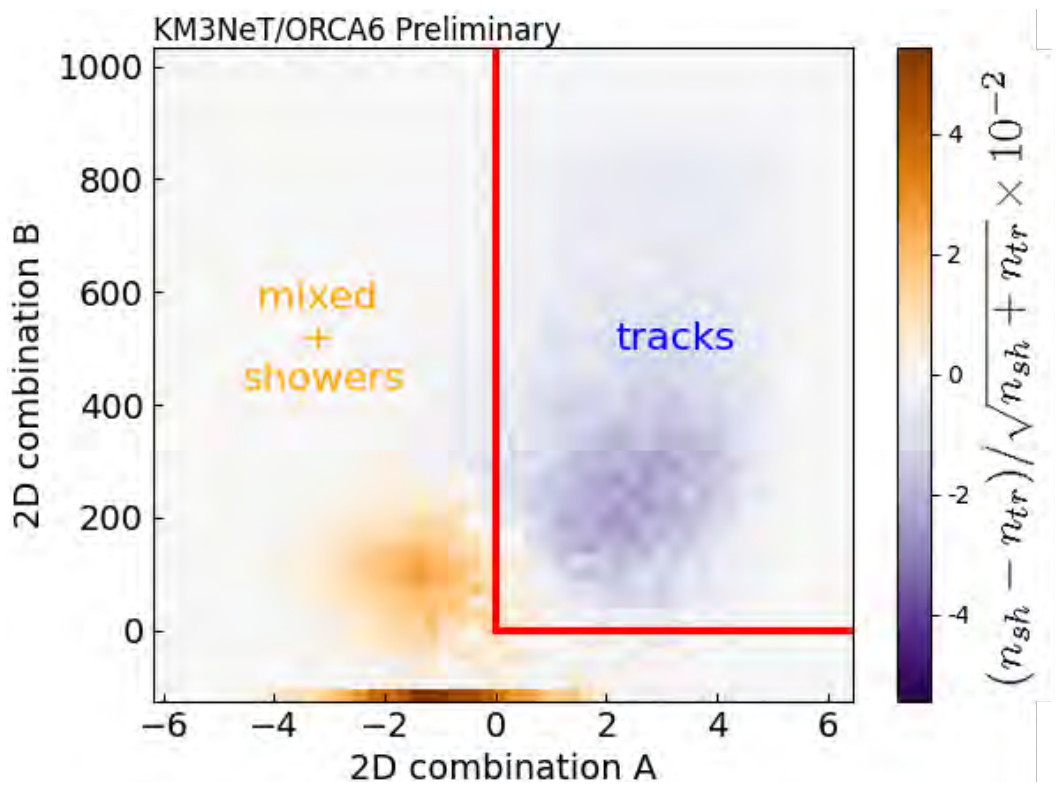

**Figure 3:** Significance of the track-shower asymmetry, with $n_{sh}$ and $n_{tr}$ the number of showers and tracks in each bin respectively. The blue region is enhanced in tracks while the orange region is enhanced in showers. Red lines are used to define the separation between the Tracks class and the Showers+Mixed classes by the RGS cuts defined in Table 2.

In the data/MC comparison for the 5 features involved in the RGS cuts, the events corresponding to all MC flavors are weighted, this allows to appreciate the good understanding of the data for each parameter that is used in the RGS class separation. For most bins the data/MC ratios are contained in a ±20% band around 1. The peak at 0 in the parameter showing the distance to the furthest Cherenkov hit correspond to events for which no hit passes the Cherenkov distance and time conditions.

| 2D combination $Z = y - (ax + b)$ | | | | | |
|---|---|---|---|---|---|
| RGS track class definition: $A\&B$ | | | | | |
| pars. | feature x | feature y | coeff a | coeff b | cut dir. |
| comb. A | n. tracks within 1° | log pre/pos fit dist. Shower Reco | -0.2356 | + 1.9124 | $Z > 0$ |
| comb. B | furthest Cherenkov hit | mean time residual of sel. hits | -5.0702 | +125.6146 | $Z > 0$ |
| RGS shower class definition:$(\bar{A} or \bar{B})$ & $(C\&D\&F)$ | | | | | |
| comb. C | log pre/pos fit dist. Shower Reco | furthest Cherenkov hit | -0.0101 | +71.1553 | $Z < 0$ |
| comb. D | log pre/pos fit dist. Shower Reco | mean time residual of sel. hits | -3.0422 | +7.4538 | $Z < 0$ |
| comb. E | mean time residual of sel. hits | log dist. Shower vs Track reco | -0.3291 | +2.503 | $Z < 0$ |

**Table 2:** Coefficients of RGS cut combination for Tracks and Showers classes definition.

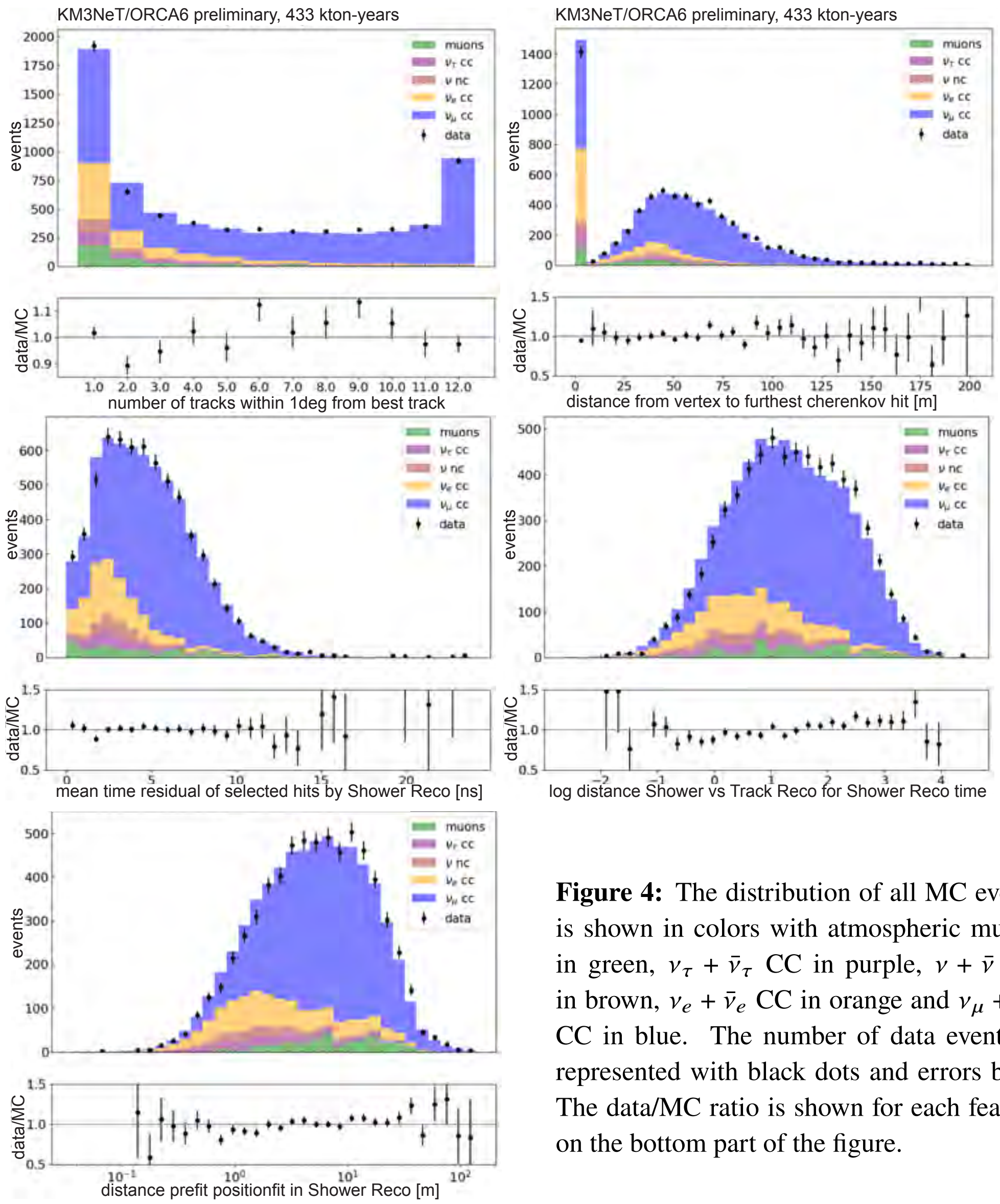

**Figure 4:** The distribution of all MC events is shown in colors with atmospheric muons in green, $\nu_\tau + \bar{\nu}_\tau$ CC in purple, $\nu + \bar{\nu}$ NC in brown, $\nu_e + \bar{\nu}_e$ CC in orange and $\nu_\mu + \bar{\nu}_\mu$ CC in blue. The number of data events is represented with black dots and errors bars. The data/MC ratio is shown for each feature on the bottom part of the figure.

## 5. Tau appearance sensitivity

After being completed, the full KM3NeT/ORCA detector is expected to measure 3000 $\nu_\tau$ per year. As a comparison IceCube DeepCore measured in 1022 days for their Analysis B, 934 $\nu_\tau + \bar{\nu}_\tau$ CC, 3368 $\nu_\mu + \bar{\nu}_\mu$ CC and 1889 atmospheric muons [5]. As the typical $\nu_\tau$ energy is close to 25GeV, above KM3NeT/ORCA energy threshold, it will already be sensitive to $\nu_\tau$ appearance at a primary stage of construction [6]. For the moment the $\nu_\tau$ are measured as an excess in the shower class. The RGS cuts presented in this work were optimised to define classes in order to have the highest possible sensitivity to $\nu_\tau$ appearance.

For each of the 3 classes defined previously the oscillation parameters are fitted in the 2D space of the reconstructed zenith angle and the reconstructed energy. The fit is accounting for various systematic effects reported in [1]. The $\nu_\tau$ normalisation is unconstrained in this study.

The sensitivity to tau appearance is shown in Figure 5. The comparison is made between the two separation methods discussed in this work, in the CC-only case where the $\nu_\tau$ normalisation affects only the events rates in the charged current interaction channel and in the CC+NC case where the $\nu_\tau$ normalisation affects event rates in both interaction channels. The sensitivity using both separation methods is similar, and indicates that the no $\nu_\tau$ hypothesis could be significantly rejected with this data set if data matches expectations.

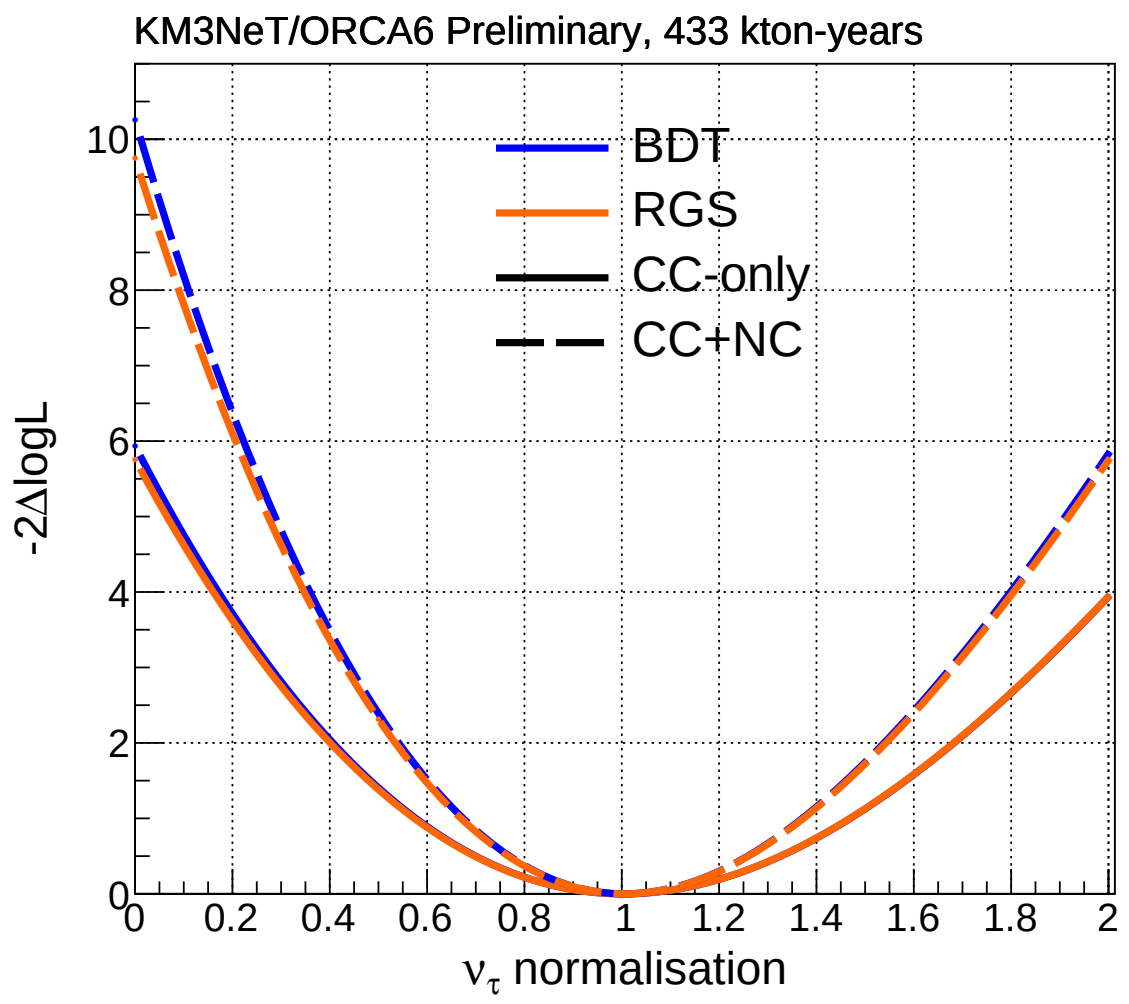

**Figure 5:** Sensitivity to $\nu_\tau$ normalisation corresponding to the ICRC23 433 kton-years data sample when using RGS (in orange) and BDT (in blue) methods for particle identification of tracks and s howers. The full line represents the $\nu_\tau$ normalisation affecting CC-only and the dotted line both CC+NC events.

Figure 6 contains the reconstructed L/E distribution which combines the two main observables, the reconstructed zenith angle and the reconstructed energy of the neutrinos. The L/E distribution offers a good visibility of the oscillation dip in both the Tracks and Showers c lasses. The Shower reconstruction is used for the Showers and Mixed classes and the Track reconstruction is used for the Tracks class from RGS and the Low/High Purity Tracks as well. The BDT classification is used on the left plots and the RGS classification for the right p lots. The top histograms show the BDT High Purity Tracks class/RGS Tracks class while the lower ones show the Showers class for both. The left histograms reveals the result of the fit with different hypothesis, free $\nu_\tau$ normalisation, fixed to 0 for the no-$\tau$ hypothesis and fixed to 1 for the nominal $\nu_\tau$ h ypothesis. The best fit lies in the middle of the two hypothesis 0 and 1. On the other side one can appreciate the L/E range between the 2 hypothesis when using the RGS classification without showing the best fi t. Data point are shown in blacks only for the official result, where the oscillation gap is clearly vi sible. More data will be needed to measure significantly the tau appearance, while high $\nu_\tau$ normalisation are already rejected by the measurement from KM3NeT [4].

From preliminary studies, using the RGS classification allows for a better agreement between the fitted model and the data, compared to the official BDT cla ssification. However, further studies are ongoing in order to investigate the differences between the two approaches and identify how to improve particle identification.

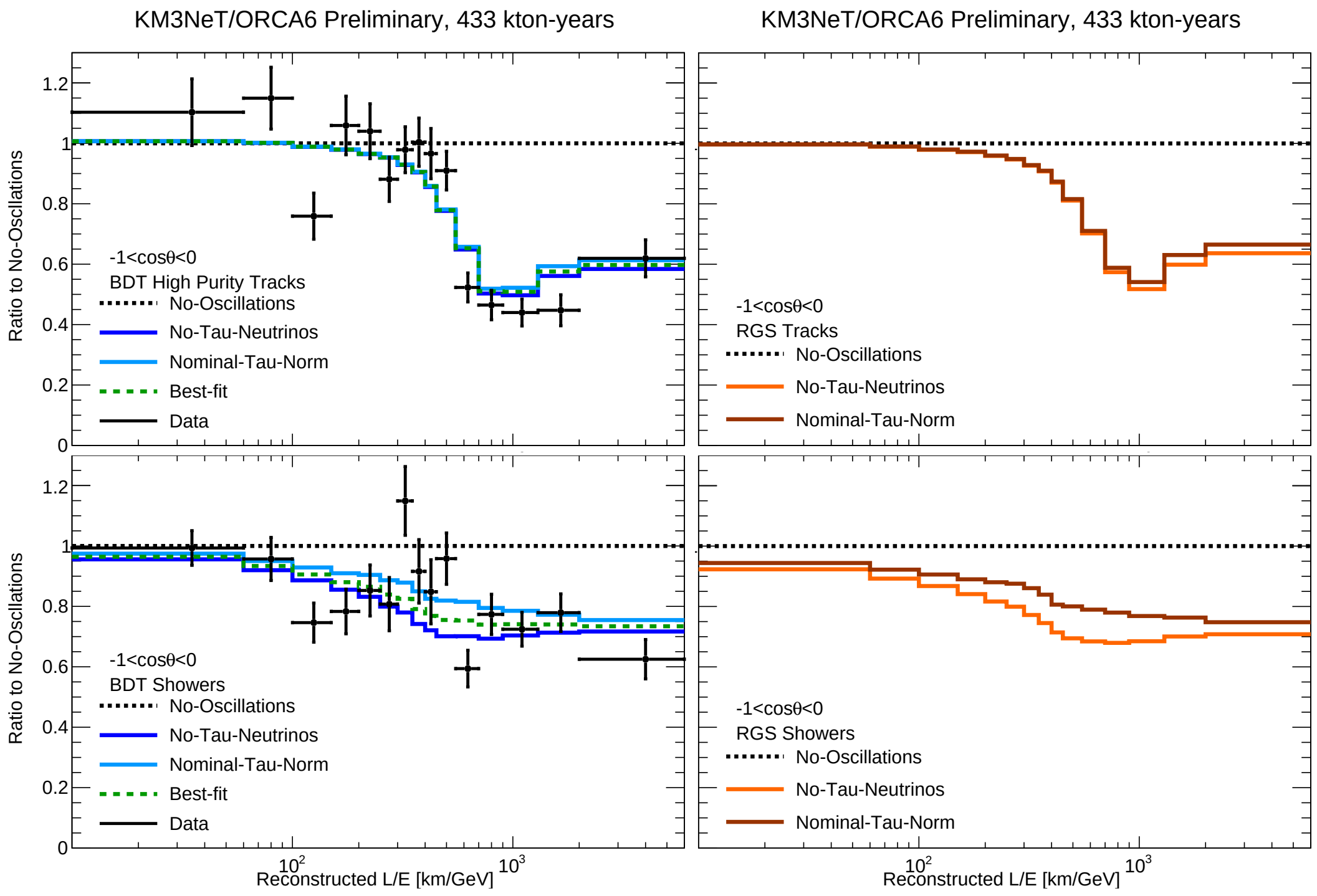

**Figure 6:** The L/E distributions of data and MC for different hypotheses are shown. The L/E using BDT classification is shown on the left side and the RGS classification on the right side. On the top the BDT High Purity Tracks class/RGS Tracks class is represented while on the lower part the BDT/RGS Showers classes are shown. One the left histograms the result of the fit with free $\nu_\tau$ normalisation is drawn in green, the no-$\tau$ hypothesis in blue and nominal $\nu_\tau$ normalisation hypothesis in cyan. In the right part the no-$\tau$ and nominal $\nu_\tau$ normalisation hypothesis are drawn in orange and brown respectively.

## 6. Conclusion

This work highlights the potential for a new particle identification method that relies on a few features and gives a similar sensitivity to $\nu_\tau$ appearance as the BDT classification. The good understanding of our data was demonstrated in the different comparisons between data and MC for the features involved in both methods. Many new possibilities in the particle identification will arise as the detector grows. One of them is the direct identification of $\nu_\tau$ CC in the Showers class.

# Selecting stopping muons with KM3NeT/ORCA

**Louis Bailly-Salins**[a,*] **on behalf of the KM3NeT collaboration**

[a] *Université de Caen Normandie, ENSICAEN, CNRS/IN2P3, LPC Caen UMR 6534, F-14000 Caen, France*

*E-mail:* baillysalins@lpccaen.in2p3.fr

The KM3NeT collaboration operates two water Cherenkov neutrino telescopes in the Mediterranean sea, ORCA and ARCA. The flux of atmospheric muons produced in cosmic ray air showers forms a background to the main objectives of KM3NeT/ORCA and KM3NeT/ARCA, respectively measuring atmospheric neutrino oscillations and detecting neutrinos from astrophysical sources. A small portion of the atmospheric muons stops inside the detector's instrumented volume. The stopping muons are 5% of the muons reconstructed using the 6 first strings deployed for ORCA. This still amounts to 1000 events per hour. We present two methods for selecting them, applied on both simulations and data. The first method uses simple cuts on a set of reconstructed variables. The second method uses a machine learning model to classify muons as "stopping" or "crossing". Both methods allow to reach a high selection purity, close to 95%. Detecting stopping muons can serve many purposes like studying muon decay via the detection of Michel electrons or estimating the flux of atmospheric muons at sea level. This work highlights the accurate reconstruction capabilities of ORCA. The median error on the reconstructed stopping point of selected muons is less than 5 meters, and the median angular deviation is 1°. This is to be compared with the 20 meters horizontal distance between strings and the 9 meters vertical distance between optical modules. Another important result is the excellent agreement between distribution of stopping muons selected in data and in simulations.

*Speaker

## 1. Introduction

The KM3NeT collaboration operates two water Cherenkov neutrino telescopes in the Mediterranean sea [1]. ORCA (*Oscillation Research with Cosmics in the Abyss*), at a depth of 2450 m, is designed to measure atmospheric neutrinos oscillations. ARCA (*Astroparticle Research with Cosmics in the Abyss*), at a depth of 3450 m, is designed to detect neutrinos from astrophysical sources. ARCA and ORCA share the same technology and detector elements. Both detectors are instrumented with photomultiplier tubes (PMTs) for the detection of Cherenkov light emitted along the path of the relativistic charged particles produced in neutrino interactions. KM3NeT detectors consist of a 3-D array of glass-spheres named Digital Optical Modules (DOMs) [2] housing 31 3-inch PMTs each. The DOMs are arranged along vertical lines called Detection Units (DU) carrying 18 DOMs each. Each DU is anchored to the seabed and remains vertical due to the buoyancy of the DOMs and of a buoy tied on its top. When completed, each detector "building block" will consist of 115 detection units arranged side by side following a cylindrical footprint. The vertical distance between DOMs and horizontal distance between DUs are different for the two detectors. ARCA is optimised to maximise its detection efficiency in the energy range 1 TeV-10 PeV, while ORCA has a denser PMT configuration to detect neutrinos in the range 1-100 GeV.

For both detectors, the vast majority of recorded events does not come from neutrinos but from atmospheric muons produced in cosmic ray air showers. Being deep underwater, the KM3NeT detectors have access to high-energy atmospheric muons [3]. Among the detected atmospheric muons, only a small portion stops inside the instrumented volume. For instance, the stopping muons are 5% of all the muons reconstructed using the 6 first strings deployed for ORCA (this partial configuration is identified in the following as ORCA6). This still amounts to more than 1000 events per hour, offering enough statistics for various purposes. First, the subsequent decay of stopping muons can be studied via the detection of Michel electrons. In addition, stopping muons can be used for tuning parameters related to the light output of the muon tracks in the simulations. Finally, reconstructing the direction and stopping point of a muon track allows to compute the range the particle has crossed through sea water, and thus to estimate the energy it had at sea level. Consequently, selecting stopping muons can lead to a measurement of the atmospheric muon flux at sea level on a given range of energy and direction (around 1 TeV for vertical muons). Given the high uncertainties in the modelling of cosmic ray-induced particle showers [4][5], specifically in the primary cosmic ray flux composition and in the hadronic interaction models, such model-independent measurement would be invaluable.

The first step to this measurement of the atmospheric muons flux at sea level with KM3NeT is to accurately select the muons stopping in the detector. This is what this contribution will describe for the ORCA6 detector.

## 2. Stopping muons in the KM3NeT detector

Of the two telescopes, ORCA is the one most suited for detecting stopping muons as it is more densely instrumented. The vertical distance between DOMs is around 9 m, and the horizontal distance between DUs is around 20 m. In the following, a stopping muon is defined as *a muon which stops inside the detector's instrumented volume*. As KM3NeT detectors do not have strict outer boundaries, there is some subjectivity in the definition of the instrumented volume. To stick to pre-existing conventions inside the collaboration, the instrumented volume is defined as *the smallest cylinder containing all the DUs*. Its limits are shown for ORCA6 on Figure 1.

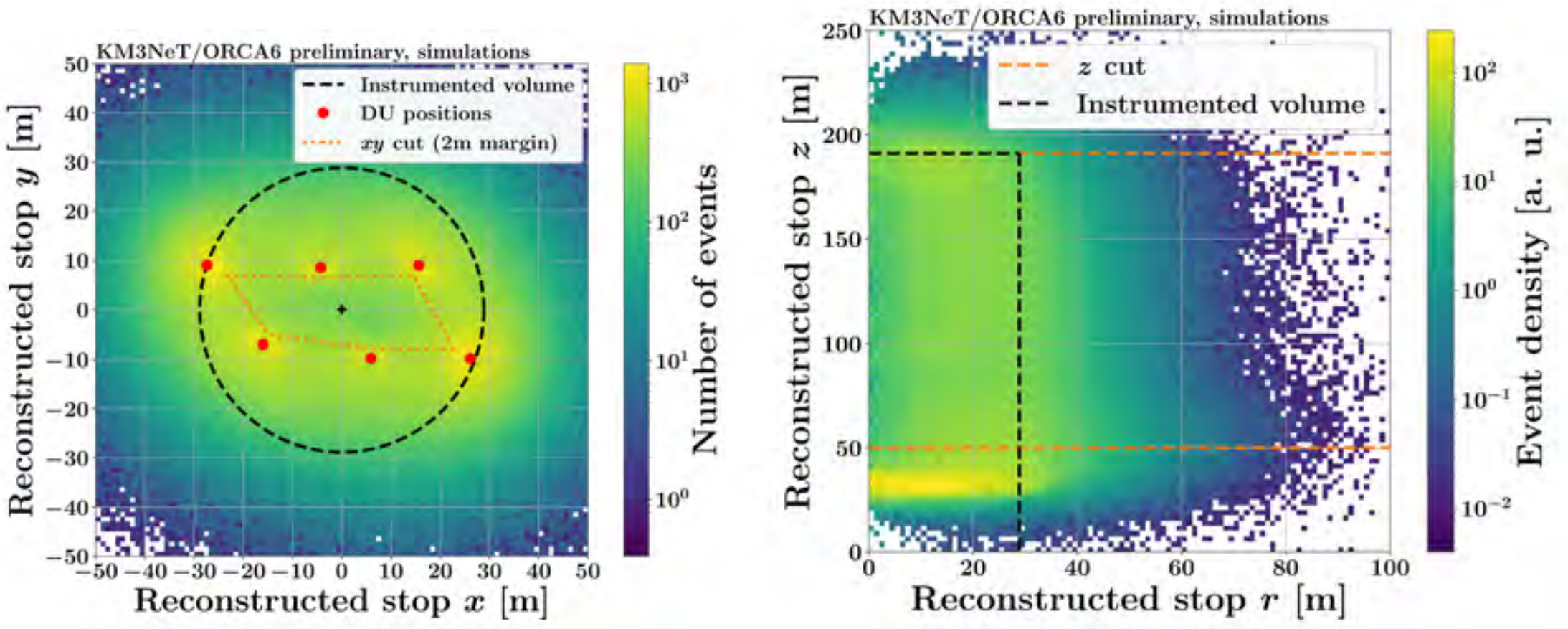

**(a)** Reconstructed stop positions in the horizontal $xy$ plane **(b)** Reconstructed stop positions in the vertical $zr$ plane

**Figure 1:** Distribution of the reconstructed stop position of all muon tracks reconstructed in ORCA6, showed with the geometrical cuts applied in the simple cuts approach.

Selecting stopping muons is more challenging than simply selecting the reconstructed muon tracks whose end point is inside the instrumented volume. Indeed, the vast majority of the muons seen by the detector do not stop inside the detector volume but keep traveling much further. Those non-stopping, or *passing* muons, represent about 95% of the muons reconstructed by ORCA6. Yet, many of the passing muons tend to have their stopping point reconstructed inside the instrumented volume, close to the DOMs (as seen on Figure 1a and Figure 1b), because the amount of light detected by the PMTs decreases quickly once the muons move away from the DOMs. Thus, a method to efficiently select true stopping muons, and reject the majority of passing muons, must be developed. Due to the abundant statistics, the methods developed to select stopping muons do not require a very high efficiency. The critical point is the purity of the selection. In the wait for an objective statistical criteria, which will come from further physics analysis, a purity close to 95% is judged satisfying for now.

## 3. Methods for selecting stopping muons

### 3.1 Simulation of atmospheric muons

The selection methods were developped using simulated atmospheric muons. The atmospheric muons Monte-Carlo (MC) simulations are done following a run-by-run approach in order to account for the variation of data taking conditions over time. For the generation of the atmospheric muon events at the surface of a virtual cylinder surrounding the active volume of the detector, the MUPAGE package [6] was used, with a parameter tuning specific to KM3NeT [7]. The propagation of generated muons and the PMT response are simulated with a custom KM3NeT software. The output of the simulation chain is a set of digitised PMT output pulses called hits. In addition to atmospheric muons, around 10% of the events in the data are noise events caused by random coincident hits on the PMTs. They mostly come from the decay of Potassium-40 in the sea water, and are simulated based on the coincidence rate in the data. Finally, the simulated events can be reconstructed under the track hypothesis with the same algorithms used for events in data.

### 3.2 Selection with simple cuts

The first approach to select stopping muons is to apply a set of cuts on chosen variables with a high separating power between stopping and passing muons. This is called the simple cuts (SC) approach. The cuts and their motivations can be summarized as follows:

- **Energy loss**: stopping muons are minimum ionizing particles (MIP), loosing energy only through ionization loss at a well-defined rate of around 0.25 GeV/m [8]. A cut on the ratio between the reconstructed energy and reconstructed track length is thus applied to remove muons which exhibit higher energy loss rates.
- **Quality of track reconstruction**: it is estimated through the likelihood resulting from the track fitting algorithms. To make that estimation independent of the length of the track, a useful variable is the normalized likelihood, obtained by dividing the likelihood by the number of hits used in the fit. As stopping muons have the end of their track contained in the instrumented volume, we expect the track reconstruction to be of good quality.
- **Number of hits**: As stopping muons have the end of their track contained in the instrumented volume, the number of hits they create on the PMTs should be higher than some threshold.
- **Reconstructed stopping point position**: the reconstructed stopping point should be inside the instrumented volume. But as many passing muons get reconstructed as stopping next to the DOMs, especially near the lowest DOMs (atmospheric muons are downgoing), the reconstructed stopping point is required to be found in an inner region of the instrumented volume. In the horizontal $xy$ plane the boundaries of this fiducial volume are 2 m from the polygon formed by the DU positions, as illustrated in Figure 1a . Along the vertical $z$ axis, to remove all the passing muons reconstructed as stopping near the lowest DOMs (at $z \approx 30$ m) only the stopping points reconstructed above $z = 50$ m are considered.
- **Zenith angle** $\theta$: only muons close to the vertical ($\cos(\theta) < -0.8$) are kept. This is because vertical muons have a longer track in the detector (due to the narrow geometry of ORCA6), making it easier to distinguish between passing and stopping muons.

With this set of cuts, the obtained performances are a **purity of 93%**[1] and an **efficiency of 5%**[2]. Although purity is satisfactory, these cuts (and the geometry of the detector) remove many true stopping muons, preventing access to stopping muons coming with a zenith angle greater than 37° ($\arccos(-0.8)$) from the vertical.

[1] The *purity* of the selection is defined as the percentage of true stopping muons among all those passing the cuts. In literature, this is sometimes referred to as the *precision* or *positive predictive value* $PPV = \frac{TP}{TP+FP}$ where $TP$ are true positive and $FP$ false positive.

[2] The *efficiency* of the selections is defined as the percentage of true stopping muons that get selected among all the true stopping muons reconstructed by the detector. In literature, this is sometimes referred to as the *sensitivity*, the *recall*, or *true positive rate* $TPR = \frac{TP}{TP+FN}$ where $FN$ are false negative.

### 3.3 Classification with a neural network

To improve efficiency and purity of the stopping muons selection, an approach based on machine learning was developed. A simple neural network (Multi-Layer Perceptron) was trained to classify muons as *stopping* or *passing*. The model which is currently used has 3 layers of 256, 128 and 64 neurons respectively. This makes a total of 48,450 trainable parameters. The input features used by the model come from the track reconstruction algorithms and no individual hit information is used, only summary information on the number of hits. 3 million simulated events were used to train the network, 750,000 to validate it during training, and 1 million to test it after training. To evaluate the dependency of the model on the training and testing sample, a 5-fold cross-validation was performed, showing very limited sample-dependency due to the sufficient sample size.

The model's output is a prediction for the muon's class under the form of a score between 0 and 1. For training, a score of 0 is associated to passing muons and a score of 1 to stopping muons. From that output score distribution, a stopping muon selection can be defined by choosing a cut value on the output score and keeping only the events with a higher score. All the possible purity and efficiency resulting from such a selection, with any chosen cut score value between 0 and 1, is shown on the precision-recall curve Figure 2. In the following, a cut value of 0.85 was chosen, as it allows to reach a **purity of 94%**. The resulting selection also reaches an **efficiency of 25%**. Thus, the machine learning approach offers the possibility to increase the efficiency by a factor 5 with respect to the simple cuts selection, while simultaneously increasing a bit the purity. Most importantly, the ML classification approach allows to access the full zenith angle range while the simple cuts selection was limited to vertical muons.

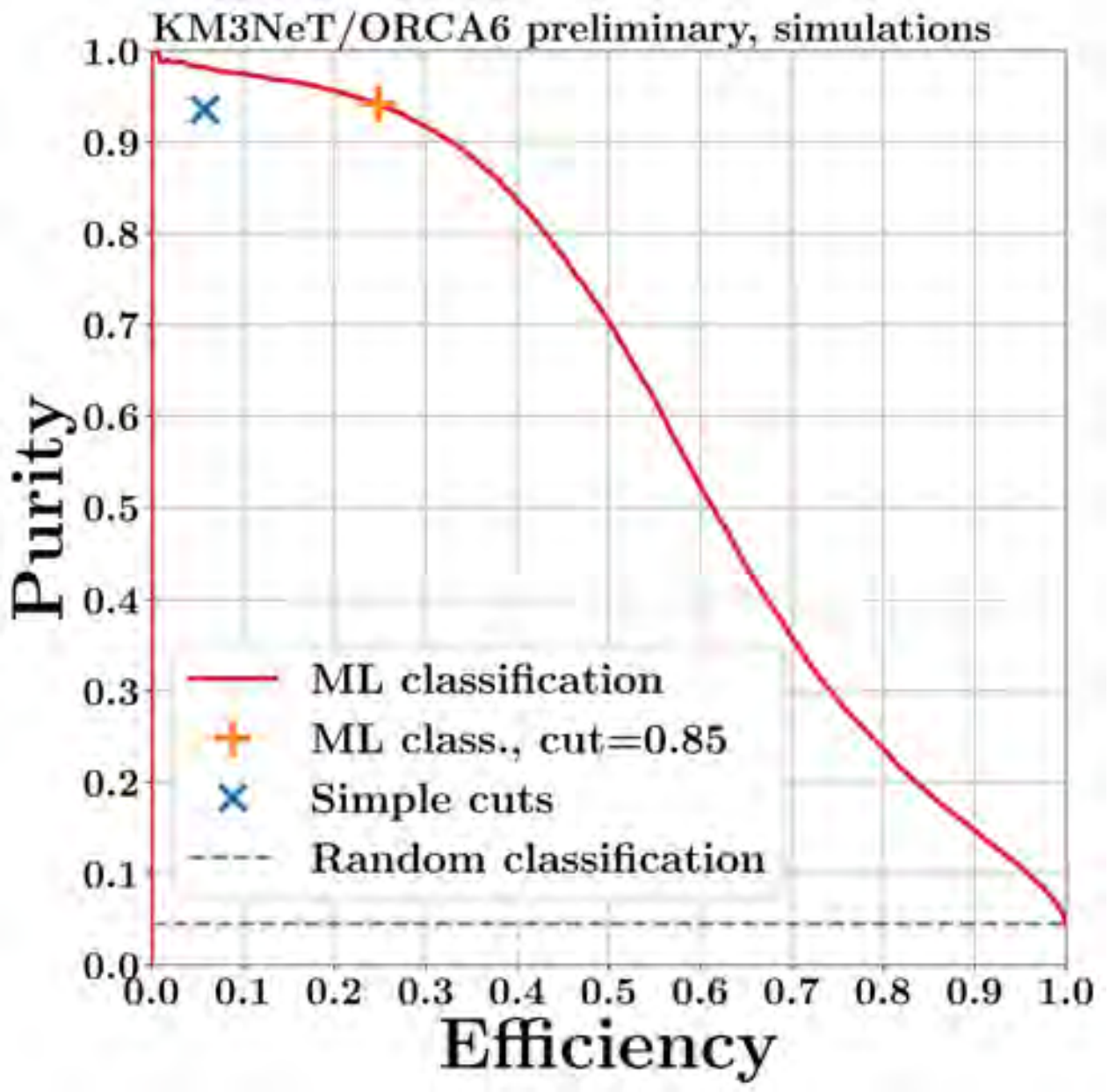

**Figure 2:** Precision-Recall curve of the machine learning classification. The performances of the two selections (ML classification with the specific cut value chosen and simple cuts) are also shown.

### 3.4 Comparison of the two selections

Figure 3 shows the comparison of purity and efficiency obtained using the two selections described previsouly as a function of the reconstructed zenith angle and the $z$ coordinate (the height with respect to the seabed) of the reconstructed stopping point. Figure 3b shows that the purity and efficiency of the two selections are close to being constant through most of the detector's height, with a loss of performance at the upper and lower limits of the instrumented volume. Figure 3a shows that the ML classification's selection efficiency decreases with the zenith angle. A possible explanation is that due to the narrow geometry of ORCA6, muons with more horizontal trajectories tend to cross a smaller portion of the instrumented volume. The simple cuts selection sees its purity quickly worsening for non-vertical muons, hence the cut at $\cos(\theta) = -0.8$.

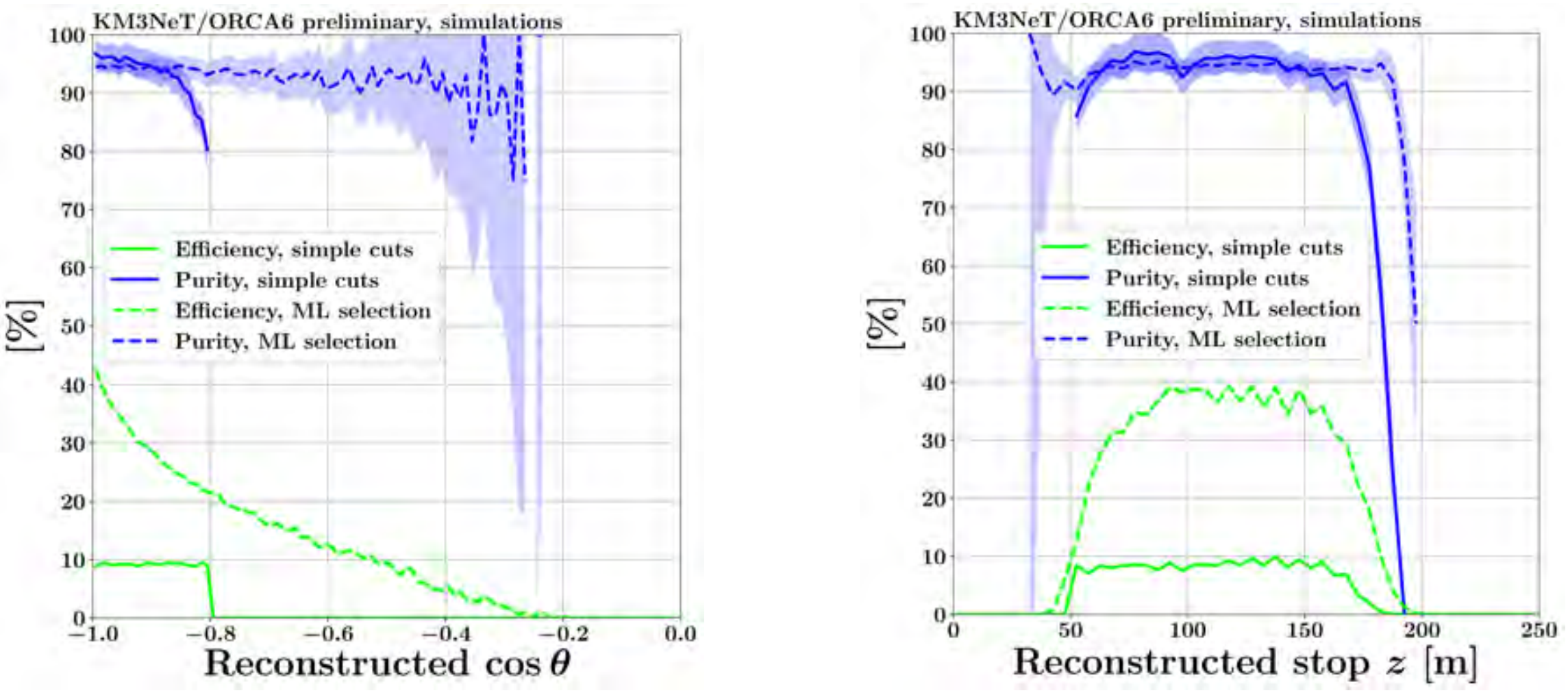

**(a)** Selection performances as a function of the reconstructed zenith angle

**(b)** Selection performances as a function of the $z$-coordinate of the reconstructed stopping point

**Figure 3:** Performance comparison between the simple cuts and machine learning-based selections.

## 4. Resolution

| Variable | Median value | |
|---|---|---|
| | Simple cuts | ML classification |
| Angular deviation [°] | 0.81 | 1.09 |
| $\Delta SP_{true/reco}$, total [m] | 3.9 | 4.9 |
| $\Delta SP_{true/reco}$, perpendicular to track direction [m] | 0.5 | 0.7 |
| $\Delta SP_{true/reco}$, parallel to track direction [m] | +3.8 | +4.6 |

**Table 1:** Median resolution values for both selections.

From selected stopping muons, the reconstruction performances of ORCA6 can be evaluated by looking at the error on the reconstructed stopping point and on the reconstructed direction. The very accurate reconstruction capabilities of ORCA6 should be highlighted: the median angular deviation on the reconstructed tracks is around 1° and the median error on the reconstructed stopping point is less than 5 meters, for both stopping muon selection methods. Table 1 shows more detailed numerical information. In this table $\Delta SP_{true/reco}$ refers to the error made on the reconstruction of the stopping point, overall or projected in the reference coordinate system of the track.

The plus signs in last line of Table 1 mean that the muon tracks tend to be reconstructed as stopping too early along the track direction. In other words, the track reconstruction algorithms miss the last 4 to 5 meters of the tracks in median. Still, it should be highlighted that these resolution values are small compared to the spacing between DOMs (9 m in the vertical direction and 20 m in the horizontal direction). Additionally, studies using stopping muons to determine the atmospheric muons flux at sea level are ongoing. In that context, it must be mentionned that the relative uncertainty on the distance a stopping muon has traveled in the sea water coming from the misreconstruction of track direction and stopping point is negligible (0.1% to 1%).

## 5. Data-MC comparison

Figure 4a shows the resulting zenith angle distribution of stopping muons from applying both selections on both MC simulations and data, for a livetime of 32.1 days (runs randomly selected over the ORCA6 operating period). The distribution of all reconstructed events, without selection, is also shown for reference. The data/MC ratio for stopping muons is very close to 1.0 overall (0.99 for simple cuts, 0.97 for ML classification) and stable over most of the phase space. Figure 4b also shows that the ratio is 1 through most of the detector's height. This excellent agreement validates the stopping muons selections, as well as the simulations.

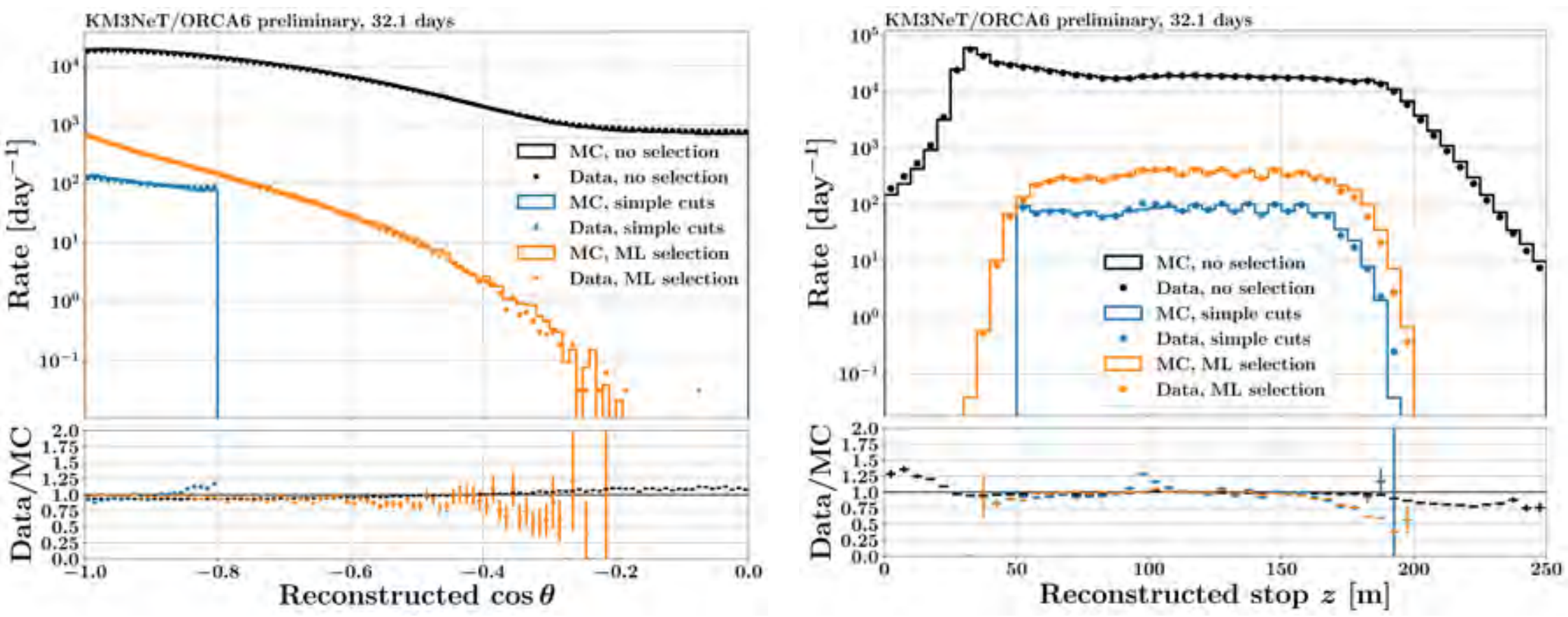

**(a)** Zenith angle distribution **(b)** Stopping point altitude distribution

**Figure 4:** Data/MC comparison for all muons and for the stopping muons selected with both methods.

## 6. Conclusion

Two methods for selecting the muons stopping inside the instrumented volume of the KM3NeT/ORCA detector with 6 detection units have been developed. The machine learning-based method allows to select five times more stopping muons than the simple cuts method with very low levels of contamination from passing muons. It also gives acces to the full range of zenith angle.

A study on the angular resolution of the selected stopping muons confirms the very good direction reconstruction capabilities of ORCA, with a median angular deviation of 1°, and shows that the median error on the reconstructed stopping point is less than 5 meters.

A very good agreement between data and simulations can be reached using both stopping muons selections, validating at the same time the selection methods and the simulations.

This preliminary study paves the way for further use of stopping muons. It must be highlighted that the results presented here were obtained with a detector made of 6 detection units only. Future configurations of ORCA, with more DUs, and a more circular footprint, will improve the resolution and the performances of the stopping muon selection methods. Work is already ongoing to use stopping muons to measure the flux of atmospheric muons at sea level on a given range of energy and zenith angle (above 1 TeV for vertical downgoing muons). Another ongoing study measures the subsequent decay of stopping muons.

# Comparison of the atmospheric muon flux measured by the KM3NeT detectors with the CORSIKA simulation using the Global Spline Fit model

**Andrey Romanov**[a,b,*] **and Piotr Kalaczynski**[c,d] **on behalf of the KM3NeT Collaboration**

[a] *Dipartimento di Fisica dell'Università di Genova, Via Dodecaneso 33, 16146 Genova, Italy*
[b] *INFN - Sezione di Genova, Via Dodecaneso 33, 16146 Genova, Italy*
[c] *National Centre for Nuclear Research, 02-093 Warsaw, Poland*
[d] *Nicolaus Copernicus Astronomical Center Polish Academy of Sciences, Particle Astrophysics Science and Technology Centre, Rektorska 4, Warsaw, 00-614 Poland*

*E-mail:* aromanov@ge.infn.it, pkalaczynski@camk.edu.pl

Atmospheric muons are the dominant component of the down-going events for the KM3NeT neutrino telescopes. Deep underwater measurements of muons provide important information about the cosmic ray properties. The KM3NeT research infrastructure includes two telescopes currently in operation while still being under construction in the Mediterranean Sea. The KM3NeT/ORCA detector is deployed at 2450 m depth near Toulon, France. The KM3NeT/ARCA telescope is located at 3500 m depth off-shore Capo Passero, Italy. In this work, the measured atmospheric muon flux is compared to the Monte Carlo simulation using the CORSIKA package with the Sibyll 2.3d model for high-energy hadronic interactions and the GSF model for mass composition. The data from both KM3NeT/ORCA and KM3NeT/ARCA telescopes are considered for this analysis. In the current configuration, KM3NeT/ORCA covers cosmic ray energy range from several TeV up to hundreds of TeV per nucleon, while for KM3NeT/ARCA the range is from several TeV up to PeV per nucleon. Systematic uncertainties considered for the analysis include that on the cosmic ray flux normalization and its composition, water properties, detector response, and high-energy hadronic interaction models.

*Speaker

## 1. Introduction

The KM3NeT research infrastructure comprises two neutrino telescopes at the bottom of the Mediterranean Sea [1]. The KM3NeT/ARCA telescope is being constructed off-shore the coast of Sicily, Italy, at a depth of ~3.5 km. Its primary scientific aim is to study High-Energy (HE) cosmic neutrinos in the TeV-PeV range. The KM3NeT/ORCA detector has a smaller and denser configuration with respect to KM3NeT/ARCA since its primary goal is to investigate atmospheric neutrino oscillations and neutrino mass hierarchy which requires a lower energy threshold (GeV) for neutrino detection. The telescope is located around 40 km away from Toulon, off the coast of France, at ~2.5 km depth. An array of 115 lines is called a building block. The final configuration of the KM3NeT/ARCA telescope will comprise two building blocks and KM3NeT/ORCA will consist of one such block. The one building block configuration is called ARCA115 and ORCA115 in the following. The analysis presented in this work was performed with data taken by the KM3NeT/ARCA and KM3NeT/ORCA telescopes with six working lines each. In the following, those configurations are denoted as ARCA6 and ORCA6.

This contribution aims to compare the atmospheric muon flux measured by ARCA6 and ORCA6 detectors to the MC simulation performed with the CORSIKA package [2] that includes the Sibyll 2.3d model [3] for description of HE hadronic interactions and the Global Spline Fit (GSF) [4] as model of the mass composition of Cosmic Ray (CR) flux.

## 2. Simulation of atmospheric muons in KM3NeT

The simulation of atmospheric muons for the KM3NeT experiment starts in the upper layers of the atmosphere and ends deep underwater. The first step is to simulate the interactions of the primary CRs with the air nuclei and the subsequent Extensive Air Shower (EAS) development. This step is performed with the CORSIKA [2] v.7.741 package.

The minimum muon energy required to reach the top part of the KM3NeT/ORCA detector (~2 km) is around 500 GeV [5]. Hence, the lower limit on the primary energy was set to 1 TeV per nucleon in the simulation. Five nuclei were used as primaries in the simulation: proton, helium, carbon, oxygen, and iron. Other primaries were taken into account by enlarging the flux weights of C, O, and Fe according to the flux of nuclei missing in the s imulation. The CR mass composition model that was used in this work is GSF [4].

The plots in Fig. 1 show the sea level flux of muon bundles that reach ORCA6 and ARCA6 detectors as a function of the bundle energy. The muon bundle energy is the sum of muon energies at sea level originating from one shower for those muons that reach the detector. The highlighted area indicates the 90% fraction of events counting from the maximum of the distribution. The energy range of the fraction spans from 0.8 TeV to 10 TeV for ORCA6 and from 1.1 TeV to 34 TeV for ARCA6. The upper limit arises due to the fast decrease of the CR flux with energy.

Therefore, the sea level energy of muons detectable by the KM3NeT experiment lays in the TeV range. This muon energy is about 3 orders of magnitude higher than for the muons detected in EAS experiments. Thus, the KM3NeT measurement is complementary to the investigations of the so-called muon puzzle [6], i.e. deficit of GeV muons detected at the ground.

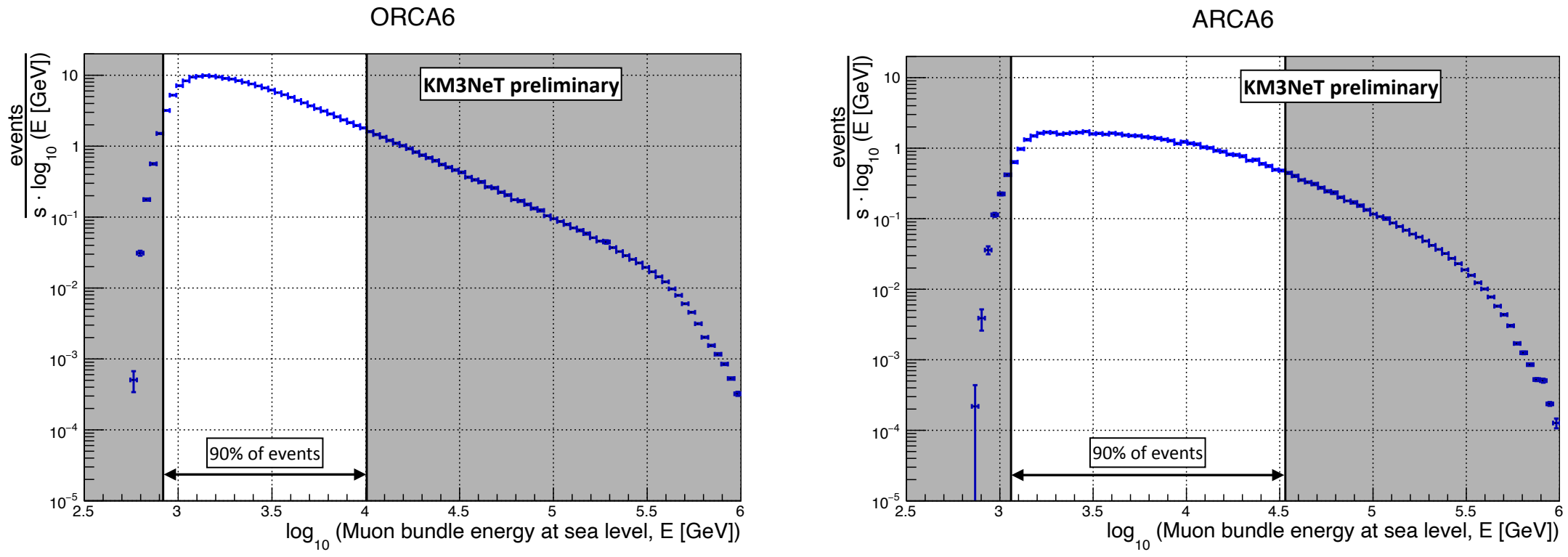

**Figure 1:** Sea level rate of generated muon bundles that reach the depth of the KM3NeT telescopes as a function of bundle energy. The left (right) plot shows the range for ORCA6 (ARCA6) detector.

The HE hadronic interaction model used in the simulation is Sibyll 2.3d [3]. Differences in the HE muon flux induced by choosing the different HE hadronic interaction models are treated as systematic uncertainties.

Fig. 2 shows pseudorapidity of muons reaching ORCA6 and ARCA6 detectors. The pseudorapidity is defined as $\eta = -\ln[\tan(\theta/2)]$, where $\theta$ is the angle between the primary particle and secondary muon at sea level. The peak of the distributions is located at $\eta \approx 9$. Therefore, muons seen by KM3NeT detectors originate from hadronic interactions laying in the very forward region of pseudorapidities. This region is not fully covered by accelerator experiments [6]. Hence, the hadronic interaction models that aim to describe the CR interactions that are seen by the KM3NeT detectors rely necessarily on the extrapolations of experimental results.

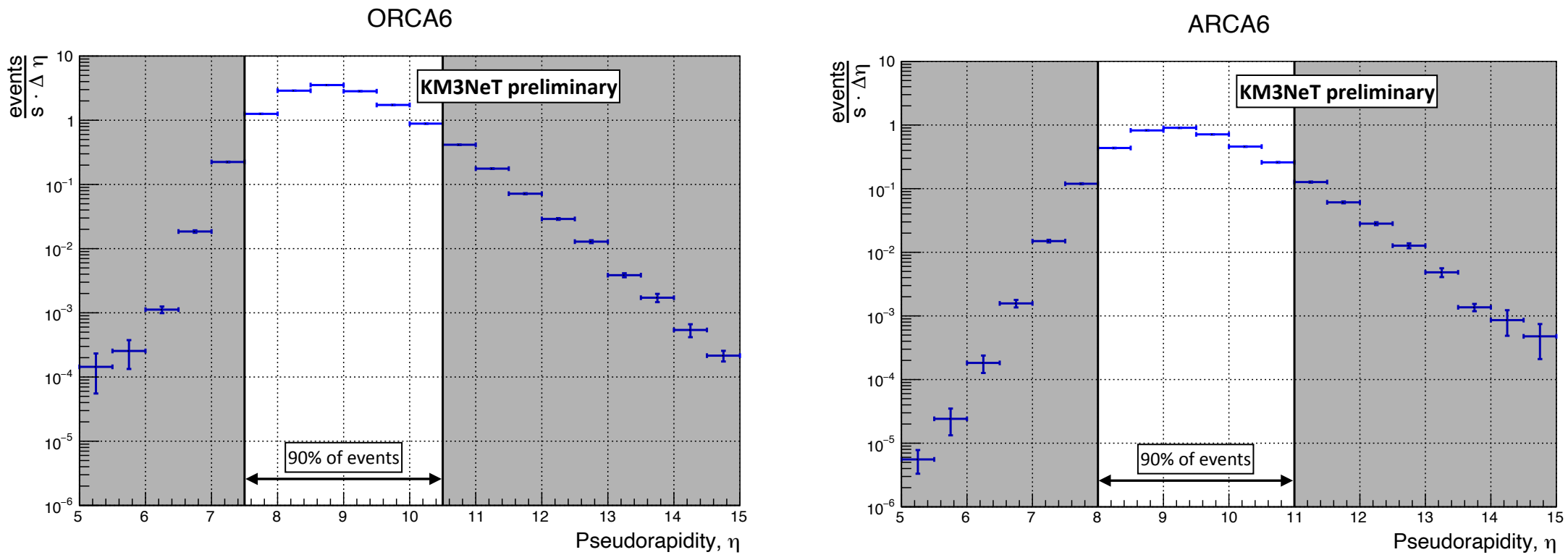

**Figure 2:** Pseudorapidity distribution of muons reaching ORCA6 (left plot) and ARCA6 (right plot) detectors. The highlighted area shows the energy range that includes a 90% fraction of events.

The propagation of muons in water down to the KM3NeT detectors is performed with the gSeaGen code [8] using PROPOSAL [9] as a muon propagator.

Simulation of the Cherenkov radiation induced by muons, its detection by the optical modules, and the muon track reconstruction is performed with the internal KM3NeT software.

The top part of Fig. 3 shows the rate of reconstructed events counted at ORCA6 and ARCA6 detectors as a function of the true primary nucleus energy. 90% fraction of the total number of events corresponds to the energy range from 3 to 316 TeV for ORCA6 and from 5 TeV to 1.3 PeV for ARCA6. The same distributions when considering ORCA115 and ARCA115 telescopes are presented in the bottom part of Fig. 3. The energy ranges for the these detector configurations are 3 TeV - 250 TeV for ORCA115 and 6 TeV - 1 PeV for ARCA115. For both detectors, the events mostly originate from p and He primaries.

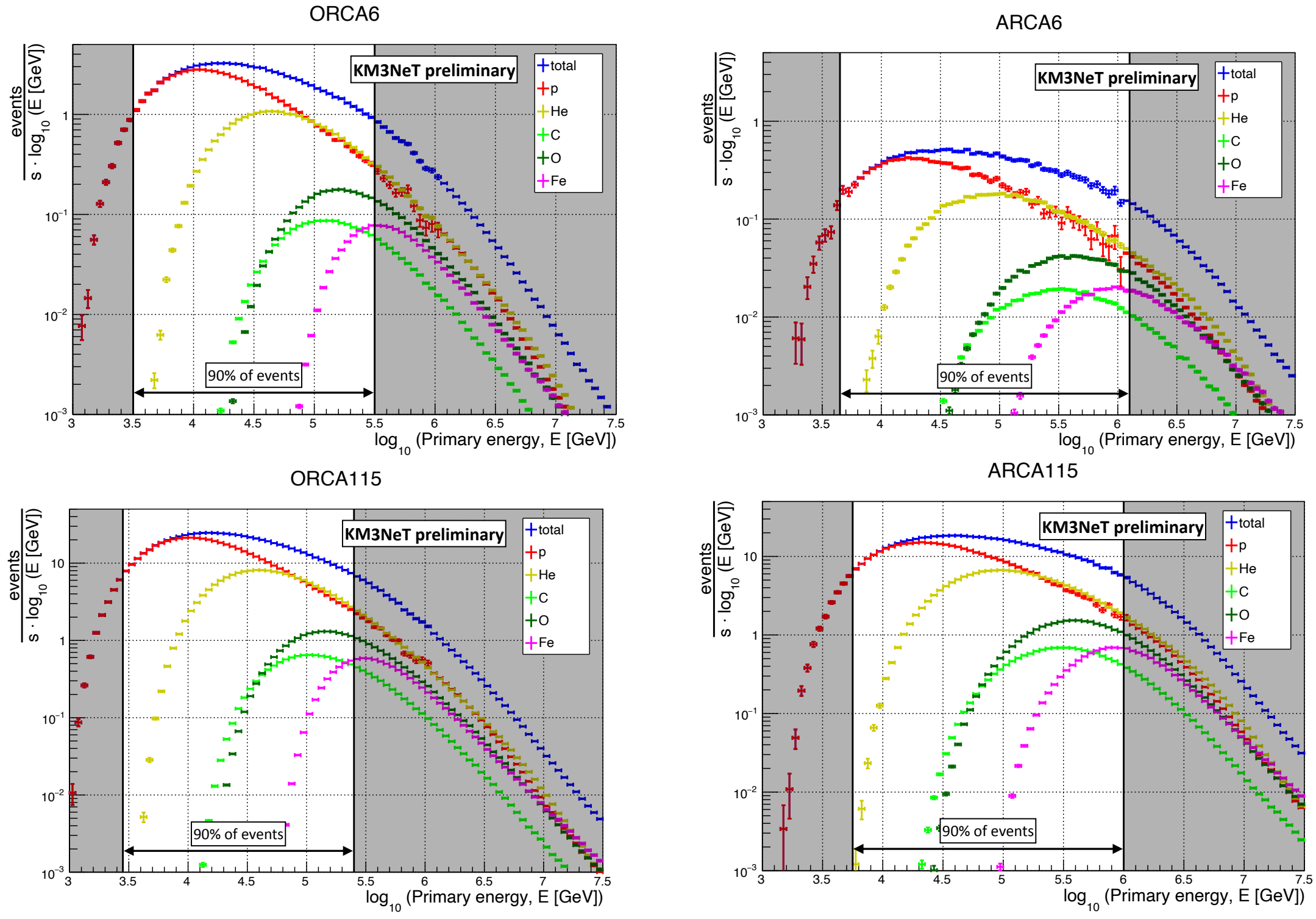

**Figure 3:** Rate of reconstructed atmospheric muon events expected with ORCA6 (top left plot), ARCA6 (top right plot), ORCA115 (bottom left plot), and ARCA115 (bottom right plot) detectors as a function of the primary energy. The total rate is shown in blue, the rates from p, He, C, O, and Fe primaries are shown as red, yellow, light green, dark green, and purple points, correspondingly.

## 3. MUPAGE tuning on CORSIKA

CORSIKA provides atmospheric muons at sea level, which can be propagated till the detector, performing the full MC simulation of EAS development through the atmosphere. The main drawback of this approach is high CPU time.

To reduce CPU time requirement, the simulation of atmospheric muons in KM3NeT is based on the fast MC generator MUPAGE [10]. It generates the muon bundle kinematics features at a certain sea depth and zenith angle based on parametric formulas. The formulas describe the flux of the single- and multi-muon bundles, the differential energy spectrum, and the lateral distance of muons from the bundle axis. Values of the parameters [11] were obtained starting from a full MC simulation performed with the HEMAS package [12] and fitting the results to MACRO [13] measurements.

With the aim of merging the advantage of a quick parameterized simulation and the details coming from the CORSIKA full simulation associated to the most recent physics models, both for hadronic interaction description (Sibyll 2.3d [3]) and for mass composition (GSF [4]), a framework was developed to adjust the MUPAGE parameters in order to reproduce CORSIKA expectations. The MUPAGE code and the parametric formulas themselves were not modified. Fig. 4 shows the muon zenith flux distributions expected with ORCA6 and ARCA6 detectors resulting from the CORSIKA full simulation and from MUPAGE with the nominal parametrization and with the modified MUPAGE tuned on CORSIKA. The distributions obtained with the *tuned* MUPAGE agree with what obtained with CORSIKA full MC within statistical fluctuations for both ORCA6 and ARCA6 detectors. Hence, the *tuned* MUPAGE can be used as a fast proxy of CORSIKA in order to obtain results that are similar to simulations with CORSIKA + Sybill 2.3d and GSF models.

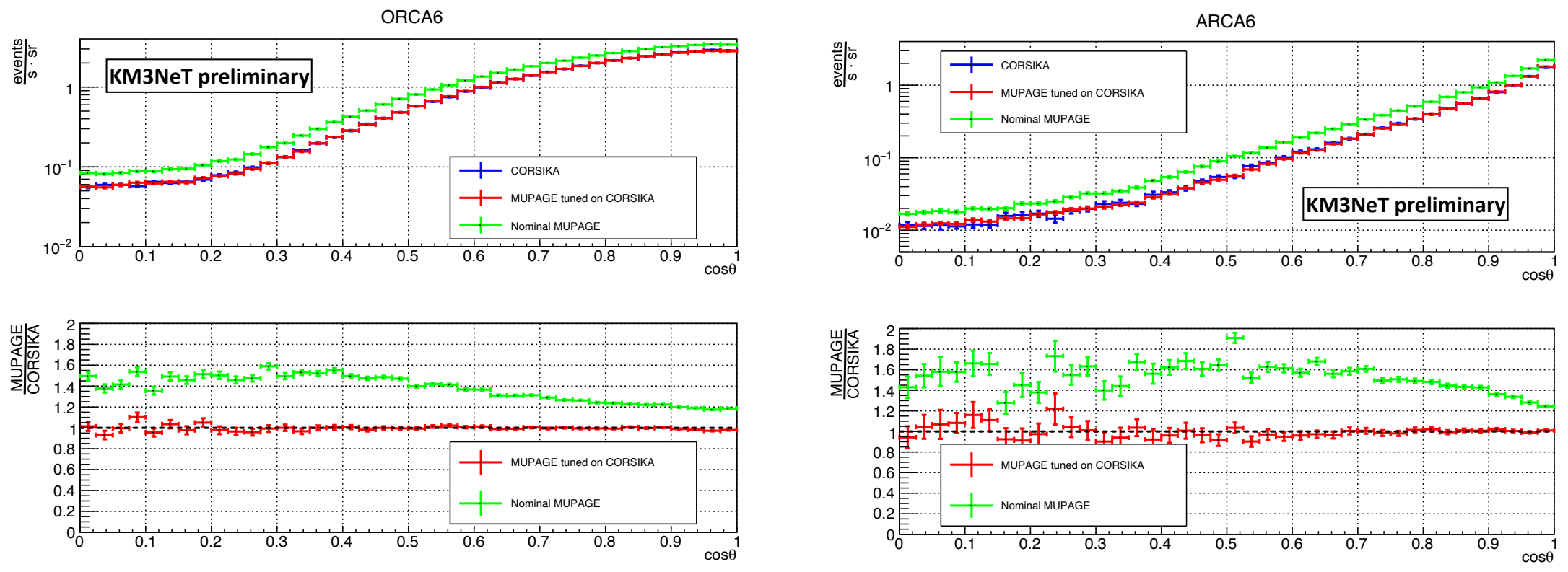

**Figure 4:** Rate of reconstructed atmospheric muons for ORCA6 (left plot) and ARCA6 (right plot) detectors as a function of the zenith angle. The blue (red) points represent the CORSIKA simulation. The nominal MUPAGE is shown as green points and the MUPAGE tuned on CORSIKA is shown in red. The ratios between MUPAGE and CORSIKA are on the bottom plots.

## 4. Reconstruction of muon arrival direction

To evaluate the direction reconstruction capabilities, the true and the reconstructed muon zenith angles, $\theta$, were used. The true zenith angle is the one of the muon bundle at the detector can, all muons are assumed to be collinear with the bundle axis in MUPAGE. The reconstructed angle is instead obtained with the KM3NeT reconstruction algorithm. In the left plot of Fig. 5, the MC true $\cos\theta$ distribution (blue points) is compared to the reconstructed one (red points) for ORCA6 detector. The discrepancy starts to emerge for events with $\cos\theta < 0.5$. The reason for this discrepancy is that the flux dependence on $\cos\theta$ is very steep and few well-reconstructed events at $\cos\theta < 0.5$ are dominated by a fraction of mis-reconstructed vertical muons. The KM3NeT angular resolution is at sub-degree level, so the fraction of mis-reconstructed events is small, however the number of vertical muons is several orders of magnitude higher. Hence, it was decided to consider only muons with $\cos\theta > 0.5$ for the final data/MC s tudies. The same plot but for ARCA6 telescope is shown in the right plot of Fig. 5. Muons with $\cos\theta > 0.6$ are considered well reconstructed.

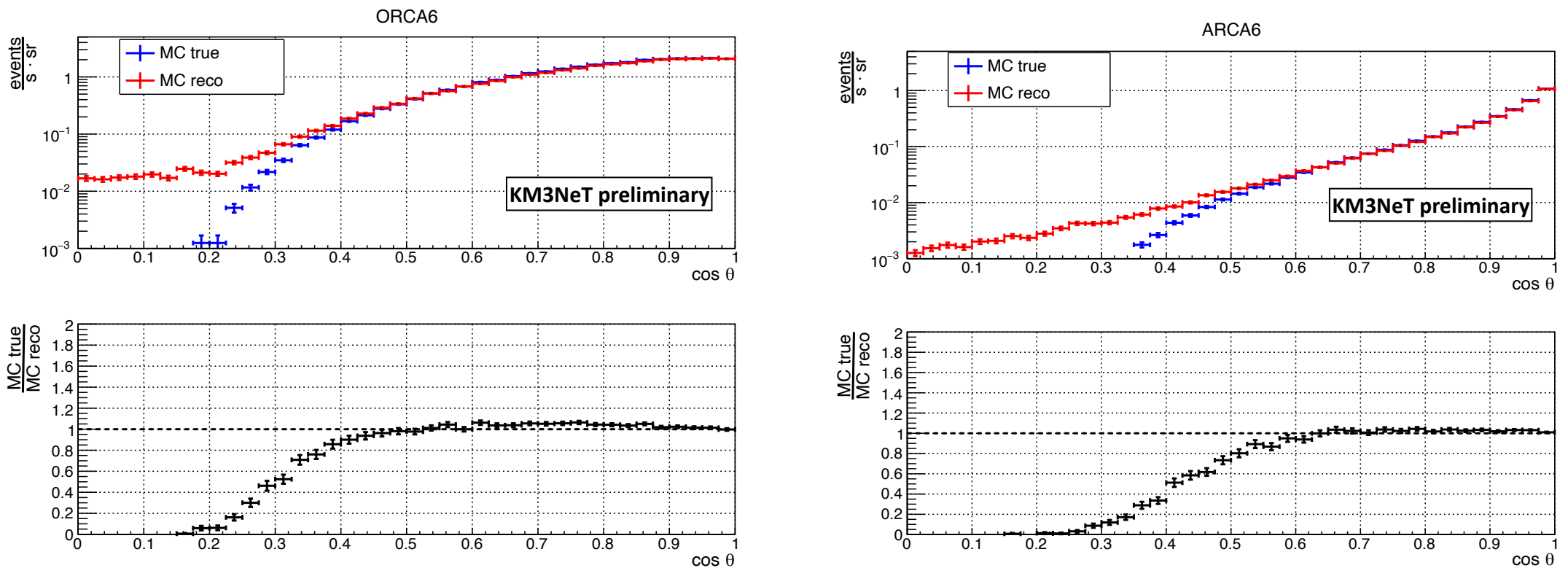

**Figure 5:** Distribution of the true muon bundle zenith angle at the can (blue points) compared to the reconstructed zenith angle (red points) for ORCA6 (left plot) and ARCA6 (right plot) detectors. The bottom plot shows the ratio of the distributions.

## 5. Data and MC with systematics

Systematic errors considered in this work include uncertainties on CR flux, light absorption length in seawater, PMT quantum efficiency, and HE hadronic interaction model.

To estimate the uncertainty on the water absorption length and PMT efficiency, ±10% variations were assumed for both quantities. The uncertainty on CR flux is evaluated using the data provided in the GSF model. The HE hadronic interaction uncertainties are estimated using the post-LHC models available in the MCEq software [14].

The final result of this analysis is the comparison of KM3NeT data with MC simulation including all systematic uncertainties mentioned above. Data and MC comparison results together with associated uncertainties are shown in Fig. 6 for ORCA6 and ARCA6 detectors. The plots below are obtained with the quality cuts on the reconstruction applied in order to remove background ($^{40}$K decays and bioluminescence).

The MC simulation underestimates data for both ORCA6 and ARCA6 detectors. The discrepancy for ORCA6 goes beyond the uncertainties considered in this work. The ratio between data and simulation is flat in the considered zenith angle range. There are ∼40% more muons in data with respect to simulation.

The shape of the ARCA6 data/MC ratio is not flat in contrast to the ORCA6 result. One of the possible explanations for the non-flat ARCA ratio and the mismatch in the ratio values between the detectors could be related to the detector response simulation uncertainties, and in particular, water properties.

Given a discrepancy between KM3NeT data and MC simulation for the muon zenith distribution underwater, it is important to investigate if it is also present for the sea level flux of HE muons. Fig. 7 shows the sea level muon flux resulting from the CORSIKA simulation, which used for the MUPAGE tuning, compared to the real data from the ground-based experiments and to the analytical models. The x-axis range of the figure covers the 99% fraction of single muon events detected by ORCA6 and ARCA6 telescopes. The data points shown in the figure are from the experiments mentioned in the overview paper [15] and from the L3+C experiment [16]. The analytical models considered are by T. Gaisser [17] and E. V. Bugaev [18]. Also, the fit of the MACRO data was added to the plot [19].

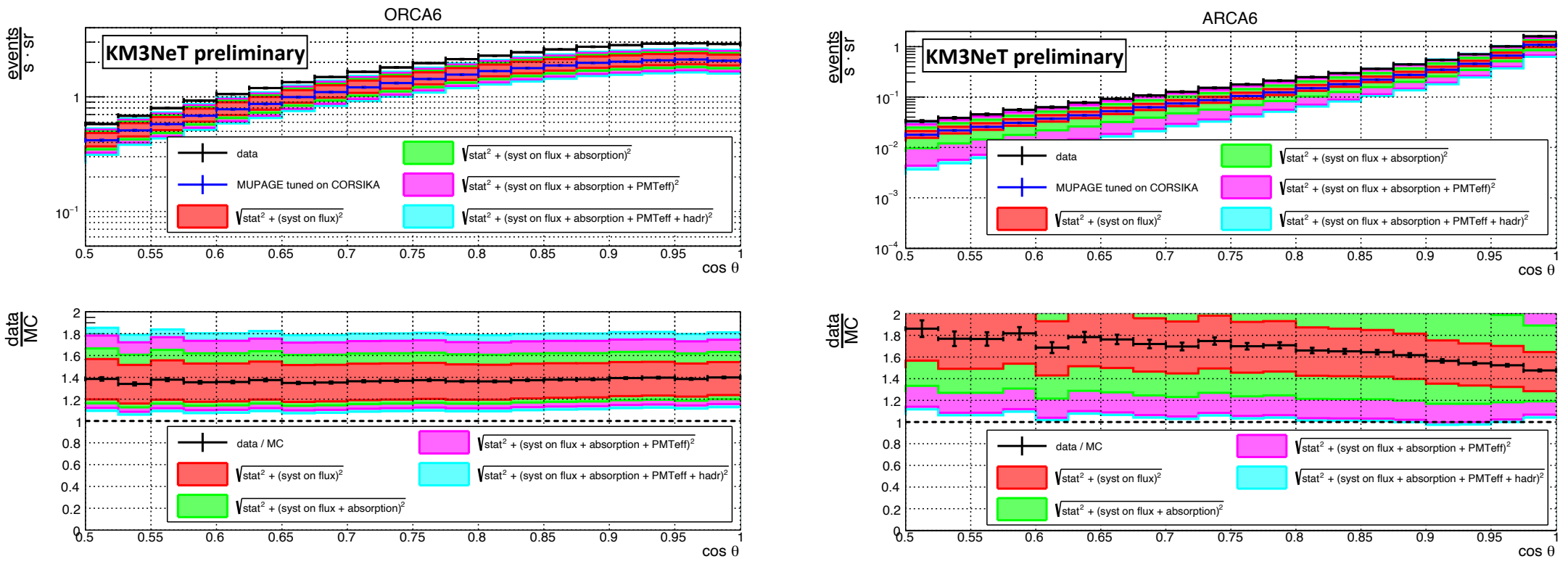

**Figure 6:** Comparison of the data (black points) and MC simulation results (blue points) for the atmospheric muons in terms of the zenith angle distribution for ORCA6 (left plot) and ARCA6 (right plot) detectors. The simulation includes statistical uncertainties that are shown as black vertical lines. The systematic uncertainties are shown as cumulative bands that include uncertainty on the CR flux (red band), the light absorption length (green band), the PMT efficiency (purple band), and the hadronic interaction models (light blue band).

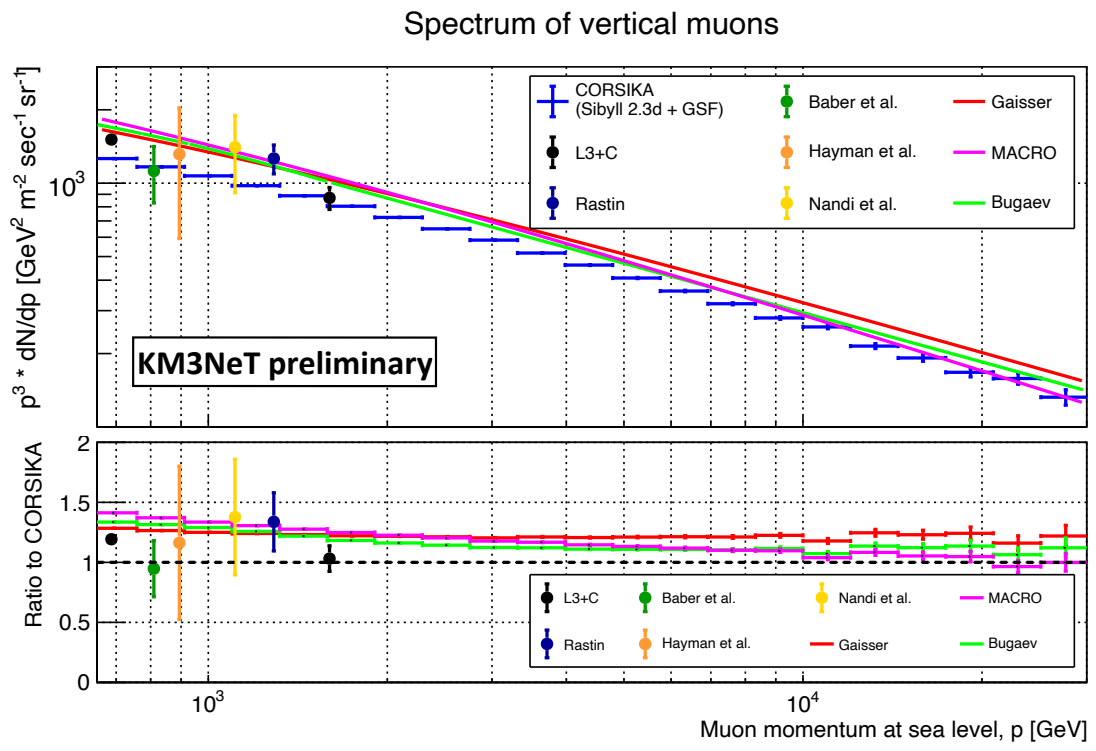

**Figure 7:** The sea level flux of muons from the CORSIKA simulation used in this work (blue points) compared to two analytical models, Gaisser [17] (red line) and Bugaev [18] (green line), to the fit of the MACRO data [13] (purple line), and to the experimental results from the ground-based experiments [15, 16]. The bottom plot shows the ratio of models and data to CORSIKA.

In general, CORSIKA expectations underestimate the sea level muon flux with respect to considered models. The discrepancy is at a level of 30%. All data points except for two exceed the CORSIKA predictions by ~20-30%.

The discrepancy between data and simulation for the underwater muon flux seen by ORCA6 telescope is around 40% as shown above. That indirectly confirms that the KM3NeT simulation describes the muon propagation in water, the light generation, the detector response, and the muon reconstruction with a precision better than 10%. The 10% disagreement between the simulation and measurements at sea level and underwater may be explained with uncertainties on light attenuation length in seawater and detector response simulation, shown as green and purple bands in Fig. 6.

The sum of the two aforementioned upper uncertainties is around 10% for the ORCA6 telescope and 20-40% for the ARCA6 detector that covers the difference.

The models used in this analysis include the post-LHC hadronic interaction model Sybill 2.3d and the GSF CR flux model which was fitted on the most recent direct CR measurements. The models provide a ~40% deficit in TeV muons with respect to the KM3NeT data. We hope this additional measurement provides new insights and the test-bench for possible solutions to the muon puzzle.

This contribution is supported by IRAP AstroCeNT (MAB/2018/7) funded by FNP from ERDF.

# Seasonal variation of the atmospheric muon flux in the KM3NeT detectors

**J. M. Mulder**[a,*] **and R. Bruijn**[a,b] **for the KM3NeT Collaboration**

[a] *University of Amsterdam, Institute of Physics/IHEF, PO Box 94216,Amsterdam, 1090 GE Netherlands*

[b] *Nikhef, National Institute for Subatomic Physics, PO Box 41882, Amsterdam, 1009 DB Netherlands*

*E-mail:* J.Mulder@nikhef.nl, rbruijn@nikhef.nl, km3net-pc@km3net.de

KM3NeT is constructing two large volume water Cherenkov detectors in the Mediterranean Sea. By instrumenting the water with photo-multipliers, neutrinos can be detected through the Cherenkov light from charged products of their interactions. The dominating signal, however, comes from muons created in extensive air-showers resulting from cosmic ray interactions in the top of the atmosphere. Despite the water column above, the highest-energy muons (> TeV) reach the detectors. Air-showers develop in a non-isotropic atmosphere where the vertical temperature profile, and thus the density, varies over time. The changing density of the atmosphere affects the balance between interaction and decay (to muons) of mesons in the air-showers. Therefore, the muon flux is expected to correlate with the seasons and short-term temperature fluctuations, as confirmed by other experiments. In this contribution, we present a first measurement of the correlation of the detected muon rate with the effective atmospheric temperature (weighted by muon production spectrum and detector response) for the KM3NeT ORCA telescope. The measured rate is compared with simulations, which model the time-dependent detector efficiency and environmental factors for a constant atmospheric muon flux.

*Speaker

## 1. Introduction

Extensive air-showers initiated by interactions of cosmic rays develop in the atmosphere. After the initial interaction, many particles are created through re-interaction of the products with molecules in the air, and through decays of unstable particles. Muons and neutrinos are created through such decays, mostly from the light mesons: pions and kaons. In regions where the density is higher such particles have a higher probability to interact. When the density is lower, the balance shifts towards decaying before re-interaction. As the atmosphere is dynamic, in the sense that the vertical temperature and thus density profile changes with time and location, the amount of muons and neutrinos created varies with time. This seasonal variation of the muon and neutrino flux is well measured by a multitude of underground detectors such as [1, 2, 5, 6, 8–10, 17, 19].

KM3NeT consists of two large-volume water-Cherenkov detectors, designed to detect neutrino interactions. The smaller (7Mt when completed) ORCA detector, anchored at 2450 meters of depth near Toulon, France, focuses on the determination of the neutrino mass ordering and oscillation parameters. The larger (1 Gt when completed) ARCA detector, anchored at 3500 meters of depth near Capo Passero, Italy, focuses on cosmic neutrino detection. Both detectors share the same technology. Pressure resistant glass spheres with a diameter of about 17 inches, house 31 light-sensitive 3-inch photo-multiplier tubes (PMTs) together with electronics for control, nanosecond timing and readout. These so-called optical modules are organised in vertical structures called detection units. These consist of two dyneema ropes along the whole length, to which 18 (buoyant) modules are connected, a backbone cable with power cables and optical fibres going to each module, and an anchor housing an electronics container. Power is provided from shore through a sea-floor network, which also houses an optical-fibre network for data transport. Both detectors are currently being expanded. Intermediate configurations of the detectors are referred to either ORCA-X or ARCA-X, where X is the amount of detection units operational.

The aforementioned studies have used variations of the same equations that describe the relation between the atmospheric temperature and muon rate. In line with that, we start here from the relation[9, 11, 15]

$$\frac{\Delta R}{\langle R\rangle} = \alpha_T \frac{\Delta T_{eff}}{\langle T_{eff}\rangle} \tag{1}$$

where $\Delta R$ is the deviation of the muon rate with respect to the average muon rate $\langle R\rangle$, and $\Delta T_{eff}$ the deviation of the *effective temperature* from the average effective temperature $\langle T_{eff}\rangle$. This relation is a simplified form, where the atmosphere is approximated as isothermal with temperature $T_{eff}$. The effective temperature takes into account that the muon production spectrum $P_\mu$ is dependent on the depth in the atmosphere $\chi$[1], while the temperature also depends on the depth. The correlation coefficient, depends on the muon energies to which a detector is sensitive, through the details of the muon production, in particular the ratio of pions and kaons in the relevant energy range. In our definition of $T_{eff}$ the energy and zenith angle $\theta$ dependent response of the detector is taken into account via the effective area $A_{eff}$. The effective temperature is in essence the depth dependent temperature profile, weighted with the energy and zenith dependent muon production spectrum and

[1]This is the column density as integrated from the top of the atmosphere.

detector response

$$T_{eff} = \frac{\int d\chi T(\chi) \left[\int dE_0 \int d\theta P_\mu(E_0,\theta,\chi) A_{eff}(E_0,\theta)\right]}{\int d\chi \left[\int dE_0 \int d\theta P_\mu(E_0,\theta,\chi) A_{eff}(E_0,\theta)\right]} \tag{2}$$

where $E_0$ is the muon energy at sea-level. For this first study in the context of KM3NeT, an analytical model based on cascade-equations from [14] is used for the muon production spectrum $P_\mu$. In this contribution, we present a new method to measure the atmospheric muon rate in the KM3NeT detectors. This method accounts for the fluctuations in the muon rate which are not due to changes in the atmosphere, but due to environmental conditions and detector effects.

## 2. Dynamic environment and detector

A particular challenge in measuring the variation of the muon rate in the KM3NeT detectors results from the fact that the detectors operate in a uncontrolled, dynamic environment. The photo-detection rate of the photo-multiplier tubes is dominated by light from $^{40}$K decays and bioluminescence. While the rate from $^{40}$K decay is rather constant, the bioluminescence is highly dynamic both in time and intensity. The baseline counting rate per KM3NeT PMT is about 7 kHz, with excursions of almost a factor 10 higher [3, 4]. Long term studies on median PMT counting rates in ANTARES [13] indicate fluctuations at various timescales including monthly and yearly. Such variations influence the muon detection rate. A moderately increased count-rate can increase the detection efficiency, while an instantaneous high count-rate may trigger a high-rate veto at the level of an optical module, suppressing the detection efficiency.

## 3. Simulation of muon flux and detector response

To simulate the muon flux at detector level, the fast parametric muon flux simulation software Mupage [12] is used. Single and multiple muon events are generated on a surface close to and surrounding the detector volume in the water. Efforts of re-tuning the parameters of Mupage on simulation and data are ongoing within KM3NeT, with the aim to improve the description and understanding of the atmospheric muon flux [ 18]. The tuning used in this analysis [16] is specific for the detector configuration. It represents a simulated flux constant in time, and thus not subject to atmospheric conditions. This is a key ingredient to this method.
The KM3NeT data taking happens in scheduled time periods of 6 hours which are called runs. The simulation scheme of KM3NeT makes use of the recorded PMT counting rates, which are stored for each PMT with a resolution of 100 µs for each run. In this so-called run-by-run simulation scheme, background photons are added to the photons resulting from the muons according to the recorded rates within a run. Also the time-dependent measured photon detection efficiencies of the PMTs are taken into account. In order to extract the effective area $A_{eff}$, Mupage simulations are used, re-weighted into a single-muon flux. As the effective area needs to be expressed in energy at sea-level, a depth and angle dependent energy loss correction is applied per muon.

## 4. Method

In order to extract the variations due to the changing muon flux, free from effects of the environment and detector response, a new method has been developed. The underlying presumption of the method is that the run-by-run simulation accounts for the influence of the ambient conditions and detector effects which are reflected in the PMT count-rates, *except* the variations caused by a change in muon flux. We define a time-dependent ratio of the recorded muon rate and the simulated muon rate:

$$R_{\frac{data}{sim}}(t) = \frac{R_{data}(t)}{R_{sim}(t)} \tag{3}$$

An average value of the of value $R_{\frac{data}{sim}}(t)$ is indicated by $\langle R_{\frac{data}{sim}} \rangle$. When the muon flux is properly described by the simulation, the average ratio should be 1. In case of the tuned Mupage muon flux parametrization for the ORCA-6 configuration and time period considered here, this is essentially the case with a value of $\langle R_{\frac{data}{sim}} \rangle = 1.015 \pm 0.0005$. We define a deviation in the muon rate ratio as

$$\Delta R_{\frac{data}{sim}}(t) = R_{\frac{data}{sim}}(t) - \langle R_{\frac{data}{sim}} \rangle \tag{4}$$

To study the variation of the muon rate variation with effective temperature, we define equivalently to eq. 1

$$\frac{\Delta R_{\frac{data}{sim}}}{\langle R_{\frac{data}{sim}} \rangle} = \alpha_{T,\frac{data}{sim}} \frac{\Delta T_{eff}}{\langle T_{eff} \rangle} \tag{5}$$

with the underlying presumption that $\alpha_{T,\frac{data}{sim}} = \alpha_T$ if all ambient and detector effects on the muon count rate are taken into account.

## 5. Data and simulation selection

The KM3NeT muon data analysed in this contribution is from the ORCA-6 detector and covers a period from 2020-01-26 to 2021-11-18, with a total livetime of 374.59 days. Some data that was excluded in this analysis would be accepted with the latest quality criteria. Selections are applied to the data to remove events that are due to neutrinos, have badly reconstructed tracks in terms of angular resolution, or are caused by ambient light. The implementation of these selections include rejecting events that come from below (through the Earth); have a reconstructed energy below 8 GeV; or have a too low likelihood value. All events in this analysis fired at least the trigger aimed at identifying muon tracks. After selection, the median zenith and angular resolution on the muon direction, evaluated from simulation, are 0.422° and 0.708° respectively. The dataset consisted of $1.648 \times 10^8$ data events with $\langle R_{\text{data}} \rangle = 5.0884 \pm 0.0056$Hz and $1.623 \times 10^8$ simulations events with $\langle R_{\text{mc}} \rangle = 5.0127 \pm 0.0045$Hz.

The atmospheric temperature profile is downloaded from the public data repository of NASA's Atmospheric Infrared Sounder (AIRS) satellite [7]. The data covers the atmospheric column above the location of the ORCA detector between 42° − 43° latitude and 5° − 7° longitude in a 1-by-1 degree grid. The dataset provides temperature measurements at 01:30 and 13:30 (so twice a day), and on 24 different pressure levels, from 1 to 1000 hPa, together with error estimates. In figure 1(left), the atmospheric temperature profile for the whole period is shown. One can notice slow variation in spring and summer, but also shorter time-scale variations in the stratosphere such as those which are very prominent for all winter periods covered. In figure 1(right) these profiles are averaged over 3 month periods for 2020 and are labeled according to the starting month. Noticeable is the different profiles for the the first and last December (each holding roughly same number of measurements). The same figure also shows the depth dependent product of the effective area with the muon production profiles, separate for pions, kaons and their sum, integrated over energy and zenith angle. These values correspond to the bracket term in eq. 2. The upper pressure layers carry the largest weights, due to $P_\mu(E_0, \theta, \chi)$ being large at low $\chi$, and thus temperature changes in the upper atmosphere are expected to be of largest influence on the muon rate. The higher frequency of data-taking runs was used to estimate the change of the muon rate during a 12 hour interval, which was used for the systematical uncertainty of the correlation. In order to match the frequency of the atmospheric temperature data, the KM3NeT dataset was re-sampled to periods of 12 hours.

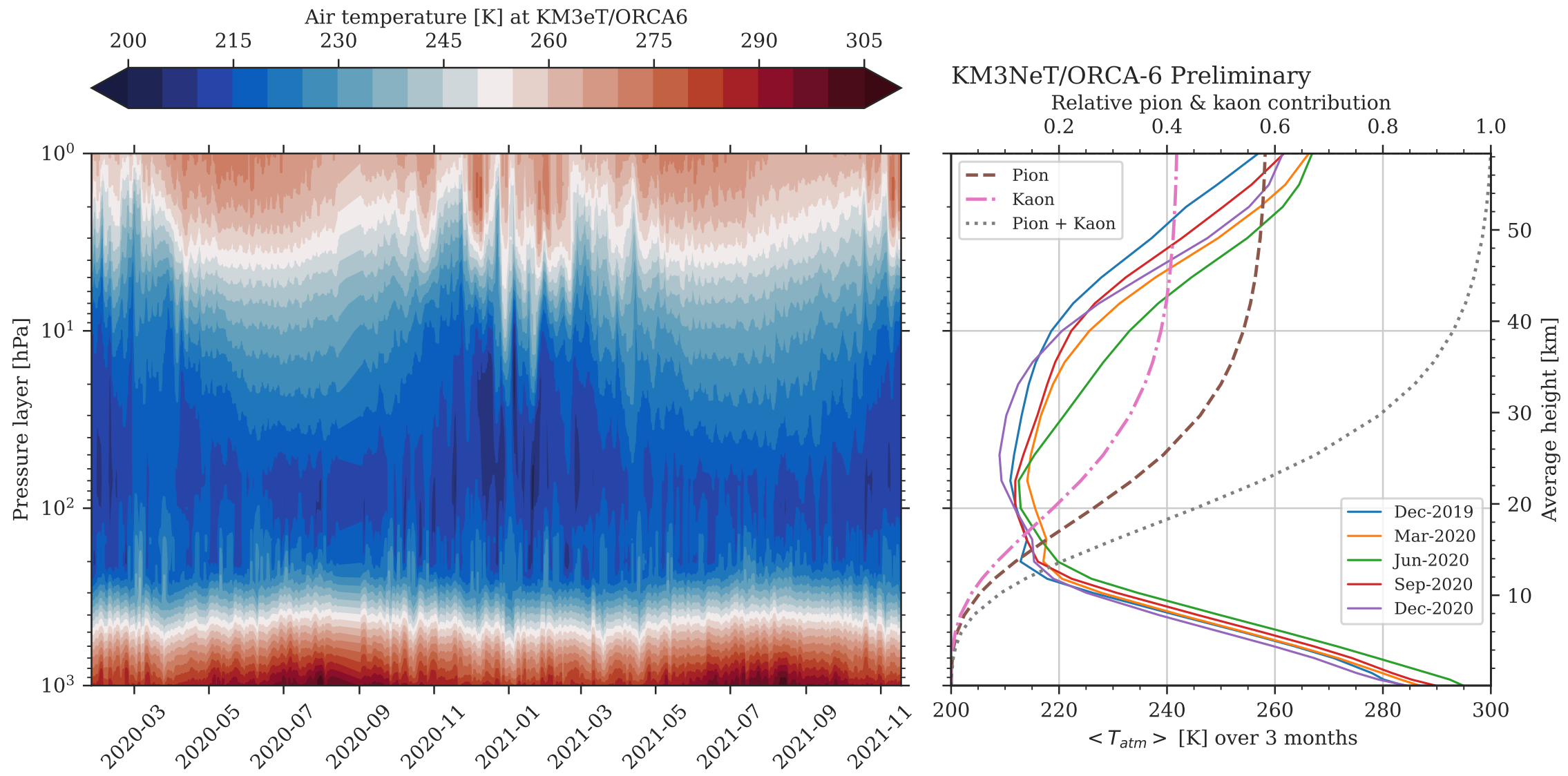

**Figure 1:** The atmospheric temperature above the ORCA detector in the period of ORCA-6 (left) and its 3-month average profiles (solid lines) for the year 2020. They share the left pressure axis which is given in 24 discrete layers[7]. The dashed lines (right) are the product of $A_{eff}$ with the pion-only, kaon-only and combined muon production spectrum $P_\mu$, integrated over energy and zenith angle. The shared right axis shows the average geometrical height that corresponds to the pressure.

## 6. Results

The challenge to extract the muon rate variations with time are illustrated in figure 3(top), where both the measured muon rates from data and the predicted muon rate from the run-by-run simulation are presented. Much of the variations in the muon rate can be attributed to the ambient conditions as they are very well reproduced with the constant flux used in the simulation. Qualitatively, the changes in muon rate are well reproduced, in particular at the shorter timescales. There seem to be some longer time-scale variations that are not seen in the simulated data.

The merit of the method introduced, where the measured rate is divided by the predicted rate $R_{\frac{data}{sim}}(t)$, per formula 4 is illustrated in figure 3(bottom). While there are still short time variations, these are much suppressed and a periodic modulation can be seen. The amplitude of this modulation is around 2 to 3 %, while the period seems consistent with a year. The highest rates are in the Northern summer of 2020. A lower peak in 2021 seems in line with an overall decrease in rate over the analysed period.

A correlation plot of the relative change in rate $\Delta R_{\frac{data}{sim}}/\langle R_{\frac{data}{sim}}\rangle$ with the relative change in effective temperature $\Delta T_{eff}/\langle T_{eff}\rangle$ for 12 hour bins is shown in figure 4. The error bars on the points are from figure 2 and 3 and include several sources of uncertainties besides the statistical errors. Included are the errors on the measurement of the temperature for each pressure layer, the uncertainty due to rate variations at time-scales smaller than 12 hours, and a conservative uncertainty on the effective area of 10 %.

There is a clear correlation between the effective temperature and the muon rate. Several clusters of correlated points suggest that there are other effects that are not accounted for. For example, the cluster of points at the right below the fit line, are all of the same period in July 2021.

Shown in the figure is a fit of the linear function

$$\Delta R_{\frac{data}{sim}}/\langle R_{\frac{data}{sim}}\rangle = \alpha_{T,\frac{data}{sim}}\Delta T_{eff}/\langle T_{eff}\rangle + \beta, \quad (6)$$

where $\alpha_{T,\frac{data}{sim}}$ and $\beta$ are free parameters. All uncertainties were taking into account. The value of the slope was $\alpha_{T,\frac{data}{sim}} = 1.005 \pm 0.041$ and with a residual variance $\chi^2/dof = 0.44$. A null-hypothesis fit (of a horizontal line) gave $\beta = 0.041 \pm 0.061$ with $\chi^2/dof = 0.74$.

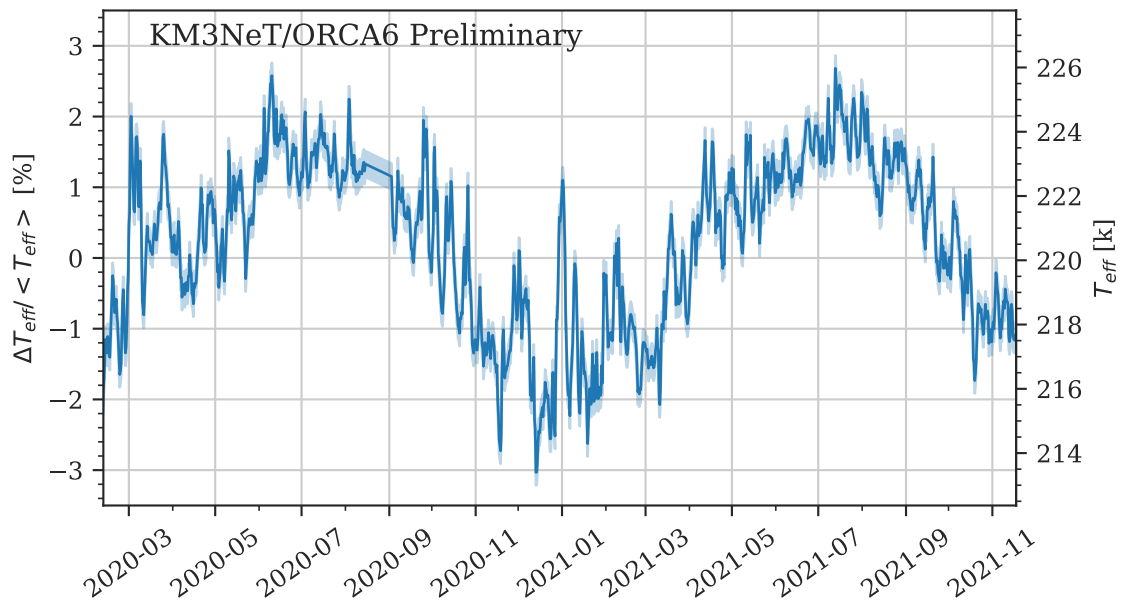

**Figure 2:** Relative variation (left y-axis) in effective temperature above the ORCA-6 detector. The right y-axis shows the absolute temperature.

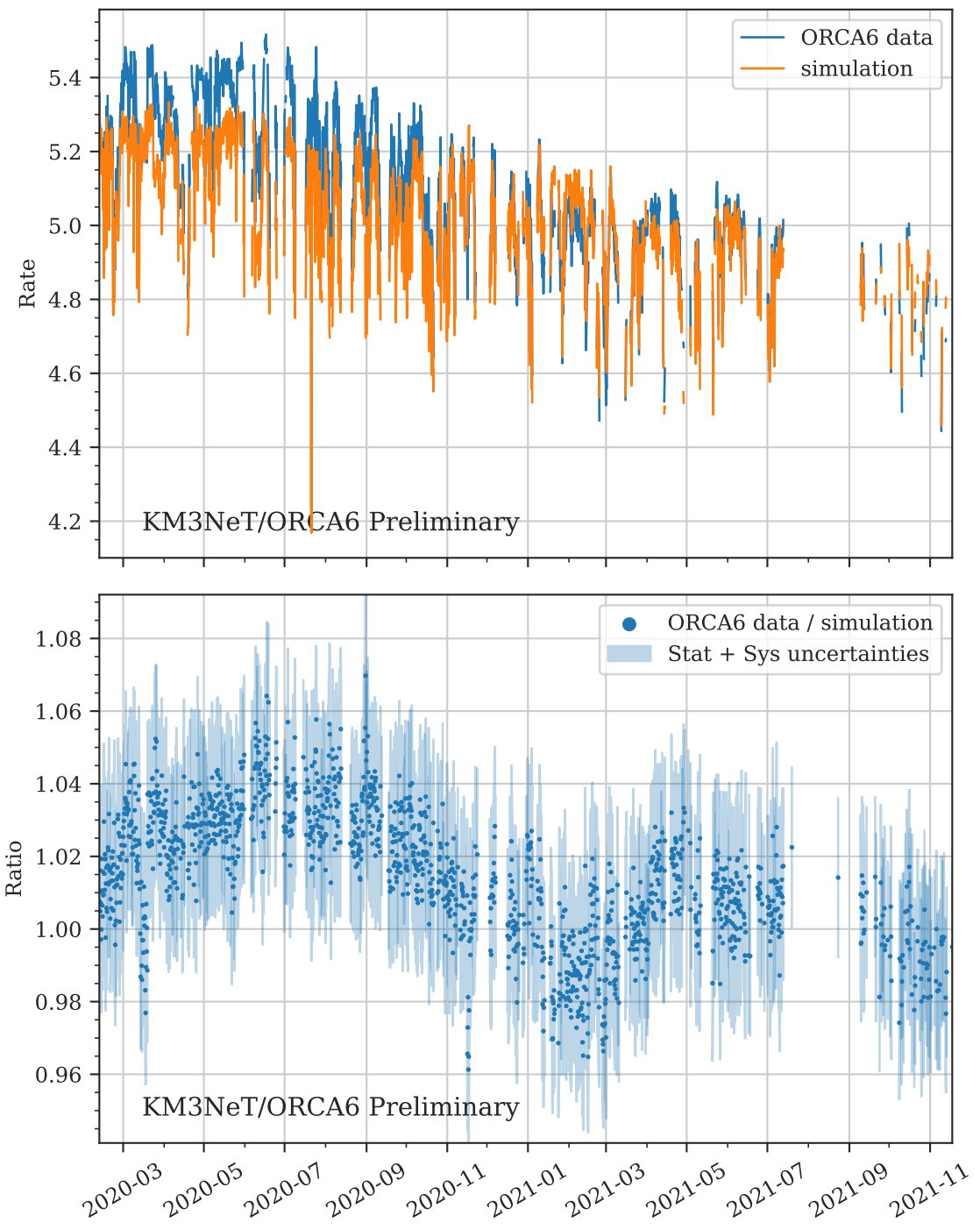

**Figure 3:** The data and simulated muon rate (top) with their ratio (bottom) for the whole ORCA-6 period. The error-bars (only shown in bottom) include statistical and systematic uncertainties are added in quadrature.

## 7. Conclusion and Outlook

In this work, a new method to measure the variation in atmospheric muon flux under changing local ambient conditions and efficiency of the KM3NeT detectors has been presented. This was applied to the KM3NeT/ORCA-6 dataset and a correlation of the high energy muon rate with the effective temperature was found. In this first study, simplified, analytic, models have been used for the muon production and propagation, combined with a parametric simulation of the muon flux in the water. Hints towards more complex time-dependent effects on the muon rate require further study, in order to understand whether they originate from the atmosphere or detector and environmental effects that are not corrected for in the simulations. In future studies a more detailed modelling e.g. by the use of full extensive air-shower simulations will be used, together with data from the growing ORCA and ARCA detectors.

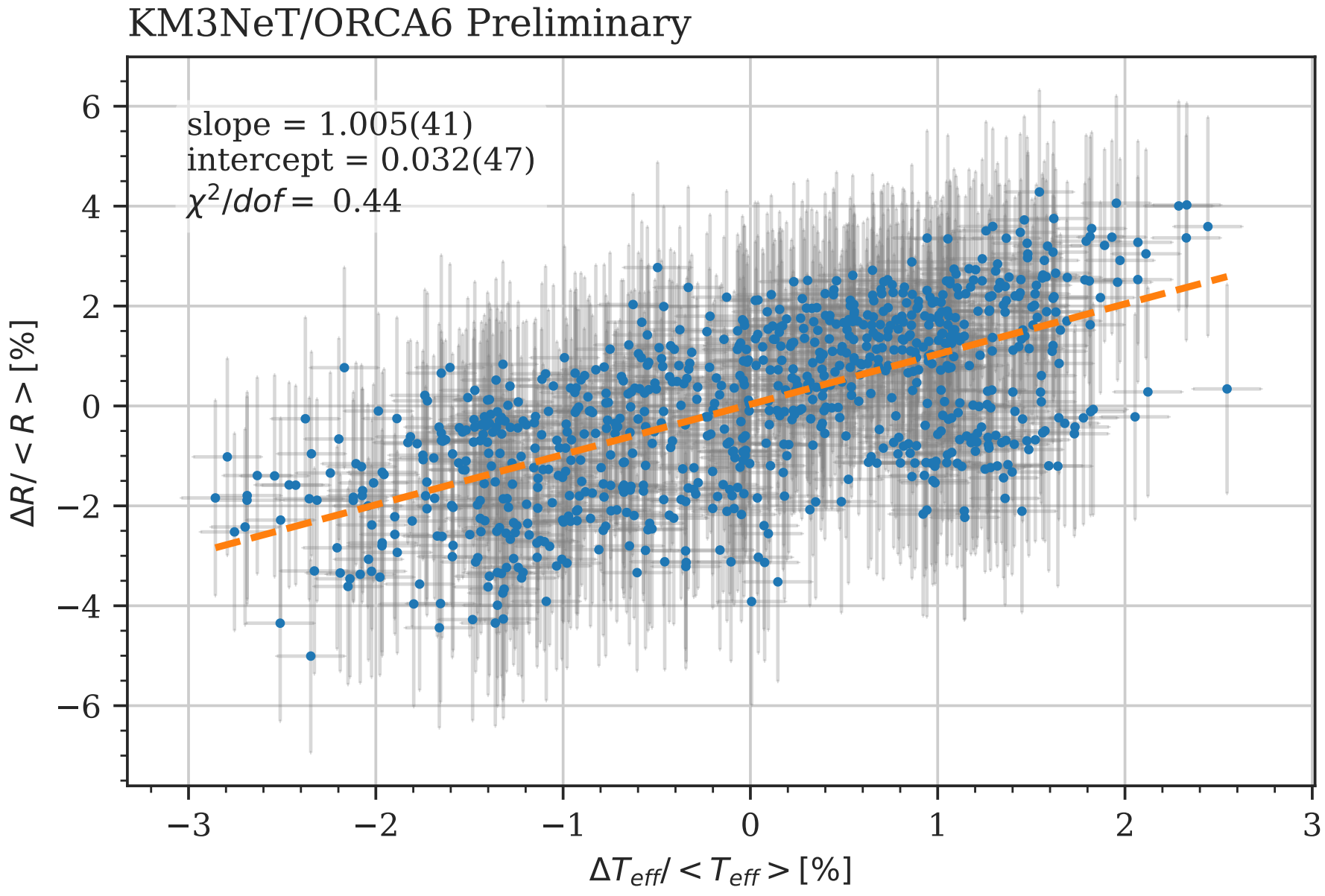

**Figure 4:** Relative variation in percentage of the effective temperature $T_{eff}$ and data/simulation ratio $R_{\frac{\text{Data}}{\text{Simulation}}}$. The orange dashed line is eq. 6 fitted to the data, using $y$ = slope $\times\, x$ + intercept. The fit takes into account both uncertainties.

# Measurement of the atmospheric muon neutrino flux with KM3NeT/ORCA6

**Dimitris Stavropoulos,**[a,b,*] **Ekaterini Tzamariudaki,**[a] **Evangelia Drakopoulou**[a] **and Christos Markou**[a] **for the KM3NeT collaboration**

[a] *NCSR Demokritos, Institute of Nuclear and Particle Physics, Ag. Paraskevi Attikis, Athens, 15310 Greece*

[b] *National Technical University of Athens, School of Applied Mathematical and Physical Sciences, Zografou Campus, 9, Iroon Polytechniou str, 15780 Zografou, Athens, Greece*

*E-mail:* dstavropoulos@inp.demokritos.gr

The KM3NeT/ORCA detector (Oscillation Research with Cosmics in the Abyss) is an array of Digital Optical Modules, spheres that host 31 photomultiplier tubes, tied together in vertical structures, the Detection Units, which are anchored on the seabed. Such an array configuration can detect neutrino events from the Cherenkov radiation emitted by the secondary particles induced by neutrino interactions in the abyssal depths of the Mediterranean Sea. The KM3NeT/ORCA detector is being deployed at a depth of 2450 m offshore Toulon, France with the determination of the Neutrino Mass Ordering being the main physics goal of the detector. The aim of this work is the study of atmospheric neutrinos with energies at the 1-100 GeV range, in order to obtain information on the atmospheric muon neutrino flux in this energy range, in which only few measurements exist by other experiments. An analysis of data collected with the 6-Detection Unit configuration of KM3NeT/ORCA (KM3NeT/ORCA6) is being presented in this contribution. The data analyzed corresponds to a time period of one and a half year. The procedure for the selection of a high-purity atmospheric neutrino sample, using a Machine Learning classifier (Boosted Decision Tree), is described. Subsequently, an unfolding scheme is used to obtain an estimation of the atmospheric muon neutrino flux in bins of energy in the region of interest.

*Speaker

## 1. Introduction

Measurements of the atmospheric neutrino flux are increasingly important for testing the models that describe the production of the cosmic rays as well as their interaction mechanisms in the atmosphere. A reliable description of the atmospheric neutrino flux is also mandatory, as atmospheric neutrinos are used from several oscillation experiments. Moreover, a precise knowledge of the atmospheric neutrino flux is critical for constraining the contribution of this irriducible background source in the search of cosmic neutrinos. The prospects of KM3NeT/ORCA to contribute at the energy range between 1 GeV and 100 GeV has been already reported in [1]. In the following, a first measurement of the atmospheric muon neutrino flux is presented using data collected with KM3NeT/ORCA. This measurement illustrates the ability of the KM3NeT/ORCA detector, even with an early-stage detector configuration, to provide experimental information at an energy region in which only few measurements exist by other experiments.

## 2. The KM3NeT/ORCA detector

The KM3NeT Collaboration is currently constructing a research infrastructure in the depths of the Mediterranean Sea [2]. The ORCA detector (*Oscillation Research with Cosmics in the Abyss*), is under construction at a location ~ 40 km offshore Toulon, France, at a depth of 2450 m. ORCA is an array of photomultiplier tubes (PMTs) capable of detecting neutrino events via the Cherenkov radiation emmited by the daughter particles. When completed, it will consist of 115 Detection Units (DUs), vertical structures with 18 Digital Optical Modules (DOMs) each. The DOM is a sphere that hosts 31 3-inch PMTs as well as the required electronics [3] (Fig. 1). The distance between the DUs in ORCA is ~ 20 m, while the vertical distance between the DOMs in the same DU is ~ 9 m, so the height of the ORCA detector is ~ 160 m. At its final form, ORCA will have a cylinrical shape with ~ 100 m radius. ORCA is currently operating using 16 Detection Units.

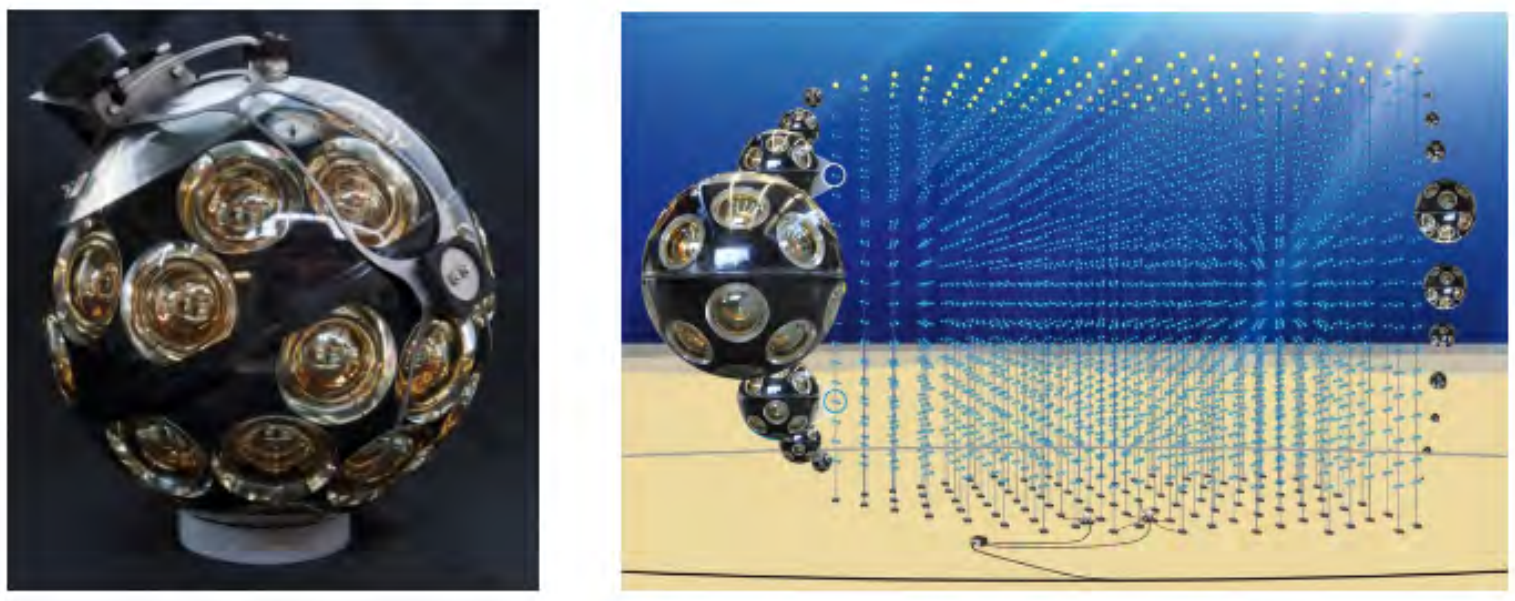

**Figure 1:** The Digital Optical Module (left) and an artistic view of the full KM3Net/ORCA detector (right).

## 3. Data and MC simulation

The data used in this analysis have been collected with the 6-DU configuration of the ORCA detector, referred to as ORCA6, which was operating from January-2020 to November-2021. The data livetime used for this analysis is equal to 555.7 days. This results in an ~ 84% time efficiency with respect to the total running time period for ORCA, as a fraction of the running time was devoted to calibration and test runs, and additional quality criteria were applied.

Monte Carlo (MC) event samples were produced to estimate the contribution of atmospheric muons and atmospheric neutrinos. The MUPAGE software has been used to generate atmospheric muon events [4]. Atmospheric neutrino events have been generated in the energy range between 1 GeV and 10 TeV using the gSeaGen software [5]. Charged current (CC) neutrino interactions have been simulated for all neutrino flavours, while the neutral current (NC) interactions of all flavours have been simulated and treated as a single component. The neutrino events have been weighted to account for the atmospheric neutrino flux and the oscillation probabilities. The HKKM14 atmospheric neutrino conventional flux model has been used [6]. For what concerns the neutrino oscillations, the parameters have been set according to NuFIT 5.2 [7], and oscillation probabilities have been computed assuming normal hierarchy. The light simulation as well as the PMT readout have been simulated using custom KM3NeT software.

The trigger algorithms that have been applied during the data acquisition as well as during the trigger level of the MC simulation, belong to the KM3NeT-specific Jpp software package [8]. The same software package has been used for the processing of triggered events in data and MC through the reconstruction chain (track and shower).

## 4. Event classification for a high-purity neutrino selection

The contribution of the random noise events, coming from optical backgrounds in sea water, is suppressed by applying a requirement on the likelihood and by requesting a minimum number of PMT hits contributing to the reconstructed track. Additionally, only events reconstructed with an upward-going direction are accepted.

The upward-going reconstructed events that survive these anti-noise cuts are further selected using a Boosted Decision Tree (BDT) algorithm. For this, the *TMVA* software is used [9]. Atmospheric $\nu_\mu + \bar{\nu}_\mu$ CC events are considered as signal, and atmospheric muons as background. 20 features are used for the BDT, and they are related to the reconstructed event position and direction, the reconstruction quality, and the event topology considering signal-like hits and charge distributions. The number of DOMs with at least one triggered hit, is also used. Dedicated MC event samples were produced to train and test the BDT algorithm.

The phase space of the BDT parameters has been scanned in order to find an optimal parameter set. For each set of parameters, the efficiency at indicative BDT score values as well as an overtraining check were performed, using an amount of $\sim 10\%$ of the data livetime. The optimal values of the chosen set of parameters, include the number of trees set to 400, the tree maximum depth to 6, and the adaptive boost parameter set to 0.3. The BDT score distributions for data and MC simulated events satisfying the anti-noise selection criteria and reconstructed as upgoing, are shown in Fig.2.

Good discrimination between the signal and background is achieved, as the former dominates at the higher score values, while the latter at the lower values. The requirement for an event to be included in the final selection is to have a BDT score greater or equal than 0.56. That leads to 4197 data events, while 4196.1 atmospheric $\nu + \bar{\nu}$ events are expected from MC simulation, from which 3214.8 are $\nu_\mu + \bar{\nu}_\mu$ CC events. The resulting contamination of atmospheric muon

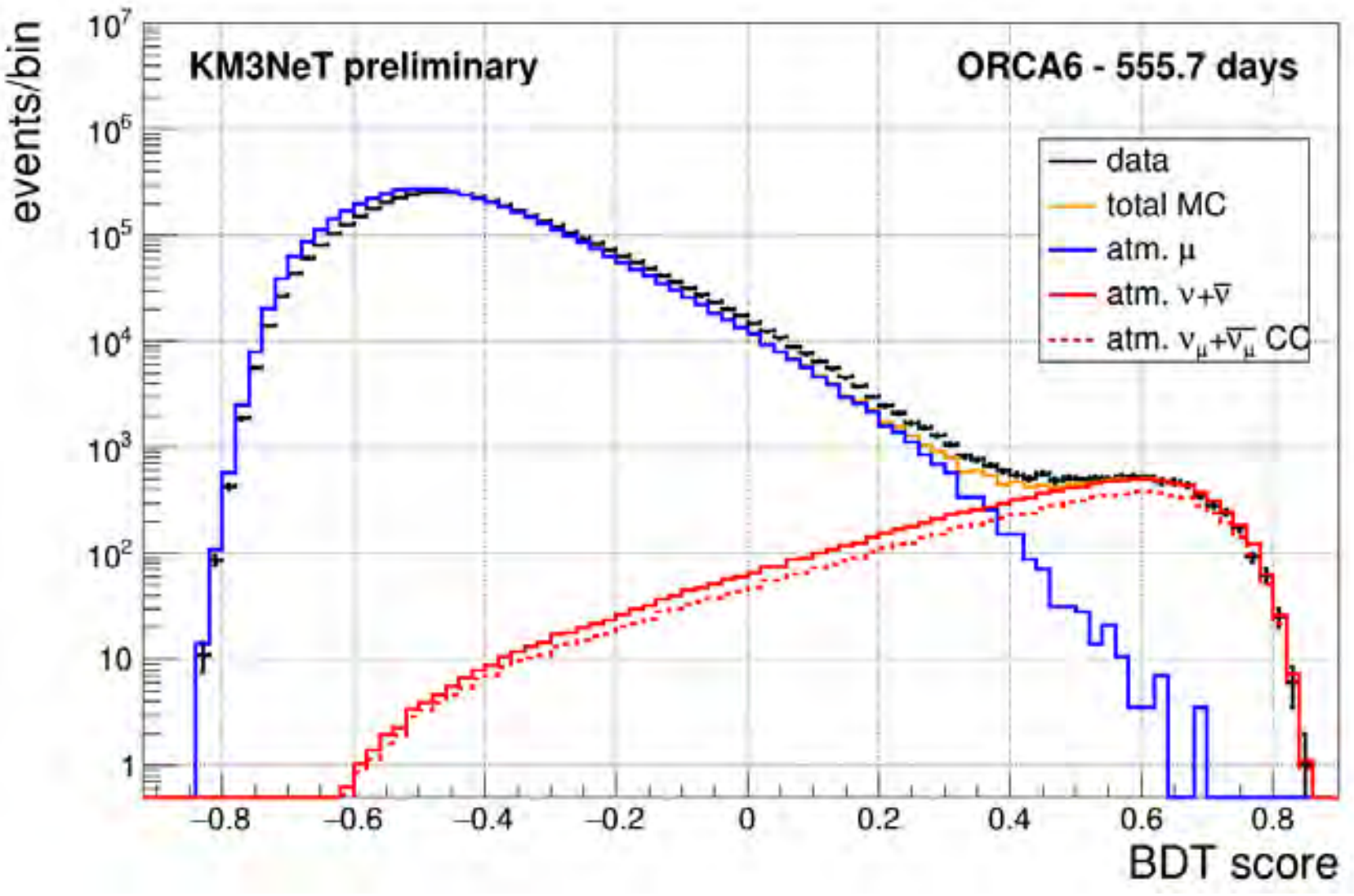

**Figure 2:** BDT score distributions for the upward-going ORCA6 events that survive the anti-noise criteria.

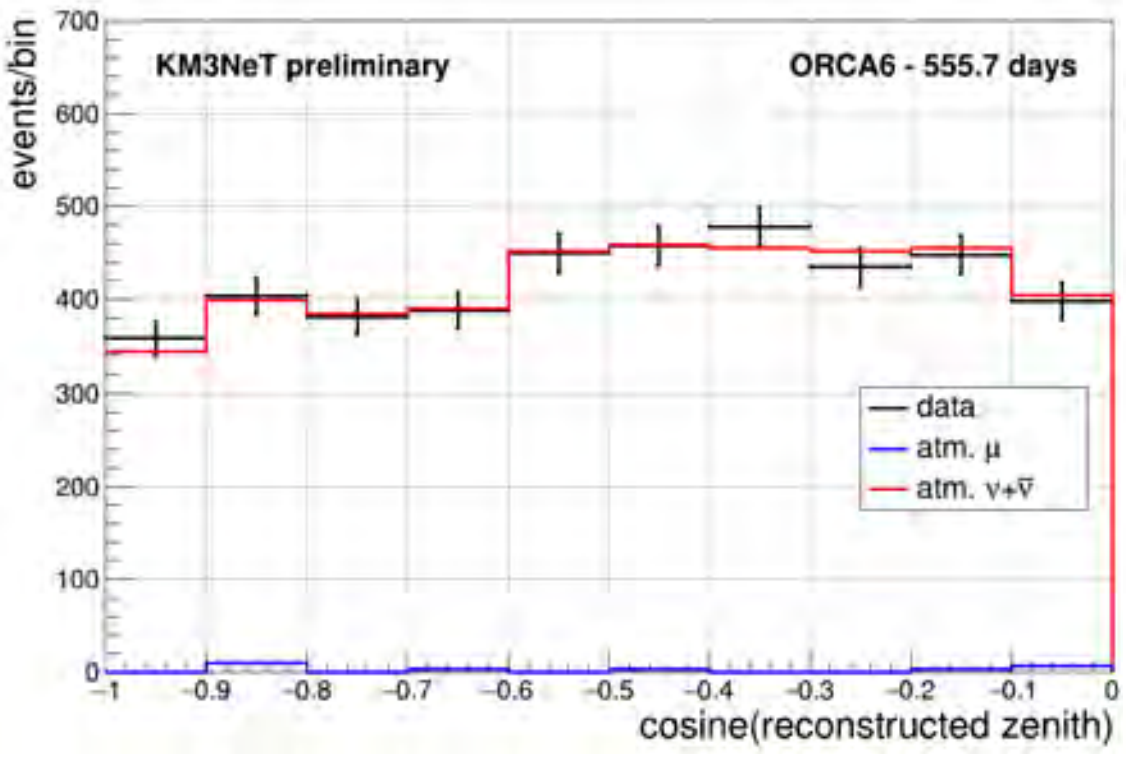

**(a)** Distribution of reconstructed cosine zenith.

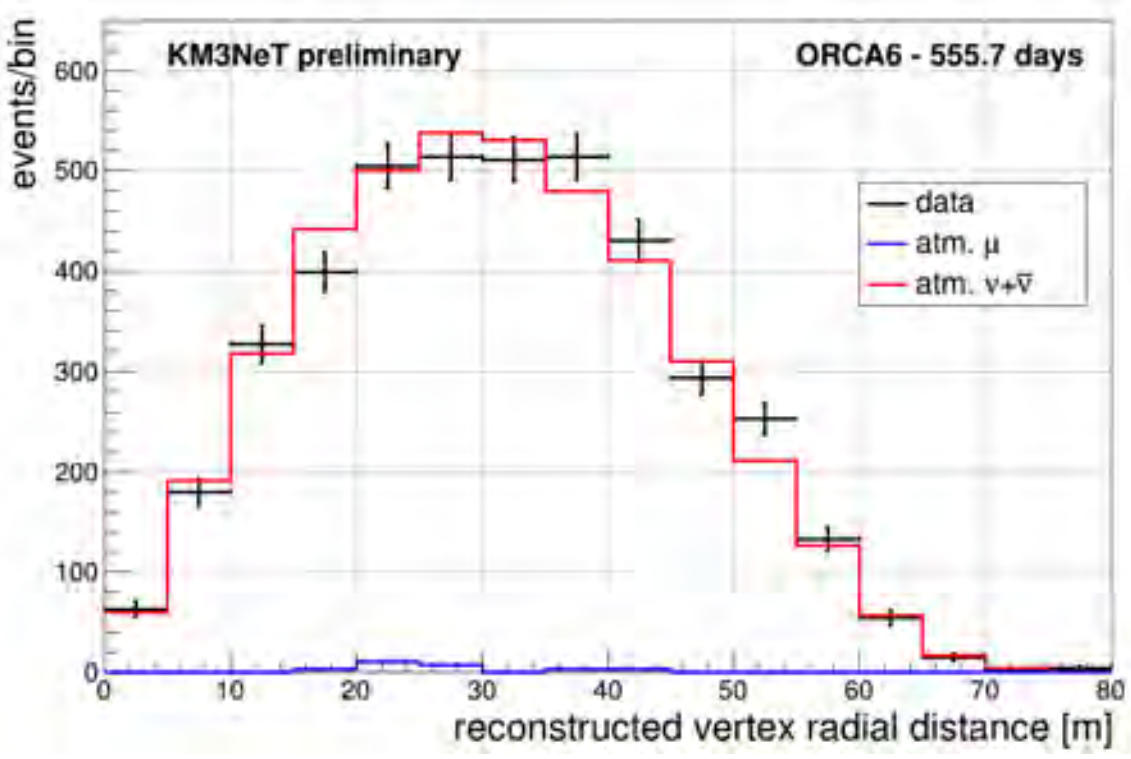

**(b)** Distribution of the reconstructed vertex radial distance.

**Figure 3:** Event variable distributions for the neutrino selection.

in the MC sample is estimated to be 28.1 events, which corresponds to 0.6% of the sample. A good agreement between data and MC simulation is obtained. The reconstructed cosine zenith distribution is presented in Fig.3a, for the BDT selected events. The distribution of the radial distance between the reconstructed vertex and the barycenter of the detector is also shown in Fig.3b.

## 5. Unfolding the $\nu_\mu + \bar{\nu}_\mu$ CC energy spectrum

An unfolding scheme is applied in order to deconvolute the $\nu_\mu + \bar{\nu}_\mu$ CC energy spectrum from the measured data energy distribution. For this, the *TUnfold* software is used [10]. In principle an unfolding scheme (with background subtraction) is described by:

$$y_i = \sum_{j=1}^{m} A_{ij} x_j + b_i, 1 \leq i \leq n \tag{1}$$

where $y$ is the energy distribution for data, $A$ the response matrix, $x$ the true distribution which is being unfolded, and $b$ the background contributions (can be more than one). The index $i$ counts over the $n$ bins of the reconstructed phase space while the index $j$ counts over the $m$ bins of the true phase space. *TUnfold* estimates the true $x$ distribution using a least square method with Tikhonov regularization and an optional constraint [10]. Due to the limited size of the ORCA6 detector configuration, the shower reconstructed energy is used for unfolding the $\nu_\mu + \bar{\nu}_\mu$ CC energy spectrum.

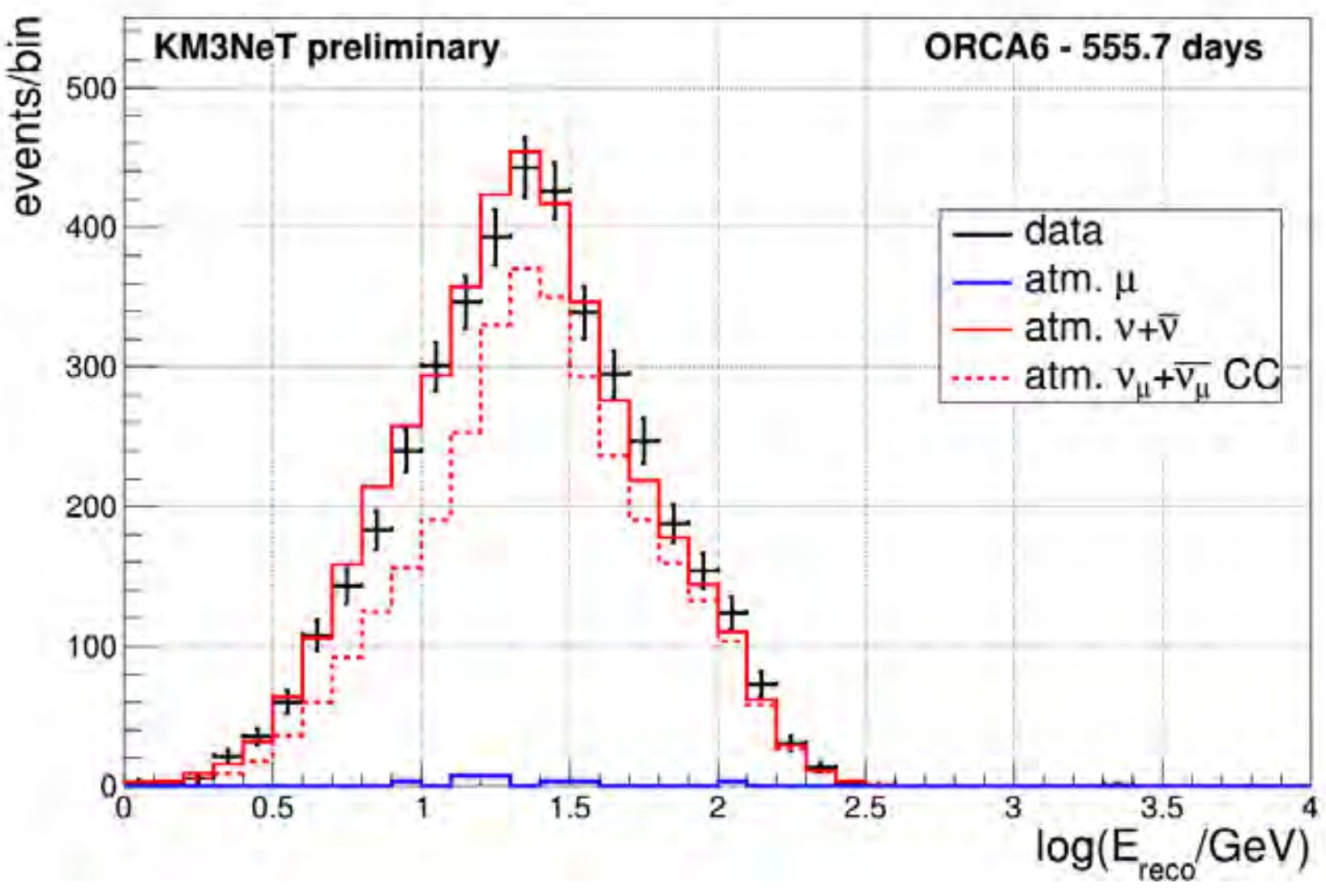

**Figure 4:** Reconstructed energy distribution for the selected events.

Only $\nu_\mu + \bar{\nu}_\mu$ CC are considered as signal events. Therefore, atmosperic muons, atmospheric $\nu_e + \bar{\nu}_e$ CC, atmospheric $\nu_\tau + \bar{\nu}_\tau$ CC and atmospheric $\nu + \bar{\nu}$ NC events are considered as background sources. In addition, atmospheric $\nu_\mu + \bar{\nu_\mu}$ CC with $E_{true} > 100$ GeV are also treated as background and their contribution is also subtracted from the reconstructed energy distribution. This approach was followed as the energy reconstruction is affected by the limited instrumented volume at this early-stage of construction. As a result, the flux measurement is performed in the range between 1 GeV-100 GeV. The consistency of the unfolding procedure was tested by replacing the reconstructed data distribution with the one for the atmospheric neutrino MC, and the robustness of the method was also checked by producing 1000 unfolding toy MC experiments. The bins used in this analysis were chosen taking the bin purity into account. The chosen binnings are $\{0.0, 0.1, 0.2, ..., 2.5, 2.6\}$ and $\{0.0, 0.8, 1.3, 1.8, 2.0\}$ for the reconstructed phase space and the true phase space respectively. The response matrix used for the unfolding procedure is shown in Fig.5.

After several tests which were performed exclusively using MC simulated events, to test the procedure and define the binning schemes, the unfolding was eventualy performed using the data reconstructed energy distribution. The unfolded $\nu_\mu + \bar{\nu}_\mu$ CC energy distribution is presented in Fig.6, while the true $\nu_\mu + \bar{\nu}_\mu$ energy distribution has been also added as a reference.

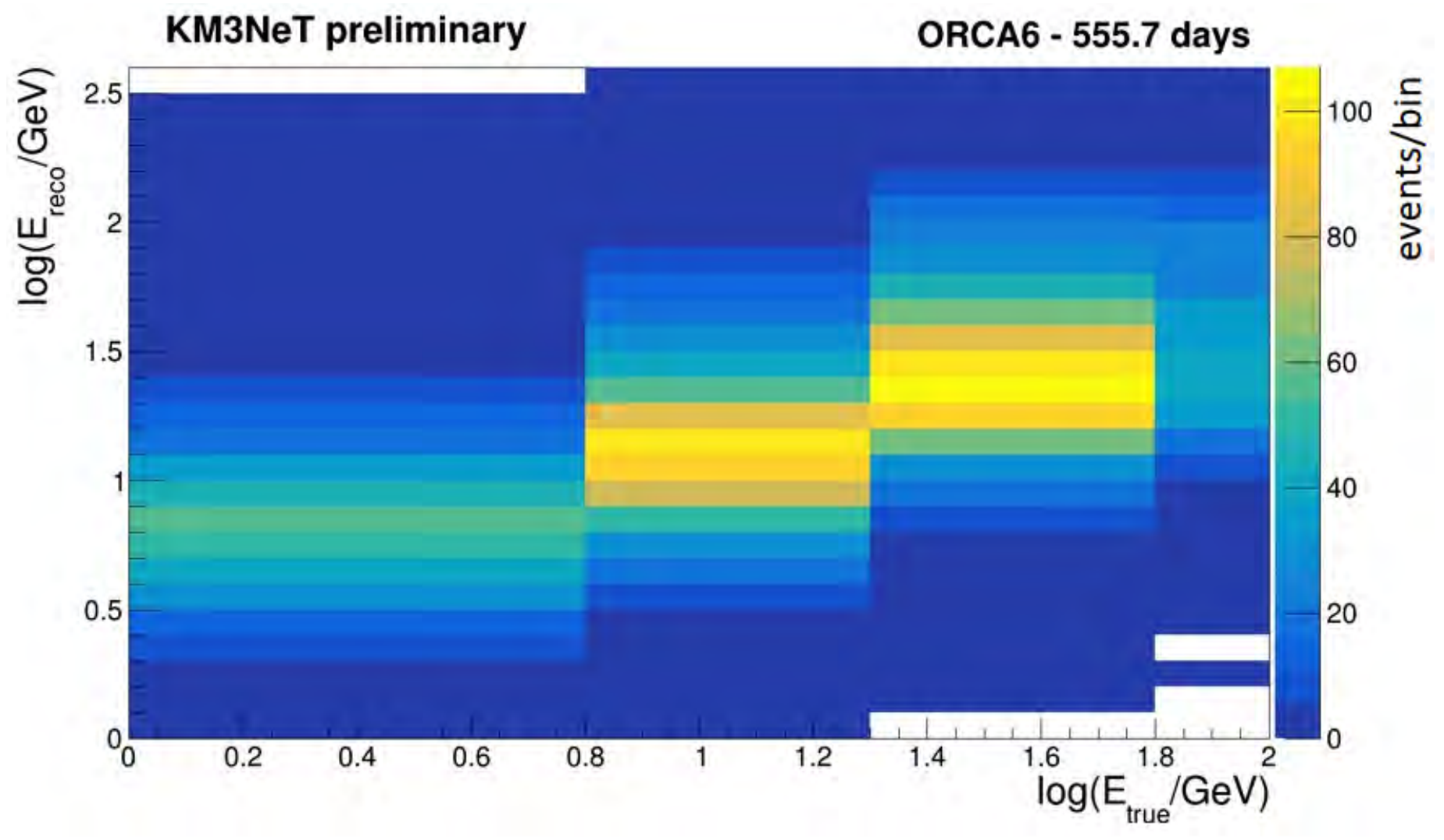

**Figure 5:** Response matrix for the event selection with the binning schemes used in the unfolding.

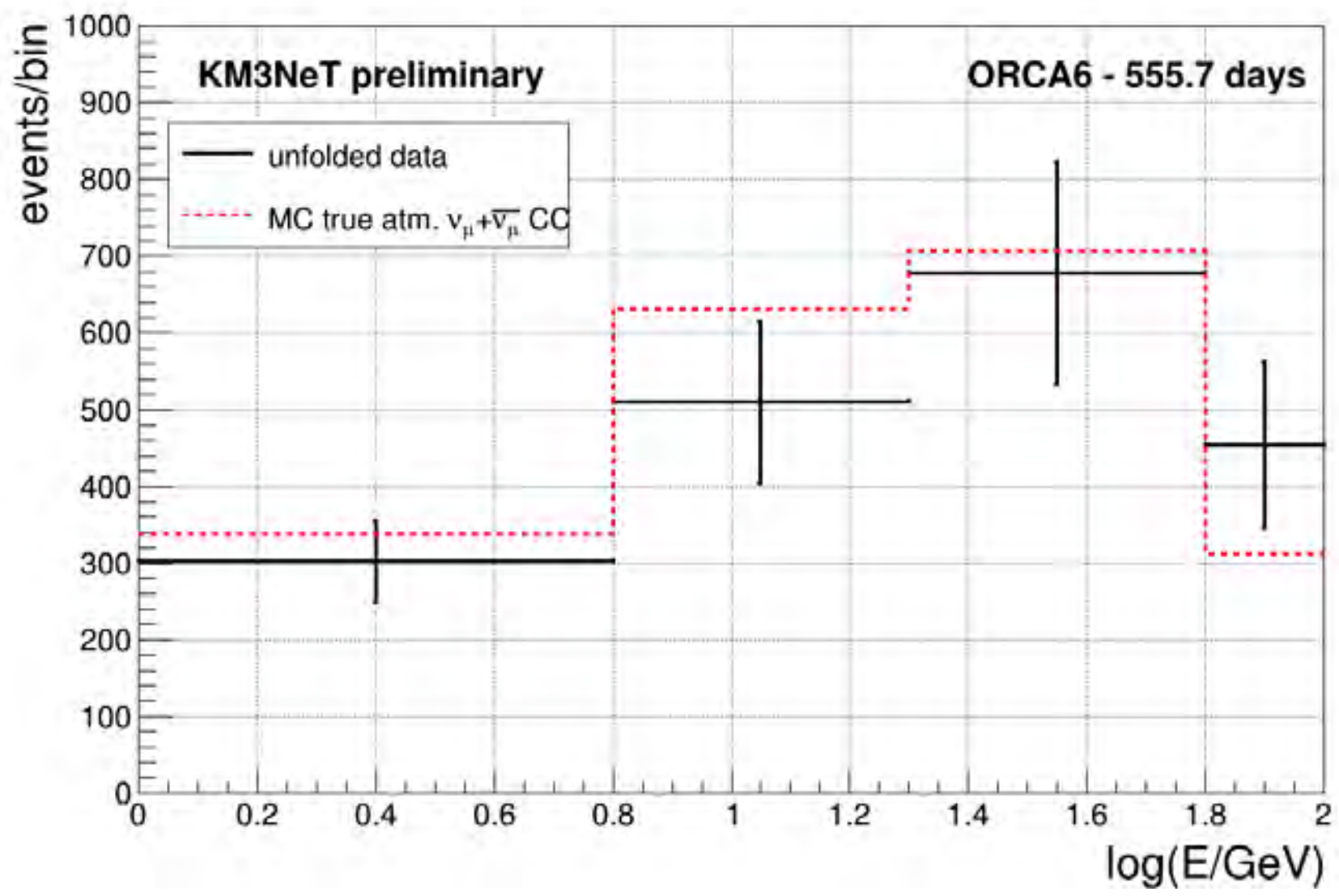

**Figure 6:** Unfolded energy distribution for the $\nu_\mu + \bar{\nu}_\mu$ CC events. The MC (true) $\nu_\mu + \bar{\nu}_\mu$ energy distribution is also shown as a reference.

## 6. Measurement of the flux

The last step is to convert the unfolded number of events per energy bin into flux values. The procedure is based on equation 2,

$$\Phi_i = \Phi_i^{MC} \cdot \frac{N_i^{unf}}{N_i^{MC}} \tag{2}$$

where $\Phi_i$ is the measured value for the bin $i$, $\Phi_i^{MC}$ is the flux value predicted by the HKKM14 atmospheric neutrino flux model [6], calculated at the weighted bin center for the bin $i$, $N_i^{unf}$ is the number of events extracted by the unfolding scheme, and $N_i^{MC}$ is the number of MC events in the bin. The value $\Phi_i^{MC}$ is calculated using the HKKM14 tabulated values are used, integrated over the azimuth angle, in order to integrate over the zenith angle according to equation 3, where $O^{\nu_i \to \nu_j}$ is the oscillation probability from $\nu_i$ to $\nu_j$, set as in [7].

$$\Phi_{MC}^{\nu_\mu+\bar{\nu}_\mu}(E_\nu) = \int_{4\pi} d\Omega \Big\{ \Phi_{MC}^{\nu_e}(E_\nu,\theta) \cdot O^{\nu_e \to \nu_\mu}(E_\nu,\theta) + \Phi_{MC}^{\bar{\nu}_e}(E_\nu,\theta) \cdot O^{\bar{\nu}_e \to \bar{\nu}_\mu}(E_\nu,\theta) + \Phi_{MC}^{\nu_\mu}(E_\nu,\theta) \cdot O^{\nu_\mu \to \nu_\mu}(E_\nu,\theta) + \Phi_{MC}^{\bar{\nu}_\mu}(E_\nu,\theta) \cdot O^{\bar{\nu}_\mu \to \bar{\nu}_\mu}(E_\nu,\theta) \Big\} \tag{3}$$

The value of $\Phi_i^{MC}$ at the equation 2 is calculated by interpolating the integrated flux of equation 3 at the energy that corresponds to the weighted bin center for each bin. The results of the measurement are shown in Table 1. The errors are statistical only and are propagated from the results of the unfolding procedure.

| $\Delta log(E_\nu/GeV)$ | $\overline{log(E_\nu/GeV)}$ | $E_\nu^2 \Phi_\nu [GeV \cdot s^{-1} \cdot sr^{-1} \cdot cm^{-2}]$ | stat. |
|---|---|---|---|
| 0.0-0.8 | 0.39 | $1.43 \cdot 10^{-2}$ | ±17% |
| 0.8-1.3 | 0.93 | $4.25 \cdot 10^{-3}$ | ±21% |
| 1.3-1.8 | 1.45 | $1.57 \cdot 10^{-3}$ | ±21% |
| 1.8-2.0 | 1.88 | $8.46 \cdot 10^{-4}$ | ±24% |

**Table 1:** From left to right: Bin energy range; weighted energy bin center; flux measurement multiplied by the weighted energy bin center; statistical error.

The KM3NeT/ORCA6 measured flux values along with measurements from other experiments ([11],[12]) are presented in Fig.7. The measurement performed by Frejus on 1995 [13] is not shown in Fig.7 as it was done before the discovery of neutrino oscillations.

## 7. Conclusion

A measurement of the atmospheric muon neutrino flux using the first KM3NeT/ORCA data has been performed. The ability of ORCA to provide significant information concerning the atmospheric neutrino flux, even with an early-stage detector configuration, is shown in Fig.7. This measurement is in good agreement with the only existing and up-to-date one below 100 GeV, from Super-Kamiokande. A study for the estimation of the systematic uncertainties in this analysis is ongoing.

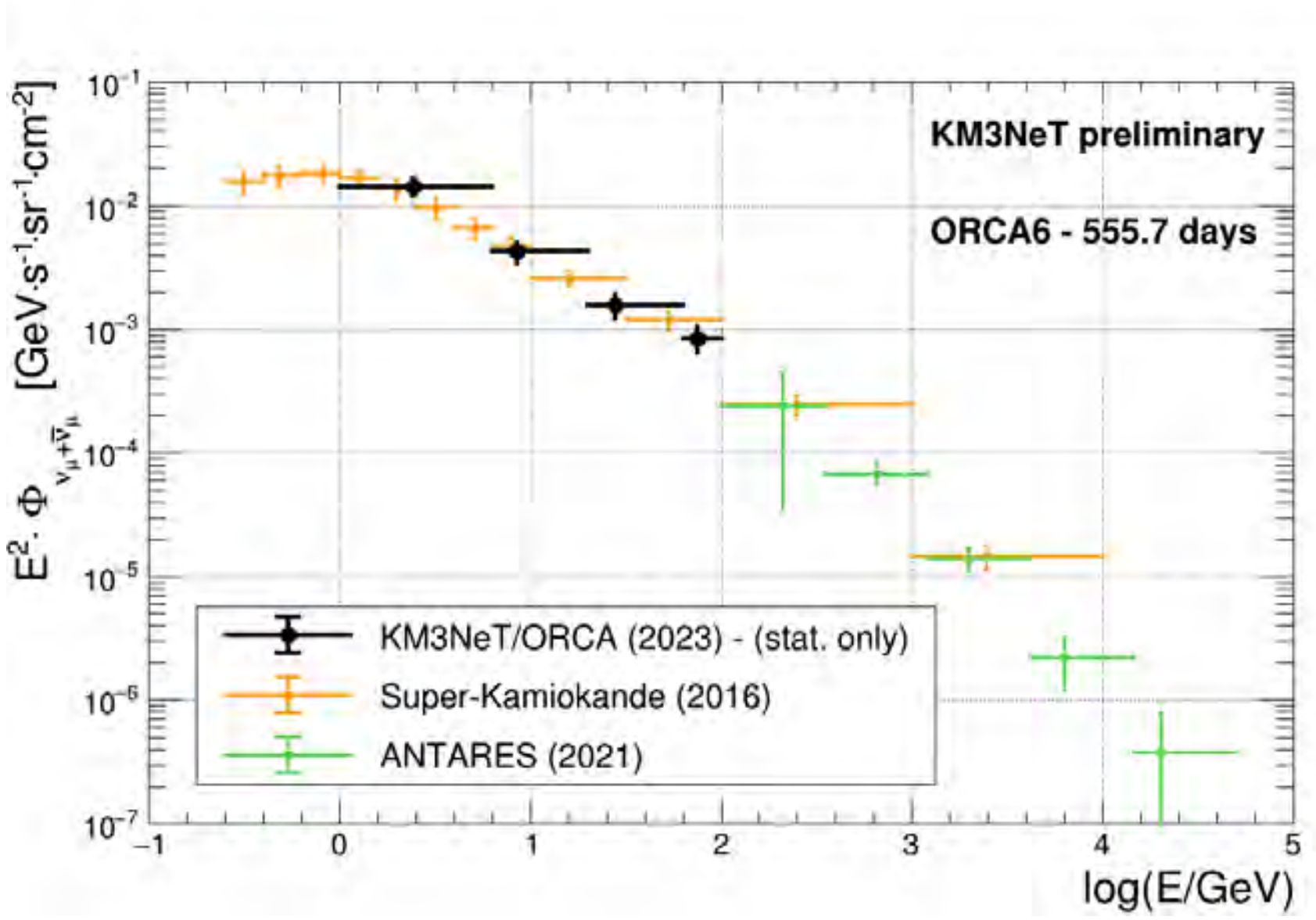

**Figure 7:** The atmospheric muon neutrino flux measurement using KM3NeT/ORCA6 data is shown along with other measurements.

# Indirect Search for Dark Matter with the KM3NeT Neutrino Telescope

**Adrian Šaina,**[a,*] **Miguel Gutiérrez,**[b] **Sara Rebecca Gozzini,**[a] **Juan de Dios Zornoza**[a] **and Sergio Navas**[b] **for the KM3NeT Collaboration**

[a] *IFIC - Instituto de Física Corpuscular (CSIC - Universitat de València), c/Catedrático José Beltrán, 2, 46980 Paterna, Valencia, Spain*

[b] *University of Granada, Dpto. de Física Teórica y del Cosmos & C.A.F.P.E., 18071 Granada, Spain*

*E-mail:* adrians@ific.uv.es, mgg@ugr.es

Neutrino telescopes aim to detect dark matter indirectly by observing the neutrinos produced by pair-annihilations or decays of weakly interacting massive particles (WIMPs). A signal excess of neutrinos resulting from the pair-annihilation of WIMPs can be detected in regions where large amounts of dark matter might accumulate. One possible source is the Sun, where WIMPs are expected to accumulate due to their scatterings in the dense core of the star. The dark matter halo of the Milky Way is another possible close dark matter container. The KM3NeT observatory is composed of two undersea Cherenkov neutrino telescopes (ORCA and ARCA) located in two sites in the Mediterranean Sea, offshore of France and Italy. The two detector configurations are optimised for the detection of neutrinos of different energies, which allows the search for WIMPs in a wide mass range, from the GeV to the TeV scale. In this contribution, searches for WIMP annihilations in the Galactic Centre and the Sun are presented. An unbinned likelihood method is used to discriminate the signal from the background in a 300-day livetime sample of the ARCA detector, and a 543-day sample of the ORCA detector. The limits on the velocity-averaged pair-annihilation cross section of WIMPs are computed for five different primary annihilation channels. For the ORCA analysis, the limits on the spin-dependent and spin-independent scattering cross sections are given for three annihilation channels.

*Speaker

## 1. Introduction

Astrophysical observations have proven that a majority of the matter budget in the Universe is composed of a cold, non-baryonic form of matter, the nature of which is unknown. Direct and indirect detection experiments assume a particle nature of dark matter, aiming to determine its properties by observing its interactions with the Standard Model particles. In the case of indirect detection, searches for the products of dark matter annihilation are performed in regions where dark matter is thought to accumulate: one such place is the Galactic Centre, as galaxy formation theory predicts the existence of galactic dark matter halos with very high densities at the centre of the object [1]. A second possible source is the Sun, where dark matter particles of the Galactic halo scatter off nuclei composing the Solar medium, causing them to be trapped in the gravitational potential of the Sun and accumulate in the centre of the object.

The neutrino flux expected at the Earth surface due to the annihilation of dark matter particles into secondary products can be formulated as:

$$\frac{d\Phi}{dE} = \frac{\Gamma}{4\pi d^2}\frac{dN}{dE}. \tag{1}$$

The parameter $\Gamma$ is the annihilation rate of WIMP particles, $d$ represents the distance to the source centre, and $\frac{dN}{dE}$ represents the number of neutrinos per unit energy emitted in one annihilation event. In order to obtain the neutrino spectra, the dark matter particle is assumed to be a weakly interacting massive particle (WIMP). A model-independent search is performed, where neutrino yields are computed for a range of WIMP masses and up to five annihilation channels:

$$\text{WIMP} + \text{WIMP} \rightarrow \mu^-\mu^+, \tau^-\tau^+, b\bar{b}, W^-W^+, \nu\bar{\nu} \rightarrow \nu\bar{\nu}.$$

The neutrino yields from the subsequent decays and emissions of the annihilation products are described using PYTHIA, and are implemented in the form of tables in the PPPC4 framework [2].

In the case of the Galactic Centre, the annihilation rate, $\Gamma$, depends on two parameters: the spatial distribution of dark matter in the target object, and the parameter we are trying to measure or constrain: the thermally-averaged cross section of WIMP annihilation, $< \sigma v >$. The spatial distribution of WIMPs is expressed in terms of a Galactic density profile, obtained with the CLUMPY program [3]. The Navaro-Frenk-White profile is used to obtain limits in this analysis [4]. This density profile is integrated along the line of sight ,l.o.s, and through the solid angle that the object subtends in the sky, $\Delta\Omega$. The equation then reads:

$$\frac{d\Phi}{dE} = \frac{1}{4\pi}\frac{< \sigma v >}{2m_{\text{WIMP}}^2}\int_{\Delta\Omega}\int_{\text{l.o.s.}} \rho^2(\theta, l)dl\, d\Omega. \tag{2}$$

In the case of the Sun, an equilibrium between capture and annihilation is assumed, which implies $\Gamma = 1/2\, C_r$, $C_r$ being the capture rate of dark matter in the Sun. The latter is related to the WIMP-nucleon scattering cross section, which can be spin-dependent or spin-independent. Therefore, a relationship between the WIMP-nucleon cross section and the flux of neutrinos can be established:

$$\sigma^{\text{SD,SI}} = K^{\text{SD,SI}}\Phi_{\nu+\bar{\nu}}, \tag{3}$$

where $\sigma^{\mathrm{SD,SI}}$ represents the spin-dependent/spin-independent cross section, and $K$ represents the so-called conversion factor, which is computed using the software package DarkSUSY [9]. The conversion factor contains information about the WIMP-WIMP annihilation channel, the mass of the WIMP, and the density and velocity distribution of halo WIMPs surrounding the Sun.

## 2. The KM3NeT neutrino telescope

The KM3NeT detector is an underwater Cherenkov detector positioned at two sites in the Mediterranean Sea [5]. The KM3NeT/ ORCA (Oscillation Research with Cosmics in the Abyss) detector, situated off the coast of Toulon, France, performs neutrino oscillation studies by detecting atmospheric neutrinos at GeV energies, whereas the KM3NeT/ARCA (Astroparticle Research with Cosmics in the Abyss) detector, placed off-shore of Sicily, Italy, attempts to detect astrophysical neutrinos at higher energies, although both detectors are well-suited for dark matter searches.

The detection principle relies on the detection of the Cherenkov light emitted by ultra-relativistic leptons that are produced when neutrinos interact in the vicinity of the detector. The light is detected by 3D arrays of Digital Optical Modules (DOMs), which are grouped in vertical Detection Units (DUs), with each DU consisting of 18 DOMs [6]. Each DOMs in itself is composed of 31 photomultiplier tubes (PMTs), along with sensors that determine the position and orientation of each DOM. The ARCA detector is currently consisting of 21 DUs. The analysis reported in this proceeding was conducted on the ARCA detector in the configuration with 6 and 8 DUs, referred to as ARCA6/8, in operation between May 2021 and June 2022. The ORCA detector currently consists of 18 DUs. The data analysed in this proceeding was taken with the configuration with 6 DUs, referred to as ORCA6, in operation between January 2020 and November 2021. The analysis was optimised on simulated events: for both detectors, the gSeaGen KM3NeT code was used to simulate neutrino interactions in water and the resulting flux of neutrinos at the detectors [7], whereas the MUPAGE package was used to simulate the atmospheric muon flux at the detectors [8]. The simulation of the light propagation and detection and the event reconstruction is handled by a custom KM3NeT software package.

## 3. Methods

Muon neutrinos interacting via the charge-current interaction produce muons that traverse the detector whilst inducing the emission of a cone of Cherenkov light. These events are referred to as track-like events. The muon produces a large amount of light while traversing through the entirety of the detector volume, which is why this event topology offers the best angular resolution. Neutral-current interactions of neutrinos produce hadronic showers which dissipate their energy in short distances, likewise for leptonic showers induced by electrons and tau leptons produced from charged-current interactions of their associated neutrino. This class of events is referred to as shower-like events. The analyses reported in this proceeding are conducted on track-like events. The largest source of background for this event topology are downgoing atmospheric muons. In order to reject them, only upgoing events, which enter the atmosphere and traverse through the Earth before arriving at the detector, are considered. Furthermore, additional selection cuts are applied to remove noise, and to remove muons that are mis-reconstructed as upgoing. A summary of the cuts applied to the data can be found in Table 1.

| | |
|---|---|
| ARCA6/8 | track likelihood > 50, $n_{\text{hits}}$ > 20, cos(zen) > 0, E(GeV) > 10, track length > x, $\beta$ < y. |
| ORCA6 | ICRC_V2 cuts, $\beta$ < a, $n_{\text{hits}}$ > b, track likelihood > c. |

**Table 1:** The cuts used to select upgoing neutrino events and reject the atmospheric muon background. Two variables were used in the optimisation procedure for the ARCA6/8 sample, the length of the reconstructed track (x ∈ 100 − 200 m), and the reconstruction angular error estimate, $\beta$ (y ∈ 0.5° − 1°). For the ORCA6 sample, the ICRC_V2 set of cuts is used in order to ensure the compatibility among data and simulations. Additionally, cuts in $\beta$ (a ∈ 0.9° − 1°), the number of hits (b ∈ 20 − 40) and the track likelihood (c ∈ 60 − 120) are used to optimise the limits on the neutrino flux.

This analysis tries to distinguish a cluster of signal events around the source centre (the signal hypothesis, $H_1$) from the null hypothesis ($H_0$), where all the events originate from the atmospheric background. Pseudo-experiments are generated with the number of signal events injected varied between 0 and 50, and an unbinned likelihood analysis is performed in order to determine the most likely number of signal events on a sky map. The probability density functions (PDFs) used in the likelihood are two-dimensional histograms in angular distance from the source centre, and the reconstructed energy. The likelihood function used is an extended log likelihood:

$$\mathcal{L} = \Sigma_{N_{\text{tot}}} \log(n_{\text{sg}} P_{\text{sg}}(\alpha, E) + (N_{\text{tot}} - n_{\text{sg}}) P_{\text{bg}}(\alpha, E)) - N_{\text{tot}}. \tag{4}$$

The variables $\alpha$ and $E$ denote the event angular distance from the source and event reconstructed energy, the parameters $P_{\text{sg}}$ and $P_{\text{bg}}$ denote the signal and background PDFs, and $N_{\text{tot}}$ and $n_{\text{sg}}$ denote the total number of events and the number of signal events. For each mock sky map generated, the number of signal events is varied in order to maximise the likelihood, and the test statistic (TS) is calculated:

$$TS = \frac{\mathcal{L}(n_{\text{sg,max}})}{\mathcal{L}(n_{\text{sg}} = 0)}. \tag{5}$$

The parameter $\mathcal{L}(n_{\text{sg,max}})$ denotes the maximised likelihood, and the parameter $\mathcal{L}(n_{\text{sg}} = 0)$ denotes the likelihood of the sky map consisting of only background events. The number of events in each pseudo-experiment is subject to Poisson fluctuations, which are accounted for by applying a transformation of the TS distribution through a Poisson function, $\mathcal{P}$ with mean $\mu$, as follows:

$$P(\text{TS}(\mu)) = \Sigma_{n_{\text{sg}}} P(\text{TS}(n_{\text{sg,max}})) \times \mathcal{P}(n_{\text{sg}}, \mu). \tag{6}$$

In order to take into account the systematic uncertainty in the number of detected signal events, a Gaussian smearing is applied with a value of 30% (15% for ORCA6): this value was obtained by simulating events in KM3NeT with a modified absorption length of photons in water, the uncertainty in this parameter exerted the largest influence on the number of detected e vents. The Neyman approach is followed to obtain an averaged upper limit on the number of signal events [11]:

the median of the background TS distribution is compared with each TS distribution with injected signal events. The sensitivity in the number of events, $n_{90}$, is obtained as the 90% confidence level upper limit for a measurement that coincides with the median of the background TS distribution. The flux sensitivity is then obtained from the number of events sensitivity, $n_{90}$, with the following equation:

$$\Phi^{90}_{\nu+\bar{\nu}} = \frac{n_{90}}{Acc \times T}. \tag{7}$$

The quantity $T$ is the livetime of the analysed dataset, whereas the parameter $Acc$ is the detector acceptance to signal events, obtained as a convolution of the detector effective area, $A_{\mathrm{eff}}$ and the WIMP annihilation spectrum:

$$Acc = \int_{E_{\mathrm{th}}}^{M_{\mathrm{WIMP}}} A_{\mathrm{eff}}(E_\nu)\frac{dN}{dE}dE_\nu. \tag{8}$$

The effective area is defined as the area of an ideal, one hundred percent efficient detector, which is computed with detector simulations from the ratio of generated and detected neutrino events. The acceptance is integrated from the minimum threshold energy of the detector, $E_{\mathrm{th}}$, up to the WIMP mass. Equations 2 and 3 are then used to convert the flux sensitivities into cross section sensitivities in the case of the Galactic Centre and the Sun.

## 4. Results: searches in the Galactic Centre

The data set taken with the ARCA6/8 configurations was analysed in search of a WIMP annihilation signal, for WIMP masses in the range 500 GeV$/c^2$ – 100 TeV$/c^2$. The event selection applied to the data was varied for each annihilation channel / WIMP mass combination, with the selected cut mask attaining the best sensitivities for that mass / channel combination. The TS of the unblinded dataset is compatible with the background hypothesis, for all combinations of WIMP masses and annihilation channels, and the limit on the thermally-averaged WIMP annihilation cross section was placed following Equation 2. The upper limits are compared to the limits obtained with the complete ANTARES dataset [10], the predecessor of the KM3NeT neutrino telescope, in Figure 1. An improvement on the ANTARES results is expected with the upcoming datasets of the expanded ARCA detector. The results are placed in context to other indirect searches in the field in Figure 2.

## 5. Results: searches in the Sun

The ORCA6 dataset has been analysed in search of dark matter in the Sun, considering WIMP masses in the range 10 GeV$/c^2$ – 10 TeV$/c^2$. As in the ARCA6+8 sample, the TS obtained for this dataset is compatible with the background hypothesis for all combinations of WIMP masses and annihilation channels. In particular, the TS of the dataset is found to be below the median of the background-only TS distribution for every test case. Consequently, the limit on the neutrino flux is set to be equal to the sensitivity. Limits to the spin-dependent and the spin-independent cross sections are obtained through Equation 3. Figures 3 and 4 show such limits.

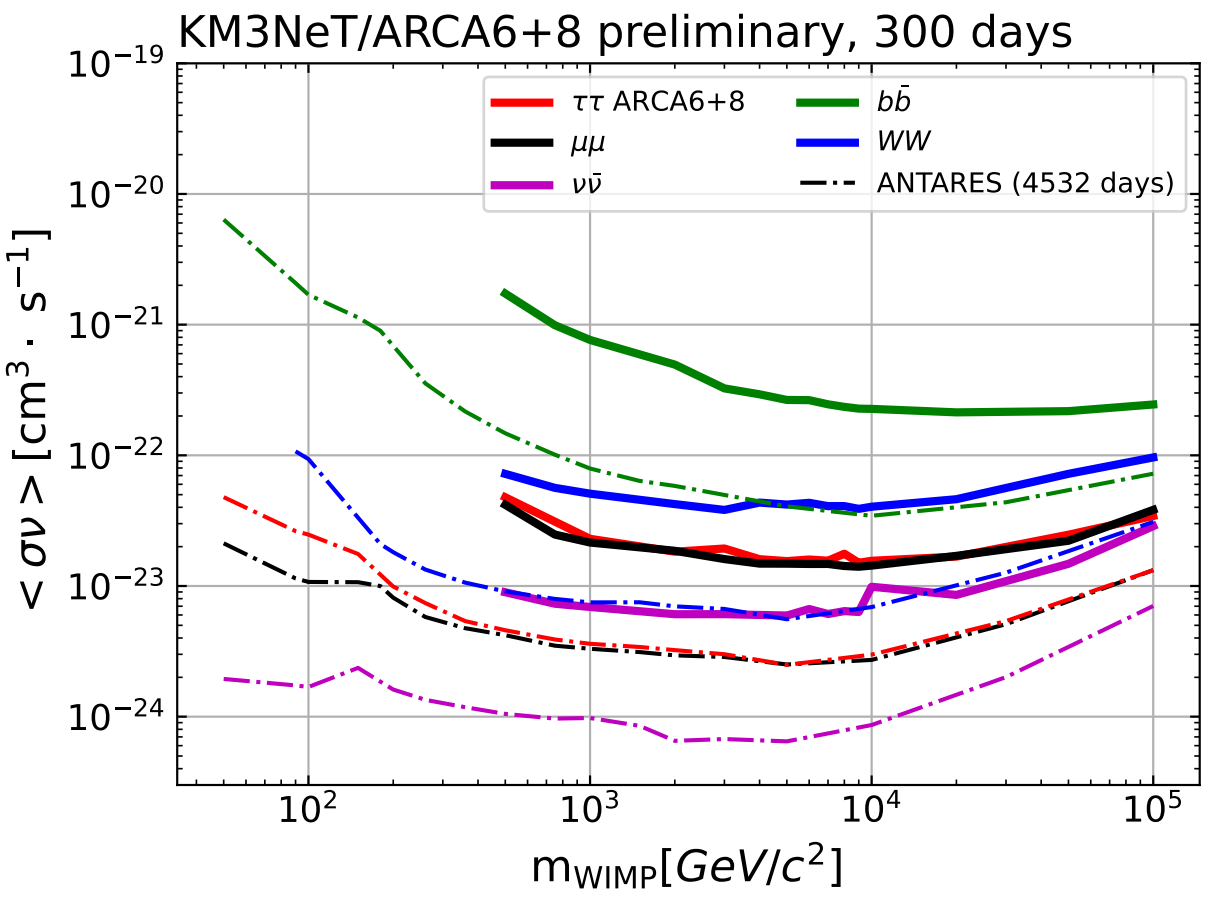

**Figure 1:** The 90% CL upper limits on the thermally-averaged WIMP annihilation cross section as a function of the WIMP mass for each of the five annihilation channels. The full lines show the results obtained in this analysis, whereas the dashed lines show the upper limits obtained with the complete ANTARES dataset [10].

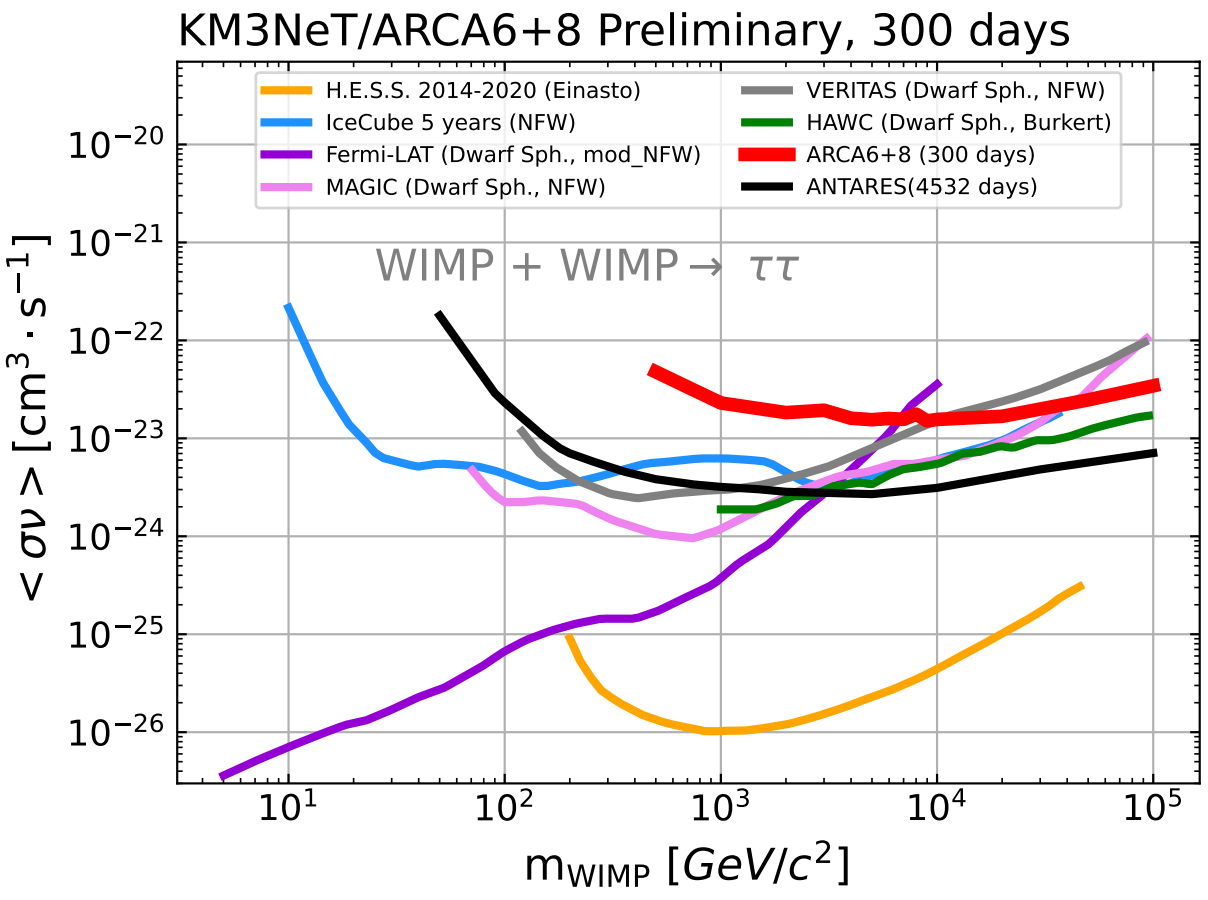

**Figure 2:** The 90% CL upper limits on the thermally-averaged WIMP annihilation cross section as a function of the WIMP mass for each of the five annihilation channels, shown along with results obtained by other experiments in the field [12–17].

## 6. Discussion

The first results on searches for dark matter annihilation signatures with the KM3NeT neutrino telescope have been presented in this contribution. Thanks to their different detector configurations, the ARCA and ORCA detectors can cover a wide range of the allowed parameter space of the WIMP mass. By searching for dark matter signatures towards the Sun and Galactic Centre, the two detectors are already providing competitive results in the field in their initial construction phase, surpassing its predecessor in certain regions of the parameter space. Follow-up searches with currently deployed and future larger detector configurations will push the boundary of dark matter searches with neutrino telescopes.

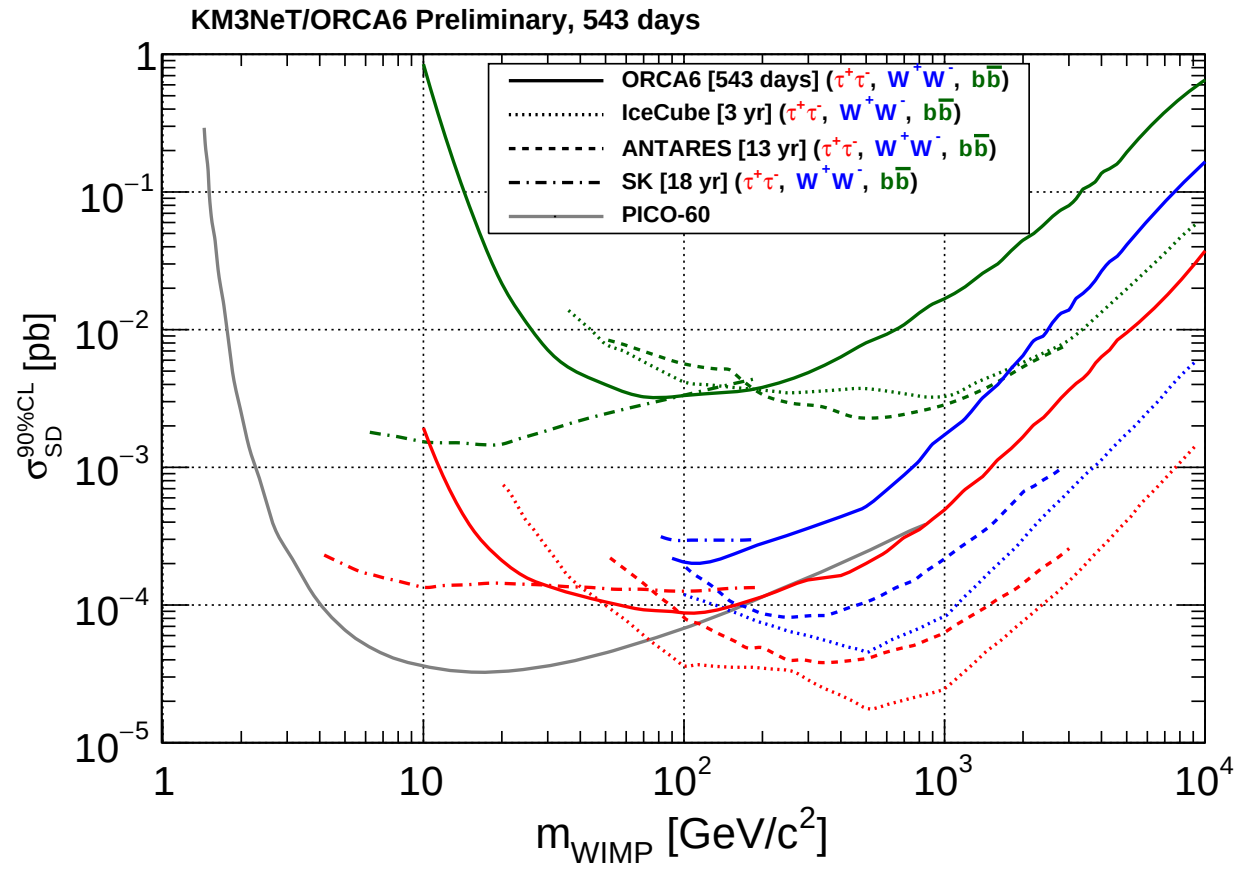

**Figure 3:** The 90% CL upper limits on the spin-dependent WIMP-nucleon cross section as a function of the WIMP mass for each of the three annihilation channels. The full lines show the results obtained in this analysis, whereas the other lines show the upper limits obtained by IceCube [18], ANTARES [19], Super-Kamiokande [20] and PICO-60 [21] (shown as a full line).

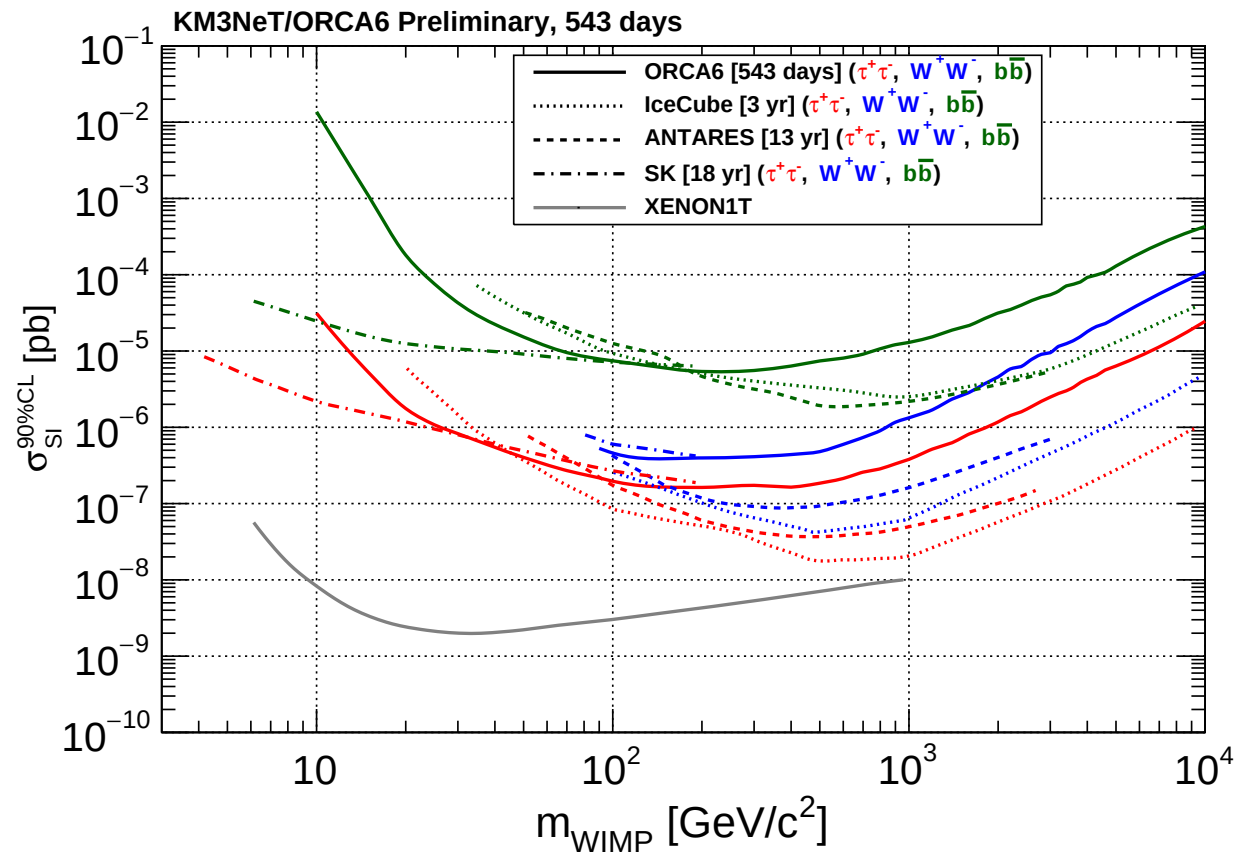

**Figure 4:** The 90% CL upper limits on the spin-independent WIMP-nucleon cross section as a function of the WIMP mass for each of the three annihilation channels. The full lines show the results obtained in this analysis, whereas the other lines show the upper limits obtained by IceCube [18], ANTARES [19], Super-Kamiokande [20] and XENON1T [22] (shown as a full line).

## 7. Acknowledgements

The authors acknowledge the support from grants Ministerio de Ciencia e Innovación (MCI): Programa Estatal para Impulsar la Investigación Científico-Técnica y su Transferencia - Subprograma Estatal de Generación de Conocimiento (reference PID2021-124591NB-C41), PID2021-124591NB-C43 funded by MCIN/AEI/10.13039/501100011033 and by "ERDF A way of making Europe", European Union; Generalitat Valenciana: Programa Santiago Grisolía (GRISOLIAP/2021/192, 2021-2025), and Program GenT (ref. CIDEGENT/2021/023).

# KM3NeT sensitivity to a flux of down-going nuclearites

**Alice Păun,**[a,b,*] **Gabriela Păvălaş**[a] **and Vlad Popa**[a] **on behalf of the KM3NeT Collaboration**

[a] *Institute of Space Science, Atomiştilor 409, Măgurele, Romania*
[b] *University of Bucharest – Faculty of Physics, Atomiştilor 405, Măgurele, Romania*

*E-mail:* alice.paun@spacescience.ro

Nuclearites are hypothetical, massive particles of Strange Quark Matter (SQM) introduced by E. Witten in 1984. They are composed of approximately equal quantities of up, down, and strange quarks. Due to the third quark flavor component which leads to a total energy lower than in the case of nuclear matter, SQM could represent the ground state of Quantum Chromodynamics (QCD). The detection and characterization of nuclearites could also bring important contributions to the Dark Matter physics. KM3NeT is a network of deep-sea neutrino telescopes placed in the Mediterranean Sea, optimized for the search for high-energy cosmic neutrinos (KM3NeT/ARCA) and the study of neutrino properties with atmospheric neutrinos (KM3NeT/ORCA). Nuclearites above a mass threshold of $10^{13}$ GeV/c$^2$ having velocities of approximately 250 km/s could reach the ground and interact in the detector through elastic collisions. A fraction of the energy generated in these collisions is dissipated as visible blackbody radiation. A customized Monte Carlo (MC) program was used to simulate the propagation and the signal of nuclearites inside the KM3NeT detector. The background considered for this study is represented by the $^{40}$K decay, which is added during the filtering stage of the MC output, and by the simulated atmospheric muons. The analysis uses selection cuts in order to remove the main background of atmospheric muons and to estimate the sensitivity of the detector. Preliminary results on the sensitivity of the KM3NeT neutrino telescope to a flux of massive down-going nuclearites will be presented.

*Speaker

## 1. Introduction

Nuclearites, or Strange Quark Matter (SQM) aggregates, are hypothetical slow (~250 km/s at the entry in the atmosphere), massive, and compact particles introduced by E. Witten in early 1984 [1, 2]. The theory assumes that they are composed of three quark flavors (up, down, and strange) [2] and their structure is similar to that of an atom (nucleus and electronic shell) [2, 3]. Due to the third quark flavor component which leads to a total energy lower than in the case of nuclear matter [4, 5], the strange quark matter could be the ground state of the Quantum Chromodynamics (QCD) [1, 6, 7]. The confirmation of this hypothesis could bring an important contribution to fundamental physics.

The detection of nuclearites could be done through the blackbody radiation emitted in the visible spectrum as a result of the elastic collision with the atoms of the transparent media along their paths [2]. Considering this feature, nuclearites could be detected in large-volume underwater neutrino telescopes.

As the Coulomb repulsion does not allow direct nuclear interactions with atoms due to the electronic shell, nuclearites interact predominantly through elastic and quasi-elastic collisions. The energy emitted in these collisions overheats the medium along the particle trajectory. In transparent media, such as water, a part of this energy is dissipated as blackbody radiation in the visible spectrum, denoted luminous efficiency ($\eta \approx 3 \cdot 10^{-5}$) [2]. The energy loss follows Equation 1 [2]:

$$\frac{dE}{dx} = -\sigma \rho v^2, \tag{1}$$

where $\sigma$ is the cross-section of the nuclearite, $\rho$ is the medium density and $v$ is the velocity of the particle. The velocity follows the Equation 2 [2]:

$$v(L) = v_0 \cdot e^{\frac{-\sigma}{M_N} \cdot \int_0^L \rho dx}, \tag{2}$$

where $L$ is the length of the path, $v_0$ is the nuclearite velocity at the entry into the atmosphere and $M_N$ is the nuclearite mass.

The cross-section of the nuclearites can be computed as a function of their mass (see Equation 3) [2]:

$$\sigma = \begin{cases} \pi \left(\frac{3M_N}{4\pi\rho_N}\right)^{\frac{2}{3}} \, cm^2, \; for \; M_N \geq 8.4 \cdot 10^{14} \; GeV \\ \pi \cdot 10^{-16} \, cm^2, \; for \; M_N < 8.4 \cdot 10^{14} \; GeV \end{cases} \tag{3}$$

where $\rho_N = 3.6 \cdot 10^{14}$ g/cm$^3$ is the density of the nuclearite.

The number of emitted photons per unit length is given by Equation 4 [2]:

$$\frac{dN_\gamma}{dx} = \eta \frac{dE/dx}{\langle E_\gamma \rangle}, \tag{4}$$

where $\langle E_\gamma \rangle \approx 3.14$ eV is the average energy of the photons emitted in the visible spectrum.

A very important aspect regarding the detection of nuclearites with such instruments is the constant background present at the detectors level [10]. The main background component for nuclearites at the detector depth is represented by atmospheric muons [8, 10], while the $^{40}$K decay and the bioluminescence (in the case of ORCA) also play an important role [8, 9].

## 2. The KM3NeT sensitivity to nuclearites

The KM3NeT detector is composed of two configurations of photomultiplier (PMT) arrays, currently under construction in the Mediterranean Sea: ARCA (Astroparticle Research with Cosmics in the Abyss) and ORCA (Oscillation Research with Cosmics in the Abyss). KM3NeT/ARCA will have two building blocks with 115 detection units (DUs) each, with the goal of identifying the high-energy cosmic neutrino sources. KM3NeT/ORCA will have a single, and more compact building block with 115 DUs. Its goal is to find a solution for the neutrino mass hierarchy problem. The two arrays are similar, both comprising DUs with 18 digital optical modules (DOMs) and 31 PMTs per DOM. However, the main differences are in the distribution of the detection instruments, size, and depth.

Massive particles of SQM that propagate through water could be detected by the light emitted from the blackbody radiation generated at the interaction of the particle with the atoms and molecules of the medium.

In this analysis, nuclearite events were generated with a dedicated MC program [11] for a mass range of $M_N \in [3 \cdot 10^{13} - 10^{17}]$ GeV/c$^2$ for both of the KM3NeT configurations. The initial flux of SQM particles considered in the simulation is 1000 events per mass. The MC production was carried out using the GRID platform hosted by the IN2P3/CNRS Computing Center.

In [11] it is shown that nuclearites that are propagated through underwater detectors, like KM3NeT or ANTARES, induce events that could be easily recognized by the large hits number and signal duration (> 1 ms).

In order to determine the feasibility of nuclearite detection with the KM3NeT underwater detector, the sensitivity of the detection instrument must be computed. This study takes into account simulated background noise due to the atmospheric muon flux at the depth of the detector. The background due to the $^{40}$K decay in sea water is also added in the processing step.

The strategy of the analysis aims to first establish the best selection criteria that could separate the nuclearite signal from other background noise present in the seawater that could be recorded by the detection instruments and then to determine the sensitivity of the detector to a flux of nuclearites. In order to make the selection, the relevant selection criteria are chosen and then a series of cuts are evaluated to establish the optimal ones for the considered selection criteria. This is done by minimizing the rejection factor (MRF method) [13]. The MRF technique consists of varying the cuts in small steps in order to obtain optimum cuts for the discrimination variables, that are leading to the maximization of the sensitivity results.

A KM3NeT analysis software has been used to separate the nuclearite signal from the background noise. The package includes a program that filters the MC data by using two muon trigger algorithms which are based on the maximum time-correlated light that can be observed on any DOM within the detector and the maximum time-correlated light that can be observed on any PMT within the detector under a track assumption.

As it was mentioned before, in the case of nuclearite detection the flux of down-going atmospheric muons is a crucial background component. To separate the muons signal from the nuclearite MC data, a MC atmospheric muon production (10 TeV and 50 TeV energy thresholds for ARCA and 1 GeV for ORCA) obtained with the MUPAGE software [12] was used.

Nuclearites are slow particles compared to atmospheric muons, which are relativistic. Thus, due to the large signal duration and strong luminous signal relative to muons, a single nuclearite can give rise to many triggered events. This nuclearite particularity can be observed also in Figure 1 and Figure 2 where it can clearly be seen that the distribution of the triggered events per timeslice in the case of nuclearites has a relatively wide range. A timeslice is a collection of all frames of the detector that correspond to the same timestamp.

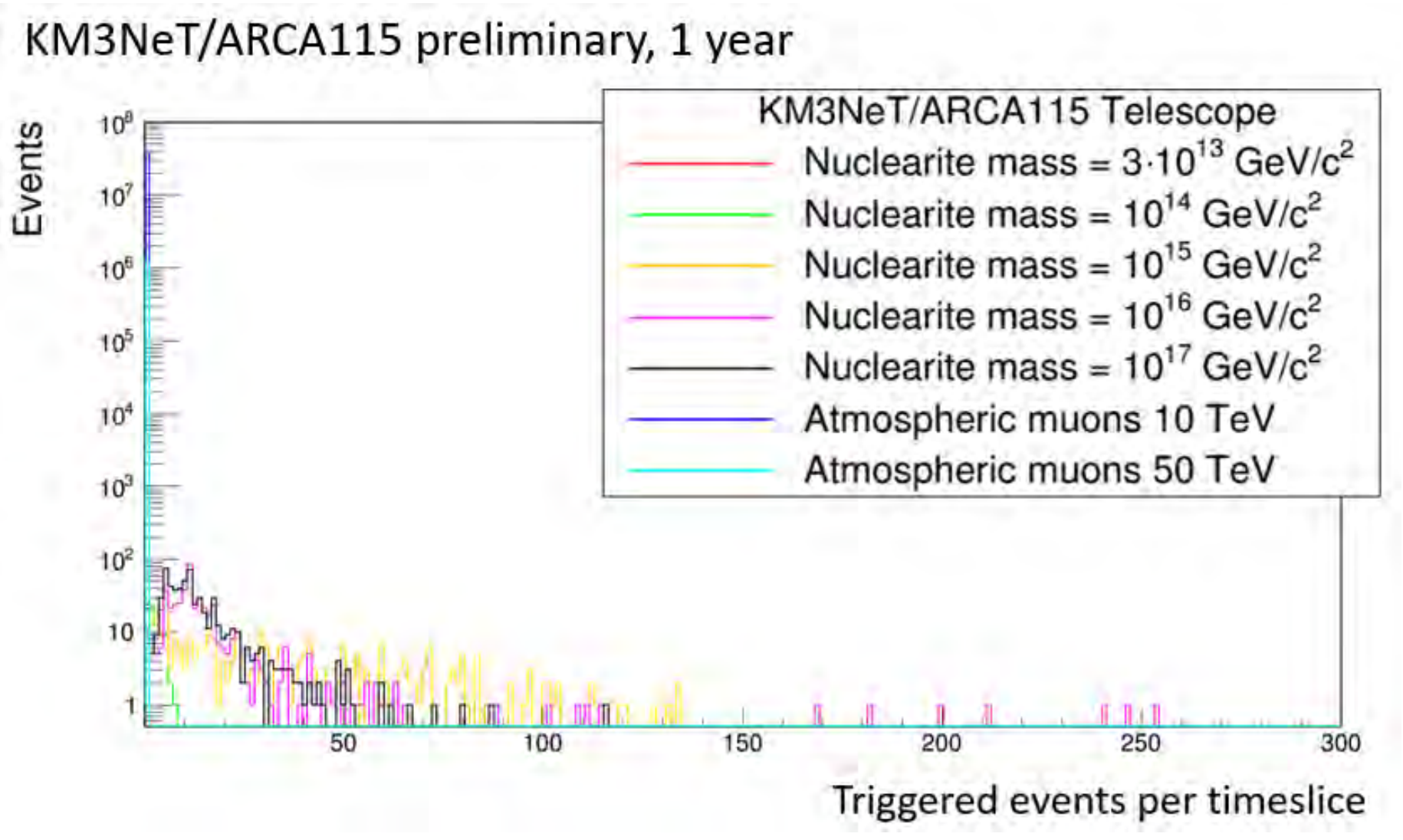

**Figure 1:** Distribution of the number of triggered events per timeslice in the case of KM3NeT/ARCA115.

Comparing Figures 1 and 2, one can observe there is an important difference regarding the multiplicity distribution between ARCA and ORCA. There are ORCA events with multiplicity values larger than for ARCA, particularly for smaller masses, as the triggered events are also more numerous. The characteristics mentioned above are a consequence of the higher PMT density of ORCA and are summarized in the first half of Tables 1 and 2.

After the filtering process, the next step is event selection. Three selection variables were considered for the nuclearite events: the triggered hits (an L1 trigger that occurs when at least two hits are detected in a given time interval at the same DOM), the snapshot hits (the raw hit information on all hits in a time interval larger than the triggered event) and the snapshot duration (the time interval between the first and the last snapshot hit: [$T_{MinTrig}$ - $T_{Extra}$, $T_{MaxTrig}$ + $T_{Extra}$], where $T_{MinTrig}$ is the time of the first triggered hit, $T_{MaxTrig}$ is the time of the last triggered hit and $T_{Extra}$ is the time offset for snapshot). The most promising discrimination variable to reduce atmospheric muon background is the snapshot duration (see Figure 3), as a larger number of nuclearite events

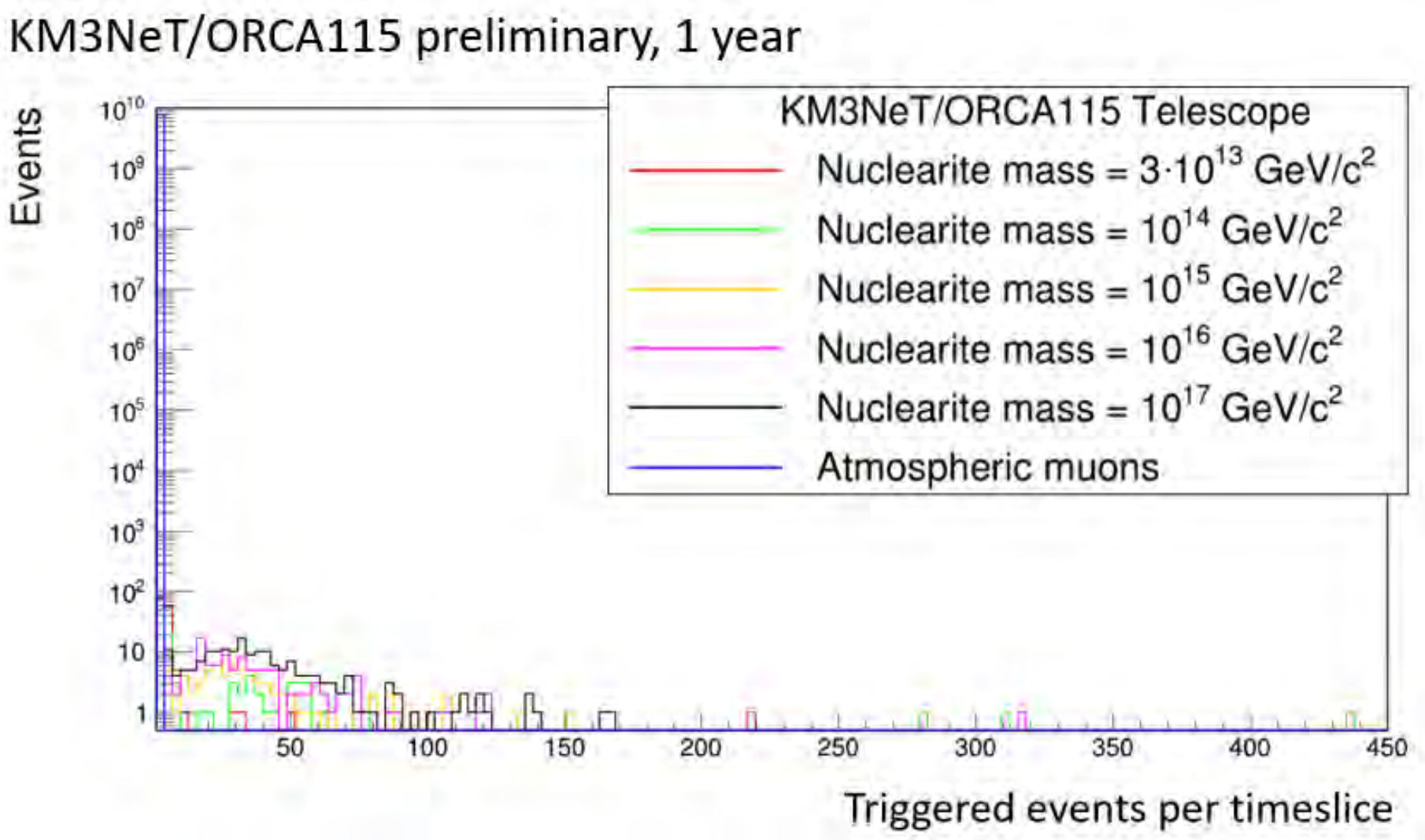

**Figure 2:** Distribution of the number of triggered events per timeslice in the case of KM3NeT/ORCA115.

pass this cut.

In the event selection of the data, it is very important to apply appropriate cuts in order to discriminate better between the noise and relevant signal. In this matter, for the optimization of the cuts, the atmospheric muon distributions were fitted and extrapolated by using a normalized form of the Gaussian function in the case of the snapshot duration variable for KM3NeT/ARCA and an exponential function for KM3NeT/ORCA. The optimal cuts for the two configurations are illustrated in Figure 3 by the red dashed lines. In this figure, the snapshot duration distribution is represented versus the MC events, considering the largest snapshot within the MC event.

| Nuclearite mass (GeV/c$^2$) | Remaining MC events | Remaining triggered events | Snapshot duration (MC) > 19000 ns |
|---|---|---|---|
| $3 \cdot 10^{13}$ | 21 | 21 | 0 |
| $10^{14}$ | 94 | 223 | 71 |
| $10^{15}$ | 344 | 11737 | 319 |
| $10^{16}$ | 468 | 8645 | 430 |
| $10^{17}$ | 609 | 8276 | 577 |
| Atmospheric muons | 9060195 (10 TeV)<br>3261452 (50 TeV) | 9060195 (10 TeV)<br>3261452 (50 TeV) | $4.92443 \cdot 10^{-6}$ |

**Table 1:** Output of the optimal cuts on the discrimination variables for KM3NeT/ARCA.

| Nuclearite mass (GeV/c$^2$) | Remaining MC events | Remaining triggered events | Snapshot duration (MC) > 6700 ns |
|---|---|---|---|
| $3 \cdot 10^{13}$ | 339 | 1062 | 1 |
| $10^{14}$ | 252 | 2992 | 29 |
| $10^{15}$ | 154 | 2832 | 62 |
| $10^{16}$ | 200 | 4651 | 101 |
| $10^{17}$ | 258 | 7792 | 144 |
| Atmospheric muons | 103935982 (1 GeV) | 103941853 (1 GeV) | $3.91748 \cdot 10^{-5}$ |

**Table 2:** Output of the optimal cuts on the discrimination variables for KM3NeT/ORCA.

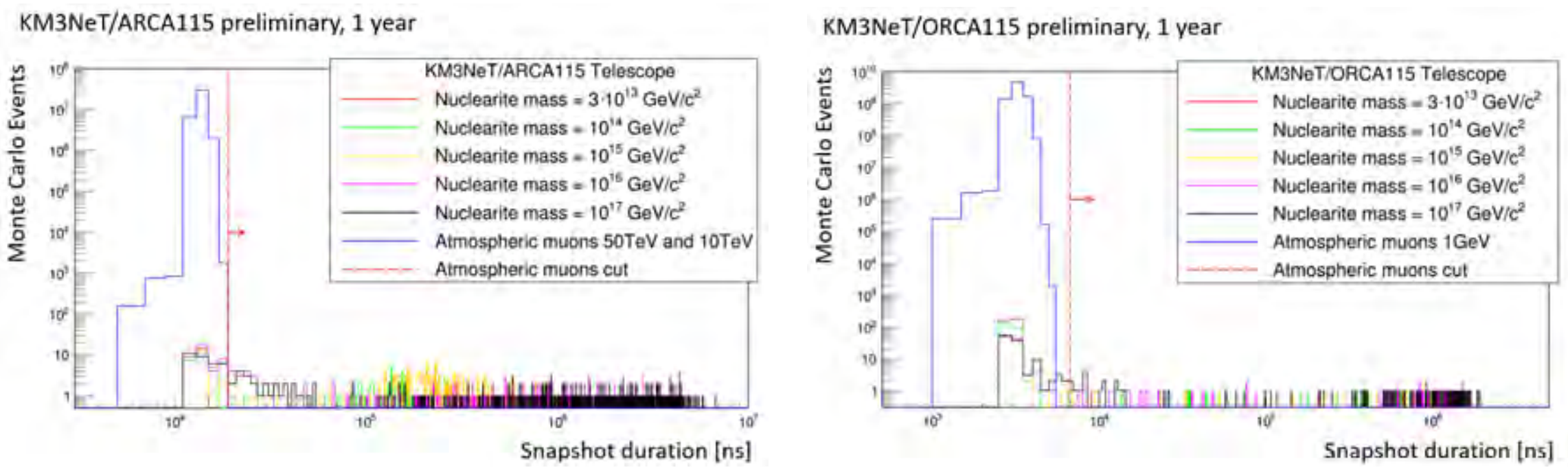

**(a)** KM3NeT/ARCA115 (cut at 19000 ns) **(b)** KM3NeT/ORCA115 (cut at 6700 ns)

**Figure 3:** Distribution of the snapshot duration for MC events, considering the largest snapshot from the MC event. The optimal cut is represented by the red dashed line.

After the evaluation of the cuts, the acceptance and sensitivity of KM3NeT/ARCA and KM3NeT/ORCA to a flux of down-going nuclearites were computed and the sensitivities were compared to the upper limits obtained for MACRO [14], SLIM [15] and ANTARES (839 days of 2009-2017) [16]. The sensitivities at 90% C.L were computed by using the Feldman-Cousins formula [17] (Equation 5), considering nuclearite events with a Poisson distribution.

$$\phi_{90} = \frac{\overline{\mu_{90}}}{A \cdot T}, \quad (5)$$

where $\overline{\mu_{90}}$ was determined from the expected background using the extrapolation of the atmospheric muon distribution, T corresponds to 1 year of live time, and A is the effective acceptance of the detector (see relation 6) [17]:

$$A = \frac{S \cdot N_{nucl}}{N_{sim}}, \quad (6)$$

where S is the area of the simulation hemisphere and $\frac{N_{nucl}}{N_{sim}}$ is the ratio of the number of nuclearite events that passed the selection cuts to the number of simulated events. The results for KM3NeT/ORCA115 and KM3NeT/ARCA230 are presented in Figure 4.

The results regarding the optimized cuts for ARCA and ORCA applied to the MC data for the snapshot duration variable are listed in Table 3 and 4 together with the background, MRF values and the sensitivity of the KM3NeT detector to massive nuclearites.

| Nuclearite mass (GeV/c$^2$) | Remaining MC events | Remaining triggered events | Background | MRF | Sensitivity 1yr (cm$^{-2}$sr$^{-1}$s$^{-1}$) |
|---|---|---|---|---|---|
| $10^{14}$ | 71 | 109 | $4.92443 \cdot 10^{-6}$ | 0.0343732 | $2.0856 \cdot 10^{-17}$ |
| $10^{15}$ | 319 | 3698 | $4.92443 \cdot 10^{-6}$ | 0.00765047 | $4.642 \cdot 10^{-18}$ |
| $10^{16}$ | 430 | 2410 | $4.92443 \cdot 10^{-6}$ | 0.00567558 | $3.4437 \cdot 10^{-18}$ |
| $10^{17}$ | 577 | 2450 | $4.92443 \cdot 10^{-6}$ | 0.00422964 | $2.5664 \cdot 10^{-18}$ |

**Table 3:** Output of the applied cut (19000 ns) on snapshot duration variable and the sensitivity estimation for KM3NeT/ARCA115 configuration to a flux of massive down-going nuclearites.

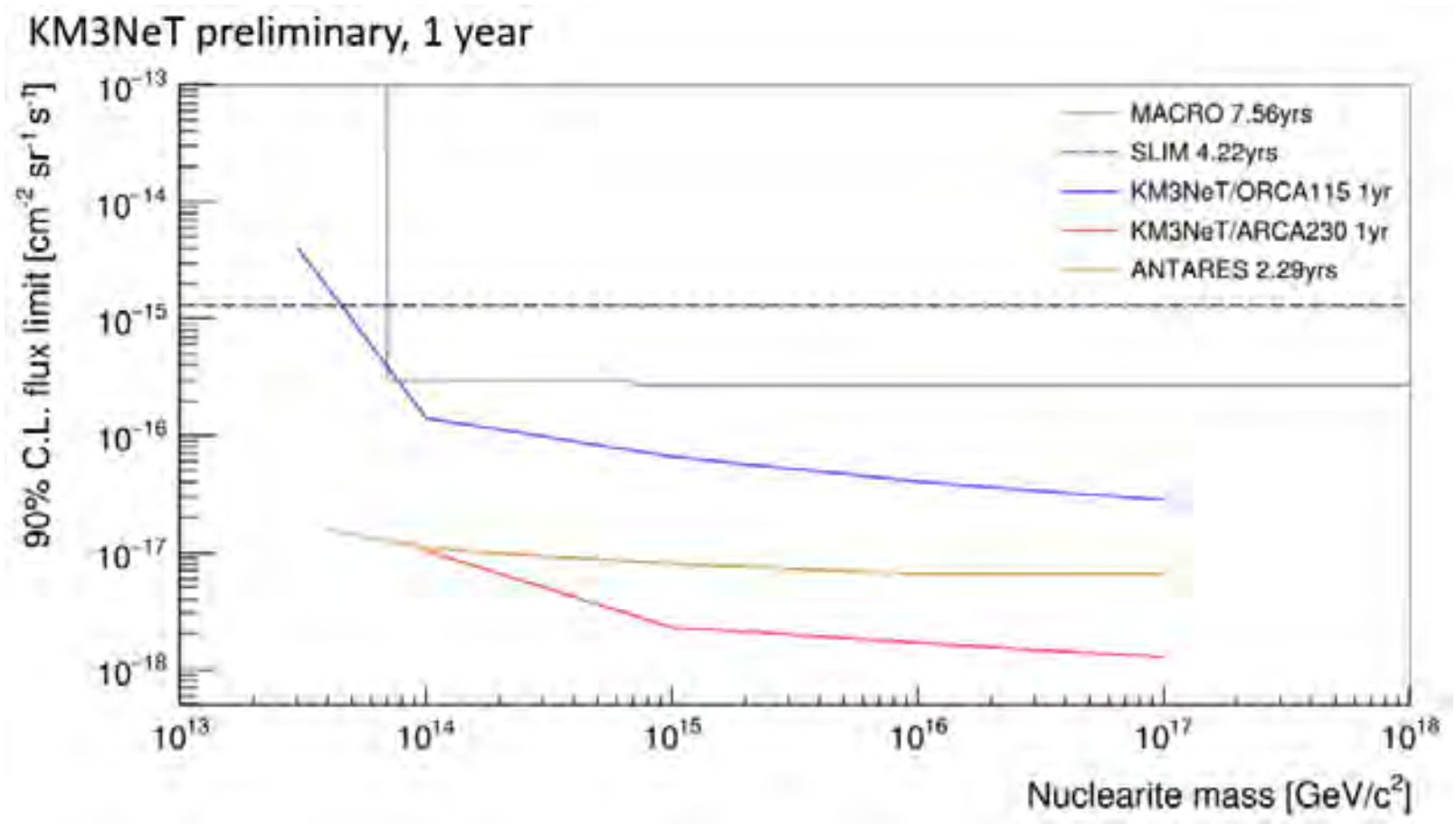

**Figure 4:** Preliminary sensitivities of ARCA and ORCA to a nuclearite flux compared to MACRO [14], SLIM [15], and ANTARES [16] upper limits.

| Nuclearite mass (GeV/$c^2$) | Remaining MC events | Remaining triggered events | Background | MRF | Sensitivity 1yr ($cm^{-2}sr^{-1}s^{-1}$) |
|---|---|---|---|---|---|
| $3 \cdot 10^{13}$ | 1 | 2 | $3.91748 \cdot 10^{-5}$ | 2.44003 | $4.10061 \cdot 10^{-15}$ |
| $10^{14}$ | 29 | 377 | $3.91748 \cdot 10^{-5}$ | 0.084139 | $1.414 \cdot 10^{-16}$ |
| $10^{15}$ | 62 | 391 | $3.91748 \cdot 10^{-5}$ | 0.0393553 | $6.61388 \cdot 10^{-17}$ |
| $10^{16}$ | 101 | 650 | $3.91748 \cdot 10^{-5}$ | 0.0241587 | $4.06 \cdot 10^{-17}$ |
| $10^{17}$ | 144 | 939 | $3.91748 \cdot 10^{-5}$ | 0.0169447 | $2.84764 \cdot 10^{-17}$ |

**Table 4:** Output of the applied cut (6700 ns) on snapshot duration variable and the sensitivity estimation for KM3NeT/ORCA115 configuration to a flux of massive down-going nuclearites.

## 3. Conclusions

The background noise is an important aspect in the detection of nuclearites with an underwater neutrino telescope. The nuclearite signal is influenced mainly by the continuous atmospheric muon flux, but also by the luminous signal produced by the decay process of $^{40}K$ always present in seawater. The signal produced by the bioluminescent organisms in the Mediterranean Sea also affects the signal of interest, but it was not considered in this analysis.

The preliminary sensitivity of KM3NeT/ARCA, determined for a flux of massive down-going nuclearites at 90% C.L. is comparable to and could improve the ANTARES upper limits obtained in [16] for 839 days of 2009-2017 data. The estimated sensitivity for KM3NeT/ORCA includes also the lightest simulated nuclearites, but is poorer than the sensitivity of KM3NeT/ARCA and ANTARES. At this point in the analysis, we can say that KM3NeT/ARCA configuration presents better characteristics for the detection and characterization of these exotic particles.

# Search for dark matter towards the Sun with the KM3NeT/ORCA6 neutrino telescope

**Miguel Gutiérrez,[a,*] Adrian Šaina,[b] Sergio Navas[a] and Sara Rebecca Gozzini[b] on behalf of the KM3NeT Collaboration**

[a] *University of Granada,*
*Dpto. de Física Teórica y del Cosmos & C.A.F.P.E., 18071 Granada, Spain*
[b] *IFIC - Instituto de Física Corpuscular (CSIC - Universitat de València),*
*c/Catedrático José Beltrán, 2, 46980 Paterna, Valencia, Spain*

*E-mail:* mgg@ugr.es, adrians@ific.uv.es

Dark matter is acknowledged to exist at different scales in the Universe. Cosmological observations strongly suggest that approximately 25% of the overall energy density of the Universe is attributed to dark matter, while roughly 5% is composed of baryonic matter. Since the neutrinos produced in pair-annihilation or decay of weakly interacting massive particles (WIMPs) could be observed by neutrino telescopes, these instruments provide an important complementarity in the quest for detecting dark matter signals. An excess of signal could be observed in regions where dark matter might accumulate, e.g., the Sun. The KM3NeT telescope is composed of two detectors, namely, ARCA (Astroparticle Research with Cosmics in the Abyss) and ORCA (Oscillation Research with Cosmics in the Abyss). The energy threshold of the latter is $\sim$ 1 GeV. Given the fact that neutrinos above 1 TeV are typically absorbed before they can escape from the Sun, the energy range of the KM3NeT/ORCA detector is optimal for the search of dark matter signals coming from our star. In this contribution, the search for WIMP annihilation signals coming from the Sun is presented. An unbinned likelihood method is used to discriminate the signal from the background in a 543-day livetime sample of data collected with the 6 first detection units of the ORCA detector. The limits on the neutrino flux and on the spin-dependent and spin-independent cross sections are given for three different annihilation channels. No evidence for a dark matter signal over the expected background has been found.

*Speaker

## 1. Introduction

Dark matter is one of the most intriguing puzzles in the modern particle physics picture. Cosmological and astrophysical observations, including the anisotropies in the cosmic microwave background radiation and the rotational velocities of galaxies, provide a solid evidence of the existence of dark matter. In particular, it is predicted that only 5% of the total energy density of the Universe belongs to baryonic matter, while 25% is attributed to dark matter, and the remaining 70%, to dark energy. A common hypothesis postulates that dark matter is made of weakly interacting massive particles (WIMPs).

While direct search experiments are designed to search for the collision of a dark matter particle with a nucleus, indirect search experiments aim to detect the neutrinos produced in a WIMP pair-annihilations. Since the average energy density of dark matter in the Galactic Halo at the location of the Sun is about $0.3\,\mathrm{GeV/cm^3}$, it is preferable to search for a signal of dark matter annihilations in places where dark matter might accumulate. Some of the most promising candidates are the Sun [1], the Earth and the Galactic Centre.

As the Sun is moving in the Milky Way, a wind of dark matter particles permeates through it. Even if most of these particles cross the Sun without interacting, a fraction of them collides with the nuclei present in it. If the collision is hard enough, the dark matter particle may lose a significant amount of energy and therefore remain gravitationally trapped in the Sun. This will increase the dark matter density in our star, which makes WIMP pair-annihilations more likely to happen.

The expected flux of neutrinos at the Earth surface due to WIMP pair-annihilations in the Sun is given by the following expression:

$$\frac{\mathrm{d}\Phi}{\mathrm{d}E} = \frac{\Gamma}{4\pi D_{\mathrm{SE}}^2}\frac{\mathrm{d}N}{\mathrm{d}E}\,, \tag{1}$$

being $\frac{\mathrm{d}N}{\mathrm{d}E}$ the so-called neutrino yield – i.e. the number of neutrinos per unit of energy that are emitted in one annihilation–, $D_{\mathrm{SE}}$ the distance from the Sun to the Earth and $\Gamma$ the annihilation rate of WIMP particles in the Sun.

The annihilation rate is strongly correlated with the capture rate, $C_r$. In particular, the following relationship can be established:

$$\Gamma = \frac{1}{2}C_r \tanh^2\left(\frac{t}{t_\odot}\right)\,, \tag{2}$$

being $t_\odot$ a characteristic time of the Sun, and $t$ its lifetime. A situation of equilibrium is often assumed, which implies that $t/t_\odot \gg 1$ and, consequently:

$$\Gamma = \frac{1}{2}C_r\,. \tag{3}$$

The capture rate of dark matter in the Sun is related to the WIMP-nucleon scattering cross section, which can be spin-dependent ($\sigma^{\mathrm{SD}}$) or spin-independent ($\sigma^{\mathrm{SI}}$). Therefore, a relationship between the WIMP-nucleon cross section and the flux of neutrinos can be established:

$$\sigma^{\mathrm{SD,SI}} = K^{\mathrm{SD,SI}}\Phi_{\nu+\bar{\nu}}\,, \tag{4}$$

where $K$ is the so-called conversion factor, which is computed using the software package DarkSUSY [2] in this analysis. The conversion factor contains information about the WIMP pair-annihilation channel, the mass of the WIMP and the energy threshold of the detector.

## 2. The KM3NeT neutrino telescope

KM3NeT [3] is a neutrino research infrastructure under construction at the bottom of the Mediterranean Sea, which consits of two different detectors: the KM3NeT/ARCA (Astroparticle Research with Cosmics in the Abyss) detector, located off-shore Sicily, Italy, and the KM3NeT/ORCA [4] (Oscillation Research with Cosmics in the Abyss) detector, located off-shore Toulon, France. The former has been designed to detect high-energy neutrinos, which makes it best suited for astronomy and astrophysics research in the TeV−PeV energy range. ORCA is optimised for the detection of atmospheric neutrinos with the primary goal of measuring neutrino properties like the oscillation parameters and the neutrino mass ordering. For this reason, the ORCA detector is optimised at lower energies, ranging from 1 GeV to 1 TeV.

Both the ARCA and ORCA detectors are composed of Digital Optical Modules (DOMs) [5], arranged in vertical strings called Detection Units (DUs), which have a segmented photo-sensitive area provided by 31 photomultiplier tubes (PMTs) housed in a 17" glass sphere. The Cherenkov light induced by the ultra-relativistic charged particles produced in the neutrino interactions nearby the detector can produce signals (*hits*) in the PMTs.

On its final configuration, the ARCA (ORCA) detector will be composed of 230 (115) detection units. The dataset analysed in this work has been taken with the 6 DUs configuration of the ORCA detector, referred to as ORCA6, in the period from January 2020 to November 2021.

## 3. Methods

In this work, the dataset of ORCA6 is analysed to search WIMP pair-annihilation signals coming from the Sun. After rejecting the runs that do not survive the quality checks, the sample analysed has a total livetime of 543 days.

Three WIMP annihilation channels are explored as benchmarks:

$$\text{WIMP} + \text{WIMP} \rightarrow \tau^+\tau^- \,, W^+W^- \,, b\bar{b} \,. \tag{5}$$

The yields for these annihilation modes are computed by the PYTHIA-based program WimpSim [6].

Based on the event topology, events can be classified as *shower-like* or *track-like*. On the one hand, shower-like events are produced in all neutral current neutrino interactions, as well as in charged current $\nu_e$ and $\nu_\tau$ interactions. They exhibit the characteristic formation of electromagnetic and hadronic showers, wherein a significant fraction of energy is promptly dissipated. On the other hand, track-like events arise mostly from charged current $\nu_\mu$ interactions. These events are distinguished by the generation of a leading muon. Being a charged particle, the muon's path through the detector materialises as a substantial emission of Cherenkov light, that produce signals in the photosensors used for the reconstruction of the neutrino direction and energy.

The analysis presented in this contribution is optimised on the selection of the track-like events, exhibiting a better angular reconstruction than the shower-like events. In addition, a set of cuts is used to reject background, mainly composed of atmospheric muons and atmospheric neutrinos, as well as to optimise the sensitivity on the flux of neutrinos. A first set of quality cuts is applied on data to reject badly reconstructed events and to reduce muon contamination, ensuring good agreement between data and Monte Carlo. In a second step, the set of cuts on $\beta$ (the estimated error in the reconstructed muon track direction), the total number of detected hits in the event, and the track likelihood (the likelihood of the track reconstruction algorithm) that optimise the sensitivity on the neutrino flux, is searched for.

The analysis method is based on the generation of pseudo-experiments (PEs), each of them consisting of a skymap populated with $n_b$ background events and $n_s$ injected signal events. Then, an unbinned likelihood analysis is performed over the skymaps to determine the most likely number of signal events within them. The ingredients of the likelihood fuction are the point spread function (PSF), which is the probability distribution of the angular distance to the Sun, and the probability density function (PDF), which is a two-dimensional probability distribution of the number of hits and the $\beta$ parameter. The corresponding signal distributions are built from Monte Carlo simulations, weighting the events by the WIMP pair-annihilation spectra. The background PDF and PSF distributions are built from the scrambled data.

The likelihood function is expressed as

$$\ln \mathcal{L}(n_s) = \sum_{i=1}^{N_{\text{tot}}} \ln \left[ n_s \mathcal{S} \left( \Psi_{\odot,i}, \beta_i, N_{\text{hits},i} \right) + n_b \mathcal{B} \left( \Psi_{\odot,i}, \beta_i, N_{\text{hits},i} \right) \right] - (n_s + n_b) \, , \tag{6}$$

where $\mathcal{S}$ and $\mathcal{B}$ denote the signal and background probability density functions, $\Psi_\odot$ denotes the angular distance to the Sun and $N_{\text{tot}} = n_s + n_b$ is the total number of events in the skymap. In order to reduce the computation time, for each skymap only the events inside a 30° cone around the position of the Sun are computed in the likelihood. The events outside this cone are treated as background: $\mathcal{S} = 0$ and $\mathcal{B} = 1$.

For each pseudo-experiment, the likelihood function is maximised with respect to $n_s$. The test statistic (TS) is calculated as

$$\text{TS} = \log_{10} \left( \frac{\mathcal{L}(\hat{n}_s)}{\mathcal{L}(0)} \right) , \tag{7}$$

where $\hat{n}_s$ is the number of signal events that maximises the likelihood. $\mathcal{L}(0)$ corresponds to the likelihood of the hypothesis that all the events originate from the atmospheric background, i.e., the null hypothesis.

A Poissonian smearing is performed over the TS distributions in order to simulate statistical fluctuations. In addition to this, a 15% Gaussian smearing is applied to include the systematic uncertainties in the number of detected events. This value was obtained by simulating events with a modified absorption length of photons: the variability in this factor had the greatest impact on the quantity of detected events. The smearings are performed as follows:

$$P(\text{TS}|\mu) = \sum_{i=0}^{n_{\text{inj}}^{\text{max}}} P(\text{TS}|i) \int_{\mu-4\sigma_\mu}^{\mu+4\sigma_\mu} P(i|\bar{\mu}) G(\bar{\mu}|\mu, \sigma_\mu) d\bar{\mu} \, . \tag{8}$$

Following the Neyman approach [7], an average upper limit on the number of signal events, $n_s$, is computed comparing each signal $P(\text{TS})$ distribution with the median of the background TS distribution. The sensitivity, $n_{90}$, is defined as the 90% CL upper limit for a measurement equal to the median of the background TS distribution. The sensitivity on the flux is obtained as

$$\Phi^{90}_{\nu+\bar{\nu}} = \frac{n_{90}}{Acc \times T}\,, \tag{9}$$

being $Acc$ the detector acceptance to the signal, and $T$ the livetime of the dataset. The acceptance is computed as

$$Acc = \int_{E_{\text{th}}}^{M_{\text{WIMP}}} A_{\text{eff}}(E)\,\frac{\mathrm{d}N}{\mathrm{d}E}\mathrm{d}E\,. \tag{10}$$

| WIMP mass range | $\beta$ | $N_{\text{hits}}$ | Track likelihood |
|---|---|---|---|
| $10\,\text{GeV} < m_\chi < 300\,\text{GeV}$ | $< 1°$ | $> 20$ | $> 120$ |
| $300\,\text{GeV} < m_\chi < 10\,\text{TeV}$ | $< 1°$ | $> 20$ | $> 60$ |

**Table 1:** Sets of cuts that optimise the sensitivity on the neutrino flux for the $\tau^+\tau^-$ channel and the $W^+W^-$ channel.

| WIMP mass range | $\beta$ | $N_{\text{hits}}$ | Track likelihood |
|---|---|---|---|
| $10\,\text{GeV} < m_\chi < 300\,\text{GeV}$ | $< 1°$ | $> 40$ | $> 120$ |
| $300\,\text{GeV} < m_\chi < 10\,\text{TeV}$ | $< 0.9°$ | $> 20$ | $> 80$ |

**Table 2:** Sets of cuts that optimise the sensitivity on the neutrino flux for the $b\bar{b}$ channel.

Here $\frac{\mathrm{d}N}{\mathrm{d}E}$ is the spectrum given by WimpSim, $E_{\text{th}}$ is the energy threshold (1 GeV for ORCA) and $M_{\text{WIMP}} \in (10\,\text{GeV}, 10\,\text{TeV})$ is the mass of the WIMP. $A_{\text{eff}}$ denotes the effective area of the detector, defined as the surface of an ideal detector with an efficiency of 100%. This quantity can only be obtained from the simulation, as the ratio between detected events and generated events.

Table 1 shows the combination of cuts that optimises the sensitivity for the $\tau^+\tau^-$ and $W^+W^-$ annihilation channels, while Table 2 shows the same for the $b\bar{b}$ channel.

Finally, Equation 4 is used to obtain the sensitivities on the cross sections for the spin-dependent and the spin-independent interactions.

## 4. Results

The KM3NeT/ORCA6 dataset has been analysed to search for dark matter in the Sun, considering WIMP masses in the range from 10 GeV to 10 TeV. After the analysis of the 543 days of livetime, the TS values obtained from this dataset are compatible with the background hypothesis for all combinations of WIMP masses and annihilation channels.

Figure 1 shows the 90% CL upper limits on the neutrino flux for the three channels explored. As neutrinos with energies above 1 TeV are absorbed before they can escape from the Sun, the spectra of the WIMP pair-annihilation for masses ranging from 1 TeV to 10 TeV are highly correlated. For this reason, the limit on the flux tends to flatten for masses in the TeV scale.

Figures 2 and 3 show the 90% CL upper limits on the spin-dependent and the spin-independent cross sections, obtained using the conversion factors (see Equation 4).

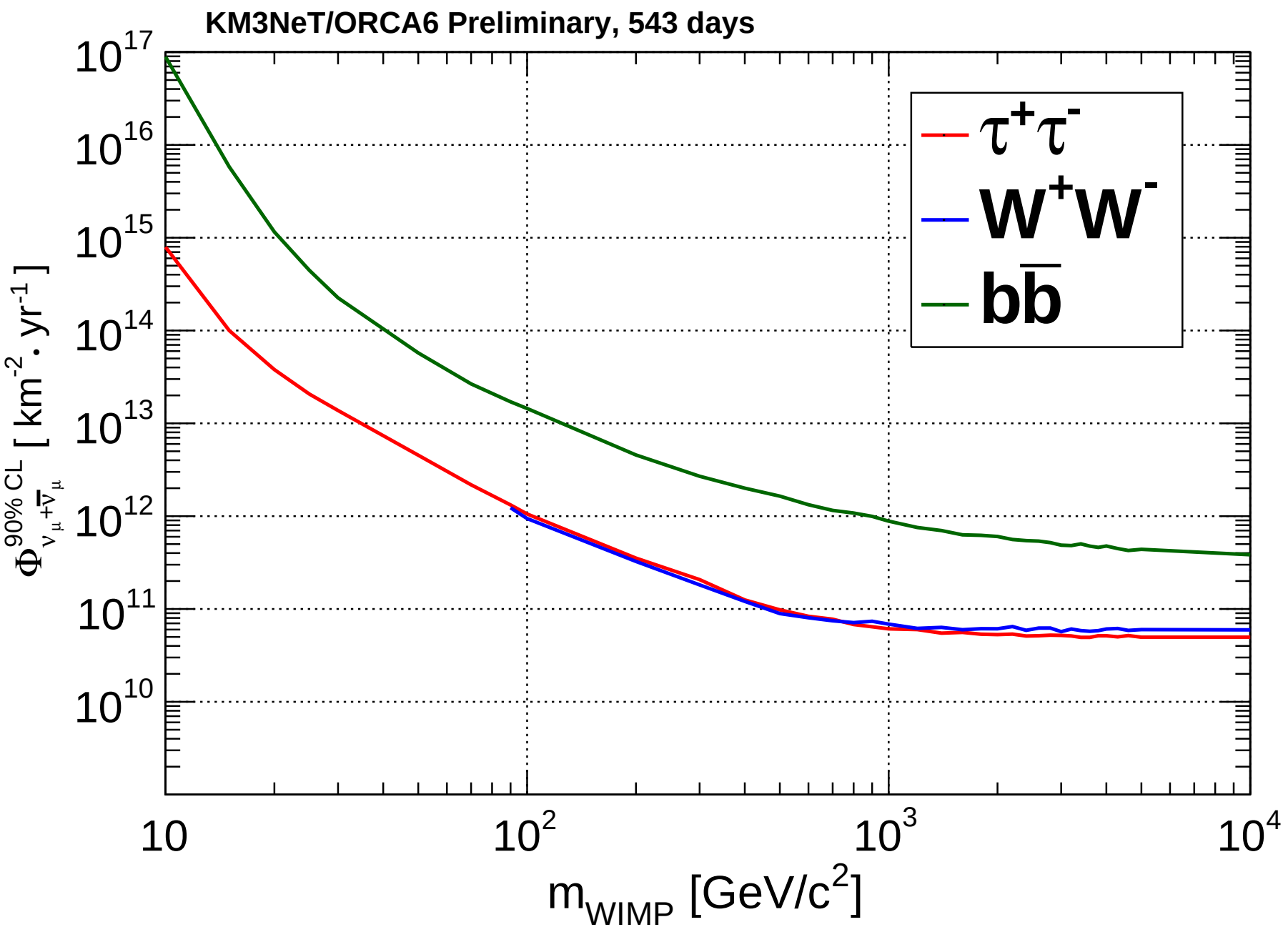

**Figure 1:** The 90% CL upper limits on the flux of neutrinos as a function of the WIMP mass for each of the three annihilation channels.

## 5. Conclusions

In this contribution, the initial findings of dark matter annihilation searches from the Sun using 543 days of data collected with the KM3NeT/ORCA detector in its 6 DUs configuration are presented. The KM3NeT/ORCA detector offers extensive coverage of the WIMP mass parameter space, namely from 1 GeV to 1 TeV. The signal has been searched for using an unbinned likelihood method in three different annihilation channels ($\tau^+\tau^-$, $W^+W^-$ and $b\bar{b}$ ). As no dark matter signal has been found in the data over the expected background, limits on the neutrino flux, and on the spin-dependent and the spin-independent cross sections have been established. This detector has already achieved competitive results during its early construction phase, even outperforming its predecessor, ANTARES [13], in some specific regions of the parameter space.

## 6. Acknowledgements

The authors acknowledge support from Grants PID2021-124591NB-C41, -C43 funded by MCIN/AEI/10.13039/501100011033 and by "ERDF A way of making Europe", European Union; Generalitat Valenciana: Programa Santiago Grisolía (GRISOLIAP/2021/192, 2021-2025); and Program GenT (CIDEGENT/2020/049).

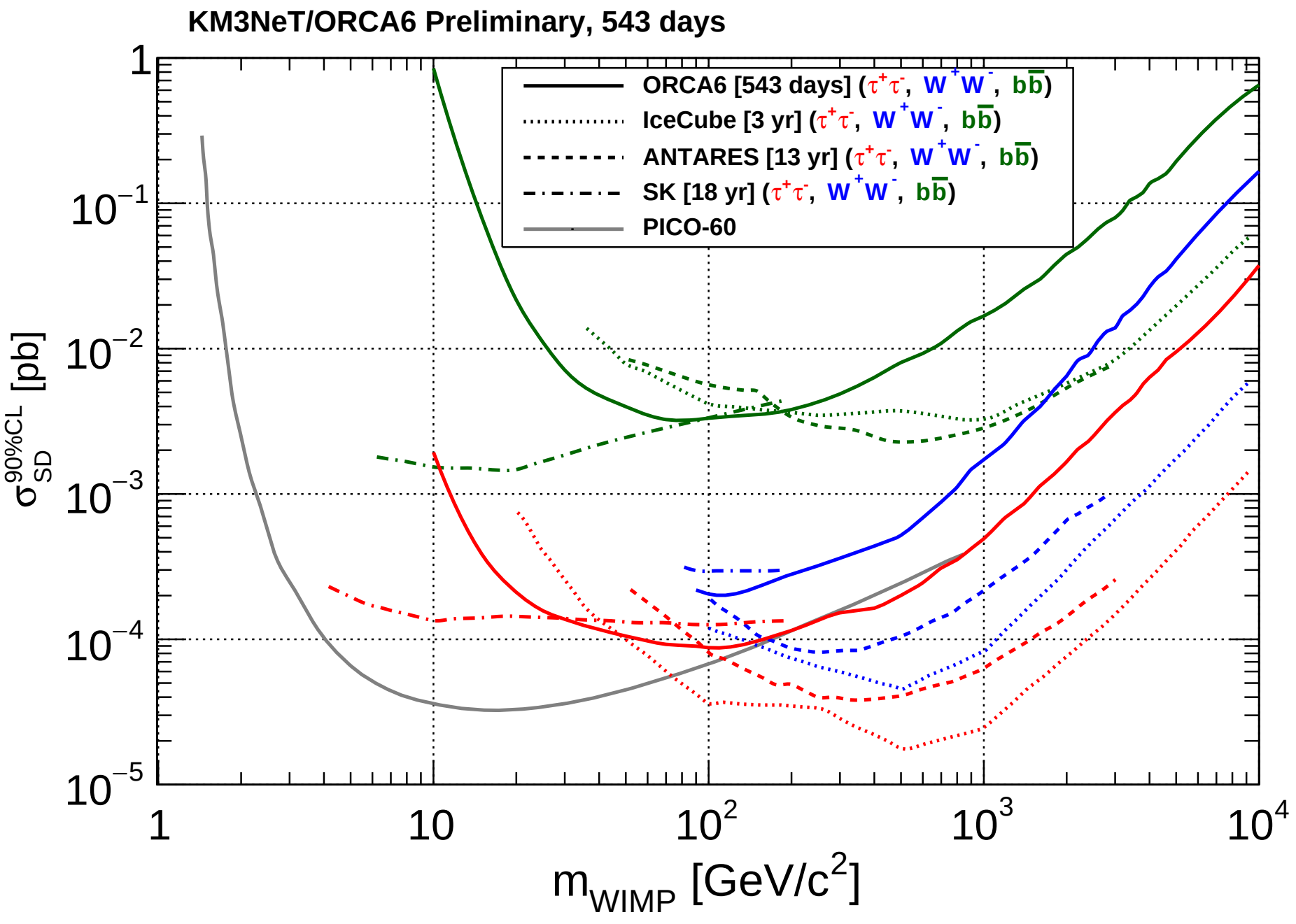

**Figure 2:** The 90% CL upper limits on the spin-dependent WIMP-nucleon cross section as a function of the WIMP mass for each of the three annihilation channels. The full lines show the results obtained in this analysis, whereas the other lines show the upper limits obtained by IceCube [8], ANTARES [9], Super-Kamiokande [10] and PICO-60 [11] (shown as a full line).

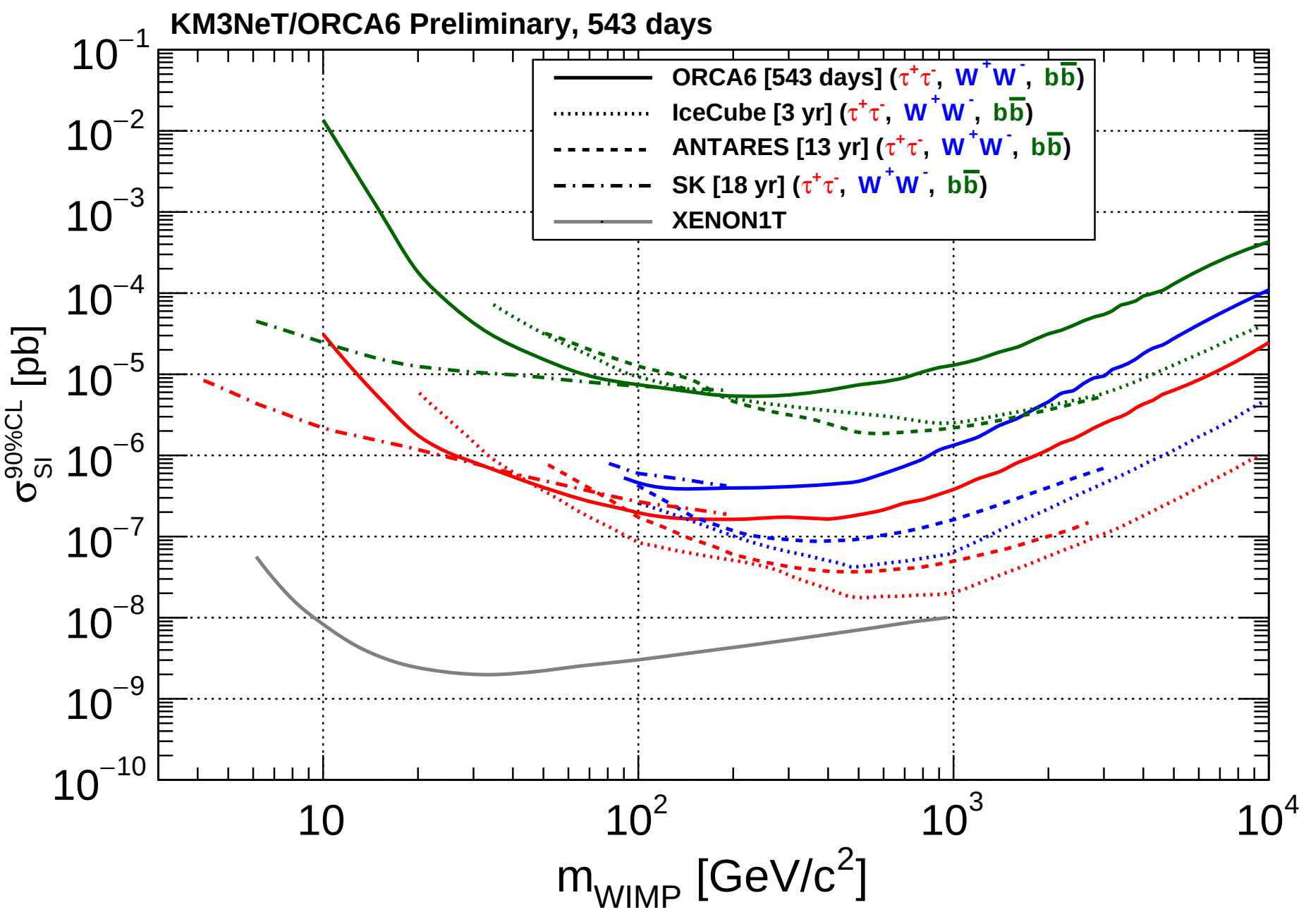

**Figure 3:** The 90% CL upper limits on the spin-independent WIMP-nucleon cross section as a function of the WIMP mass for each of the three annihilation channels. The full lines show the results obtained in this analysis, whereas the other lines show the upper limits obtained by IceCube [8], ANTARES [9], Super-Kamiokande [10] and XENON1T [12] (shown as a full line) .

# Time, position and orientation calibration using atmospheric muons in KM3NeT

**Louis Bailly-Salins**[a,*] **on behalf of the KM3NeT collaboration**

[a] *Université de Caen Normandie, ENSICAEN, CNRS/IN2P3, LPC Caen UMR 6534, F-14000 Caen, France*

*E-mail:* baillysalins@lpccaen.in2p3.fr

The KM3NeT collaboration operates two Cherenkov neutrino telescopes in the deep Mediterranean sea, ORCA and ARCA. Both detectors consist of an array of light-sensitive detectors called Digital Optical Modules (DOMs) assembled along vertical strings anchored to the seafloor. Although the abundant muon flux from cosmic ray air showers is a background for the main scientific objectives of KM3NeT/ORCA and KM3NeT/ARCA, it can be exploited in various ways for calibration purposes. In this contribution, the methods implemented within the KM3NeT calibration workflow which exploit the muon track reconstruction are presented. For the latter, a likelihood fit of a track hypothesis to a set of observed hits on the DOMs is performed. For calibration purposes, the optimal position, time reference and orientation of each string of the detector can be found as the parameters maximizing the overall likelihood of reconstructed muon tracks. The muon track quality method is shown to reach the desired accuracy in time and position, allowing for the determination of the relative time offsets between strings. It also represents an important tool to cross-check position and orientation calibrations obtained by other means. Another muon-based calibration method is used to determine the relative time offsets between DOMs. It is based on the evaluation of the difference between the measured hit time and the one predicted from the fitted muon track's position.

*Speaker

# 1. Introduction

The KM3NeT collaboration [1] operates two Cherenkov neutrino telescopes in the deep Mediterranean sea. ORCA (*Oscillation Research with Cosmics in the Abyss*) is designed to measure atmospheric neutrinos oscillations, while ARCA (*Astroparticle Research with Cosmics in the Abyss*) is designed to detect neutrinos from astrophysical sources. ARCA and ORCA share the same technology and detector elements. Both detectors are instrumented with photomultiplier tubes (PMTs) for the detection of the Cherenkov light emitted by the relativistic charged particles produced in neutrino interactions. KM3NeT detectors consist in a 3-D array of glass-spheres named Digital Optical Modules (DOMs) housing 31 3-inch PMTs each. The DOMs are arranged along vertical strings called Detection Units (DU). Each DU hosts 18 DOMs, is anchored to the seabed and remains vertical due to the buoyancy of the DOMs and to a buoy tied on its top. When completed, each detector "building block" will consist of 115 DUs arranged side by side following a cylindrical footprint. The geometry of ARCA is optimised to maximise its detection efficiency in the energy range 1 TeV-10 PeV, while ORCA is built to detect neutrinos in the range 1-100 GeV. This translates into a different spacing between the DOMs: the vertical and horizontal distances between the DOMs of ORCA are respectively 9 meters and 20 meters, while for ARCA, the corresponding distances are 36 and 90 meters.

To be able to reconstruct accurately the direction of neutrino events, the PMTs need to be synchronized with a 1 ns precision, and their position determined with an accuracy of 20 cm. The orientation of the optical modules must also be known with a precision of 3°. To calibrate the detector in time, position and orientation, several procedures using data from dedicated instruments located inside the DOMs [2], on DU bases or on the seafloor, as well optical data from the PMTs, are used.

# 2. Calibration of KM3NeT detectors

## 2.1 Position and orientation calibration

The KM3NeT positioning system [3] relies on a set of acoustic emitters and receivers. The emitters are beacons anchored on the seabed near the detector. There are two sets of acoustic receivers: hydrophones located on the DU bases, and piezoelectric sensors glued at the South pole of the DOMs. The position of each DOM is determined using triangulation of the acoustic signals, constrained by a mechanical model for the DUs. For the orientation calibration, an attitude and heading reference system (AHRS), referred to as "compass", is used. It consists of a set of accelerometers and magnetometers mounted on the electronics boards of each DOM. An important specificity of KM3NeT detectors is that the DOMs move and rotate over time due to sea currents. Thus, dynamic position and orientation calibration procedures have been developped [4], updating the positions and orientations of the DOMs every 10 minutes. The expected accuracies of those dynamic position and orientation calibrations are less than 10 cm and a few degrees, respectively.

## 2.2 Time calibration

Because of delays arising at different levels of the infrastructure, the time calibration of KM3NeT detectors must be achieved at various scales [5] to synchronize all PMTs:

- inter-PMT (or intra-DOM): synchronize the individual photomultiplier tubes inside a given DOM. This is done using the light resulting from the $\beta^-$ decay of Potassium-40 naturally present in sea water, which can be detected as coincident hits on a few neighboring PMTs.
- inter-DOM: synchronize the optical modules of a given detection unit. A preliminary calibration is done before deployment of the DU in the sea, using a laser source. In-situ, light pulse emitters located on the DOMs, called nanobeacons and developped by the collaboration [6], are used to adjust the inter-DOM time calibration[7]. Atmospheric muons can also be used for cross-checks (see Section 3.4).
- inter-DU: synchronize the detection units forming the building block. There is currently no dedicated instrument for this inter-DU time calibration. Instead, the muon track quality method, using the reconstructed tracks of the detected atmospheric muons, has been developped. This method will be described in the following sections of the current contribution.

## 3. Calibration with reconstructed atmospheric muon tracks

### 3.1 Atmospheric muons in KM3NeT

Cosmic rays are charged particles arriving on the Earth atmosphere with very high kinetic energy. They interact with air nuclei in the upper atmosphere and produce showers. Among the charged secondary particles produced, muons are the most penetrating. The atmospheric muons with enough energy can reach KM3NeT/ORCA and KM3NeT/ARCA [8]. Although forming a background for the main physics goals of KM3NeT, atmospheric muons can be used for cosmic ray physics, as well as for checking the detector performance [9] and for calibration.

### 3.2 Muon track reconstruction

During data acquisition, a *hit* is produced when a photon reaching a PMT induces an electrical signal above a defined threshold. The hit information includes a time stamp, a time-over-threshold (not relevent in the following), and the geometrical properties of the PMT detecting the hit: position and orientation. A series of causally-connected hits forms an *event*. Muon track reconstruction algorithms process an event by fitting its observed hits with the hypothesis of a straight track of a muon emitting Cherenkov light at a fixed angle along its trajectory. Fitting a muon track to observed hits on PMTs in a non-linear problem. In KM3NeT, an approach with several consecutive steps is employed. The principal step of the fit, determining the direction and vertex position of the track, adopts a maximum likelihood approach. The likelihood is a quantity which describes the agreement between the track hypothesis and the properties of the observed hits. In other words, it describes the *quality* of the track hypothesis. The *best track* is the one maximizing the likelihood.

### 3.3 Track quality method

The principle of the track quality method (TQM) [10] is to add, on top of the track reconstruction process during which the best track properties are found, a step where the best hit properties are found. Indeed, the likelihood of a reconstructed track depends both on the track properties and on the observed hit properties. As the hit properties are the properties of the PMTs being hit (reference time, position and orientation), finding the best hit properties is equivalent to calibrating the detector. Thus, the optimal calibration can be determined by reconstructing the same data with different values of some detector parameters and monitoring the quality of the reconstructed tracks.

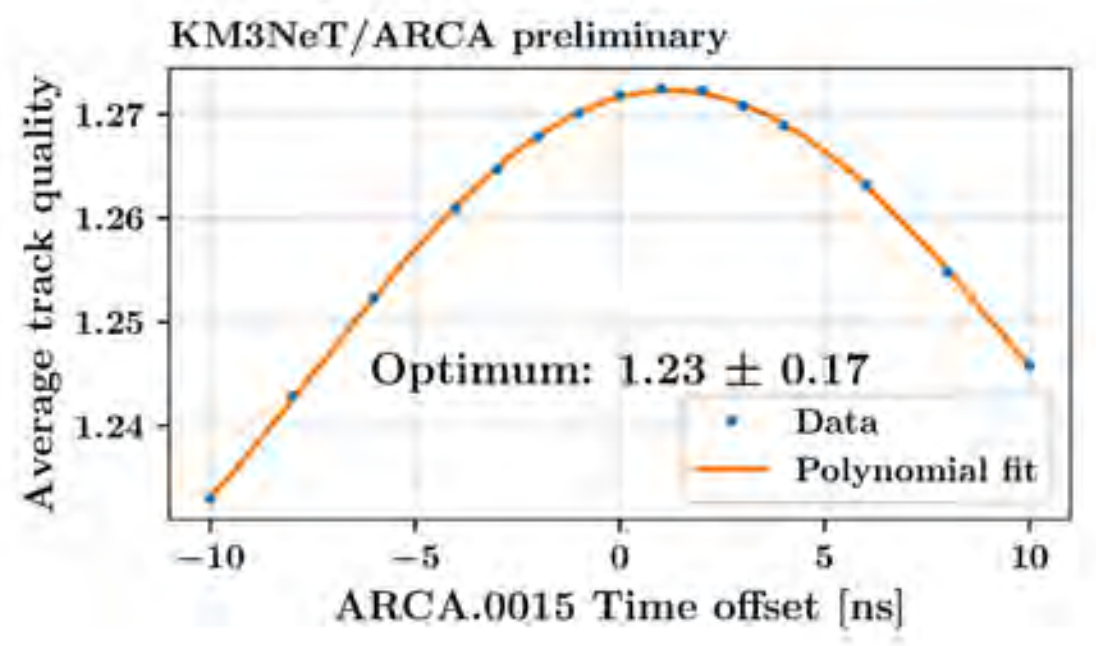

**Figure 1:** Track quality scan for the time offset of the DU 15 of ARCA. The metric for track quality is the likelihood resulting from the fit of the track divided by the number of hits used in the fit.

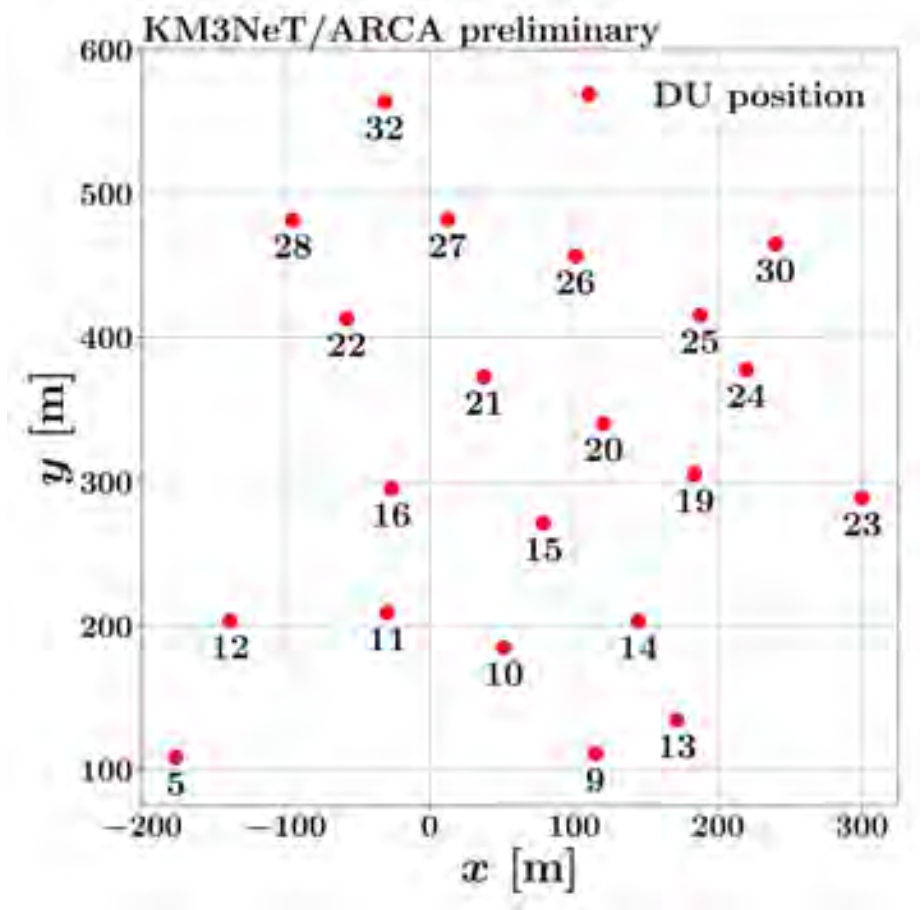

**Figure 2:** Footprint of the ARCA detector in its 21-DU configuration.

The current implementation of the TQM works at the Detection Unit scale. One-dimensional track quality scans are performed over some parameters of an individual DU (keeping the parameters of other DUs fixed): the time reference of the DU; its position along the $x$,$y$, and $z$ axis; and its orientation around the $z$-axis. An example is shown on Figure 1 for the time calibration of a DU of ARCA. The time offset is relative to a given nominal time calibration. It is applied to all the DOMs of the DU considered. Here, the optimal offset obtained with the muon track quality method is around 1.2 ns, meaning that the reference time of that DU should be shifted by 1.2 ns. The TQM is currently the standard way to obtain the time calibration at the inter-DU scale. For position and orientation calibrations, the TQM is an important tool to cross-check results from the nominal calibration obtained with acoustic data and compasses (see Section 2.1).

It should be highlighted that the TQM is only a relative calibration method. The quality of reconstructed tracks will only be affected by relative shifts in the time, position and orientation of an individual DU with respect to the rest of the detector. For instance, if all the DUs are shifted by 1 m in a given direction, then the reconstructed position of the track will be shifted by 1 m, but its quality will not change. In addition, the accuracy of the method will be worse for DUs which are more isolated from the others, like DU 5, 23 or 32 of ARCA (see Figure 2).

### 3.4 Hit time residuals method

Another calibration method using reconstructed muon tracks relies on the hit time residuals. Once a muon track is reconstructed, the expected arrival time of a photon on each PMT can be computed under the hypothesis of non-scattered Cherenkov photon. The expected hit time can then be compared with the observed hit time: the difference is called the *hit time residual* (HTR). This method can be used to measure the inter-DOM time offset by fitting the distribution of HTR for each DOM: the difference in HTR value between DOMs is the inter-DOM offset to correct for.

## 4. Results of the muon track quality method

### 4.1 Position calibration

The track quality method is important to check the results from the position calibration resulting from acoustic data. In particular, it highlights the improvement in positionning coming from the use of a dynamic calibration, where the position of the DOMs is updated every 10 minutes, rather than a static calibration where the fit of acoustic data is done only once and the movement of the DOMs is neglected. To make that comparison, the same track quality scan as the one shown in Figure 1 can be done on consecutive sets of events. That way, the evolution over time of the optimal offset found with the muon track quality, with respect to the nominal calibration relying on acoustics, can be monitored. An example is shown in Figure 3 for one string of ARCA, covering a period of 86 h where the sea currents were among the highest ever observed on the site. With the static calibration, an evolving x-position offset of several meters is seen with the TQM with respect to the acoustic positioning: this means that due to sea currents, the DU is moving relative to other DUs. This displacement is not accounted for by static calibration. Using dynamic positioning, the offset found with the TQM is very close to zero and almost constant over time, showing that dynamic calibration indeed corrects for the displacement of the strings.

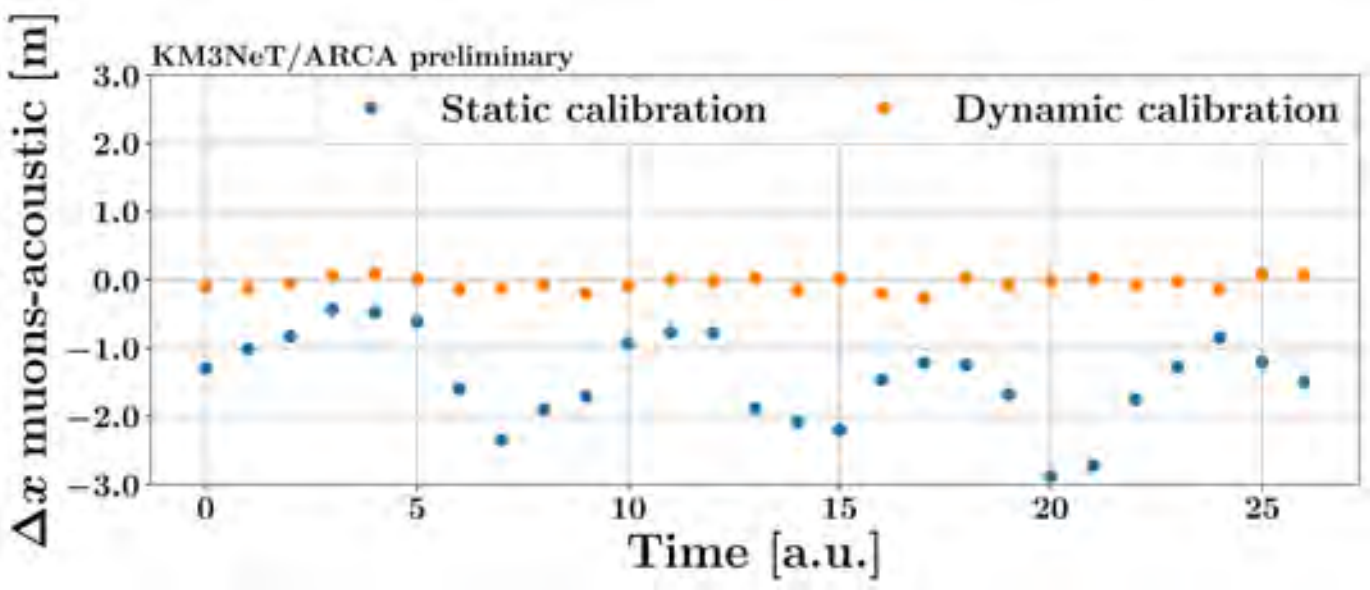

**Figure 3:** Evolution of the optimal x-position offset with respect to the acoustic positionning obtained with the track quality method, for the DU 15 of ARCA, over a period of 86 h. Each point corresponds to the optimal offset on a set of 5000 consecutive events.

The distribution of the optimal offsets obtained with the TQM for the 21 currently deployed DUs of ARCA is shown in Figure 4. These are obtained for the same period as in Figure 3, both with a static and dynamic position calibration. The same effect is seen for all DUs: when using dynamic calibration, the offsets seen from the muon tracks are much lower, and the spread of that offset over time drastically reduces. This shows that the movement of the DOMs is correctly accounted for.

For ORCA, the relative displacements of each DU with respect to the rest of the detector are much smaller, even in periods of high sea currents, because the DUs are much closer to one another (20 meters appart) than in ARCA (90 meters apart), so they move in currents of similar strength and direction. This is illustrated in Figure 5, where the y-position offset obtained with the TQM with respect to the acoustic position calibration is already very small (around 10 cm on average) with a small spreading of values. As 10 centimeters is essentially the accuracy of the track quality method for ORCA [10], using dynamic calibration does not bring a substantial improvement.

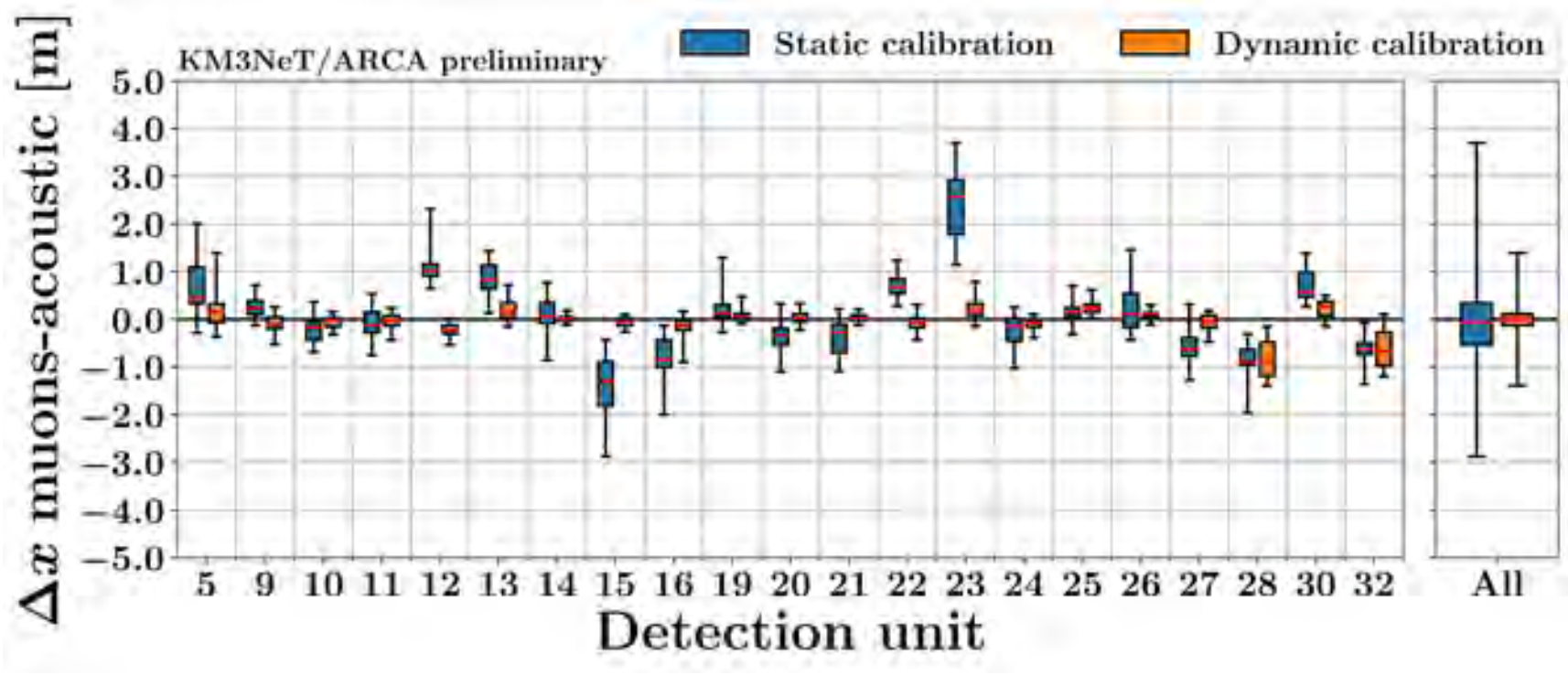

**Figure 4:** Distribution of the optimal x-position offset obtained with the track quality method with respect to the position calibration from acoustics, shown for each individual DU of ARCA and for all DUs. One entry is an optimal offset on a set of 5000 consecutive events. The boxes contain 50% of the DU entries (between the first and third quartiles). The whiskers show the minimum and maximum entry values.

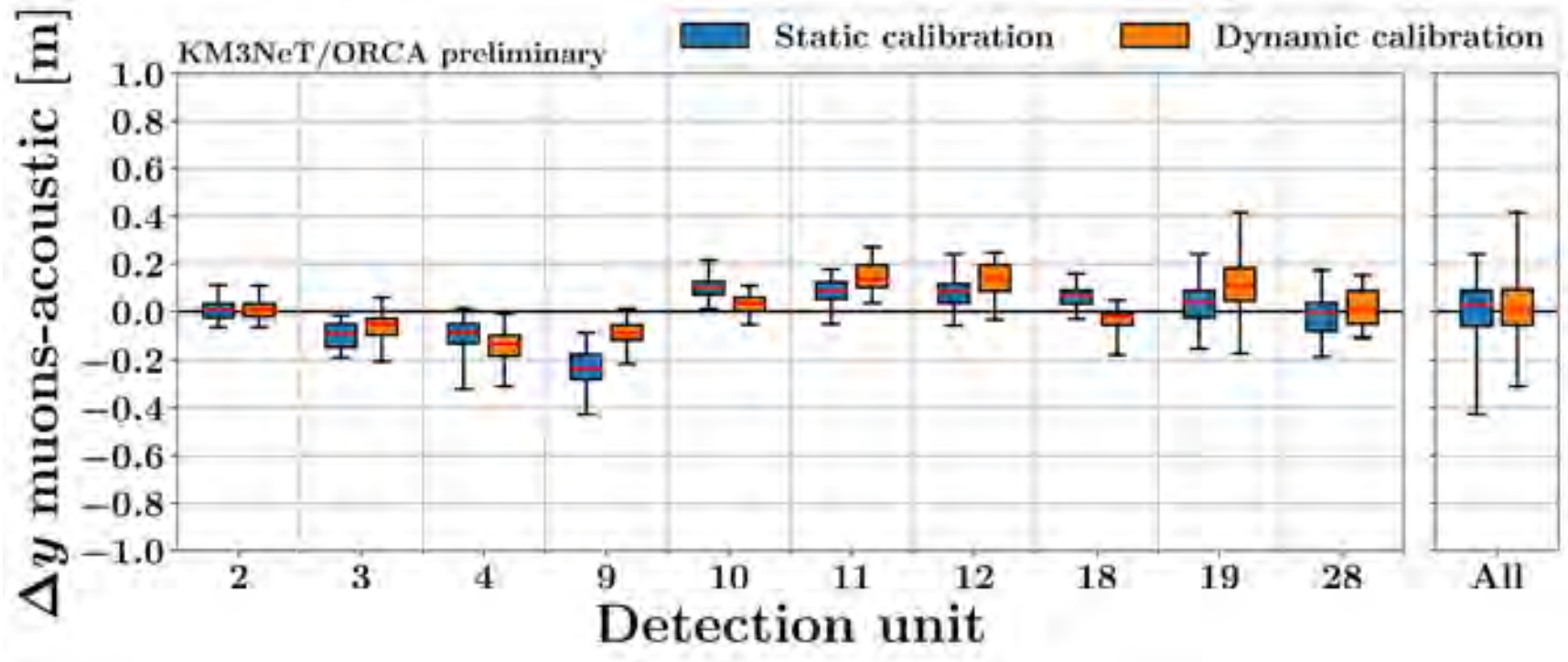

**Figure 5:** Distribution of the optimal y-position offset obtained with the track quality method with respect to the position calibration from acoustics, shown for each individual DU of ORCA and for all DUs. One entry is an optimal offset on a set of 5000 consecutive events. See Figure 4 for the definition of boxes and whiskers.

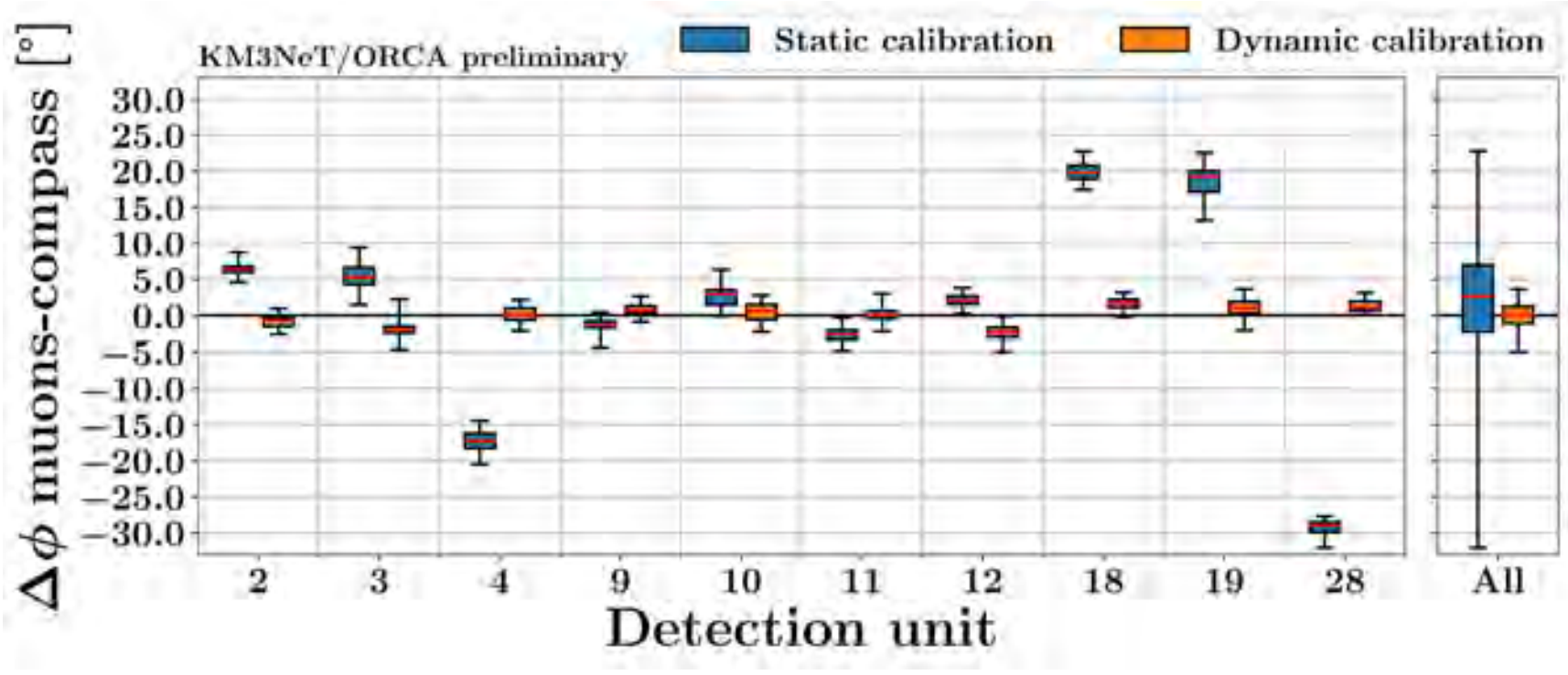

**Figure 6:** Distribution of the optimal orientation offset obtained with the track quality method with respect to the orientation calibration from compasses, shown for each individual DU of ORCA and for all DUs. One entry is an optimal offset on a set of 5000 consecutive events. See Figure 4 for the definition of boxes and whiskers.

### 4.2 Orientation calibration

For the ORCA detector, the muon track quality method agrees with the dynamic orientation calibration within a few degrees, as depicted in Figure 6, obtained with the same period used in Figure 5. This validates the calibration obtained with the compasses. Similar results are obtained with ARCA.

### 4.3 Time calibration

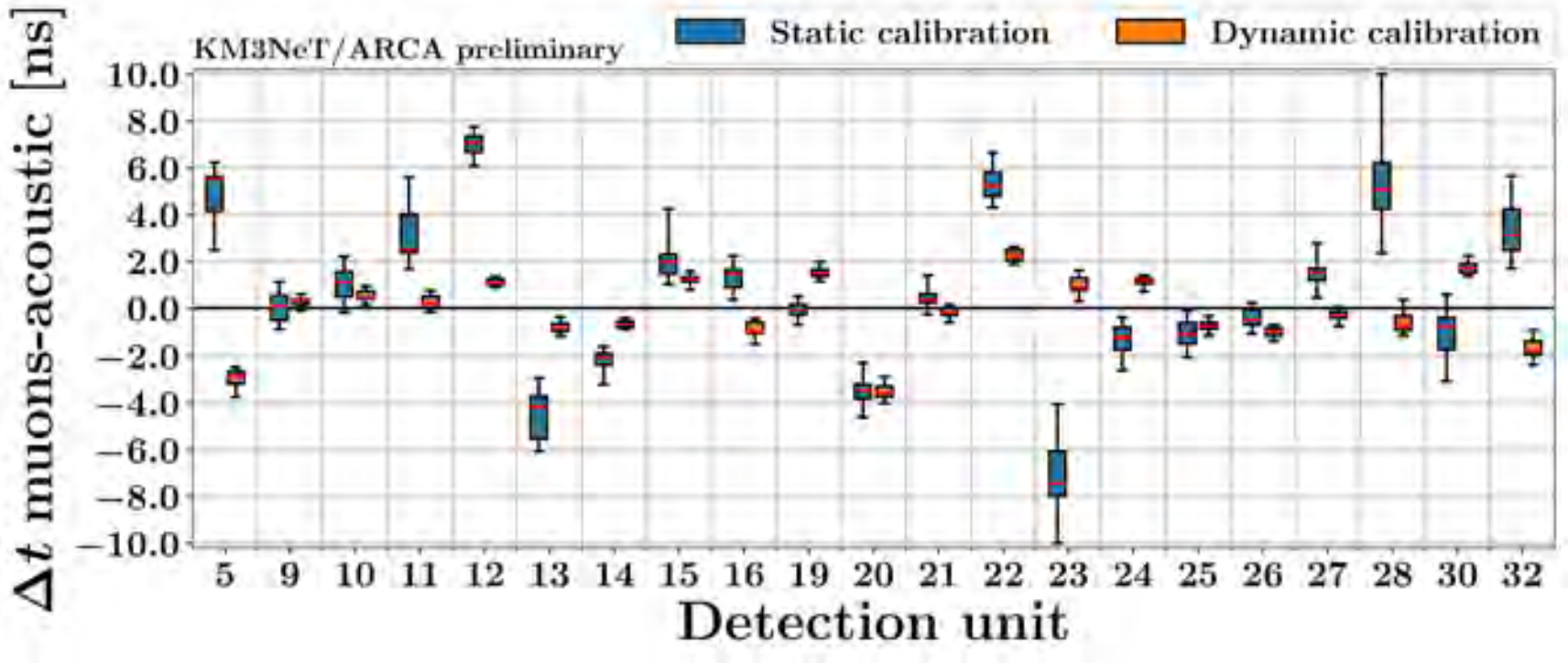

**Figure 7:** Distribution of the optimal time offset obtained with the track quality method with respect to the pre-existing calibration, for ARCA. One entry is an optimal offset on a set of 5000 consecutive events. See Figure 4 for the definition of boxes and whiskers.

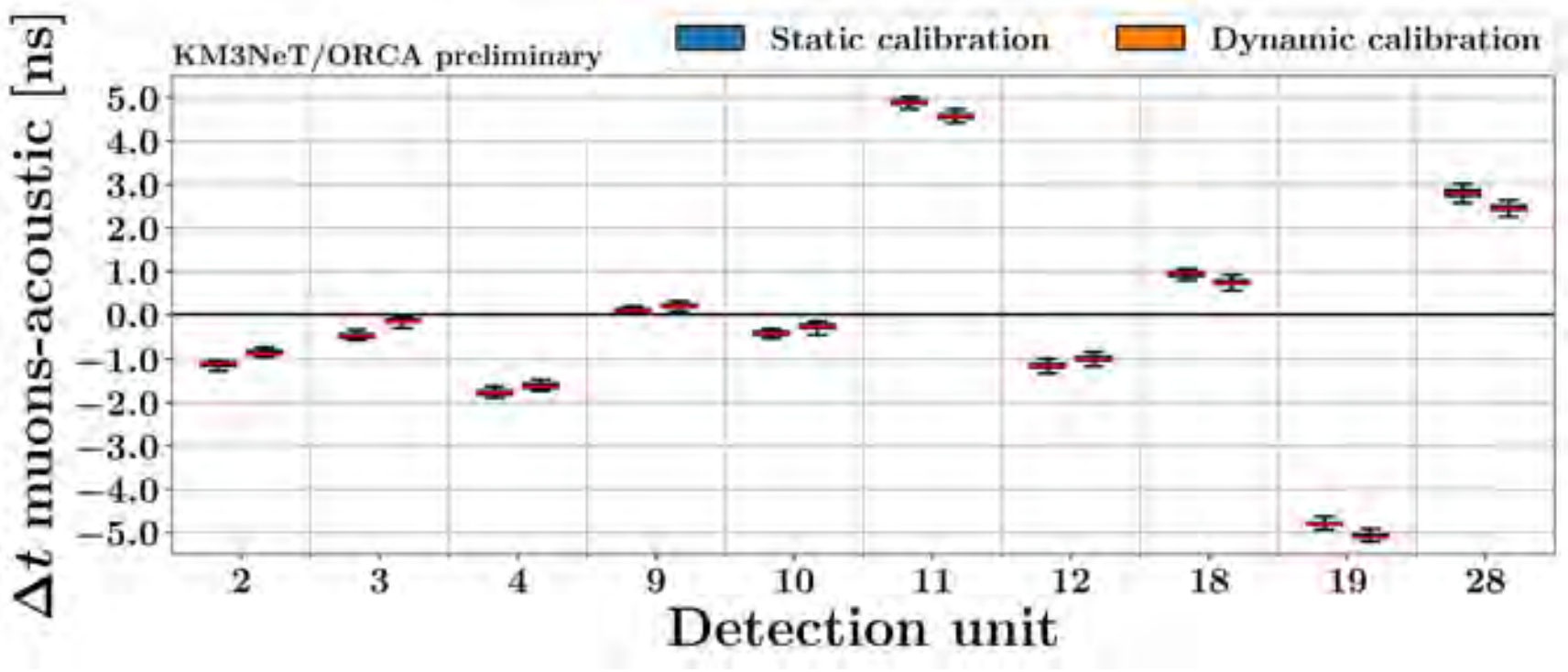

**Figure 8:** Distribution of the optimal time offset obtained with the track quality method with respect to the pre-existing calibration, for ORCA. One entry is an optimal offset on a set of 5000 consecutive events. See Figure 4 for the definition of boxes and whiskers.

Even though the time calibration does not result from the acoustic calibration procedure, the time offset obtained with the track quality method is still dependent on the acoustic calibration due to the degeneracy between the displacement of the DUs and the light arrival time. This is why a spreading in the optimal time offset value is seen in Figure 7. It is more pronounced when using static calibration for the same reason as for the position (DU movements not accounted for).

The offsets are also non-zero for some DUs, even in the dynamic case. This comes from remaining miscalibrations in the reference time of some DUs. More specifically, DUs 4, 11, 18, 19, and 28 of ORCA, which are the ones with the higher time offsets, were newly deployed DUs for the configuration studied here. It is thus expected that addiditional corrections to their time calibration are needed. The corrections to apply are the mean values of the offsets displayed in Figures 7 and 8.

The spread of the values when using dynamic calibration (thus correcting for DU movements) gives an idea of the accuracy of the muon track quality method for determining inter-DU time offsets. If we take the inter-quartile difference as reference metric (height of the boxes in Figure 7 and 8), it is around 0.3 ns for ARCA and 0.1 ns for ORCA. These values are consistent with the resolution values obtained from simulations [10], and satisfy the sub-nanosecond precision requirement on the time calibration. The better accuracy for ORCA is expected from the smaller distance between optical modules leading to a more precise muon track reconstruction.

## 5. Conclusion

Reconstructed muon tracks are used in KM3NeT for the inter-DU time calibration, achieving the required sub-nanosecond precision. The track quality method also allows to cross-check position and orientation calibrations, confirming the accuracy of the dynamic positioning and orientation procedure.

# Dynamical position and orientation calibration of the KM3NeT telescope

**Félix Bretaudeau,**[a] **Clara Gatius Oliver,**[b,*] **Maarten de Jong**[b,c] **and Lilian Martin**[a]

[a] *Subatech, IMT Atlantique, IN2P3-CNRS, Université de Nantes, 4 rue Alfred Kastler - La Chantrerie, Nantes, BP 20722 44307 France*

[b] *Nikhef, National Institute for Subatomic Physics, PO Box 41882, Amsterdam, 1009 DB Netherlands*

[c] *Leiden University, Leiden Institute of Physics, PO Box 9504, Leiden, 2300 RA Netherlands*

*E-mail:* cgatius@km3net.de

KM3NeT is an underwater neutrino telescope which detects the Cherenkov radiation created by the products of neutrino interactions. To accurately reconstruct neutrino events, a precise determination of the position and orientation of the optical modules, which detect the Cherenkov radiation, is required. As the detector elements sway with the deep sea currents, a continuous tracking of the positions and orientations is necessary. A network of acoustic emitters and receivers is used to position the optical modules. Their orientation is determined by compasses placed in each optical module. This contribution presents the methods to perform the position and orientation calibration of the KM3NeT telescope. The positions of the optical modules need to be resolved with an accuracy of better than 20 cm in order to achieve the envisaged angular resolution of the KM3NeT/ARCA telescope of 0.05 degrees. The orientations of the optical modules need to be resolved with an accuracy of about 3 degrees in order to not compromise the quality of the event reconstruction.

*Speaker

## 1. Introduction

KM3NeT is an underwater neutrino telescope located at two sites in the Mediterranean sea. Neutrinos are indirectly detected by the products of their interactions in the sea water, which produce Cherenkov radiation. The ORCA[1] detector, off the coast of Toulon, is used to measure atmospheric neutrino oscillations, and the ARCA[2] detector, off the coast of Sicily, is used to search for neutrinos from astrophysical sources.

Cherenkov radiation is measured by photo-multiplier tubes (PMTs) placed in pressure-resistant glass spheres called Digital Optical Modules (DOMs). Each DOM houses 31 PMTs, the front-end and readout electronics and calibration devices, which are relevant to this work [1]. Eighteen optical modules are attached to two vertical Dyneema® ropes via a titanium collar, forming the detection units (DUs). DUs are attached to an anchor to ensure a fixed position at the sea bed, while at the top a buoy is placed to reduce its movement. Due to the different science goals of ARCA and ORCA, the detectors are build to be sensitive to higher energy (TeV-PeV) and lower energy (GeV-TeV) neutrinos, respectively. This is achieved by a different inter-DU and inter-DOM spacing, which results in about 700 m long detection units for ARCA and 200 m for ORCA [2].

To accurately reconstruct neutrino events, a precise determination of the position and orientation of the optical modules is required. However, the detection units tilt with the sea current, causing a displacement of the optical modules, and twist around their vertical axis, modifying the orientation of the PMTs. Hence, a continuous tracking of the positions and orientations of the optical modules is necessary. The optical module positions are determined by an acoustic positioning system, while their orientation is determined by compasses.

In this contribution we present the methods used to determine the fixed or static parameters in the system, as well as the dynamic position and orientation calibration of the optical modules. To achieve the envisaged angular resolution of the ARCA detector of 0.05 degrees, the positions of the optical modules need to be resolved with an accuracy of better than 20 cm, which corresponds to a hit time uncertainty of around 1 nanosecond in water. The optical modules orientation need to be resolved to an accuracy of about 3 degrees, to not compromise the event reconstruction quality [2]. The present calibration method concerns the relative positions and orientations of the optical modules, but not the absolute position and orientation of the detector. The second has to be determined with a different approach, as dedicated sea operations or studying the atmospheric muons created by cosmic rays shadowed by the Moon or Sun [3].

## 2. Acoustic position calibration

### 2.1 Acoustic positioning system

Piezo-electric acoustic sensors are housed within the optical modules and external piezo sensors are located at the base modules of the detection units. The latter are not yet used for the position calibration. The emitters are placed in autonomous tripod structures on the sea bed, on some of the anchors of the detection units and on some of the electrical junction boxes where the detection units are connected [4]. In Fig.1 the footprint of the current configuration of the ARCA and ORCA detectors are shown, indicating the location of the detection units and the emitters.

[1]Oscillation Research with Cosmics in the Abyss
[2]Astroparticle Research with Cosmics in the Abyss

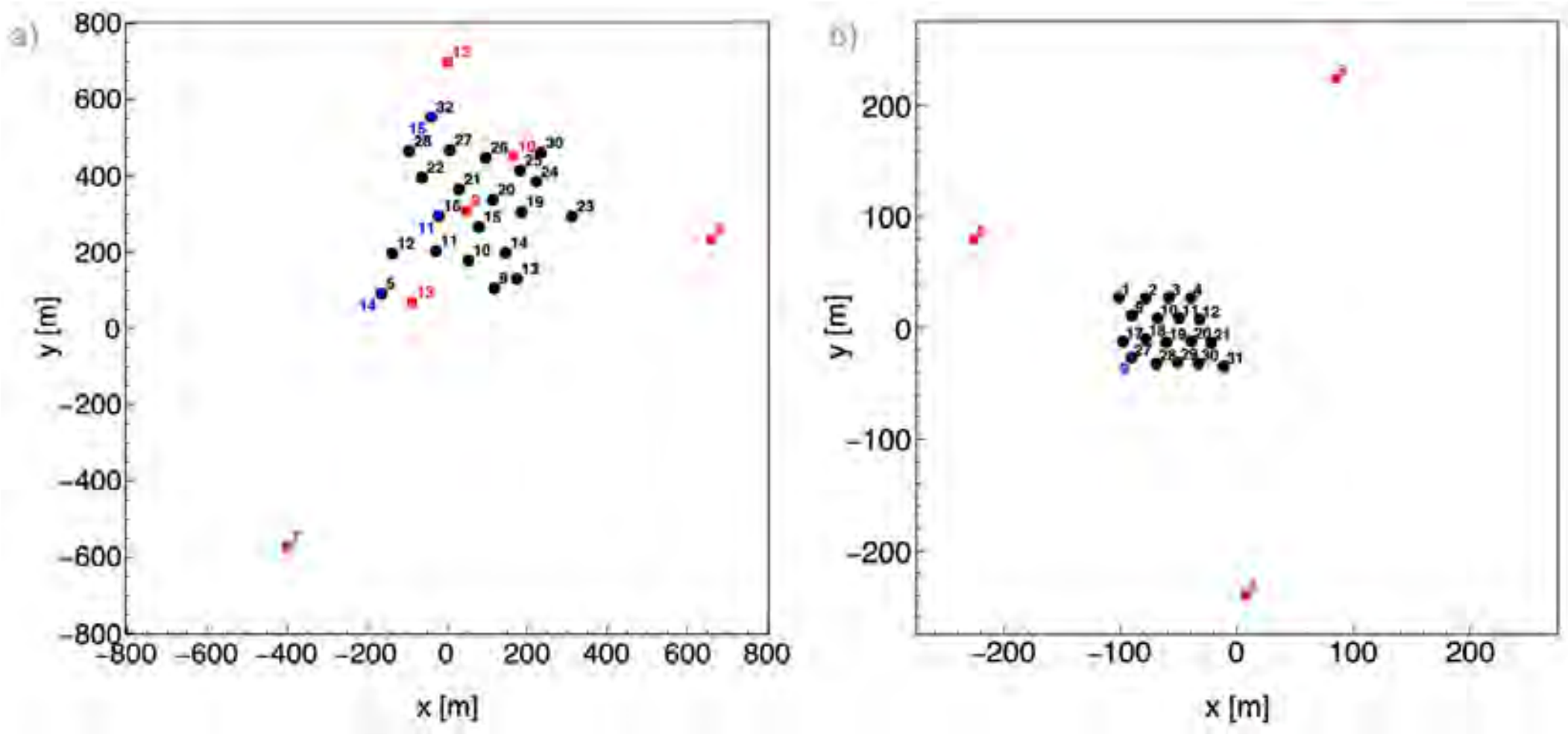

**Figure 1:** Footprint of the current (a) ARCA and (b) ORCA detector configurations. The x and y direction correspond to easting and northing with respect the detectors reference frame. Detection units are represented by black dots, autonomous acoustic emitters by red squares and emitters on the base modules by blue triangles.

The emitters are distributed within the detector footprint, ensuring a proper positioning of the optical modules in the horizontal plane, as well as their height. The emission pattern of the autonomous emitters, which are battery powered, is of the order of 10 consecutive emissions or pings every 10 minutes. The emitters which are connected to the base modules or junction boxes, emit continuously, approx. 1 ping every 30 seconds. Each of the emitters has an specific frequency, which is recognised by an acoustic data filter. The filter matches the corresponding waveform of the signal with the raw data, giving as output, among others, a time-of-arrival and an emitter identifier.

### 2.2 Acoustic fit

From each recorded time-of-arrival a preliminary estimation of the time-of-emission is computed assuming the nominal position of the optical modules that recorded the signal. If enough optical modules measure an emission within a certain time window, an acoustic event is triggered. The acoustic fit takes as input a set of acoustic events and fits a model of the detector geometry to the acoustic data. The set consists on acoustic events within a time window of 10 min, during which detector elements are assumed to not move significantly. This ensures that pings from all emitters are used for the fit.

The model describing the detector geometry is parameterised as a function of fixed or static parameters and dynamic parameters, the latter fitted every 10 min. The fixed parameters consist of the anchor position of the DUs; the height of each DOM in the DU; the piezo-sensor position in the base module, which depends on the anchor orientation; and the emitter positions on the sea bed. For the fit are also required two fixed mechanical model parameters describing the DU shape, which are computed beforehand, and the sound velocity at the detector depths, which has been measured. The dynamic parameters are 2 tilt angles and their second order corrections; a dynamical stretching factor, which describes the creep and stretching due to the tilt, of the Dyneema® ropes holding the DU together; and the time-of-emission of each ping.

The equation relating the times of flight with the detector geometry is the following:

$$t_A^c[i,j] = t_E^c + \mid \vec{x}_0[i] + \Delta\vec{x}[i,j] - \vec{x}^c \mid v^{-1} \tag{1}$$

Where $t_A^c[i,j]$ is the time-of-arrival at DU $i$ DOM $j$ of emitter $c$, $t_E^c$ is its time-of-emission, $\vec{x}_0[i]$ is the position of the DU anchor, $\Delta\vec{x}[i,j]$ is a vector from the anchor of the DU to DOM $j$ [3], $\vec{x}^c$ is the position of emitter $c$ and $v$ is the average speed of sound in between emitter and receiver.

The vector $\Delta\vec{x}[i,j]$ is described as a function of the mechanical model of the detection unit and its tilt and stretching. The mechanical model of the detection unit describes its curvature with respect to the vertical considering the buoyancy and the drag force of all elements [5]. In the following equation the effective height, given the nominal height of the DOM $z_0$, the phenomenological parameters $a$ and $b$ describing the curvature of the DU and a stretching factor $(1+\alpha)$, is shown.

$$z' = (1+\alpha)\ z_0 + b\log(1 - a\ (1+\alpha)\ z_0) \tag{2}$$

The tilt is represented by a unit vector $\hat{T} = (T_x, T_y, (1 - T_x^2 - T_y^2)^{1/2})$, where $T_x$ and $T_y$ are the gradients of the x and y positions in the z' direction. The components of the $\Delta\vec{x}[i,j]$ vector in Eq.1 can be expressed in the following way:

$$\Delta x[i,j] = T_x[i]z'[i,j] + T_x^{(2)}[i](z_0[i,j])^2 \tag{3}$$

$$\Delta y[i,j] = T_y[i]z'[i,j] + T_y^{(2)}[i](z_0[i,j])^2 \tag{4}$$

$$\Delta z[i,j] = f^{-1}(\ (1+\alpha[i])z_0[i,j]\ ) \tag{5}$$

where $z'[i,j]$ is the effective height and follows Eq.2, considering the dependence of the parameters of each DOM or DU. Second order corrections of the tilt $(T_x^{(2)}, T_y^{(2)})$ are taken into account to compute the horizontal displacement of the DOMs. These are introduced to correct for possible features not taken into account by the mechanical model, for example, the effect of varying sea currents. The vector component $\Delta z[i,j]$ is obtained by computing the length of the DU, $(1+\alpha)z_0$, as a function of the height z and inverting the equation. This is computed from the horizontal displacements and effective height.

### 2.2.1 Minimisation method

The fit minimises the $\chi^2$, which is the sum of differences between the measured and modelled times of arrival normalised by an assumed resolution, chosen to be 50 $\mu s$ ($\sim$ 7.5 cm). Each measured time of arrival involves 6 dynamic parameters: 2 tilt angles and their second order corrections, a stretching factor and the time-of-emission. Given that a 10 min data set is used as input to the fit, the number of free parameters amounts to $n_f = \sum_{i=1}^{M} n_i + 5N$, where $M$ is the number of emitters, $n_i$ the number of emissions within 10 min per emitter, and N the number of DUs. The number of data points corresponds to $n_p = \sum_{i=1}^{M} n_i \cdot N \cdot 18$, where 18 is the number of optical modules in a DU. As a result, the number of degrees of freedom is very large $NDF \equiv n_p - n_f$, increasing with each new emitter, receiver or detection unit. The fit uses a Lorentzian M-estimator to mitigate the effect of possible outliers.

[3] To ease the calculation, the third component of $\vec{x}_0[i]$ is set to 0 and the nominal height $z_0$ is incorporated in $\Delta\vec{x}[i,j]$.

#### 2.2.2 Determination of the fixed parameters

The values of the fixed parameters are determined after the deployment of each new detection unit or acoustic emitter. For this, a fit of multiple fits is required, not only to determine the dynamic parameters, but also the fixed ones. The fixed parameters amount to 4 per DU (position and initial stretching), 1 per DOM (height) and 3 for each emitter (position). A set of static parameters is used to fit the dynamic ones, and a conjugate gradient method is implemented to determine the fixed parameters minimising the $\chi^2$. The less constrained fixed parameters are varied first.

### 2.3 Dynamic calibration

The dynamic parameters are fitted every 10 min, updating the tilt and stretching of each DU, hence, the displacement of the optical modules. In Fig.2a and 2b, the tilt ($\hat{T}$) amplitude and orientation are shown for four months of the ORCA detector, with a configuration of six detection units. A coherent movement between the detection units is observed in both plots. In Fig.2d and 2e, the square of the sea current speed and the direction are displayed, measured at the detector location during the same period as the tilts. From the mechanical model of the detection units, the square of the sea current speed should be proportional to the tilt. As can be seen, the two quantities have a coherent behaviour, as well as the tilt orientation and sea current direction.

In Fig.3, the tilt amplitude and orientation for few days of the current ARCA detector, consisting of 21 lines, is shown. The tilt reaches values up to 100 mrad ($\sim$ 5.7 deg), highlighting the importance of the position calibration to achieve the envisaged angular resolution. The creep and dynamic stretching of the detection units is shown in Fig.4, for the period of the ARCA detector with a configuration of 6 detection units. The oldest detection unit, DU 9, shows a constant stretching, since it had already been deployed for several months, while the other detection units show a creep due to their adaptation to the medium.

In Fig.5, the residuals of the fit for one detection unit during two different periods of the ARCA detector are shown. The residual distributions in Fig.5a correspond to the period shown in Fig.3, during which extremely high tilts are fitted. The residuals expand beyond $\pm$ 100 $\mu s$ ($\sim$ 15 cm), specially for the top optical modules of the detection unit. Instead, the distributions in Fig.5b correspond to a period with low tilts, and are well contained within a range of $\pm$100 $\mu s$.

## 3. Orientation calibration

### 3.1 Compass system

A magnetometer and accelerometer are integrated in each optical module, referred to as "compass", which can provide the yaw, pitch and roll of the module [1]. The compass data is continuously recorded every 10 sec and calibrated with in-lab compass measurements. To convert the calibrated data to yaw, pitch and roll, a correction for the magnetic declination and the meridian convergence angle needs to be applied, which is different in the two sites of the detector. The compass data is converted to quaternions, which are a compact way to describe rotations: $Q \equiv (\cos(\theta/2), \sin(\theta/2)\hat{u})$, where $\theta$ is the rotation angle around the axis $\hat{u}$.

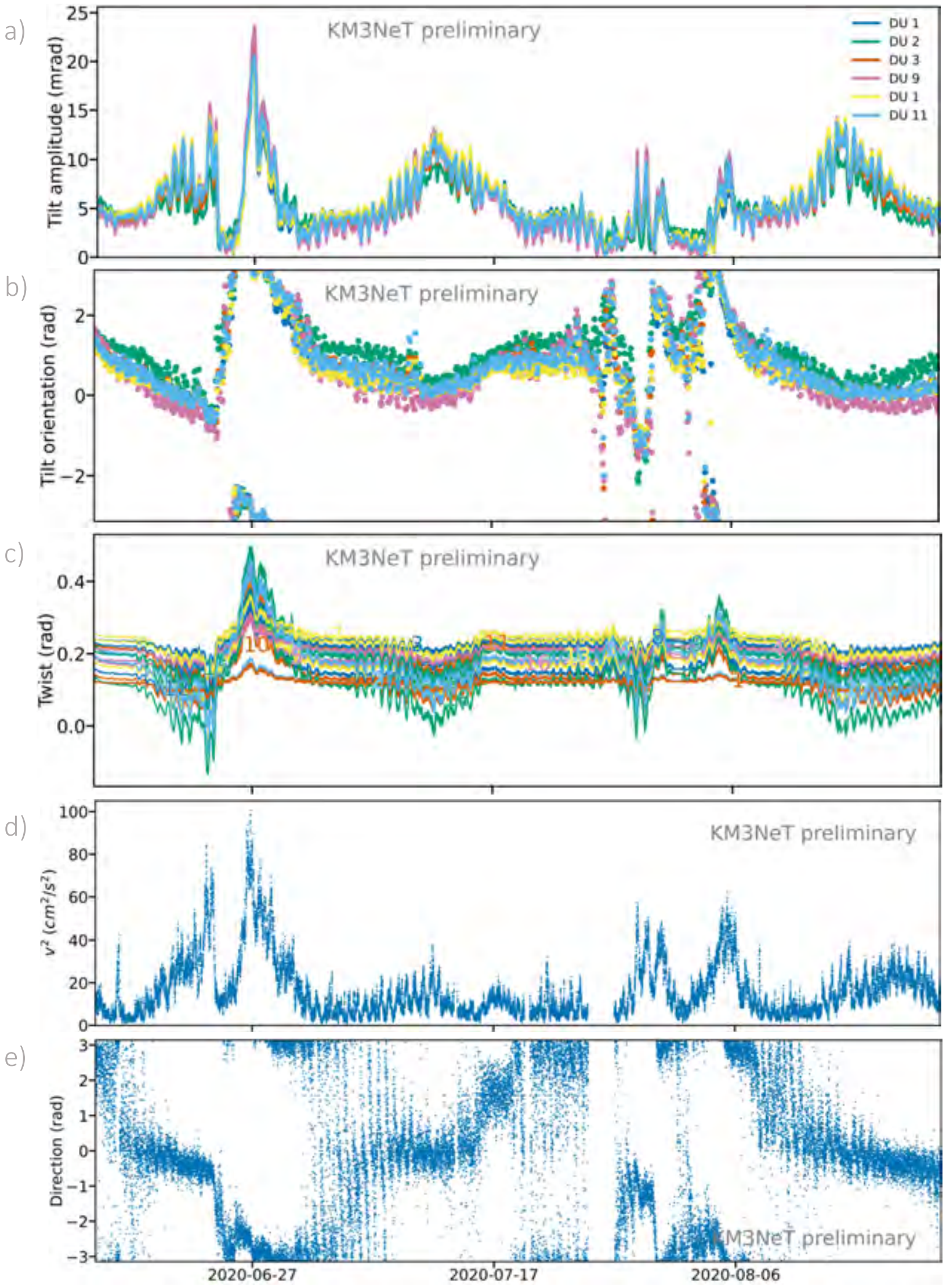

**Figure 2:** (a) Tilt amplitude for each detection unit. (b) Tilt orientation for each detection unit. (c) Twist of the optical modules from detection unit 11. The different modules are indicated by a number corresponding to their position along the DU; 1 for the lowest module, 18 for the top one. (d) Square of the measured sea current velocity. (e) Measured sea current direction. All plots are for the same period, which comprises four months of the ORCA detector with a configuration of six detection units.

### 3.2 Static calibration

An *in-situ* calibration of the optical module orientations is done, which consists of the alignment of the modules for each detection unit. This is applied by fitting the compass data within a time window of 10 min, to a model of the DU twist around the vertical axis. The model is a polynomial function, where $Q_0$ represents the tilt of the DU and $Q_1$ represents its twist around the vertical axis, which depends on the height of each module $z_i$. The model ensures a continuous change of the twist along the DU.

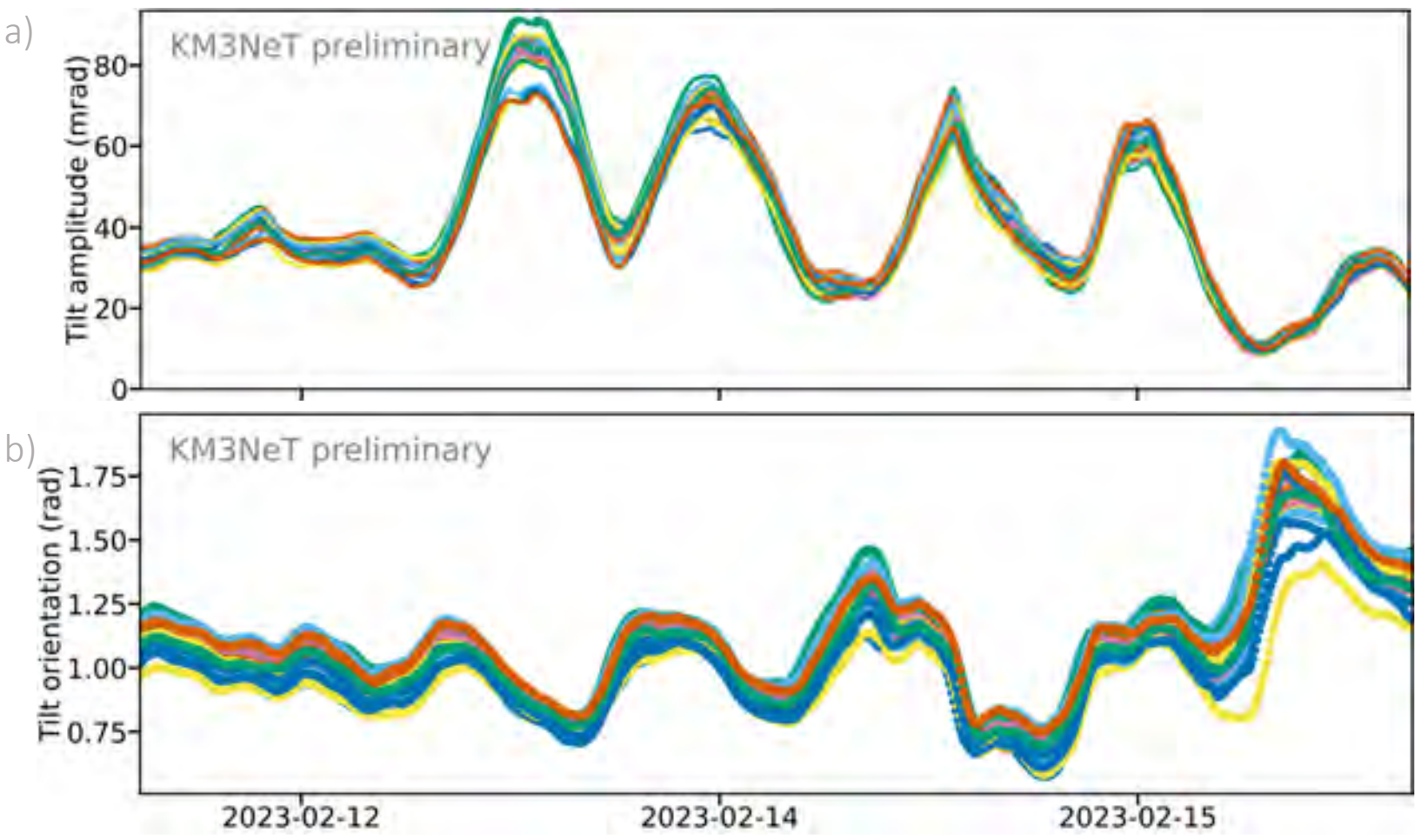

**Figure 3:** a) Tilt amplitude and b) tilt orientation, for a few days period of the current ARCA detector, with a 21 detection units configuration. The different lines indicate different detection units.

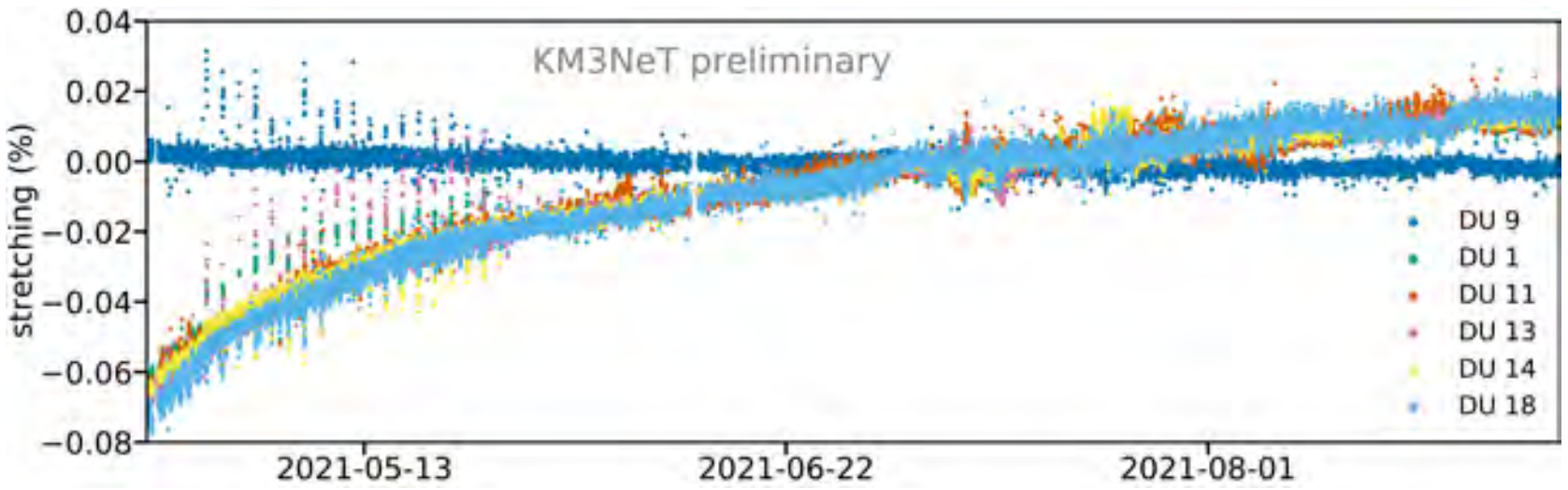

**Figure 4:** Stretching as a function of time for the ARCA detector, during the period it had a 6 detection units configuration.

$$Q = Q_0 \, Q_1^{z_i} \tag{6}$$

Multiple fits are performed during a period in which the DUs do not move significantly. The fit minimises the $\chi^2$, defined as the square of the shortest angle between the measured and modelled quaternion of the DOMs within a DU, normalised by an assumed resolution of 1 deg. The average residual between the modelled and measured quaternion constitutes the alignment of the DOM.

### 3.3 Dynamic calibration

The orientation of the optical modules are updated every 5 min, using the continuous recording of compass data. After applying the lab calibration and the alignment of the compasses, the data is filtered by removing measurements that deviate more than 5 deg from a local interpolation of the time-ordered data. In case a DOM does not record data, an interpolation with the neighbouring DOMs measurements is applied to find its orientation. In Fig.2 the dynamic orientation of the DOMs of the detection unit 11, during four months of the ORCA detector with a configuration of six detection units, is shown. The DOMs move coherently among themselves as well as with the tilt amplitude and orientation variations derived from the acoustic data.

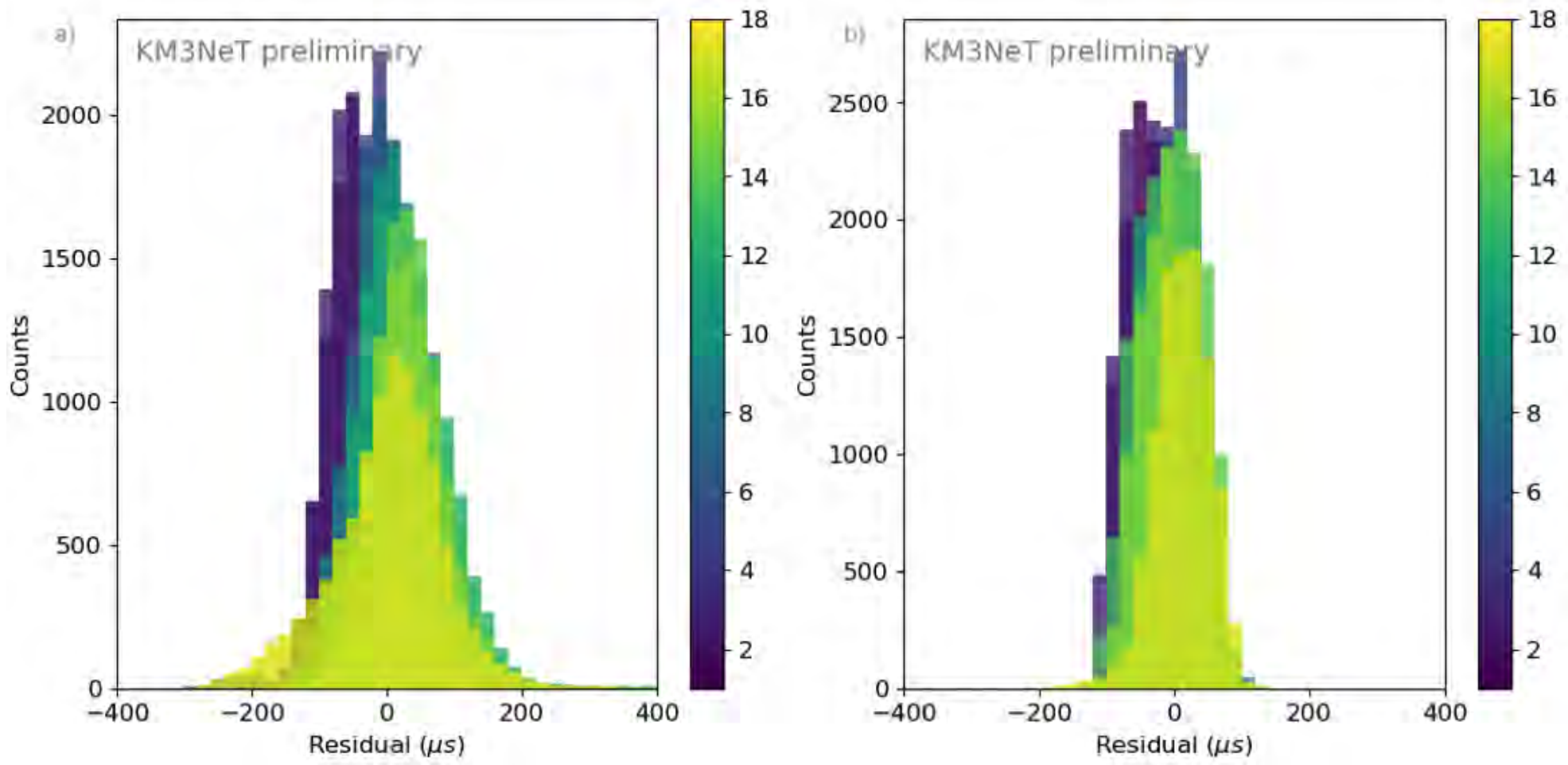

**Figure 5:** a) Residuals of the fit for a detection unit of the current ARCA detector, during the period shown in Fig. 3. b) Residuals during a period with small values of the tilt. The colour scheme indicates the floor of the modules; 1 for the lowest module, 18 for the top one.

## 4. Conclusion

Acoustic positioning and orientation methods are developed to calibrate the static and dynamic positions and orientations of the optical modules. These are cross-checked by a muon calibration technique, which exploits the muon track reconstruction to find the optimal orientation, position and time reference of the optical modules. The positions are found to agree with the muon calibration within a range of ±10 cm while the orientations show an agreement within a range of less than ±3 deg [6]. The agreement is within the required specifications to achieve the envisaged angular resolution of the KM3NeT telescope.

# KM3NeT Time calibration with Nanobeacons

**Agustín Sánchez Losa,[a,*] Juan Palacios González,[a] Francisco Salesa Greus,[a] Juan Zúñiga Román,[a] Diego Real Máñez[a] and David Calvo Díaz-Aldagalán[a] for the KM3NeT collaboration**

[a] *Instituto de Física Corpuscular (CSIC-UV), Parque Científico, C/ Catedrático José Beltrán, 2. Valencia, Spain.*

*E-mail:* Agustin.Sanchez@ific.uv.es

The KM3NeT Collaboration is building a neutrino telescope in the Mediterranean Sea. The detector is expected to achieve an angular resolution better than 0.1 degrees for energies above 10 TeV. This is critical for attaining one of the key goals of the experiment, i.e. the identification of cosmic neutrino sources. In order to achieve a good angular resolution, the detector requires a relative time calibration of the order of 1 ns.

The Nanobeacon is a cost-effective time calibration device developed by the KM3NeT Collaboration to synchronise *in situ* the detector photomultipliers with a nanosecond relative accuracy. In this contribution we will describe the design and operation of the Nanobeacon. Moreover, we will present the results of data taken in real sea-conditions and we will show how they are used to validate the time calibration parameters measured onshore.

*Speaker

## 1. The KM3NeT detectors

The Kilometre Cube Neutrino Telescope (KM3NeT) is a research infrastructure currently under construction in the Mediterranean Sea [1]. It comprises two detectors situated at different locations. The first one, known as ARCA (Astroparticle Research with Cosmics in the Abyss), is being installed off the coast of Capo Passero in Italy at a depth of 3.5 km. The second detector, ORCA (Oscillation Research with Cosmics in the Abyss), is located off the coast of Toulon in France at a depth of 2.5 km. The main scientific goal of ARCA is the detection of high-energy cosmic neutrinos, while ORCA focuses on studying neutrino properties, including neutrino oscillations and neutrino mass hierarchy.

Both detectors share the same hardware, differing only in their layout, which determines the energy range to which they are sensitive. KM3NeT relies on a network of Digital Optical Modules (DOMs) [2], each equipped with 31 photomultiplier tubes (PMTs) measuring 3 inches in diameter [3]. These DOMs are arranged in strings called detector units (DUs), forming a 3D matrix that optimally captures the Čerenkov radiation induced by high-energy neutrinos interacting with the matter surrounding the detector.

Each DOM is controlled by a Central Logic Board (CLB) [4], where all registered data in such DOM is formatted and sent to shore. In particular, when photons are detected by a PMT a "hit" entity is generated. This entity contains the time of detection plus a proxy of the amount of light converted in the PMT photocathode. The latter, called ToT (Time-over-Threshold), is the time the PMT charge has been integrated above a certain detection threshold. The former, is the time when such threshold was exceeded.

In this work, it is discussed how the LED beacon housed in the DOM is used to calibrate one of the corrections that must be applied to the hit time.

## 2. Time calibration in KM3NeT

Accurate time calibration is a crucial aspect in the operation of neutrino telescopes, playing a fundamental role in event reconstruction, which relies on both, the measurement of the arrival times of the Čerenkov light on the different sensors of the detector and their position estimate at the moment of the light detection. For a detailed description of the determination of the KM3NeT sensor positions see contribution [5].

Čerenkov light propagation through the detector medium introduces unavoidable uncertainties on individual photon detection, partially compensated statistically when, after its propagation, enough light is detected. This contribution is limited by the optical medium, in practice water or ice for neutrino telescopes, and therefore to its scattering and absorption lengths, being better those of water. The remaining uncertainties reside only in the capability to assign a correct time for the photon detection in the PMT photocathode.

Due to various factors, such as signal propagation delays and instrument response time, achieving precise time calibration to millisecond events along a kilometer scale detector poses significant challenges. Absolute time accuracy is limited by the GPS reference used in computers on shore, which in practice implies a few tens of nanosecond accuracy, always below the millisecond, typically enough for astrophysical purposes. On the other hand, the necessary relative time accuracy to achieve angular resolutions below the degree in a neutrino telescope is at the nanosecond level on the photon detection.

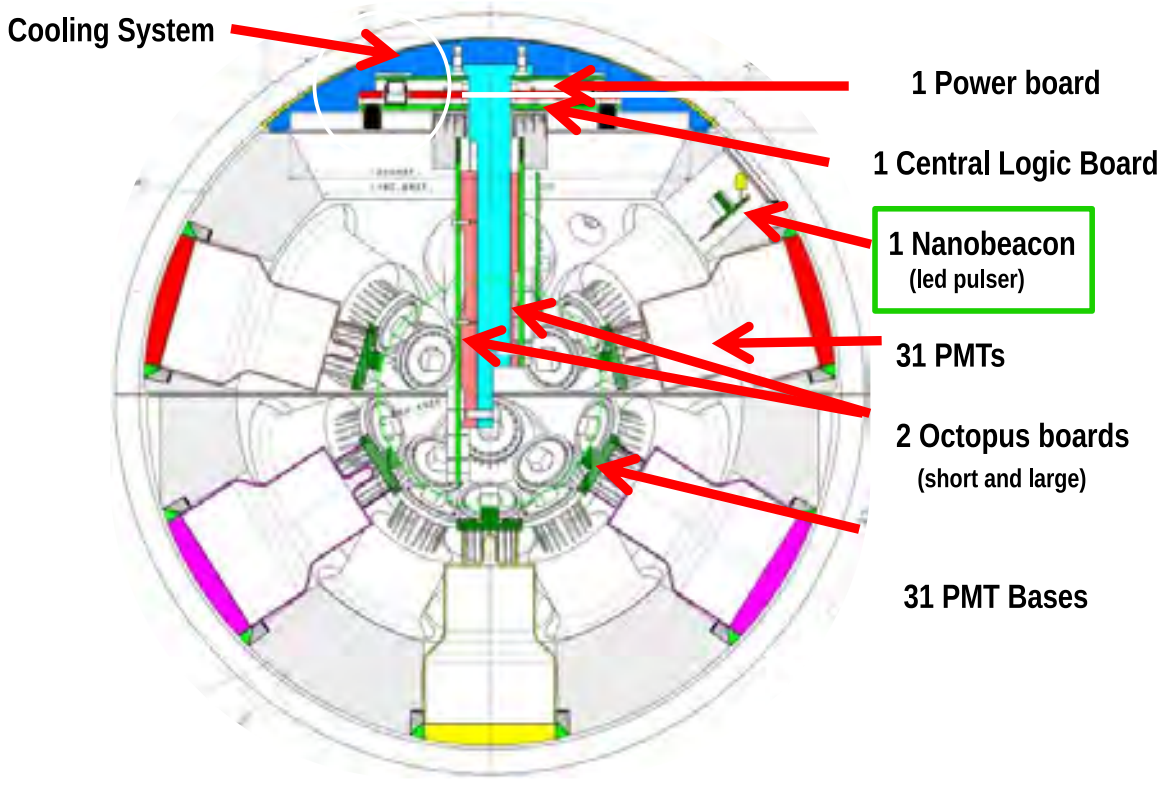

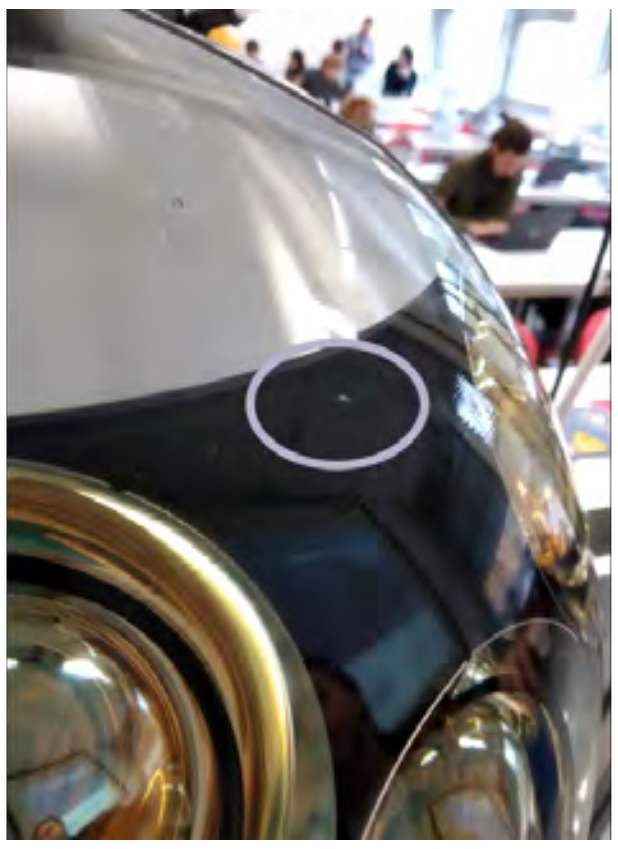

**Figure 1:** Left: diagram of a KM3NeT DOM, showing the position of the NB LED. Right: A picture of part of the top hemisphere of a DOM with the position of the NB highlighted by a circle. The three closest PMTs used for the flash trigger are partially visible. From [7].

An intrinsic uncertainty contribution comes from the Transit Time Spread of the PMTs, which for KM3NeT is around 2.5 ns [3]. This contribution is compensated statistically in the event reconstruction when the number of detected photons increases.

The time of detection of a hit is assigned by the DOM CLB according to the clock signal propagated along the whole detector infrastructure from the shore computers. This delay is partially estimated during the DU integration in the dark room plus multiple calibrations on parts of the infrastructure. As a result there are three levels of relative time calibration in KM3NeT:

- Intra-DOM: this one reefers to the relative time synchronization of all the PMTs within each DOM. This one can be achieved *in situ* using the $K^{40}$ decays in the sea and looking for the offsets that maximize the coincidences.
- Inter-DOM: that corresponds to the relative timing of the DOMs along the same DU. It is firstly estimated in the dark room using a laser that simultaneously illuminates two reference PMTs in each DOM and can be checked *in situ* using muon tracks and optical beacons, in particular the Nanobeacons.
- Inter-DU: that determines the relative offsets between the different DUs of the detector. It is firstly estimated via multiple calibrations on parts of the infrastructure and during the dark room calibration and can be checked *in situ* using muon tracks (see contribution [6]) and optical beacons, like a laser beacon.

In this work we present the application of the Nanobeacons to the inter-DOM calibration in KM3NeT plus its prospects.

## 3. Nanobeacon devices

The Nanobeacon (NB) system [7] is an evolution of the LED beacon system [8] mainly used for time calibration purposes in the ANTARES neutrino telescope. The NB has two main components:

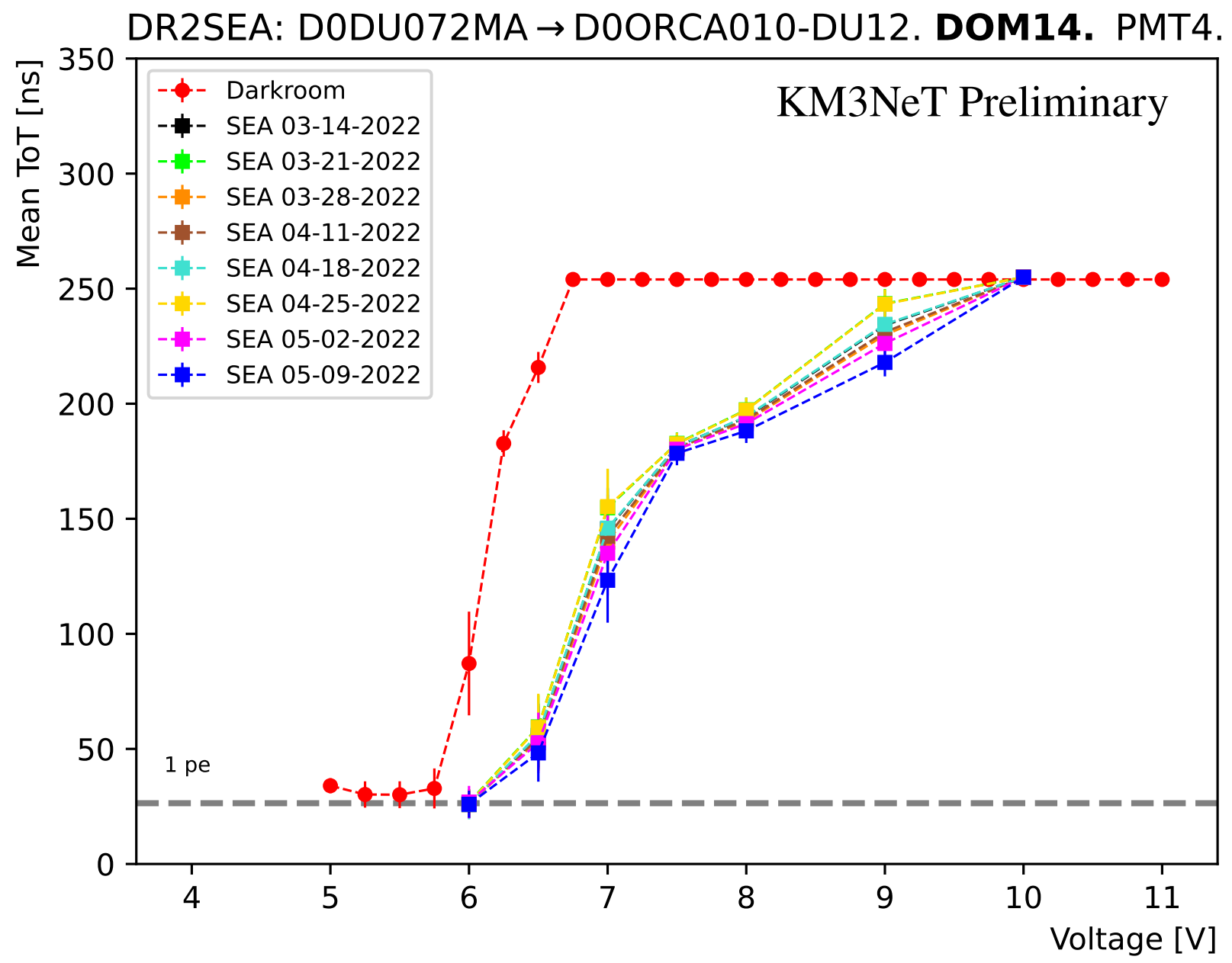

**Figure 2:** Sigmoid plot for a particular DOM and reference PMT in ORCA.

the pulser and the DC/DC converter. The pulser consists of a LED and an electronics board responsible for generating the trigger signal for the optical pulse. On the other hand, the DC/DC converter is integrated in the power board inside the DOM and delivers the power supply to the NB pulser.

The inherent simplicity of the NB system not only makes it a cost-effective solution but also allows for a smooth integration inside each DOM, providing a high-redundancy calibration of the detector.

In order to mitigate the effects of sedimentation and biofouling on the glass sphere of the DOM, the NB is strategically positioned ~45° off the axis from the top of the DOM, as shown in Figure 1.

The selected LED model was the HLMP-CB1A-XY0DD produced by Broadcom [9]. This model can provide light intensities that are one order of magnitude higher than the intensities provided by the LEDs used in ANTARES.

The NB emits short-duration pulses with a width of approximately 5 ns (FWHM) and a rise time of around 3 ns (from 10% to 90% of full amplitude). The intensity of the pulse is controlled by a configurable voltage that can range between 4.5 and 30 V.

To perform the calibration of the KM3NeT PMTs using the NB system, a reference time is required to accurately establish the moment when the NB flash is emitted. This reference time is obtained by using the three closest PMTs to the NB within the DOM. To prevent data-taking saturation and confirm the NB's emission, a voltage scanning procedure is conducted in a dark room before deploying the DOMs. This scanning helps determine the optimal voltage at which the NB will operate effectively.

The optimal voltage for operating the NB system is expected to vary depending on the medium in which it operates. This is due to the different reflection indices of air (in the dark room) and water (in the sea), resulting in different signals collected by the reference PMTs. Since voltage scanning after deploying the DOMs would significantly reduce detector uptime, the scanning process in the sea can only be afforded for a limited number of lines. The characteristic sigmoid-shaped curves resulting from an example of these scanning done in the dark room and in the sea are shown in Fig. 2.

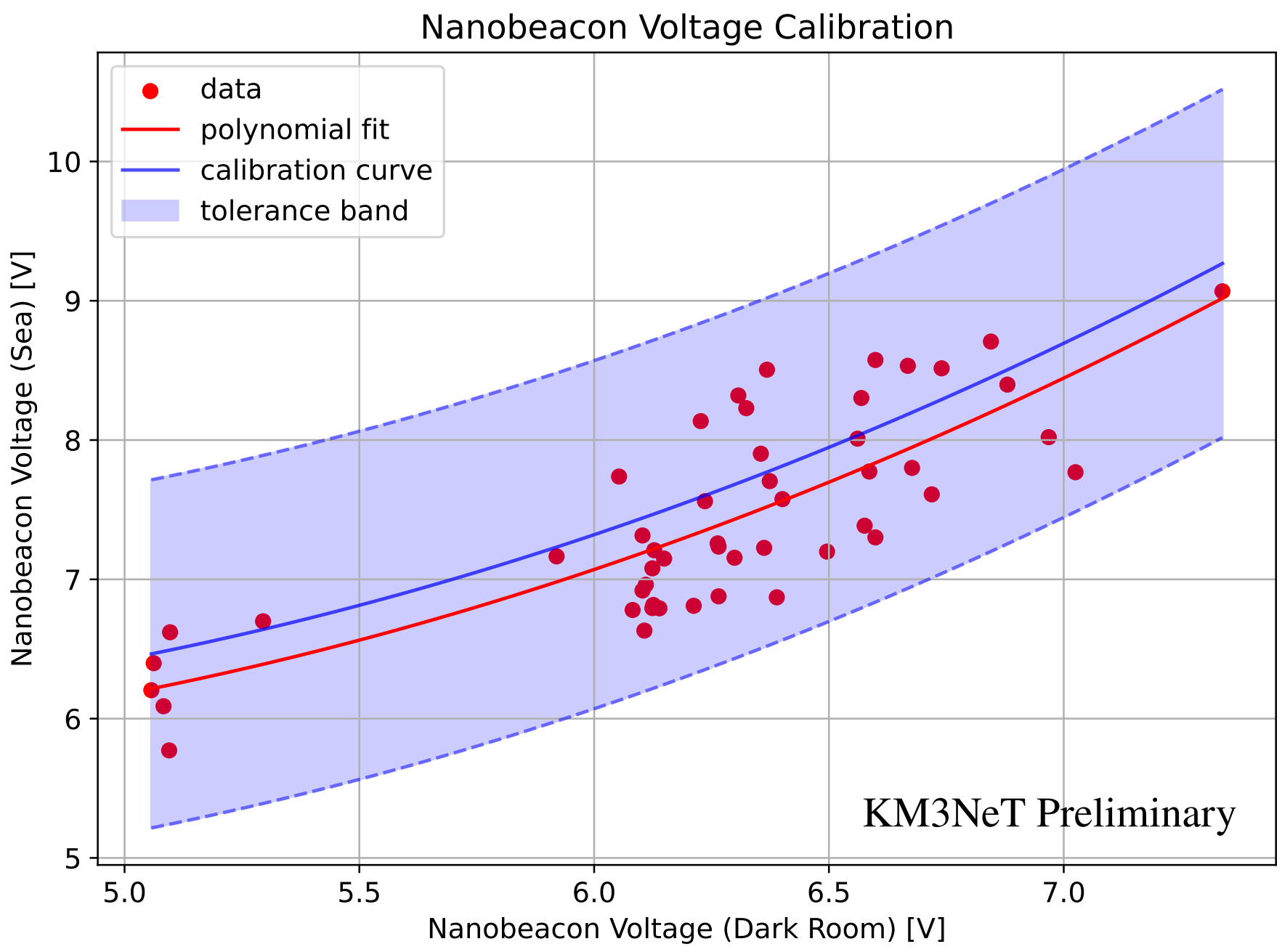

**Figure 3:** Resulting calibration curve for the NB voltage.

From the sigmoid curve, it is possible to calculate the cumulative distribution and determine the voltage at which the $50^{\text{th}}$ percentile, very stable and robust to determine, is reached. Matching the percentiles for the dark room and sea sigmoids gives the data points shown in Fig. 3. These points exhibit a clear trend which can be fitted to a 2-degree polynomial (red line). To account for fluctuations along the fitted line and minimize cases where the assigned V may not allow proper light emission from the LED, a systematic shift of 0.25 V has been introduced (blue line). This blue line will ultimately be used to determine the appropriate voltage for operating the NB in the sea.

## 4. Calibration methodology

It is possible to find the relative time offsets between DOMs by illuminating simultaneously various of them with the same NB. This can be done via the detection time of the NB flash measured on each DOM relative to the flash emission corrected by the time it takes the NB light pulse to reach the corresponding DOM. Due to the NB emplacement at the top of the DOM and its PMT distribution, it is intended to be used to illuminate most efficiently and directly the lower PMTs of the DOMs above the flashing NB (see Fig. 4 ). These DOM pairs, with the most direct light illumination, will be considered for the calibration.

The observed light distribution in the illuminated DOMs follows a Gaussian like distribution, due to the direct light, with a tail due to the scattered light (see Fig. 5). This corrected time difference can be denoted as $\Delta t_0(i,j)$ and under a perfect calibration it will be null. If any of the implied DOMs has a miscalibrated offset, it will then be $\Delta t_0(i, j) = t_0^j - t_0^i$, where $t_0^i$ would be the miscalibrated time offset of the flashing DOM and $t_0^j$ the one of the illuminated DOM. It is worth to mention that this method will not be sensitive to determine any global absolute offset of the whole DU, that should be determined with an inter-DU calibration, so only DOM relative corrections can be provided.

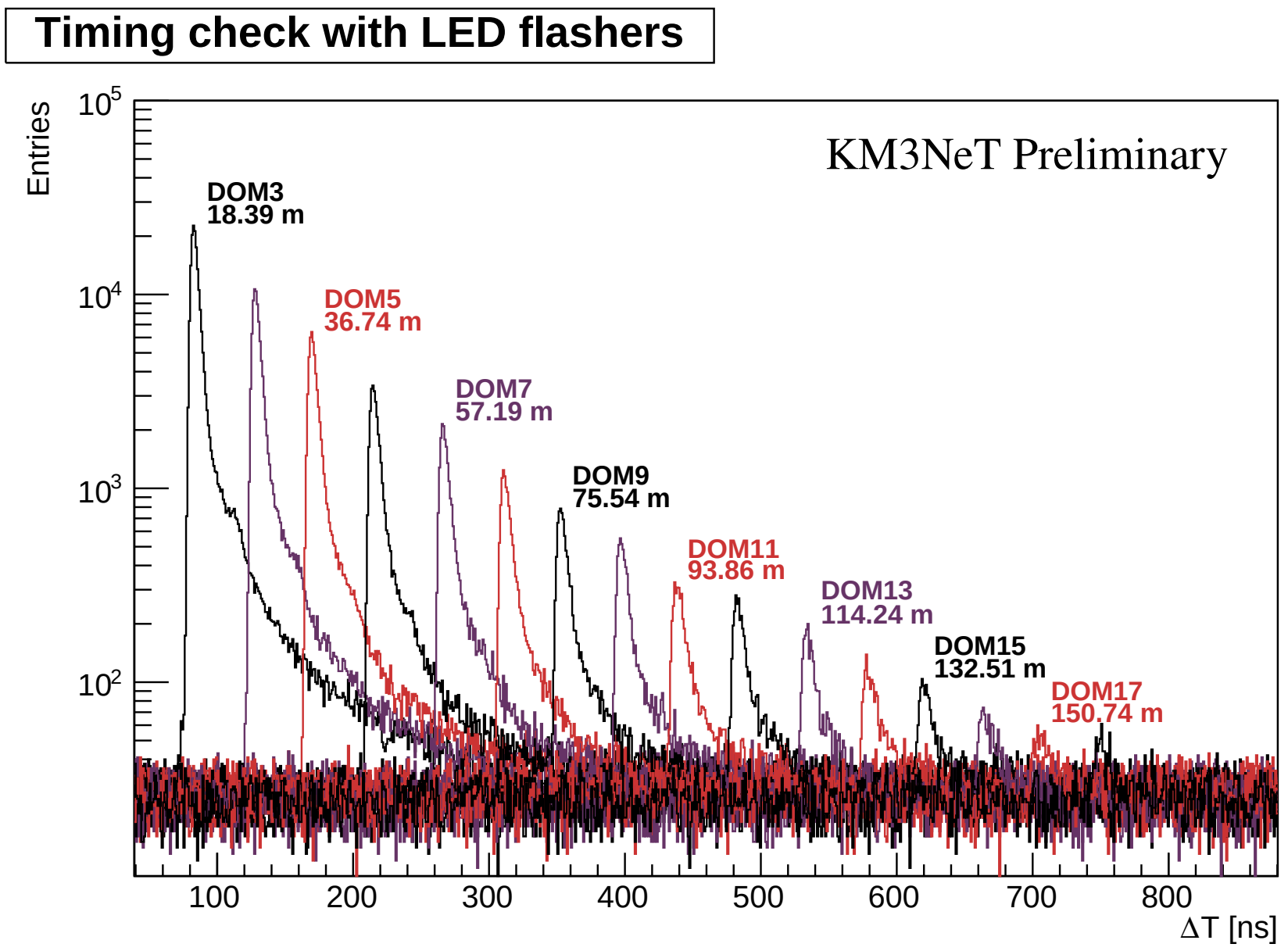

**Figure 5:** NB illumination of ORCA.0011 line.

Being a NB on each DOM, it is possible to have a significant redundancy by considering all the possible pairs of flashing-illuminated DOMs, allowing a robust *in situ* calibration system. For 18 DOMs it is possible to define a system of up to $\frac{1}{2}18(18-1) = 153$ equations, with 17 unknown offsets (one DOM offset can be fixed, e.g. $t_0^1 = 0$, since the method is not sensitive to any DU global offset) which can be summarized as follows:

$$\begin{pmatrix} 1 & 0 & 0 & \cdots & 0 & 0 \\ -1 & 1 & 0 & \cdots & 0 & 0 \\ -1 & 0 & 1 & \cdots & 0 & 0 \\ \vdots & \vdots & \vdots & \ddots & \vdots & \vdots \\ 0 & 0 & 0 & \cdots & -1 & 1 \end{pmatrix} \begin{pmatrix} t_0^1 \\ t_0^2 \\ t_0^3 \\ \vdots \\ t_0^{18} \end{pmatrix} = \begin{pmatrix} 0 \\ \Delta t_0(1,2) \\ \Delta t_0(1,3) \\ \vdots \\ \Delta t_0(17,18) \end{pmatrix} \tag{1}$$

or more compactly:

$$M_{mn} \cdot (T_0)_n = (\Delta T_0)_m \tag{2}$$

with $m = 1, \ldots, 154$ and $n = 1, \ldots, 18$.

Each available $\Delta t_0(i, j)$ pair can be estimated from a Gaussian fit to the time difference distribution. This fit has an associated error $\sigma_\mu^{i,j}$ that can be introduced in the least squares solutions by means of a weighting covariance matrix $W$:

$$W_{mm} = diag\left(\frac{1}{x^2}, \frac{1}{(\sigma_\mu^{1,2})^2}, \frac{1}{(\sigma_\mu^{1,3})^2}, \cdots, \frac{1}{(\sigma_\mu^{17,18})^2}\right) \tag{3}$$

This will be helpful since sometimes the $\Delta t_0(i,j)$ fits can be not optimal nor very reliable, weighting them out when solving the system in eq. 2 via:

$$W_{mm} \cdot M_{mn} \cdot (T_0)_n = W_{mm} \cdot (\Delta T_0)_m \tag{4}$$

leading to the system solution:

$$(T_0)_n = \left(M_{nm}^T \cdot W_{mm} \cdot M_{mn}\right)^{-1} \cdot M_{nm}^T \cdot W_{mm} \cdot (\Delta T_0)_m \tag{5}$$

A covariance matrix $\mathbb{V}$ for the time offsets can also be defined:

$$\mathbb{V} = \left(M_{nm}^T \cdot W_{mm} \cdot M_{mn}\right)^{-1} \tag{6}$$

allowing to evaluate the goodness of the found offsets.

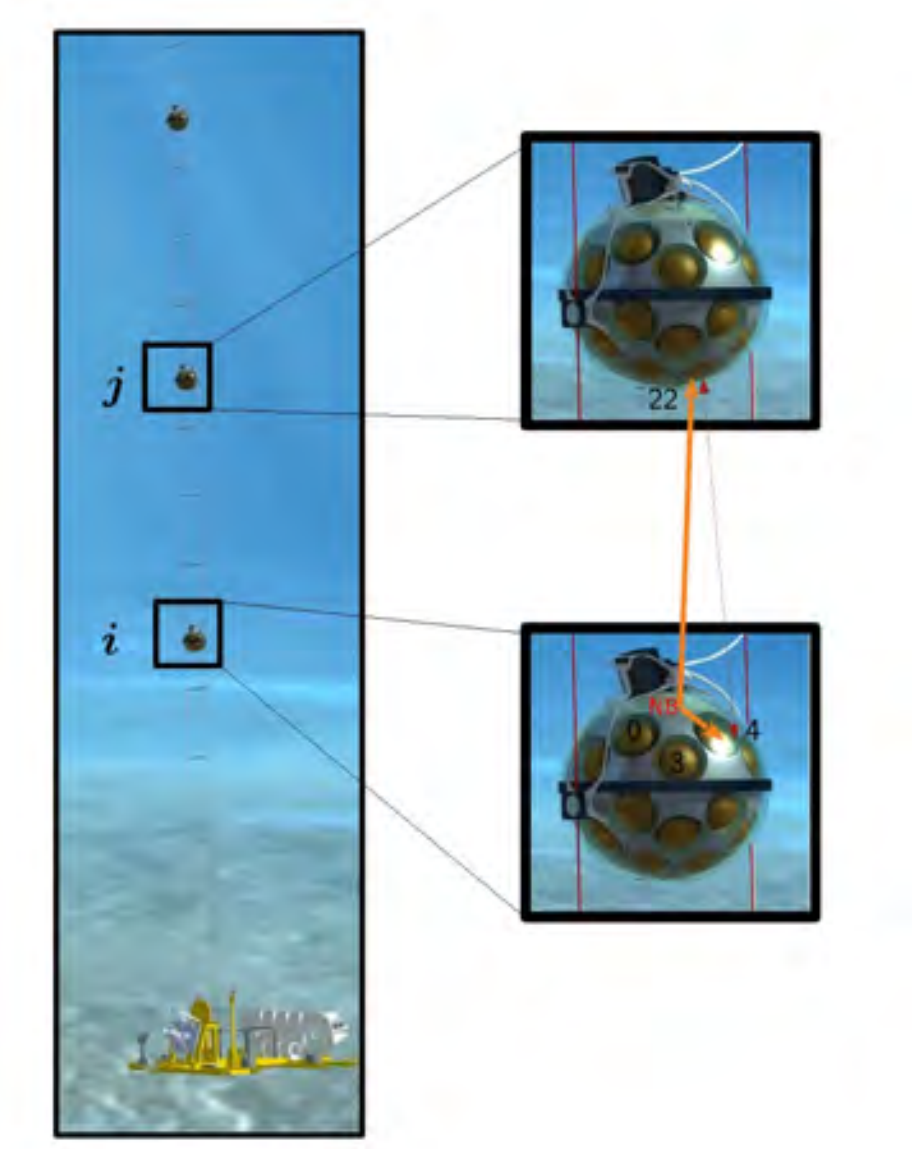

**Figure 4:** Concept of NB time calibration. DOM $i$ NB flash is detected by the closest PMTs and observed mostly by those at the bottom of DOM $j$.

## 5. Results

This calibration method has been applied to *in situ* NB data (see an example in Fig. 6) assuming a preliminary fixed geometry of the detector and mostly confirming the goodness of the dark room inter-DOM calibration. Further checks with a revised geometry are on-going. It will be used to determine the inter-DOM offset of the few DOMs that, due to technical issues, were not possible to calibrate during the dark room phase. For the future it is planned to design a sustainable NB data taking to scale reasonably up to the whole detector including the optimally tuned NB voltages described in this work.

## 6. Acknowledgments

The authors acknowledge the financial support from Ministerio de Ciencia, Innovación, Investigación y Universidades (MCIU): reference PGC2018-096663-B-C41; Ministerio de Ciencia e Innovación (MCI): reference PID2021-124591NB-C41; and Generalitat Valenciana: references CIDEGENT/2018/034 and CIDEGENT/2020/049. Also, this research was supported by an FPU grant (Formación de Profesorado Universitario) from the Spanish Ministry of Universities (MIU) to Juan Palacios González (reference FPU20/03176).

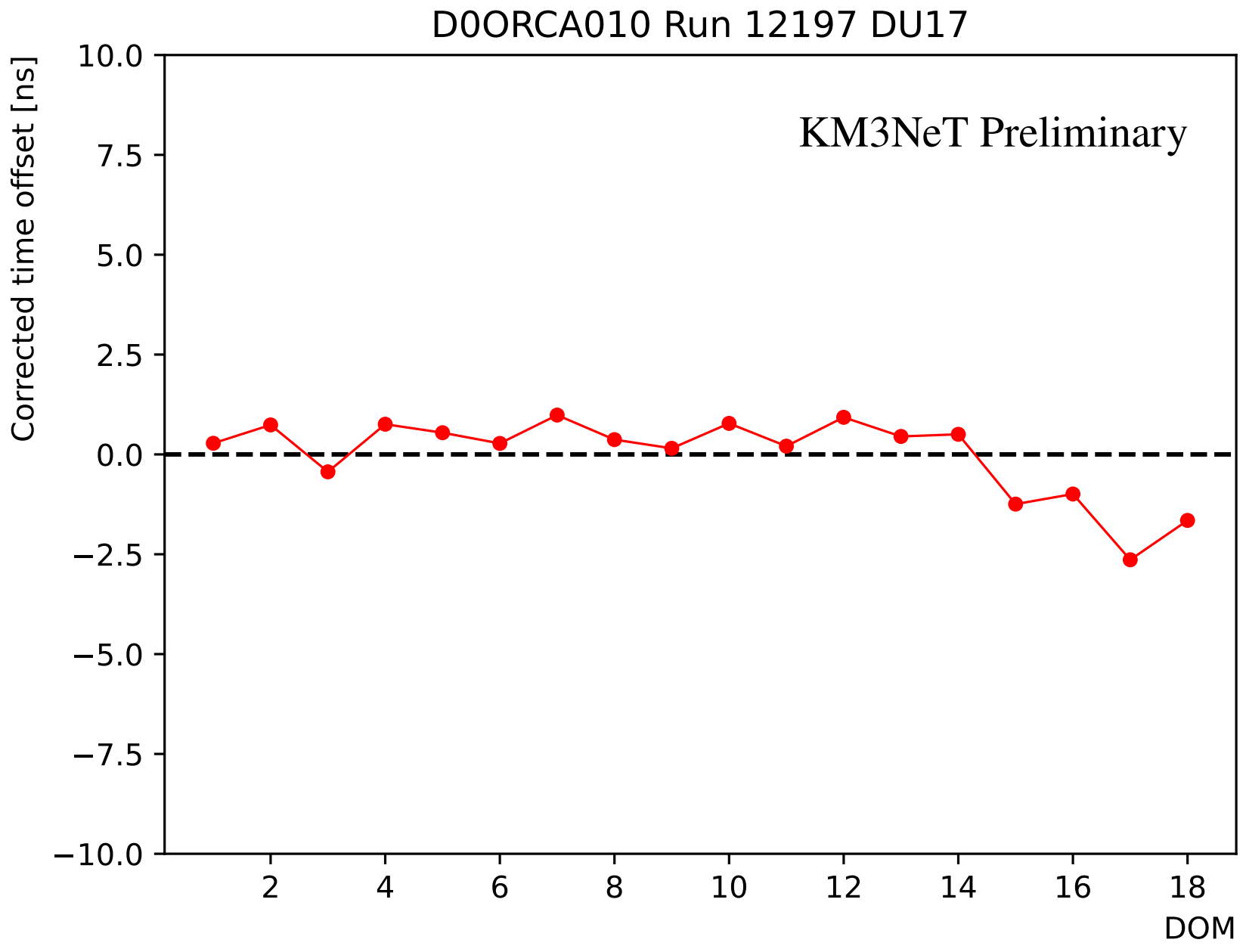

**Figure 6:** NB time calibration.

# The KM3NeT/ARCA Calibration Unit

**M. Anghinolfi,[c] F. Benfenati,[a,b,*] R. Cereseto,[c] R. Cocimano,[d] K. Leismuller,[d] C. D'Amato,[d] S. Dalle Fabbriche,[d] S. Mastroianni,[e] P. Migliozzi,[e] C.M. Mollo,[e] C. Nicolau,[f] S. Ottonello,[c] G.Pellegrini,[a] G. Riccobene,[d] L. Roscilli,[e] A. Rovelli[d] and S.Viola[d] for the KM3NeT collaboration**

[a] *Istituto Nazionale di Fisica Nucleare - Sezione di Bologna, Viale Berti-Pichat 6/2, 40127 Bologna, Italy*

[b] *Università di Bologna, Dipartimento di Fisica e Astronomia, Viale Berti Pichat 6/2, 40127*

[c] *Istituto Nazionale di Fisica Nucleare - Sezione di Genova, Via Dodecaneso 33, 16146 Genova, Italy*

[d] *Istituto Nazionale di Fisica Nucleare - Laboratori Nazionali del Sud Via S. Sofia 62, 95125 Catania, Italy*

[e] *Istituto Nazionale di Fisica Nucleare - Sezione di Napoli Via Cintia, 80126 Napoli, Italy*

[f] *Istituto Nazionale di Fisica Nucleare - Sezione di Roma Piazzale Aldo Moro 2, 00185 Roma, Italy*

*E-mail:* benfenat@bo.infn.it

The KM3NeT/ARCA calibration unit is a dedicated calibration system designed to improve the accuracy of the acoustic positioning system of the detector optical modules and monitor the water column physical properties. The calibration unit is composed of a calibration base and an instrumentation unit, connected with an electrical inter-link cable. The deployment of one calibration unit for each of the two building-blocks, in which the ARCA detector is subdivided, is foreseen, with the first one to be deployed in 2024. The calibration base is made of an anchoring structure, connected for power supply and communication to a junction-box, where an acoustic beacon and a hydrophone used for the positioning system are mounted and which hosts a pressure vessel containing the required electronics. The instrumentation unit consists of an anchoring base and a 750m-long inductive line, kept vertical by a top buoy and equipped with oceanographic sensors. The base is linked to the calibration base for power and readout and hosts an absolute pressure gauge used as depth reference and a vessel containing the electronics for managing sensor communication. The line hosts two sound velocimeters and two conductivity-temperature-depth probes equipped with dissolved oxygen sensors, to measure sound velocity and allow for the determination of acoustic wave speed, and two Doppler current sensors to provide information on sea current speed and direction, further improving the accuracy of the positioning system.

*Speaker

## 1. Introduction

The KM3NeT underwater neutrino telescope consists of two detectors, ORCA and ARCA [1], implemented as large-volume 3D arrays of Digital Optical Modules (DOMs) [2] hosting photomultipliers tubes (PMTs) and placed at a depth of ~2450 m off-shore Toulon, France (ORCA) and of ~3500 m off-shore Sicily, Italy (ARCA). Both detectors are made of vertically aligned Detection Units (DUs), each hosting 18 DOMs. In their final configuration, ARCA will have 230 DUs and ORCA 115 DUs. Each DOM contains 31 3-inch PMTs, calibration and positioning instrumentation including a light emitter and readout electronic boards named Central Logic Boards (CLBs). Each DU is equipped with a top buoy to provide vertical support and is anchored to the sea-floor with a heavy anchor frame which also incorporates the DU Base Module (DU-BM) and calibration devices such as hydrophones, lasers, acoustic beacons. The DU-BM hosts the electronics for powering the DU and a CLB for instrument control and data readout. In ARCA, DUs are connected to submarine Junction-Boxes (JBs), which act as electrical and optical fiber distribution systems, distributing data and power from the main electro-optical cables to the DUs and vice-versa. JBs are also endowed with hydrophones, lasers and acoustic beacons. While both ARCA and ORCA feature the same detection elements, they have distinct layouts designed to suit their respective scientific objectives; in particular, ARCA is designed to study cosmic neutrinos from the TeV to PeV scale and their possible astrophysical sources. The PMT array detects Cherenkov light emitted when relativistic charged particles traverse the detector volume. The recorded data is then utilised to determine the direction and energy of the incoming neutrino, which is responsible for generating these particles. In order to reconstruct neutrino direction with a precision better than 1°, DOMs need to be synchronised with nanosecond accuracy [3], and their exact location determined with an accuracy < 20 cm [4]. Having an impact on sound propagation, sea water properties must be continuously monitored because they affect the positioning calibrations.

To meet these goals, KM3NeT is going to deploy several dedicated Calibration Units (CUs). Measurements of the environmental parameters will be also exploited for Earth and marine sciences studies, providing the capability for continuous online monitoring of the marine ecosystem for extended durations. Due to the different specific requirements of ARCA and ORCA detectors and their sizes, the design of the CUs differs between the two sites. This proceeding provides a comprehensive description of the features, the objectives and the current status of the first ARCA CU that will be depolyed shortly; for the ORCA CU, refer to [5].

## 2. Positioning calibrations

Under the effect of deep sea currents, DOMs tend to float around the vertical position which must hence be continuously monitored as well as their orientation, provided by internal compass data. This is done by means of a relative acoustic positioning system (APS) which relies on an auto-calibrating Long-Baseline (LBL) system of emitters and receivers displaced in the detector volume, located on the detector elements and on additional autonomous tripods. The accuracy in the determination of the DOMs position with respect to the LBL reference system, required to reach the desired detector angular resolution, is ~10 cm [7] CUs are going to improve the APS in several ways: adding permanent acoustic beacons and hydrophones in the sea-floor network, providing measurements of sea currents along water column and allowing the characterization of sound velocity in the medium with precise measurements to be used in combination with data from the APS.

## 3. KM3NeT/ARCA Calibration Unit

The Calibration Unit is divided into two main components, connected by a 700m-long electrical ROV-operable inter-link cable: the Calibration Base (CB) and the Instrumentation Unit (IU). The CB will be connected to a JB via a 300m-long standard DU electro-optical inter-link for power and communication with the shore station. Detailed descriptions of the CB and the IU are in Sec. 3.1 and Sec. 3.2. Their locations at the ARCA site are shown in Fig. ?? along with the foreseen dates of deployment.

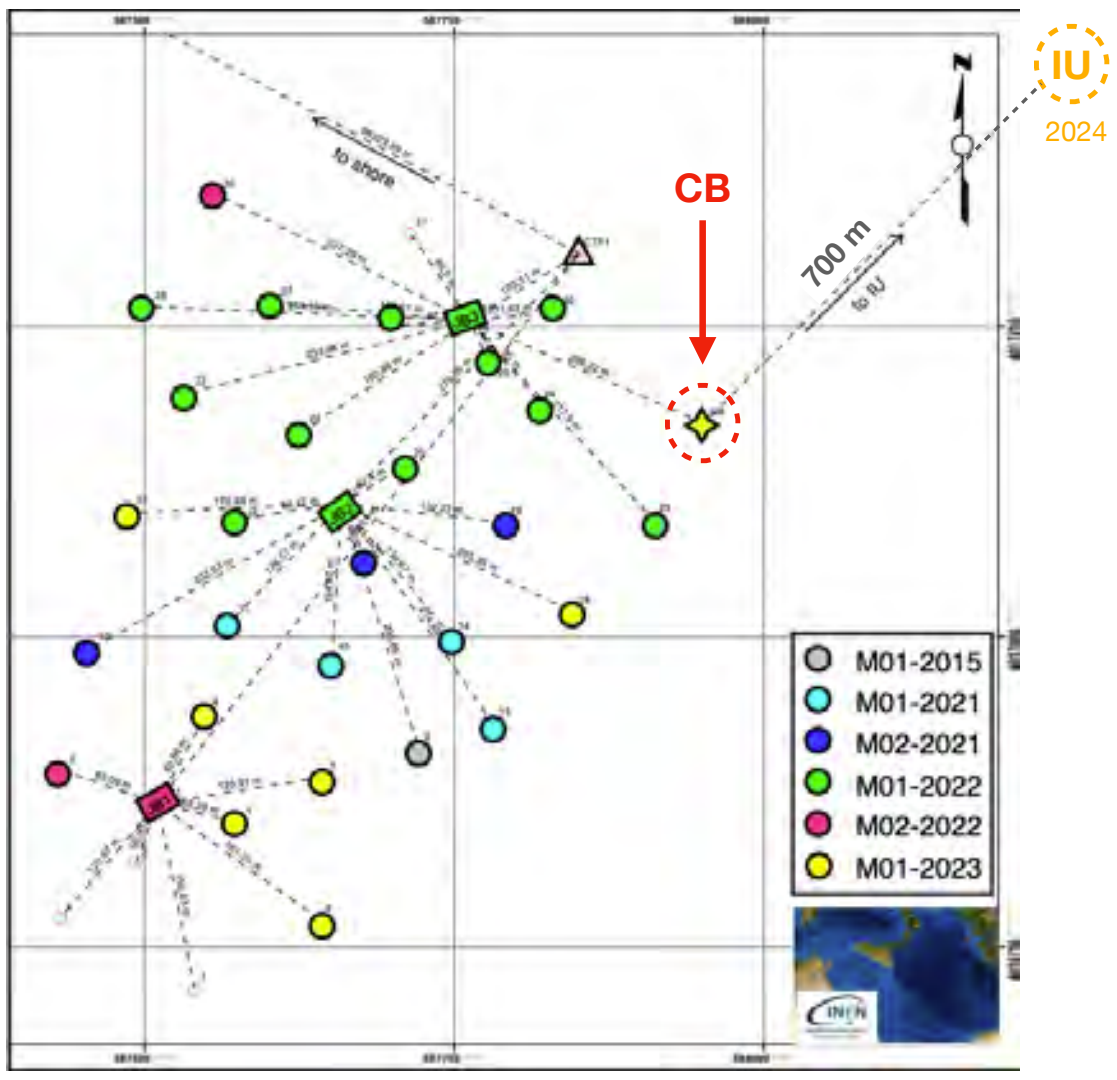

**Figure 1:** The current ARCA sea-floor map; coloured circles (boxes) indicate installed and next-to-be-deployed (Sep. 2023) DUs (JBs), while the straight lines indicate the inter-link cables connecting the DUs to the JBs. The colour scheme follows the sequence of marine campaigns indicated by a progressive number and the year. Highlighted are the positions of the CB and of the IU. The IU position is chosen to be in safe distance from the DU array and, according to the dominant current measured at site, to minimise the risk of line drifting into the field during deployment and recovery operations.

### 3.1 KM3NeT/ARCA Calibration Base

The main role of the CB is to contribute to the detector APS by adding a LBL hydrophone and an acoustic beacon to the network. These instruments are mounted on the CB frame and are connected to a BM which contains the electronics for managing the power and communication both with the shore station and with the IU. The main components of the CB are described in the following subsections and are shown in Fig 2a.

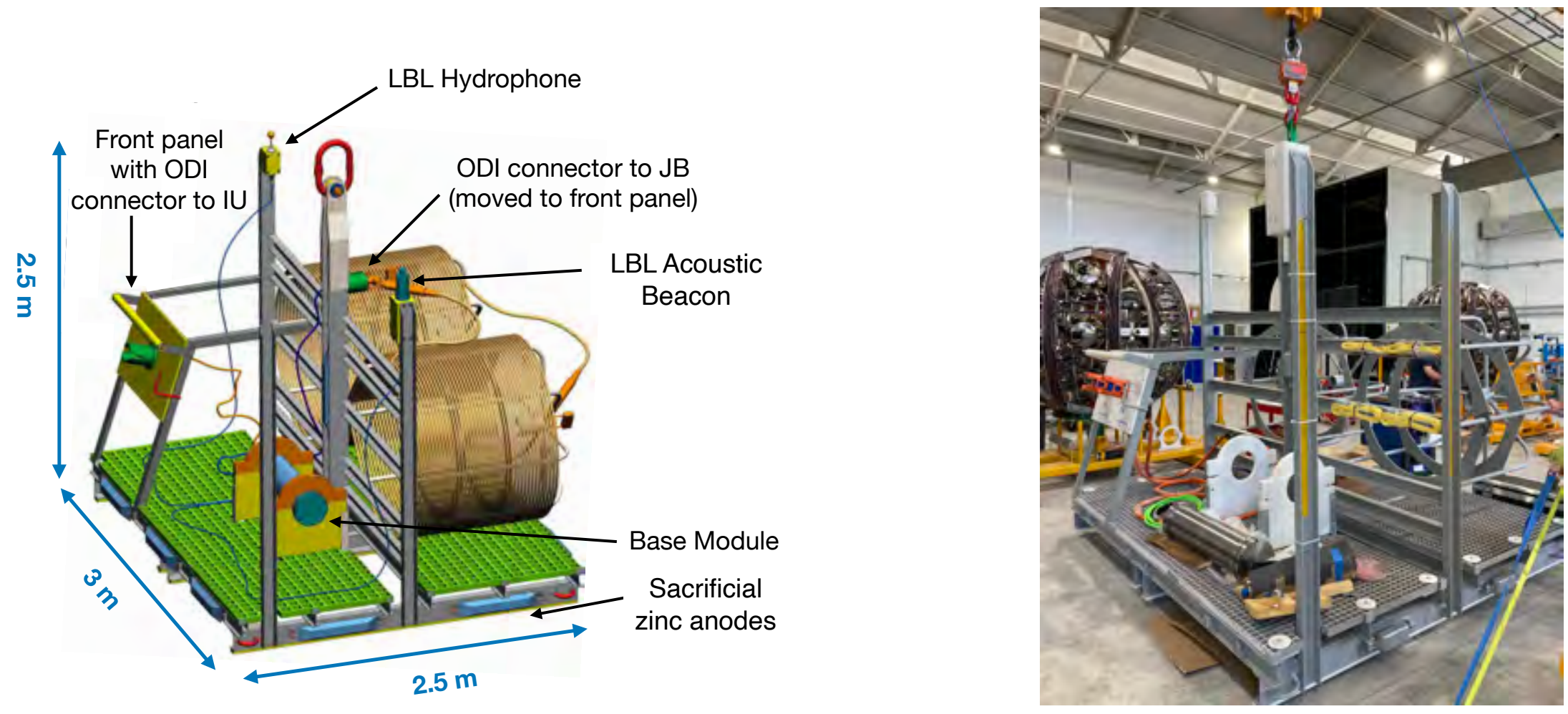

**(a)** **(b)**

**Figure 2:** (a): mechanical structure of the Calibration Base: components and instrumentation are highlighted. (b): the Calibration Base during integration.

### 3.1.1 Anchoring frame

The CB is a compact structure made of primary and secondary iron crossbars welded together in a configuration which is expected to be effective against corrosion ph enomena. It has a central load-bearing beam and two lateral beams to support the hydrophone and the acoustic beacon, and it features a high density polyethylene (PEHD 500) front panel where the ODI ROV-operable plugs [9] for the inter-links towards the JB and the IU are integrated. PEHD 500 is also used for the supports that fix the BM, the hydrophone and the acoustic beacon to the s tructure. All the bolts and screws used to fix the components with plastic supports are made of t itanium. In order to guarantee protection against corrosion for long-term deployment, all the metal surface is coated with anti-corrosion painting (standard NORSOK M-501 [8]) and a cathodic protection system with eleven ~12 kg sacrificial zinc anodes located on the bottom sides and on the central beam is applied. The base includes fiberglass grids easing the assembly and preventing the anti-corrosion painting from scratching during integration. Initially, the structure was built with a support for hosting the inter-link towards the JB and the corresponding ODI connector was foreseen to be integrated on a panel adjacent to the central beam; since this inter-link has been separately deployed in advance, the connector has been moved to the front panel to facilitate ROV operations.

### 3.1.2 Instruments

The selected hydrophone is the DG1330, a digital omni-directional hydrophone specifically designed and produced by Colmar S.r.l. [10] for KM3NeT to be operated at 3500m depth. It consists of a spherical piezo-ceramic element, read-out by an analogue board splitting the signal in two lines with different gains (+46 dB and +26 dB). The low gain channel has been implemented in order to prevent signal saturation due to the acoustic emission from the beacon in close proximity (~3 m), while the high gain channel is used for analyzing data received from distant beacons, spanning up to a few kilometers, as well as for studying faint acoustic signals, such as those related to bio-acoustics, environmental noise, and acoustic neutrino detection [11]. It includes an analogue signal high-pass filtering stage at 700 Hz to reject the low frequency ambient sea noise and improve the signal to noise ratio in the detection of beacon pulses range (20-40 kHz). The sampling frequency is 195.3 kHz, and the acceptance frequency range is 5-90 kHz. The two streams are sampled with a stereo 24 bit commercial ADC (CS-4270) and converted into AES/EBU protocol using a digital interface transmitter. The acoustic beacon is the Mediterraneo Senales Maritimas [12] MAB 100, endowed with a FFR SX30 acoustic transducer. The electronic boards are contained in a shielded titanium case resisting up to 4000m of depth. The beacon is programmed to autonomously emit every 30s its unique modulation signature carried by a sweep signal ranging from 40kHz to 36kHz, but it can also emit with external triggers. Both instruments are connected to the CB-BM via a common GISMA [13] MCIL6M connector, and linked through it to the CLB via the FMC board (see Sec. 3.1.3). Connection for power and communication is done via RS-485 with a RJ45 connector for the hydrophone, while the acoustic beacon power is taken directly from the BPS board (see Sec. 3.1.3) and communication uses RS-232 lines connected to the FMC. The clock from the CLB, synchronised to the master clock in the shore station, is exploited to timestamp data retrieved by the hydrophone and can be used to emit synchronised triggers to the acoustic beacon.

#### 3.1.3 Base Module

The mechanical container of the ARCA CB-BM is made of titanium and it is identical to the standard ARCA DU-BM, except for the interface flange where the connectors for the CB instruments and for the two jumper cables towards the plugs for the IU and the JB inter-links are located. The internal frame is also identical, excluding the electrical and optical elements used to connect the DU-BM to DOMs. The BM hosts the following electronics boards:

- The Base Power Supply (BPS) board receives 375VDC power from the JB and converts it into low voltage to power other electronic boards installed in the BM, the CB instruments and the IU.
- The Central Logic Board is the core of KM3NeT front-end electronics, installed inside each DOM and BM. Regardless of the detector element where they are integrated, all CLBs are identical except for the installed firmware. The CLB in the CB-BM receives commands and the common clock from the shore station and, through the FMC board, it interfaces the BPS, the CB instruments and the IU. Communication between CLB and shore station is established using an SFP laser integrated into the CLB.
- The FPGA Mezzanine Card (FMC) is a piggy pack board mounted on the CLB. It enables CLB communication with BPS, CB instruments and IU.

### 3.2 KM3NeT/ARCA Instrumentation Unit

The role of the IU is to acquire data regarding sea water properties. It is composed of the Instrumentation Line (IL) and the recoverable frame, shown in Fig. 3a. The IL is an inductive line, kept vertical by a top buoy, which both provides support for the oceanographic sensors and acts as transmission medium of the inductive transducers used to communicate with them, avoiding additional conductors. The frame consists in a metallic anchoring structure which hosts the Base Container (IU-BC) where the electronic boards are located, a front panel with the ROV bulkhead

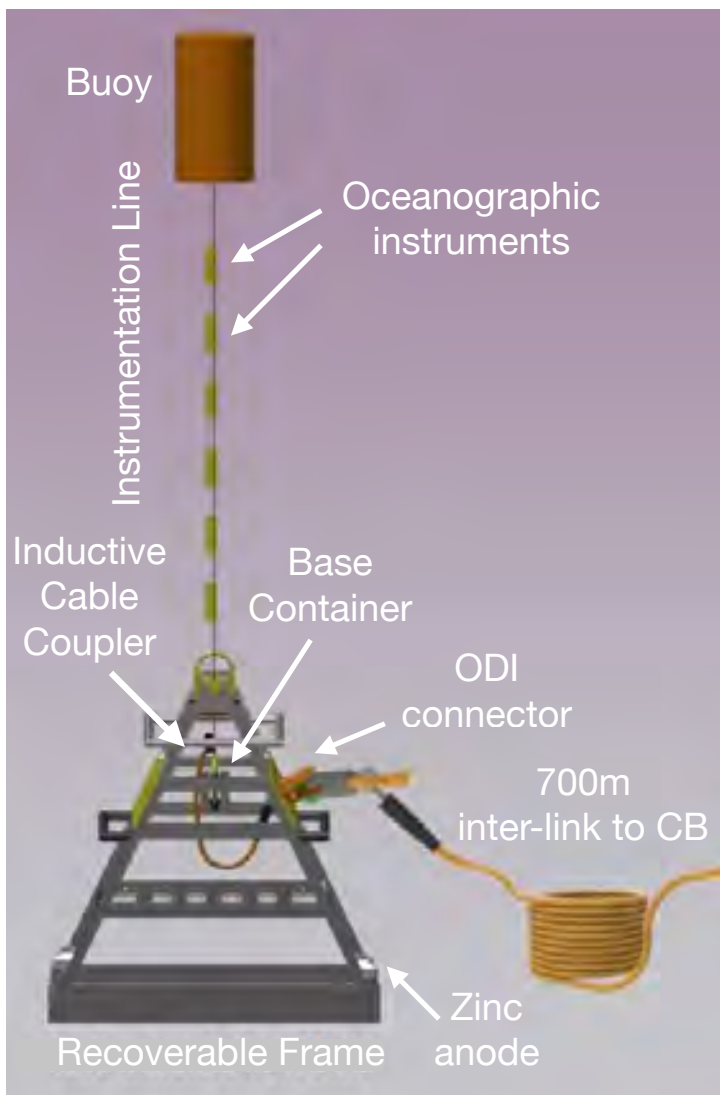

**(a)**

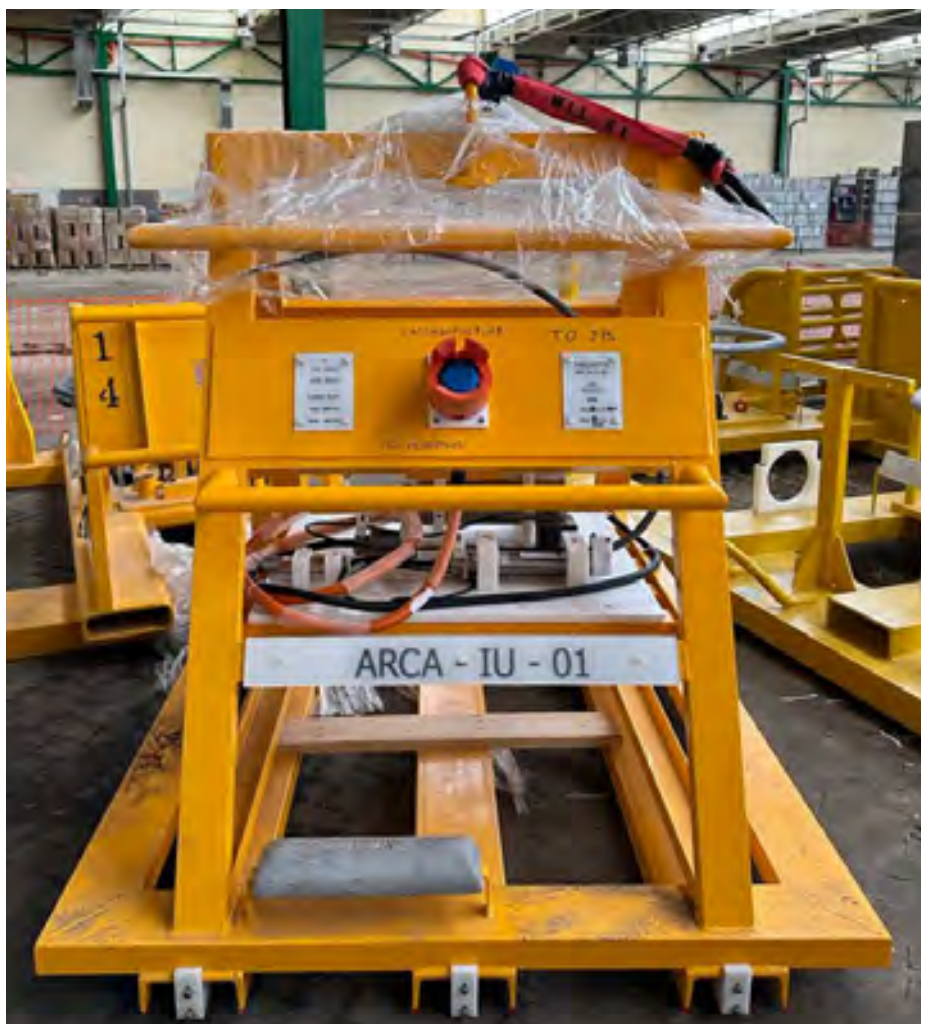

**(b)**

**Figure 3:** (a): ARCA IU main components. (b): the IU recoverable frame after integration.

plug for connecting the IU-BC to the CB, an absolute pressure gauge and the inductive coupler for the IL. Communication between the CB and IU instrumentation occurs through inductive modems. Two types of modem are present: a Seabird [14] Inductive Modem Module (IMM) hosted inside the IU-BC to allow the communication between the CLB in the CB-BM and inductive sensors, and Seabird SBE-44 Underwater Inductive Modems to interface non-inductive serial sensors to the line. The SBE-44 provides also battery power to connected sensors through a wired power link. A Seabird Inductive Cable Coupler, located on the frame, couples the IMM to the inductive cable.

### 3.2.1 Recoverable frame

The recoverable frame structure is made of stainless steel coated with anti-corrosive painting (standard NORSOK M-501) and cathodic protection is granted by two $\sim$12 kg welded sacrificial zinc anodes. The front panel and the support floor for the IU-BC and the other devices are built in PEHD 500. The frame has been designed in order to be recovered every two years, together with the IL, for batteries change and instrument re-calibration. To preserve the inter-link cable to the CB during recovery operations, a long-term cable parking terminal frame will be used.

## 3.3 Base container

The IU Base Container is a titanium pressure vessel containing an electronic board that receives the power and communication lines from the CB. The IMM is mounted on this board, from which it receives 12V power and which also includes a serial converter from RS-422 to RS-232 standards. The board is connected to the CB through a DMS connector on one of the vessel flange via a jumper cable to the ROV connector on the front panel. On the other flange, another DMS connector is used for the cable coupler connection to the board. The floor and the supports which fix the container and the instruments are made of PEHD 500, while titanium screws are used.

#### 3.3.1 Instrumentation Line

The inductive cable is a 750m-long, 3x19-strands galvanised jacketed wire rope with swaged socket terminations to allow grounding with seawater. The IU is equipped with the following instruments:

- Two native inductive conductivity-temperature-depth probes with pressure and optical dissolved oxygen sensors. They measure conductivity, temperature and depth and allow indirect calculations of sound velocity in water as a function of temperature, pressure and salinity through Chen and Millero [16] or Del Grosso's [17] formulas. They will be used to improve the fits of the APS. Measurements of oxygen and salt concentration (the latter inferred from conductivity) in water, as well as its temperature, will be used for oceanographic studies;
- Two non-inductive sound velocity sensors. They perform a direct measurement of the sound velocity in water, thus contributing to improve APS fits.
- Two non-inductive Doppler current sensors. They perform measurements of the sea current along the 3 directions and will be used to improve the mechanical DU line fit model [15].
- One non-inductive pressure gauge, used as absolute depth reference to improve the accuracy of the APS; its measurements will also allow geo-oceanographic studies on the variation of sea depth and can be used to monitor earthquakes and tsunamis.

Collected data will be made available to European oceanographic observatories and organizations such as ESONET-EMSO [18]. The instruments will be distributed along the water column as shown in Tab. 1, where models are also indicated.

| **Height from seabed (m)** | **Instrument** | **Measurement** |
|---|---|---|
| 650 | Valeport [20] mini-SVS | Sound velocity |
| 600 | Seabird SBE-37 IMP-ODO MicroCAT CTD | Conductivity, temperature, pressure |
| 550 | AAndera [19] ZPulse Doppler Current Sensor 4520R | Current velocity |
| 200 | Valeport mini-SVS | Sound velocity |
| 150 | Seabird SBE-37 IMP-ODO MicroCAT CTD | Conductivity, temperature, pressure |
| 100 | AAndera ZPulse Doppler Current Sensor 4520R | Current velocity |
| 0 | Paroscientific [21] Digiquartz Depth Sensor 8CB4000-I | Pressure |

**Table 1:** Oceanographic sensors mounted on the Inductive Line, and respective indicative height from the seabed.

## 4. Outlooks and conclusions

The first ARCA CB is fully tested and ready to be deployed, while the IU is under finalisation: the recoverable frame is integrated, and instruments have to be configured and mounted on the inductive cable; a final communication test with the integrated IU will be done before preparation for deployment, foreseen in 2024. Firmware for CLB and software for communication with CB and IU instrumentation have been tested. The software update to the detector control system required for interfacing to the IU instruments and to store its data in the KM3NeT database is under development.

# KM3NeT acquisition electronics: status and upgrade

**D. Calvo,[a,*] D. Real,[a] P. Jansweijer,[b] V. van Beveren,[b] J.W. Schmelling,[b] G. Pellegrini,[c] A. Diaz,[d] S. Colonges,[e] C. Champion,[e] F. Benfenati,[c] F. Filippini,[c] T. Chiarusi,[c] P. Musico,[f] P. Litrico,[g] J.D. Zornoza[a] and R. Gozzini[a] on behalf of the KM3NeT Collaboration**

[a] *IFIC - Instituto de Física Corpuscular (CSIC - Universitat de València), c/ Catedrático José Beltrán, 2, 46980 Paterna, Valencia, Spain*

[b] *Nikhef, National Institute for Subatomic Physics, PO Box 41882, Amsterdam, 1009 DB, Netherlands*

[d] *University of Granada, Dept. of Computer Architecture and Technology/CITIC, 18071 Granada, Spain*

[e] *APC, Université Paris Diderot, CNRS/IN2P3, CEA/IRFU, Observatoire de Paris, Sorbonne Paris Cité, 75205 Paris, France*

[c] *INFN, Sezione di Bologna, v.le C. Berti-Pichat, 6/2, Bologna, 40127, Italy*

[f] *INFN, Sezione di Genova, Via Dodecaneso 33, Genova, 16146, Italy*

[g] *INFN, Laboratori Nazionali del Sud, Via Santa Sofia,62 , CT - 95123, Catania, Italy*

*E-mail:* david.calvo@ific.uv.es, real@ific.uv.es

The KM3NeT Collaboration is building and operating two deep-sea neutrino telescopes at the bottom of the Mediterranean Sea. The telescopes consist of lattices of light detectors housed in pressure-resistant glass spheres, the so-called digital optical modules, which house 31 3-inches of diameter photomultipliers, as well as the acquisition electronics. The so-called detection units are vertical strings along which digital optical modules are installed. For the first phase of the construction of the telescopes, several tens of detection units have been produced, out of which almost 40 have already been deployed with more than 20,000 photomultipliers installed and taking data. Once finished, the two telescopes will have installed more than ten thousand acquisition nodes, completing one of the more complex networks in the world in terms of operation and synchronisation. This work presents the current status of the acquisition electronics, including the upgrade of the principal components such as the central logic board, the signal collecting board, switching core board and Glenair back plane.

*Speaker

## 1. Introduction

The KM3NeT Collaboration is engaged in the construction and operation of two deep-sea neutrino telescopes located at the bottom of the Mediterranean Sea. These telescopes are designed to detect neutrino interactions by using lattices of light detectors contained within pressure-resistant glass spheres known as Digital Optical Modules (DOMs) [1]. Each DOM houses a cluster of 31 3-inches of diameter photomultiplier tubes (PMTs) distributed on the surface of the glass sphere, and the necessary acquisition electronics inside. DOMs are installed along vertical strings called Detection Units (DUs). There are 18 DOMs per DU. In the initial phase of telescope construction, a considerable number of DUs have been manufactured, and almost 40 of them have already been deployed, incorporating over 20,000 installed PMTs that are actively collecting data. Once fully completed, the two telescopes will have an intricate network comprising over ten thousand acquisition nodes, establishing one of the most sophisticated operational and synchronised networks worldwide. This paper provides an overview of the current status of the acquisition electronics in the KM3NeT neutrino telescopes [2–4]. Emphasis is placed on the ongoing upgrade of key components, the so-called: central logic board [5, 6], signal collecting boards, switching core board and Glenair back plane. The upgraded electronics will enhanced data collection capabilities, improving the overall performance and efficiency of the data acquisition system of the telescopes. The work presented herein highlights the progress made so far in the electronic boards of the development of the KM3NeT neutrino telescopes.

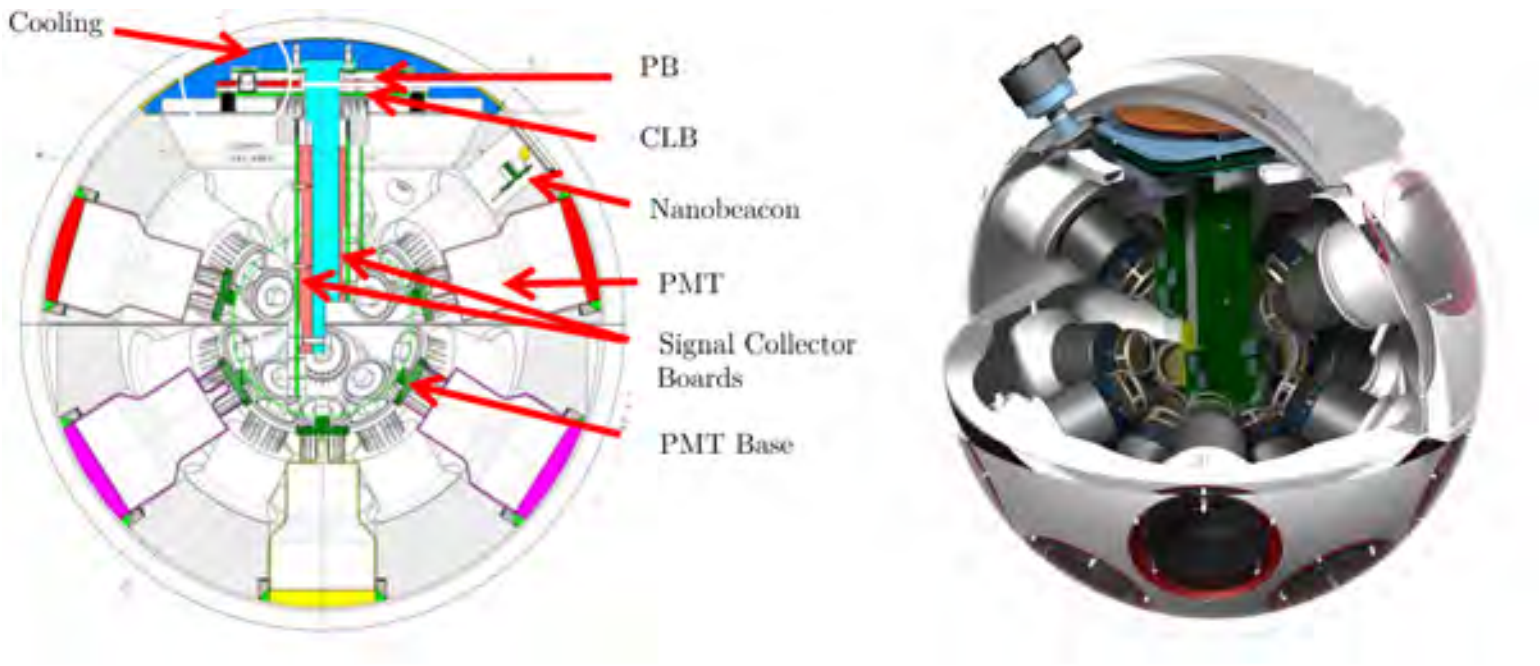

**Figure 1:** Left: 2D vertical cross section of the DOM with the main components indicated. Right: 3D representation of the DOM.

## 2. KM3NeT hardware

The DOM is the main element in the KM3NeT telescope operation. The 31 PMTs and all the necessary electronics to control the operation of the DOMs are located inside. The principal electronic board is called the Central Logic Board (CLB), which manages and controls all the functionalities of the DOM. Connected to the CLB there is the Power Board (PB), whose mission is to generate all the voltages required by the DOM. The PMTs are connected to boards called bases, which are responsible for generating the power supply voltage for the PMT, as well as digitising its electrical signal to be sent to the CLB. The signals from the PMTs are sent to the CLB for processing by means of two boards known as the Signal Collector Boards (SCBs). Finally, all the DOMs incorporate a device called Nanobeacon for time calibration, which uses a pulsed LED [7]. Figure 1 shows a representation of the DOM with the position of all its componentes.

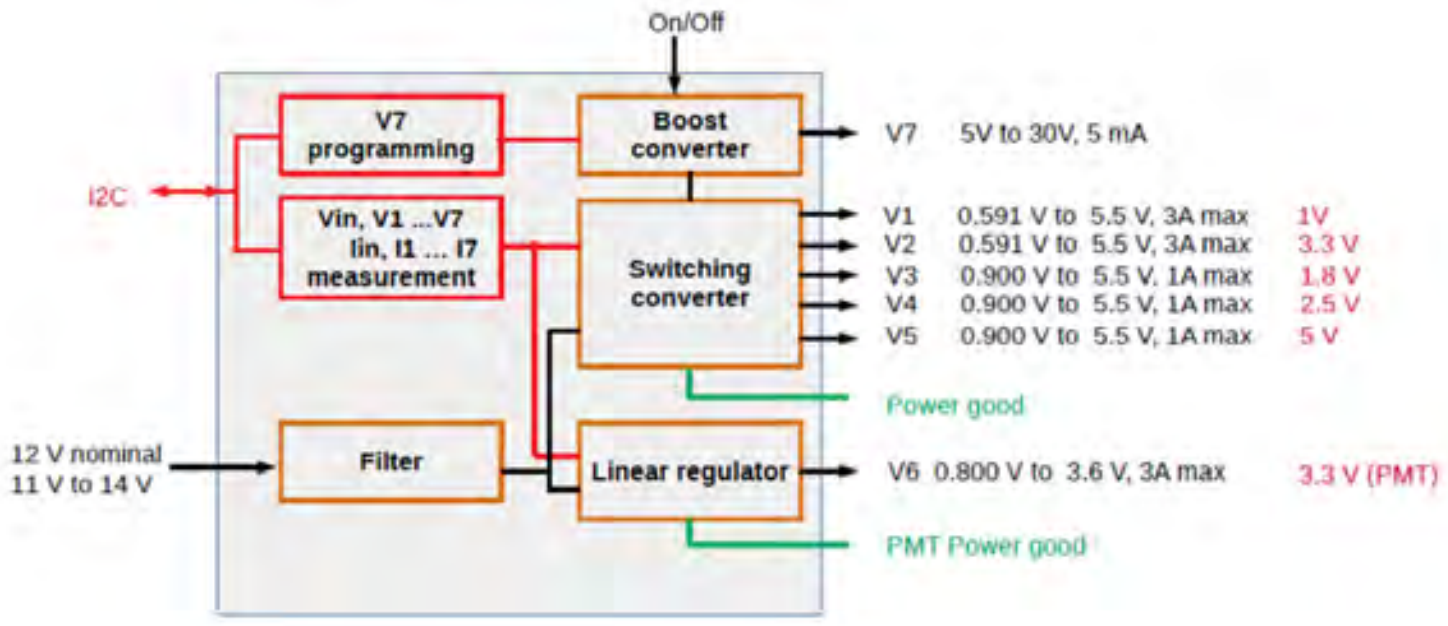

**Figure 2:** Block diagram of the power board. From a 12 V input voltage, all the necessary voltages for the operation of the DOM are generated using voltage regulators.

### 2.1 Central Logic Board

The CLB, [8, 9] is a multi-layered electronic board that controls the functioning of the DOM, including acquisition, synchronisation, instrumentation, and various communication interfaces. The control of all the components of the electronic board is implemented on a Field-Programmable Gate Array (FPGA) Kintex-7 from Xilinx Inc. The FPGA is responsible for the acquisition and processing of the optical signals generated by the PMTs, and the acoustic signals collected by the piezoelectric sensor of the positioning system. The CLB includes the acquisition firmware and the embedded software [10, 11]. All events, both optical and acoustic, are digitised with a timestamp provided by the White Rabbit (WR) synchronisation protocol. This protocol enables the synchronisation of all the KM3NeT nodes with an accuracy of 1 ns.

### 2.2 Power Board

At the entrance of each DOM, there is a voltage converter that transforms the voltage signal of 375 – 400 V DC from the high-voltage power supply system of KM3NeT into a 12 V DC voltage. The PB receives these 12 V and generates all the voltages required by the DOM: 1 V, 1.8 V, 2.5 V, two 3.3 V (one for the CLB and another for the PMT power supply), 5 V, and an adjustable voltage between 5 – 30 V for the calibration systems [12]. If all the voltages have been generated correctly, a signal called Power Good is activated. Two analog-to-digital converters (ADCs) are used to read all voltage and current values. Figure 2 shows a descriptive diagram of the PB with all its power supplies and configuration interfaces.

A previous study was conducted to accurately select the voltage converters, taking into consideration the current requirements for each power supply rail. This allowed for the optimisation of the board's efficiency. The total power consumption of a DOM is approximately 6 W. The PB incorporates a sequential start-up of the voltages, ensuring proper power supply to the FPGA. This prevents excessive currents during start-up that could damage the system.

### 2.3 Signal Collector Board

The Low voltage Differential Signals (LVDSs) generated by the PMTs bases are sent to the CLB by two SCBs. The first one, called SCB Large, sends the signals of the 19 PMTs located in the lower hemisphere of the DOM, and the second one, called SCB Small, sends the signals of the 12 PMTs located in the upper hemisphere. The SCBs are also responsible for sending configuration and monitoring commands from the CLB to the PMTs. The main components of the SCBs include a Molex 754332104 connector, a Xilinx Coolrunner Complex Programmable Logic Device (CPLD) chip for controlling and monitoring the PMT bases through current sensors, current limiters, a multiplexer for different Inter-Integrated Circuit($I^2C$) lines, and a link that allows communication between the acoustic piezoelectric sensor and the CLB.

## 3. Upgrade and new developments

### 3.1 Central Logic Board

The CLB has been designed to include a total of 12 layers, comprising 6 signal layers, 2 power planes, and 4 ground planes. These layers are arranged symmetrically around the 2 power planes. The positioning of the ground planes alongside the signal layers was done to enhance signal integrity. Special attention was given to routing the differential pairs in order to maintain a time difference of less than 100 ps between different PMT signals and less than 20 ps between clock s ignals. A reliability analysis was conducted using the FIDES method [15], which indicated an estimated failure risk of less than 10% over a span of 15 years. Various signal integrity simulations were performed on different signals present on the board, demonstrating a high level of discrimination. Several modifications have been made regarding the original board. One of the most important has been the provision of a secondary clock circuit. Since KM3NeT is an isolated system that cannot be repaired or modified when in operation, if the clock system fails, the DOM would be completely lost. This is why it has been decided to include a backup clock system. Figure 3 shows a diagram of the complete clock scheme of the upgraded CLB. Another challenge faced by an experiment whose construction can last several years is component obsolescence. Several components in the CLB have already become obsolete, being the most important one the Molex connector that connects the CLB with the signal collector boards. A redesign of the electronic board has been necessary, replacing the Molex connectors with another combination of SAMTEC connectors specifically designed for KM3NeT. Figure 4 shows a 3D representation of the upgraded CLB design.

### 3.2 Signal Collector Boards

Also, the SCBs have been affected by the obsolescence of components, which has led to a redesign of the electronic board. Similarly to the CLB, the main obsolescence problem in the SCBs has arisen with the Molex connector that connects them to the CLB. A change to SAMTEC connectors with a specific configuration for KM3NeT has also been necessary for these boards. A model with the CLB and the signal collection boards integrated in the mechanical frame of the DOM is shown in Figure 5.

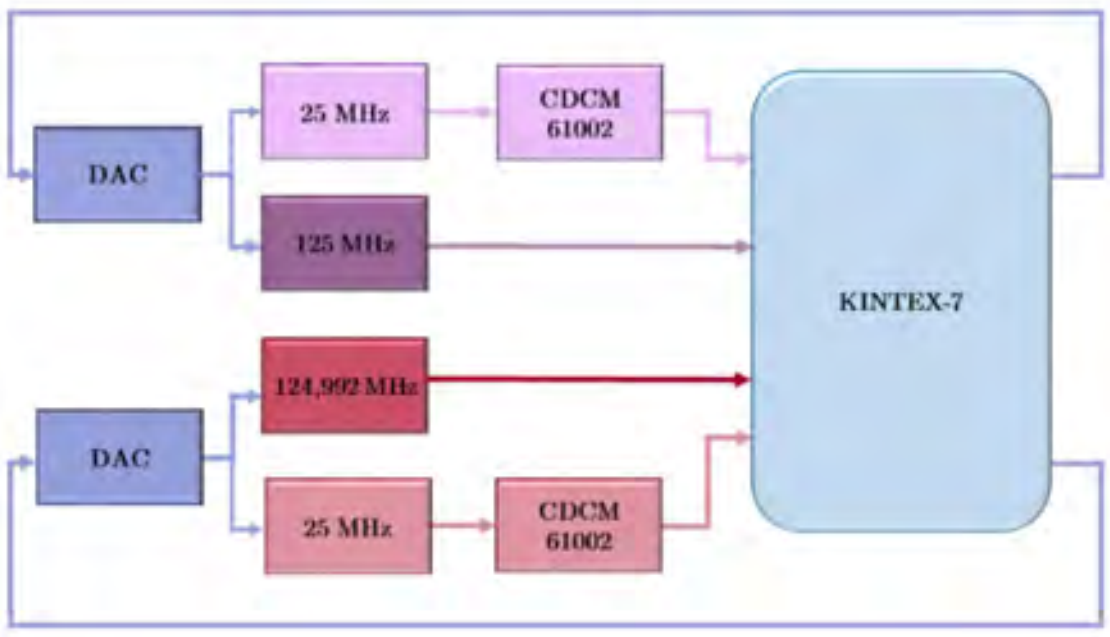

**Figure 3:** Block diagram of the clock system of the upgraded version of the CLB, including the main clocks and the backup circuit based on 25 MHz and CDCM61002.

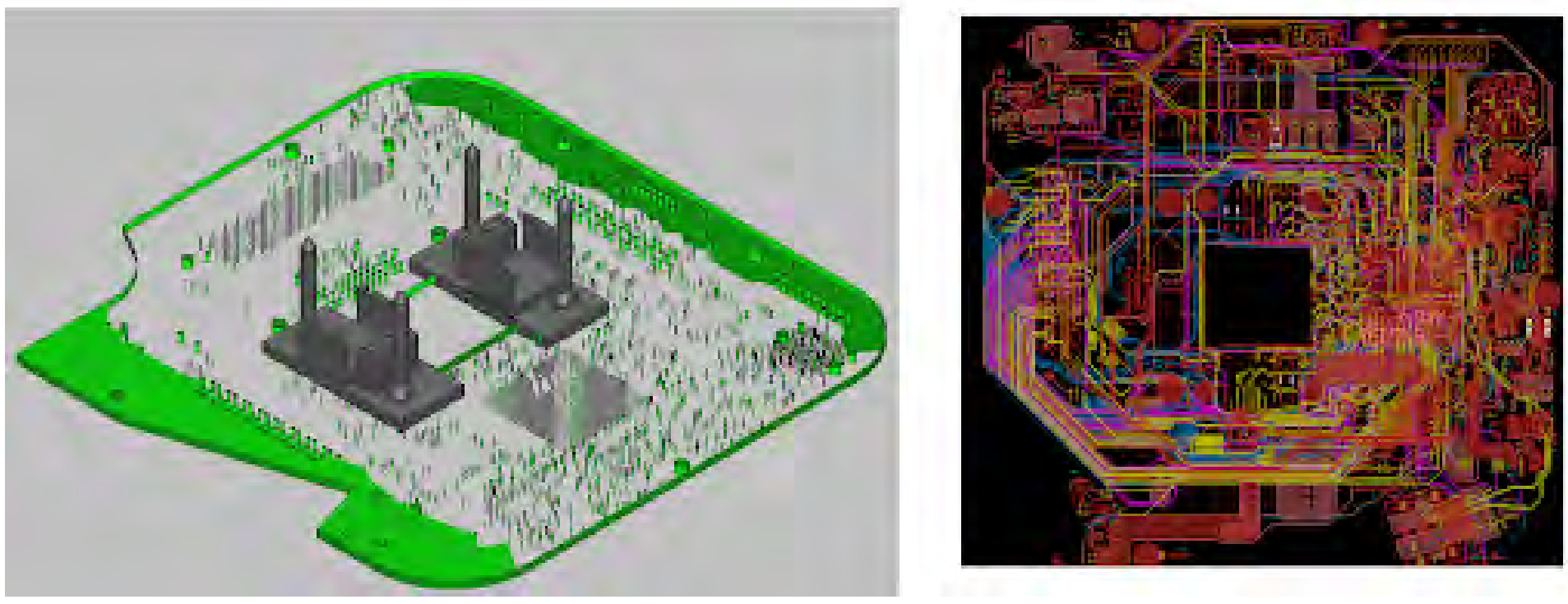

**Figure 4:** 3D model (left) and layout (right) of the upgraded CLB.

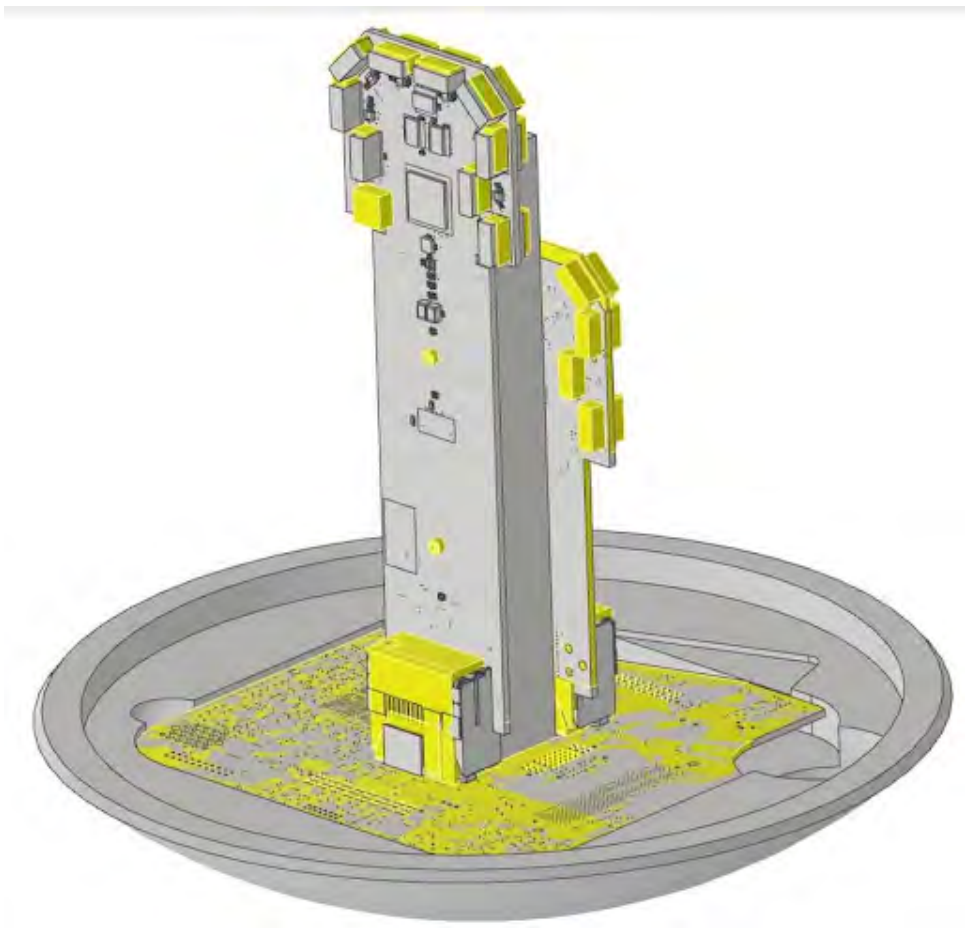

**Figure 5:** Mechanics model where the upgraded versions of the CLB and the SCBs are integrated into the mechanical support of the DOM.

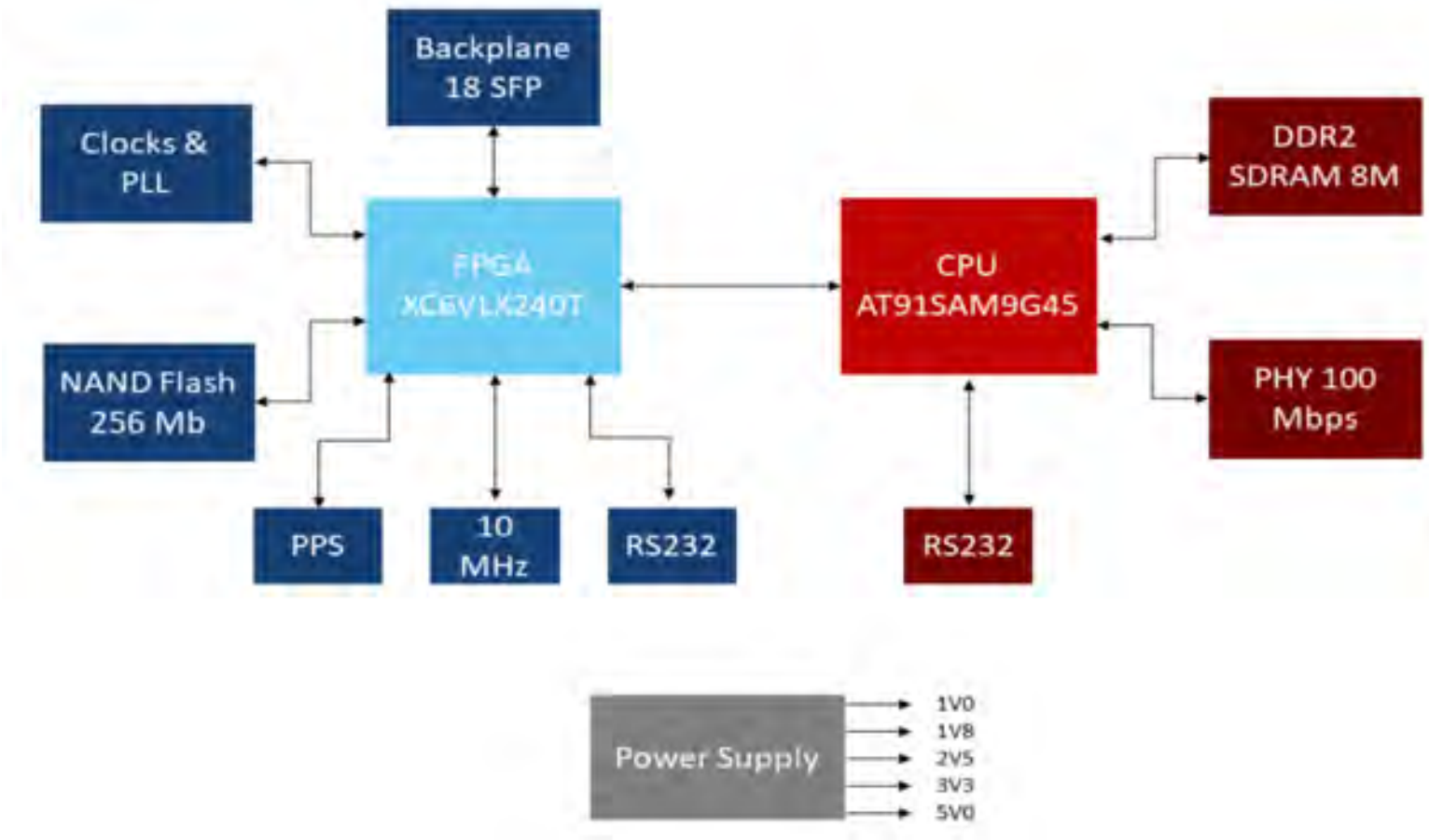

**Figure 6:** Diagram of the main blocks that form the so-called White Rabbit Switch.

### 3.3 White Rabbit Switch

The KM3NeT experiment is an infrastructure composed of thousands of DOMs, which are responsible for detecting and processing the Cherenkov radiation induced by the charged particles produced in the neutrino interactions, to reconstruct the trajectory and energy of these neutrinos. The DOMs, separated by tens of meters, have their own acquisition electronics and their own temporal scale, so it important to synchronise the clocks of all DOMs with high precision. To achieve this, the White Rabbit (WR) protocol has been used [13, 14]. WR forms a hierarchical network topology designed on top of the Ethernet physical layer, based on bidirectional timeTransmitter - timeReceiver links, where the timeReceivers nodes are the DOMs and the timeTransmitters are the White Rabbit Switches (WRSs). This device allows to combine both synchronisation and data packets on the same physical connection medium. It consists of two electronic boards: the SWitching Core Board (SWCB) and the back plane. Figure 6 shows a diagram of the WRS with its main functional blocks.

#### 3.3.1 Switching core board

The main hardware of the WRS is the switching core board. It is implemented in the microTCA standard, which is a compact standard with a redundant power supply, remote control, and designed to support high data rates. Due to the compact size of the microTCA standard, the optical connectors of the 18 ports are incorporated on another card that is attached using two QSS-048-01-LD-DP connectors. A reliability study was conducted based on the FIDES method, which revealed that the components that contributed the most to increasing the device's failure rate, known as FIT (Failures In Time), were the decoupling capacitors. As a result, an upgrade of the board was performed by selecting new capacitors that improved the ratio between their maximum voltage and the voltage at which they operate in the WRS. This new component selection will reduce the failure rate by a 66%. The testing of the produced switching core boards has also been enhanced. The main goal is to improve the reliability. The testing tool developed for the automatic test of the switching core boards is shown in Figure 7.

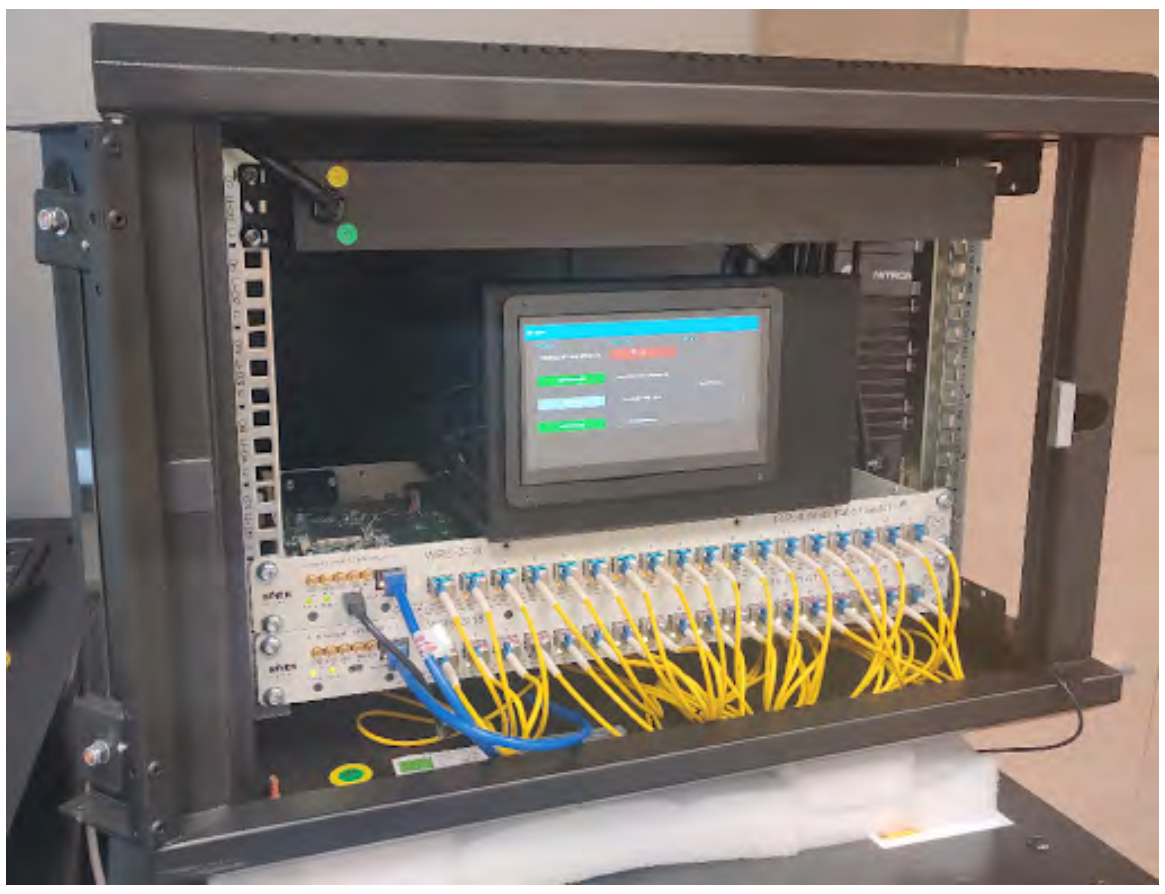

**Figure 7:** System test for the functional test of the switching core boards. The automatisation of the production is of crucial importance for increasing the reliability of the boards to be used in the experiment.

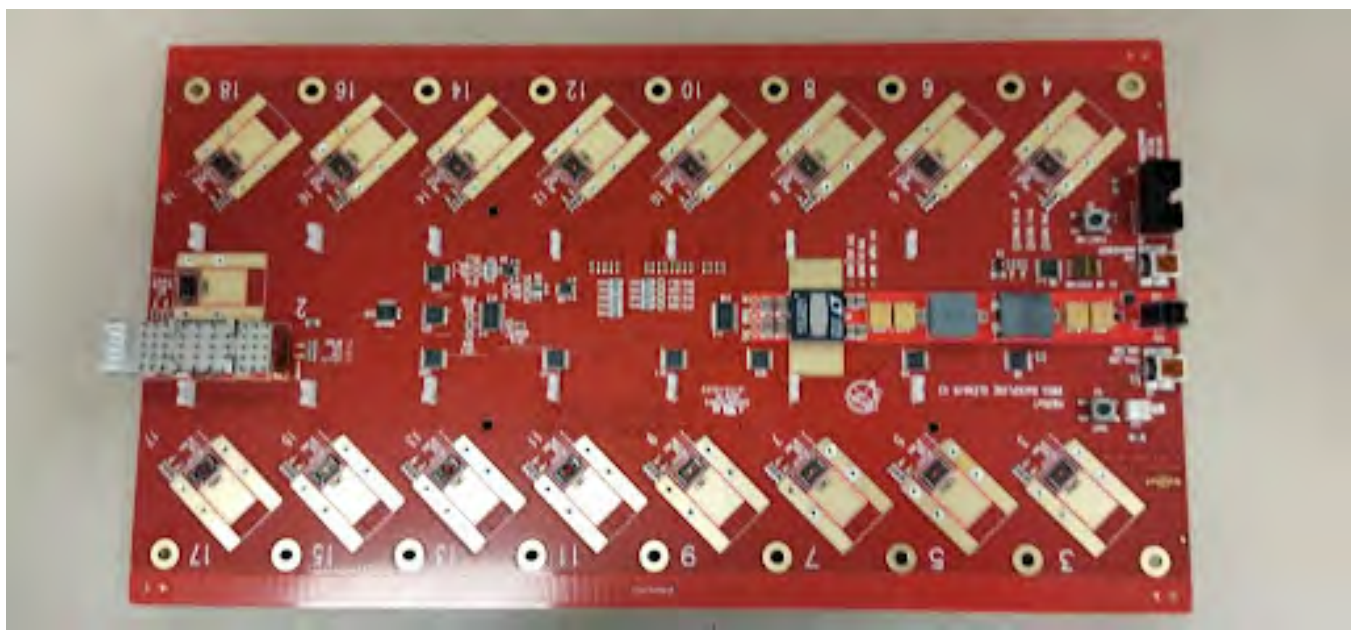

**Figure 8:** New design of the Glenair back plane adapting the geometry to KM3NeT requirements and including the new optical transceivers from Glenair.

#### 3.3.2 Glenair back plane

The other electronic board that forms the WRS is called the back plane. It incorporates 18 Small Form-factor Pluggable (SFP) optical fiber transceivers. In the case of KM3NeT, due to space constraints, power consumption, and reliability requirements, a new design was necessary. The board was designed with a different geometry and with different optical transceivers from the Glenair company, which provide improved reliability and power consumption. This new board, called the Glenair back plane, is shown in Figure 8.

## 4. Conclusions

The main electronic boards of the KM3NeT experiment and their basic operation within the experiment have been presented. The main upgrades of these boards have been described, as well as the main redesign criteria such as higher efficiency, improved reliability, or component obsolescence requirements. The modifications and enhancements implemented have significantly improved the overall reliability and performance of the KM3NeT experiment. These advancements have been adapted to obsolescence ensuring the performance and minimising failure rates.

## 5. Acknowledgements

The authors acknowledge the support from Ministerio de Ciencia e Innovación: Programa Estatal para Impulsar la Investigación Científico-Técnica y su Transferencia (ref. PID2021-124591NB-B-C41), and Generalitat Valenciana: Programa de Planes Complementarios I+D+I (refs. ASFAE/2022/023).

# The new generation of the Data Acquisition System of KM3NeT

**M. Circella[a,*], N. Battista[a], F. Benfenati[b,c], R. Bruno[d], E.-J. Buis[e,f], D. Calvo[g], T. Chiarusi[b], R. Cocimano[h], S. Colonges[i], A. D'Amico[f], A.F. Díaz[j], F. Filippini[b,c], S. Henry[k], P. Jansweijer[f], M. Lindsey Clark[i], M. Mongelli[a], C. Nicolau[l], C. Pastore[a], G. Pellegrini[b], S. Pulvirenti[h], D. Real[g], D. Samteleben[f], J.-W. Schmelling[f], I. Sgura[a], F. Tatone[a], C. Valieri[b], V. Van Beveren[f], D. Vivolo[m,n],on behalf of the KM3NeT Collaboration**

[a]*Istituto Nazionale di Fisica Nucleare – Sezione di Bari, Via Orabona 4, Bari, 70125 Italy*

[b]*Istituto Nazionale di Fisica Nucleare – Sezione di Bologna, V.le C. Berti-Pichat 6/2, Bologna, 40127 Italy*

[c]*Università di Bologna, Dipartimento di Fisica e Astronomia, V. le C. Berti-Pichat 6/2, Bologna, 40127 Italy*

[d]*Istituto Nazionale di Fisica Nucleare – Sezione di Catania, Via Santa Sofia 64, Catania, 95123 Italy*

[e]*TNO, Technical Sciences, PO Box 155, Delft, 2600 AD the Netherlands*

[f]*Nikhef, National Institute for Subatomic Physics, PO Box 41882, Amsterdam, 1009 DB the Netherlands*

[g]*IFIC - Instituto de Física Corpuscular (CSIC - Universitat de València), c/Catedrático José Beltrán, 2, 46980 Paterna, Valencia, Spain*

[h]*Istituto Nazionale di Fisica Nucleare – Laboratori Nazionali del Sud, Via S. Sofia 62, Catania, 95123 Italy*

[i]*Université Paris Cité, CNRS, Astroparticule et Cosmologie, F-75013 Paris, France*

[j]*University of Granada, Dept. of Computer Architecture and Technology/CITIC, 18071 Granada, Spain*

[k]*Aix Marseille Univ., CNRS/IN2P3, CPPM, Marseille, France*

[l]*Istituto Nazionale di Fisica Nucleare – Sezione di Roma, Piazzale Aldo Moro 2, Roma, 00185 Italy*

[m]*Istituto Nazionale di Fisica Nucleare – Sezione di Napoli, Complesso Universitario di Monte S. Angelo, Via Cintia ed. G, Napoli, 80126 Italy*

[n]*Università degli Studi della Campania "Luigi Vanvitelli", Dipartimento di Matematica e Fisica, V.le Lincoln 5, Caserta, 81100 Italy*

*E-mail:* marco.circella@ba.infn.it

[*]Speaker

The KM3NeT Collaboration is installing two neutrino telescopes at the bottom of the Mediterranean Sea, ARCA and ORCA. These detectors are built in the form of very large 3D arrays of photomultipliers (PMTs) detecting the Cherenkov light induced by relativist particles propagating in the sea water. The PMTs are installed inside pressure-resistant glass spheres, the Digital Optical Modules (DOMs), each containing 31 PMTs together with the electronics for data taking and communication. Eighteen DOMs and a base module are comprised in a detection unit (DU), a structure standing on the sea bottom with a height of almost 700 m for ARCA (200 m for ORCA). For optimal performance of the detectors, a good synchronization of all parts of the detectors is required. The data acquisition system of KM3NeT has been designed based on the White Rabbit (WR) protocol. In a first phase of construction of the detectors, this has been developed in a tailored design based on a "broadcast" approach: in such scheme, the communications from the shore station to the detectors is performed by distributing the same data stream to all underwater nodes, while in the opposite path each single DOM sends data to shore through its own communication channel. For the next phase of construction, a new system has been developed, based on the standard WR protocol. This approach has some significant advantages, in terms of maintainability and scalability of the system, which are obtained with a non-negligible cost, however, since it became necessary to design tailored WR components which would meet the KM3NeT requirements and to modify the full design of the apparatus consequently. In this paper we will present and compare these two DAQ system designs.

## 1. Introduction

KM3NeT is the next-generation neutrino telescope project, being implemented in the Mediterranean Sea [1]. KM3NeT comprises ARCA (Astrophysics Research with Cosmics in the Abyss) and ORCA (Oscillation Research with Cosmics in the Abyss), which are under construction at almost 3,500 m depth, 80 km offshore Capo Passero at the southern tip of Sicily (Italy), and at about 2,500 m depth, 40 km offshore Toulon (France), respectively. The two detectors share the same technology and neutrino detection principle. They are built as very large, 3D arrays of photosensors, called digital optical modules (DOMs) [2], capable to detect the Cherenkov light induced during their propagation in the sea water by relativistic particles emerging from neutrino interactions. The DOMs are made with pressure-resistant glass spheres, each housing a set of 31 photomultipliers (PMTs) of 3" photocathode diameter, the electronics for data taking and transmission, and calibration devices. The DOMs are arranged on detection units (DUs), which are anchored on the sea bottom and stand due to the buoyancy of the DOMs and a top buoy. Each DU comprises 18 DOMs connected through a backbone cable to a base module (BM), which is located on the DU anchor. The BM is connected to a submarine network of Junction Boxes (JBs) and cables, which is connected to the shore with a long-distance cable, the MEOC (Main Electro-Optical Cable). Due to the size of the detectors and the distance from shore, all data communications take place on an optical fibre system.

There are significant geometrical differences between ARCA and ORCA. ORCA has been optimized for studying neutrino oscillations and mass hierarchy with atmospheric neutrinos in the few-GeV energy region. In ORCA the DOMs are spaced by ~9 m, and the distance between the closest DUs is of ~20 m. ARCA, which aims at the search of high-energy neutrinos from astrophysical sources, is considerably larger, with a DOM spacing of 36 m and a distance between the closest DUs of about 90 m.

The challenges for implementation of the data acquisition system for such detectors are manifold. KM3NeT is based on the approach of sending all data to shore, with minimal filtering performed offshore. In this scenario, all signals detected by all PMTs above a set threshold (which can be controlled from shore and is typically set at the level corresponding to 0.3 of the average signal expected for a single photoelectron) are sent to a cluster of computers onshore where the data are filtered according to various filtering algorithms and dispatched for all usages (including an online alert distribution system, a prompt data analysis system for swift analyses to be done in response to alerts and to a data storage system for offline analysis). The DAQ system has therefore to ensure the necessary throughput to cope with the PMT signal rate. Due to the optical background, which is unavoidable in the sea water, due to the decays of $^{40}K$ and bioluminescence, each PMT has a signal rate of about 7 kHz, which can increase occasionally due to bioluminescence bursts. In the design of the DAQ system a target throughput corresponding to about twice the expected PMT signal rate has been conservatively assumed. The system has also to provide the necessary reliability for long-term operation (10 years minimum). Redundancy, when possible, is a plus. Finally, for optimal detector performance, a good synchronization and time calibration of the large number of PMTs included in the detectors, which are distributed in very large volumes and at a considerable distance from the control station, is required.

The DAQ system of KM3NT was designed taking into account all such requirements. In particular, in order to satisfy the requirement about synchronization, an approach based on the White Rabbit (WR) protocol [3] has been adopted. This was implemented in two different designs: a "broadcast" system, which has been used in the first phase of construction of ARCA and ORCA, and a WR-standard system, which will be used next. These two systems are illustrated in the next sections.

## 2. The "broadcast" system

In the "broadcast" system, used so far in the construction of ARCA and ORCA, the White Rabbit protocol is used in a non-standard way, since the same data stream is distributed from shore to all offshore nodes. The distribution of such "broadcast" channel requires therefore no Ethernet or WR switching and is implemented directly inside the optical system, whose architecture is shown in Figure 1. Long-distance data communications take place thanks to Small-Factor Pluggable (SFP) electro-optical transceivers; these have been used in two versions: either transmitting at a predefined wavelength or with a wavelength tunable by remote control. The purpose is to implement a DWDM (Dense Wavelength Division Multiplexing) approach in which the data streams from 72 DOMs (i.e., 4 DUs) are injected into a single fibre of the MEOC and demultiplexed onshore.

In such "broadcast" system, the BMs work as "slaves" of the WR "master" onshore, which is synchronized, through a "Grand master" switch to GPS; this allows to use the WR approach to dynamically reconstruct a time offset between the downstream and upstream clocks at the level of the BMs. The DOMs instead communicate with shore over a standard Ethernet protocol. The time offsets between the master clock onshore and the local clocks running in the different DOMs cannot be determined through WR. Consequently time calibration procedures, which include the usage of optical calibration sources as well as data analyses performed on coincident signals detected by the PMTs due to the optical background and particle events, have been developed.

The "broadcast" system was designed in order to rely as much as possible on solutions which could be more easily implemented, considering that a fully WR-compliant design would require a significant effort, as explained in the next section. This system has been proven to be fully functional. At the moment of writing this text, in total almost 40 DUs are in operation in ARCA and ORCA.

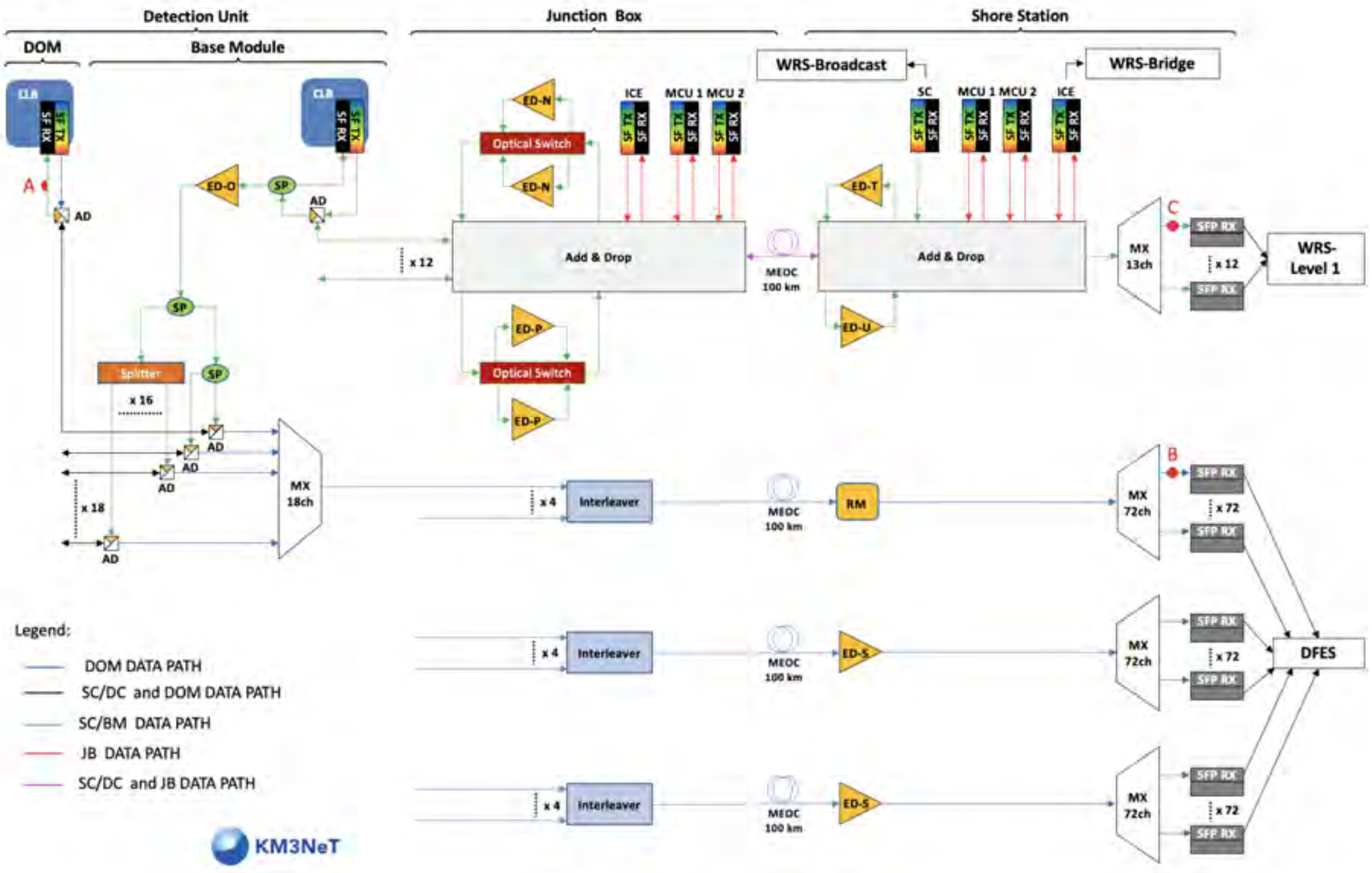

**Figure 1:** The optical network for the "broadcast" DAQ system of KM3NeT ARCA (the design for ORCA is similar, but not the same). The system requires the usage of Erbium-doped optical amplifiers (ED-x) both onshore and offshore. For additional details on this scheme see the text and [4].

## 1. The upgraded WR system

A non-standard WR system like the "broadcast" system illustrated above, leads to some disadvantages on the long term, in particular for maintaining the system compatible with the latest WR releases. This requires a continual effort from the Collaboration. Furthermore, the complexity of the network increases significantly with the size of the detector. Due to these reasons, a new implementation of the DAQ system has been designed, following a standard WR approach. It should be noted that, even if this development pays in terms of maintainability and scalability of the project, it comes with a significant cost, requiring several actions not only for redesign of the DAQ system itself but also in the following areas:

- Electronics: the necessary circuitry for implementing the WR system offshore had to be designed specifically, as none of the already available solutions could fit in the KM3NeT detectors.
- Power system: the new electronics requires increased power consumption, in particular inside the BM. The power system has therefore to be able to deliver such power with the required reliability.
- Optical system: this has been upgraded, starting with an adequate choice of the electro-optical transceivers to use inside the DOMs and BMs and redefining the optical network.
- Mechanics: the design of the DOMs and the BMs of the DUs had to be modified in order to accommodate the changes in the electronics and optical system. This was quite a demanding task in particular for the BM, due to the increased size of the instrumentation to be located inside it and the significant increase of power consumption.
- Calibration: the calibration procedures need to be updated in order to include the additional functionalities provided by the new DAQ system

The architecture of the new system is illustrated in Figure 2. The main feature of the new system is that the BM has been upgraded so as to provide a WR switching capability. This has been done by implementing tailored electronics boards equipped with the switches and the electro-optical transceivers for communication with the DOMs of the DU (for more details on such implementations and other recent upgrades of the electronics see [5]). Inside a BM there are two of such switches, each equipped with 18 ports. One of such ports is needed for the long-range communication with the shore station, another one is used for the control board inside the BM. A port on each switch is used for establishing a direct communication between the two switches, for redundancy. Therefore, a total of 12 ports, extra to the 18 needed for communication with the DOMs, remain available on the two switches. They are used for performing redundant (cold) communication channels with selected DOMs of the DUs.

The DOMs have been upgraded as follows: the central logic board has been modified in order to host the electro-optical transceiver for communication with the BM and the firmware which is run inside the DOM is a standard WR release.

The optical network in the submarine part as well as onshore has also been upgraded: each DU now communicates with shore with just two communications channels, and this allows a different DWDM scheme to be implemented, in which the communication channels for 7 DUs are multiplexed into a single optical fibre inside the submarine junction boxes. In this new system the circuitry of the shore station is fully WR-compatible.

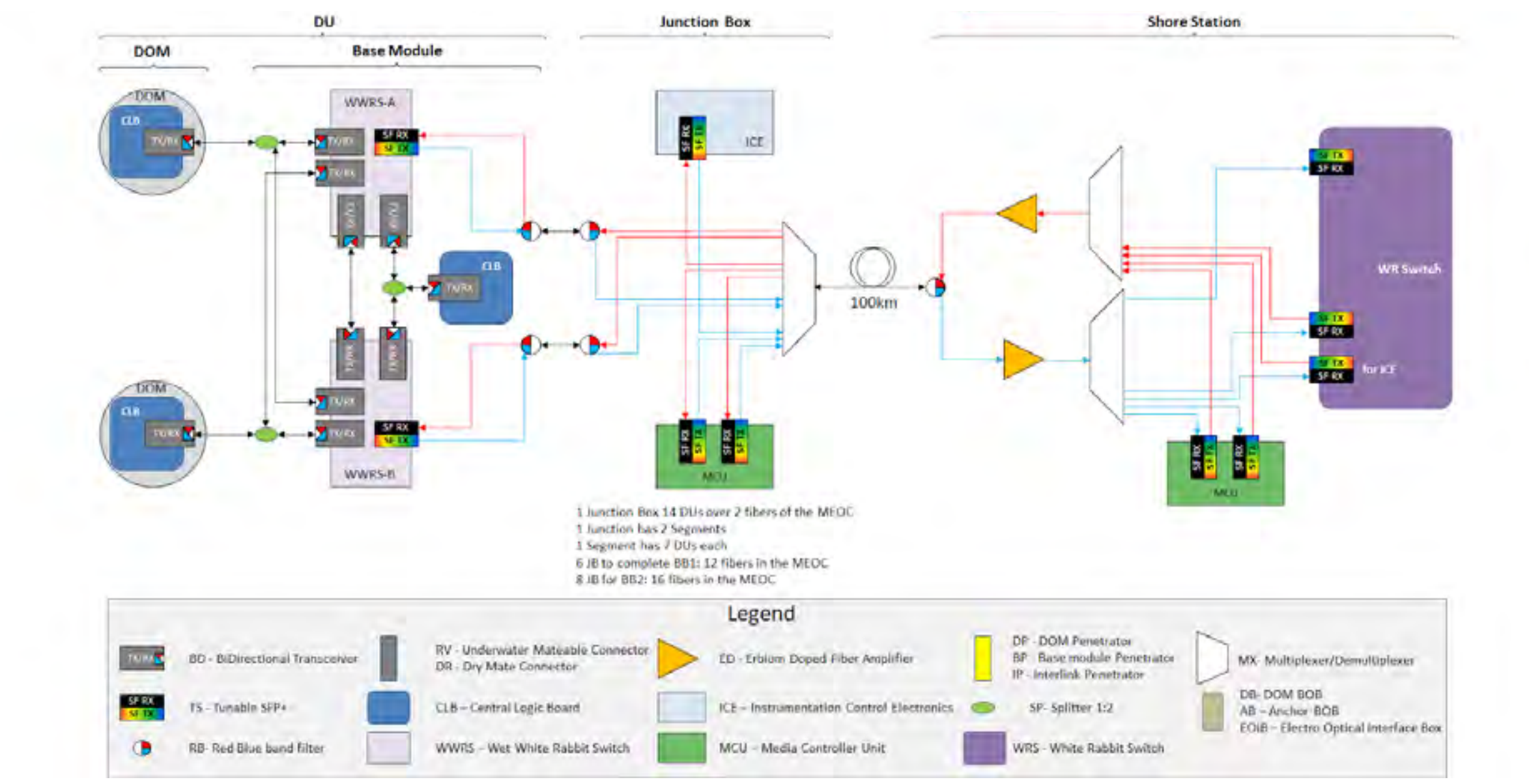

**Figure 2:** The architecture of the new DAQ system of KM3NeT ARCA (the design for ORCA is similar, but not the same). BB stands for Building Block, a set of 115 DUs (two BBs are foreseen in ARCA, one in ORCA).

## 2. Conclusions

The implementation of large, deep, submarine detectors, like ARCA and ORCA comprised in KM3NeT, requires highly reliable solutions. The DAQ system in particular has to satisfy demanding requirements as to data throughput and the need of synchronization at the level of the nanoseconds of detection elements distributed over very large volumes and at a very large distance from the control station. The KM3NeT Collaboration has implemented two layouts of the DAQ system, based on the White Rabbit protocol. The first sectors of ARCA and ORCA have been built with a DAQ system implementing a “broadcast” approach, i.e. a system in which the same data stream is distributed from onshore to all offshore nodes. A new system, implementing a standard WR network has been later implemented for the purposes of increased maintainability and scalability. Construction of new DUs equipped for such DAQ system and of the corresponding sector of the submarine network is starting.

# Art and Astrophysics in Conversation with KM3NeT Deep in the Mediterranean Sea

**Donald Fortescue,[a] Gwenhaël de Wasseige,[b] Jonathan Mauro,[b,*] Paschal Coyle[c] and Alin Ilioni[d] on behalf of the KM3NeT Collaboration**

[a] *California College of the Arts, San Francisco California*
[b] *Centre for Cosmology, Particle Physics and Phenomenology - CP3, Universite Catholique de Louvain, B-1348 Louvain-la-Neuve, Belgium*
[c] *Aix-Marseille Université, CNRS/IN2P3, CPPM, 163 Avenue de Luminy, Case 902, 13288 Marseille Cedex 09, France*
[d] *Université de Paris, CNRS, Astroparticule et Cosmologie, F-75013 Paris, France*
*E-mail:* dfortescue@cca.edu, gwenhael.dewasseige.@uclouvain.be, jonathan.mauro@uclouvain.be

We present the result of a cross-disciplinary collaboration between Donald Fortescue of the California College of the Arts in San Francisco and researchers from the KM3NeT Collaboration. The project involved creating an analog sound producing instrument which was installed within a standard KM3NeT pressure resistant glass sphere housing along with sound and video recording equipment. The instrument, titled "Bathysphere", was deployed in the sea for the first time adjacent to the ORCA array of KM3NeT off the coast of France on September 23, 2021.
One outcome of this project is the video work "Below the Surface". It incorporates sound and video recorded from within the "Bathysphere" as it floated on the surface of the Mediterranean and then as it dived down to 300m depth. As the "Bathysphere" dived, the analog instrument it contained quietened. The sound in the dive portion of the video is created from the sonification of data from the KM3NeT array that was recorded during the deployment of the "Bathysphere".
"Below the Surface" highlights the extraordinary environment in which the KM3NeT array is being created and operates. The data sonification illustrates the potential of this method of data representation to connect with viewers in a deeply physical way and offers new perspectives on the data collected by KM3NeT.

*Speaker

## 1. Introduction

We present the result of a cross-disciplinary collaboration between Donald Fortescue and researchers from the KM3NeT Collaboration [1]. The project builds on the successful art/science research collaboration between Fortescue and De Wasseige, initiated during Fortescue's US National Science Foundation funded residency with the IceCube Neutrino Observatory [2] at the South Pole in the austral summer of 2016/17. The outcomes of this previous collaboration were presented at the 36th International Cosmic Ray Conference (ICRC2019), in Madison and were detailed in the proceedings of that conference [3].

KM3NeT is a European particle astrophysics research infrastructure located at several sites deep in the Mediterranean Sea, illustrated in Fig. 1. When fully completed, KM3NeT will search for neutrinos from distant astrophysical sources like supernova remnants, gamma-ray bursts, supernovae or colliding stars and will be a powerful tool in the search for dark matter in the universe. It also houses instrumentation for monitoring the deep-sea environment used by other sciences such as marine biology, oceanography and geophysics. KM3NeT consists of two discrete detector arrays: ORCA situated off the coast of Marseille, France, and ARCA situated off the coast of Sicily, Italy.

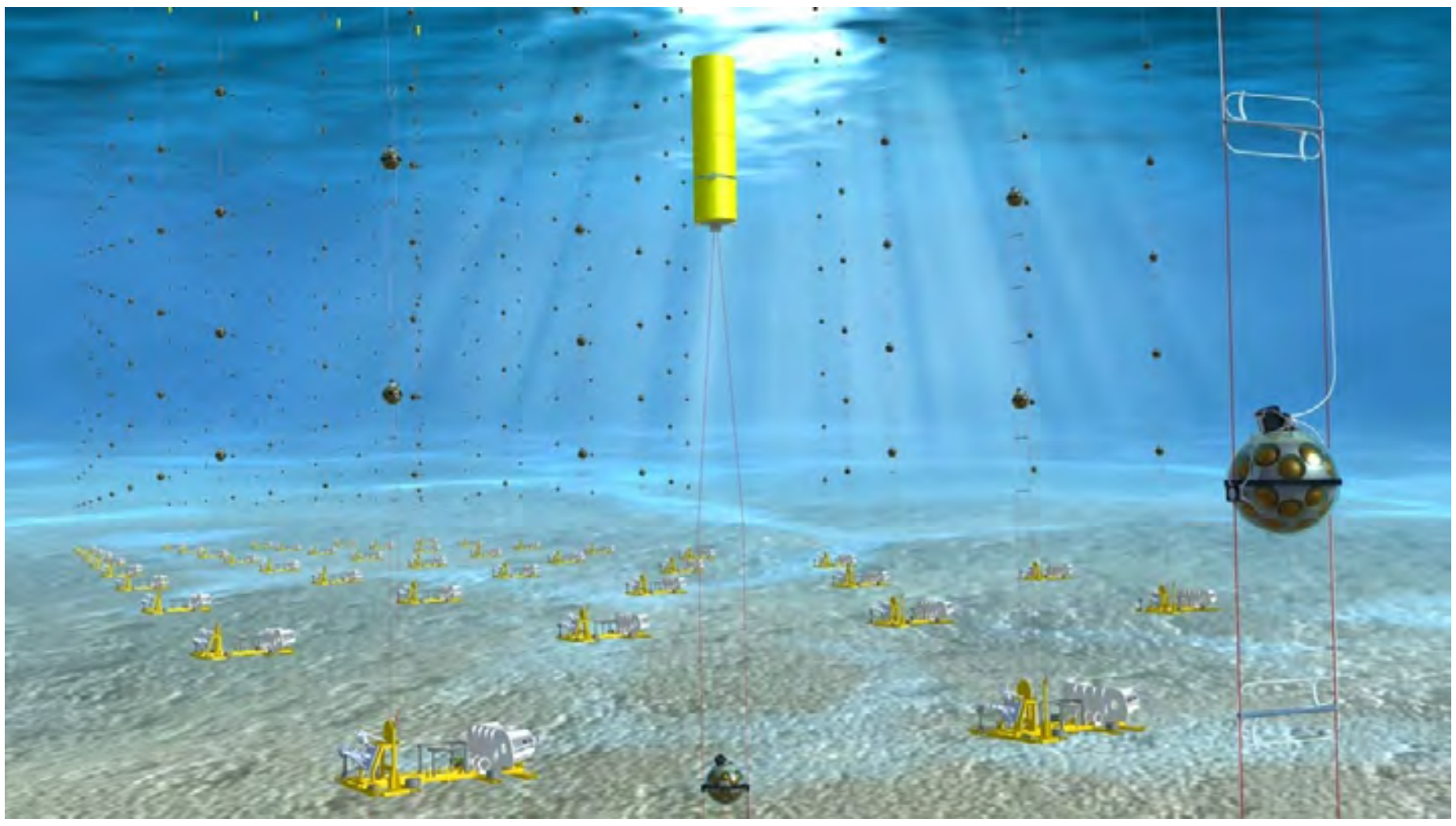

**Figure 1:** Artist's Rendering of the ORCA Array.

The current project involved Fortescue creating an analog, sound producing, instrument which was installed within a standard KM3NeT pressure resistant glass sphere housing along with sound and video recording equipment. This instrument, titled "Bathysphere"[1], was deployed for the first time, adjacent to the ORCA array of KM3NeT, off the coast of Marseilles, on September 23, 2021.

One outcome of this project is the video work "Below the Surface". The work incorporates sound and video recorded from within the "Bathysphere" as it floated on the surface and as it dived down to 300 m depth. As the "Bathysphere" dived, the analog instrument it contained quietened. The sound in the dive portion of the video was created from the sonification of data from the ORCA array that was captured during the dive of the "Bathysphere". "Below the Surface" is publicly available at https://vimeo.com/844369324

[1]Named in honor of the Bathysphere developed by Otis Barton and deployed by Barton and William Beebe during the first deep-water dives in Bermuda in the early 1930's.

## 2. Process

### 2.1 The Bathysphere

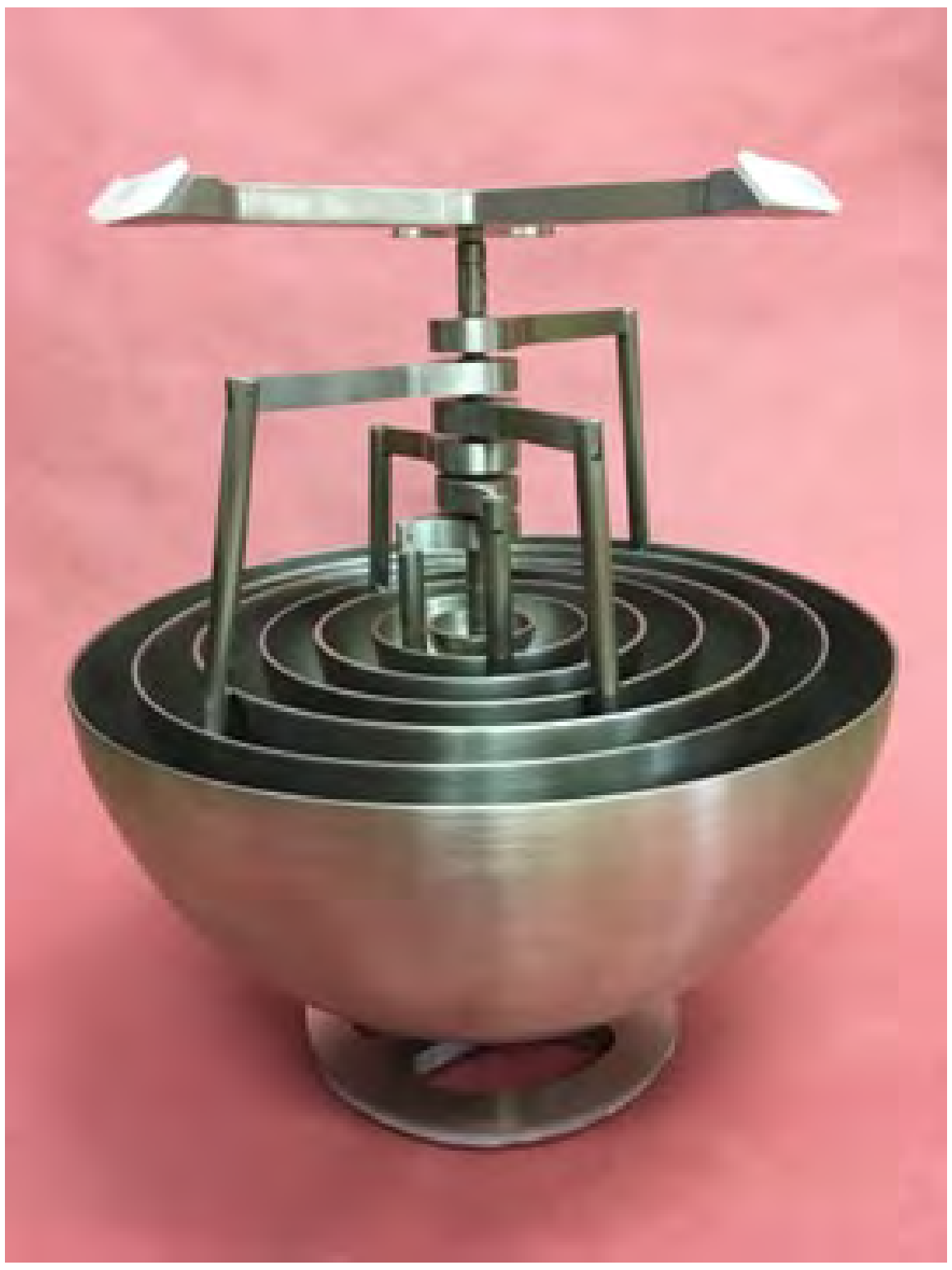

**Figure 2:** The "Bathysphere" internal mechanism.

The completed "Bathysphere" is an idiophonic instrument that operates on the surface and at depth to transduce water oscillations into sound. The analog, sound producing component of the bathysphere consists of seven, concentric, stainless steel, hemispheric bells attached to a central axis. These bells are activated by six stainless steel hammers which are free to rotate around the central axis, as shown in Fig. 2. Sounds are produced by the hammers hitting the bells and by them running around the rim of the bells, much like the action employed in the playing of Himalayan prayer bells. A standard GoPro Hero 8 and Sony Digital Audio Recorder are attached to the upper platform of the "Bathysphere" to record video and the sound produced by the "Bathysphere".

Deployment involved lowering the "Bathysphere" with its attached ballast weight and connected flotation buoy to the sea's surface (Fig. 3). The flotation buoy for the ballast was connected to a line running to the deployment vessel and the "Bathysphere" was connected to another line with depth markers every 5 m. The "Bathysphere" was allowed to float on the surface some 50 m from the deployment vessel for about 15 minutes (Fig. 4), then the flotation buoy was released and retrieved back to the ship. With the release of the flotation buoy, the ballast sank and towed the "Bathysphere" along with it. Once the “Bathysphere” reached approximately 300 m depth, it was hauled back to the surface and retrieved. Three separate dives were possible during the first Mediterranean deployment of the “Bathysphere”.

Video and sound were recovered from the recorders. Recordings from the third dive were edited and used to create the video work “Below the Surface”, which incorporates both recorded “Bathysphere” sounds and sonified data from the ORCA array.

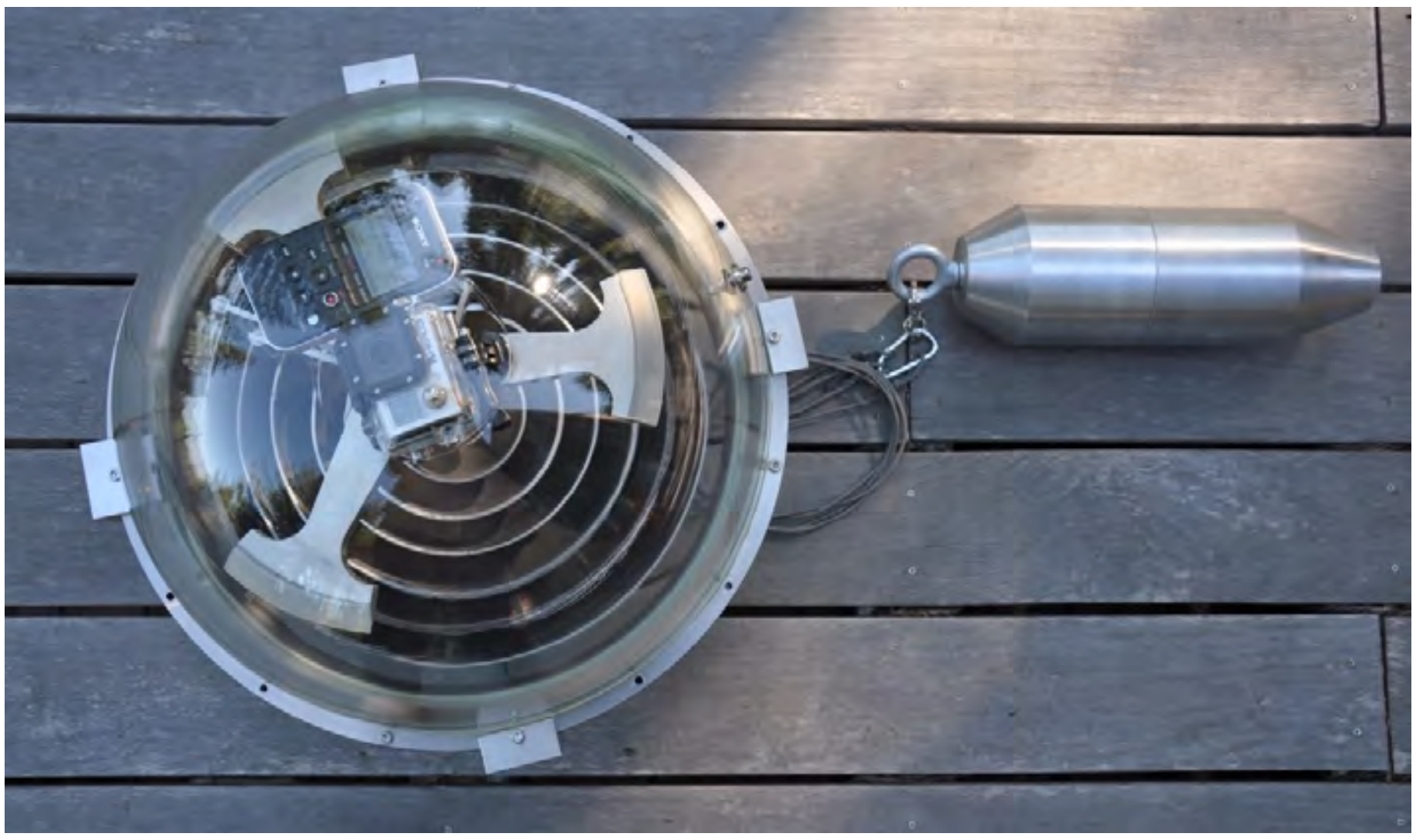

**Figure 3:** The “Bathysphere” ready for deployment.

### 2.2 Sonification

Two terms, used by both scientists and sound artists working with converting data to sound, are ’audify’ and ’sonify’. Audification is the making audible of an inaudible sound through amplification or through transposing frequencies into the range that humans can hear, this is the process that makes whale song audible to us. Audification is essentially sound transcription. Sonification is the process of rendering other forms of data into sound, e.g., electromagnetic, particle flux or gravitational wave signals. Sonification is the transduction of non-audio data into audible sound. “Below the Surface” incorporates the sonification of digital signals derived from the detection of photons by the ORCA array, deep in the in the Mediterranean Sea, into sounds we can hear.

There are innumerable ways that data can be transduced into sound. Each requires the allocation of sound frequency, volume, duration and timing values to map the event rate, deposited energy, photon flux and location values in the data. Some mappings are readily suggested by the shared characteristics of sound and light waves, such as frequency, intensity, and duration. Others, such as which events are sampled, the sampling rate and playback speed, need to be selected to both reflect the underlying physics and to satisfy aesthetic considerations, that is, to ensure the resulting work will be engaging. As well, the mode or “instrument” chosen to play the sounds is open for creative interpretation. We selected a digital instrument based on the glockenspiel for the “Below the Surface” sonification as it was a close analog to the sounds produced by the “Bathysphere”.

During the "Bathysphere" deployment, six Detection Units (DU) were active in the ORCA array. Currently, there are 18 operational DUs after three successful deployments since 2022. Each DU holds 18 Digital Optical Modules (DOMs) and each DOM houses 31 Photomultiplier Tubes (PMTs) [4].

The sonification process we developed involved assigning four discrete notes to each DOM by dividing each DOM into four regions (upper, mid-north, mid-south, and lower), each with either 8 or 7 PMTs. Each region was assigned a note so that groups of three adjacent DOMs on the same DU cover one octave on the chromatic scale. Within each octave, individual DOMs have been tuned to avoid strong dissonances in adjacent groups. This was done because luminous events happening close to a DOM will likely be detected by multiple PMTs belonging to different groups. Without appropriate tuning, strong dissonances would characterise the overall sound of the track as an artifact of the grouping of PMTs. In a manner similar to tuning a musical instrument, each of the three DOM has been given a unique sound given by the choice of the four notes that it covers.

The four-note groups for each of the three DOMs that form an octave are as follows (in the order: upper, mid-north, mid-south, lower): (1) B, E, G, C; (2) F, Bb, Eb, Gb; (3) Ab, Db, D, A.

Each DU has a range of 6 octaves, going from C1 to B6. Lower octaves are mapped to the bottom of the detector, while higher octaves are found at the top. This allows us to reproduce the dimensionality of the detector and the geometrical distribution of the sonified signal. A filter was used at the DOM level on low level data from the detector. A cut on the minimum cumulative charge detected by signals in coincidences of 100 ns on the same DOM was applied to get rid of most of the random coincidences caused by incoherent noise and to focus more on coherent environmental and atmospheric activities. These signals are expected to be produced by the decay of sea salt ($^{40}$K), and from bioluminescent species (fish and plankton), as well as Cherenkov photons from the interaction of atmospheric muons and neutrinos. The detected charge is used to modulate the volume of the notes, such that higher charge corresponds to a louder volume. The timing of detection is scaled and directly mapped onto the timing of the notes in the track, coincident notes with the same pitch coming from the same DOM are merged into a single note with volume level given by the sum of the original volumes.

The data that we selected for the sonification came from the duration of the dive that was recorded and edited for the video portion of this project The video "Below the Surface" is 6 minutes and 6 seconds long (6:06), and is played on a continuous loop. The dive portion of the video runs for 4 minutes and 40 seconds. The "Bathysphere" was underwater for exactly 30 minutes (1800 seconds) during its third dive. To produce the track in "Below the Surface" around 0.4 seconds of data were used, starting from September 23rd, 2021, 16:34:00 UTC, which coincides with the time the "Bathysphere" reached its maximum depth during the dive.

## 3. Outcomes and Perspectives

"Below the Surface" was shown for the first time at I CRC2023. It will be presented at both science and art venues in Australia, the USA, and Europe over the coming months and years. It will bring awareness of the research of the KM3NeT Collaboration to new audiences and provide an original perspective on leading astrophysical research, which can be difficult for lay audiences to engage with.

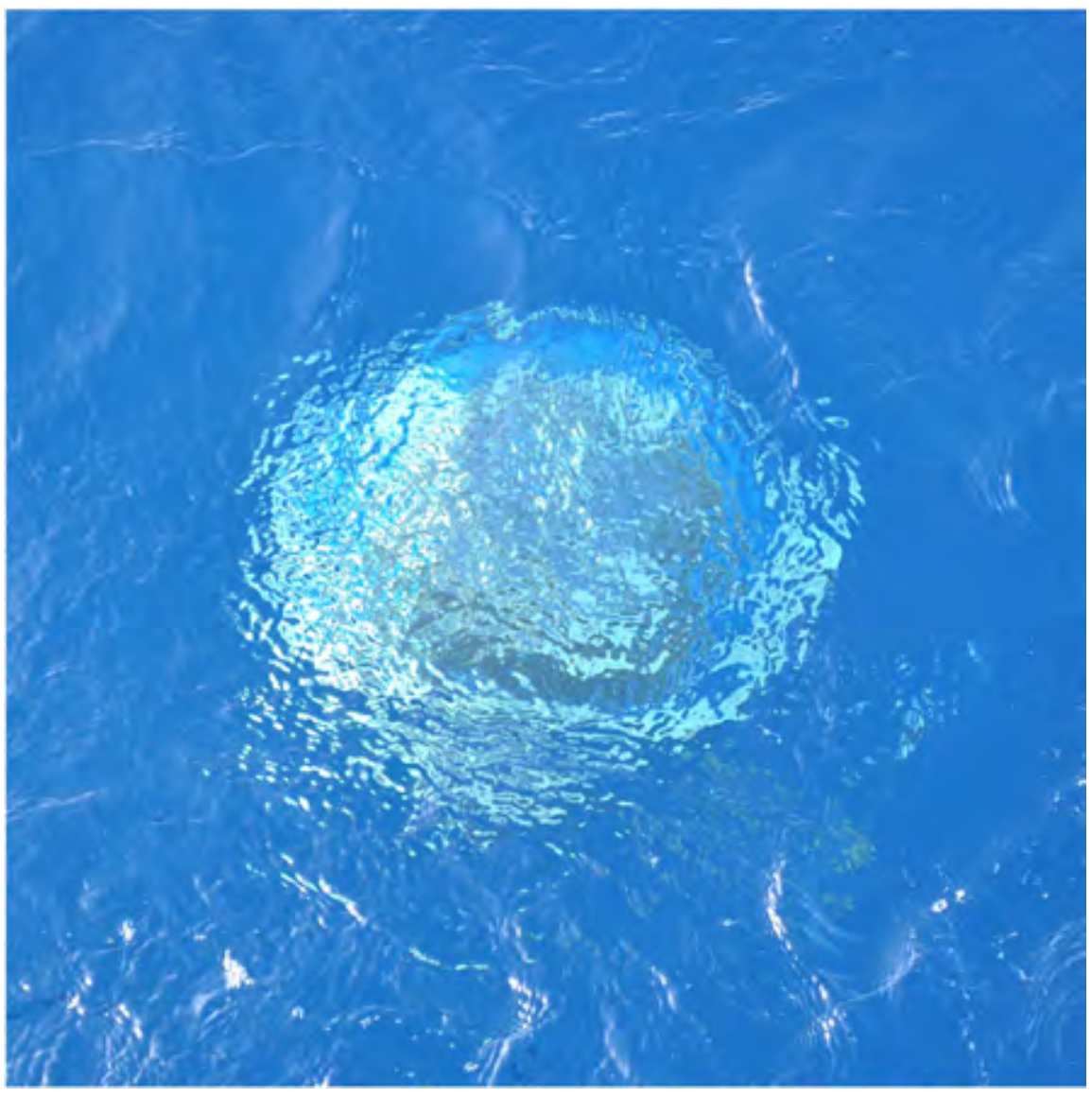

**Figure 4:** The "Bathysphere" submerged.

Cutting-edge contemporary astrophysical research instruments, like KM3NeT, are often located in challenging and inaccessible locations, such as the Antarctic, the high deserts of South America, deep in the waters of the Mediterranean Sea and Lake Baikal. The locations in themselves are sources of wonder. "Below the Surface" aims to highlight the extraordinary collaborative effort and technology required to create a complex instrument to be deployed deep in the ocean, as well as the unique natural environment in which it is located.

The craftsmanship involved in creating the unique, specifically designed and calibrated, analog instrument contained within the "Bathysphere" echoes the incredible technical craftsmanship that has been developed for and is continuing to evolve as the deep-sea array of KM3NeT is created and deployed. This technical ingenuity and sophisticated craftsmanship is often overlooked by the wider public and in the case of neutrino telescopes is hidden in the vast, inaccessible natural spaces in which they are deployed.

### 3.1 Objective approaches to art making

Viewers may assume that the underwater audio for "Below the Surface" was composed specifically for the video. In a sense it was, but the individual notes and their timing were not selected by the work's authors. Instead, it was established an objective set of constraints on the data set selected and the way this data was then transduced into sound. This is analogous to the way a scientific study uses a blind procedure, that is, developing the analysis on sample data, selecting relevant 'signal' data and then processing that data for final a nalysis. Such an objective approach to art making might, at first glance, seem to deny the creativity of the artist. However, this approach has a strong precedent in contemporary art. An objective methodology where distinct rules are established to constrain artists' choices was first established by the minimalist and conceptual artists of the 1960's and 70's who were striving to reduce artistic agency in reaction to the dominant paradigm of Abstract Expressionism. The renowned conceptual artist Sol LeWitt explained in his Paragraphs

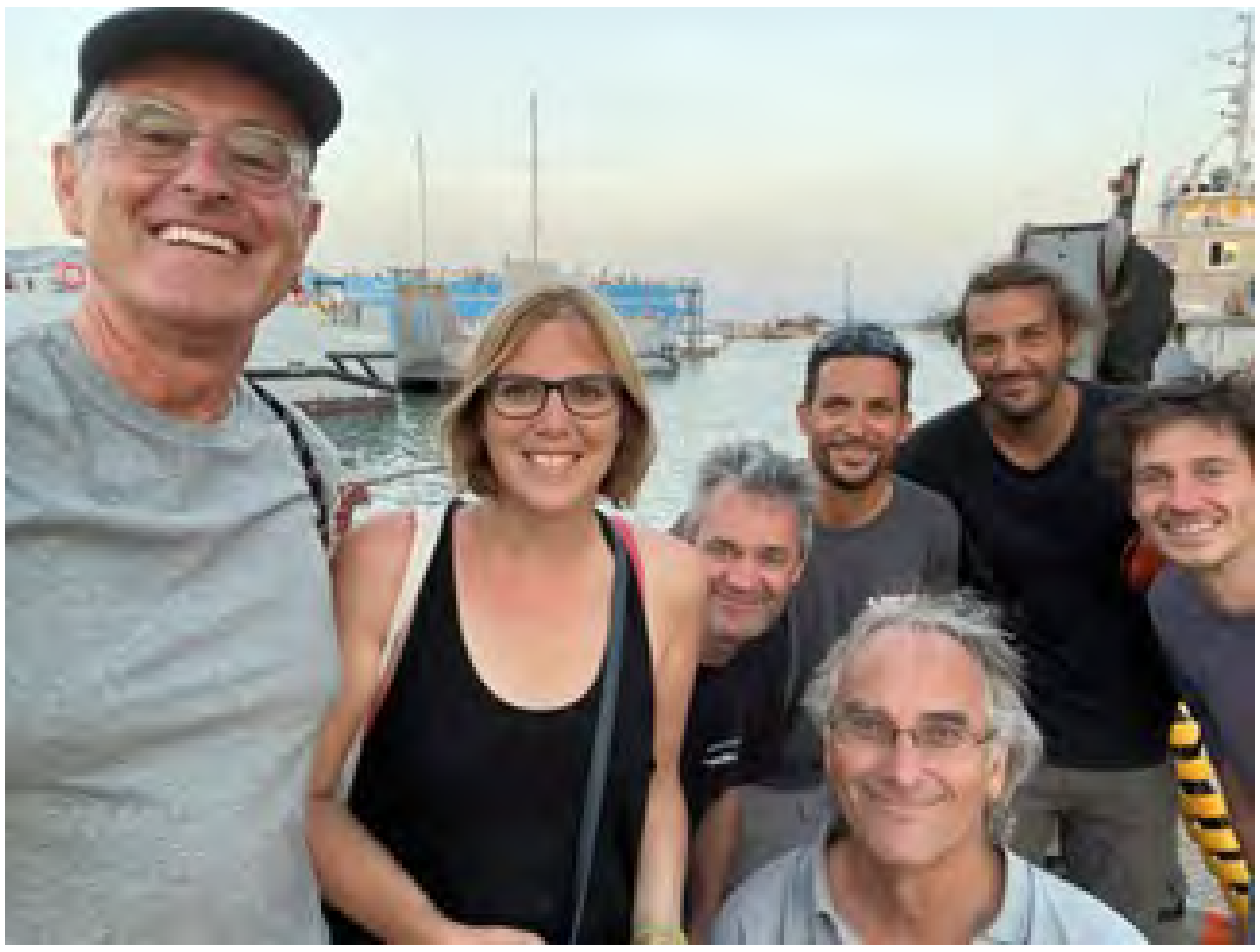

**Figure 5:** The team who participated in the successful deployment of the "Bathysphere". From L to R. Prof. Donald Fortescue, Prof. Gwenhaël De Wasseige, Dr. Vincent Bertin, Conrad Bertrand, Alexandre Cornillon, Alin Ilioni, and (foreground) Dr. Paschal Coyle

on Conceptual Art: *"To work with a plan that is preset is one way of avoiding subjectivity. It also obviates the necessity of designing each work in turn. The plan would design the work. Some plans would require millions of variations, and some a limited number, but both are finite. . .. In each case, ... the artist would select the basic form and rules that would govern the solution of the problem. After that the fewer decisions made in the course of completing the work, the better. This eliminates the arbitrary, the capricious, and the subjective as much as possible" [5].*

The objective approach provides a strong link between the methodological practices of science and art and can help delineate the required conditions for effective art/science collaborations, that is, the common ground where the two methodologies may coalesce.

### 3.2 Value of sonification

Sonification is a methodology that is generating wide interest in contemporary data science, and for this there are several reasons. We readily discern sound as occurring in three dimensions, so it would seem that 'listening' to data that derives from interactions across three-dimensional space could be more intuitively informative than 'looking' at graphic analysis or animated renderings on a 2D screen. Also, sonification permits access to data sets by researchers for whom vision might not be their primary sense, thus enabling a broader diversity of scientific researchers. And finally, more general audiences have a direct visceral connection to sound which can facilitate a deeper direct connection with the phenomena being studied by leading research institutions.

### 3.3 Collaboration

Artists are rarely considered key components of collaborative science research teams. They are often seen as addenda to the core science, as illustrators or designers who can help with posters or presentations for public outreach. The durable art/science collaboration developed by the authors over many years provides a model for imbedding artists within science teams to facilitate new perspectives and audiences.

As the KM3NeT array is still under development, the current collaboration provides the potential for ongoing art/science projects as the one described in these proceedings (Fig. 5). For example, future sonified works from the array could become more richly complex. The sonification for "Below the Surface" could be considered as just one instrument playing a motif that has the potential to grow into a full orchestral work, over time, as the array attains its full functionality.

## 4. Conclusions

"Below the Surface" highlights the extraordinary environment in which the KM3NeT array is being created and operates. The craftsmanship of the "Bathysphere" echoes the technical sophistication of the unique instrumentation of KM3NeT. The data sonification created for "Below the Surface" illustrates the potential of this method of data representation to connect with viewers in a deeply physical way and offers new perspectives on the data collected by KM3NeT. The durable art/science collaboration developed by the authors over many years provides a successful model for imbedding artists within science teams to facilitate new perspectives and audiences.

## Acknowledgements

Special thanks to Dr. Véronique Van Elewyck, Lawrence LaBianca, Paolo Salvagione, and Dr. Sandra Kelch for assistance with the development of the "Bathysphere". Thanks, are also due to the other team members who so generously contributed to the successful deployment of the "Bathysphere" in September 2021 - Dr. Vincent Bertin, and Conrad Bertrand and Alexandre Cornillon aboard the MV Foselev Onyx. Fortescue's research is supported by the California College of the Arts and by KM3NeT. Fortescue and De Wasseige received a Materials Based Research Grant from the Center for Contemporary Craft in Asheville NC, USA to develop this project. The data used was generously provided by and used with the permission of the KM3NeT Collaboration.